\documentstyle[twoside]{article}

\input ijmpa.sty
\input psfig
\def\Journal#1#2#3#4{{#1} {\bf #2}, #3 (#4)}
\def\RMP{\em Rev. Mod. Phys.}
\def\NIM{\em Nucl. Instrum. Methods}
\def\NIMA{{\em Nucl. Instrum. Methods} A}
\def\CPC{\em Comp. Phys. Comm.}
\def\NPB{{\em Nucl. Phys.} B}
\def\NPA{{\em Nucl. Phys.} A}
\def\PLB{{\em Phys. Lett.}  B}
\def\MPLA{{\em Mod. Phys. Lett.}  A}
\def\PRL{\em Phys. Rev. Lett.}
\def\PRD{{\em Phys. Rev.} D}
\def\PRC{{\em Phys. Rev.} C}
\def\ZPC{{\em Z. Phys.} C}
\def\JPG{{\em J. Phys.} G.}
\def\JETP{\em Soviet Phys. JETP}
\def\SJNP{\em Soviet J. Nucl. Phys.}

\newcommand{\MSB}{\mbox{$\overline{\rm MS}$}}
\newcommand{\leqsim}{\,\raisebox{-0.6ex}{$\buildrel < \over \sim$}\,}
\newcommand{\geqsim}{\,\raisebox{-0.6ex}{$\buildrel > \over \sim$}\,}
\begin{document}

\runninghead{Structure functions $\ldots$} {Structure functions $\ldots$}

\normalsize\textlineskip
\thispagestyle{empty}
\setcounter{page}{1}

\rightline{OUNP-97-10}
\rightline{DESY 97-226}

\vspace*{0.88truein}

\fpage{1}
\centerline{\bf STRUCTURE FUNCTIONS OF THE NUCLEON}
\vspace*{0.035truein}
\centerline{\bf AND THEIR INTERPRETATION}
\vspace*{0.37truein}
\centerline{\footnotesize A M COOPER-SARKAR \& R C E DEVENISH}
\vspace*{0.015truein}
\centerline{\footnotesize\it Physics Department, University of Oxford,}
\baselineskip=10pt
\centerline{\footnotesize\it Nuclear \& Astrophysics Laboratory,}
\baselineskip=10pt
\centerline{\footnotesize\it Keble Road, Oxford OX1 3RH, England}
\vspace*{0.37truein}
\centerline{\footnotesize A DE ROECK}
\vspace*{0.015truein}
\centerline{\footnotesize\it DESY,}
\baselineskip=10pt
\centerline{\footnotesize\it Notkestrasse 85,}
\baselineskip=10pt
\centerline{\footnotesize\it 22607 Hamburg, Germany}
\centerline{\footnotesize Presently on leave at CERN Geneva, Switzerland}
\vspace*{0.225truein}
\centerline{\footnotesize November 1997}
\vspace*{0.225truein}
\centerline{\footnotesize Submitted to International Journal of Modern Physics A}

\vspace*{0.21truein}
\abstracts{The current status of measurements of the  nucleon 
structure functions and their understanding is reviewed. The fixed
target experiments E665, CCFR and NMC and the HERA experiments H1 and
ZEUS are discussed in some detail. The extraction of parton momentum
distribution functions from global fits is described, with particular 
attention paid to much improved information on the gluon momentum
distribution. The status of $\alpha_s$ measurements from deep
inelastic data is reviewed. Models and non-perturbative approaches for the
parton input distributions are outlined. 
The impact on the phenomenology of QCD of the 
data at very low values of the Bjorken $x$ variable is discussed in detail.
Recent advances in the understanding of the transition from deep 
inelastic scattering to photoproduction are summarized. Some 
brief comments are made on the recent HERA measurements of the ep NC and CC 
cross-sections at very high $Q^2$.}{}{}

%
%
\vspace*{1pt}\textlineskip	
\section{Introduction}	
\vspace*{-0.5pt}
\noindent
There have been two main strands of interest in experiments on deep inelastic 
scattering (DIS) since the initial observation of Bjorken scaling. 
Firstly, they are
used to investigate the theory of the strong interaction and secondly, they
are used to determine the momentum distributions of the partons within the
nucleon.

The observation of Bjorken Scaling established that the quark-parton
model (QPM) is a valid framework in which to interpret the data and thus 
that the 
deep inelastic structure functions can be used to measure the nucleon's 
parton distributions. The later observation of
logarithmic scaling violations indicated that the non-Abelian
gauge theory of Quantum Chromodynamics (QCD) might be the
correct theory of the strong interactions. Parton distributions may still be
measured, but one must account for their evolution with $Q^2$~\fnm{a}
\fnt{a}{~$Q^2$ and $x$ are defined in Sec.~\ref{sec:lorinv}}.

In the early 1980's arguments raged, as to whether the observed $Q^2$ 
dependence of structure functions was uniquely described by perturbative 
QCD (pQCD), or whether alternative
theories (e.g. Abelian theories, fixed point theories) provide an equally good
description of data. The consensus came down in favour of QCD as more and more
accurate data, from a variety of physical processes, were able to establish some
of the crucial features of QCD. For example, direct evidence for the 
existence of spin-1 gluons came from the observation and properties of three
jet events in $e^+e^-$ scattering at PETRA. Evidence for the 
running of the strong coupling constant, $\alpha_s$ with $Q^2$, came from 
measurements of production rates for high transverse momentum 
hadrons and jets, in 
many different processes at different scales. Evidence for the
non-Abelian nature of the gluon-gluon coupling came from the observation that
the fraction of the momentum of the nucleon carried by the quarks decreases
as $Q^2$ increases and from studies of four jet events at LEP.

These observations made the theory of QCD part of the Standard Model (SM) of
particle physics, and the variety of different processes which can be
successfully described by pQCD is now  detailed in many 
textbooks. Hence the interest in structure function measurements 
from the mid to late eighties lay in extracting accurate parton
distributions taking the framework of pQCD for granted. These parton 
distributions are of interest since we still have only
a limited understanding of the non-perturbative physics involved in 
confinement. Accurate parton distribution functions are also of vital 
importance as the input for calculations of high energy scattering 
processes which might involve physics beyond the Standard Model.

With the advent of DIS data from HERA we have entered into a new phase of 
interest in structure function measurements, where we are again using the 
data to investigate the properties of pQCD, in a new kinematic regime where 
our (by now) conventional calculations may not be adequate. Firstly because
the conventional treatment does not account for contributions to the 
cross-section which are leading in $\alpha_s\ln(1/x)$ and we are now making
measurements at very low $x$, and secondly because 
the parton densities, in particular the gluon, are becoming large and we
may need to develop a high density formulation of QCD.
\textheight=7.8truein
\setcounter{footnote}{0}
\renewcommand{\thefootnote}{\alph{footnote}}
It is an appropriate time to take stock of what has been achieved in the five
years since HERA started operation in 1992. The first high statistics 
measurements of $F_2^p$ have been published by H1 and ZEUS and during the
same period the fixed target experiments NMC, E665 and CCFR have also published
their complete structure function data sets. Activity on the theoretical
side has also increased and there is much to report on, even though we are 
still some way from a complete understanding of QCD at low $x$. 
We cover published data up to the end of June 1997 and we include 
some preliminary data shown at the Symposium on Lepton and Photon 
Interactions, Hamburg 1997~\cite{lpham97} and at
the EPS Conference, Jerusalem 1997~\cite{jer97}.
Material on theoretical interpretations is covered
up to the end of November 1997.

This review is concerned with the unpolarized proton and
neutron structure functions and their interpretation. It does not deal with
the interesting questions raised by DIS data on nuclear targets, such
as the EMC effect and nuclear shadowing. We also do not not discuss the
field of polarized structure functions and spin dependent
effects. It has been established that some $10\%$ of the cross-section
for deep inelastic events consists of events with a significant rapidity gap.
These are generally thought to be mediated by diffractive processes, discussion
of such processes is beyond the scope of the present review, see 
refs.~\cite{gallo97,eich97} for recent review talks. We note
that such events are implicitly included in the inclusive measurements from
which the structure functions and parton densities are derived.
In more detail: section 2 contains a collection of kinematic 
definitions and standard formulae for deep inelastic neutral-current (NC)
and charged-current (CC)
scattering cross-sections; section 3 provides an pedagogical outline 
of the pQCD
improved quark-parton model framework; section 4 gives an account
of the fixed target and collider detectors relevant to this review and a brief
discussion of experimental methods and event reconstruction; section 5 is a
review of the structure function data from the NMC, E665, CCFR, H1 and ZEUS
experiments; section 6 discusses the extraction of 
parton distribution functions from this data and the related extraction of 
the value of $\alpha_s$; section 7
provides a broad survey o of theoretical and phenomenological 
approaches to understanding structure 
function data at small $x$; section 8 considers the transition region from
deep inelastic data to photoproduction: 
section 9 contains a brief account of the
measurements by H1 and ZEUS of the NC and CC cross-sections at very large $Q^2$
and the observation of a possible excess of events above the Standard Model 
expectation; finally section 10 contains a summary and outlook.

\section{Formalism}
\label{sec:formalism}
\noindent
We give definitions of the commonly used Lorentz invariants and the 
formulae for the neutral and charged current cross-sections in terms of
the structure functions. The latter are put in context by reference to
the quark-parton model. The details of how the expressions are derived
may be found in many books, for example Halzen and Martin~\cite{Halzen}
or Roberts~\cite{roberts}.  

\subsection{Lorentz invariants}\label{sec:lorinv}
\noindent
The basic process is
{\small \begin{equation}
               l N \rightarrow l' X
\end{equation}}
where $l,l'$ represent leptons~\fnm{b}\fnt{b}{~Lepton is taken to include 
antileptons, unless it is necessary to distinguish them.}, $N$ represents 
a nucleon and $X$ represents the
hadronic final state particles. Such processes were studied using a lepton beam
and a fixed nucleon target until the advent of HERA and the terminology of 
lepton as probe and nucleon as target is still widely used.

\begin{figure}[ht]

\centerline{\psfig{figure=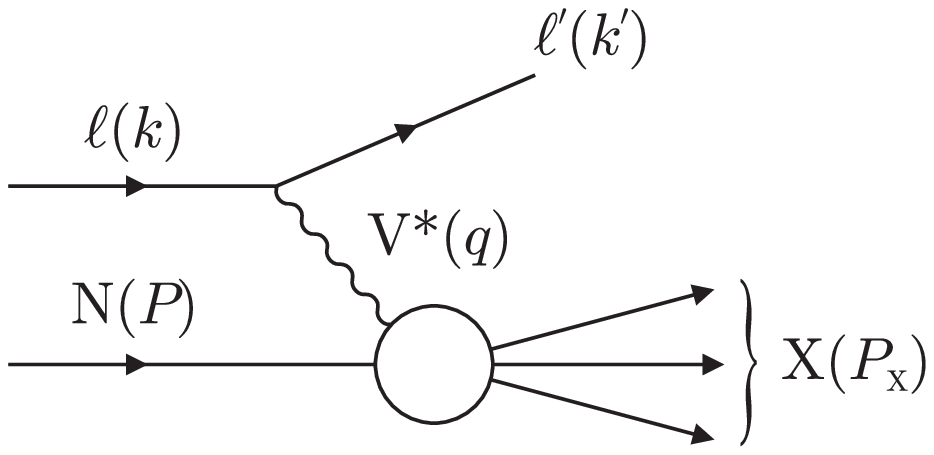}}
\label{scatter}
\fcaption{Schematic diagram of lepton-hadron scattering via Vector-Boson 
Exchange}
\label{fig:lpscat}
\end{figure}
\noindent
The associated four vectors are $k,k'$ for the
incoming and outgoing leptons respectively, and $P$ for the target 
(or incoming)
nucleon. The process is mediated by the exchange of a virtual vector boson, 
$V^*$($\gamma, W$ or $Z$), see Fig.~\ref{fig:lpscat}. This boson has four 
momentum given by
{\small \begin{equation}
               q = k - k'
\end{equation}}
and the four vector $p_X$ of the hadronic final state system $X$ is given by
{\small \begin{equation}
              P_X = P + q.
\end{equation}}
Various Lorentz invariants are useful in the description of the kinematics of 
the process:
{\small \begin{equation}
                   s = ( P + k )^2
\end{equation}}
the centre of mass energy squared for the $lp$ interaction,
{\small \begin{equation}
                   Q^2 = -q^2
\end{equation}}
the (negative of) the invariant mass squared of the virtual exchanged boson,
{\small \begin{equation}
                   x = Q^2/2P.q
\end{equation}}
the Bjorken $x$ variable, 
which is interpreted in the quark-parton model as the fraction 
of the momentum of the incoming nucleon taken by the struck quark, 
{\small \begin{equation}
                   W^2 = ( P + q )^2
\end{equation}}
the invariant mass squared of the hadronic system $X$, and
{\small \begin{equation}
                   y = P.q/P.k
\end{equation}}
a measure of the amount of energy transferred between the lepton and the hadron
systems. 
  
Note that (ignoring masses),
{\small \begin{equation}
                   Q^2 = s x y,\ \  W^2 = Q^2 (1/x - 1), 
\end{equation}}
so that only three of these quantities are independent. We shall give formulae
appropriate for $Q^2 \gg M_N^2$, where $M_N$ is the nucleon mass, unless
otherwise stated.

\subsection{Neutral current cross-sections}\label{sec:NCxsec}
\noindent
The general form for the differential cross-section for charged 
lepton-nucleon scattering, mediated by the neutral current at high energy, 
is given in terms
of three structure functions, $F_2, F_L, xF_3$, as,
{\small 
\begin{equation}
\frac {d^2\sigma (l^{\pm}N) } {dxdQ^2} =  \frac {2\pi\alpha^2} {Q^4 x}  
\left[Y_+\,F_2^{lN}(x,Q^2) - y^2 \,F_L^{lN}(x,Q^2)
\mp Y_-\, xF_3^{lN}(x,Q^2) \right],
\label{eq:NCxsec}
\end{equation}
} 
where $\displaystyle Y_\pm=1\pm(1-y)^2$ and we have ignored mass terms as 
appropriate at high $Q^2$.
For $Q^2$ values much below that of the $Z^0$ mass squared, the parity 
violating structure function $xF_3$ is 
negligible and the structure functions $F_2, F_L$ are given purely by 
$\gamma^*$ exchange. We begin by considering this region in more detail.
The differential cross-section is 
conveniently rewritten in terms of $R = \sigma_L/ \sigma_T$, 
the ratio of the longitudinally to transversely polarized virtual photon
absorption cross-sections, as follows
{\small 
\begin{equation}
\frac {d^2\sigma (l^{\pm}N) } {dxdQ^2} =  \frac {4\pi\alpha^2} {Q^4 x}  
\left[ 1 - y  - \frac{M^2_N x^2 y^2}{Q^2} + 
\frac{y^2}{2}\frac{1+4M^2_N x^2/Q^2}{1+R} \right]F_2^{lN}(x,Q^2),
\label{eq:Rxsec}
\end{equation}
}  
where we have specified the terms in $M^2_N$ which are important at 
low $Q^2$, and we have used the relationship $R = F_L/ 2xF_1$, 
where $2xF_1 = F_2(1 + 4M^2_N x^2/Q^2) - F_L$ relates the structure functions 
$2xF_1$, $F_2$ and $F_L$.

 It will aid understanding of the significance of the structure 
functions if we specify their interpretation in terms of the partons within
the nucleon as follows:
{\small \begin{equation}
        F_2^{lN}(x,Q^2) = \Sigma_i e_i^2*(xq_i(x,Q^2) + x\bar q_i (x,Q^2)),
\end{equation}}
a sum over the quark, $xq_i$, and antiquark, $x\bar q_i$, momentum 
distributions, contained in the
nucleon, multiplied by the corresponding quark charge squared $e_i^2$ 
(where $e_i$ is conventionally understood to be the fraction 
(quark charge/positron charge)). 
In writing this formula we have extended the naive quark-parton model 
interpretation to include
$Q^2$ dependence following the conventional manner of implementing first order
pQCD~\fnm{c}\fnt{c}{~At second order the 
same definition of
$F_2$ may be used, but only for the DIS renormalization scheme.}.
The spin-1/2 nature of the quarks also implies that $\sigma_L = 0$, i.e. 
there is no longitudinal absorption cross-section for virtual photon scattering
on quarks, and thus that $R = 0$ and $F_L = 0$, from which the 
Callan-Gross relationship~\cite{cg}, $2xF_1 = F_2$, follows
provided that $Q^2 \gg 4 M^2_N x^2$.
These relationships are also preserved in first order pQCD.

The differential cross-section is then related directly to $F_2(x,Q^2)$ by the
simple relationship,
{\small \begin{equation} 
\frac {d^2\sigma (lN) } {dxdQ^2} = \frac {2\pi\alpha^2} {Q^4 x} 
   Y_+\,F_2^{lN}(x,Q^2).
\end{equation}} 
and thus the lepton-nucleon scattering process has been used extensively
to measure quark distribution functions, and to investigate their $Q^2$
dependence.

Nucleon parton distribution functions (PDFs) are {\bf defined} to be those
for the proton. By default we shall mean the momentum distribution $xq(x,Q^2)$
(also called the momentum density) when we use this term. The number 
distribution (or density) is given by $q(x,Q^2)$. For ease of reference,
we shall specify the flavours entering into the sum for lepton probes
on proton, neutron and isoscalar targets.
For charged lepton scattering on protons  
{\small \begin{eqnarray}
  F_2^{lp}(x,Q^2) & = & \frac {4} {9} x(u(x,Q^2) + \bar u (x,Q^2) + c(x,Q^2) + 
\bar c (x,Q^2))\nonumber\\
& &\mbox{\hspace{1cm}} + \frac {1} {9}
 x(d(x,Q^2) + \bar d (x,Q^2) + s(x,Q^2) + \bar s (x,Q^2)) 
\label{eq:f2lp}
\end{eqnarray}}
whereas for a neutron target strong isospin swapping gives
{\small \begin{eqnarray}
  F_2^{ln}(x,Q^2) & = & \frac {4} {9} x(d(x,Q^2) + \bar d (x,Q^2) + c(x,Q^2) + 
\bar c (x,Q^2))\nonumber\\ & &\mbox{\hspace{1cm}} + \frac {1} {9}
 x(u(x,Q^2) + \bar u (x,Q^2) + s(x,Q^2) + \bar s (x,Q^2)) 
\label{eq:f2ln}
\end{eqnarray}}
and an isoscalar target is treated as an average of these two. We have assumed
that there is no significant bottom or top quark content in the nucleon. 
(We note that not all authors define a parton distribution for a heavy quark
such as the charm quark. The treatment of heavy quarks is discussed in
Sec.~\ref{sec:heavyq})
\vspace{3mm}

At values of $Q^2$ comparable to the $M_Z^2$, the formula for
$F_2$ must be extended to account for $Z^0$ exchange and $\gamma-Z^0$ 
interference as follows
{\small \begin{equation}
 F_2^{lN}(x,Q^2) = \Sigma_i A_i^{L,R}(Q^2)*(xq_i(x,Q^2) + x\bar q_i (x,Q^2)),
\label{eq:f2lNhiq2}
\end{equation}}
where, for lepton scattering,
{\small \begin{equation}
        A_i^{L,R}(Q^2) = e_i^2 - 2e_i e_l(v_l \pm a_l)v_i P_Z + (v_l \pm a_l)^2
(v_i^2 +a_i^2) P_Z^2
\end{equation}}
The coupling of the fermions to the currents now depends on whether the
polarization of the lepton beam is left ($L$) or right ($R$) handed. The
notation $e_l$ specifies the incoming lepton's charge such that 
$e_l= \pm 1$. The 
vector and axial-vector couplings of the fermions are given by
{\small \begin{equation}
 v_f = (T_{3f} - 2 e_f \sin^2\theta_W),\ \ a_f = T_{3f}
\end{equation}}
where the definition holds good for any fermion, whether lepton or quark;
$T_{3f}$ is the weak isospin, and $\theta_W$ is the Weinberg 
angle~\fnm{d}\fnt{d}{~Neutrinos and charged leptons of the same 
family form weak 
isospin doublets with $T_3 = 1/2,-1/2$ respectively; and the  
quarks form similar weak isospin doublets, within the families 
$(u,d),(c,s),(t,b)$, with $T_3 = 1/2, -1/2$ respectively.}.
The term  $P_Z$ accounts for the $Z_0$ propagator
{\small \begin{equation}
           P_Z = \frac {Q^2} {Q^2 + M_Z^2} \frac {1} {\sin^2 2\theta_W}.
\end{equation}}

Furthermore, the parity violating structure function $xF_3$ is no longer 
negligible, and is given by
{\small \begin{equation}
  xF_3^{lN}(x,Q^2) = \Sigma_i B_i^{L,R}(Q^2)*(xq_i(x,Q^2) - x\bar q_i (x,Q^2)),
\label{eq:xf3ln}
\end{equation}}
where, for lepton scattering,
{\small \begin{equation}
        B_i^{L,R}(Q^2) =  \mp 2e_i e_l(v_l \pm a_l)a_i P_Z \pm 2(v_l \pm a_l)^2
v_i a_i P_Z^2.
\end{equation}}
\vspace{3mm}
The corresponding cross-sections for antilepton scattering are given by
swapping $L \rightarrow R$, $R\rightarrow L$ in the expressions for $F_2$ and
$xF_3$ given in Eq.~\ref{eq:f2ln} and Eq.~\ref{eq:xf3ln}.

\subsection{Charged current cross-sections}\label{sec:CCxsec}
\noindent
The charged lepton-nucleon differential cross-sections mediated by the charged
current $W^{\pm}$ (where the final state lepton is a neutrino) are 
given by
{\small \begin{equation}
\frac {d^2\sigma^{CC}(l^\pm N) } {dxdQ^2} =  \frac {G_F^2} {4\pi x} \frac {
 M^4_W} { (Q^2 + M_W^2)^2 }\left[  
 Y_+\,F_2(x,Q^2) - y^2 \, F_L(x,Q^2) \mp Y_-\, xF_3(x,Q^2) \right]
\label{eq:CCxsec}
\end{equation}}
and the correspondence to the neutral current case can be seen easily if we 
express the Fermi coupling constant $G_F$ as
{\small \begin{equation}
        G_F = \frac{\pi \alpha} {\sqrt 2 \sin^2\theta_W M^2_W}
\end{equation}}
Then, using the predictions of first order pQCD, we again have $F_L=0$, and
the differential cross-section for lepton scattering becomes
{\small \begin{equation}
\frac {d^2\sigma^{CC}(l^-N) } {dxdQ^2} = (1 - P)\frac {G_F^2} {2\pi x} \frac 
{ M^4_W} { (Q^2 + M_W^2)^2 }   \Sigma_i \left[ 
 xq_i(x,Q^2) + (1-y)^2 x\bar q_i (x,Q^2) \right]
\end{equation}}
whereas for antilepton scattering we have
{\small \begin{equation}
\frac {d^2\sigma^{CC}(l^+N) } {dxdQ^2} = (1 + P)\frac {G_F^2} {2\pi x } \frac 
{ M^4_W} { (Q^2 + M_W^2)^2 }   \Sigma_i \left[ 
(1-y)^2 xq_i(x,Q^2) +  x\bar q_i (x,Q^2) \right]
\end{equation}}
where the sums contain only the appropriate quarks or antiquarks for the 
charge of the current and the polarization of the lepton beam, $P = 
\frac{N_R - N_L}{N_R + N_L}$.

We specify the flavours entering into the quark sums for ease of 
reference. 
For $l^- p\rightarrow \nu X$ and left handed polarization we have
{\small \begin{eqnarray}
     F_2 &=& 2x(u(x,Q^2) + c(x,Q^2) + \bar d (x,Q^2) + \bar s (x,Q^2)),
           \nonumber \\
    xF_3 &=& 2x(u(x,Q^2) + c(x,Q^2) - \bar d (x,Q^2) - \bar s (x,Q^2))
\label{eq:lhp}
\end{eqnarray}}
and $F_2 = xF_3 = 0$, for right handed polarization.

Whereas for $l^+ p\rightarrow \bar \nu X$ and right handed polarization 
we have 
{\small \begin{eqnarray}
     F_2 &=& 2x(d(x,Q^2) + s(x,Q^2) + \bar u (x,Q^2) + \bar c (x,Q^2)),
           \nonumber \\
    xF_3 &=& 2x(d(x,Q^2) + s(x,Q^2) - \bar u (x,Q^2) - \bar c (x,Q^2)),
\label{eq:rhp}
\end{eqnarray}}
and $F_2 = xF_3 = 0$, for left handed polarization.
Again we have assumed that there is no significant top or bottom quark content
in the nucleon and we have also assumed that we are considering energies 
above threshold for the
production of charmed quarks in the final state~\fnm{e}\fnt{e}{~Otherwise 
we need to 
multiply $\bar d $ by $\cos^2\theta_c$ and $\bar s $ by $\sin^2\theta_c$ in 
Eq.~\ref{eq:lhp},
and $d$ by $\cos^2\theta_c$  and $s$ by $\sin^2\theta_c$  in 
Eq.~\ref{eq:rhp}.}.

For charged lepton scattering on neutron targets these relationships undergo
isospin swapping $u \rightarrow d,d \rightarrow u, \bar u  \rightarrow 
\bar d ,
\bar d \rightarrow \bar u $, and if we are considering isoscalar targets we 
must take the average $(n+p)/2.$

\vspace{3mm}
To date charged current reactions have been studied most extensively in 
the reverse neutrino and antineutrino scattering processes, 
{\small \begin{equation}
               \nu N \rightarrow \mu^- X,\ \ \bar \nu  N \rightarrow \mu^+ X
\end{equation}}
at $Q^2$ values far below the mass of the $W$ squared. 
The polarizations of the (anti)neutrino probes are automatically 
(right)left-handed, and these cross-sections have conventionally been 
written as
{\small \begin{equation}
\frac {d^2\sigma(\nu,\bar \nu ) } {dxdy} = \frac {G_F^2 s} {4\pi } \left[ 
Y_+\,F_2(x,Q^2) - y^2 \,F_L(x,Q^2) \pm Y_-\,xF_3(x,Q^2) \right]
\end{equation}}
and using the results of first order pQCD we have for neutrino scattering 
{\small \begin{equation}
\frac {d^2\sigma(\nu) } {dxdy} = \frac {G_F^2 s} {\pi } \Sigma_i \left[ 
 xq_i(x,Q^2) + (1-y)^2 x\bar q_i (x,Q^2) \right]
\end{equation}}
and for antineutrino scattering,
{\small \begin{equation}
\frac {d^2\sigma(\bar \nu) } {dxdy} = \frac {G_F^2 s} {\pi } \Sigma_i \left[ 
(1-y)^2 xq_i(x,Q^2) +  x\bar q_i (x,Q^2) \right]
\end{equation}}
Hence antineutrino and neutrino data have been 
combined to extract quark and antiquark momentum distributions and data from
proton and isoscalar targets have been combined to achieve flavour separation.

We specify here the flavours entering into these sums in terms of the
structure functions for ease of reference.
{\small
\begin{eqnarray}
    F_2^{\nu p} &=& 2x(d(x,Q^2) + s(x,Q^2) + \bar u (x,Q^2) + \bar c (x,Q^2)),
         \nonumber \\
   xF_3^{\nu p} &=& 2x(d(x,Q^2) + s(x,Q^2) - \bar u (x,Q^2) - \bar c (x,Q^2)),
\label{eq:nup}
\end{eqnarray}
\begin{eqnarray}
    F_2^{\nu n} &=& 2x(u(x,Q^2) + s(x,Q^2) + \bar d (x,Q^2) + \bar c (x,Q^2)),
         \nonumber \\
   xF_3^{\nu n} &=& 2x(u(x,Q^2) + s(x,Q^2) - \bar d (x,Q^2) - \bar c (x,Q^2)),
\end{eqnarray}
\begin{eqnarray}
     F_2^{\bar \nu p} &=& 2x(u(x,Q^2) + c(x,Q^2) + \bar d (x,Q^2) 
+ \bar s (x,Q^2)), \nonumber \\
    xF_3^{\bar \nu p} &=& 2x(u(x,Q^2) + c(x,Q^2) - \bar d (x,Q^2) 
- \bar s (x,Q^2)),
\label{eq:nubarp}
\end{eqnarray}
\begin{eqnarray}
     F_2^{\bar \nu n} &=& 2x(d(x,Q^2) + c(x,Q^2) + \bar u (x,Q^2) 
+ \bar s (x,Q^2)), \nonumber \\
    xF_3^{\bar \nu n} &=& 2x(d(x,Q^2) + c(x,Q^2) - \bar u (x,Q^2) 
- \bar s (x,Q^2))
\end{eqnarray}}

Again we are assuming no top or bottom quarks in the nucleon targets and 
that we are working above threshold for production of charmed quarks in the 
final state~\fnm{f}\fnt{f}{~Below the charm threshold, one has to multiply $d$ 
by $\cos^2\theta_c$ and $s$ 
by $\sin^2\theta_c$ in Eq.~\ref{eq:nup} and $\bar d$ by $\cos^2\theta_c$  and 
$\bar s$ by $\sin^2\theta_c$ in
Eq.~\ref{eq:nubarp}, where $\theta_c$ is the Cabibbo mixing angle. Similar 
modifications would also be necessary for the neutron reactions.}.

Having set up the formalism we now proceed to outline the theoretical framework
of the pQCD improved quark-parton model.
%
%

\section{Theoretical framework}\label{sec:Thframe}
\subsection{The naive quark-parton model}\label{sec:qpm}
\begin{figure}[ht]
\centerline{\psfig{figure=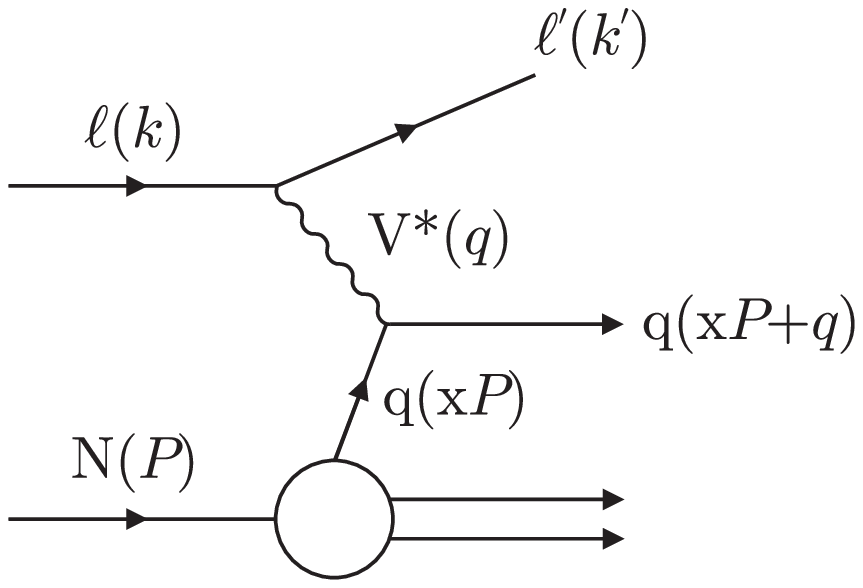,width=6cm,height=5cm}}
\fcaption{Schematic diagram of lepton-hadron scattering in the 
quark-parton model}
\label{qpm}
\end{figure}

\noindent
The quark-parton model (QPM) grew out of the attempt by Feynman~\cite{feynman}
to provide a simple physical picture of the scaling that had been predicted 
by Bjorken~\cite{bjorken} and observed in the first high energy deep 
inelastic electron scattering experiments at SLAC~\cite{nobel1990} where
$F_2$ was observed to be independent of $Q^2$ for $x$ values around
$x \sim 0.3$.
The model states that the nucleon is full of point-like
non-interacting scattering centres known as partons. The lepton-hadron 
reaction cross-section is approximated by an incoherent sum of
elastic lepton-parton scattering cross-sections, see Fig.~\ref{qpm}. 
In the infinite momentum
frame it is then easy to show that the variable $x$ is identified with the
fraction of the nucleon's momentum involved in the hard scattering.
\noindent

The parton model had to be reconciled with the static
quark model which pictures a nucleon and other baryons as made of three 
constituent quarks which give them their flavour properties.
The reconciliation was effected in the QPM by 
considering the nucleon as made up of valence quarks, which give it its 
flavour properties, and a `sea' of quark antiquark pairs which have no 
overall flavour. Both the valence quarks and the sea quarks and 
antiquarks are identified as partons. Hence the antiquark
distributions within a nucleon are purely sea distributions, whereas the 
quark distributions have both valence and sea contributions
{\small \begin{equation}
   xq(x) = xq_v(x) +xq_s(x),\  x\bar q(x) = x\bar q_s(x),
\end{equation}}
and clearly
{\small \begin{equation}
         xq_s(x) = x\bar q_s(x),
\end{equation}}

To substantiate this idea, early (anti)neutrino data~\cite{fisk_sciulli}
were used to take a first look at the shapes of 
the sea and valence quark momentum distributions in the nucleon. 
The structure functions for (anti)neutrino scattering on 
isoscalar targets are given by
{\small \begin{equation}
F_2^{\nu N}(x) = x(u(x) + d(x) + \bar u(x) + \bar d(x) + 2s(x) + 
2\bar c(x))
\end{equation}
\begin{equation}
F_2^{\bar \nu N}(x) = x(u(x) + d(x) + \bar u(x) + \bar d(x) + 2\bar s(x)
 + 2c(x))
\end{equation}}
 and
{\small \begin{equation}
xF_3^{\nu N}(x) = x(u(x) + d(x) - \bar u(x) - \bar d(x) + 2s(x) 
- 2\bar c(x) )
\label{eq:nucs}
\end{equation}
\begin{equation}
xF_3^{\bar \nu N}(x) = x(u(x) + d(x) - \bar u(x) - \bar d(x) - 2\bar s(x) 
+ 2 c(x) )
\label{eq:anucs}
\end{equation}}
Hence such data may be combined to yield, 
{\small \begin{equation}
xF_3^{\nu N}(x) \simeq xF_3^{\bar \nu N}(x) \simeq x(u_v(x) + d_v(x)) 
= xV(x)
\end{equation}}
where $xV(x)$ represents a purely valence, or non-singlet, distribution 
and
{\small \begin{equation}
 F_2^{\nu N}(x) = F_2^{\bar \nu N}(x) = x(u_v(x) + d_v(x) + 2 \bar u(x) +
2 \bar d(x) + 2\bar s(x)) = xV(x) + xS(x)
\end{equation}}
represents a combination of valence and sea~\fnm{g}\fnt{g}{~In early 
analyses one often assumed an SU(3) symmetric sea, i.e. 
$\bar u(x) =\bar d(x)= \bar s(x), \bar c(x) = 0 $, 
though it now appears that $ \bar s(x) \approx 0.5*(\bar u(x) + 
\bar d(x))/2$.} which is a singlet distribution. The terminology 
non-singlet/singlet refers to flavour exchange/non-exhange and is explained 
in context in Sec.~\ref{sec:pdfhilo}.  

Experimental information on valence and sea distributions may also be 
obtained from charged lepton-nucleon scattering. $F_2$ on an isoscalar target 
may be written as
{\small \begin{equation}
F_2^{lN} = \frac{5}{18} x(u(x) + \bar u(x) + d(x) + \bar d(x) ) + 
\frac{1}{9} x(s(x) + \bar s(x)) + \frac{4}{9} x(c(x) + \bar c(x))
\end{equation}}
which is a combination of valence and sea quarks
and, assuming $\bar u(x) =\bar d(x)$,
{\small \begin{equation}
 F_2^{lp} - F_2^{ln} = \frac {1}{3} x (u_v(x) - d_v(x))
\end{equation}}
is a pure valence distribution. Thus the shapes of the 
valence and sea distributions may be extracted separately and furthermore, 
the neutrino and lepton structure functions may be related as follows. Assuming
$s(x) = \bar s(x), c(x) = \bar c(x)$,
{\small \begin{equation}
F_2^{lN} = \frac{5}{18} F_2^{\nu N} - \frac{1}{3} xs(x) + \frac{1}{3} 
xc(x) \approx \frac{5}{18} F_2^{\nu N} 
\end{equation}}
The observation that the charged lepton scattering and the
neutrino scattering structure functions are related by $\sim 5/18$, 
taken together with the approximate verification of the Callan-Gross 
relationship was a triumph for the QPM.

Further support came from considering sum rules. Since $xq(x)$ gives the 
quark momentum distribution, $q(x)$ must give the quark number distribution.
Then if we take the quark-parton model seriously, we predict
{\small \begin{equation}
 \int_0^1 dx u_v(x) = 2 ,\int_0^1 dx d_v(x) = 1
\end{equation}}
Of course these relations cannot be directly verified, but their 
consequences
can be, for example the Gross Llewelyn-Smith sum rule~\cite{GLS}
{\small \begin{equation}
 \int_0^1 dx F_3^{\nu N} \approx \int_0^1 dx (u_v(x) +d_v(x)) = 3
\end{equation}}
was verified in early neutrino data on isoscalar targets. 
Similarly, the Adler sum rule~\cite{Adler}
{\small \begin{equation}
 \int_0^1 \frac{dx}{x}\,\bigl(F_2^{\bar \nu p} - F_2^{\nu p}\bigr) = 
2\int_0^1 dx  ((u_v(x) - d_v(x)) = 2  
\end{equation}}
was verified using neutrino data on hydrogen targets.
The Gottfried sum rule~\cite{Gottfried}
{\small \begin{equation}
 \int_0^1 \frac{dx}{x}\,\bigl(F_2^{l p} - F_2^{l n}\bigr) \approx 
 1/3 \int_0^1 dx (u_v(x) - d_v(x)) = 1/3 
\label{eq:gott}
\end{equation}}
was approximately verified in early data, but more recent data 
shows that it is 
actually violated since the approximation in its derivation requires the 
assumption of an SU(2) symmetric sea.

A sum rule can also be applied to the sum over the momenta of all types of 
quarks and antiquarks in the nucleon. Denoting the singlet distribution by
{\small \begin{equation}
x\Sigma(x) = x (u(x) + \bar u(x) + d(x) + \bar d(x) + s(x) + \bar s(x) + c(x)
+ \bar c(x) ) = F_2^{\nu N} 
\end{equation}}
we have the momentum sum rule (MSR),
{\small \begin{equation}
\int_0^1 dx x\Sigma(x) = 1 
\end{equation}}
if quarks and antiquarks carry all of the momentum of the nucleon. This
was not confirmed, the experimental measurement of $\sim 0.5$ 
implied that there was more momentum in the nucleon than that carried by the
quarks and antiquarks and gave 
impetus to the development of the theory of QCD, in which the deficit in
momentum is carried by the gluons.

\subsection{Parton distributions at low and high $x$}
\label{sec:pdfhilo}
\noindent
The shapes 
extracted for valence and sea quark distributions were 
roughly of the form 
{\small \begin{equation}
    xV(x) = A x^\alpha (1-x)^\beta   
\label{eq:alfbet}
\end{equation}}
where $\alpha \simeq 0.5$, $\beta \simeq 3$ for the valence distribution, 
and
{\small \begin{equation}
    xS(x) = B (1-x)^\gamma
\label{eq:sea}
\end{equation}}
where $\gamma \simeq 7$ for the sea distribution. Thus the simple
picture of a separation between valence and sea quarks contributions was
verified~\cite{Abramowicz}. 

There are theoretical guidelines 
for both the low $x$ and the high $x$ behaviour of these distributions.
The high $x$ behaviour has been predicted from the 
dimensional counting rules~\cite{Farrar}.
As $x \rightarrow 1$ there can be no momentum left for any of the partons
apart from the struck quark, thus they become `spectators' and the prediction 
is that
$xq(x) \rightarrow (1-x)^{2n_s - 1}$, where $n_s$ is the minimum number of 
spectators. Thus, for valence quarks one has $n_s=2$, 
$xV(x) \rightarrow (1-x)^3$, and for sea quarks one has $n_s=4$, $xS(x) 
\rightarrow (1-x)^7$, where it may be noted that the counting
of spectators represents a very naive view of the 
interaction. Nevertheless 
these rules gave a first rough indication of the differing behaviour of 
valence and sea quarks at high 
$x$\fnm{h}\fnt{h}{~For completeness we also note for future reference 
that for gluons one has $n_s=3$, $xg(x) \rightarrow (1-x)^5$.}.

The low $x$ behaviour is predicted from Regge theory. At small $x$ we have 
large $P.q$ and hence large centre of mass energy, $W$, of the virtual 
boson-nucleon system.
Thus we are in the Regge region for the virtual boson-nucleon subprocess. 
Regge phenomenology gives predictions for the scattering amplitudes in high
energy hadron-hadron interactions based on considering the process 
$ a b \rightarrow a b$ as mediated by the exchange of a `trajectory' of
virtual particles which have the correct quantum numbers to be exchanged. The
prediction is that the imaginary part of the scattering amplitude, 
$Im A(ab \rightarrow ab)$, depends on the centre of mass energy of the 
process squared, $s(ab)$ as (ignoring spin)
{\small \begin{equation}
    Im A(ab \rightarrow ab) \sim \Sigma_i \beta_i s(ab)^{\alpha_i}
\end{equation}}
\noindent
where $\alpha_i$ is the intercept of the Regge trajectory and $\beta_i$ is 
an (unknown) residue function. 

\begin{figure}[ht]

\centerline{\psfig{figure=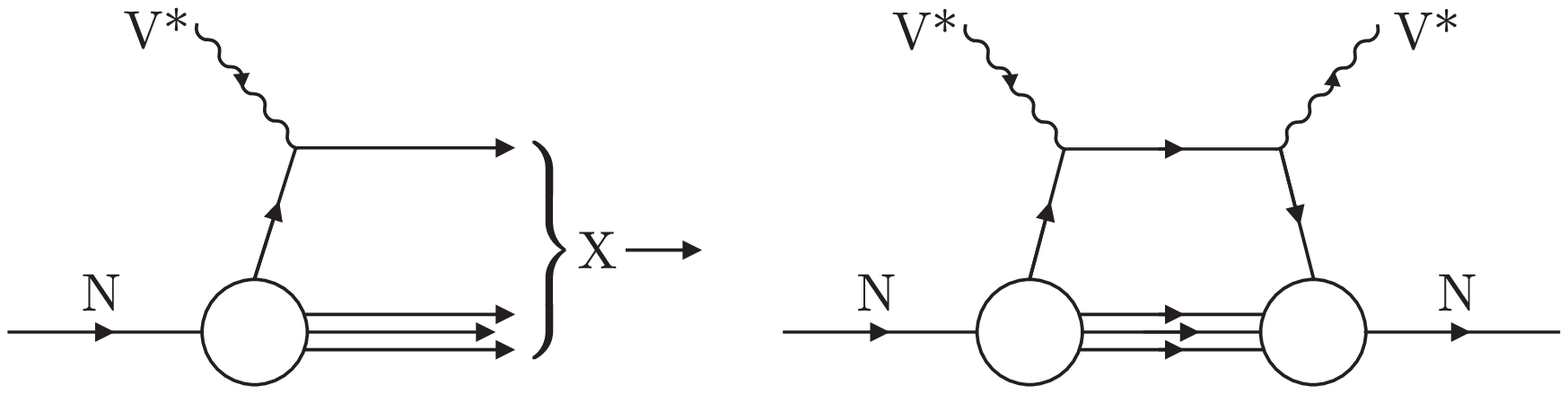,width=10cm,height=5cm}}
\fcaption{Illustration of the Optical Theorem}
\label{optical}
\end{figure}

\begin{figure}[ht]

\centerline{\psfig{figure=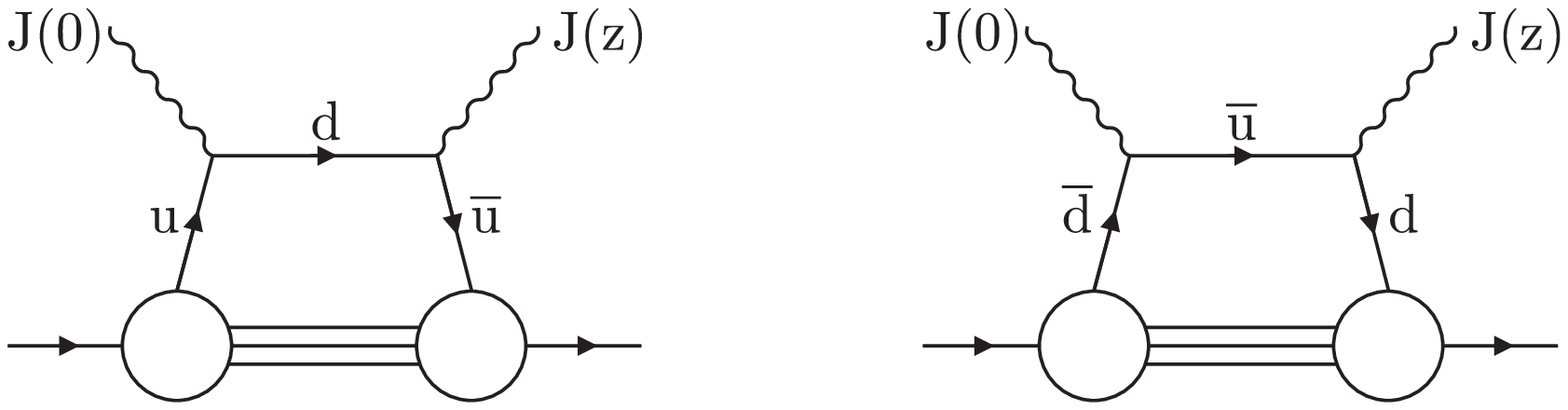,width=10cm,height=5cm}}
\fcaption{Flavour structure of t channel exchanges for the handbag diagram in
$\nu, \bar \nu$ scattering. The notations $J(z)$, $J(0$ refer to the 
space-time structure of the vector boson currents.}
\label{handbag}
\end{figure}

To gain information on the behaviour of total cross-sections we use the 
Optical Theorem,
which relates the inclusive cross-section to the imaginary part of the 
forward elastic scattering amplitude as follows
{\small \begin{equation}
 \sigma( a b \rightarrow X) = \frac {1}{s(ab)} Im(a b \rightarrow a b)
\end{equation}}
\noindent
If we apply the Optical Theorem and Regge theory to the virtual boson-nucleon 
total cross-section (as 
illustrated in Fig.~\ref{optical}) we predict an $s(V^* N)$ dependence 
{\small \begin{equation}
 \sigma( V^* N \rightarrow X) \sim \frac {1}{s(V^* N)} 
\Sigma_i s(V^* N)^{\alpha_i} \sim \Sigma_i s(V^* N)^{\alpha_i -1}
\end{equation}}
\noindent
where the sum is dominated by the appropriate trajectory with the largest
$\alpha$. This dependence on $s(V^*N)$ translates back to an $x$ dependence
of $x^{(1-\alpha_i)}$ for the deep inelastic scattering process. 
\noindent
The appropriate Regge trajectory depends on 
whether we consider the contribution of $F_2$ or $xF_3$ 
to the total cross-section, since these pick out different flavour exchanges.
For $\nu$,$\bar \nu$ scattering the exchange in the $(V^*N \rightarrow V^*N)$
process has the possible flavour combinations $ u \bar u + d \bar d$ and  
$ u \bar u - d \bar d$, corresponding to $F_2$ and $xF_3$ respectively
(see Fig.~\ref{handbag}). The
latter is flavour non-singlet (i.e. flavour IS exchanged) and hence the 
leading Regge trajectory is the $\rho/A_2$ trajectory, with intercept 
$\alpha_R \sim 0.5$.
The former is flavour singlet and the leading trajectory is called the 
Pomeron, with intercept $\alpha_P \sim 1.0 + \lambda$, where
$\lambda \sim 0.08$. The exchange of the Pomeron trajectory is
considered responsible for the 
slowly increasing behaviour of hadron-hadron total cross-sections with
increasing energy~\cite{DoLa}. 
\noindent
Hence we have the predictions that the shape of $xF_3$ as 
$x \rightarrow 0$ is given by $x^{0.5}$, whereas the shape of $F_2$ as 
$x \rightarrow 0$ is given by $x^{-\lambda}$, 
 i.e. it is flattish or very slowly increasing as $x$ decreases.
These predictions were born out by the early data, which 
were mostly taken at moderate $Q^2 (\sim 10\,$GeV$^2$) and $x$
values of around $x \geqsim 0.01$. It is also true that the real photon 
nucleon cross-section 
(for which $Q^2 = 0$) obeys the Regge prediction for the Pomeron.
However since we now know that parton distributions do not exhibit exact
Bjorken Scaling, but evolve with $Q^2$, one may ask in what region of
$Q^2$ should these predictions be relevant? Since Regge phenomenology is 
an essentially
non-perturbative approach we might expect it to be most relevant at 
low $Q^2$. However, the new HERA data allow us to probe
very low $x$ values ($x \leqsim 10 ^{-4}$) 
and we shall see that there are significant deviations from the
Regge predictions for Pomeron exchange, 
even for $Q^2$ values as low as $Q^2 \simeq 2\,$GeV$^2$. 
This has led to theoretical developments such that the conventional Pomeron
referred to above is now called the `soft' Pomeron to distinguish it from
hard Pomeron-like behaviour which can be predicted in the framework of 
pQCD~\cite{ForRoss}.
We discuss this further in Secs.~\ref{sec:lowx},~\ref{sec:lowq2}.

\subsection{Parton distributions with $Q^2$ dependence: QCD evolution}
\label{sec:qcdevol}
\noindent
The QPM must be modified since partons cannot be non-interacting. They are
confined within nucleons by the strong interaction.
QCD is a non-Abelian gauge theory of the strong interactions between quarks
and gluons, which allows us to reconcile short distance freedom with long
distance confinement. This comes about because the
strong interaction's strength is variable. 
\newpage
\subsubsection{The running coupling constant}

\begin{figure}[ht]
\centerline{\psfig{figure=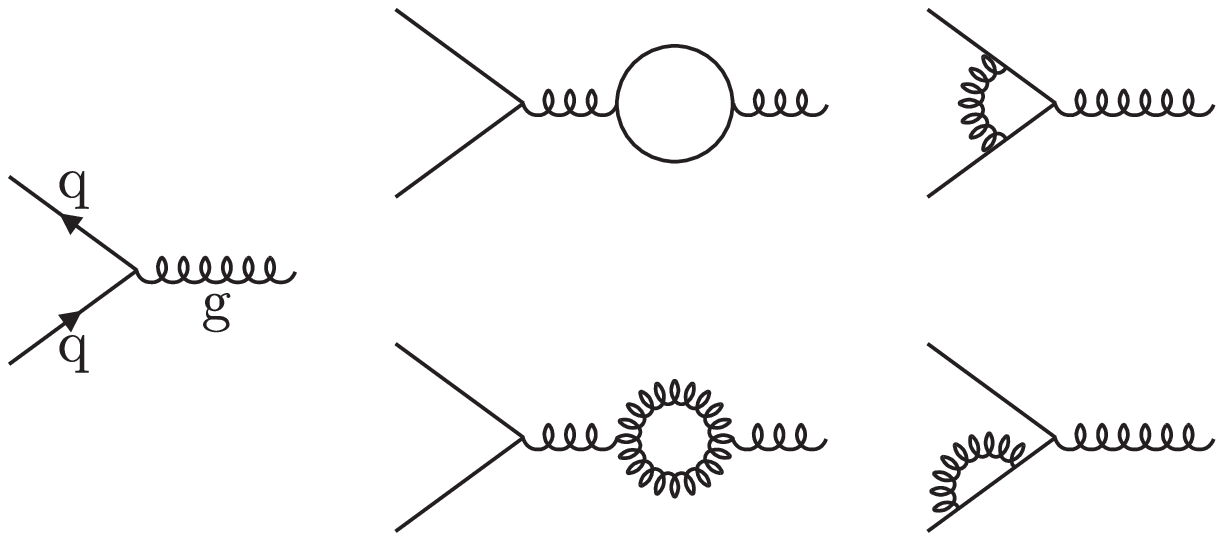,width=9cm,height=5cm}}
\fcaption{Schematic diagram of the $q\bar qg$ vertex diagram plus virtual loop
corrections}
\label{vertex}
\end{figure}

This is best understood by 
considering the behaviour of the strong coupling `constant',  $\bar g$, 
which is defined as the value of the $q\bar q g$ vertex diagram 
(see Fig.~\ref{vertex})
\noindent
The value of $\bar g$ which one measures in strong interactions must include all 
the virtual loop diagrams (only the 1-loop corrections are illustrated in 
Fig.~\ref{vertex}). If one tries to calculate this series of diagrams one 
obtains infinities and 
these are controlled by a renormalization procedure, in which one 
defines the coupling to be finite at some scale $\mu^2$, and 
expresses $\bar g(Q^2)$ at any
other scale in terms of this fixed value. Renormalization can be 
performed in various
different ways or schemes. It is clear that the results for 
physical quantities cannot
depend on the arbitrary scale $\mu^2$ and this independence is 
expressed in terms of
a Renormalization Group Equation, which can be solved to give 
the dependence of 
the coupling $\bar g$ (or indeed of any Green's Function) on the external 
scale in terms of
a calculable $\beta$ function
{\small \begin{equation}
 2\frac{d}{dt}\bar g(t) = \beta (\bar g) = -\beta_0 \frac{\bar g^3}{16\pi^2}  
-  \beta_1 \frac{\bar g^5}{(16\pi^2)^2} +\dots
\end{equation}}
\noindent
where $t = ln(Q^2/\mu^2)$, $\beta_0 = 11 - 2 n_i/3$, 
$\beta_1 = 102 - 38 n_i/3$ , and 
$n_i$ is the number of quarks participating in the interaction at the scale 
$Q^2$. The term in $\beta_0$ gives the 1-loop results for the 
$\beta$ function, the 
term in $\beta_1$ gives the 2-loop result etc. The 1-loop solution 
of this equation 
 is often expressed in 
terms of the `running coupling constant' $\alpha_s(Q^2) = \bar g^2(Q^2)/(4\pi)$ as 
{\small \begin{equation}
\alpha_s(Q^2) = \frac{4\pi}{\beta_0 ln(Q^2/\Lambda^2)}
\label{eq:alphas1}
\end{equation}}
\noindent
where $\Lambda$ is now a parameter of QCD, 
which depends on the renormalization scale and scheme and also on the number
of active flavours $n_i$ at the scale $Q^2$ we are working at.

The fact that the coupling constant actually depends on the 
external scale $Q^2$
is true of all field theories including QED, where it manifests itself
as charge screening - one does not see the full charge until one probes
sufficiently close to the source. However, the non-Abelian nature
of the gluon-gluon coupling in QCD leads to anti-screening - the
closer one probes the less strong the charge appears!. Thus when $Q^2$ is 
fairly large, (say $Q^2>4\,$GeV$^2$ for deep inelastic scattering), 
$\alpha_s$ is small and the quarks are `asymptotically free'.
It is in this kinematic region that we may use perturbation theory to 
perform calculations within QCD. At large distances the coupling constant 
increases, and quarks are confined within hadrons. We need
non-perturbative techniques to work in this region. The present 
section is concerned 
only with perturbative QCD, we consider the transition to low $Q^2$ in 
Sec.~\ref{sec:lowq2}.

The first order, or 1-loop, expression for $\alpha_s$ is not 
adequate, since
 the parameter $\Lambda$ cannot be unambiguously defined.
 Changing the value of $\Lambda$,
$\Lambda \to \Lambda/k$, changes the expression for $\alpha_s$ only by 
a term of 
order $\alpha_s^2$. Since a change in $\Lambda$ is clearly
equivalent to a change in the scale $Q^2$, we therefore cannot define 
the scale of the
theory at 1-loop. We must compute $\alpha_s$ to 2-loops. This is given by 
{\small \begin{equation}
 \ln \frac{Q^2}{\Lambda^2} = \frac {4\pi}{\beta_0 \alpha_s} -\frac{\beta_1}
{\beta_0^2} \ln \left[ \frac {4\pi}{\beta_0\alpha_s} + \frac{\beta_1}
{\beta_0^2}\right]
\label{eq:alphas2}
\end{equation}}
\noindent
and a change of scale $\Lambda \to \Lambda/k$ now produces a change in
$\alpha_s$ at the same order. Practically, it is also true that we cannot avoid
the need to consider higher order effects, since much of the 
available data is in the moderate $Q^2$ range, $5 < Q^2 < 100\,$GeV$^2$,
where $\alpha_s^2$ is not negligible. 

It is also clear that, when we are working to finite order, $\alpha_s$ depends
on the renormalization scheme in which we are working, since $\Lambda$
depends on the renormalization scale $\mu$. The choice of scheme was much 
debated in the early days of 
pQCD~\cite{Buras},~\cite{PMS}. One wishes to chose a scheme in 
which higher
order corrections are progressively smaller than the lower order results,
i.e. a scheme in which calculations converge rapidly, and at the same time one
wishes to chose a scheme which will preserve this property in other
perturbatively calculable processes, such as Drell Yan production and $e^+e^-$
scattering). The consensus
of the community has settled on the $\MSB$ scheme although the DIS
scheme (which maintains the identification of structure functions in terms 
of a simple weighted sum over quark distributions) is extensively used in 
Monte Carlo simulation codes. Values of $\Lambda$ and $\alpha_s$ which we quote
will refer to the $\MSB$ scheme unless otherwise stated.
We shall indicate how the choice of scheme affects the results
as we discuss how pQCD modifies the simple results of the QPM.
  
\subsubsection{$Q^2$ dependence of parton distributions functions: first order}

\begin{figure}[ht]
\centerline{\psfig{figure=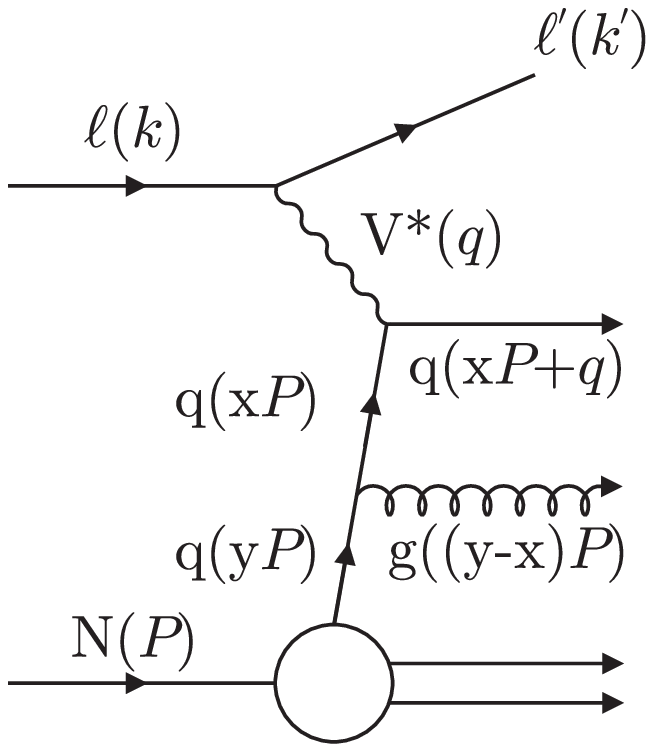,width=5cm,height=6cm}}
\fcaption{Schematic diagram of the QCD Compton (QCDC) process}
\label{qcdc}
\end{figure}

\begin{figure}[ht]
\centerline{\psfig{figure=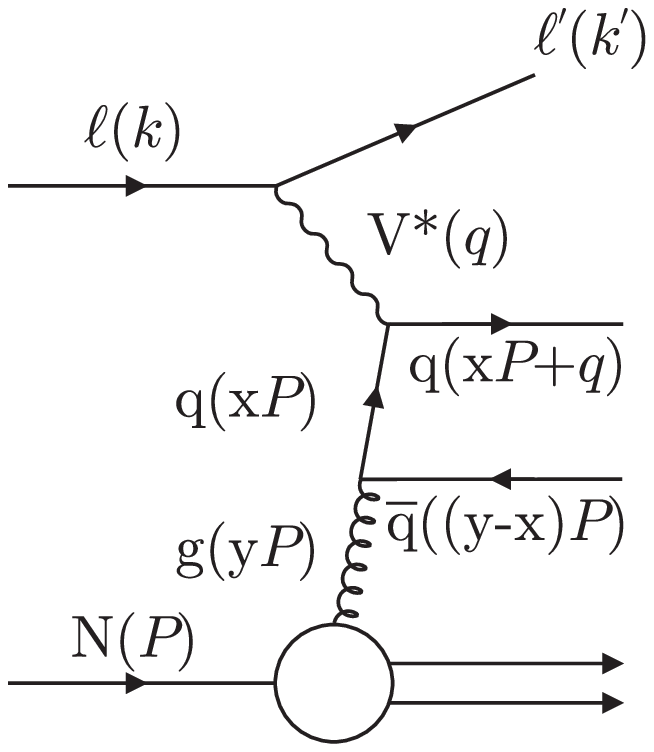,width=5cm,height=6cm}}
\fcaption{Schematic diagram of the boson-gluon fusion (BGF) process}
\label{bgf}
\end{figure}

Firstly, the effect of the quark-gluon interaction is to make the
quark momentum distributions, and thus the structure function, depend on
$Q^2$, (the momentum distributions are said to `evolve' with $Q^2$) 
such that $F_2(x,Q^2)$ rises with $Q^2$ 
at small values of $x$, and falls with $Q^2$ at large values of $x$.
To understand this $Q^2$ evolution we refer to Fig.~\ref{qcdc} and 
Fig.~\ref{bgf}.
The struck quark
may have a history before it interacts with the vector boson. It could radiate 
a gluon as in Fig.~\ref{qcdc} (the QCD Compton (QCDC) process), and thus, 
although the quark which is struck 
has momentum
fraction $x$, the quark originally had a larger momentum fraction, $y>x$.
Alternatively, as in Fig.~\ref{bgf}, it may be that a gluon with momentum 
fraction $y$ 
produced a $q \bar q $ pair and one of these became the struck quark of 
momentum fraction $x$ (the boson-gluon fusion (BGF) process). 
Thus quark distributions, $q(y,Q^2)$ for all momentum
fractions $y$ such that $ x < y < 1$, contribute to the process shown in 
Fig.~\ref{qcdc}, 
and the gluon distribution $g(y,Q^2)$, for all momentum fractions $y$ such 
that $x < y < 1$, contributes to the process shown in Fig.~\ref{bgf}.

To summarize, the parton being probed may not be an `original' constituent,
but may arise from the strong interactions within the nucleon. The smaller 
the wavelength of the probe (i.e. the larger the scale $Q^2$) the more of such
quantum fluctuations can be observed and hence the amount of $q \bar q$ pairs 
and gluons in the partonic `sea' increases. Although these sea partons carry 
only a small fraction of the nucleon momentum, their increasing number leads to 
a softening of the valence quark distributions as $Q^2$ increases.
Thus $F_2$, which contains both valence and sea quark distributions, will rise
with $Q^2$ at small $x$, where sea quarks dominate, and fall with $Q^2$ at 
large $x$, where valence quarks dominate.

We may quantify these effects using the DGLAP~\cite{DGLAP} formalism
which expresses the evolution of the quark distribution by
{\small \begin{equation}
\frac {d q_i(x,Q^2)} {d \ln Q^2} = \frac {\alpha_s(Q^2)} {2\pi} 
\int_x^1 \frac {dy}{y} \left[\sum_j
q_j(y,Q^2) P_{q_iq_j}(\frac{x}{y}) + g(y,Q^2) P_{q_ig} (\frac{x}{y}) \right]
\label{eq:DGLAPq}
\end{equation}}
\noindent
and the corresponding evolution of the gluon distribution 
(due to contributions from
the diagram in which a quark radiates a gluon
and from a diagram in which a gluon can split into two gluons) by
{\small \begin{equation}
\frac {d g(x,Q^2)} {d \ln Q^2} = \frac {\alpha_s(Q^2)} {2\pi} 
\int_x^1 \frac {dy}{y} \left[
\sum_j q_j(y,Q^2) P_{gq_j}(\frac{x}{y}) + g(y,Q^2) P_{gg} (\frac{x}{y}) \right]
\label{eq:DGLAPg}
\end{equation}}
\noindent
where the `splitting function' $P_{ij}(z)$ represents the probability of 
a parton
(either quark or gluon) $j$ emitting a parton $i$ with momentum fraction $z$
of that of the parent parton, when the scale changes from 
$Q^2$ to $Q^2 + d \ln Q^2$. These splitting functions contribute to the
evolution of the parton distributions at order $\alpha_s$, $\alpha_s^2$, etc.
{\small \begin{equation}
 P_{qq} (z) = P_{qq}^0(z) + \frac{\alpha_s(t)}{2\pi} P_{qq}^1(z) + \dots
\end{equation}}

The diagrams of Fig.~\ref{qcdc} and Fig.~\ref{bgf} give the 
splitting functions which contribute at order $\alpha_s$
{\small \begin{equation}
 P_{qq}^0(z) = \frac {4}{3} \frac {1 + z^2}{1-z}
\label{eq:pqq}
\end{equation}
\begin{equation}
 P_{qg}^0(z) = \frac {1}{2} \left [ z^2+ (1-z)^2 \right]
\label{eq:pqg}
\end{equation}
\begin{equation}
 P_{gq}^0(z) = \frac {4}{3} \frac {1 + (1-z)^2}{z}
\label{eq:pgq}
\end{equation}
\begin{equation}
 P_{gg}^0(z) = 6 \left[\frac{z}{1-z} + \frac{1-z}{z} +  z(1-z) \right]
\label{eq:pgg}
\end{equation}}
\noindent
where the poles at $z=1$ can be regularized by including virtual gluon
diagrams, see reference~\cite{Halzen}. To this order there is no quark flavour
mixing since $P_{q_i q_j} = 0$, unless $i=j$.

Secondly, after we have specified the evolution of the parton 
distributions we must still
relate these parton distributions to the measurable cross-sections and 
structure
functions. Consider, for simplicity, $V^*N$ scattering with singlet exchange.
For the QPM we may calculate the cross-section in terms of a convolution of
the point like $V^*q$ scattering (see Fig.~\ref{qpm}) and the 
parton distribution function
{\small \begin{equation}
 \frac{F_2(x)}{x} = \int dydz\ \delta(x-zy)\ \sigma^{point}(z) \ q(y)
\end{equation}}   
and, in electroproduction for example, we have
{\small \begin{equation}
 \sigma^{point}(z) = e_i^2\ \delta(1-z)
\end{equation}} 
so that
{\small \begin{equation}
 \frac{F_2(x)}{x} = e_i^2\ q(x)
\end{equation}} 
\noindent
where clearly one has to sum over all relevant parton flavours to obtain 
the full result.
  
Now when this is modified in pQCD it amounts to adding to the pointlike
parton cross-section further terms which allow for processes such as 
$\gamma^* q \to g q$ scattering (see Fig.~\ref{qcdc}), so that for 
electroproduction,
{\small \begin{equation}
 \frac{F_2(x)}{x} = \int dydz\ \delta(x-zy)\ q(y) \left[ e_i^2
\delta(1-z) + \sigma^{\gamma^*q\to gq}\right]
\end{equation}}
\noindent 
giving,
{\small \begin{equation}
 \frac{F_2(x)}{x} = \int_x^1 \frac{dy}{y}  q(y) \left[ e_i^2
\delta(1-x/y) + \sigma^{\gamma^*q\to gq}(\frac{x}{y},Q^2)\right]
\label{eq:kernel}
\end{equation}}
\noindent 
where 
{\small \begin{equation}
 \sigma^{\gamma^*q\to gq}(x/y,Q^2) = e_i^2 \frac{\alpha_s}{2\pi}
\left[ P_{qq}(z) \ln \frac{Q^2}{Q^2_0}\right] 
\label{eq:xsec}
\end{equation}}
\noindent 
where $Q^2_0$ is a low momentum cut-off for the integration over the quark
propagator which mediates the process $\gamma^*q \to gq$ (see for 
example~\cite{Renton}) and $\alpha_s$ is the relevant $q\bar qg$ 
coupling constant. 
Thus
{\small \begin{equation}
 \frac{F_2(x)}{x} = e_i^2 \int^1_x \frac{dy}{y} \left[ q(y) + 
\Delta q(y,Q^2) \right]\delta(1-\frac{x}{y})
\end{equation}} 
\noindent
so that
{\small \begin{equation}
 \frac{F_2(x)}{x} = e_i^2 \left[ q(x) + \Delta q(x,Q^2) \right]\ 
=\ e_i^2\ q(x,Q^2)
\end{equation}}
\noindent 
where we have transferred the $Q^2$ dependence in the parton cross-section
into the parton distribution
function $q(x) \to q(x,Q^2)$. We think of $q(x,Q^2)$ as the effective 
distribution seen by the vector boson as it explores a wider range of 
$p_t^2$ within the nucleon when it has larger $Q^2$. We have almost
obtained the form of the DGLAP equations since
{\small \begin{equation}
 \Delta q(x,Q^2) = \frac{\alpha_s}{2\pi} \ln \frac{Q^2}{Q^2_0} \
\int^1_x \frac{dy}{y} q(y) P_{qq}(\frac{x}{y})
\label{eq:dq}
\end{equation}}
\noindent
can be written as 
{\small \begin{equation}
 \frac{d q(x,Q^2)}{d\ln Q^2} = \frac{\alpha_s}{2\pi} \int^1_x \frac
{dy}{y}q(y,Q^2) P_{qq}(x/y)
\label{ldglap}
\end{equation}}
\noindent
and although this result lacks any contribution from  the gluon 
distribution the
extension of the derivation to include the cross-section 
$\sigma(V^*g\to q\bar q)$ follows similarly. Finally,
we have so far treated $\alpha_s$ as a constant in the calculation of the
cross-section $\sigma(V^*q\to gq)$,
all that remains to complete the DGLAP formalism
is to substitute $\alpha_s \to  \alpha_s(Q^2)$.

\subsubsection{$Q^2$ dependence of parton distributions: second order}

First order pQCD introduces $Q^2$ dependence into the 
parton distributions while
preserving the simple expressions for the structure functions in terms of the
parton distributions.
However these expressions require some modification when calculations are 
made to second order. Such calculations require us to use the 2-loop
expression for $\alpha_s$ and the $\alpha_s^2$ contributions from the
splitting functions ($P^1_{qq}(z)$ etc.). The main new feature is that
the separation which we made at first order, such that the $Q^2$ dependence
of the parton cross-section was transferred into the parton distribution
function, cannot be maintained at second order because it cannot be done in 
the same way for all processes. Consider the cross-section for $\gamma^* q$
scattering at second order. 
Equivalent to the kernels of Eqs.~\ref{eq:kernel},~\ref{eq:xsec}
we now have
{\small \begin{equation}
e_i^2 \left[ \delta(1-z) + \frac{\alpha_s}{2\pi} P_{qq}(z) \ln \frac{Q^2}
{Q^2_0} + \alpha_s f^r_{2,3}(z)\right]
\end{equation}}
\noindent
which contains terms $f(z)$ which depend on the symmetry properties of the
particular structure function (e.g.$F_2, xF_3$) being considered and on the
renormalization scheme in which the calculations are being performed. Hence
we should now write 
{\small \begin{equation}
 \frac{F_2(x,Q^2)}{x} = \int dydz\ \delta(x-yz)\ \sigma(z,Q^2,\alpha_s^r)\ 
q^r(y,Q^2)
\end{equation}}
\noindent
where the dependence of $\sigma$ and $q$ on the renormalization scheme which
defines $\alpha_s$ is denoted by the superscript $r$.

The equations which identified the structure functions as sums
over quark distributions  must be modified accordingly to give expressions like
{\small \begin{equation}
\frac{F_2(x,Q^2)}{x} = \int^1_x \frac{dy}{y} \left[ \sum_i C_2\left(
\frac{x}{y},\alpha_s\right)\ q_i(y,Q^2) + C_g\left(\frac{x}{y},\alpha_s\right) 
\ g(y,Q^2)\right]
\label{eq:coeff}
\end{equation}}
\noindent
where the sum is over the appropriate flavours of quarks and 
the coefficient functions 
$C$ represent the appropriate parts of the $V^*$-parton scattering 
cross-sections,
{\small \begin{equation}
C_{2} \left(\frac{x}{y},\alpha_s\right) = 
\sigma_2\left(\frac{x}{y},\alpha_s\right) = e_i^2\left[ \delta
(1-\frac{x}{y}) + \alpha_s(Q^2) f_2(\frac{x}{y})\right]
\label{eq:coeff2}
\end{equation}}
\noindent
and
{\small \begin{equation}
C_{g} \left(\frac{x}{y},\alpha_s\right) = 
\sigma_g\left(\frac{x}{y},\alpha_s\right) = 
\left[ \alpha_s(Q^2 ) f_g(\frac{x}{y})\right]
\label{eq:coeffg}
\end{equation}}
\noindent
Similar expressions obtain for $xF_3$ in terms of $f_3$, but in this case 
the gluon makes no contribution. One may chose to include the $f_2$ term in the
definition of the parton distribution so that the coefficient function 
$C_2$ remains
as a delta function (this is done in the DIS scheme) but then the coefficient
function $C_3$ would have to be more complicated.  
The quark distributions, coefficient functions and splitting functions are
all renormalization scheme dependent, and although certain combinations of
these quantities are not scheme dependent, in general the predictions
for physical quantities such as structure functions  ARE
renormalization scheme dependent when calculated to finite order. 
Of course the result summed to all orders cannot depend on the scheme, 
but we have
to live with finite order calculations and thus
one must define the scheme in which one is working in order to use the same 
parton distributions in different physical processes.

One very important consequence of the fact that at second order
gluon radiation can no longer be accounted for by making the 
quark distributions scale dependent, is that we can no longer picture 
the target 
purely as a sum of spin 1/2 quarks and thus the Callan-Gross relationship, 
$2xF_1 = F_2$, is violated at second order.
A consequence of this violation is that longitudinal structure 
function $F_L$ is no longer zero. It is given in terms of $F_2$
and the gluon distribution as
{\small \begin{equation}
F_L(x,Q^2) = \frac{\alpha_s}{\pi}\left[\frac{4}{3}\int^1_x \frac{dy}{y}\left(
\frac{x}{y}\right)^2
F_2(y,Q^2) + 2 c \int^1_x \frac{dy}{y}\left(\frac{x}{y}\right)^2 
(1-\frac{x}{y}) yg(y,Q^2)\right]
\label{eq:flqcd}
\end{equation}}
\noindent
where $c = \sum e^2_i$ for charged lepton scattering and $c = 4$ for neutrino 
scattering.    
At small $x$ ($x \leqsim 10^{-3}$) the dominant contribution comes from the 
glue regardless of the exact shape of the gluon distribution. In fact the
weighting function in the integral over the gluon distribution  
approximates to a $\delta$ function~\cite{AMCS}, 
such that a measurement of $F_L(x,Q^2)$
is almost a direct measure of the gluon distribution $yg(y,Q^2)$ for $y
\simeq 2.5x$.

\subsubsection{Connection to the moment approach}
\label{sec:moments}

The theory of QCD is more formally derived from the 
Operator Product Expansion and
the Renormalization Group Equation to give predictions in terms of
the moments of the structure functions. It is not the purpose of the 
current review to cover this formal approach.  Nice expositions 
are given, for example, in references~\cite{Buras,roberts,yndurain}. 
We shall require only some concepts and terminology for our future reference. 

\begin{figure}[ht]
\centerline{\psfig{figure=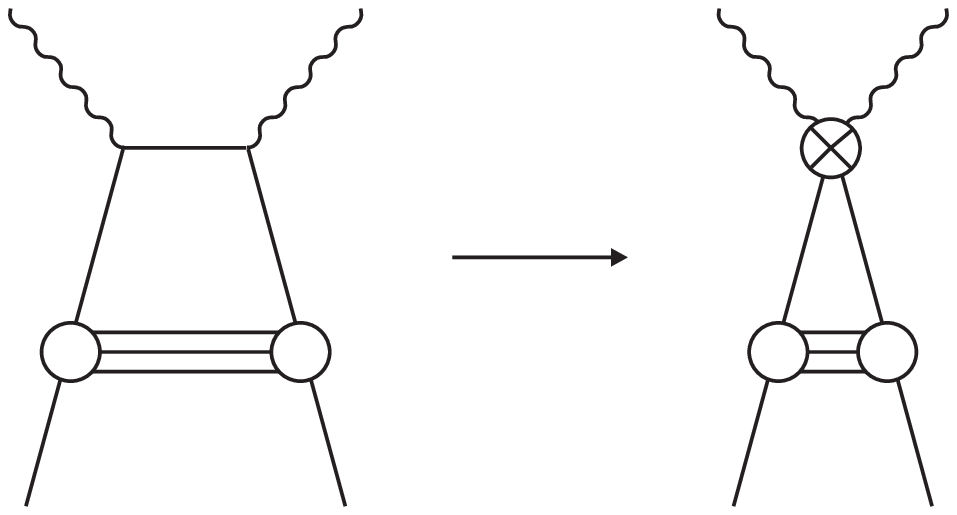,width=9cm,height=5cm}}
\fcaption{Schematic diagram of the relationship between the handbag diagram
and its operator structure}
\label{operatorQPM}
\end{figure}

\begin{figure}[ht]
\centerline{\psfig{figure=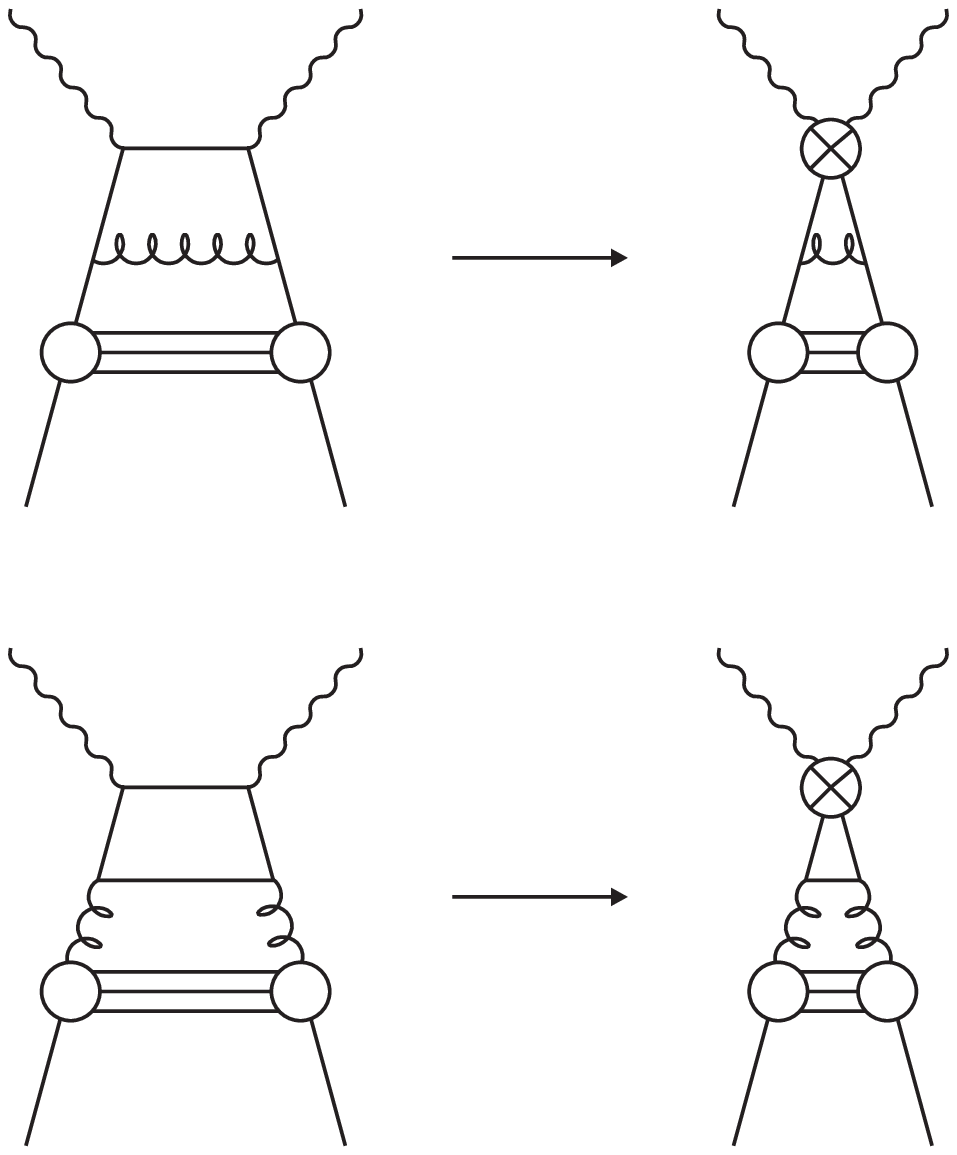,width=9cm,height=10cm}}
\fcaption{Schematic diagram of extensions to the handbag diagram from pQCD
and their relationship to quark and gluon operators}
\label{operatorQCD}
\end{figure}

The deep inelastic scattering process is calculated by using the Optical
Theorem to express it in terms of the elastic process $V^*N \to V^*N$.
This involves a product of vector boson currents (see Fig.~\ref{handbag}). 
The Operator Product Expansion consists of expanding this current product
$J(z)J(0)$ in terms of a sum over operators of
different spin and type (e.g. quark or gluon, singlet or non-singlet). 
For example for the electromagnetic current we have terms such as
{\small \begin{equation}
 \frac{2}{3} \bar \psi_u \gamma_{\mu}  \psi_u
\end{equation}}
\noindent
So that the product $J(z)J(0)$ contains terms like
{\small \begin{equation}
 \frac{4}{9} \left[ \bar \psi_u(z) \gamma_{\mu}  \psi_u (z) \bar \psi_u(0)
\gamma_{\mu} \psi_u(0)\   + ... \right] 
\end{equation}}
\noindent
which contract to
{\small \begin{equation}
 \frac{4}{9} \left[ \bar\psi_u(z) \gamma_{\mu}\psi_u(0) \right] 
\end{equation}}
\noindent
which have the form of a sum of coefficients (like $q_u^2=4/9$)
and operators (like $\bar \psi\gamma_{\mu}\psi$). These operators can
be calculated just like any other Green's Function.
Rather than trying to follow the development mathematically 
we illustrate it
schematically in Fig.~\ref{operatorQPM} and Fig.~\ref{operatorQCD} 
which show how the handbag diagram of Fig.~\ref{handbag} is expressed 
in terms of operators, and how this concept is extended when the handbag 
diagram is modified in pQCD. 
\noindent
\noindent
To obtain the deep inelastic scattering cross-section we must take the 
nucleon matrix element of the current product, and hence of the operator 
product expansion.
The operator matrix elements are not calculable within perturbative QCD, 
however their coefficient functions are, and their $Q^2$ dependence is given
by a renormalization group equation. 

The results of this formal approach show that the $N$th 
order moment of the structure function projects out the spin $N$ contribution
to the operator product expansion, e.g. for $F_2$ 
{\small \begin{equation}
 M_N(Q^2)\ =\ \int^1_0 x^{N-1}\ \frac{F_2(x,Q^2)}{x} = \sum_a\  
 A_N^a(Q^2_0)\ C_N^a(\frac {Q^2}{Q^2_0},\alpha_s)
\label{eq:moment}
\end{equation}}
\noindent
where the sum is over the types of operator which can contribute (gluon,
(non)-singlet quark). The notation $A_N^a$ represents the operator matrix
elements and $C_N^a$ represents their coefficient functions. 
In the case when $a$
represents a non-singlet quark operator the $Q^2$ dependence of the coefficient
function may be written quite simply as
{\small \begin{equation}
 C_N^a (\frac{Q^2}{Q^2_0},\alpha_s) = C_N^a(1,\alpha_s)\ \exp \left[
-\int^{\alpha_s(Q^2)}_{\alpha_s(Q^2_0)} d\alpha \frac{\gamma_N^a(\alpha)}
{\beta(\alpha)} \right]
\end{equation}} 
\noindent
where $\gamma_N^a$ is a function known as the anomalous dimension 
of the non-singlet
quark operator, $\beta$ is the QCD $\beta$ function and $Q^2_0$ is 
an arbitrary 
starting point. In the case of singlet quark and gluon operators 
we have mixing, so
that the above equation must be formulated as two coupled equations relating to
a 2 by 2 matrix of anomalous dimensions $\gamma^{qq}, \gamma^{qg}, \gamma^{gq},
\gamma^{gg}$. 

This formalism may be related to the DGLAP approach as follows.
The matrix element $A_N^a$ of the spin $N$ operator is identified with the 
$N$th moment of the corresponding parton distribution ($q_N^a$) 
evaluated at the same scale
($Q^2_0$) and the $Q^2$ dependence of the coefficient function $C_N^a$ 
is transferred 
to the parton distribution such that
{\small \begin{equation}
 q_N^a(Q^2) =q_N^a(Q^2_0) exp \left[ -\int^{\alpha_s(Q^2)}_{\alpha_s(Q^2_0)} 
d\alpha \frac{\gamma_N^a(\alpha)}{\beta(\alpha)}\right]
\end{equation}}
\noindent
for non-singlet quarks (and two similar coupled equations are necessary to 
express the mixing of singlet quarks
and gluons). Then Eq.~\ref{eq:moment} becomes
{\small \begin{equation}
 M_N(Q^2)\ =\ \sum_a\ q_N^a(Q^2)\ C_N^a(1,\alpha_s)
\end{equation}}
\noindent
and this is precisely what we will get if we take moments on either side of 
Eq.~\ref{eq:coeff}, provided that $C_N^a$ denotes the moments of the $x$ space 
coefficient functions and $q_N^a$ denotes moments of the appropriate 
type of parton 
distributions (quark-singlet or gluon). The $Q^2$ dependence of the 
moments of the 
parton distributions can then be obtained by taking moments of the 
DGLAP equations
{\small \begin{equation}
\frac {dq_N(Q^2)} { d\ln Q^2} = \frac {\alpha_s}{2\pi}\left[ 
\frac{-\gamma_N^{qq}}{4} 
\ q_N(Q^2) + \frac{-\gamma_N^{qg}}{4}\ g_N(Q^2) \right]
\end{equation}} 
{\small \begin{equation}
\frac {dg_N(Q^2)} { d\ln Q^2} = \frac {\alpha_s}{2\pi}\left[ 
\frac{-\gamma_N^{gq}}{4} 
\ q_N(Q^2) + \frac{-\gamma_N^{gg}}{4}\ g_N(Q^2) \right]
\end{equation}} 
\noindent
 where
{\small \begin{equation}
  \frac{-\gamma_N^a}{4} = \int_0^1 dz\ z^{N-1} P_{a}(z) 
\end{equation}}
\noindent
defines  the relationship of the  anomalous dimension function 
to the corresponding splitting function.

\subsection {Higher twist}\label{sec:hitwist}
\begin{figure}[ht]

\centerline{\psfig{figure=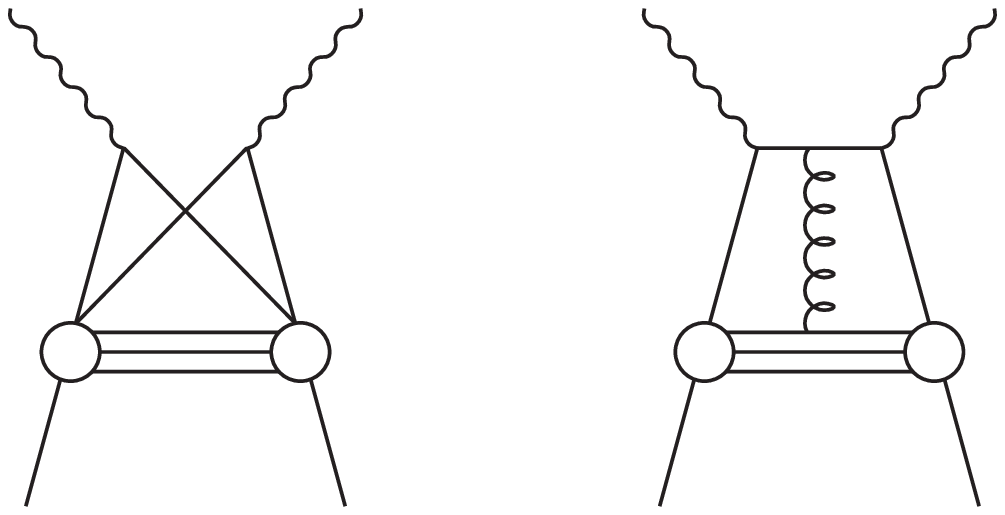,width=9cm,height=5cm}}
\fcaption{Alternatives to the handbag diagram for higher twist operators}
\label{operatorht}
\end{figure}

\noindent
So far we have only considered the predictions of QCD at
leading twist. Twist refers to the (dimension - spin)
of the operators entering into the operator product expansion. The dominant 
contributions are twist = 2 and involve the sort of operators depicted in
Fig.~\ref{operatorQPM} and Fig.~\ref{operatorQCD}. 
Higher twist ($\geq 4$) operators relate to diagrams like those shown in
Fig.~\ref{operatorht}. They are suppressed by powers of 
$1/Q^2$ in comparison to the leading twist diagrams and so 
they may become important at low $Q^2$, as we approach the region where 
perturbative QCD becomes inapplicable (since the coupling constant $\alpha_s$
becomes too large). In the same
kinematic region target mass effects become important. The identification 
of $x$ with the fraction of the proton's momentum taken by the struck
quark cannot be maintained when $Q^2 \simeq M_N^2$, and corrections to
the formulae are necessary~\cite{tm}. Since these involve powers of 
$1/Q^2$ they are often called kinematic higher twist effects, 
whereas terms coming
from operators of higher twist are called dynamic higher twist effects.

\noindent
Conventional higher twist effects are usually only important at 
higher $x$, for the following reason.
At twist=2 there are only 2 operators for any value,
$N$, of the spin, but at twist=4 there are $N$ quark-gluon operators and
$N^2$ 4-quark operators. Hence, because there are at least $N$ times as many 
operators contributing to the $N$th moment of the structure function, we expect
its $Q^2$ dependence to be modified by a multiplicative factor of the form
$( 1 \ + \ C \frac{N}{Q^2})$
and correspondingly the structure function would be modified by a 
factor of the form
$( 1 \ + \ C \frac{1}{Q^2}\frac{1}{(1 - x)})$.
Thus such higher twist effects can  be important at low $Q^2$ and
large $x$, i.e. when $W^2$ is small and the contributions of specific exclusive
processes to the inelastic cross-section become important.

These dynamic higher twist effects have been estimated 
only for some specific cases ~\cite{1987}. Their full calculation awaits a 
solution to the problem of confinement since they clearly involve 
reinteraction of the struck quark with the proton remnant. We would then
be able to calculate the 
nucleon matrix elements of all operators and assess their relative
contributions in different kinematic regions. 
Present calculations~\cite{Miramontes,GLSnew} have included 
contributions to the GLS and Bjorken 
sum rules and to the longitudinal structure function $F_L$.
The latter is particularly interesting since only one twist=4 operator is
involved and hence the higher twist contribution is not associated only with
high $x$, see ref.~\cite{Harindranath}.

There has been some progress on calculating higher twist terms recently, from
the renormalon approach~\cite{webberon,das}. When calculating physical
quantities within QCD the perturbation series will eventually breakdown. 
Typically such a divergence may come from renormalon graphs - chains of vacuum
polarization bubbles on the gluon line. However, the perturbative series which
describes such a graph does give a good approximation to the true value if it
is truncated appropriately, and one may make an estimate of the error, or
ambiguity, on the truncated result. This ambiguity decreases like $Q^{-2p}$ 
with increasing $Q^2$, where $p$ depends on the particular quantity being 
calculated. This implies that the perturbative prediction should be
supplemented by non-perturbative information from the higher twist 
contributions. 
Thus by studying the onset of the breakdown in perturbation theory via 
renormalon graphs, one gains information on the $Q^2$ dependence of the
 non-perturbative terms which are necessary. This
is very useful when studying processes for which one has no handle on such
terms from the operator product expansion. Even for DIS where the OPE 
controls the form of the higher twist terms, this approach may enable us to 
go further and predict the $x$ dependence of the dominant higher twist 
contributions. Such predictions give successful descriptions of higher twist
contributions to $F_2$ and $xF_3$ (see Sec.~\ref{sec:pdf}).

Recently it has been realised that higher twist terms may also be important
at very low $x$, this will be considered further in Sec.~\ref{sec:lowx}.

\subsection{The leading log approximation: LLA, NLLA, DLLA}\label{sec:LLA+}
\noindent
There is a very important aspect which has been glossed over in the 
preceding subsections. When we make the replacement 
$\alpha_s \to \alpha_s(Q^2)$ to make the DGLAP equations equivalent to the 
formal theory we are effectively making the virtual boson-parton 
cross-sections (like $V^*q \to qg$) take a $\ln(\ln Q^2)$ dependence 
rather than a $\ln Q^2$ dependence.
Thus the contribution of these cross-sections to the parton distribution 
(which we absorbed into its $Q^2$ dependence c.f. 
Eqs.$~(\ref{eq:xsec}-\ref{eq:dq})$) is given by 
{\small \begin{equation}
\Delta q(x,Q^2) \sim\ \alpha_s\ ln(Q^2)
\end{equation}}
\noindent
but from  Eq.~\ref{eq:alphas1} we know that 
$\alpha_s \sim 1/\ln Q^2 $, so the
contribution is of order $O(1)$, rather than of order $O(\alpha_s)$.
Hence we must go beyond first order and sum terms of the type 
$\alpha_s^n (\ln Q^2)^n$ (the `leading logs') to all orders. 
This is what is actually done by the Renormalization Group Equation in
the formal approach and by the DGLAP equations when making the substitution
$\alpha_s \to \alpha_s(Q^2)$. 
It is known as the Leading Log Approximation (LLA). The extension to 
second order includes terms of the type
$\alpha_s^n (\ln Q^2)^{n-1}$ and is known as the Next-to-Leading Log 
Approximation (NLLA). 

The extension of first order results by the LLA is
often referred to as leading order (LO) and the extension of second order 
results by the NLLA is often referred to as next-to-leading order (NLO), 
however one has to be careful about which quantity is being calculated.
An LO result for the structure functions
$F_2$ or $xF_3$ is obtained by using the zeroth order result for the 
coefficient functions $C_2, C_3$ (i.e. that they are delta
functions) together with the order $\alpha_s$ contribution from the splitting
functions ($P^0_{qq}$ etc.). An NLO result is obtained by using the first
order result for the coefficient functions (see 
Eqs.~\ref{eq:coeff2},~\ref{eq:coeffg}) together with the order $\alpha_s^2$
contributions from the splitting functions ($P^1_{qq}$ etc.). This is what we 
shall normally mean by the terminologies LO and NLO. However the
structure function $F_L$, for example, 
involves coefficient functions which are zero at
zeroth order so that the lowest order result for $F_L$ involves calculations
to the same orders in $\alpha_s$ as the NLO result for $F_2$.
Authors differ as to whether such a result for $F_L$ is termed LO or NLO. 
In this review we shall term such a result lowest order and specify  
the order
of $\alpha_s$ involved in the calculation of the relevant coefficient 
function if  ambiguity arises. 

\begin{figure}[ht]
\centerline{\psfig{figure=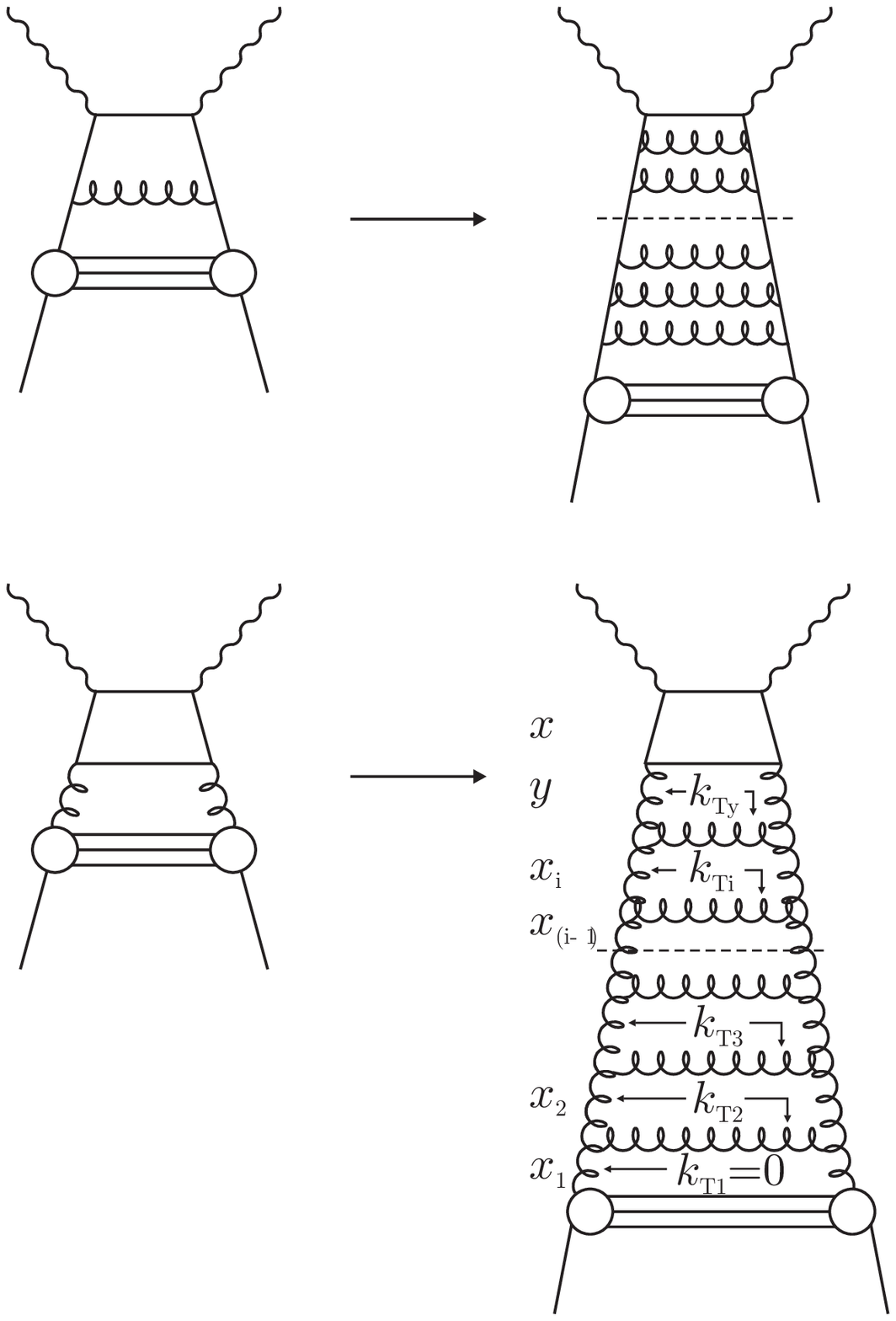,width=9cm,height=11cm}}
\fcaption{Schematic representation of ladder diagrams contributing to 
$V^*N\to V^*N$ in the Leading Log Approximation}
\label{ladder}
\end{figure}

\noindent
Diagrammatically we are extending the 
diagrams of Fig.~\ref{operatorQCD} to include many `rungs' of gluons as 
shown in Fig.~\ref{ladder}. The dominant contributions to these ladder 
 diagrams come from ordered momenta so that the virtual boson sees 
successive layers of off-mass shell partons all of which contribute to 
the cross-section. Thus if we label the t-channel partons (struts) from 
nucleon to boson as
$1$ to $i$, we have $x_1 > x_2...> x_i$ for longitudinal momentum fractions 
and $k_{T,1}^2 < k_{T,2}^2 ...< k_{T,i}^2$ for transverse momenta as we go up 
the ladder.

At large $Q^2$ the dominant contribution (LLA) has strong ordering in $k_T^2$,
 $k_{T,1}^2 << k_{T,2}^2 ...<< k_{T,i}^2$, as the ladder 
diagrams become dominated by collinear gluon emission~\fnm{i}\fnt{i}{~The 
first parton of the chain 
with longitudinal momentum fraction $x_1$ is considered to be collinear with 
the proton so that it has $k_{T,1}=0$, and the second parton of the chain
has $k_{T,2}$ equal to the $k_T$ of the first emitted parton rung. When the
ladder becomes dominated by collinear gluon emission one can think of all
successive emissions as taking place off partons which are collinear with
the proton so that the $k_T$'s of the t-channel gluons (struts) are equal
to those of the emitted partons (rungs) as indicated.}.
The NLLA contributions 
correspond to the case when a single pair of gluons are emitted without strong
$k_T$ ordering, and hence give a power of $\alpha_s$ unaccompanied by 
$\ln Q^2$.

\begin{figure}[ht]
\centerline{\psfig{figure=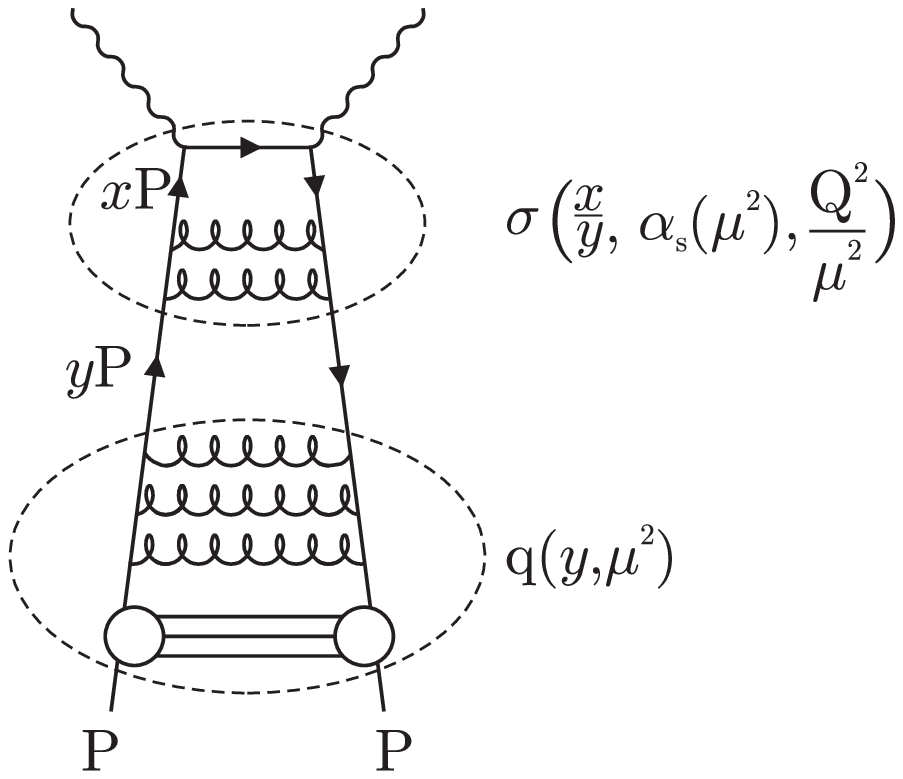,width=7cm,height=6cm}}
\fcaption{Schematic representation of collinear factorization in the
gluon ladder diagram}
\label{dickladder}
\end{figure}

We may generalize Eqs.~\ref{eq:coeff},~\ref{eq:coeffg} to express 
collinear factorization
{\small \begin{equation}
 F(x,Q^2) = \int^1_x \frac{dy}{y} \sigma(\frac{x}{y},\alpha_s(\mu^2),
\frac{Q^2}{\mu^2})\ q(y,\mu^2)
\end{equation}}
\noindent
where $F(x,Q^2)$ is a generalized structure function, $q(y,\mu^2)$ is a 
generalized parton density, and $\sigma(z,Q^2,\mu^2)$
is the cross-section for that parton scattering elastically from a 
virtual boson of virtuality $Q^2$. The meaning of the arbitrary scale $\mu$ is 
illustrated by Fig.~\ref{dickladder}. 
It represents the point in the ladder where we chose to separate the parton
distributions from their interaction cross-sections. This choice is a part of 
the renormalization scheme dependence, which is known as the factorization 
scheme. It essentially defines the point at which we consider that the 
proton `ends'. Collinear logarithmic singularities which arise from gluon
emission in the partonic subprocess are absorbed into the parton densities
(which then run with $Q^2$). The factorization scheme defines the point at 
which this happens.
 So far we have made the choice $\mu^2=Q^2$, taking the factorization at
the top of the gluon chain. Of course the observable 
structure function must be independent of this choice.

At small $x$ and at large $x$ there are large corrections to the conventional
leading $\ln Q^2$ calculations which imply that $\mu^2 = Q^2$ may not be the 
best choice of factorization scale. In both cases this arises when there is a 
second large scale in consideration. At large $x$, $Q^2 \gg W^2$, but $Q^2$ and
$W^2$ can both still be large. There are large corrections of the form
$\alpha_s(Q^2)\ln(1-x)$, since as $x \to 1$ there is less and less phase space 
for the emission of real gluons ($Q^2(1-x_i)$ is the upper limit on the 
invariant mass of the emitted gluon) and thus real gluon emission cannot cancel
virtual gluon contributions. It has been suggested that these corrections may 
be accounted for by replacing $\alpha_s(Q^2)$ by $\alpha_s(Q^2(1-x))$ or by 
$\alpha_s(W^2)$~\cite{roberts}. There is interesting new work in this area by
Sterman and collaborators~\cite{Ster97}. We do not discuss such modifications 
further since there is little data at very high $x$ on nucleon targets.

At small $x$, $W^2 \gg Q^2$, but $Q^2$ and $W^2$ can both still be large. 
There 
are large corrections of the form $\alpha_s(Q^2)\ln(1/x)$. In conventional
calculations the ordering in $x$ along the gluon ladder  becomes strong 
$x_1>>x_2...>>x_i$, and we
then obtain the Double Leading Log Approximation (DLLA) where the 
cross-sections are dominated by terms in $\ln(\ln Q^2 )\ln(1/x)$. 
At very low $x$
it may be necessary to go beyond the conventional LLA, NLLA or DLLA and
consider terms which are leading in $\ln(1/x)$ regardless of 
whether or not they
are leading in $\ln Q^2$. This is the kinematic
region where HERA has probed for the first time and we will consider 
such extensions
to conventional pQCD in detail in Sec.~\ref{sec:lowx}.

\subsection{Heavy quarks}
\label{sec:heavyq}
\noindent

Firstly we consider the behaviour of $\alpha_s$ across flavour thresholds. At
second order $\alpha_s$ is defined in terms of $\beta_0$ and $\beta_1$ through
Eq.~\ref{eq:alphas2}. These in turn depend on the number of flavours, $n_i$. 
Thus, for a given value of
$\Lambda$, the running coupling $\alpha_s(Q^2)$ should reflect 
the appropriate
number of active flavours. For $Q^2 \ll m_c^2$ the function $\alpha_s(Q^2)$
follows a curve appropriate to 3-flavours, whereas for 
$m_b^2 \gg Q^2 \gg m_c^2$ it
should follow a curve appropriate to 4-flavours, 
and for $m_t^2 \gg Q^2 \gg m_b^2$ it
should follow a curve appropriate to 5-flavours. 
In the threshold regions, $Q^2 \sim
m_c^2$ and $Q^2 \sim m_b^2$ one must make a smooth transition between different
curves. The widely used prescription of Marciano~\cite{Marciano} 
is basically to 
match the values of $\alpha_s$ at the thresholds 
$Q^2 = m_c^2$ and $Q^2 = m_b^2$.
Explicitly, if one uses $\Lambda$ for 4 flavours as our 
reference parameter,
then defining $\alpha_s(Q^2,n_i)$ to be the solution of 
Eq.~\ref{eq:alphas2} for 
$n_i$ flavours, we have
{\small \begin{equation}
\alpha_{s,4}(Q^2) = \alpha_s(Q^2,4)
\end{equation}}
\noindent 
for 4 flavours and
{\small \begin{equation}
\alpha_{s,5}^{-1}(Q^2) = \alpha_s^{-1}(Q^2,5) + \alpha_s^{-1}(m^2_b,4) 
- \alpha_s^{-1}(m^2_b,5)
\end{equation}}
\noindent 
for 5 flavours and
{\small \begin{equation}
\alpha_{s,3}^{-1}(Q^2) = \alpha_s^{-1}(Q^2,4) + \alpha_s^{-1}(m^2_c,4)
- \alpha_s^{-1}(m^2_c,3)
\end{equation}}
\noindent 
for 3 flavours. One could chose to express $\alpha_s$ in terms of 
$\Lambda$ for 5 flavours or 3 flavours, or one could
chose different values for the thresholds, such as 
$Q^2 =4m_c^2, 4m_b^2$. It is necessary to specify the procedure being used when
comparing results. Values of $\Lambda$ quoted will refer to 4 flavours unless
otherwise stated.

Secondly we consider the theoretical issues to be resolved when considering 
heavy quark production in DIS.
These are concerned with the non-trivial question of when can a heavy
quark be treated as a massless parton~\cite{hws_cc}.
A range of options has been explored. We  
consider the contribution of the charmed quark to the structure function, the
contributions of bottom and top may be treated similarly. At one
extreme charm has been treated as a massless parton. For example, the MRS 
team~\cite{MRSD,MRSA,MRS96} assume that
$c(x,Q^2) = 0$ for $Q^2\leq \mu^2_c$ (where $\mu_c = O(m_c)$) and 
generate the charm
parton distribution for larger $Q^2$ by the splitting $g \to c \bar c$, 
using the usual NLO DGLAP equations for massless
partons. This may be termed a zero mass variable flavour number scheme 
(ZM-VFN). One has $n_i = 3 + \theta(Q^2-\mu^2_c)$ and the charm 
contribution to
the structure function is given by 
$F_2^{c\bar{c}}=\frac{8}{9}xc(x,Q^2)$ at LO. The
value of $\mu_c$ was adjusted to give a satisfactory description of the 
EMC charmed
structure function data. It turns out that the shape of the charmed sea 
generated 
by this procedure is similar to that of the non-charmed sea, 
with normalization 
given by $2\bar c(x,Q^2_0) = \delta S(x,Q^2_0)$, $\delta=0.02$ at 
$Q^2_0 = 4\,$GeV$^2$~\fnm{j}\fnt{j}{~The latest MRS fits~\cite{MRS96} 
have a starting value of $Q^2$ 
which is less than $\mu^2_c$ and thus require an additional phenomenological 
threshold factor to ensure that the charm density turns on smoothly.}. 

Obviously, such a procedure cannot give a good description of the 
charm contribution 
in the threshold region. In fact one can create a $c\bar c$ pair by 
boson-gluon 
fusion (BGF), when $W^2 \geq 4m_c^2$, and since $W^2=Q^2(1-x)/x + M_N^2$, 
this can be well below the $Q^2$ threshold,
$Q^2 \ll m_c^2$,  at small $x$. Hence, at the other extreme, the GRV 
team~\cite{GRV91,GRV94,GRS}
treat charm as a heavy quark, dynamically generated by the BGF process. 
This means that there is no concept of a charmed
parton distribution we have a fixed flavour number scheme (FFN) 
with $n_i = 3$. 
The contribution to $F_2$ comes  through
BGF as embodied in the DGLAP equations with massive quark coefficient 
functions. In this picture the lowest order result for $F_2^{c\bar c}$
arises at order $\alpha_s$ in the coefficient functions~\cite{bgf_lw}. 
However calculations have now been made to next-to-lowest order 
(order $\alpha_s^2$ in the coefficient functions) by Laenen et al~\cite{nlocc}.
$F_2^{c\bar{c}}$ is given in terms of the massless parton distributions
by
\begin{eqnarray}
F_2^{c\bar{c}}(x,Q^2)&=&{Q^2\alpha_s\over 4\pi^2m^2_c}\int_{ax}^1
{dy\over y}\left[c^{(0)}_{2,g}+4\pi\alpha_s\left\{c^{(1)}_{2,g}+
\bar{c}^{(1)}_{2,g}\ln{\mu^2\over m^2_c}\right\}\right]
e^2_c yg(y,\mu^2) \nonumber \\
& & \mbox{}+{Q^2\alpha_s^2\over \pi m^2_c}\int_{ax}^1{dy\over y}
\left[c^{(1)}_{2,q}+\bar{c}^{(1)}_{2,q}\ln{\mu^2\over m^2_c}\right]e^2_c
\sum_i yq_i(y,\mu^2) \nonumber \\
& & \mbox{}+{Q^2\alpha_s^2\over \pi m^2_c}\int_{ax}^1{dy\over y}d^{(1)}_{2,q}
\sum_i e^2_i yq_iy,\mu^2),
\label{eq:bgf_nlo}
\end{eqnarray}
where $a = 1 + 4m_c^2/Q^2$, 
$e_i (i=u,d,s,c)$ are the quark charges in units of the proton charge, $m_c$
is the charm quark mass and $\mu$ is a mass factorization scale (which
has been put equal to the renormalization scale). The coefficient
functions, $c^{(\ell)}_{2,k},~\bar{c}^{(\ell)}_{2,k},~d^{(\ell)}_{2,k}$ 
($\ell=0,1; k=g,q$), are calculated in the $\MSB$ scheme and are 
functions of the variables 
$\xi=\frac{Q^2}{m^2_c},~\eta=\xi\frac{1-z}{4z}-1$, as 
specified in ref.~\cite{nlocc}. 
The strong coupling $\alpha_s$ is taken as a function of $\mu^2$. 
\par
\begin{figure}[htbp]
\vspace*{13pt}
 \centerline{\psfig{figure=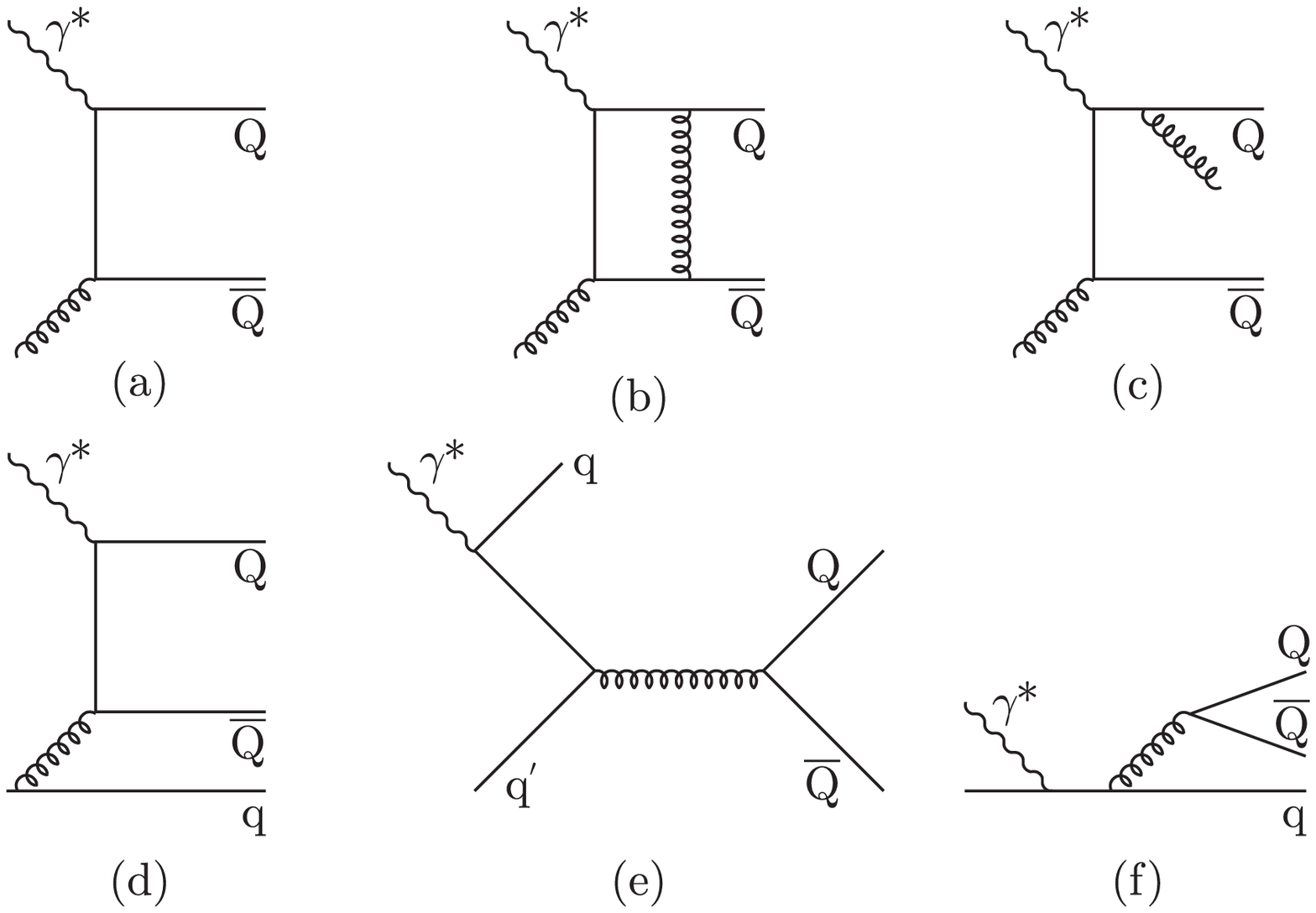,width=7cm,height=6cm}}
\fcaption{The DIS boson-gluon fusion process: (a) order $\alpha_s$ diagram;
(b) one loop gluon diagram at order $\alpha_s^2$, 
there are another 7 similar diagrams 
plus the one loop quark insertion into the gluon propagator; (c) gluon
bremsstrahlung at order $\alpha_s^2$ -- plus another 3 similar diagrams; 
(d),(e) and (f)
are the $\alpha_s^2$ diagrams which involve the light quarks in the target.}
\label{fig:bgf}
\end{figure}

\noindent
In Eq.~\ref{eq:bgf_nlo} the terms have been grouped according to their
origin. At first order in $\alpha_s$ only the gluon distribution function
is involved through the coefficient function $c^{(0)}_{2,g}$ 
(see Fig.~\ref{fig:bgf}{a}). The gluon
one loop corrections and bremsstrahlung diagrams (see 
Fig.~\ref{fig:bgf}(b) and (c)) are brought in at order $\alpha_s^2$ 
through the coefficient
functions $c^{(1)}_{2,g},~\bar{c}^{(1)}_{2,g}$. At this order c\=c production
also occurs from light quarks through the processes shown in
Fig.~\ref{fig:bgf}(d)(e)(f): in (d) the virtual photon produces a 
c\=c pair with rate given by the sum of the light quark densities and 
the coefficient functions $c^{(1)}_{2,q},~\bar{c}^{(1)}_{2,q}$; in
(e) and (f) the virtual photon first interacts with a light quark which later
produces a c\=c pair via a gluon with rate given by the
charge-weighted sum of the light quark densities and the function  
$d^{(1)}_{2,q}$.  
\noindent
\par\noindent
A number of numerical studies of Eq.~\ref{eq:bgf_nlo} have been made by
the original authors~\cite{nlocc}, by Vogt~\cite{vogt_cc} and by Laenen
et al~\cite{hws_cc} for the 1996 HERA workshop. What all these studies
show is that c\=c production in DIS is largely determined by the gluon
density in the proton and hence is largest at small $x$ -- at 
$Q^2\sim 100\,$GeV$^2$ and $x\sim 0.0001$, $F_2^{c\bar{c}}$ is more than
30\% of $F_2$. The scale dependence of calculations performed 
to next-to-lowest order is under control: allowing
$\mu$ to vary in the range $m_c\leq \mu \leq 2\sqrt{Q^2+4m_c^2}$, gives
variations in $F_2^{c\bar{c}}$ less than 10\% for $x\leq 0.01$ compared to
more than 20\% for calculations made to lowest order.
The largest theoretical uncertainty comes from the
value of $m_c$, a $10$\% change in $m_c$ produces a $15-25$\%
change in $F_2^{c\bar{c}}$ at a $Q^2$ of $10\,$GeV$^2$ (the magnitude
of the effect diminishes with increasing $Q^2$). If a scale of
$\mu=\sqrt{Q^2+4m_c^2}$ is chosen, the gluon is sampled over a rather 
narrow region in the integral at a fixed $(x,Q^2)$. All this points to
$F_2^{c\bar{c}}$ being rather a good experimental handle on the gluon
density. This is discussed further in Sec.~\ref{sec:gluon}.

Fixed flavour number schemes are well suited to the description of
differential distributions for charm production just above threshold.
As $Q^2$ increases into the asymptotic region, $Q^2 \gg 4m_c^2$, 
one expects that 
the charmed quark can be treated as a massless parton. However, the
coefficient functions of Eq.~\ref{eq:bgf_nlo} do not switch smoothly 
from a heavy quark description to a massless partonic description. 
The problem is the presence of 
large logarithms, $\ln{Q^2\over m_c^2}$, in the NLO BGF coefficient 
functions. In the usual NLO DGLAP calculation of
the splitting functions the quarks are taken to be massless and the
corresponding large logarithms are summed and absorbed into the definition
of the parton distributions (thus giving rise to their scale dependence).
If we want to have a consistent description of heavy quark production from the
threshold region to the asymptotic region we must find a way to include the 
heavy quark mass in the DGLAP splitting functions and
coefficient functions in such a way as not to destroy the 
partonic interpretation, 
so that one can define universal heavy quark distributions which can be 
used in 
different processes. 

There has been progress in this direction recently by Lai and 
Tung~\cite{LaiTung}
and Martin et al (MRRS)~\cite{MRRS}. 
The contribution of charm to $F_2$ at NLO is 
given by a convolution of parton distributions and coefficient functions as
{\small \begin{equation}
\frac{F_2^{c \bar{c}}}{x} = \frac{8}{9}\int_x^1 \frac{dy}{y}
\left[C_c(\frac{x}{y},
Q^2,\mu^2) c(y,\mu^2) + C_g(\frac{x}{y},Q^2,\mu^2) g(y,\mu^2)\right]
\end{equation}}
\noindent
where $C_c = C_c^0 + \frac{\alpha_s}{4\pi} C_c^1$ and 
$C_g = \frac{\alpha_s}{4\pi} C_g^1$. In the fixed flavour number scheme which 
we outlined above one would have $c(y,\mu^2) = 0$ by definition and 
$C_g^1 = C_g^{BGF}$ given by the
boson-gluon fusion cross-section. Instead of this, the new approaches use
general mass variable flavour number schemes (GM-VFN), 
in which, for $Q^2 \leqsim m^2_c$,
$c(x,Q^2) = 0$ and $F_2^{c\bar{c}}$ is described
completely by BGF via the $C_g \otimes g$ convolution (flavour creation), 
but for $Q^2 \geq m_c^2$ a charm distribution is generated from the $P_{cg}$ 
splitting $g \to c \bar c$, and an additional contribution to $F_2^{c\bar c}$ 
from $\gamma c$ scattering via the $C_c \otimes c$ convolution 
(flavour excitation) results, and rapidly becomes 
dominant.
Once the charm distribution has been created
 part of the BGF cross-section is automatically generated by the evolution
of the charm distribution. To avoid double
counting one must subtract from $C_g$ the contribution which is generated 
this way
such that, $C_g^1 = C_g^{BGF} - \Delta C_g$, where 
$\Delta C_g \sim P_{cg}^0 \otimes C_c^0$. Thus the troublesome
$\ln(Q^2/m_c^2)$ terms in $C_g^{BGF}$ 
are cancelled by similar terms in $\Delta C_g$ and the analysis is
applicable from the threshold region into the asymptotic region.
 
Lai and Tung implement a GM-VFN to LO as first proposed by 
Aivasiz et al~\cite{varflav} (ACOT), in such a way that the massive terms are
incorporated in the coefficient functions so that the charmed partons 
still evolve like massless partons. There has been recent progress on the NLO
implementation by Schmidt~\cite{ScmidtC} and Buza et al~\cite{Buza}.
MRRS chose to calculate massive splitting functions which generate
partons which may be used with the conventional $\MSB$ 
coefficient functions at NLO. These two approaches are compared by Olness and
Scalise~\cite{Olsca}. Although the detailed implementations differ, 
the resulting parton distributions are
very similar when evolved to $Q^2 = 25\,$GeV$^2$. There is not yet a consensus
on the definitive way to implement a GM-VFN scheme. Roberts and Thorne 
have recently made substantial progress~\cite{DickRob}.


\section{The experiments}
\begin{figure}[htbp]
\vspace*{13pt}
\begin{center}
\psfig{figure=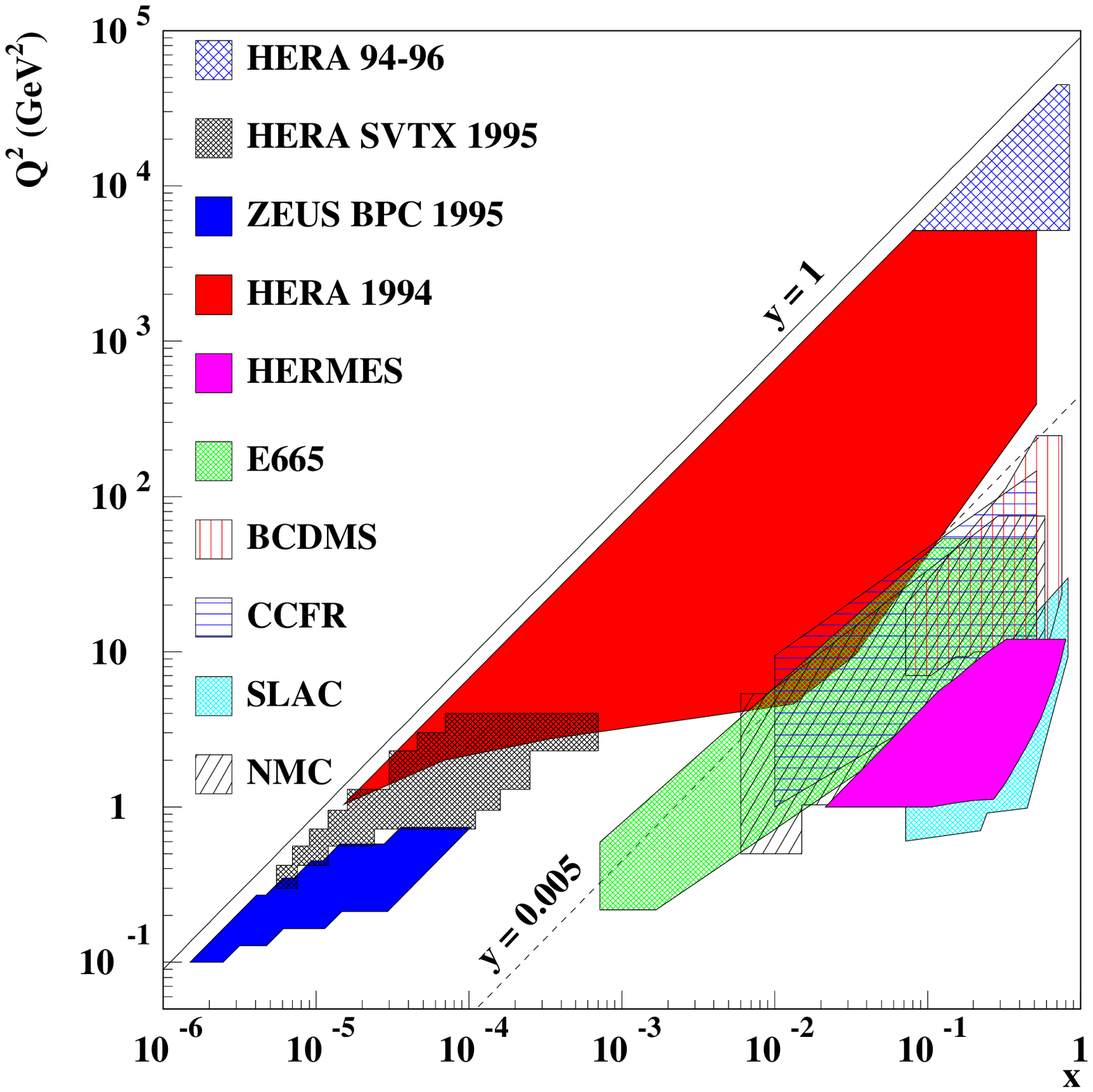,bbllx=-40pt,bblly=160pt,bburx=605pt,bbury=645pt,height=9cm} 
\fcaption{Regions in the ($x,Q^2$) plane covered by DIS experiments.}
\label{fig:kinreg}
\end{center}
\end{figure}

\noindent 
In this section we 
outline the detectors, data selection and extraction of 
structure functions from the scattering data.
We discuss in some detail the NMC, E665 
and CCFR fixed target experiments and the two
HERA collider experiments H1 and ZEUS.   
Many excellent reviews cover the earlier 
deep inelastic experiments from SLAC~\cite{nobel1990}, 
FermiLab~\cite{mishra_sciulli} and CERN~\cite{sloan} that have contributed 
so much to our understanding of the nucleon structure and strong 
interactions. The $(x,Q^2)$ regions covered by DIS experiments are 
shown in Fig.~\ref{fig:kinreg}.

In essence the measurement of the double differential inclusive 
deep inelastic scattering cross-section is straight forward. 
Assuming that the kinematic variables are to be reconstructed 
from the scattered lepton only, one needs precise measurements 
of the momentum of the
beam and scattered leptons, accurate knowledge of the event trigger and
reconstruction efficiencies, accurate knowledge of the detector 
acceptance, good control of backgrounds and radiative corrections 
and a precise measurement of the luminosity. As a number of the 
corrections, particularly the radiative corrections, depend 
on the structure functions, an iterative procedure is used to 
extract $F_i(x,Q^2)$ from the raw data.
Although simple in principle, to get systematic errors below
5\% requires a very thorough knowledge of the detector and 
careful attention to many details during both data-taking and 
analysis. As we shall see below there are quite considerable 
differences in the problems faced
by the fixed target and collider experiments.   

\subsection{Fixed target experiments}\label{sec:fixedT}
\noindent
First we outline the determination of the variables $x$ and $Q^2$ 
from the measurement of the scattered lepton in fixed target 
experiments. Assume that the beam energy, $E$, is known and that 
the lepton beam direction defines the axis from which the 
scattered lepton angle, $\theta$, is measured. $E^\prime$ is 
the energy of the scattered lepton
and $M_N$ is the nucleon mass. Then, ignoring the lepton masses, 
\begin{equation}
s=2M_N E~~~~~~~{\rm and}~~~~~y=(E-E^\prime)/E,
\end{equation}
\begin{equation}
Q^2 = 4EE^\prime\sin^2{\theta\over 2},
\end{equation}
\begin{equation}
x = {2EE^\prime\sin^2{\theta\over2}\over M_N(E-E^\prime)},
\label{eq:fixtx}
\end{equation}
definitions of the Lorentz invariants $x$, $y$, $s$ and $Q^2$ are 
given in Sec.~\ref{sec:lorinv}. It is also useful to define the
energy transfer in the lab frame $\nu=E-E^\prime$. 
In the ($x,Q^2$) 
plane lines of constant scattering angle $\theta$ are given by 
$Q^2=sxD/(x+D)$ where $D=2E\sin^2{\theta\over 2}/M_N$ and lines of 
constant $E^\prime$ are given by straight lines $Q^2=2M_N(E-E^\prime)x$.
The need for a well reconstructed scattered lepton requires minimum
values of $\theta$ and $E^\prime$ which then limit the 
region in which data can be analysed. Further restrictions also 
follow from how the resolutions
of $x$ and $Q^2$ depend on the resolutions of $\theta$ and 
$E^\prime$. This can be seen be looking at the contributions 
to the relative errors
\begin{equation}
\begin{array}{rcl}
{\delta Q^2\over Q^2}&=&
{\delta E^\prime\over E^\prime}\oplus 
\cot\bigl({\theta\over 2}\bigr)\delta \theta\\
&&\\
{\delta x\over x}&=&{1\over y}{\delta E^\prime\over E^\prime}
\oplus \cot\bigl({\theta\over 2}\bigr)\delta\theta
\end{array}
\label{eq:kinerr}
\end{equation}
where $\oplus$ indicates addition in quadrature.
From these equations we see that $Q^2$ is well determined except 
at very small scattering angles. The same is also true of $x$, but in 
addition, the error in the energy gets magnified at small values of $y$.
Looking at Eq.~\ref{eq:fixtx} one sees that errors in the beam
energy ($E$) will also get magnified at small $y$. For fixed target
experiments accurate control of $\delta E$, $\delta E^\prime$ and
the relative calibration of $E$ and $E^\prime$ are very important.
Together the restrictions on resolution and acceptance 
give the regions for the fixed target experiments in the ($x,Q^2$) 
plane that are shown in the lower right corner of Fig.~\ref{fig:kinreg}. 
 
\subsubsection{NMC}\label{sec:nmcdet}
\begin{figure}[htbp]
\vspace*{13pt}
\begin{center}
\begin{tabular}{c}
\psfig{figure=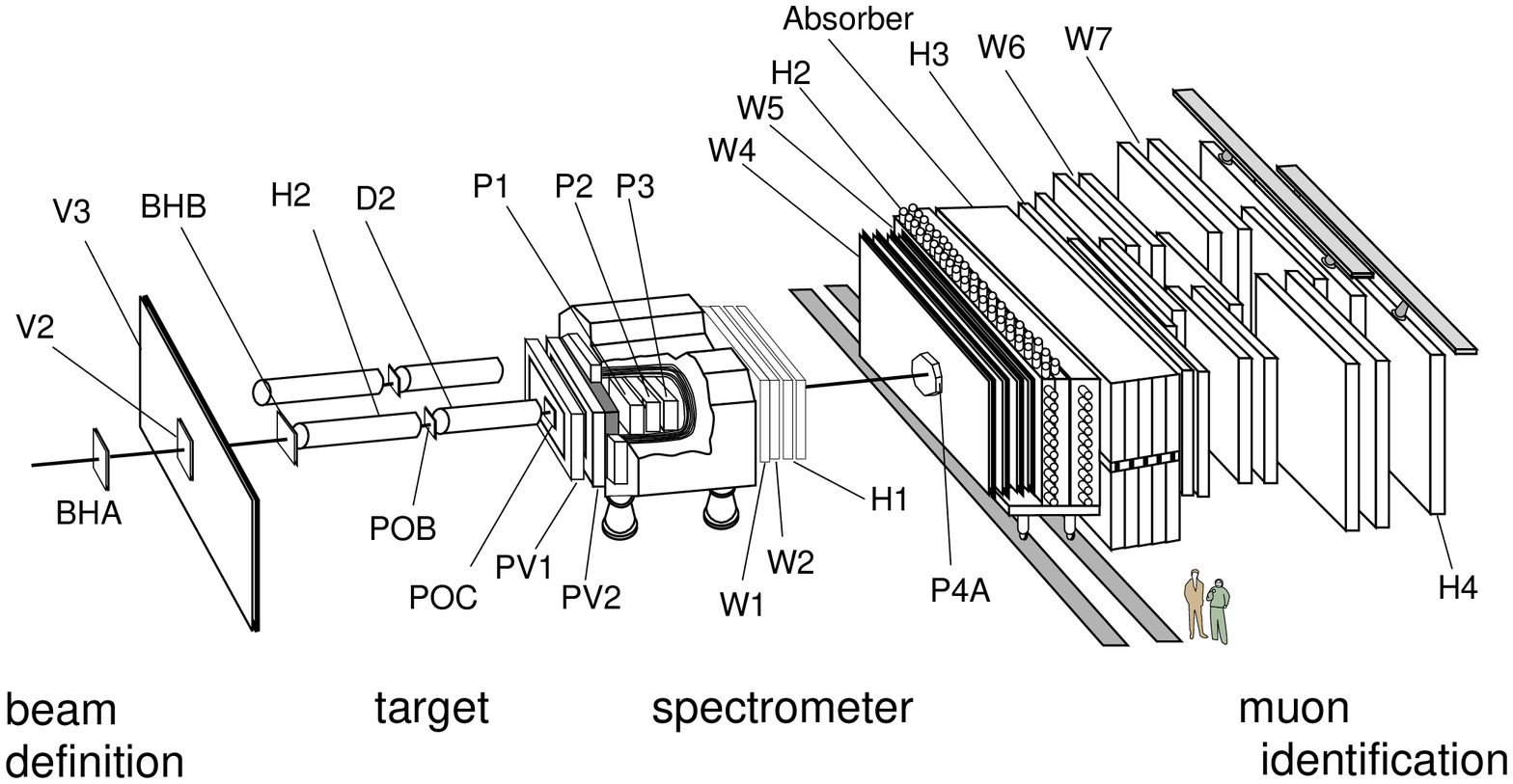,width=14cm,height=7cm} \\
\psfig{figure=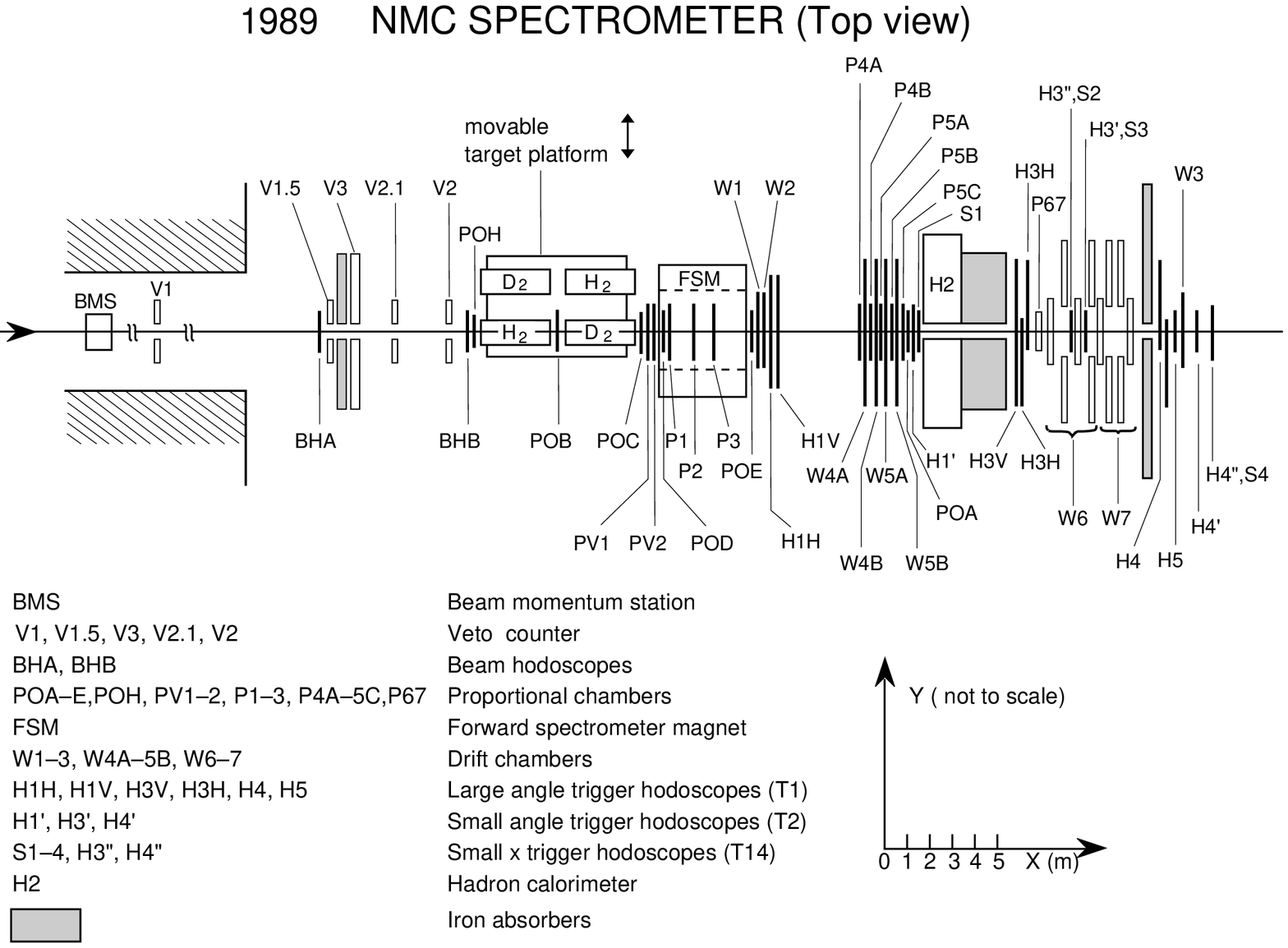,width=14cm,height=9cm} 
\end{tabular}
\vspace*{13pt}
\fcaption{The NMC detector.}
\label{fig:nmc}
\end{center}
\end{figure}

\noindent
The New Muon Collaboration (NMC) experiment~\cite{nmcdet} at 
CERN was an extension and improvement of the European 
Muon Collaboration (EMC) experiment. 
Originally designed to provide better data on nuclear effects in 
DIS, particularly the EMC effect, it has also provided
accurate data on $F_2^p$ and $F_2^d$. The NMC experiment was
situated on the M2 beam line of the CERN SPS and took data
at muon energies of 90, 120, 200 and $280\,$GeV at various
times during 1986, 1987 and 1989. A schematic diagram 
of the NMC apparatus is shown in Fig.~\ref{fig:nmc}. 

The important features of the NMC detector for structure function
measurements are the following:
\smallskip
\par\noindent
{\bf BMS.} The beam momentum station measures the deviation of beam
muons from a central assumed momentum. It achieves an rms resolution
of about $1\,$GeV at $280\,$GeV.
\smallskip
\par\noindent
{\bf FSM.} The forward spectrometer magnet, associated upstream
proportional chambers and large area downstream drift chambers
provide accurate track reconstruction and momentum determination for
charged particles with a typical momentum resolutions of $\Delta p/p
\approx 10^{-4}$. Scattered muons are identified by  large area
drift chambers placed after the electromagnetic calorimeter and a 
$2\,$m thick steel absorber.
\smallskip
\par\noindent
{\bf BCS.} The beam calibration spectrometer provides absolute calibration
of beam muons for the BMS. It is situated $36\,$m behind the FSM.
Special
runs at 90, 200 and $280\,$GeV gave 0.2\% accuracy.
\smallskip
\par\noindent
{\bf Targets.} The target setup has been designed to reduce systematic
effects, particularly for cross-section ratio measurements. For the
$p$ and $d$ runs it consisted of two similar pairs of $3\,$m long
target cells exposed alternately to the beam. In one pair the 
upstream target was liquid hydrogen and the downstream target liquid
deuterium, while in the other pair the order was reversed. The 
acceptance of the main spectrometer is significantly different for the
upstream and downstream target positions, giving two determinations
of the structure functions for each material. 
\smallskip
\par\noindent
{\bf Trigger.} Various triggers can be constructed by using signals
from the trigger hodoscopes in the forward spectrometer and from the 
beam halo veto system. 
Two physics triggers were used:
\begin{description}
\item[T1.] The large angle physics trigger. This requires a scattered
muon pointing at the target with a minimum scattering angle of $10\,$mrad
together with suppression of events with large energy
transfer $\nu$ at small scattering angles (almost real photoproduction).
\item[T2.] The small angle trigger. This overlaps with T1, but uses some
additional small hodoscopes to cover small scattering angles in the
range $5-15\,$mrad.
\end{description}
\smallskip
\par\noindent
Many other triggers were used for beam and beam halo studies and for
beam normalization.

As we have noted above, it is very important to have accurate absolute
and cross calibrations of the beam and forward spectrometers. This was 
achieved in a series of dedicated runs at all beam energies with specially
installed silicon microstrip detectors. The calibration of the FSM was
also checked to an accuracy of 0.2\% against the measured masses of 
$K^0$ mesons at 90 and $120\,$GeV and $J/\psi$ mesons at 200 
and $280\,$GeV. 

The integrated incident muon flux was measured by 
 random sampling of the beam~\cite{mount},
and by reconstructing beam tracks using
prescaled triggers on two planes of the beam hodoscope. 
The second method gave a statistical precision of 1\% in a few hours of data
taking. 

Selection cuts were applied to the data according to the trigger and
the muon beam energy, $E_\mu$ as given in Table~\ref{tab:nmccuts}.
The scattered muon momentum, $p_\mu$, was  required to be above a
minimum value to suppress muons from pion and kaon decays. Events with
small $\nu$, where the spectrometer resolution is poor, were rejected.
Regions with rapidly varying acceptance were excluded by requiring 
minimum muon scattering angles, $\theta_{min}$.
Cuts on the maximum values of $y$ and the mass squared of the
hadronic final state, $W^2$, excluded the kinematic domain 
where higher order electroweak processes dominate. 
The position of the reconstructed vertex was constrained to be 
within one of the targets. Finally, at each value of $x$, data in 
regions of $Q^2$ where the
acceptance was less than 30\% of the maximum at that $x$ were removed.

\begin{table}
\tcaption{Cuts applied to the NMC data, as explained in the text.
Different values of $\theta_{min}$ were used for the upstream and downstream
targets, as indicated. $N_p$ and $N_d$ are the total number of events for
protons and deuterons, respectively, after applying all cuts.
}
\centerline{\footnotesize\smalllineskip
\begin{tabular}{cccccccccc}\\
 \hline
 Trigger & $E_\mu$ & $p_\mu(min)$ & $\nu_{min}$ & $\theta^{up}_{min}$ &
 $\theta^{down}_{min}$ & $y_{max}$ & $W^2_{max}$ & $N_p$ & $N_d$ \\
    & ~[GeV]   & [GeV]      & [GeV] & [mrad]              &
[mrad]                &     & [GeV$^2$] & [$10^3$]  &  [$10^3$] \\
 \hline
 T1 &  90 & 15 &  7 & 13 & 15 & 0.9 & 130 & 255 & 533 \\
    & 120 & 20 & 10 & 13 & 15 & 0.9 & 150 & 103 & 215 \\
    & 200 & 35 & 20 & 13 & 15 & 0.9 & 250 &  56 & 114 \\
    & 280 & 40 & 30 & 13 & 15 & 0.9 & 400 & 141 & 406 \\
 \hline
  & $E_\mu$ & $p_\mu(min)$ & $\nu_{min}$ & $\theta^{up}_{min}$ &
 $\theta^{down}_{min}$ & $y_{max}$ & $y_{max}$ & $N_p$ & $N_d$ \\
\hline
 T2 & 200 & 35 & 15 & 6-6.5 & 6.5-7 & - & 0.8 &  78 & 162 \\
    & 280 & 40 & 30 & 6-7 & 6-7.5 & 0.2 & 0.8 &  97 & 207 \\
\hline\\
\end{tabular}}
\label{tab:nmccuts}
\end{table}

\subsubsection{E665}
\label{sec:e665det}
\noindent

\begin{figure}[t]
\vspace*{13pt}
\begin{flushleft}
\psfig{figure=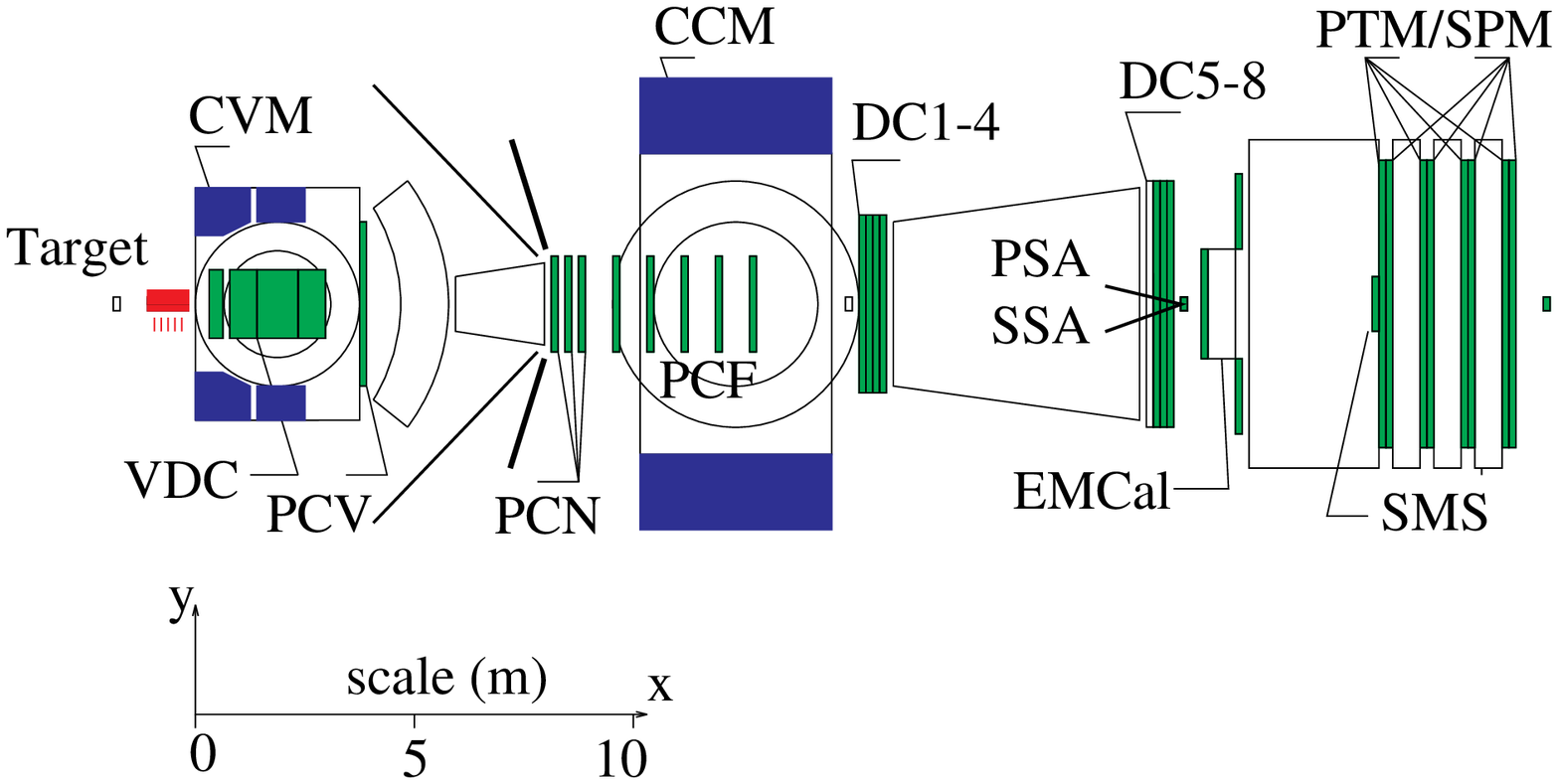,width=13cm} 
\fcaption{The E665 detector.}
\label{fig:e665}
\end{flushleft}
\end{figure}

The Experiment E665~\cite{e665det} was located at the end of the NM
beamline 
at Fermilab. The NM beam provided muons of average energy $470\,$GeV with 
a spread of $50\,$GeV. The experiment took data in 1987-88, 1990 and 
1991-92. The detector, shown in Fig.~\ref{fig:e665}, was designed to
measure beam and scattered muons with high precision and to provide
a good measurement of charged and neutral particles in the final state.
Components important for structure function measurement are:
\smallskip
\par\noindent
{\bf Beam spectrometer.} This was positioned between the end of the NM
beamline and the main detector. It consisted of 4 measuring stations
each equipped with beam hodoscopes and multiwire proportional chambers,
two before a dipole magnet (NMRE) and two after,
resulting in a  resolution on the beam
momentum of
$\delta(p^{-1})\sim~8\times~10^{-6}\,$GeV$^{-1}$.

\smallskip
\par\noindent
{\bf Target.} The target assembly was placed in the
field free region in front of the first spectrometer magnet (CVM). It
consisted of three identical target cells on a precision table that 
moved the targets laterally into the beam following a regular cycle.
The three target cells were identical and of active length $1\,$m, 
two were filled with liquid hydrogen and liquid deuterium respectively. 
The third was empty and was used to provide data for subtraction
of off-target scatters. 
\smallskip
\par\noindent
{\bf Main spectrometer.} The charged particle spectrometer was constructed
around two large magnets with reversed polarities positioned
so that the 
position of the scattered muon at a focussing plane
(where the muon
detector was situated) was independent of momentum and depended only
on the scattering angle. Tracking was performed by wire planes placed
inside both magnets and before and after the CCM. For tracks that
traversed the full length of the spectrometer a momentum resolution
of $\delta(p^{-1})\sim 2\times 10^{-5}\,$GeV$^{-1}$ was achieved, 
which gave resolutions of about 5\% on $x$ at low $x$ and about 4\% 
on $Q^2$.  Muons were identified
by four sets of wire planes placed behind a $3\,$m iron absorber, with
the sets separated by $1\,$m thick concrete absorbers. 
\smallskip
\par\noindent
{\bf EMcal.} The electromagnetic calorimeter was placed just in 
front of the muon absorber and was a 20 radiation length 
lead gas-sampling device. In addition to measuring photons from
neutral hadron decays it also provided information on elastic 
muon-electron scatters.
\smallskip
\par\noindent
{\bf Trigger.} Various triggers were used in the experiment, for
beam normalization, for detector and beam monitoring and for physics.
Three classes of physics triggers were used: 1) calorimeter -- which
used calorimeter signals to identify muon interaction events without 
using signals from behind the absorber; 2) large-angle  --
which required a muon to be identified in the muon chambers behind
the absorber; 3) small-angle (SAT) -- which used only the veto hodoscope 
to indicate the absence of an unscattered muon, and allowed 
events with muon scattering angles as small as $1\,$mrad to be triggered.

The integrated incident muon flux was measured by a variation on
the EMC technique~\cite{mount}. It assumes that
the beam spectrometer response is the same for random and physics
triggers. The number of usable beam muons is determined by 
counting the number of random beam triggers with a good muon and
multiplying by the pre-scale factor. The latter is determined by
comparing the number of random beam triggers with the actual
number counted by scalers. 

The calibration and resolution of the E665 spectrometers has been
checked using the following techniques. Primary protons at 
$800.6\pm2\,$GeV (determined from the Tevatron magnet currents) 
were directed through the beam and main spectrometers. The forward
spectrometer measured $800.5\pm0.14\,$GeV and an error estimate of
0.3\% was found for the momentum calibration of the beam spectrometer.
The relative calibration of the beam and main spectrometers was 
checked using non-interacting muons and comparing the difference
of the momenta measured in the two spectrometers in data and
simulation. This lead to an estimate of 0.13\% relative error at the
nominal muon beam energy of $470\,$GeV. Elastic muon-electron 
scattering events also provided information on the calibration and 
resolution of the spectrometer. Using signals from the EMcal, the 
events have a distinctive signature and are constrained by kinematics to
have $x=m_e/M_N$. From the measured $x$ distribution one finds the
absolute value correct to 0.1\% with a resolution $\delta x/x$ of
5.5\%. Finally an uncertainty of 0.35\% in the momentum calibration
of the forward spectrometer was estimated from $K^0_S$ decays.

Using only the small-angle trigger, $159,853$ H$_2$, $100,648$ D$_2$
and $31,796$ empty target events were collected with beam 
muon energy in the range $350-600\,$GeV. For the structure function
sample further cuts were applied: a single scattered muon from within 
the target with a minimum momentum of $100\,$GeV; a muon energy 
loss, $\nu$, of at least $35\,$GeV with calculated relative error 
less than 50\%.
 
\subsubsection{Neutrino scattering experiments}
\label{sec:ccfr_det}

\noindent
Two high statistics neutrino scattering experiments were performed in
the 1980s, CDHSW at CERN and CCFR at Fermilab. 
Both have been important for providing tests of pQCD and for helping to
untangle parton distributions. More details on these and earlier
neutrino experiments are given in the reviews by Fisk and 
Sciulli~\cite{fisk_sciulli} and Mishra and Sciulli~\cite{mishra_sciulli}.
However, because the greatest precision 
has been achieved by CCFR their results for $F_2^{\nu N}$ and 
$xF_3^{\nu N}$ are the ones that continue to be used in global parton
determinations. We give a brief account of the CCFR detector.

\begin{figure}[t]
\vspace*{13pt}
\begin{center}
\psfig{figure=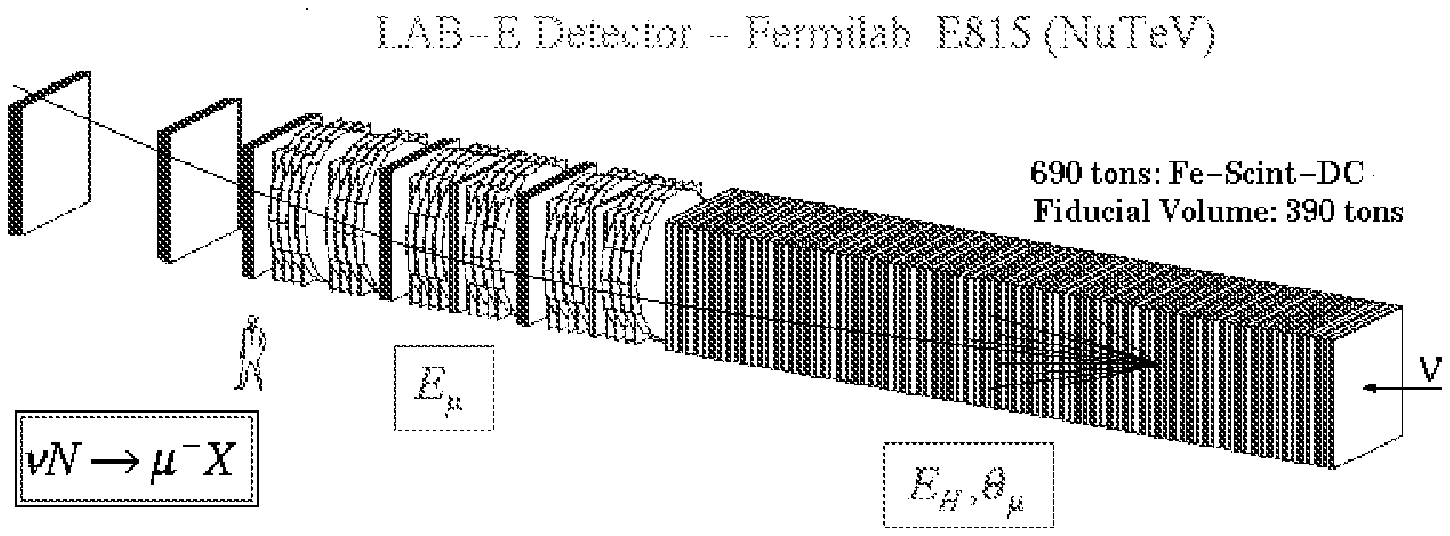,bbllx=50pt,bblly=250pt,bburx=540pt,bbury=500pt,width=.95\textwidth} 
\fcaption{The CCFR detector.}
\end{center}
\end{figure}

The CCFR neutrino scattering data were collected in two Fermilab
experiments, E740 in 1984 and E770 in 1987. The detector was
exposed to the Tevatron Quad-Triplet wide-band neutrino beam, 
which was composed of neutrinos with average energy $185\,$GeV and
antineutrinos of average energy $143\,$GeV. The maximum beam energy
was about $600\,$GeV and the ratio of $\nu:\bar{\nu}$ was about
$2.5:1$. The CCFR detector~\cite{ccfrdet}
~\fnm{k}\fnt{k}{~The detector
 is also the basis of that for the NuTeV experiment.}
consists of a $17.7\,$m long
690 ton unmagnetized steel-scintillator target calorimeter, which
is instrumented with drift chambers for muon tracking. The hadronic
energy resolution of the calorimeter is $\sigma/E=0.85/\sqrt{E({\rm GeV})}$.
The calorimeter energy scale was calibrated to 1\% using momentum analysed
hadron beams with energies between 15 and $450\,$GeV. The target
is followed by a $17.8\,$m long solid iron toroidal magnetic
spectrometer for muon identification and momentum measurement. The
spectrometer was calibrated to about 0.5\% using a momentum analysed muon
beam with energies of $50,75,120$ and $200\,$GeV. The muon momentum
resolution is $\Delta p/p=0.11$ and it is limited by multiple Coulomb
scattering in the iron. The detector provides measurements of the
visible hadronic energy $E_{had}$, the momentum, $p_\mu$,
and angle with respect to the neutrino beam line, $\theta_\mu$, of
the scattered muon. 
The relative neutrino flux at different energies and the relative
$\bar{\nu}/\nu$ flux are obtained from 
events with low hadron energy, $E_{had}<20\,$GeV. The absolute
normalization is determined 
so that the total $\nu N$ cross-section equals the average value 
for Fe from the CHDSW and CCFR experiments of 
$\sigma^{\nu N}/E=(0.677\pm 0.014)\times 10^{-42}\,$m$^2$GeV$^{-1}$
per nucleon~\cite{ccfr_norm}.

In neutrino scattering, the incident neutrino energy cannot be measured
directly but has to be inferred from the final state measurements. The
kinematic variables $x$ and $Q^2$ are first estimated from the
measured quantities by:
\begin{equation}
Q^2_{vis}=4E_{vis}E_\mu\sin^2{\theta_\mu\over 2},~~~~~~~
x_{vis}={Q^2_{vis}\over 2M_N E_{had}},
\end{equation} 
since the lepton energy loss $\nu_{vis}=E_{had}$. Average values of
physical quantities are calculated by using Monte Carlo simulation
to correct distributions of the corresponding visible quantity.

The data sample consists of 950,000 $\nu$ and 170,000 $\bar{\nu}$ 
events after the fiducial and kinematic cuts: $p_\mu>15\,$GeV;
$\theta_\mu<0.15\,$rad; $E_{had}>10\,$GeV and
$30<E_{vis}<360\,$GeV. The cuts are designed to select the region
of high efficiency and small systematic errors.


\subsection{HERA Experiments}
\label{sec:heraexp}
\noindent
HERA is the first $ep$ collider and consists of two separate rings
of circumference $6.3\,$km, one a warm magnet electron
(or positron) ring with maximum energy $30\,$GeV and the other a
superconducting magnet proton ring of maximum energy $820\,$GeV.
The rings are brought together at four intersection regions now 
occupied by the experiments H1, ZEUS, HERA-B (proton fixed wire
target B physics) and HERMES (polarized-electron -- polarized nuclear
gas-jet target). Although not optimized for the task, HERMES
 can in principle also measure unpolarized structure functions 
in a similar kinematic region to the SLAC experiment.

HERA can operate with up to 220 bunches in each
ring. In 1994 (1995) the collider operated with 153 
(174) colliding bunches of
$27.5\,$GeV positrons and $820\,$GeV protons. Additional 15 (6) positron
 and 17 (15) proton  bunches were used to study beam related 
backgrounds. Data were collected in 1992 and 1993, but we shall
concentrate on the high statistics measurements from 1994 and some
 results from the 1995 data.

The interval between bunch crossings in HERA is $96\,$ns
which necessitates sophisticated multi-level trigger systems. 
For ZEUS and H1, at the first level data is stored temporarily 
(`pipelined') while
hardware specific trigger processors, operating synchronously
with HERA beam crossings, arrive at a decision in about $2-5\,\mu$s. 
The higher trigger levels operate asynchronously and involve
more sophisticated calculations, culminating in an `event filter'
which uses a fast version of the offline reconstruction code 
running on a farm of RISC processors. 
 The overall reduction achieved is from
a raw interaction rate of many $100\,$KHz (which is
very sensitive to beam conditions in HERA) to about $5-10\,$Hz written
to tape. 

The H1 and ZEUS detectors are nearly hermetic multi-purpose 
devices designed  to investigate all aspects of high energy $ep$ 
collisions. In particular both 
the scattered electron and the hadronic system in a hard $ep$ 
interaction are measured, the latter allowing one to  estimate
the energy and angle of the struck quark. Together with 
the information from the scattered electron one has an 
overdetermined system and there are 6 possible ways~\cite{bek} to 
reconstruct $x$ and $Q^2$. We will only discuss
those that have been used in $F_2$ analyses. 

\begin{figure}[htbp]
\vspace*{13pt}
\begin{center}
\psfig{figure=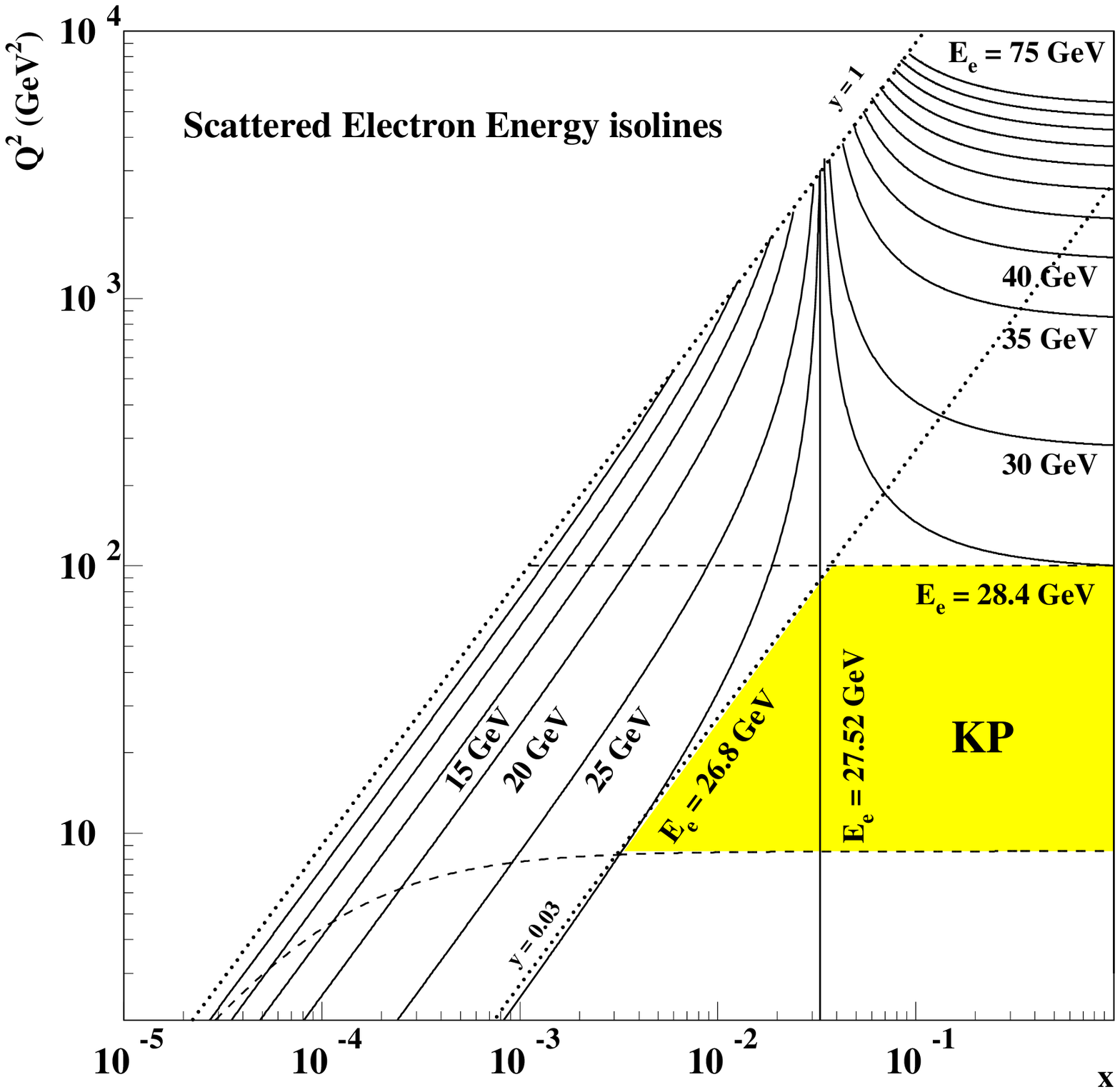,bbllx=-70pt,bblly=160pt,bburx=575pt,bbury=645pt,height=8cm} 
\fcaption{Contours of constant energy of the 
scattered lepton in the ($x,Q^2$) plane for the HERA $ep$ collider, together
with the kinematic peak (KP) region.}
\label{fig:herakp}
\end{center}
\end{figure}

First the electron 
(E) method. The proton axis is conventionally taken to define 
the positive $z$ direction from which all angles are measured. At 
low $Q^2$ this means that the electron scattering angle approaches 
$180^\circ$. If $E_e$, $E_p$ are the electron and proton beam 
energies, $E^\prime$ and $\theta_e$ the energy and angle of the 
scattered electron, then:
\begin{equation}
s=4E_eE_p~~~~~~~{\rm and}~~~~~
y_e=\bigl(1-{E^\prime\over E_e}\sin^2{\theta_e\over 2}\bigr),
\label{eq:hk1}
\end{equation}
\begin{equation}
Q_e^2 = 4E_eE^\prime\cos^2{\theta_e\over 2} = 
{E^{\prime 2}\sin^2\theta_e\over 1-y_e},
\label{eq:hk2}
\end{equation}
\begin{equation}
x_e = {E_eE^\prime(1-\sin^2{\theta_e\over2})\over E_p(E_e-
E^\prime\sin^2{\theta_e\over 2})}.
\label{eq:hk3}
\end{equation}
Note that if $E^\prime=E_e$ then $x=x_0\equiv E_e/E_p$, the 
position of the so-called kinematic peak. In the ($x,Q^2$) plane 
lines of constant
scattering angle $\theta$ are given by $Q^2=sxC/(sx+C)$ where 
$C=4E_2^2\sin^2{\theta_e\over 2}$ and lines of constant 
$E^\prime$ 
are given by $Q^2=(4xE_e(E_e-E^\prime))/(x_0-x)$.
 Fig.~\ref{fig:herakp} shows 
contours of constant energy of the scattered lepton.
The large region of phase space around the kinematic peak, in 
which $E^\prime\approx E_e$, is evident. Kinematic peak
events, which can be selected to be almost independent of the 
structure function samples, are used for calibration purposes
by both H1 and ZEUS. A typical selection is shown by the shaded
region in the figure. As in fixed 
target experiments, $E^\prime$ is required to be greater than a 
minimum value and $\theta_e$ less than a maximum angle if it is to 
be measured in the rear EM calorimeters of the collider detectors. 
These considerations limit the accessible region of $x$ and $Q^2$. 
Although there are differences of detail, the formulae for the
resolutions of $x$ and $Q^2$ are similar to those of Eq.~\ref{eq:kinerr}
and the message is the same, 
that at small $y$ (large $x$) the resolution in $x$ deteriorates. 
The other potential problem with the electron method is that radiative
corrections can be large.

In the double angle (DA) method, only angles are used to reconstruct $x$ and 
$Q^2$. In addition to $\theta_e$ one constructs the angle $\gamma$ from
the hadronic energy flow using:
\begin{equation}
\cos\gamma = {(\sum_h p_x)^2+(\sum_h p_y)^2-(\sum_h(E-p_z))^2\over
                (\sum_h p_x)^2+(\sum_h p_y)^2+(\sum_h(E-p_z))^2}
\label{eq:gamh}
\end{equation}
where $\sum_h$ runs over all energy deposits (with momentum vectors
$(p_x,p_y,p_z)$) not assigned to the scattered electron. In the QPM
$\gamma$ is the direction of the struck quark. The variables $x$ and 
$Q^2$ are then determined by:
\begin{equation}
Q^2_{DA}=4E_e^2{\sin\gamma(1+\cos\theta_e)\over
                \sin\gamma+\sin\theta_e-\sin(\gamma+\theta_e)}
\end{equation}
\begin{equation}
x_{DA}=x_0{\sin\gamma+\sin\theta_e+\sin(\gamma+\theta_e)\over
             \sin\gamma+\sin\theta_e-\sin(\gamma+\theta_e)}
\end{equation}

The method is insensitive to hadronization and, to first order, is 
independent of the detector energy scales~\cite{bek}.
At small values of $\theta_e$ or $\gamma$ the resolution in $x_{DA}$ and
$Q^2_{DA}$ worsens. In order that
the hadronic system be well measured, it is necessary to require a
minimum of hadronic activity away from the beampipe. A suitable
quantity for this purpose is the hadronic estimator~\cite{jb} of $y$:
\begin{equation}
y_{JB}={\sum_h(E-p_z)\over 2E_e}
\label{eq:yjb}
\end{equation}
\noindent

At low $y$ ($<0.04$) and low $Q^2$, when there is low hadronic activity
in the detector the DA method becomes sensitive to noise in the calorimeter.
The double angle method is also used to cross calibrate the energy 
scale of the detectors.

Two other methods, which use a mixture of electron and hadronic 
information, will be considered with the relevant experiment below.

Because of the interest in measuring the behaviour of $F_2$ or the 
total $\gamma^*p$ cross-section as $Q^2\to 0$, both ZEUS and H1 have 
improved and extended their detectors to be able to measure 
$\theta_e$ very close to $180^\circ$. In addition they have taken 
data from special runs
of the HERA collider in which the primary interaction vertex is 
shifted by $70\,$cm in the proton beam direction. The interaction
is thus further from the rear calorimeters which are used to 
identify the scattered electron and hence smaller angles can be
reconstructed. Such data are referred to as {\it shifted vertex data}.
Another way to reach very low $Q^2$ is to select a sample of events
in which the incoming electron (or positron) radiates a hard 
collinear photon and hence provides a lower interaction energy, 
this technique has been used by both experiments and will be
referred to as ISR. 

At HERA luminosity is measured using the very small $Q^2$ Bethe-Heitler 
process $ep \to ep\gamma$, for which the cross-section is large and
known very accurately. ZEUS uses lead-scintillator 
calorimeters and H1 uses KRS-15 (TlCl/TlBr) crystal cerenkov counters
situated very close to the electron beam pipe in order to 
detect scattered beam electrons and bremsstrahlung photons. 
Photons emerging from the electron-proton 
interaction point (IP) at angles $\theta_\gamma \le 0.5$ mrad with 
respect to the electron beam axis hit the photon calorimeter at 
107 m  (103 m) from the IP
for ZEUS (H1). Electrons emitted from the IP at scattering
angles less than or equal to 6 mrad and with energies  
$0.2 E_e <E_e^{\prime} < 0.9 E_e$ are deflected by beam magnets and            
hit the electron calorimeter placed 35~m (33 m) from the IP. The
systematic error on the luminosity measurement is typically less than 2\%.

The regions in the $(x,Q^2)$ plane covered by the
various H1 and ZEUS data sets are shown in Fig.~\ref{fig:kinreg}.
At high $Q^2$ the measurements are limited by event statistics. 
The integrated HERA luminosity collected to date is
insufficient to allow measurement of structure functions for
$Q^2 > 10\,000\,$GeV$^2$. However first measurements of the NC
cross-section at very high $Q^2$ have been published and have
generated great interest. They will be discussed further in 
Sec.~\ref{sec:hiq2}.

\subsubsection{H1}
\label{sec:h1dat}
\begin{figure}[htbp]
\vspace*{13pt}
\begin{center}
\psfig{figure=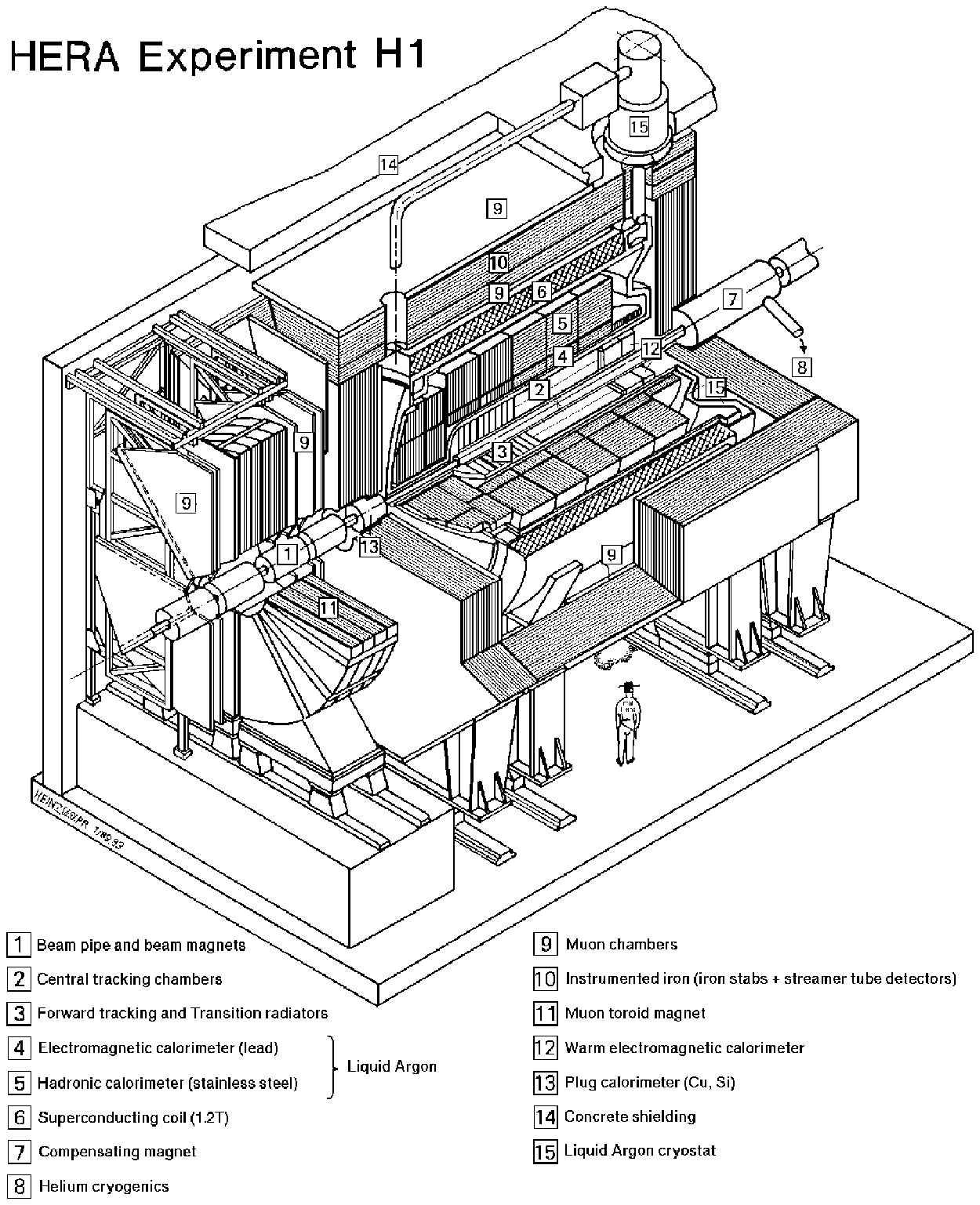,bbllx=40pt,bblly=90pt,bburx=420pt,bbury=570pt,height=18cm}
\fcaption{The H1 detector in its configuration until end 1994.}
\label{fig:h1det}
\end{center}
\end{figure}

\noindent
The H1 detector~\cite{h1det} for data taken until 1994 is shown in 
Fig.~\ref{fig:h1det}. The structure function measurements  have 
relied on the inner tracking system and on the backward and liquid 
argon calorimeters. The inner trackers and the argon calorimeter are
 surrounded by a large superconducting
solenoid which provides a uniform magnetic field of $1.15\,$T.
The 1995 detector upgrade will also be briefly mentioned.

\smallskip
\par\noindent
{\bf Inner Tracker.} The inner tracker consists of the forward tracker
(FT) modules, a backward proportional chamber (BPC) and the central jet 
chamber (CJC). The CJC consists of two concentric drift chambers
covering a polar angle range of $15^\circ-165^\circ$. For tracks
crossing the CJC the resolution in transverse momentum is
$\delta p_T/p_T<0.01p_T$ ($p_T$ in GeV). The FT is used to determine
the vertex for events which leave no track in the CJC and it covers
the polar angle range of $7^\circ-20^\circ$. The BPC provides a
space point, with a resolution of about $1.5\,$mm in the plane
perpendicular to the beam, for charged particles entering the 
backward EM calorimeter and has a polar angle acceptance of 
$151^\circ-174.5^\circ$.  
\smallskip
\par\noindent
{\bf BEMC.} The backward electromagnetic calorimeter covers the polar
angle region $155^\circ-176^\circ$ and is of lead-scintillator 
construction with a depth of 22 radiation lengths, giving an energy
resolution of $10\%/\sqrt{E({\rm GeV})}$. 
\smallskip
\par\noindent
{\bf LAR.} The liquid argon calorimeter is used to measure the
hadronic final state and scattered electrons at $Q^2>120\,$GeV$^2$.
The LAR covers the angular range $3^\circ-155^\circ$ and has an
EM section with lead absorber plates of $20-30$ radiation lengths
and a hadronic section with steel absorber plates giving a total
depth of $4.5-8$ interaction lengths. For EM particles the energy
resolution is $12\%/\sqrt{E({\rm GeV})}$, and for hadrons 
$50\%/\sqrt{E({\rm GeV})}$. 
\smallskip
\par\noindent
{\bf 1995 upgrade.} The H1 detector was upgraded in 1995~\cite{h1up}
with the replacement of the BPC and BEMC by drift chambers and an
improved calorimeter. For data with an
interaction  vertex shifted by 70 cm from 
the nominal position the
backward drift chambers (BDC) cover the angular range 
$152^\circ-178.3^\circ$ and provide track segments for charged particles
entering the rear calorimeter. The BEMC was replaced by a lead
scintillating-fibre calorimeter (SPACAL) which covers the range
$153^\circ-178.5^\circ$ with high granularity giving an angular
resolution of $1-2\,$mrad. A preliminary value for the relative
energy resolution, measured during experimental operations, is
$7.5\%/\sqrt{E({\rm GeV})\oplus 2.5\%}$. For EM energy the absolute energy
scale
is presently known to $1-3$\%.
 
Both the BEMC and LAR electromagnetic 
energy scales were determined by using the
observed shape of the kinematic peak and the double angle method to
refine the results. The BEMC energy scale was checked to 1\% and the LAR
energy scale was checked to 3\%.
 The calibration of the LAR hadronic energy
scale relies on test beam measurements and a comparison of the
balance between the transverse energy of electrons and hadrons in DIS
events. This gives a 4\% uncertainty in $p_T^h/p_T^e$.

For DIS structure function measurements two triggers were used: 
{\it low $Q^2$} required a BEMC energy cluster with 
$E^\prime > 4\,$GeV which was in time with the beam crossing;
 {\it high $Q^2$} required
an electromagnetic cluster in LAR with either $E^\prime > 8\,$GeV or
$E^\prime > 6\,$GeV plus a track trigger. The samples collected
in the 1994 HERA run were $2.2\,$pb$^{-1}$ at the nominal vertex position 
and $58\,$nb$^{-1}$ with the shifted vertex.  Simplifying
somewhat~\cite{h1n94}, after offline reconstruction DIS NC events were 
further required to satisfy $E^\prime>11\,$GeV, to be fully 
contained by either the BEMC ($Q^2<120\,$GeV$^2$) or LAR 
($Q^2>120\,$GeV$^2$) and to have a primary vertex within $\pm 30\,$
of the nominal interaction point.

For the 1995 shifted vertex sample~\cite{h1v95} of $114\,$nb$^{-1}$ 
the new SPACAL allowed a 
lower cut on $E^\prime$ of $7\,$GeV. 

To improve the resolution of $x$ at small $y$ and to reduce radiative
corrections H1 use the $\Sigma$ method of reconstruction~\cite{sigma}.
Returning to Eq.~\ref{eq:yjb}, $\sum_h(E-p_z)$ is
referred to as $\Sigma$, and the essence of the idea is to remove
explicit dependence on $E_e$ from the kinematic reconstruction of
Eqs.~\ref{eq:hk1}-\ref{eq:hk3}. This is done by noting
that energy and momentum conservation give 
$\displaystyle 2E_e=\Sigma+E^\prime(1-\cos\theta_e)$ so one defines
\begin{equation}
y_\Sigma={\Sigma\over \Sigma+E^\prime(1-\cos\theta_e)}~~~~~~~~~
Q_\Sigma^2={E^{\prime 2}\sin^2\theta_e\over 1-y_\Sigma}.
\label{eq:sigma}
\end{equation}
The variable $x$ is then calculated from $Q^2=sxy$. Note that
the equation for $y_\Sigma$ may also be written
$y_\Sigma=y_{JB}/(1+y_{JB}-y_e)$.
Apart from allowing an extension of the measurements to lower 
$y$ values than the E method, the $\Sigma$ method has a reduced
sensitivity to radiative corrections.

\subsubsection{ZEUS}
\begin{figure}[htbp]
\vspace*{13pt}
\begin{center}
\psfig{figure=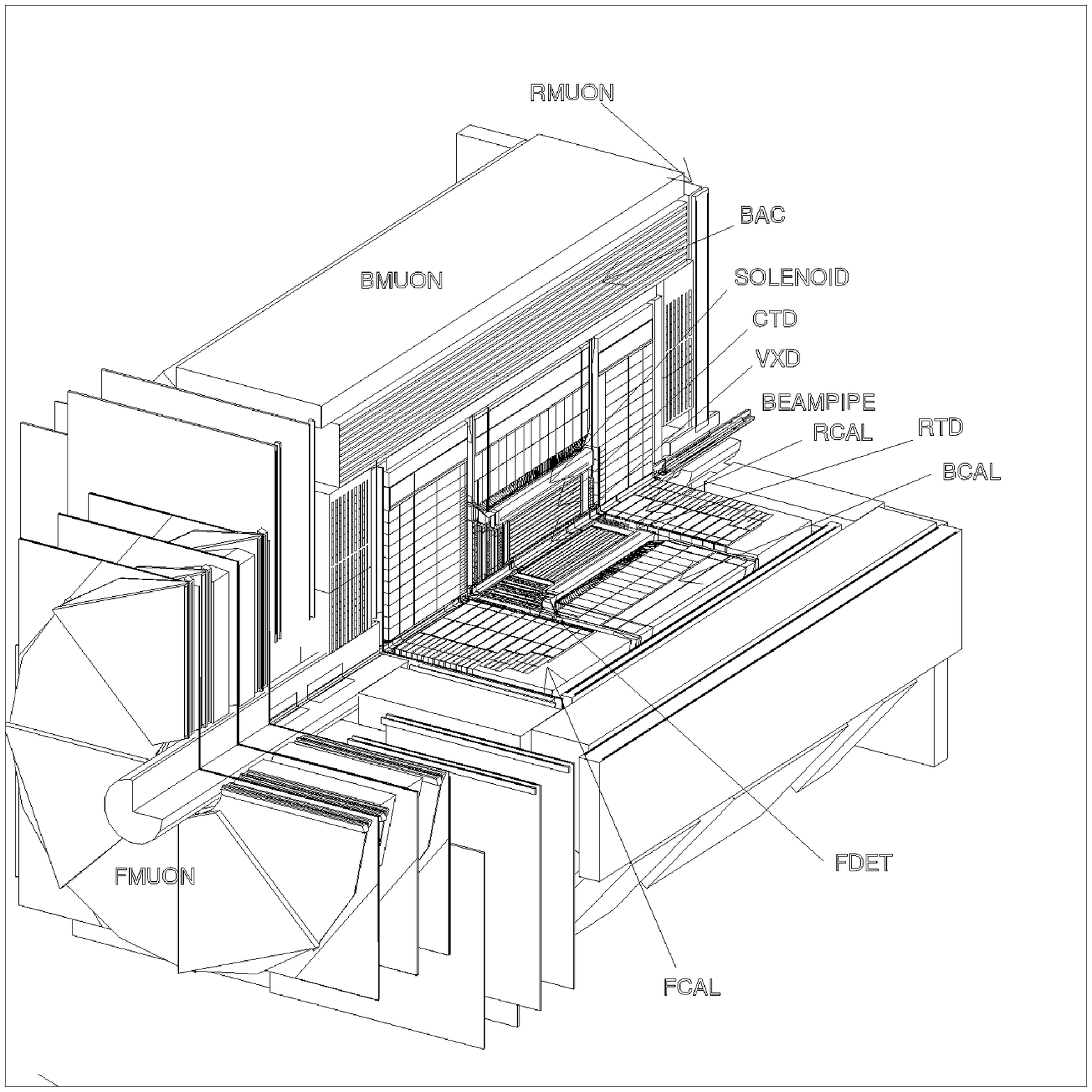,bbllx=30pt,bblly=30pt,bburx=580pt,bbury=720pt,height=18cm} 
\fcaption{The ZEUS detector.}
\label{fig:zeus}
\end{center}
\end{figure}

\noindent
ZEUS is a multipurpose detector~\cite{zeusdet} shown
in Fig.~\ref{fig:zeus}. We give 
a brief description of those parts of the detector relevant for 
structure function measurements.  

\smallskip
\par\noindent
{\bf Tracking.} Charged particles are tracked by the inner tracking 
detectors which operate in a magnetic field of 1.43 T provided by a 
thin superconducting coil. Immediately surrounding the beampipe is 
the vertex detector (VXD), which is a cylindrical drift chamber. Surrounding 
the VXD is the central tracking detector (CTD) which consists of 
72 cylindrical drift chamber layers, giving a resolution in 
transverse momentum for full length tracks of 
$\sigma_{p_T}/p_T=0.005p_T\oplus 0.016$ (for $p_T$ in GeV). The
interaction vertex is measured with a typical resolution of $0.4\,$cm
along the beam direction and for tracks with momentum above $5\,$GeV
the extrapolated position on the inner face of the calorimeter is
$0.3\,$cm.
\smallskip
\par\noindent
{\bf UCAL.} The solenoid is surrounded by a high resolution 
uranium-scintillator calorimeter covering
 the polar angle region 
$2.6^\circ - 176.2^\circ$.
The resulting 
solid angle coverage is $99.7\%$ of $4\pi$.  Under test beam
conditions the UCAL has an energy resolution of 
$\sigma_E = 0.35\,\sqrt{E({\rm GeV})}$ for hadrons and 
$\sigma_E = 0.18\, \sqrt{E({\rm GeV})}$ for electrons. The UCAL also provides
a time resolution of better than 1 ns for energy deposits greater than
4.5 GeV, which is used for background rejection.    
\smallskip
\par\noindent
{\bf SRTD.} The position of positrons scattered at small angles 
to the positron beam direction
is measured using the small angle rear tracking detector (SRTD).
The SRTD consists of two planes of scintillator strips, 1 cm wide and 
0.5 cm thick, arranged in orthogonal directions and read out 
via optical fibres and photo-multiplier tubes. 
The SRTD signals resolve single minimum ionizing particles 
and provide a position resolution of 0.3~cm.
The time resolution is less than 2 ns for a minimum ionizing particle. 
\smallskip
\par\noindent
{\bf BPC.} The beam-pipe calorimeter is a small tungsten-scintillator
sampling calorimeter located about $5\,$cm from the beam line just 
upstream of the RCAL ($294\,$cm from the interaction point). It is
designed to measure scattered electrons at $\theta_e$ very close to
$180^\circ$ with high precision. 
Position measurements are made with
horizontal
and vertical scintillator strips $8\,$mm wide. The energy resolution is 
$17\%/\sqrt{E({\rm GeV})}$ as measured in a test beam and confirmed by
use of kinematic peak events. The absolute energy scale and 
strip-to-strip calibrations were also determined using kinematic peak
events and confirmed using quasi-elastic $ep \to ep\rho^0$ events. 
The BPC was installed for the 1995 HERA physics run.  

\noindent

The absolute calibration error of the UCAL is estimated to be about
2\% based on test-beam measurements and it is maintained by 
monitoring the level of natural uranium radioactivity. The main problem
in the accurate determination of $E^\prime$ is to account properly
for energy loss in dead material in front of the UCAL, typically
1.5 radiation lengths. Events selected in the kinematic peak have been
used to calibrate the correction procedure for 
$E^\prime\approx 27.5\,$GeV and $\theta_e>135^\circ$. QED Compton
events ($ep \to e\gamma p$) and DIS $\rho^0$ production also provide
checks on the calibration at lower $E^\prime$ and smaller $\theta_e$.
Together these methods give a resolution on $E^\prime$ of
$\sigma/E=(20-27)\%/\sqrt{E({\rm GeV})}$ depending on $\theta_e$.

The ZEUS pipelined trigger~\cite{zcalflt} was designed to operate with 
minimum deadtime. The primary trigger signals were:
electromagnetic (EMC) energy above 2.5 - 4.8 GeV, depending on the
position, and proper timing for an $ep$ collision measured with
 upstream veto counters and the SRTD.
To reject photoproduction events, and proton beam-gas events 
originating within the detector, events were required to satisfy
$\delta\equiv\sum_i E_i(1-\cos\theta_i)>(25-2E_\gamma)\,$GeV, where
$E_i$ and $\theta_i$ are the energies and polar angles of UCAL cells
and $E_\gamma$ is the energy deposited in the luminosity photon 
calorimeter. Simplifying~\cite{zn94} somewhat, selected DIS NC events
are required to have a positron candidate with a fully contained 
shower and a track match for $\theta_e<135^\circ$. The electron
must also satisfy $E^\prime > 10\,$GeV and $y_e<0.95$. In addition
the events must satisfy $38<\delta<65\,$GeV. The 1994 sample 
corresponds to $2.5\,$pb$^{-1}$.

For analysis of their 1994 data ZEUS have used a variation on the
$\Sigma$ method, the PT method, to extend coverage to 
smaller $y$ and hence to
make an overlap with the fixed target data possible. The basic idea
is first to use the balance in transverse momentum between the scattered 
positron and the hadronic system to correct $y_{JB}$ for hadronic
energy loss, either in the beam pipe or in dead material. 
 A correction function
calculated from a Monte Carlo simulation (both of lepton-nucleon physics 
including hadronization  and of detector response~\cite{zn94}) is used.
The $\Sigma$ method is then used to improve $y$ over the whole 
kinematic region.
Finally $x$ and $Q^2$ are calculated using the double angle formula
but with $\gamma$ calculated from
\begin{equation}
\cos\gamma_{PT}={p_T^{e 2}-4E_e^2y^2\over
                 p_T^{e 2}+4E_e^2y^2}.
\end{equation} 
For the PT method to work it is necessary to limit the loss of
hadronic transverse energy in the forward beam pipe and the cut
$\displaystyle {p_T^h\over p_T^e}>0.3 - 0.001\gamma$, where $\gamma$
is calculated from Eq.~\ref{eq:gamh}, is applied to the data.

\subsection{Structure function extraction}
\noindent
There are three general problems to be solved in the extraction of
structure functions from the observed data. We consider the
following in turn for the case of NC scattering:
theoretical corrections for terms that cannot be measured directly;
radiative corrections to the Born-level cross-section; 
detector resolution and acceptance corrections. 

\subsubsection{Corrections for $F_L$ and $Z^0$ exchange}
\label{sec:rcorr}
\noindent

The full equation relating the structure functions $F_2, F_L$
and $xF_3$ to the Born cross-section for NC $\ell N$ scattering is 
given in Eq.~\ref{eq:NCxsec}. Without measurements from both e$^-$ 
and e$^+$ beams at high $Q^2$ the contribution of $xF_3$ cannot
be isolated. In addition $Z^0$ exchange terms contribute to $F_2$.
Corrections for both contributions must be made
in order to extract $F_2^{em}$.  These corrections are only significant
for the HERA experiments at $Q^2>1000\,$GeV$^2$ and they are
calculated using the measured Standard Model electroweak couplings of the $Z^0$
and parton distributions extracted from lower $Q^2$ data. The corrections
are not very sensitive to the choice of parton distribution.
Ignoring $xF_3$, the cross-section may be written in terms of $F_2$ and
 $R$ as in Eq.~\ref{eq:Rxsec}
respectively. $R$ (or $F_L$) is difficult to 
measure since its contribution to the cross-section is
suppressed by the $y^2$ factor. Measurements are discussed in more detail in
Sec.~\ref{sec:rdata}, but note that in order to measure $F_L$ cross-section
data at two different centre of mass energies are required. 
In the QPM and leading twist pQCD $R$ is predicted to be small for most of the
kinematic range, and generally the contribution of $R$ (or $F_L$) to the 
cross section is small except for $y$ close to $1$. However, for experiments
in which it has not been measured, it represents  an 
important correction. Indeed, the treatment of the $F_L$ correction
was the source of some of the apparent discrepancies
between earlier fixed target $F_2^{\mu N}$ data sets. 
Corrections have usually been made by parametrizing $R$ and using 
Eq.~\ref{eq:Rxsec} to extract $F_2$. Various strategies have been adopted:
$R=0$ (naive QPM for $Q^2 \gg M^2_N$); $R=\,$constant; $R=R_{QCD}$ (i.e. 
approximating $R = F_L/(F_2 - F_L)$ with $F_L$ given by Eq.~\ref{eq:flqcd}); 
and $R=R_{SLAC}$ where
the last is a parametrization~\cite{whitlow90} of data on $R$. This
parametrization will be discussed in more detail in Sec.~\ref{sec:rdata}.
As the first three methods do not the describe the data on $R$ 
at low $Q^2$ they have been discarded and all fixed target groups now
use $R_{SLAC}$ (unless $R$ is measured in the same experiment).
At larger $Q^2$ and large centre of mass energies, pQCD should be
applicable and $R_{QCD}$ is used. To calculate $R_{QCD}$ groups
have used either a recent set of parton distributions or the results
of their own NLO fit to the $F_2$ data. Experiments now quote the 
prescription used for $R$ such that their values of $F_2$ may be recalculated
appropriately if a different prescription is favoured by future data.

\subsubsection{Radiative corrections}
\noindent
Radiative corrections are large for `classic' DIS measurements based
on the reconstruction of kinematics from the scattered lepton only.
Expressions for the fixed target case involving only virtual photon 
exchange were first worked out by Mo and Tsai~\cite{motsai}
and refined by the Dubna group~\cite{dubna}. 
In preparation for HERA, calculations were performed 
by a number of groups~\cite{radcor} and a number of computer codes
developed. Recently a new code, HECTOR~\cite{hector}, based on 
HELIOS~\cite{helios} and TERAD91~\cite{terad91},
 has been 
developed for $ep$ colliders, but is also applicable to $\mu N$ 
scattering. It provides a
framework for semi-analytical calculations using a variety of 
kinematic reconstruction schemes. Another important code for
corrections at HERA is HERACLES~\cite{heracles}. HERACLES
may also be interfaced through DJANGO~\cite{django} to codes 
simulating the DIS final state such as LEPTO~\cite{lepto} or
ARIADNE~\cite{ariadne}.

\begin{figure}[htbp]
\vspace*{13pt}
\begin{center}
\psfig{figure=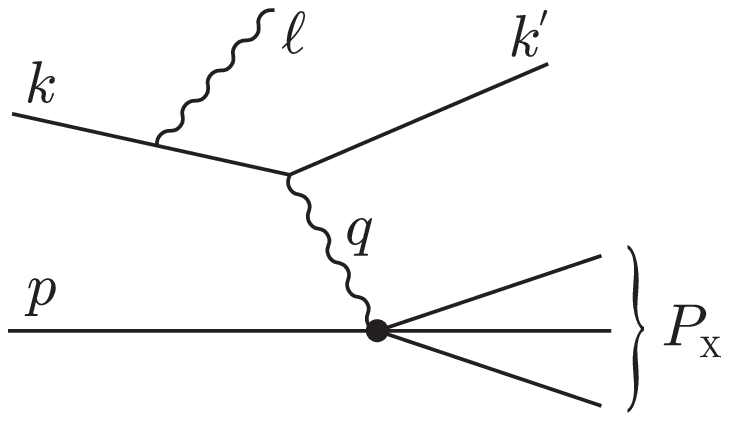,bbllx=140pt,bblly=360pt,bburx=350pt,bbury=480pt,height=5cm} 
\fcaption{Schematic diagram for radiative lepton-nucleon scattering.}
\label{fig:gamrad}
\end{center}
\end{figure}

We will not consider the full electroweak calculations here, but will
concentrate on the important features of QED corrections for
NC reactions. The notation for single photon emission from the 
charged lepton lines is shown in Fig.~\ref{fig:gamrad}. Schematically 
the cross-section including radiative effects is related to the Born
cross-section by an expression of the form
\begin{equation}
{d\sigma\over d{\bf v}}^{rad}=\int d{\bf v^\prime} K({\bf v},
{\bf v^\prime}){d\sigma\over d{\bf v^\prime}}^{Born}
\label{eq:conv1}
\end{equation}
where ${\bf v}$ and ${\bf v^\prime}$ are 2-dimensional vectors
representing the kinematic variables $(x,Q^2)$, and 
$K({\bf v},{\bf v^\prime})$ is the radiative kernel.

The corrections are of two sorts, virtual corrections and the infra-red 
part of real photon emission which give an overall modification to
the Born cross-section and non-infra-red real photon emission in which
the photon is not identified. The radiative process $ep\to e\gamma X$
is dominated by photon emission from the initial and final lepton. In
principle there is also interference between the two. Consider first the
situation when the kinematics is reconstructed using the scattered
lepton only and the radiation is from the initial lepton (ISR). The
standard definitions of the kinematic variables are (where $\ell$
indicates that they have been calculated from the lepton side),
\begin{equation}
Q^2_\ell=-(k-k^\prime)^2~~~~x_\ell={Q^2_\ell\over 2p.(k-k^\prime)}~~~~
y_\ell={p.(k-k^\prime)\over p.k}.
\end{equation} 
The effect of the radiation of a photon of 4-momentum $l$, is to
lower the actual centre of mass energy in the $ep$ interaction from
$s=(k+p)^2$ to $s^\prime=(k-l+p)^2$. The true kinematic variables
are correctly reconstructed from the hadron side
\begin{equation}
Q^2_h=-(k-k^\prime-l)^2~~~~x_h={Q^2_h\over 2p.(k-k^\prime-l)}~~~~
y_h={p.(k-k^\prime-l)\over p.(k-l)}.
\end{equation}
 
\begin{figure}[htbp]
\vspace*{13pt}
\begin{center}
\psfig{figure=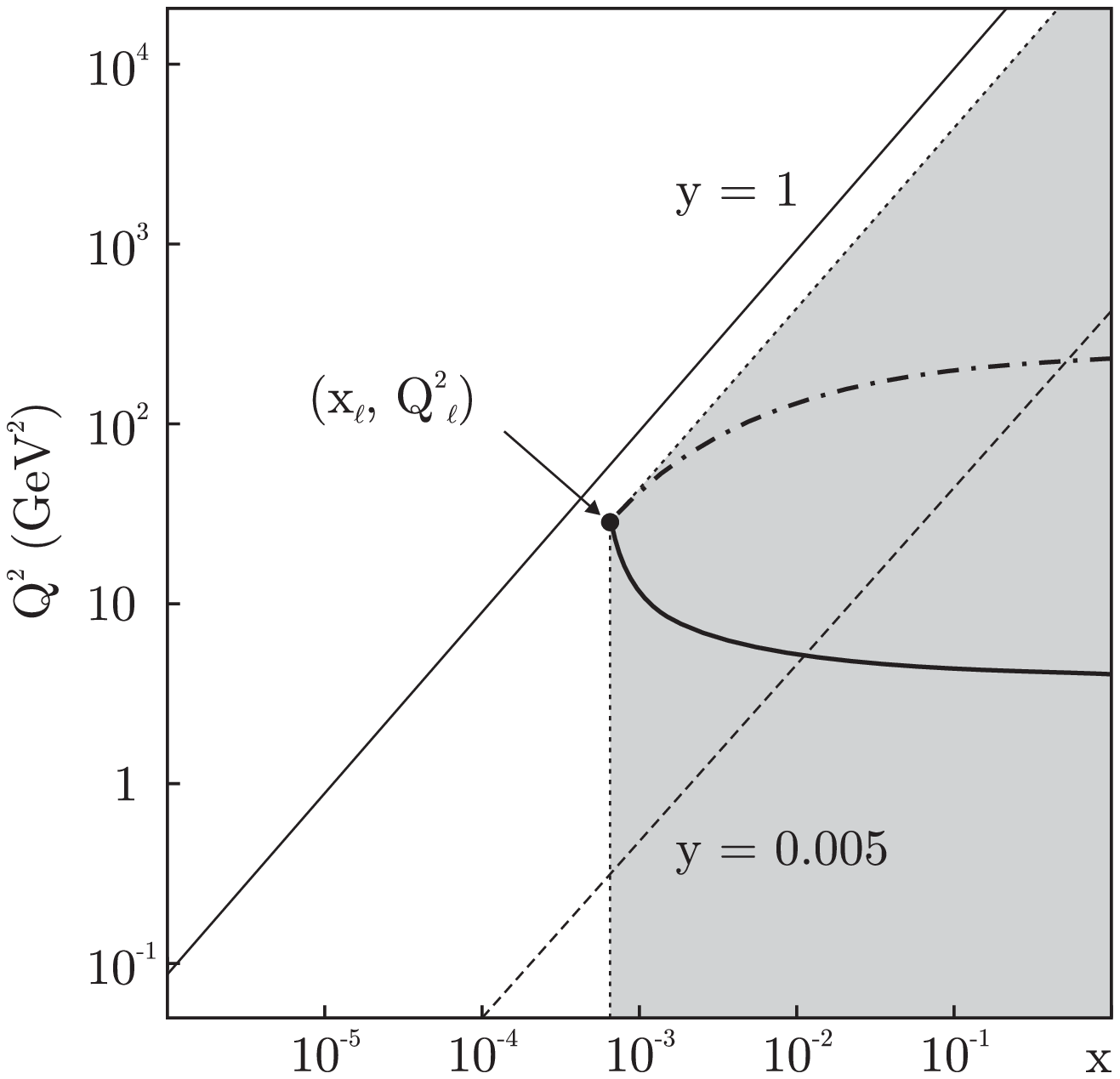,bbllx=-70pt,bblly=180pt,bburx=405pt,bbury=625pt,height=7cm} 
\fcaption{Phase space for one-photon emission at fixed 
$(x_\ell,Q^2_\ell)$. The full curve shows the s-peak (photon collinear
with incoming lepton) and the dash-dotted curve the p-peak (photon
collinear with the outgoing lepton) as the emitted photon energy ranges 
from zero to its kinematically allowed maximum.}
\label{fig:radxq2}
\end{center}
\end{figure}

\noindent
For collinear ISR the lepton side and true variables are related by
\begin{equation}
Q^2_h=z_i Q^2_\ell~~~~~~~~x_h={x_\ell y_\ell z_i\over y_\ell+z_i-1}
\label{eq:isrxq}
\end{equation}
where $z_i$ is the fractional energy loss from the initial state lepton
$z_i=(E-E_\gamma)/E$. 
Similar expressions can be deduced in the case of collinear radiation
from the final state lepton (FSR).

In terms of the variables $x$ and $y$, the radiative kernel $K$
 has to be integrated over the domain $x_\ell\le x\le 
1,~0\le y\le y_\ell$ (shaded region in Fig.~\ref{fig:radxq2}). 
Particularly for electrons, the radiative kernel is
large in three regions: (i) the `s-peak' region corresponding to 
collinear ISR and following a curve in the ($x,Q^2$) plane given by 
Eq.~\ref{eq:isrxq} and shown as the full curve in the figure; 
(ii) the `p-peak' region corresponding to collinear FSR and shown as 
a dash-dotted line; (iii) the `t-peak' or Compton region which occurs 
as $Q^2\to 0$ and may be viewed as Compton scattering from the 
almost real exchange photon.
The size of the contribution from the `peak' regions is characterised
by $\ln(m^2_\ell/Q^2)$ and is thus less pronounced for muon scattering.
In order to calculate the radiative corrections, one needs to know the
structure functions from the point measured down to very small values 
of $Q^2$ as the shaded region in Fig.~\ref{fig:radxq2} shows. Not
only is data poorly measured in parts of the integration domain, but it may
also include some of the region that is being measured in the same
experiment. The procedure adopted for fixed target experiments~\cite{bbkk}    
is to use a phenomenological parametrization of $F_2(x,Q^2)$ chosen to give
a good representation of the data over a wide range of $Q^2$ and 
particularly at very small values. The fit parameters may be improved
during unfolding. One of the biggest uncertainties in this procedure
comes from lack of knowledge of $R$. The radiative corrections are
sizeable for large $y$ and small $x$, for example in the lowest $x$ bins
of the E665 experiment the corrections reach 40\% at the largest $Q^2$
values.\cite{kotwal}

At HERA, even though the lepton beam is composed of electrons or positrons,
the radiative corrections turn out to be less severe. There are two reasons
for this. The first is that since the scattered electron is measured by 
calorimetric methods the energy lost through FSR will usually be deposited 
in the same calorimeter cell as the electron itself. The second and more
important is that kinematic information is measured from both the electron 
and hadron sides. Even if the hadron information is not used in the final
reconstruction of $x$ and $Q^2$, requiring that, for example, 
$|y_e-y_{JB}|$ is less than some small value will limit the energy
lost through hard radiation. In practice the variable that is often used
by the HERA experiments is $\delta=\sum_iE_i(1-\cos\theta_i)$ where
the sum runs over all calorimeter cells and $\theta_i$ is the angle 
with respect to the proton beam direction. For a completely hermetic
detector and in the absence of ISR, $\delta=2E_e$, where $E_e$ is the
HERA $e^\pm$ beam energy. The measured values of $\delta$ are lowered
by losses in rear beam-pipe (electron side), particularly
real or nearly real photoproduction events, and ISR. Thus requiring
a minimum value of $\delta$ removes both ISR events and photoproduction
background. Over the region of the $(x,Q^2)$ plane used for $F_2$
measurements at HERA, radiative corrections do not exceed 10\% and are
typically a factor 2 smaller. A summary of radiative corrections at high
$y$ for different reconstruction methods at HERA, calculated using the
HECTOR code, is given in Bardin et al~\cite{bardin}.

\subsubsection{Extracting $F_2$ from the observed events}
\noindent

Before $F_2$ values can be extracted from the measured cross-section
residual backgrounds must be subtracted. Off-target scattering in the 
fixed target case or beam-gas interaction at HERA is estimated from 
empty target runs (fixed target) or the  use of the non-colliding 
bunches (HERA). In the case of HERA, background from real photoproduction
processes in which a false scattered electron is identified is a
serious background at high $y$. This is estimated from the shape of
the $\delta$ distribution, from Monte Carlo simulations and using tagged
photoproduction events. 

Although there are considerable differences in detail, all experiments
follow much the same procedure in principle. $F_2$ is extracted from the
observed distribution of events by unfolding
\begin{equation}
N^{obs}={\cal L}\int_{bin} d{\bf v_m} \int d{\bf v} {\cal A}_K({\bf v_m},
{\bf v})F_2({\bf v})
\label{eq:conv3}
\end{equation}
where $N^{obs}$ is the number of observed events in a bin (after
background subtraction) and 
${\cal A}_K$ is a convolution of the radiative kernel $K$ and
the acceptance function $A$. The quantities ${\bf v}$ and ${\bf v_m}$
are 2 dimensional vectors representing the true and
measured kinematic variables ($x$,$Q^2$) respectively. With the advent
of DIS Monte Carlo event generators incorporating radiative effects
${\cal A}_K$ is usually determined in one step. For some of the older
fixed target experiments, the radiative corrections were calculated
semi-analytically and only the acceptance and resolution smearing function
$A$ was determined from the detector Monte Carlo code. In all cases an
iterative approach is followed as corrections depend on the shape
of the cross-section that is being measured.

A wide variety of unfolding methods have been proposed and more details
are given in the papers by Blobel~\cite{blobel},
D'Agostini~\cite{dag1} and Zech~\cite{zech}. The more complicated
methods are analogous to matrix inversion and using such methods
one is less sensitive to the Monte Carlo simulation and the unfolded data
tends to be `smoother'. Large Monte Carlo samples are needed to 
guarantee the necessary stability of the result.
However, in practice, if the Monte Carlo
data from which the unfolding matrix is generated describes the data
reasonably well the different methods give very similar results for the 
unfolded data~\cite{aqthesis}.

We give a very brief account of the simplest method -- known as 
bin-by-bin unfolding.
Starting from a recent set of parton density functions, 
giving $F_2^{input}$, DIS events are generated with the help 
of a lepton-nucleon physics
simulation code such as LEPTO~\cite{lepto}, ARIADNE~\cite{ariadne}
or GMUON~\cite{gmuon} (and as we have mentioned radiative effects may also be
included in the generation). 
These Monte Carlos not only describe the lepton-nucleon
interaction at the parton level using QCD matrix elements, they also deal
with the hadronization of the final state partons. The hadronization
prescriptions differ between the Monte-Carlos - the interested reader is 
referred to Sec.~\ref{sec:hfs} where this is discussed in context of 
measurements of the structure of the hadron final state. The events are then
passed through a Monte Carlo simulation of the detector (usually the
most time-consuming task) and subjected to the same set of cuts as the
real DIS data. Normalising the Monte Carlo data to the measured
luminosity gives the predicted array $N^{MC}(x_m,Q^2_m)$. This is now used
to correct the input structure functions using
\begin{equation}
F_2^{new}(x,Q^2)={N^{obs}(x_m,Q^2_m)\over N^{MC}(x_m,Q^2_m)}F_2^{old}(x,Q^2),
\label{eq:f2unf}
\end{equation}
where $F_2^{old}=F_2^{input}$ initially.
The $F_2^{new}$ values are then `smoothed' in some way, either fitting a
suitable function of $(x,Q^2)$ or most recently by a NLO QCD fit.
The smoothed structure function is then used to re-weight the Monte Carlo 
events such that $F_2^{new} = F_2^{input}$ for a second iteration of the
procedure, which produces a new array $N^{MC}(x_m,Q^2_m)$. The procedure is 
repeated, usually only 2-3 times, until a convergence criterion is satisfied
(typically that the change in $F_2$ is less than $0.2-0.5$\%). To account
properly for the migration between bins caused by finite detector 
resolutions, the simulated data must describe well the key
measured quantities such as the angle and energy of the scattered electron.

Strictly speaking $F_2$ is measured as an average over the bin area, 
$\Delta x\Delta Q^2$. However it is much more convenient to present
the data at a predetermined $(x,Q^2)$ point in the bin rather than
at $\langle x,Q^2\rangle_{bin}$. This requires a correction, known
as {\it bin-centering}, which is usually performed with the function used
to smooth the data during unfolding. This
introduces a small uncertainty in the quoted value of $F_2$.

\subsubsection{Systematic errors}
\noindent

One of the most important and time consuming tasks in the extraction
of structure function data is the estimation of systematic errors.
This requires a good knowledge of the detector and the methods used
for event reconstruction. Some uncertainties, such as those in beam
flux or luminosity will affect all data by a common factor, other
changes, for example energy scales or cut values, may affect the
distribution of the data in a correlated way. To estimate these latter
effects, a change is made and the complete structure function
extraction repeated so that the changes $\delta F$ can be determined. 
Most experimental teams now try
to group the systematic errors into classes that are more or less 
independent and some have tried to parameterize the correlated errors.
The actual values of the full correlated systematic errors are usually
available from the experimental groups. The phenomenological fits 
done by experimental groups take full account of their own
systematic errors but the problem of how to do this between experiments
and for global parton density  determinations, such as those discussed in
Sec.~\ref{sec:mpdf}, is still a subject of
active discussion~\cite{syserrs}.

\section{The data}
\label{sec:data}

In this section we consider the structure function data in some detail,
covering the range and quality of the data and the consistency between
the different experiments. A summary is given in Table~\ref{tab:datsum}.
All experiments measure $F_2$, those indicated have measured $R$ and 
only the CCFR experiment (of the ones we consider) has measured $xF_3$.
We note in passing that all the fixed target experiments have measured
the structure functions for nuclear targets as well as $p$ and $d$. Apart 
from the CCFR experiment we do not consider such data in this review.

\begin{table}[htb]
\tcaption{Summary of data from recent structure function experiments.
 An entry in the $R$ column indicates that $R$ was measured.
All the data referred in this table are available from 
the Durham HEPDATA database, at 
http://durpdg.dur.ac.uk/HEPDATA on the world wide web.
}
\centerline{\footnotesize\smalllineskip
\begin{tabular}{llllcccc}\\
 \hline
 Beam(s) & Targets & Experiment  & $Q^2$ (GeV$^2$) & 
 $x$ & $R$ & $Ref.$ \\
 \hline
 $e^-$ & p,d,A & SLAC  & $0.6-30$ & $0.06-0.9$ & $\surd$ & 
 \cite{whitlow92,whitlow_thesis}\\
 $\mu$ & p,d,A & BCDMS  & $7-260$ & $0.06-0.8$ & $\surd$ & 
 \cite{BCDMS,bcdmserr}\\
 $\mu$ & p,d,A & NMC  & $0.5-75$ & $0.0045-0.6$ & $\surd$ & 
 \cite{nmcdata1,nmcdata2}\\
 $\mu$ & p,d,A & E665  & $0.2-75$ & $8\cdot10^{-4}-0.6$ & - & 
 \cite{e665data,kotwal}\\ 
 $\nu,\bar{\nu}$ & Fe &CCFR & $1.-500.$ & $0.015-0.65$ & $\surd$ &
 \cite{ccfr_lambda,ccfr_gls,ccfr_seligman} \\
 $e^\pm$, p& - & H1  & $0.35-5000$ & $6\cdot10^{-6}-0.32$ & - & 
 \cite{h1n94,h194r,h1v95}\\
 $e^\pm$, p& - & ZEUS  & $0.16-5000$ & $3\cdot10^{-6}-0.5$ & - & 
 \cite{zv94,zn94,zbpc95}\\
\hline\\
\end{tabular}}
\label{tab:datsum}
\end{table}

When there is no ambiguity we shall refer to a dataset simply by the 
collaboration acronym or code name (SLAC, BCDMS, E665). For NMC 
and CCFR datasets we qualify the acronym by year of {\it publication}
of the data, e.g. NMC(95), but note that the data may sometimes appear
in global fits a year earlier. For HERA datasets we qualify the
acronym by the year in which the data was {\it collected}, 
thus H1(94) etc.

\subsection{$F_2^{\mu N}$ and $F_2^{eN}$}
\label{sec:f2lepto}

The overall situation on the data for $F_2$ is rather pleasing. The 
fixed target programme is complete and 
the final results published. Most of the fixed target results at low 
and moderate $Q^2$ are systematics limited and the level of
understanding achieved (better than 5\% systematic errors in many cases)
is a tribute to the hard work over many years of the experimental 
teams. The major inconsistencies between different experiments, 
particularly EMC versus BCDMS, have been resolved~\cite{emc_rev}
(in favour of BCDMS). 
The other major advance in the field is the addition of
data from HERA and in particular the first high statistics data from
the 1994 data taking period. 
The reach in the $(x,Q^2)$ has improved by two orders
of magnitude at both low $x$ and high $Q^2$. The data from H1 and
ZEUS are also nicely consistent with each other and with 
fixed target data. 
The use of improved $x$ and $Q^2$ reconstruction in the 
analysis of the 1994 data has given a small region of overlap in 
which a direct check on the relative normalisation of the collider 
and fixed target results may be made. A large overlap region will only
be possible if HERA is run at a lower centre of mass energy. 
Until recently fits to determine global parton distribution functions
 and QCD parameters were dominated by the high statistics 
fixed target data, particularly BCDMS and NMC. Now the HERA data have a
statistical weight that is beginning to match that of the other experiments
and as systematic errors diminish, the data are starting
to have an influence outside the HERA kinematic region.

\noindent

\subsubsection{Fixed target $F_2^{\mu N}$ data}
\label{sec:ftf2mu}
In addition to the experiments covered in detail in this review we 
note that structure function data from two of the earlier 
fixed target experiments are still used in the determination 
of parton distributions.
They are the SLAC experiments, following the re-analysis by 
Whitlow et al~\cite{whitlow92} in 1992, and the high statistics 
BCDMS experiment. 

Taken together the 8 SLAC $ep$ 
and $ed$ experiments provide data on $F_2^p$ and $F_2^d$ in the 
ranges $0.06\leq x\leq 0.90$ and $0.6\leq Q^2 \leq 30\,$GeV$^2$
(SLAC dataset). 
The normalisation uncertainties of the SLAC data are $\pm2.1$\%
for proton data and $\pm1.7$\% for deuteron data. Relative 
normalization between the different SLAC experiments is $\pm1.1$\%
and the other systematic uncertainties are at the level of $0.5-
2.0$\% from radiative corrections and $R$. Full details are given
by Whitlow~\cite{whitlow_thesis}.

BCDMS~\cite{BCDMS} studied $\mu p$ and $\mu d$ scattering at CERN 
using muon beam energies of $100, 120, 200$ and $280\,$GeV giving 
data on $F_2^p$ and $F_2^d$ in the ranges $0.06\leq x\leq 0.80$ and 
$7\leq Q^2 \leq 260\,$GeV$^2$ (BCDMS dataset). The systematic 
uncertainty in the absolute cross-section normalisation is 
$\pm3.0$\% and the uncertainty in the relative normalisation 
between data taken at
different beam energies is $\pm1$\%. The relative uncertainty in
the finite resolution of the spectrometer is 5\% and other
uncertainties are less than 1\%. More details are to be found in
ref.~\cite{bcdmserr}.

\begin{figure}[t]
\vspace*{13pt}
\begin{center}
\begin{tabular}[t]{ll}
\psfig{figure=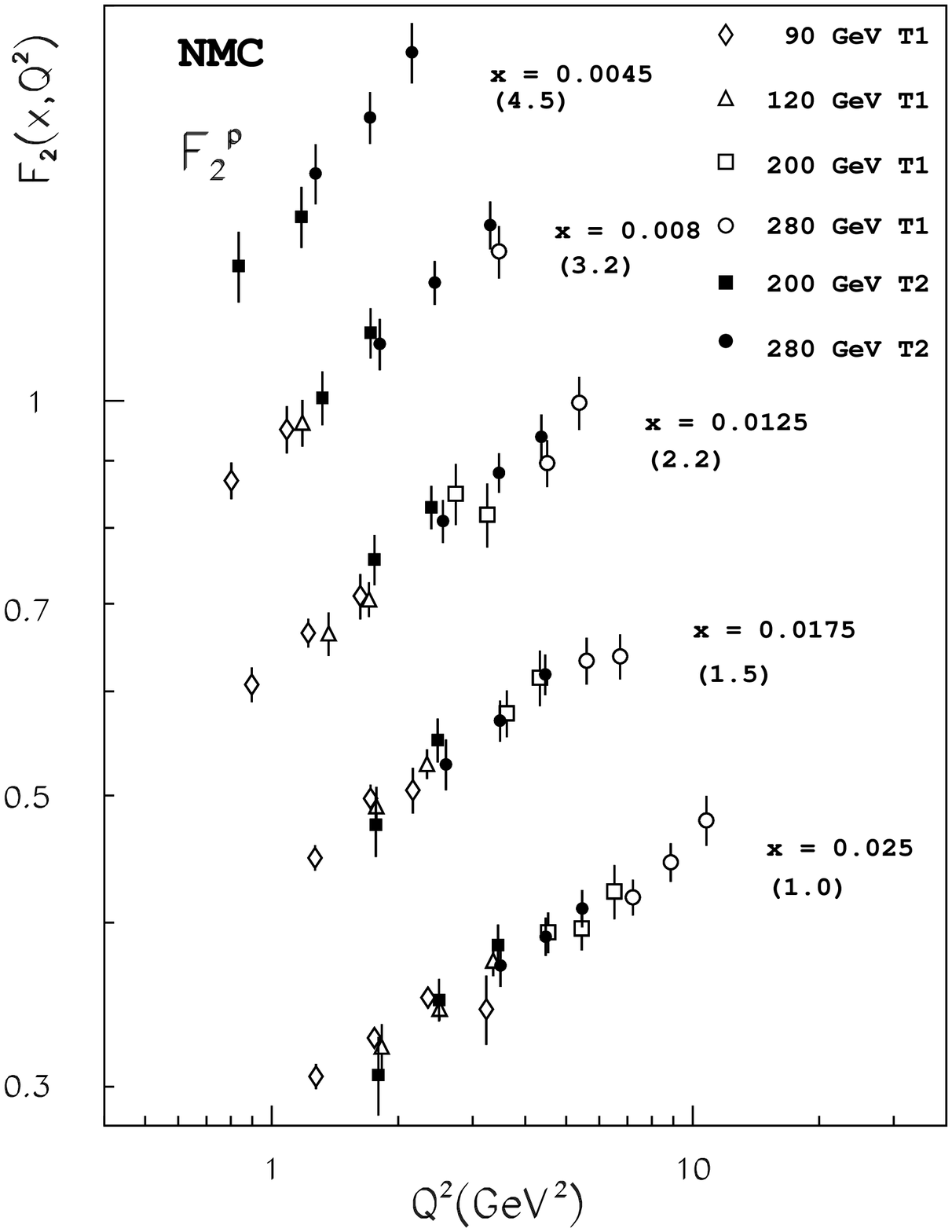,width=.45\textwidth} &
\psfig{figure=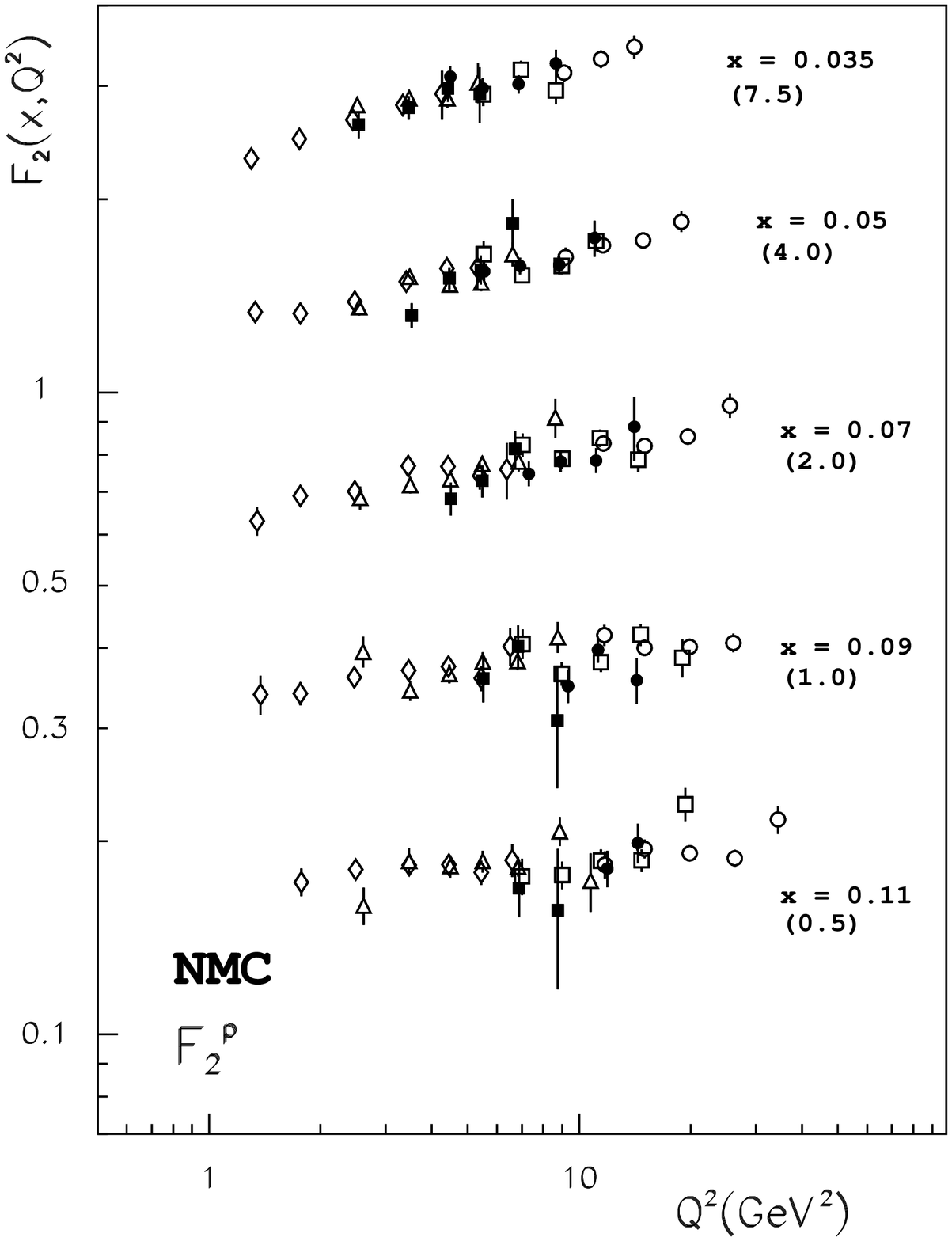,width=.45\textwidth} 
\end{tabular}
\fcaption{NMC(97) $F^p_2$ data at different muon beam energies and triggers
before averaging.
The inner error bar shows the statistical error and the full error bar
the total error (statistical error and systematic error 
added in quadrature). The data in each $x$ bin are scaled by the factors 
indicated in brackets.}
\label{fig:nmcdat1}
\end{center}
\end{figure}

The NMC collaboration has published data~\cite{nmcdata1} at all 
muon beam energies (90, 120, 200 and 280 GeV) using the large 
angle T1 trigger (NMC(95) dataset). For this data 
$R_{SLAC}$ was used in the extraction of $F_2$.
More recently NMC has presented results from the small angle T2 
trigger at beam energies of 200 and 280 GeV~\cite{kabuss,nmcdata2}. 
The small angle trigger data set covers the kinematic range 
$0.002<x<0.14$ and $0.5<Q^2<25\,$GeV$^2$. The enlarged data 
set permits the determination of $R$ as well as $F_2$ 
for both the proton and deuteron. 
The full range in $x$ and $Q^2$ now spanned by NMC $F_2$ data is
$0.002<x<0.6$ and $0.5<Q^2<75\,$GeV$^2$, and this is referred to as the 
NMC(97) dataset. These are the data shown in figures unless otherwise stated.  
The consistency of the data taken with different triggers
and beam energies is shown for $F_2^p$ in Fig.~\ref{fig:nmcdat1}. The 
figure also shows the characteristic features of pQCD scaling violations,
with the slope  ${\partial F_2\over \partial\ln Q^2}$ increasing
as $x$ decreases. To give the final NMC(97) data, data taken with 
different triggers and beam energies are averaged where appropriate. 
 
\begin{figure}[htbp]
\vspace*{13pt}
\begin{center}
\begin{tabular}[t]{ll}
\psfig{figure=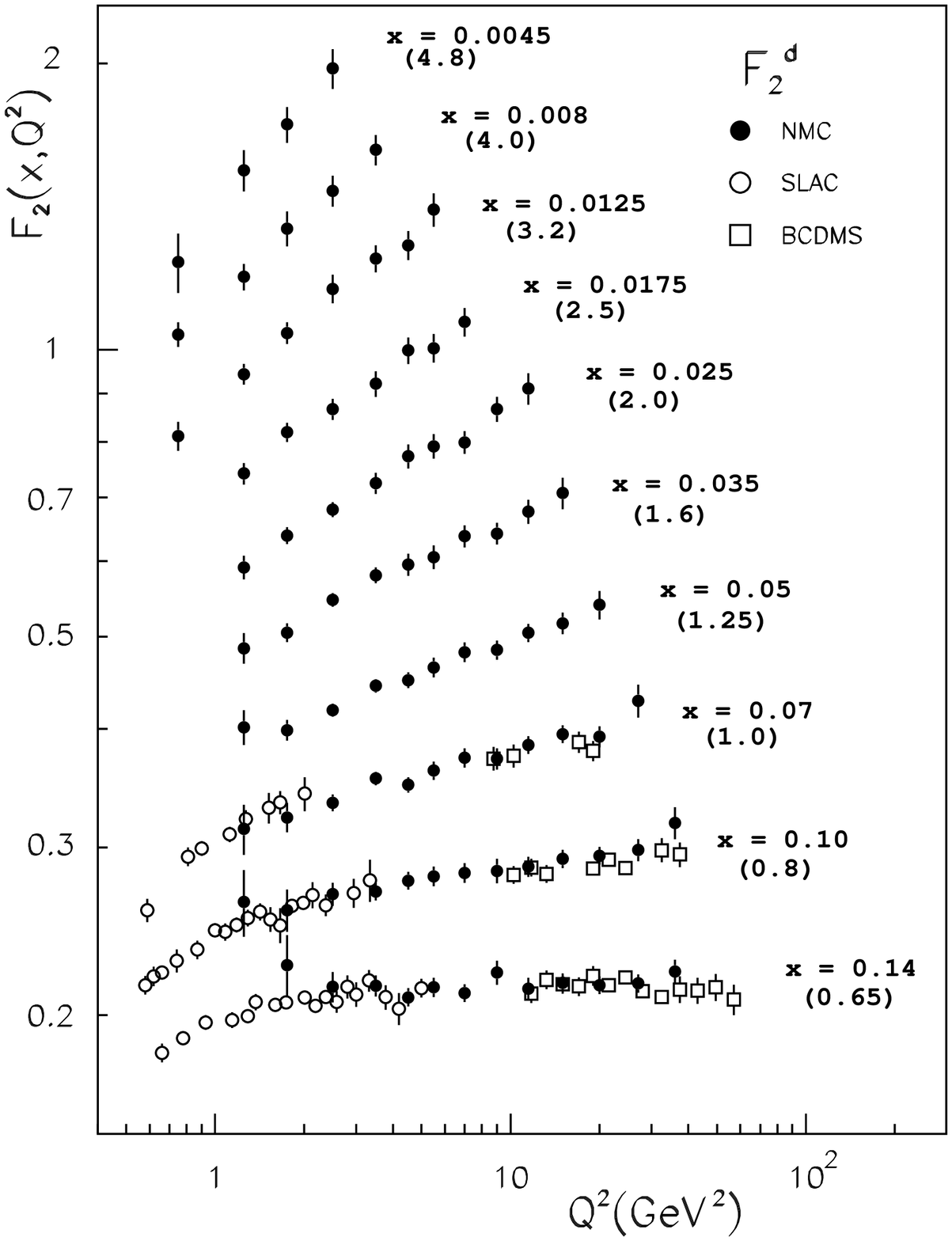,width=.45\textwidth} &
\psfig{figure=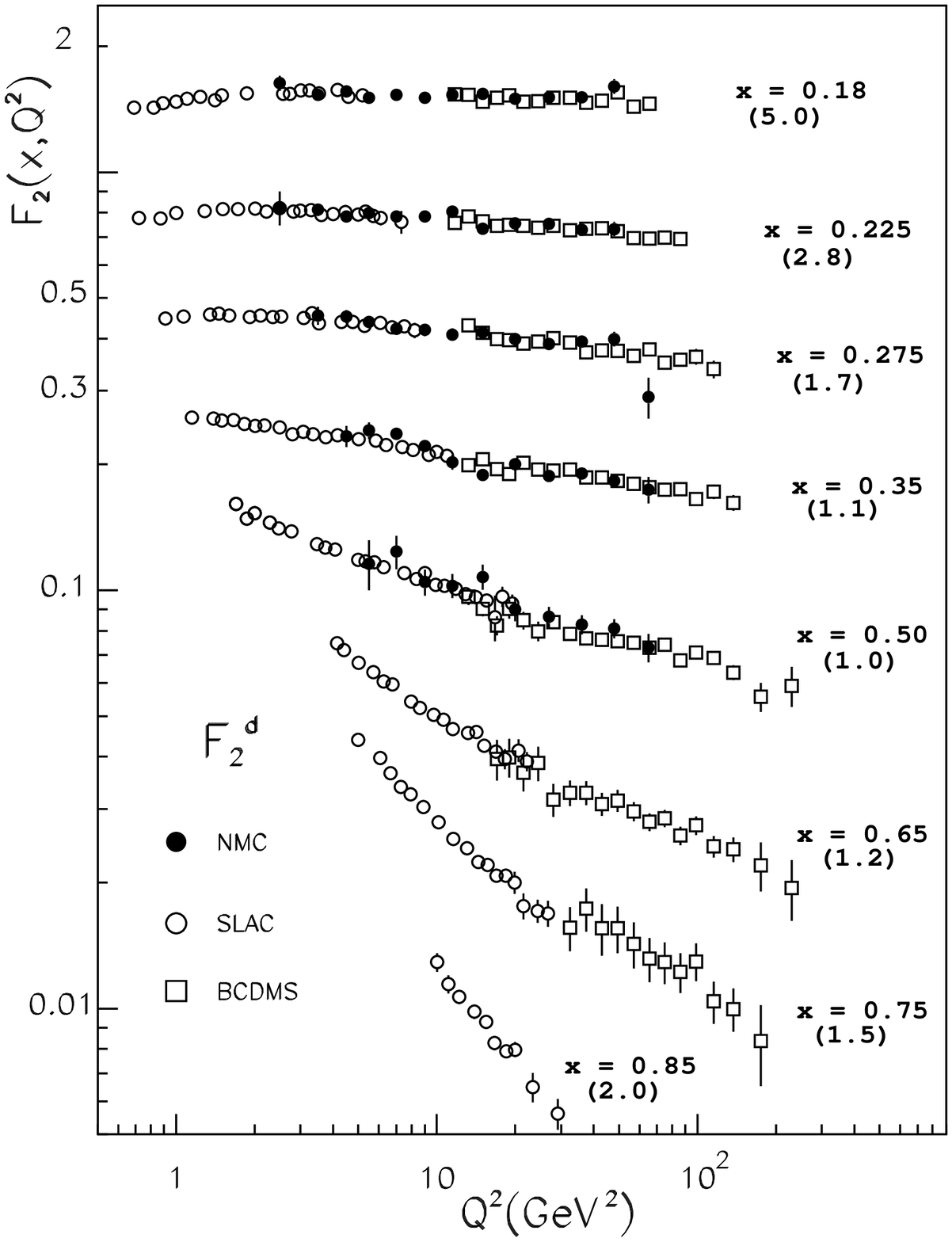,width=.45\textwidth} 
\end{tabular}
\fcaption{A comparison of $F^d_2$ from the NMC(97), SLAC and BCDMS
datasets. The SLAC and BCDMS data were rebinned to the NMC $x$ bins.
 The data in each $x$ bin are scaled by the factors indicated in brackets. 
The error bars
represent the total error, apart from the overall normalization. Relative
normalization is discussed in the text.}
\label{fig:nmcdat2}
\end{center}
\end{figure}

Two methods were used to extract $F_2$ values from the data. Method A used
$R_{SLAC}$~\cite{whitlow90} and in method B
$F_2$ and $R$ were both extracted from the data. More details on the second
method will be given in Sec.~\ref{sec:rdata} below. 
The overall normalization uncertainty is $\pm2.5$\%. Other
systematic errors are divided into five independent classes
(incident beam energy, scattered muon energy, acceptance, radiative
corrections and reconstruction) which are then added in quadrature 
to give the total systematic error on each $F_2$ measurement. Radiative
corrections are calculated using the formalism of the 
Dubna group~\cite{dubna}. Typically
the total systematic error is less than 2\% but it can rise to 4\%
on the boundaries of the measured region in ($x,Q^2$) at a given energy.
To estimate precise relative normalizations between their own data and those
from SLAC and BCDMS, the NMC group use a convenient
parametrization of $F_2^p$ and $F_2^d$ (and their uncertainties)
of the form~\cite{milsztajn}
\begin{equation}
F_2(x,Q^2)=A(x)\left({\ln(Q^2/\Lambda^2)\over 
\ln(Q_0^2/\Lambda^2)}\right)^{B(x)}(1+{C(x)\over Q^2})
\label{eq:nmcfit} 
\end{equation}
where $Q_0^2=20\,$GeV$^2$, $\Lambda=0.250\,$GeV and $A(x),B(x),C(x)$ 
are simple functions of $x$ with a total of 15 parameters. To 
determine the parameters the data points were fit using their statistical
errors and adjusting the normalization of each experiment within its
quoted range. The total uncertainties in the parameterized structure
functions were determined by taking into account the systematic errors
of the data, including correlations. The normalization shifts are 
given in Table~\ref{tab:nmcnorm}.  The data from the three different 
experiments are well
consistent with each other as can be seen from Fig.~\ref{fig:nmcdat2}. 

\begin{table}[htb]
\tcaption{
Normalization shifts for SLAC, BCDMS and NMC data from the fit using 
Eq.~\ref{eq:nmcfit}. The quoted overall normalization uncertainties are 
2\%, 3\% and 2\% for the SLAC, BCDMS and NMC experiments respectively.
}
\centerline{\footnotesize\smalllineskip
\begin{tabular}{lccc}\\
 \hline
 Data set & proton &  & deuteron \\
          &        & both & \\ 
 \hline
 SLAC & -0.4\% & & +0.9\% \\
 BCDMS & -1.8\% & & -0.7\% \\
 NMC 90 GeV & & -2.7\% & \\
 NMC 120 GeV & & +1.1\% & \\
 NMC 200 GeV T1 & & +1.1\% & \\
 NMC 280 GeV T1 & & +1.7\% & \\
 NMC 200 GeV T2 & & -2.9\% & \\
 NMC 280 GeV T2 & & +2.0\% & \\
\hline\\
\end{tabular}}
\label{tab:nmcnorm}
\end{table}

\begin{figure}[ht]
\vspace*{13pt}
\begin{center}
\psfig{figure=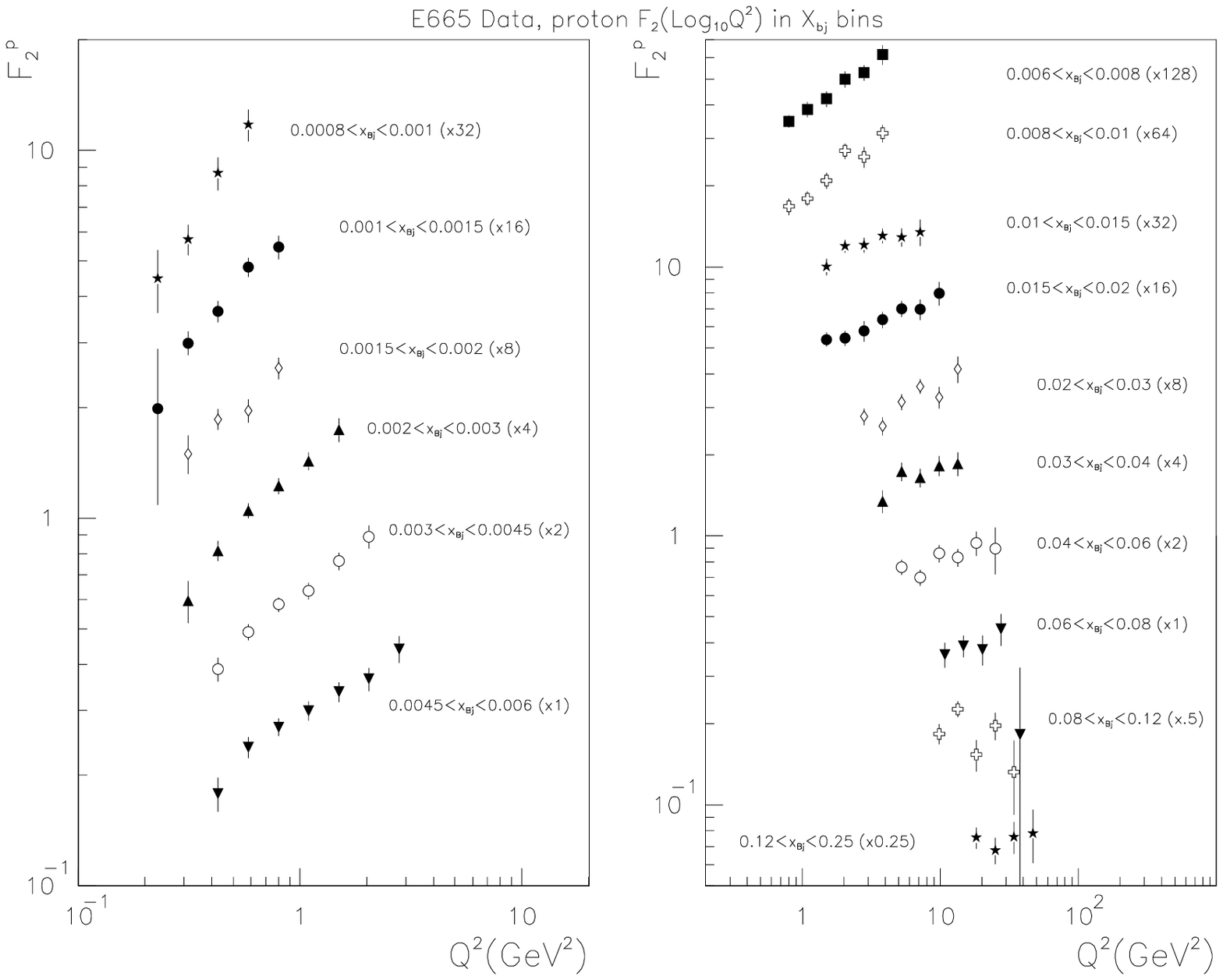,bbllx=7pt,bblly=310pt,bburx=523pt,bbury=680pt,height=7.0cm} 
\fcaption{$F_2^p$ data from the E665 experiment in bins of $x$ as
a function of $Q^2$. The errors shown are the sum in quadrature
of the statistical and systematic errors. The data in each $x$ bin
are scaled by the factors indicated in brackets.}
\label{fig:e665dat1}
\end{center}
\end{figure}

\begin{figure}[htbp]
\vspace*{13pt}
\begin{center}
\psfig{figure=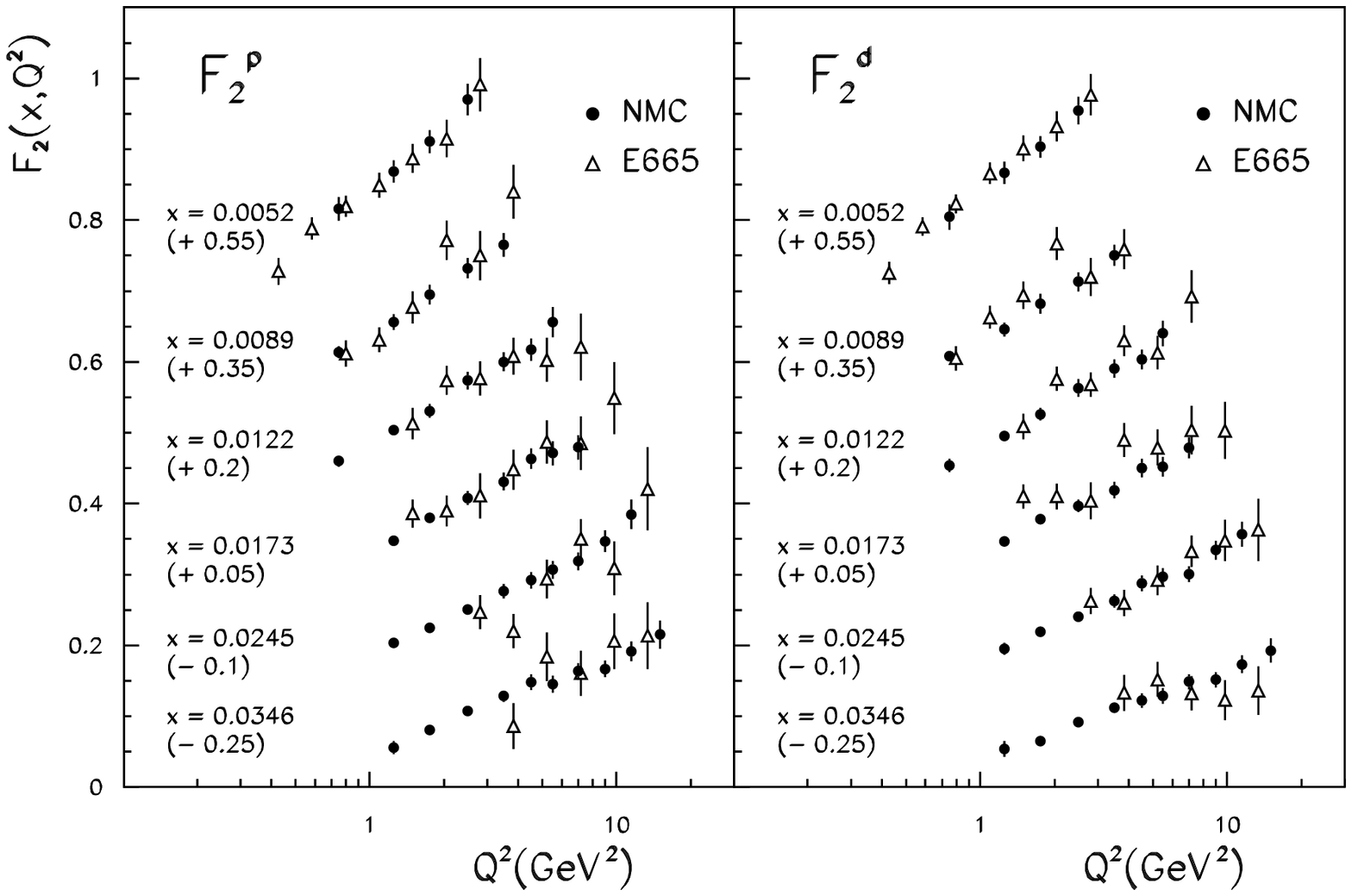,height=7.0cm} 
\fcaption{A comparison of $F^p_2$ data from the E665 and NMC(97) datasets.
The data in each $x$ bin are shifted  by the constants indicated
in brackets.}
\label{fig:e665dat2}
\end{center}
\end{figure}

The E665 collaboration has recently published the full analysis
of data on $F_2^d$ and $F_2^p$ at a mean muon beam energy of 
$470\,$GeV~\cite{e665data,kotwal}.
The data were collected using only the small angle trigger (see 
Sec.~\ref{sec:e665det}) and give a coverage of $0.0008<x<0.6$ and
$0.2<Q^2<75\,$GeV$^2$ (E665 dataset). To extract $F_2$ from 
the measured cross-sections,
$R_{SLAC}$ was used, but $R=0$ and $R=R_{QCD}$ were also used in the
estimate of systematic errors. Radiative corrections were calculated
using the same formalism as used by the NMC experiment. The uncertainties
in the overall normalizations are $\pm1.8$\% and $\pm1.9$\% for the
proton and deuterium data respectively. Other systematic errors are
considered in seven categories (trigger, reconstruction, absolute
energy scales, relative energy scales, radiative corrections, $R$,
bin centering). The total systematic error is formed by adding the
errors from the seven sources in quadrature. The largest sources of
uncertainty are from the trigger and reconstruction at low $x$ and
$Q^2$. In this region the total systematic error can reach 50\% at
the boundaries of the measured region. Away from this region the
typical total systematic error is around 3\% and is often comparable
to the statistical error. Fig.~\ref{fig:e665dat1} shows the data
for $F_2^p$ as a function of $Q^2$. Very large scaling violations at
small $x$, already seen in the NMC data, are manifest.

\noindent
There is a large overlap in the $(x,Q^2)$ regions explored by E665 and
NMC and in this common region the data from the two experiments agree
well both in shape and normalization -- this is shown in 
Fig.~\ref{fig:e665dat2}. 
Because of the very small $Q^2$ values reached by E665, this data has been
important in testing models for $F_2$ as $Q^2\to 0$. 

\subsubsection{HERA $F_2^{eN}$ data}

The first $F_2$ measurements at HERA (HERA(92))
were made using 1992 data~\cite{hera92}
and gave the first indication of the steep rise of $F_2$ as $x$ decreases.
The rise was confirmed by the much improved data from the 1993 
run (HERA(93))~\cite{hera93}. We shall concentrate here on the 
first really large samples provided by the 1994 run. In 1992-93 
HERA collided electrons and protons. From 1994 positrons have 
been used instead of
electrons as larger currents could be kept with longer 
lifetimes. To exploit the 
full potential of HERA at very high $Q^2$ large samples of both 
$e^- p$ and $e^+ p$ data with and without
polarization will eventually be required.

The ZEUS collaboration has published their 1994 results in two papers,
the first~\cite{zv94} contains the low $Q^2$ data from $58\,$nb$^{-1}$
of shifted vertex data (SVX) and initial state radiation events (ISR)
and the second~\cite{zn94} the analysis of the large nominal vertex 
sample (NVX), together giving the ZEUS(94) dataset. The low $Q^2$ data 
covers the kinematic region 
$1.5<Q^2<15\,$GeV$^2$, $3.5\cdot 10^{-5}<x<0.004$ and events are 
reconstructed using the E method. $R$ is calculated using $R_{QCD}$ 
(as calculated to lowest order from Eq.~\ref{eq:flqcd})
with parton distributions determined from the QCD NLO fit used to smooth the
data during unfolding. 
Sources of systematic error considered are:
positron finding and energy scale; positron angle; photoproduction
background; primary vertex reconstruction; varying the $\delta$ cut.
The total systematic error varies between 4 and 14\% (compared to the
statistical error which is in the range $4-9$\%). The overall normalization
error from the trigger and luminosity is 3\%. The ISR analysis used the
full 1994 sample and events were selected with a modification of the 
DIS selection to require a photon with energy satisfying
$6<E_{\gamma}<18\,$GeV tagged
in the luminosity photon calorimeter. The $\delta$ cut was replaced by
a cut on $\delta^\prime=\delta+2E_\gamma$. 
The main source of 
background for this 
technique, which  is the random coincidence of a DIS or photoproduction event
with a bremsstrahlung event $ep\to ep\gamma$,   is estimated
from a sample of events selected without the tagged photon to which
a photon is added with an energy selected from a random sampling
of the photon energy spectrum of bremsstrahlung events.
The level of this background is
below 10\% except in the lowest $x$ bins where it rises to a maximum
of 24\%. 

\begin{figure}[htbp]
\vspace*{13pt}
\begin{center}
\psfig{figure=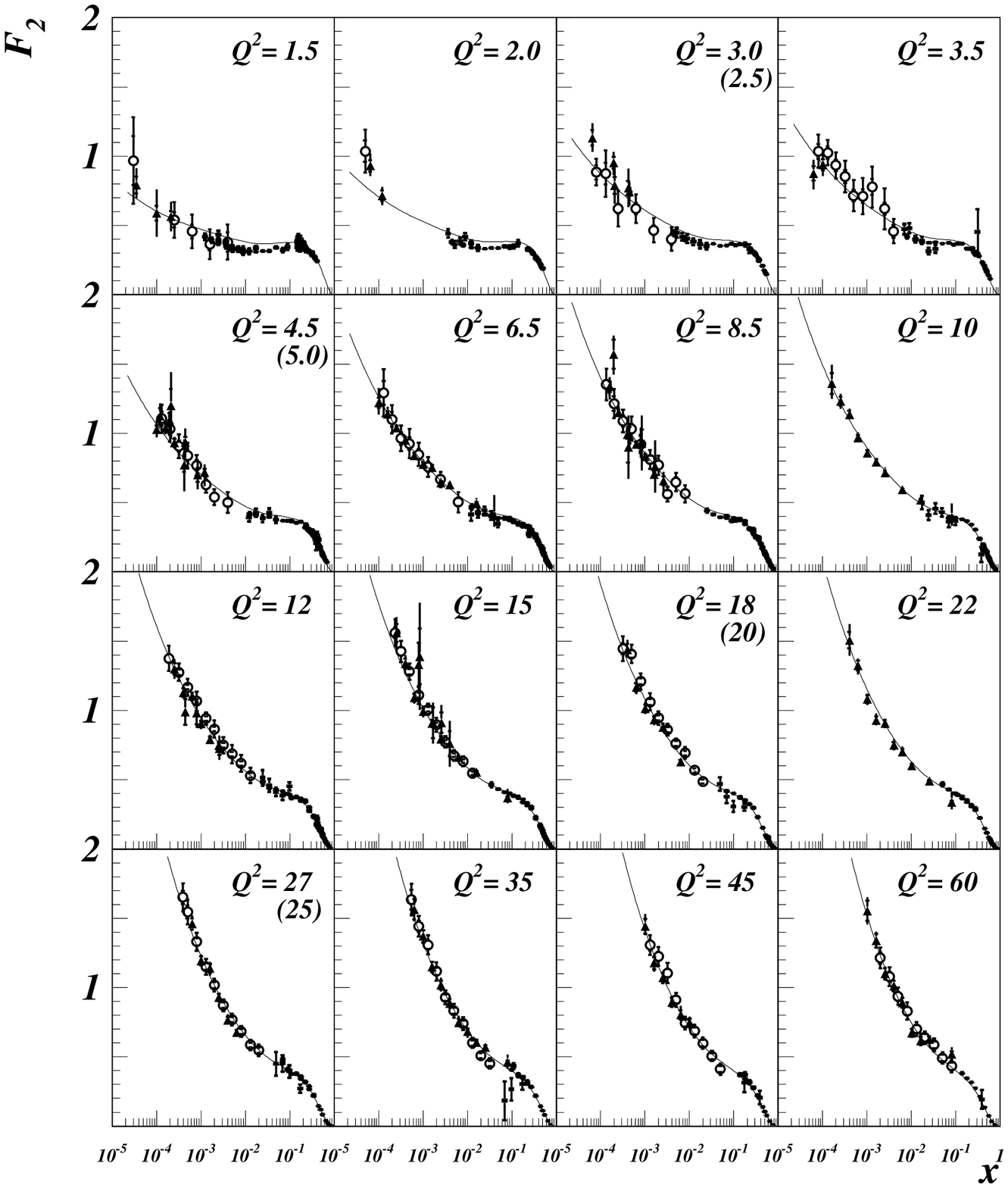,bbllx=50pt,bblly=70pt,bburx=650pt,bbury=770pt,height=.95\textheight} 
\fcaption{ZEUS(94) and H1(94) $F_2^p$ data at fixed $Q^2 \le 60\,$GeV$^2$. 
The curves shown are the NLO DGLAP QCD fit used to smooth the data during 
unfolding.}
\label{fig:zdat1}
\end{center}
\end{figure}

\begin{figure}[htbp]
\vspace*{13pt}
\begin{center}
\psfig{figure=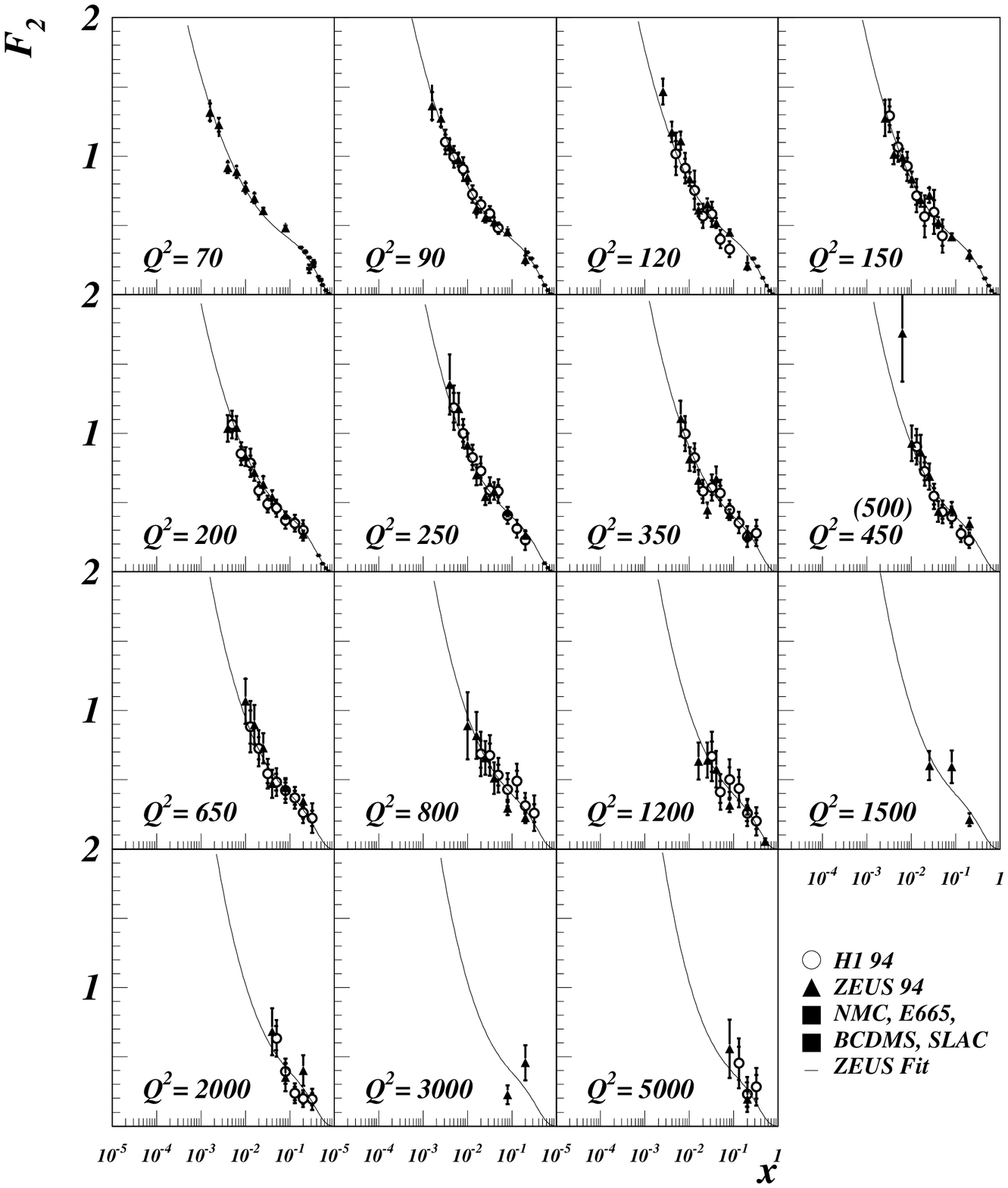,bbllx=50pt,bblly=70pt,bburx=650pt,bbury=770pt,height=.95\textheight} 
\fcaption{ZEUS(94) and H1(94) $F_2^p$ data at fixed $Q^2 \ge 60$ GeV$^2$. 
The curves shown are the NLO DGLAP 
QCD fit used to smooth the data during 
unfolding.}
\label{fig:zdat2}
\end{center}
\end{figure}

The ZEUS NVX sample~\cite{zn94} covers the region $3.5<Q^2<5000\,$GeV$^2$, 
$6.3\cdot 10^{-5}<x<0.5$, and events were reconstructed using the 
PT method. 
First order radiative effects were included in event simulation using
HERACLES. Higher order QED corrections, such as soft photon exponentiation
not included in HERACLES, were calculated using the programme HECTOR. These
additional corrections are small ($0.2-0.5$\%) and vary smoothly with $Q^2$. 
Systematic errors are grouped into six categories: positron finding;
positron angle; positron energy scale; hadronic energy scale and noise;
hadronic energy flow and photoproduction background. For the majority
of bins the total systematic error is below 5\% but increases to around
10\% as $y$ increases above $0.5$. The overall normalization uncertainty
is 2\%. $R$ is calculated from $R_{QCD}$ as in the ZEUS SVX analysis
and at high $Q^2$ the data are corrected to give $F_2$ as explained
in Sec.~\ref{sec:rcorr}. The ZEUS data are shown in Figs.~\ref{fig:zdat1}
and ~\ref{fig:zdat2},
which show the strong rise in $F_2$ as $x$ decreases.

H1 have published results from their 1994 sample using NVX, SVX and ISR 
data~\cite{h1n94} (H1(94) dataset).
 The E and $\Sigma$ kinematic reconstruction methods were
used at low ($<0.15$) and high ($> 0.15$)
$y$ respectively to give an overall $x$ resolution better
than 20\%. The coverage in $(x,Q^2)$ from all methods is 
$1.5<Q^2<5000\,$GeV$^2$ and $3\cdot10^{-5}<x<0.32$. 
For the ISR data a photon with energy $E_{\gamma} > 4 $ GeV was required,
and the minimum electron energy was lowered to $E_e = 8$ GeV.
To separate signal from background events overlapping with bremsstrahlung
events, the quantity $(E_{\gamma}-E_e(y_e-y_h))$ was required to be less 
than 0.5 (with $y_e$ and $y_h$ defined in Eqs.~\ref{eq:hk1} and~\ref{eq:yjb}).
The remaining background amounts to an average of 8\%, with at most 
15\% in one ($x, Q^2$) bin, and it was statistically subtracted.
For the value of $R$ the QCD prediction was taken with the GRV parton
distributions as input. 
The following 
sources of systematic errors were considered: positron energy scale; 
hadronic energy scale; positron angle; vertex reconstruction; radiative
corrections; photoproduction background. The typical total systematic error 
is around 5\%. For the ISR analysis there is an additional uncertainty
of around 2\% from the geometrical acceptance and relative energy calibration
of the luminosity photon tagger. For the NVX and SVX samples the overall
normalization uncertainties (from trigger and luminosity measurement) are
1.5\% and 3.9\% respectively. First order radiative effects were included
in the Monte Carlo simulation events using the HERACLES code. These were
checked using the HECTOR programme. 
The data are shown in Figs.~\ref{fig:zdat1} and ~\ref{fig:zdat2}.
The data points include also the high $y$ measurements made by 
H1~\cite{h194r}, 
primarily used to extract $R$, as discussed in Sec.~\ref{sec:rdata}.

\begin{figure}[htbp]
\vspace*{13pt}
\begin{center}
\psfig{figure=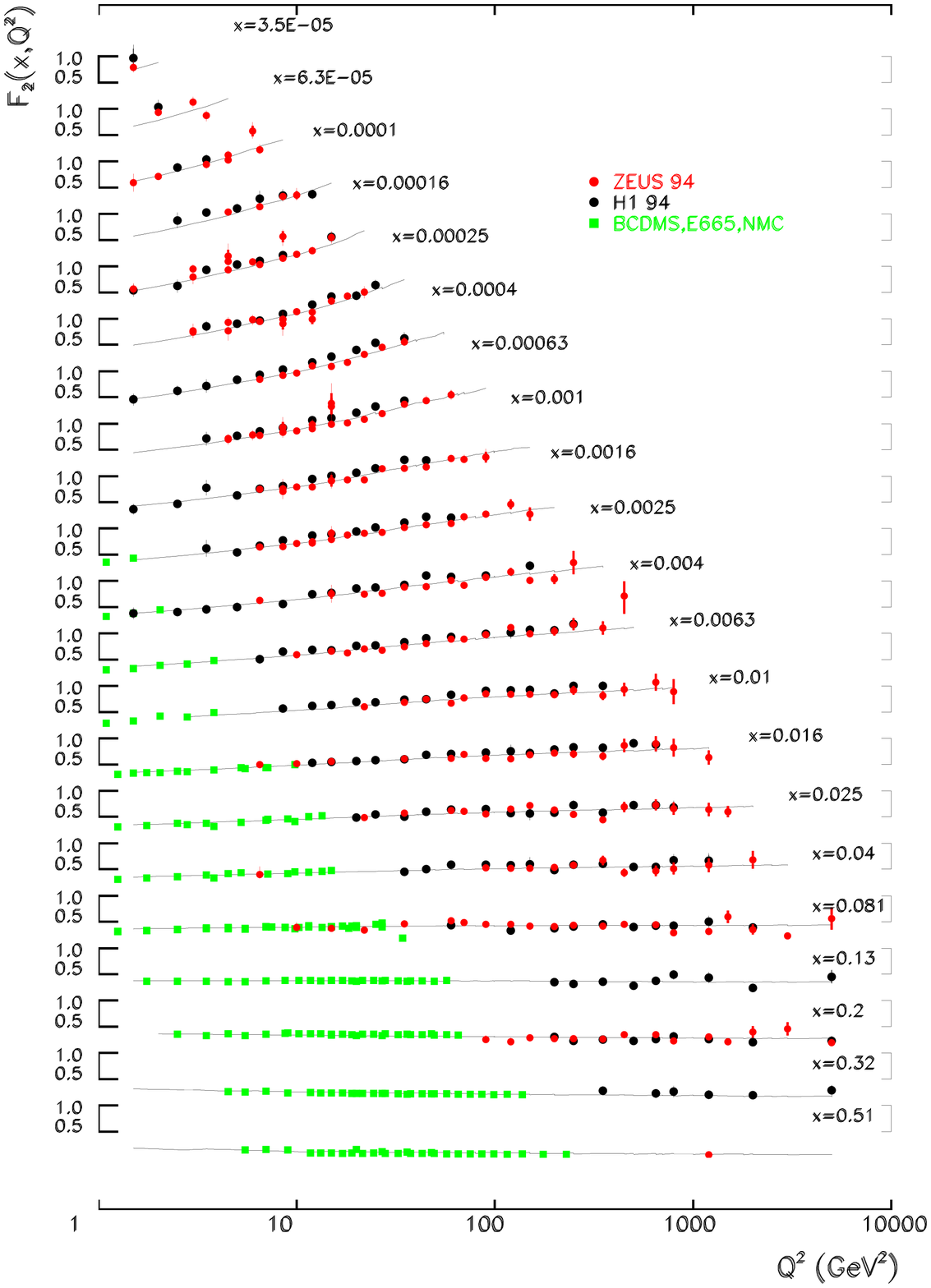,bbllx=0pt,bblly=0pt,bburx=570pt,bbury=840pt,height=.95\textheight} 
\fcaption{$F_2^p$ data from HERA(94) and fixed target experiments at fixed
$x$ as a function of $Q^2$. The curves shown are the NLO DGLAP 
QCD fit used to smooth the data during 
unfolding.}
\label{fig:zeusall_q2}
\end{center}
\end{figure}

\begin{figure}[htbp]
\vspace*{13pt}
\begin{center}
\psfig{figure=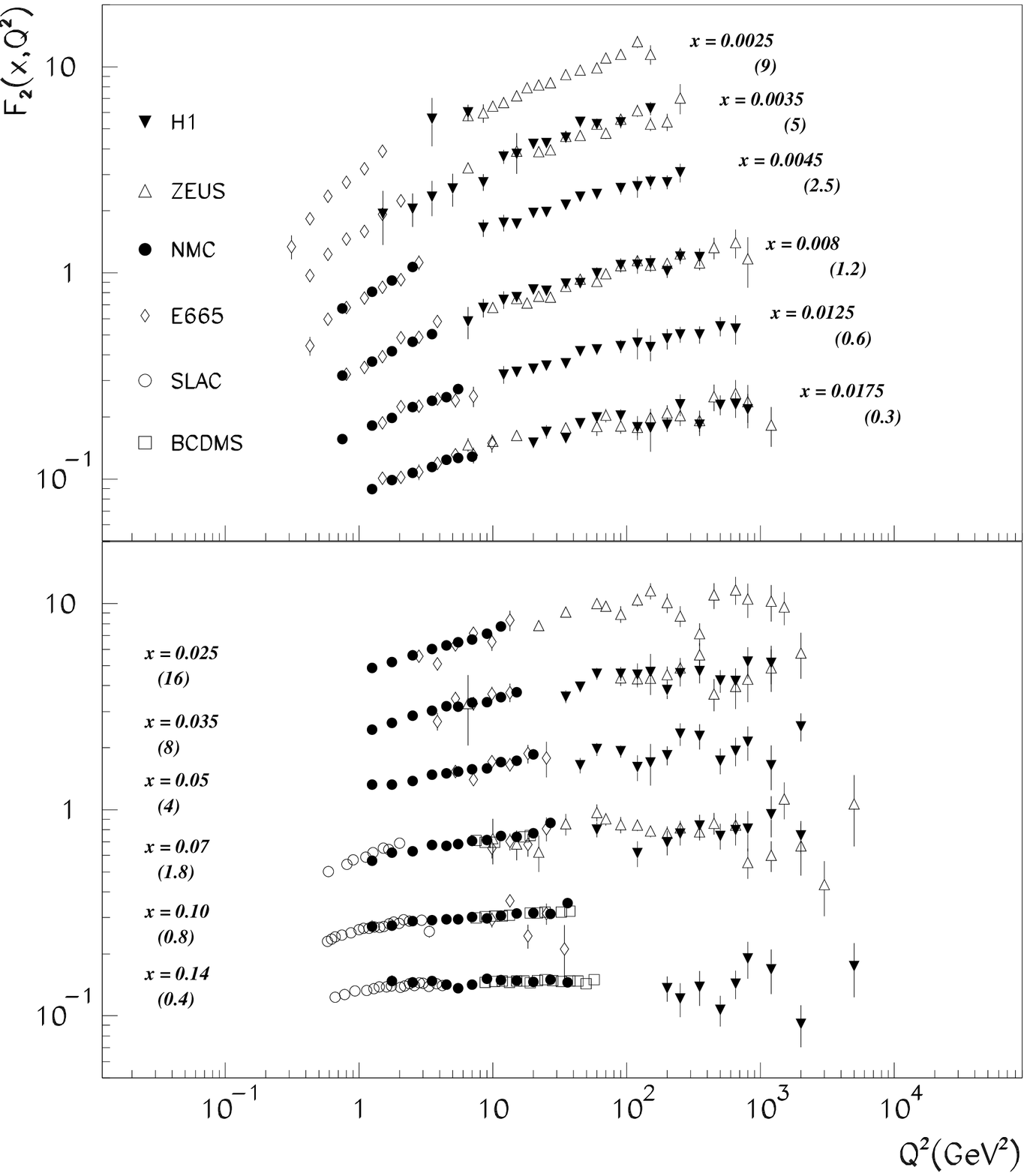,bbllx=20pt,bblly=0pt,bburx=690pt,bbury=750pt,height=.95\textheight} 
\fcaption{$F_2^p$ data from HERA(94) and fixed target experiments at fixed
$x$ as a function of $Q^2$. The data in each $x$ bin are scaled by the factors
indicated in brackets.
Here the NMC data are those from the NMC(95) dataset.}
\label{fig:f2_fixed}
\end{center}
\end{figure}

The $F_2^p$ data from the two HERA experiments are nicely consistent with
each other and appear to connect smoothly to data from the fixed target
experiments. This is shown in Fig.~\ref{fig:zeusall_q2}, which also shows a
NLO DGLAP QCD fit. Using the extended kinematic reconstruction methods for 
HERA data now gives a small region of overlap with the fixed target regime
in both $x$ and $Q^2$.
Fig.~\ref{fig:f2_fixed} shows the comparison in more detail for the 
region $0.0025<x<0.14$.
 The absolute normalizations of the 
data from the two types of experiment are in good agreement. This is a
significant achievement and is of considerable importance when it comes
to combining fixed target and HERA data for further physics analysis.

The major new information on structure functions from HERA is the very
strong rise of $F_2$ as $x$ decreases. This has now been measured with
considerable precision. The slope ${\partial F_2\over \partial x}$ does
decrease as $Q^2$ decreases, but there is still a noticeable rise at
$Q^2$ values as low as $1.5\,$GeV$^2$. The apparent success of NLO QCD
fits in describing the data over the $Q^2$ range $1.5-5000\,$GeV$^2$ 
and for $x$ values down to $3.5\times 10^{-5}$ as
shown in the figures above, is also somewhat surprising  and will be 
discussed in greater detail in Sec.~7 below.

As $Q^2\to 0$  $F_2\to 0$ and at
some smallish value of $Q^2$ the pQCD description afforded by the DGLAP
evolution must breakdown. The surprise at HERA is that 
the size of the transition region, from photoproduction to the deep inelastic
regime, is so narrow. To measure structure functions at 
very small $Q^2$ requires a detector that can be triggered by, and  can
reconstruct,  the scattered muon or electron at very small scattering 
angles. Of the fixed target experiments E665 has this capability and 
has published $F_2$ data down to $Q^2=0.2\,$GeV$^2$. 
At HERA the first attempt to explore this region was by use of SVX data, 
but that still had a lower limit of about $1.5\,$GeV$^2$. In 1995, to cover 
the transition region  $0<Q^2<1.5\,$GeV$^2$,
ZEUS~\cite{zbpc95} added a special detector near the rear beam-pipe 
and the new rear detectors of H1~\cite{h1v95} gave them improved coverage
at small electron scattering angles.

The  ZEUS BPC data (ZEUS(95)BPC data set) 
covers the region $0.11<Q^2<0.65\,$GeV$^2$, 
$2\cdot 10^{-6}<x<6\cdot 10^{-5}$ and is based on $1.65\,$pb$^{-1}$
data collected in 1995. The largest 
systematic errors are the position and energy scale of the BPC and the 
estimate of the photoproduction background. The total systematic error 
varies from 6\% to 11\% at the highest $y$ values. In addition there is 
an overall normalization uncertainty of 2.4\%.
 
ZEUS announced preliminary low $Q^2$ data from the 1995
shifted vertex run (ZEUS(95)SVX)~\cite{zsvx95} at the DIS97 meeting.
 The method used was
essentially the same as that described above for the 1994 SVX analysis,
the primary interaction point was shifted $70\,$cm in the proton beam
direction. The increased statistics from $236\,$nb$^{-1}$ allowed
$F_2$ to be extracted in seven $Q^2$ bins between 0.65 and $4.64\,$GeV$^2$.

\begin{figure}[htbp]
\vspace*{13pt}
\begin{center}
\psfig{figure=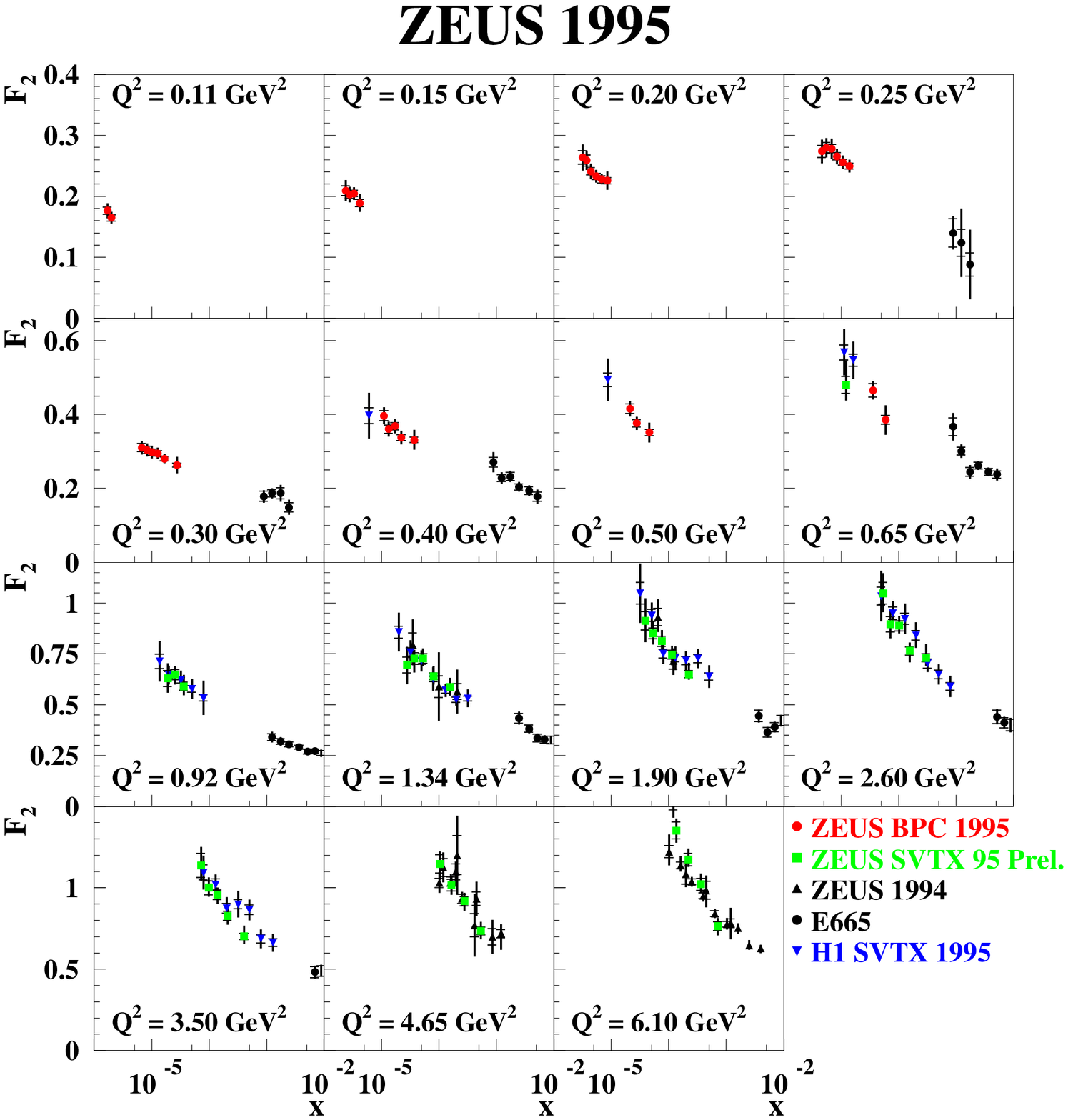,bbllx=50pt,bblly=70pt,bburx=650pt,bbury=770pt,height=.95\textheight} 
\fcaption{$F_2^p$ at very small values of $Q^2$ from the E665, ZEUS(94), 
ZEUS(95)BPC, ZEUS(95)SVX (preliminary) and H1(95)SVX data sets}
\label{fig:loqf2}
\end{center}
\end{figure}

Using their new rear detector and the 1995 SVX sample of $114\,$nb$^{-1}$, 
H1 cover the region $0.35<Q^2<3.5\,$GeV$^2$ and $x>6\cdot 10^{-6}$ (H1(95)SVX
data set).
The  estimate of systematic errors are in the range $5-10$\%, 
from positron energy and angle determination, radiative corrections and 
photoproduction background. There is an overall 3\% normalization 
error from the luminosity measurement. The very low $Q^2$ $F_2$ data 
are shown in Fig.~\ref{fig:loqf2}. The phenomenology of this data will
be discussed in Sec.~\ref{sec:lowq2}.

\subsection{$F_2^{\nu N}$ and $xF_3^{\nu N}$}
\label{sec:xf3dat}

\begin{figure}[htbp]
\vspace*{13pt}
\begin{center}
\psfig{figure=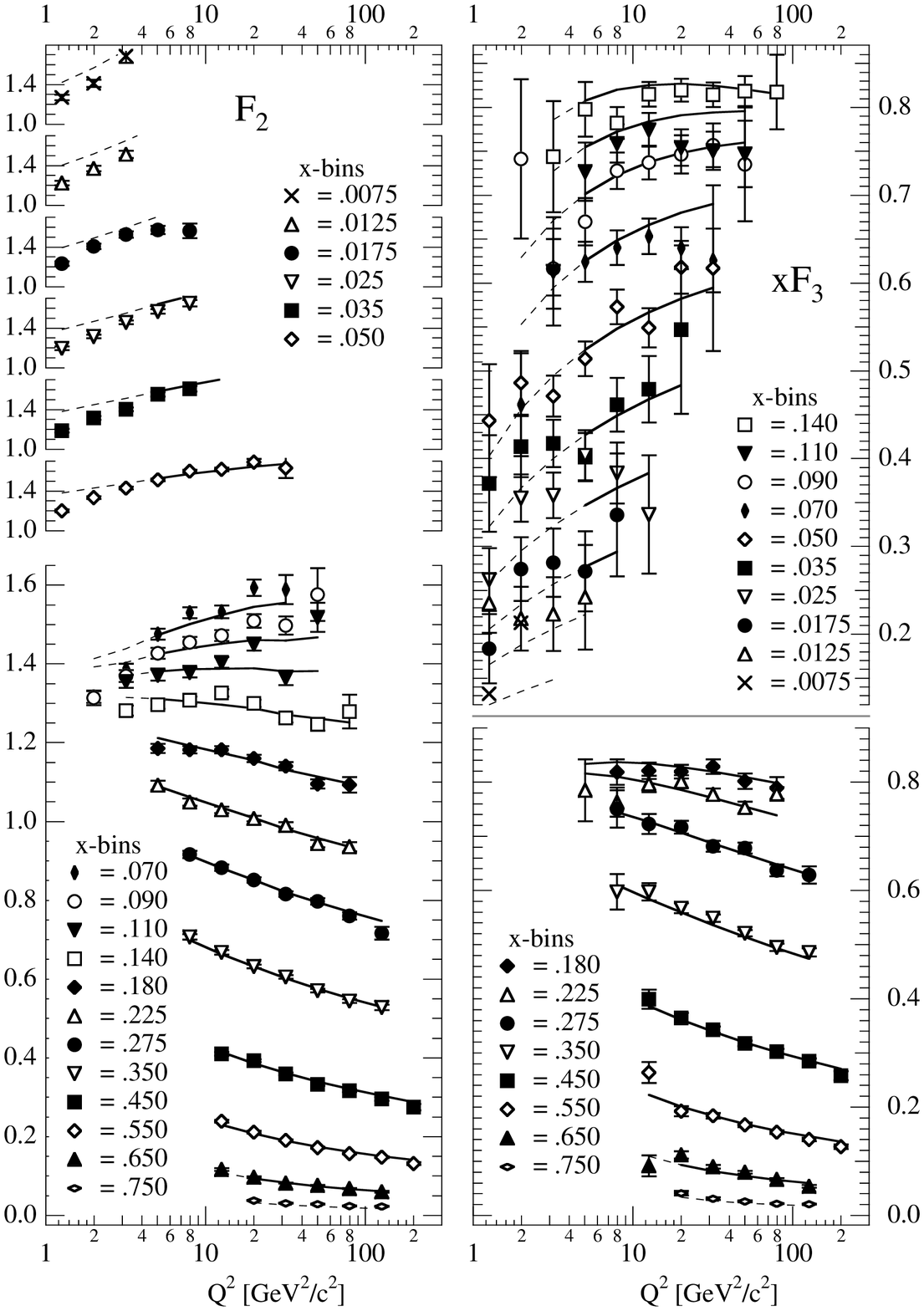,bbllx=-20pt,bblly=30pt,bburx=630pt,bbury=760pt,height=14cm} 
\fcaption{CCFR(97) $F_2$ and $xF_3$ data as functions of $Q^2$ at fixed $x$.
The results of a NLO QCD are also
shown (full line), the dashed line is the extrapolation of the fit to lower 
$Q^2$.}
\label{fig:ccfrdat1}
\end{center}
\end{figure}

Accurate $\nu$Fe structure function data have 
been available for some time from the high statistics CCFR 
experiment~\cite{ccfr_lambda,ccfr_gls} (CCFR(93) dataset). 
Details of the detector and the 
calibration procedures have been given in Sec.~\ref{sec:ccfr_det}. 
Recently the analysis has been
reassessed and improved~\cite{ccfr_seligman} to give $F_2$ and $xF_3$ 
values for $x$ between $0.0075$ and $0.75$ and $Q^2$ between
$1.3$ and $126\,$GeV$^2$ (CCFR(97) dataset). Improvements in 
the analysis include
a new determination of the muon and hadron energy calibrations from 
test beam data and a more complete treatment of 
radiative corrections~\cite{ccfr_radcor1}.
Generally the estimates of systematic errors have been refined. Other
corrections applied are: 
an isoscalar correction for the $6.8$\% excess of neutrons
over protons in Fe; a correction for the charm quark threshold using the
slow-rescaling model with $m_c=1.34\,$GeV and a correction for the mass of
the $W$-boson.  The $R_{SLAC}$ parametrization was used and nuclear 
effects have not been corrected for. 
The largest sources of systematic errors are the 1\% errors in the muon
and hadron energy scales. The error in the ratio 
$\sigma^{\bar{\nu}}/\sigma^\nu$ is also important. The world average
value of the ratio from $\nu$Fe DIS experiments of $0.499\pm0.007$ was
used. Other sources of systematics include the flux determination
and variations in the physics models used in the Monte Carlo simulation. 
The structure function data are shown in Fig.~\ref{fig:ccfrdat1}.

\begin{figure}[htbp]
\vspace*{13pt}
\begin{center}
\psfig{figure=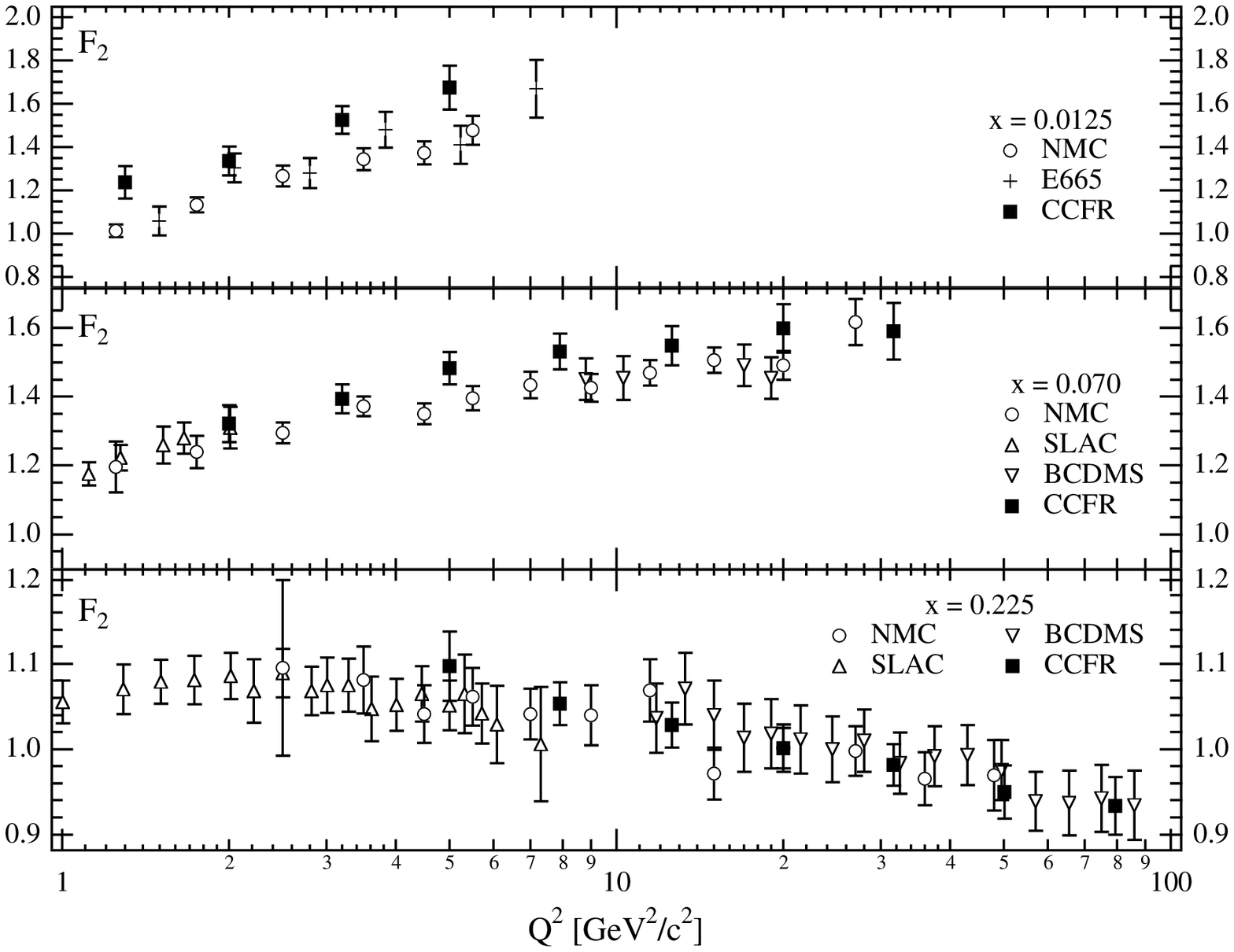,bbllx=-65pt,bblly=330pt,bburx=535pt,bbury=750pt,width=10cm,height=8cm} 
\fcaption{Comparisons of CCFR(97) $F_2$ values for 
$\nu Fe$ with those for $\ell D$ from SLAC, NMC(97) and BCDMS converted 
using the `5/18' relation as described in the text).}
\label{fig:ccfrdat2}
\end{center}
\end{figure}

As we have noted in Sec.~3.1 one expects 
$F_2^{\ell N}\approx {5\over 18}F_2^{\nu N}$ and this was a very important
and early success of the QPM. CCFR have made a more exact
comparison by using 
$\displaystyle F_2^{\ell N}= {5\over 18}\left(1-
{3\over 5}{s+\bar{s}-c-\bar{c}\over q+\bar{q}}\right)F_2^{\nu N}$ 
with the strange sea extracted from their own dimuon data~\cite{ccfr_dimuon}. 
The comparison is made with deuterium 
data from SLAC, NMC and BCDMS corrected to Fe using the 
$F_2^{\ell Fe}/F_2^{\ell d}$ ratio as measured by SLAC and NMC, 
and is shown in Fig.~\ref{fig:ccfrdat2}. 
The $F_2$ values generally agree well except in
the low $x$ bin ($0.0125$) in which the CCFR values lie about 15\% above 
those of NMC. The discrepancy is larger than the quoted systematic errors
of the two experiments. It cannot be explained by increasing the
size of the strange sea as this is limited by CCFR dimuon data, however
it has been suggested that its distribution may be more complicated than
usually assumed~\cite{brodsky-ma}. Further possibilities are that the nuclear
corrections are different for neutrinos and charged leptons or that
the treatment of the charm threshold in 
the CCFR analyses~\cite{CCFRth} is not fully correct to NLO~\cite{GKR}.
Recently Caldwell~\cite{caldwell97} has compared the CCFR $F_2$ data,
corrected to single protons, with that from ZEUS and NMC and concluded
that the trend of the CCFR data at small $x$ may be in better agreement with
the HERA data than that from NMC. It is possible that the HERMES 
experiment at HERA could measure
$F_2$ in this overlap region and help to resolve the discrepancy.

The IHEP-JINR neutrino detector collaboration have recently published 
data on $F_2$ and $xF_3$ from three exposures of their detector to the
wide band $\nu(\bar{\nu})$ beams at the Serpukhov U70 
accelerator~\cite{barabash}. Events were selected to satisfy 
$W^2>1.7\,$GeV$^2$ and $6<E_{\nu(\bar{\nu})}<28\,$GeV. Samples of
741 $\bar{\nu}$ events with $\langle Q^2\rangle=1.2\,$GeV$^2$ and 
5987 $\nu$ events with $\langle Q^2\rangle=2.3\,$GeV$^2$. Structure 
functions have been extracted in 6 bins with $0.052<x<0.563$ and
$0.55<Q^2<4.0\,$GeV$^2$ with $R=0$ and $R=0.1$. Corrections have been
made for acceptance, Fermi motion smearing, radiative effects and target
non-isoscalarity. The overall normalization errors from the neutrino
flux determination are 4\% and 11\% for $F_2$ and $xF_3$ respectively.
\smallskip

Given the size of the CCFR data sample it remains the definitive neutrino
nucleon DIS experiment.

\subsection{Data on $R$}
\label{sec:rdata}

In this section we will discuss the methods that have 
been used to measure $R$ or $F_L$ and summarise the existing data.
Consider charged lepton induced NC scattering. It is convenient to
write the double differential cross-section in terms of the cross-sections
for the scattering of transverse and longitudinal virtual 
photons~\cite{roberts}
\begin{equation}
{d^2\sigma\over dxdQ^2}=\Gamma
\left[\sigma_T(x,Q^2)+\varepsilon\sigma_L(x,Q^2)\right]
\label{eq:sigTL}
\end{equation}
where the virtual photon cross-sections are related to $F_2$ and $F_L$ by
\begin{equation}
4\pi^2\alpha F_2={\nu Q^2 K\over (\nu^2+Q^2)}(\sigma_T+\sigma_L) ~~~~~
4\pi^2\alpha F_L={\nu Q^2 K\over \nu^2}\sigma_L
\label{eq:relate}
\end{equation}
and
\begin{equation}
\Gamma={\alpha\over\pi}{M_N K \over s^2x^2(1-\varepsilon)},\\
~~~\varepsilon^{-1}=1+2\left(1+{\nu^2\over Q^2}\right)\left[
{s(s-2M_N\nu)\over M_N^2Q^2}-1\right]^{-1}
\label{eq:epsgam}
\end{equation} 
with $\nu=Q^2/(2M_N x)$.
In these expressions the factor $K$ gives the flux of virtual photons which 
may be defined according to the conventions of Gilman~\cite{Gilman} 
($K = \sqrt{\nu^2 + Q^2}$) or Hand~\cite{hand} ($K = \nu(1-x)$) 
such that it
should be equal to that of a real photon beam ($K \to \nu$) as $Q^2 \to 0$.
At small $x$ and/or large $Q^2$ we may drop terms in 
$Q^2/\nu^2 (=4 M^2_N x^2/Q^2)$ and 
the above expressions simplify greatly. In particular,
\begin{equation}
\varepsilon={2(1-y)\over 1+(1-y)^2}
\label{eq:vareps}
\end{equation}
and
\begin{equation}
4\pi^2\alpha F_2= Q^2(\sigma_T+\sigma_L) ~~~
4\pi^2\alpha F_L=Q^2 \sigma_L  
\label{eq:relatesim}
\end{equation}
At this level of approximation one may rewrite 
Eqs.$~\ref{eq:sigTL}-\ref{eq:epsgam}$ to read
\begin{equation}
{d^2\sigma\over dxdQ^2}={2\pi\alpha^2\over Q^4 x}Y_+
\left[{1+\varepsilon R\over 1+R}\right]F_2(x,Q^2),
\end{equation}

To measure $R$ directly at a fixed value of $(x,Q^2)$ requires data at 
two different values of $y$ and hence at two different centre of mass energies 
(because of the constraint $Q^2=sxy$), then
\begin{equation}
R={\tilde{\sigma_2}Y_{+1}-\tilde{\sigma_1}Y_{+2}\over 
2\tilde{\sigma_1}(1-y_2)-2\tilde{\sigma_2}(1-y_1)}
\end{equation}
where $\displaystyle\tilde{\sigma}= {Q^4x\over 2\pi\alpha^2}
{d^2\sigma\over dxdQ^2}$. This equation shows why $R$ is 
difficult to measure accurately, not only does one 
need very small statistical errors on $\tilde{\sigma_i}$ but also very good
control of the systematic errors at the two different energies. From
the form of the expression for $\varepsilon$ as a function of $y$ it can
be seen that one gets the largest variation in $\varepsilon$, for a given
change in $\sqrt{s}$, at large $y$. This condition is often difficult
to achieve, particularly at HERA. If one has
data at many different energies then $\sigma_L$ can be extracted from
the slope of the total cross-section as a function of $\varepsilon$ -- the
`Rosenbluth plot' of low energy nuclear and nucleon-electron scattering.

Early fixed target measurements of $R$ were made by EMC~\cite{emcr},
BCDMS~\cite{BCDMS}, CDHSW~\cite{cdhswr} and SLAC following the
re-analysis of the data by Whitlow et al~\cite{whitlow90}.
The most extensive results are those of SLAC, 
$R^p$ and $R^d$ measured for $0.1\leq x\leq0.9$, 
$0.6\leq Q^2\leq 20\,$GeV$^2$, and BCDMS, $R^p$ and $R^d$ 
measured for $0.07\leq x\leq0.65$, $15\leq Q^2\leq 50\,$GeV$^2$.
The data satisfied $R^p=R^d$ but the 
behaviour of $R$ as a function of $x$ and $Q^2$ did not follow any
model or theory, including pQCD. The need, none the less, for a good
description of the behaviour of $R$ led the authors of 
ref.~\cite{whitlow90} to produce a convenient parametrization of their
own data and data from EMC, BCDMS and CDHSW. It is
\begin{equation}
R_{SLAC}={b_1\over \ln(Q^2/\Lambda^2)}\Theta(x,Q^2)+{b_2\over Q^2}+
         {b_3\over Q^4+(0.3)^2}
\end{equation}
where
$\displaystyle
\Theta(x,Q^2)=1+12\left({Q^2\over Q^2+10.}\right)\left({(0.125)^2\over
(0.125)^2+x^2}\right)$,
$\Lambda=0.2\,$GeV and the best fit results for the parameters
$b_i$ ($\chi^2$ of 110 on 139 data points) are $b_1=0.06347,b_2=0.57468,
b_3=-0.35342$. $R_{SLAC}$ is not valid for $Q^2<0.35\,$GeV$^2$.

More recently $R$ has been measured by a new SLAC experiment~\cite{e140x},
by CCFR~\cite{ccfr_r} and
NMC~\cite{nmc_r,nmcdata2}. The new data from SLAC are for hydrogen, deuterium
and beryllium targets and cover the ranges $0.1\leq x\leq 0.7$, 
$0.5\leq Q^2\leq 7\,$GeV$^2$. The advantage of the new SLAC data is that
$R$ is extracted from a single experiment which leads to a reduction of the
systematic error from relative normalizations. The total systematic error
is at the level of 10\%. 

\begin{figure}[htbp]
\vspace*{13pt}
\begin{flushleft}
\psfig{figure=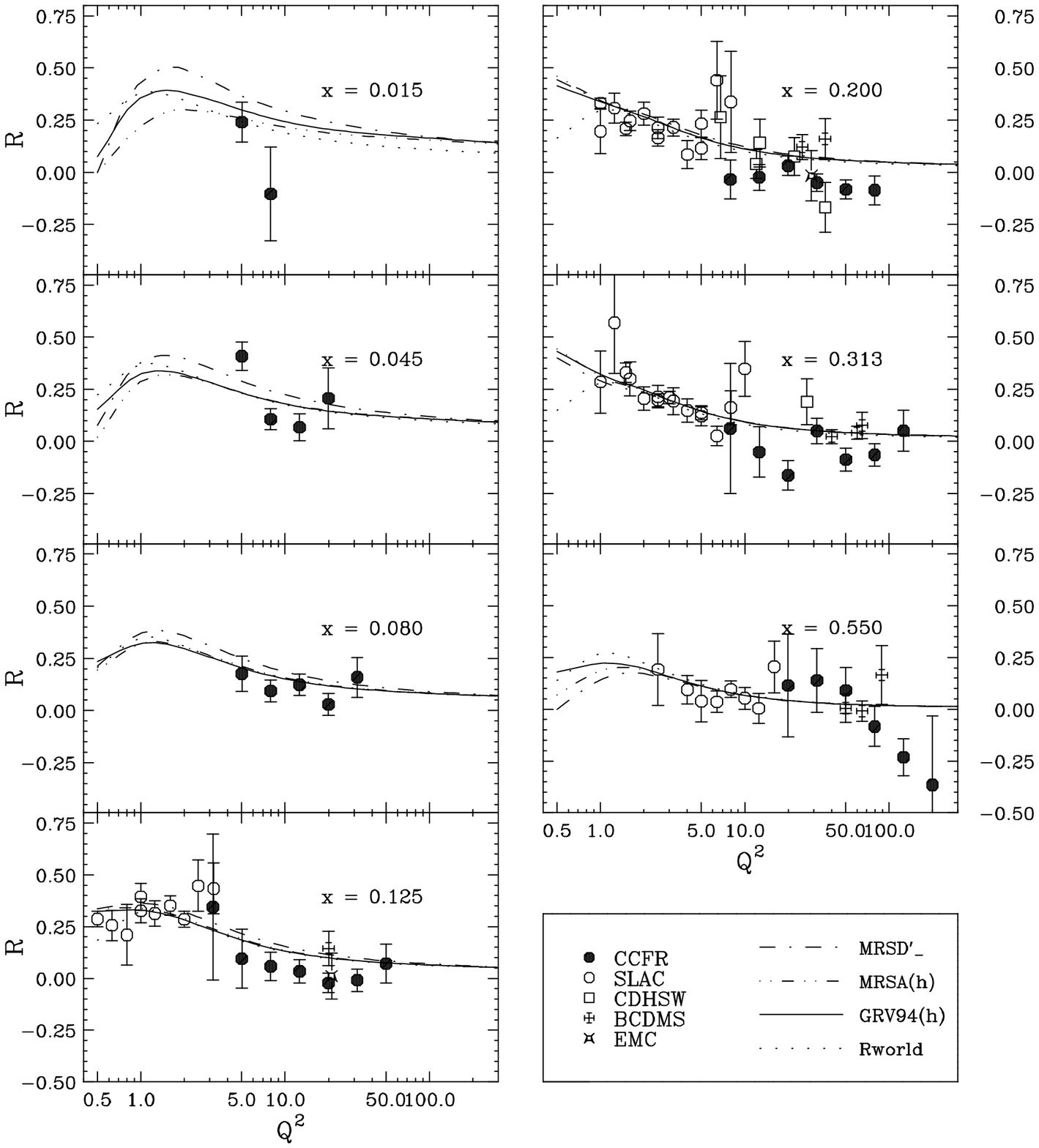,width=13cm} 
\fcaption{$R$ at fixed $x$ vs $Q^2$. Data from CCFR is shown
together with data from EMC, CDHSW, BCDMS and SLAC. The curves are
from Bodek, Rock and Yang~\cite{bry} (see text).}
\label{fig:rdat_ccfr}
\end{flushleft}
\end{figure}

The CCFR $R$ measurements are from the same
data sample discussed in Sec.~\ref{sec:xf3dat} above.
$R$ is extracted from a linear fit to the averaged 
cross-section data for $\nu Fe$ and $\bar{\nu}Fe$ versus $\varepsilon$. The
measurements cover the ranges $0.01<x<0.6$, $4<Q^2<300\,$GeV$^2$. The 
method assumes that $xF_3^\nu=xF_3^{\bar{\nu}}$ which is not exact at 
low $x$. Corrections are made for the differences caused by the strange
sea and charm mass effects (see Eqs.~\ref{eq:nucs},~\ref{eq:anucs}). 
The CCFR data, together with some of the 
earlier $R$ data are shown in Fig.~\ref{fig:rdat_ccfr}.
To describe the $R$ data at both low and high $Q^2$ Bodek, Rock and 
Yang~\cite{bry} have developed a model which includes non-pQCD effects
such as higher twist terms and target mass effects at low $Q^2$ but which
matches smoothly to the next-to-lowest order expression for $R$ 
(worked out to order $\alpha_s^2$ in the coefficient functions~\cite{nnlor})
 at large $Q^2$. The results of this model using various different parton
distribution functions
as input are shown in the figure. The dependence on the choice of PDF is 
not strong and the model gives a reasonable description of all the $R$
data.

\begin{figure}[t]
\vspace*{13pt}
\begin{center}
\begin{tabular}[t]{ll}
\psfig{figure=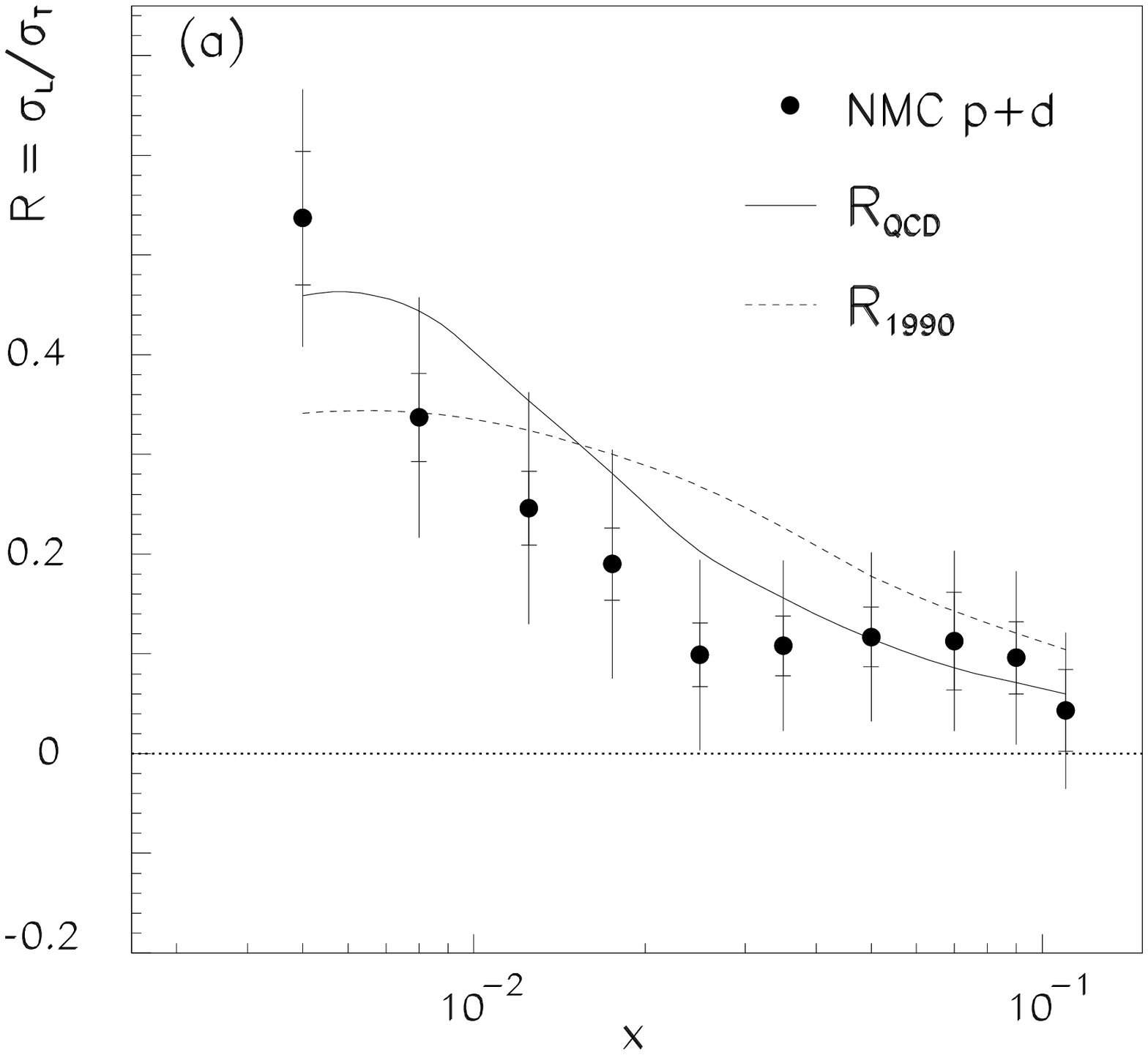,width=.475\textwidth} &
\psfig{figure=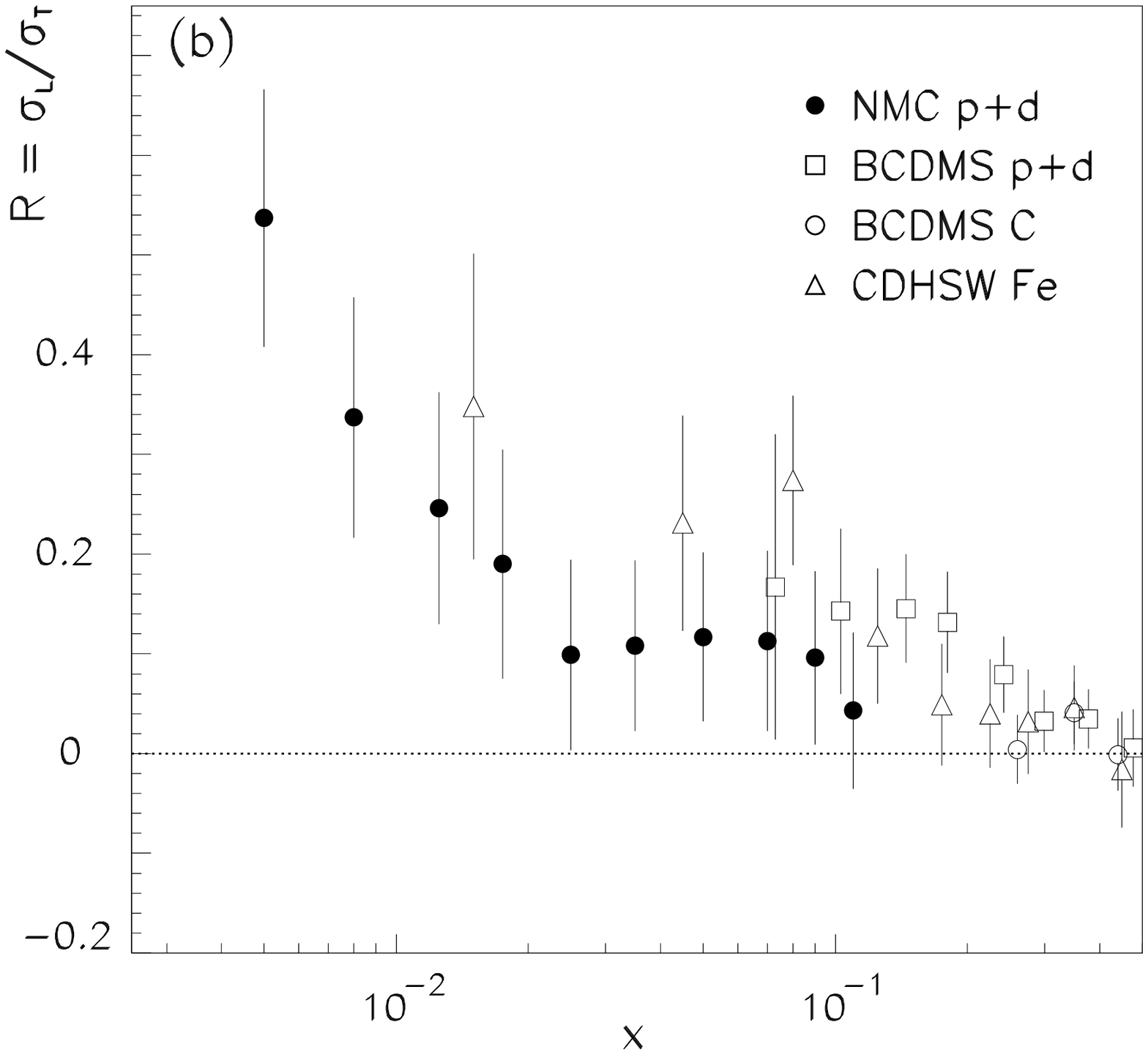,width=.475\textwidth} 
\end{tabular}
\fcaption{Left: $R$ vs $x$, data from NMC. The full curve
is from a QCD fit and the dotted curve is from $R_{SLAC}$. Right: data on
$R$ from NMC, BCDMS and CDHSW.}
\label{fig:rdat_nmc}
\end{center}
\end{figure}

The NMC results on $R$ come from the T1 trigger data sample already 
published and the more recent data from the small angle 
trigger T2. 
The data on $R^p$ and $R^d$ cover the ranges $0.002 < x < 0.12$ and
$\langle Q^2\rangle=1.4 - 20.6\,$GeV$^2$.  The largest source of 
systematic error is normalization uncertainty. The total systematic
errors are 1-5 to 3 times larger than the statistical errors. This
is shown Fig.~\ref{fig:rdat_nmc}(a), in which the inner error bar is
statistical and the outer the sum of statistical and systematic errors
in quadrature. The data are in good agreement with other measurements 
at large $x$ and provide a considerable increase in data for 
$x<0.1$, where $R$ is
seen to rise as one would expect from its sensitivity to the gluon
distribution function. This can be seen in Fig.~\ref{fig:rdat_nmc}(b).

\begin{figure}[htbp]
\vspace*{13pt}
\begin{center}
\begin{tabular}[t]{ll}
\psfig{figure=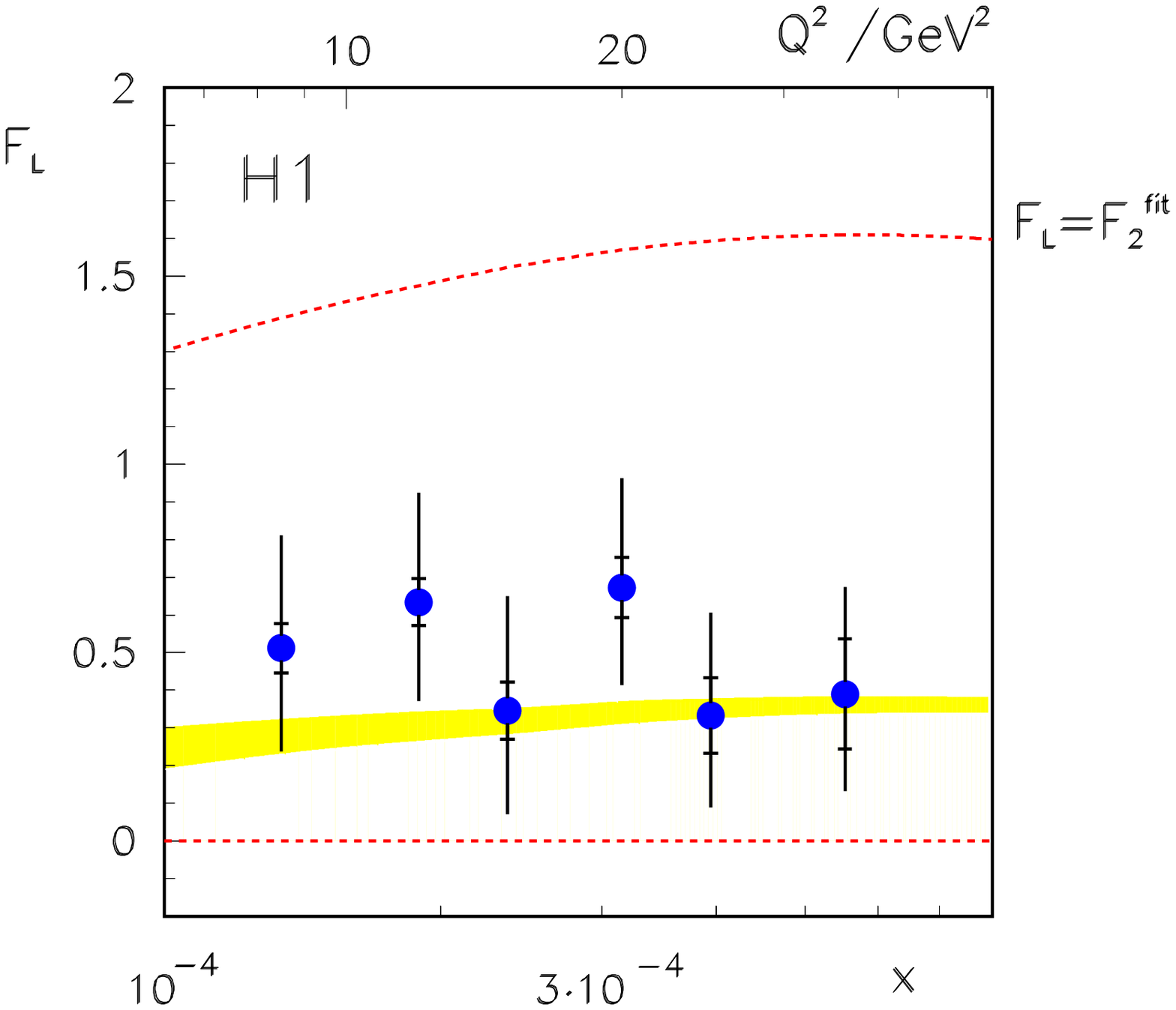,bbllx=40pt,bblly=160pt,bburx=525pt,bbury=645pt,width=.475\textwidth}
&
\end{tabular}
\fcaption{Measured longitudinal structure function $F_L$ by H1 for 
$y = 0.7$ as functions of $x$ and $Q^2$ for $Q^2= 8.5, 12, 15, 20 ,25$
and $35$ GeV$^2$. The error band represents the uncertainty of the
calculation
of $F_L$ using the gluon and quark distributions, as determined from the
NLO
QCD analysis of the H1 data for $y\le 0.35$ and the BCDMS data.
The dashed lines define the allowed range of $F_L$ values from $F_L=0$ to
$F_L = F_2$ where $F_2$ is given by the QCD fit.}
\label{fig:h1fl}
\end{center}
\end{figure}

At HERA no direct measurements of $R$ have yet been attempted as the
collider has operated essentially at a fixed CM energy of around
$300\,$GeV. 
In the meantime the H1 collaboration~\cite{h194r} have estimated $R$ 
by using their high statistics 1994 data and assuming that $F_2$
is well described by NLO QCD. As we have noted a number of times 
the contribution of $F_L$ to the double differential cross-section
is suppressed by the factor $y^2$. The idea of the H1 method is first
to determine $F_2$ by applying a NLO QCD fit to data satisfying 
$y<0.35$, i.e. in a region where the effect of $F_L$ in 
the cross-section is small. 
The fit is then extrapolated to the high $y$ region and 
used to subtract the contribution of $F_2$ from the cross-section at 
$y= 0.7$, leading to an estimate of $R$.
Although it has been checked that extrapolations based on some other
models~\cite{royon,dewolf} give the same result at $y = 0.7$ within 
a few per cent, the extrapolation remains the most uncertain part 
of this analysis.\fnm{l}\fnt{l}{~Thorne~\cite{thornefl} has argued in
some detail that the extrapolation error may be considerably larger.}
The data sample for this study  is essentially that
of the `nominal vertex' sample H1(94)
but with the minimum cut on the energy of the scattered positron lowered
from 11 to $6.5\,$GeV, thus allowing higher values of $y$ to be reached.
 The
data cover the ranges $0.6<y<0.78$ and $7.5<Q^2<42\,$GeV$^2$. A careful
study of radiative corrections was made using the HECTOR code and they are 
stable within about half the statistical error of the cross-sections.
Systematic errors are highly correlated, sources studied are 
$y$ dependent effects such as
cut values and the photoproduction background subtraction, the 
relative uncertainty in the luminosities and the error from the 
subtraction of the fitted $F_2$. Averaging over all the data
gives the result $F_L=0.52\pm0.03(stat)^{+0.25}_{-0.22}(sys)$
for $\langle Q^2\rangle=15.4\,$GeV$^2$ and 
$\langle x\rangle=2.43\times10^{-4}$, 
well compatible with pQCD calculations using recent parton distribution 
functions. The result is shown in Fig.~\ref{fig:h1fl}.

An alternative approach to the measurement of $R$ at HERA has been
proposed by Krasny, Placzek and Spiesberger~\cite{kps}. The idea
of the method is to use hard initial state radiation as a way of 
varying the incident e$^\pm$ beam energy thus allowing the 
cross-section to be measured for different values of $\varepsilon$, but
fixed $(x,Q^2)$ -- rather analogous to a broad-band neutrino beam. 
Experimentally one uses the luminosity monitor to
measure the energy of the hard ISR photon, $E_\gamma$. Integrating
over photon emissions angles inside a cone $\theta_\gamma\geq \pi-\theta_a$
and neglecting infrared and collinear finite terms, the cross-section
for $ep\to e\gamma X$ is
\begin{equation}
{d^3\over dxdQ^2dz}=\alpha^3P(z){1+(1-y(z))^2\over xQ^4}
                     \left[F_2(x,Q^2)-(1-\varepsilon(z))F_L(x,Q^2)\right],
\label{eq:isrfl}
\end{equation}
where $P(z)$ is the  function
\begin{equation}
P(z)={1+z^2\over 1-z}\ln\left({E_e^2\theta_a^2\over m_e^2}\right)-{z\over 1-z}
\end{equation}
and 
\begin{equation}
\varepsilon(z)={2(1-y(z))\over 1+(1-y(z))^2}~~~~y(z)={Q^2\over xzs},
\end{equation}
\noindent
where $z=(E_e-E_\gamma)/E_e$.
The variables $x$ and $Q^2$ are the {\it true} values and if the electron
method is used for their reconstruction the formulae given in 
Eq.~\ref{eq:isrxq} must be used to take account of the radiated energy.
$F_L$ or $R$ is extracted from the dependence of the cross-section on
$\varepsilon$.
A measurement of $F_L$ for $0.0006<x<0.02$,
$15<Q^2<120\,$GeV$^2$ at HERA could be made using ISR events, but would
need a total luminosity of the order of $200\,$pb$^{-1}$.

The method is a very elegant one in principle, but it has not been possible
to exploit it yet in practice.  The major problems are the small
cross-section, the stringent requirements on resolution of the kinematic
variables (particularly $x$) and control of the bremsstrahlung background.
Improvements have been suggested by
 by Frey~\cite{frey},  using a more restricted kinematic range, and 
 by Favart et al~\cite{favart}, using a modified analysis method.
The latter exploits the knowledge of $F_2$
and extracts $R$ from its influence on the shape of the $\varepsilon$
distribution of events. However the method does also require a very
low value of $E^\prime_{min}$, well below $5\,$GeV and, 
if possible, as small as $2\,$GeV. 
Although none of the direct experimental methods is likely to 
give a measurement of $R$ to better that $20-30$\%, it is important to 
have a direct check on this quantity at the high values of $W$ and small
values of $x$ available at HERA. The uncertainty in $R$ could eventually
be one the limiting systematic errors on the extraction of $F_2$ at
low $x$. $F_L$ is also a very interesting quantity in its own right, because
of its very direct dependence on the gluon density as shown by 
Eq.~\ref{eq:flqcd}.

Recently the possibility of measuring $R$ at HERA by reducing 
the beam energies has been studied~\cite{bauerdick}.
The conclusion was that $R$ could be measured with a precision
of about 15-20\% at low $x$ and $Q^2< 100 $ GeV$^2$, by running
at proton beam energies of 450, 350 and $250\,$GeV. A data sample of 
10 pb$^{-1}$ would be required at each energy. One lower beam 
energy data sample would be sufficient to give a measurement with 
approximately 30\% precision in a large part of this kinematic region.

\subsection{Open charm production and $F_2^{c\bar{c}}$}
\label{sec:cdata}
\noindent 
Heavy quark production in DIS is thought to occur predominantly through
the boson-gluon fusion process (BGF) -- see Fig.~\ref{fig:bgf}(a)
 --
at relatively small $x$. Other 
mechanisms such as diffractive production (which is important for
$J/\psi$ production) have much smaller cross-sections. The evidence
for intrinsic charm in the proton is somewhat equivocal and in 
any case such events would be produced at large $x$. 

The first measurement of $F_2^{c\bar{c}}$
was made by the BFP collaboration~\cite{bfp} who used a $209\,$GeV
muon beam at FermiLab and a multimuon spectrometer to extract
$F_2^{c\bar{c}}$ from dimuon events over a range $0.0013<x<0.13,
~0.63<Q^2<63\,$GeV$^2$. Later the EMC 
collaboration~\cite{emcf2cc} measured $F_2^{c\bar{c}}$, in the 
range $0.0042<x<0.422$, $1<Q^2<70\,$GeV$^2$, from
di- and trimuon events produced from $250\,$GeV muon scattering on an iron 
target. The dimuon data were corrected for acceptance effects using the
lowest order 
BGF model of Leveille and Weiler~\cite{bgf_lw} and a mean semileptonic
branching ratio for the $D$ meson to muons of 8.2\% was used to correct
to the $c\overline{c}$ cross-section.
The results from both experiments were found to be well represented by the
BGF production mechanism. However because of the rather low cm energies of
around $20\,$GeV, their results could not rule out other explanations.

The advent of HERA, with an order of magnitude increase in $W$ and a 
reach to much smaller $x$ values, together with the sophisticated
detectors H1 and ZEUS capable of measuring many details of the DIS
final state gives the prospect of accurate data on $F_2^{c\bar{c}}$
over a much wider range of $x$ and $Q^2$. The second major step
forward since the early measurements is the calculation of the massive 
quark coefficient
functions to order $\alpha_s^2$ by Laenen et al~\cite{nlocc,fastbgf} 
and the attempts to
provide consistent treatments of heavy quark production from the
threshold region $Q^2\sim m_q^2$ to the asymptotic region
$Q^2 \gg m_q^2$ by MRRS~\cite{MRRS} and CTEQ~\cite{LaiTung} - see
Sec.~\ref{sec:heavyq}. In addition the 
differential distributions for the rapidity and transverse momentum
of the charm quark have also been worked out to next-to-lowest 
order~\cite{nlocc_pty}.

Both the H1 and ZEUS experiments have searched for $D^*$ 
production in DIS using
the well-established technique which exploits the accurately known 
mass difference 
$\Delta M=M(K\pi\pi_s)-M(K\pi)$~\fnm{m}\fnt{m}{~The PDG value is 
$145.42\pm0.05\,$MeV.} (from the decay chain
$D^{*+}\to D^0\pi^+_s\to K^-\pi^+\pi^+_s$) as the primary signal. The $D^*$ 
is only just above threshold and its decay produces a `slow' pion labelled 
as $\pi_s$. Although the signal is a very clean one, the combined branching 
ratio for the $D^*\to K\pi\pi$ decay is only $2.62\pm0.10$\%~\cite{pdg96}, 
which means that this measurement requires high luminosity.

\begin{figure}[htbp]
\vspace*{13pt}
\begin{center}
\psfig{figure=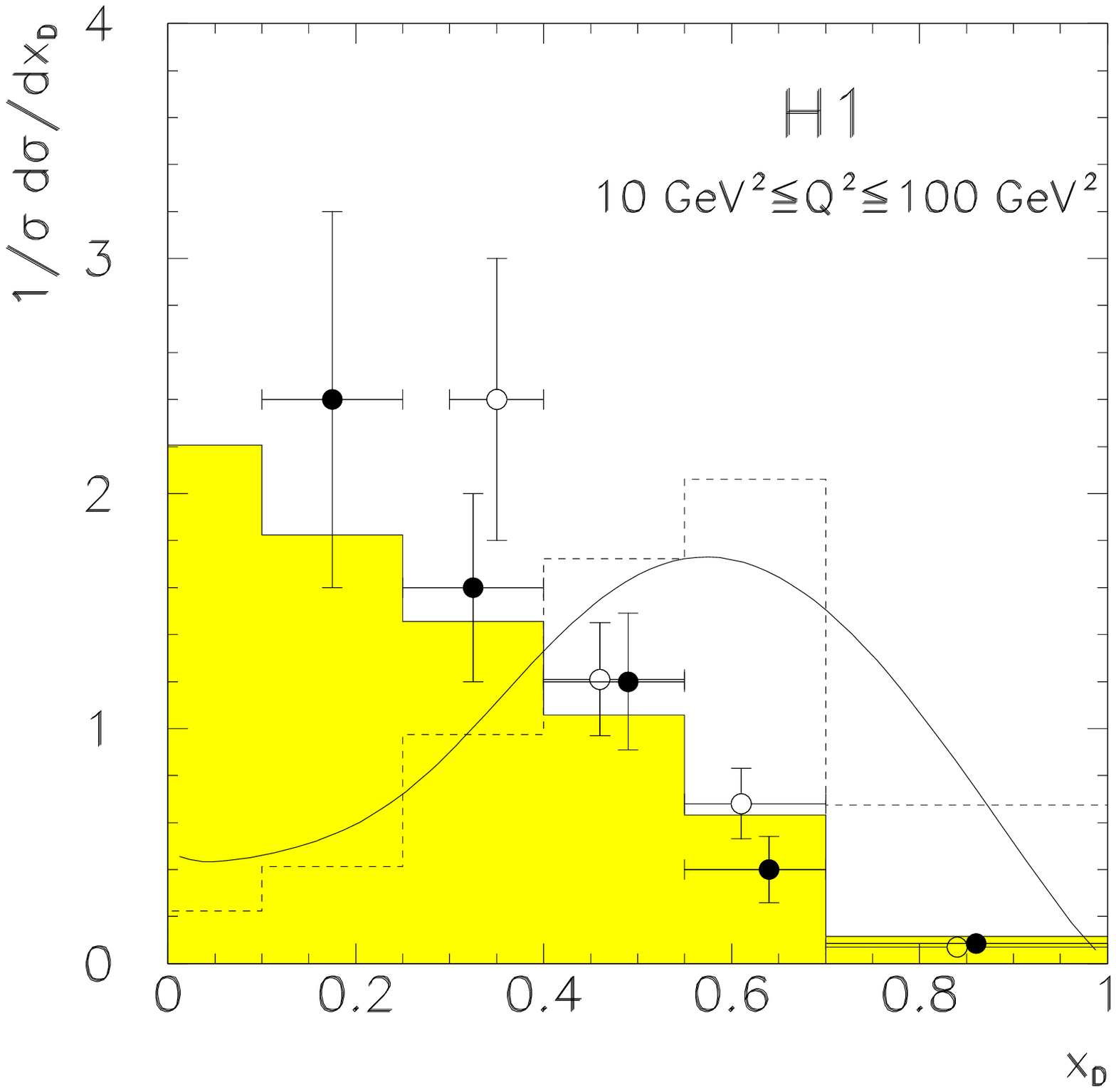,bbllx=-200pt,bblly=0pt,bburx=360pt,bbury=560pt,width=7cm,height=7cm} 
\fcaption{Distribution of $x_D=2|{\bf p}_D|/W$ measured by H1 in
the $\gamma^*p$ frame (full points $D^*$ data and open points $D^0$ data), 
compared to the BGF (shaded histogram) and charm sea
(dashed histogram) expectations.}
\label{fig:h1_xd}
\end{center}
\end{figure}

The H1 collaboration~\cite{h1_d*} has made a detailed study of 
DIS charm production
using both $D^*$ and $D^0$ channels. 
The analysis is based on $2.97\,$pb$^{-1}$ of 1994 DIS events, 
restricted to $10<Q^2<100\,$GeV$^2$ and $y<0.53$. Charged particles  
were reconstructed in the H1 central tracker
 and must have at least $0.25\,$GeV
momentum transverse to the beam line. Particle identification is not
used but charged particle are ranked in order of the magnitude of
their momentum in the $\gamma^*p$ frame, to take advantage of the fact
that the fragmentation function for charmed mesons is `harder' than that
for the light mesons and so the decay products will have large momenta.
Using only highly ranked particles to construct the $D^0$ reduces
the combinatorial background. Charm events were selected by using the
$D^*\to D^0\pi$ ($\Delta M$ method) and also
a direct inclusive $D^0\to K^-\pi^+$ search. In both analyses the
$D^0$ candidate was required to satisfy 
$|\eta(D^0)|<1.5$~\fnm{n}\fnt{n}{~The pseudo-rapidity $\eta=
-\ln(\tan(\theta/2))$ where $\theta$ is the polar angle with respect
to the proton beam direction.}. The $D^*$ and
$D^0$ signals are extracted by fitting the $\Delta M$ and $K\pi$
mass distributions, respectively, with Gaussians and suitably shaped
background functions. The resulting yields of signal events are
$103\pm13$ $D^*$ events and $144\pm19$ events for the inclusive $D^0$ 
analysis, with $20\pm5$ events in common. 
The data are corrected for acceptance and resolution effects,
using the AROMA Monte Carlo~\cite{aroma}, to the full phase space 
in $p_T(D)$ 
and $\eta(D)$ and the region $10<Q^2<100\,$GeV$^2$, $y<0.7$. 
The efficiencies for
reconstructing $D^*$ and $D^0$ mesons, including acceptance, are
about 16\% and 6\% respectively. The cross-sections for $ep\to eDX$
and $ep\to eD^*X$ are $20.4\pm2.7^{+2.7+1.6}_{-2.4-1.2}\,$nb
and $7.8\pm1.0^{+1.2}_{-1.0}\pm0.6\,$nb respectively, where the last 
error refers to model dependent uncertainties. The charm production
mechanism has been studied by measuring the the $x_D=2|{\bf p}_D|/W$ 
distribution in the $\gamma^*p$ frame. The normalized $x_D$ distribution
for an average $\gamma^*p$ CM energy of $125\,$GeV and $|\eta_D|<1.5$
is shown in Fig.~\ref{fig:h1_xd} compared to calculations of the BGF
mechanism using the AROMA Monte Carlo simulation and to the expectation
from the charm sea only (using LEPTO/MEPS).  From these comparisons, H1 has 
set an upper limit of 5\% 
(at 95\% CL) on charm production from the sea. 
A similar result has been
found by the ZEUS collaboration from the DIS $D^*$ analysis described 
below. 

To calculate the inclusive charm production 
cross-section one needs to know the probability that the charm quark
will fragment into a $D^0$ or $D^*$ meson, $P(c\to D)$ and to correct 
for the small fraction, $\xi$, of indirect $D$ mesons produced through 
$B$ meson decay or fragmentation,
\begin{equation}
\sigma(ep\to ec\bar{c}X)={1\over 2}{\sigma(ep\to eDX)\over
P(c\to D)(1+\xi)}.
\end{equation}
From measurements at $e^+e^-$ colliders,
$P(c\to D^0)\sim 50$\%, $P(c\to D^{*\pm})\sim 25$\% and $\xi$ is about
2\%. The final result averaging over both $D^*$ and $D^0$ channels is
$\sigma(ep\to ec\bar{c}X)=17.4\pm1.6(stat)\pm1.7(sys)\pm1.4(model)\,$nb, 
where the last error
is from model dependent uncertainties in the extrapolation to the full
phase space. The systematic error is dominated by
the mass resolution and the separation of signal and background, other
contributions come from tracker efficiency, luminosity measurement and 
radiative corrections. H1 has compared their measured cross-section
with next-to-lowest order 
calculations using MRS and GRV gluons and $m_c$ between $1.3$ and
$1.7\,$GeV and find that the calculated cross-sections are in the range
$8.7-11.4\,$nb, well below the measured value. Using a 
gluon distribution from a NLO QCD fit to their own 
1994 $F_2$ data the $m_c=1.5$, H1
predict a value of $13.6\pm1.0\,$nb.

\begin{figure}[ht]
\vspace*{13pt}
\begin{center}
\psfig{figure=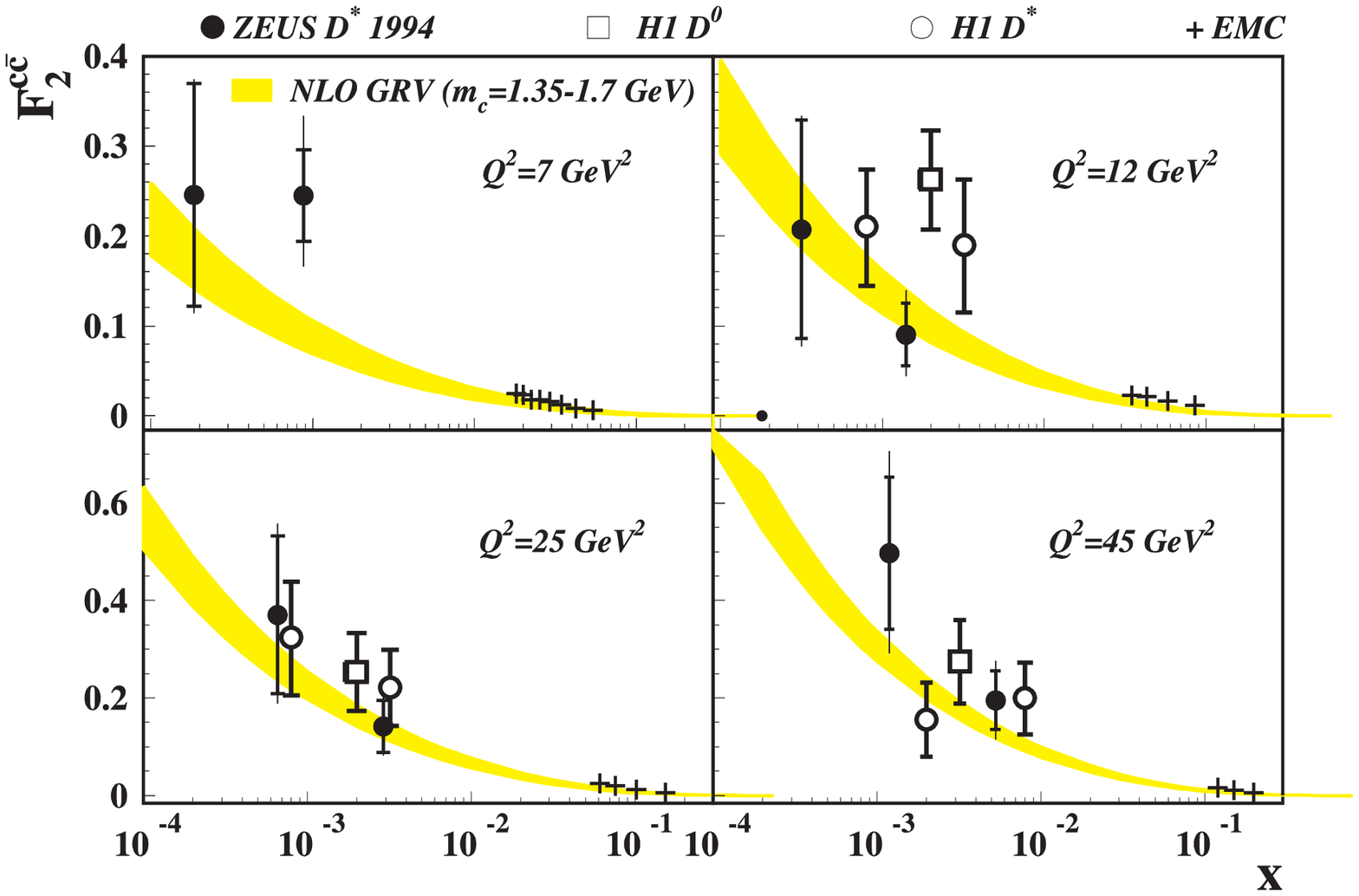,bbllx=-40pt,bblly=245pt,bburx=510pt,bbury=610pt,width=10cm} 
\vspace*{13pt}
\fcaption{$F_2^{c\bar{c}}$ from the ZEUS and H1 analyses of the HERA 
1994 data. Earlier results from EMC are also shown. The shaded band
corresponds to a next-to-lowest order QCD calculation using the GRV94 
gluon distribution
and a range of charm quark masses (upper edge $1.35\,$GeV, lower edge
$1.7\,$GeV). }
\label{fig:f2cc_all}
\end{center}
\end{figure}

Assuming that $R=0$, $F_2^{c\bar{c}}$ is related to the $ep\to ec\bar{c}X$
cross-section by
\begin{equation}
{d^2\sigma(c\bar{c})\over dxdQ^2}={2\pi\alpha^2\over Q^4x}
(1+(1-y)^2)F_2^{c\bar{c}}(x,Q^2).
\label{eq:f2cc}
\end{equation}
H1 has binned the $D^*$ and $D^0$ data into 9 bins with mean values of
$x$ and $Q^2$ of $0.008, 0.0020, 0.0032$ and $12,25,45\,$GeV$^2$, 
respectively. The conversion to $c\bar{c}$ cross-sections follows the
procedure described above and $F_2^{c\bar{c}}$ is extracted using 
Eq.~\ref{eq:f2cc}. The results are shown in Fig.~\ref{fig:f2cc_all}. 
Averaging over all bins,
H1 finds a value $\displaystyle 0.237\pm0.021^{+0.043}_{-0.039}$
for the ratio $F_2^{c\bar{c}}/F_2$.
\smallskip

The ZEUS collaboration has also studied DIS $D^*$ production~\cite{z_d*}. 
DIS events from $2.95\,$pb$^{-1}$ of the 1994 sample satisfying 
$y<0.7,~5<Q^2<100\,$GeV$^2$ were chosen. Charged particle tracks,
reconstructed in the ZEUS central tracker, had
to have a transverse momentum $p_T>0.125\,$GeV and a polar angle 
satisfying $20^\circ<\theta<160^\circ$. The $K\pi$ invariant mass from
opposite sign particles had to lie in the range $1.4 - 2.5\,$GeV and
if adding a third track (with opposite sign to the $K$ track) gave 
$\Delta M < 180\,$MeV, the three charged particles formed a $D^*$ candidate.
$D^*$ candidates were required to satisfy $1.3<p_T(D^*)<9.0\,$GeV
and $|\eta(D^*)|<1.5$.
Combinatorial background was subtracted using an average of estimates
from side-bands in the $\Delta M$ distribution and that from wrong-sign
particle combinations. In the given kinematic region $122\pm17$ signal
events were found above a background of $95\pm8$. The corrections for 
detector effects and acceptance were 
calculated using the DIS heavy flavour production simulation programme
AROMA. For the kinematical region selected by the cuts, the mean
event selection efficiency is about 75\% and
mean $D^*$ reconstruction efficiency is about 38\%. The total systematic error
on the cross-section is 15\% with the major contributions coming from
the signal selection and background subtraction, Monte Carlo corrections,
radiative effects and the overall normalization.

\noindent

Integrating the cross-section over
the region $5<Q^2<100\,$GeV$^2$ and $y<0.7$ ZEUS finds the result
$\sigma(ep\to eD^*X)=5.3\pm1.0\pm0.8\,$nb for 
$1.3<p_T<9.0\,$GeV and $|\eta|<1.5.$
Using the GRV94 gluon density and $m_c=1.5\,$GeV, the next-to-lowest order 
calculated 
cross-section(for the same restricted kinematic region) is  
$4.15\,$nb, about one sigma below the measured cross-section. ZEUS have
followed a similar procedure to that described above for H1 to extrapolate the
$D^*$ cross-section to the full phase space. The next-to-lowest order 
calculation of 
Harris and Smith~\cite{nlocc_pty} was used for this purpose and it is
found that about 50-65\% of the cross-section is outside the measured region. 
 $F_2^{c\bar{c}}$ is extracted for two $x$ bins at each of
four values of $Q^2$ and the results are shown in Fig.~\ref{fig:f2cc_all}
together with those from H1 and EMC. This figure also shows a calculation 
from the GRV94 
parton distribution functions~\cite{GRV94} for a range of charm quark masses
between $1.35$ and $1.7$ GeV. It should also
be noted that the recent CTEQ4F3~\cite{LaiTung} or MRRS~\cite{MRRS} 
gluon densities give results within 5\% of the GRV94 parametrization 

The first results from HERA on $F_2^{c\bar{c}}$ are very encouraging. 
The charm signal can be identified cleanly, backgrounds are
under control and the measurements indicate that the BGF process is dominant.
However high precision results await high luminosity from HERA, though
the use of silicon micro-vertex detectors could improve the detection
efficiency by an order of magnitude. For the future there is still much
to do. Is the apparent discrepancy between next-to-lowest order 
calculation and the measured
cross-section at low $Q^2$ real? Will better data help to choose between
the various approaches to $c\bar{c}$ production and the quark versus
parton question at large $Q^2$? Will precise data provide more constraints
on the gluon and possibly shed light on the necessity or
otherwise of terms beyond standard DGLAP in the QCD evolution equations
(see Sec.~\ref{sec:summary})? 

\subsection{$F_2^d/F_2^p$ and $R^d-R^p$}

The ratio $F_2^d/F_2^p$ is directly related to $F_2^n/F_2^p$ which 
gives information on the ratio of up and down quarks in the nucleon.
Because the $Q^2$ dependences of $F_2^d$ and $F_2^p$ are similar
one expects the ratio to be almost independent of $Q^2$, the slight
residual dependence can be calculated from pQCD. Similarly the
difference $R^d-R^p$ is expected to be small and is sensitive to
differences in the gluon distributions in the proton and neutron.
 
The NMC collaboration has recently published~\cite{nmcdata3}
accurate and extensive measurements of the ratio $F_2^d/F_2^p$ 
and the difference $R^d-R^p$. The ratio data cover the range
$0.001<x<0.8$ and $0.1<Q^2<145\,$GeV$^2$. They are derived from the
previously mentioned large and small angle trigger 
data~\cite{nmcdata1,nmcdata2} and an additional small angle trigger T14.
T14 uses only the central part of the muon beam and is derived from small 
scintillators just above and below the muon beam. Measurement of the ratio
exploits the special feature of the NMC target arrangement 
(see Sec.~\ref{sec:nmcdet}) which has two pairs of targets with H$_2$ 
and D$_2$ up and down stream, respectively, in one and with the 
order reversed in the other. By alternating the target pairs in the beam,
the measured cross-section ratio does not depend on either the
incident muon flux or the detector acceptance. The systematic errors
can thus be kept to a minimum and the region in which the ratio 
is measured is larger than that for $F_2^p$ or $F_2^d$ alone.

If $R_d=R_p$ then $F_2^d/F_2^p$ is given by the cross-section ratio
$\sigma^d/\sigma^p$ in a given $(x,Q^2)$ bin -- see Eq.~\ref{eq:Rxsec}.
Given that the difference $\Delta R=R^d-R^p$ is small, the
group use the following approximate relation
\begin{equation}
{\sigma^d\over \sigma^p}\approx{F_2^d\over F_2^p}\left(1-
{1-\varepsilon\over (1+\bar{R})(1+\varepsilon\bar{R})}\Delta R\right),
\label{eqn:f2r_ratio}
\end{equation}  
where $\bar{R}=(R^d+R^p)/2$. The results are extracted in two steps.
First $\Delta R$ averaged over $Q^2$ is extracted for each $x$-bin 
by fitting the data in a bin to Eq.~\ref{eqn:f2r_ratio} using 
4 parameters ($\Delta R, \bar{R}, a_1, a_2$) where 
$F_2^d/F_2^p=a_1+a_2\ln Q^2$. The values of $\Delta R$ are all
small and $\Delta R$ shows no significant $x$ dependence. Averaging 
over $x$ gives $\Delta R=0.004\pm0.012(stat)\pm0.011(sys)$ 
at an average $Q^2$ of $5\,$GeV$^2$ and $0.003<x<0.35$.
Having established that $\Delta R$ is compatible with zero, the second
step in the extraction of the structure function ratio data is to take
$\Delta R$ as zero so that $F_2^d/F_2^p=\sigma^d/\sigma^p$.
Neglecting nuclear effects in the deuteron 
$F_2^n/ F_2^p=2\sigma^d/ \sigma^p-1$. 

The E665~\cite{e665_ratio} collaboration collected data on the 
ratio $\sigma^d/\sigma^p$  in 
1991. In addition to the small angle trigger (SAT) used for structure function 
measurements, they also collected data with the EM calorimeter trigger which 
allowed them to get to $Q^2$ as small as $0.001\,$GeV$^2$. The calorimeter 
trigger recognizes very low $Q^2$ inelastic scattering events by the large 
spread of energy deposits and in particular can reject elastic $\mu e$ events 
which are a significant background. The ratio $F_2^n/F_2^p$ is extracted 
from the cross-section ratio ignoring any nuclear effects. Overall systematic 
errors are in the range $2-3.5$\%, depending on the trigger, dominated by 
uncertainty in the relative acceptances. No significant $Q^2$ dependence is 
found but at low $x$ the ratio is less than 1, more precisely for 
$x\leq0.01, ~F_2^n/F_2^p=0.935\pm0.008\pm0.034$ indicating that there may be 
some nuclear shadowing in this region. 

\begin{figure}[htbp]
\vspace*{13pt}
\begin{center}
\psfig{figure=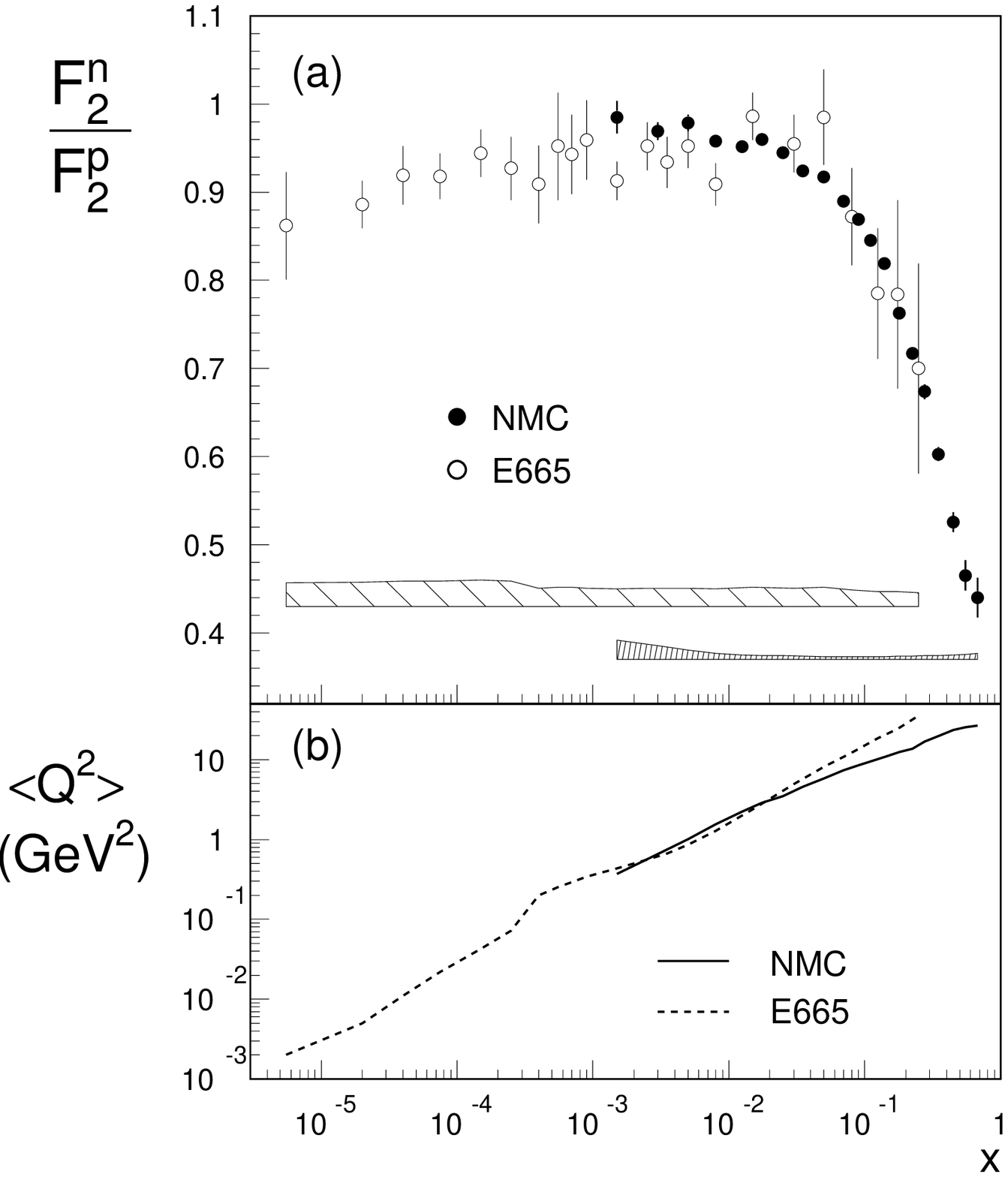,bbllx=-230pt,bblly=150pt,bburx=270pt,bbury=700pt,width=6cm,height=8cm} 
\fcaption{$F_2^n/F_2^p$ averaged over $Q^2$ as a function of $x$ from the
NMC and E665 experiments.}
\end{center}
\label{fig:nmcf2ratio}
\end{figure}

The results for $F_2^n/F_2^p$ as a function of $x$ averaged over $Q^2$ 
are shown in Fig.~\ref{fig:nmcf2ratio} for both NMC and E665.
 One should be warned that
the $F_2^n/F_2^p$ as defined above may deviate from the free nucleon
ratio, though such effects are expected to be small for deuterium.

\vfill
\vskip1cm
\subsection{Sum rule data}
\label{sec:gott}

We summarise fairly recent data on the Gottfried sum rule. Recent
evaluations of the Gross Llewellyn-Smith sum rule will be 
discussed in Sec.~6.2 in the context of measurement of $\alpha_S$.

From the QPM relations given in Eqs.~\ref{eq:f2lp} and ~\ref{eq:f2ln} 
and valence quark sum rules one finds
\begin {equation}
S_G=\int_0^1(F_2^{\mu p}-F_2^{\mu n}){dx\over x}={1\over 3}+{2\over 
3}\int_0^1(\bar{u}-\bar{d})dx.
\end{equation}
Early data supported the assumption that the $q\bar{q}$ sea was SU(2) 
symmetric, i.e. $\bar{u}=\bar{d}$. 
Using the relation
\begin{equation}
F_2^p-F_2^n=2F_2^d{(1-F_2^n/F_2^p)\over (1+F_2^n/F_2^p)}=
2F_2^d{(1-F_2^d/F_2^p)\over F_2^d/F_2^p}
\end{equation}
the integral can be evaluated by using ratio data and $F_2^d$.
In 1991 NMC published an evaluation of the sum rule using their early data 
on the ratio $F_2^d/F_2^p$ at 90 and $280\,$GeV muon beam momentum 
and $F_2^d$ from a global fit. This result showed that the integral was 
significantly below 1/3 and that  $\bar{u}\not=\bar{d}$.
NMC made a more exact evaluation at $Q^2=4\,$GeV$^2$ using their own 
$F_2^d$ data in 1994~\cite{nmc_gsr} giving 
\begin{equation}
S_G(0.004-0.8)=0.221\pm0.008(stat.)\pm0.019(sys.).
\end{equation}
The final NMC data referred to in the previous section gives 
$0.2281\pm0.0065$ consistent with the 1994 result. 
To complete the integral, the contributions from the unmeasured regions at 
high and low $x$ have to be estimated. The contribution from $x>0.8$ is 
consistent with zero ($0.001\pm0.001$). For the region $x<0.004$ a 
Regge like parametrization, $ax^b$, was fit  to the data in the region 
$0.004<x<0.15$ and extrapolated to give a contribution to $S_G$ of 
$0.013\pm0.005$. The final result for $S_G$ is $0.235\pm0.026$, where 
the error is the sum in quadrature of the statistical and systematic 
contributions.
The asymmetry $\bar{u}-\bar{d}$ is not something that can be generated 
by pQCD evolution and must be inserted by hand in the parametrizations of 
the parton distribution functions at the starting scale $Q^2_0$. We discuss 
this further in Sec.~\ref{sec:pdf}.

\section {Determining parton distribution functions and $\alpha_s$}
\noindent

The success of pQCD, as worked out in the conventional framework outlined in
Sec.~\ref{sec:Thframe}, has been well documented in the textbooks of 
Roberts~\cite{roberts} and, more recently, 
Ellis, Stirling and Webber~\cite{esw}. 
In the present section we consider how the conventional framework has been
used to determine parton distributions and the strong coupling
constant $\alpha_s$ (it is now conventional to
quote the value of $\alpha_s$ at the scale $M_Z^2$, and this is what we shall
mean by the notation $\alpha_s$ when we quote numerical values, unless
otherwise specified). We discuss the extraction of parton distributions in
Sec.~\ref{sec:pdf}. We pay particular attention
to the extraction of the gluon distribution since most cross-sections 
at hadron colliders, present and future, are dominated 
by gluon induced processes, but the gluon distribution is the least 
accurately determined of the parton distributions.
We summarize determinations of $\alpha_s$ from deep inelastic scattering, 
and compare with evaluations from  other processes in Sec.~\ref{sec:alphas}. 
Finally we consider the non-perturbative techniques which have been used 
to predict the form of parton distributions in Sec.~\ref{sec:modelpdfs}.

\subsection{Extraction of parton distributions}
\label{sec:pdf}

Ideally one would like to find analytic parametrizations of parton 
distributions,
or structure functions, which are consistent with pQCD. The problem
is to perform a Mellin inversion of the exact predictions of pQCD for the
moments of structure functions in order to find suitable analytic expressions
for the structure functions themselves. Such parametrizations have been given by many
authors, see for example~\cite{BurGam,BB,OR,Yndurainold}. However, if one requires consistency with pQCD
beyond leading order, one cannot find exact analytic expressions which are valid for more than
a limited $x,Q^2$ range. For this reason, 
the most commonly used method of extracting parton distributions is to
perform a direct numerical integration of the DGLAP equations at NLO.
The technique is broadly as follows. 
An analytic shape for the parton distributions (valence, sea and gluon) is
assumed to be valid  at some starting value of $Q^2 = Q^2_0$. This starting
point is arbitrary, but should be large enough to ensure that 
$\alpha_s(Q^2_0)$ is small enough for perturbative calculations to be
applicable. Then the DGLAP equations are used to evolve the 
parton distributions up to a different $Q^2$ value, where they  are 
convoluted with coefficient functions, appropriate to the chosen 
renormalization scheme, 
in order to make predictions for the structure functions. These predictions 
are then fitted to the data, for whatever $x,Q^2$ points have been measured.
The fit parameters are those necessary to specify the input analytic shape, 
and the
QCD $\Lambda$ parameter (though one should note that this is fixed in some 
parametrizations). The input analytic form assumed for the parton 
distributions is only valid at the starting scale $Q^2_0$. For other $Q^2$ 
values the distributions must be interpolated from values at grid points, 
such as those provided in 
the parton distribution function library PDFLIB~\cite{PDFLIB} (see also
HEPDATA). 
 
\begin{figure}[ht]

\centerline{\psfig{figure=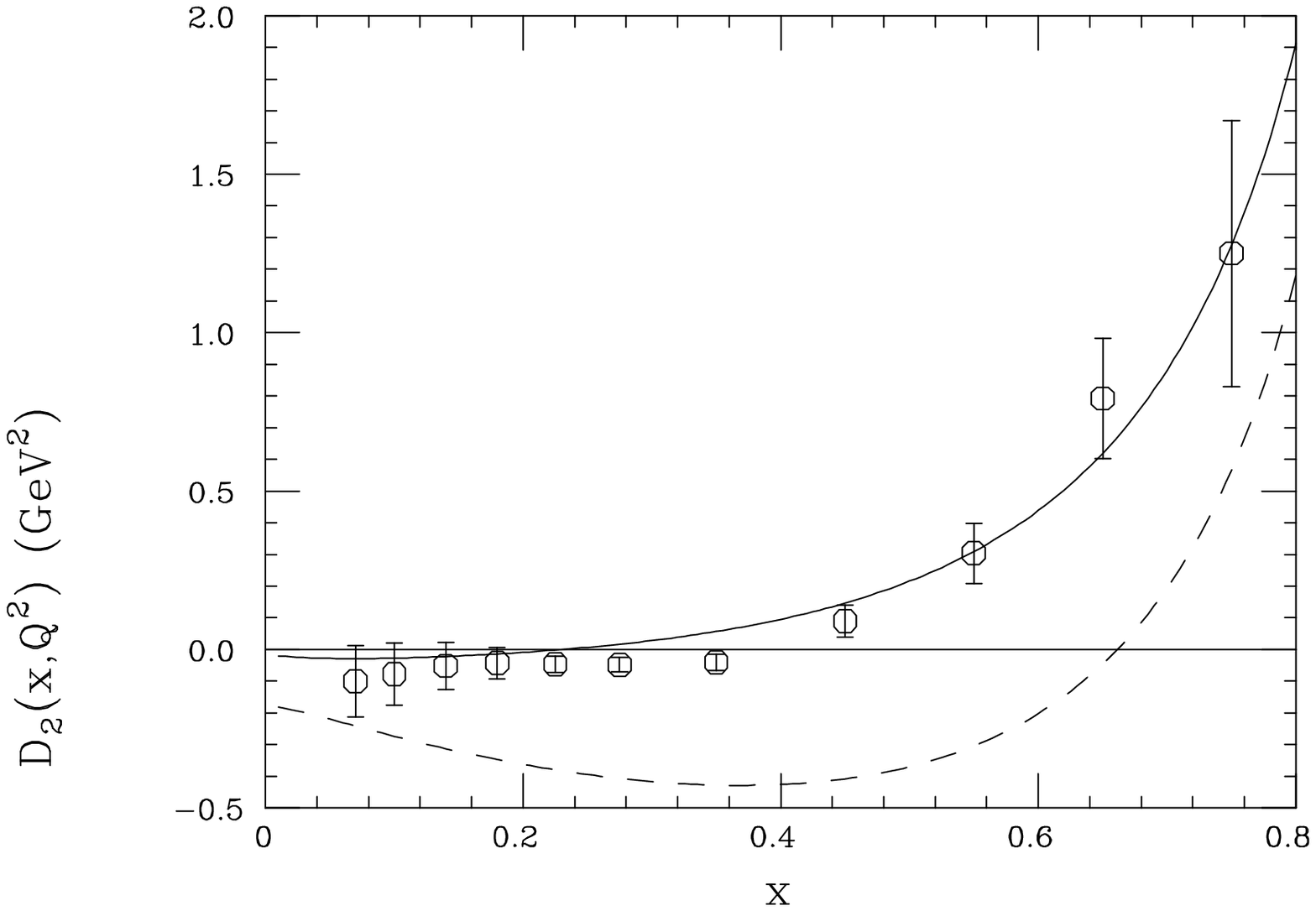,bbllx=30pt,bblly=200pt,bburx=570pt,bbury=630pt,width=8cm,height=7cm}}
\fcaption{Renormalon predictions for the twist-4 contributions to $F_2$ 
(full line) and $xF_3$ (dotted line). The SLAC and BCDMS data from the analysis
of Virchaux and Milsztajn are shown for comparison.}
\label{renormalon}
\end{figure}

The DGLAP equations embody the predictions of pQCD in the NLLA 
at leading twist only.
Thus one must be sure that the data which are fitted are not likely to 
be subject to strong higher twist corrections.
An analysis of the need for higher twist terms 
 was made by Virchaux and Milsztajn~\cite{VirMil} on 
SLAC and BCDMS hydrogen and deuterium charged leptoproduction data. They
accounted for target mass effects and included dynamical 
higher twist terms by allowing the leading twist form of the
structure function to be multiplied by $( 1 + \frac{C}{Q^2})$,
where the parameter $C$ is fitted separately in each $x$ bin, so that no
specific function of $x$ is imposed on the data. The values for $C$ are
found to be non-zero and positive only for $x>0.5$ and 
$Q^2 \leqsim 10\,$GeV$^2$, see Fig.~\ref{renormalon}.

An analysis of data on the ratio $F_2^{\mu n}/F_2^{\mu p}$ by the NMC 
Collaboration~\cite{NMCrat} concluded that small higher twist terms were 
necessary to describe
data in the kinematic region, $0.1 < x< 0.3$, $1 < Q^2 < 10\,$GeV$^2$,
 but a more recent analysis of the latest NMC ratio data and comparison with
SLAC and BCDMS data~\cite{nmcdata3}, indicates that
this ratio has only a small logarithmic $Q^2$ dependence compatible with 
leading twist pQCD predictions.

Thus higher twist contributions seem to be important only for high $x$ and 
low $Q^2$. It is interesting to note that the form of these  
contributions as measured experimentally is in agreement with 
the recent renormalon predictions~\cite{das} of the $1/Q^2$ contribution to 
structure functions. These predictions differ for $F_2$ and
$xF_3$, the predictions for $xF_3$ being small and negative for $x < 0.65$
(see Fig.~\ref{renormalon}). This may explain longstanding differences in 
higher twist terms between charged lepton and neutrino data, whereby the 
analysis of EMC/SLAC proton target data~\cite{EMC} found large 
positive values of $C$ 
for $x > 0.5$, whereas analysis of WA59 (anti)neutrino data~\cite{Varvell}, 
found small negative values for most of the measured $x$ range. 
Unfortunately, the WA59 data did not distinguish $F_2$ and $xF_3$ for the 
higher twist analysis, and these data were taken on a heavy target without 
nuclear corrections. Thus these observations
were not conclusive. However some analyses of the much more precise CCFR(93) 
and CCFR(97) $xF_3$ 
data~\cite{Sidarov1,KKSP97} and of a combination of world $xF_3$ 
data~\cite{Sidarov2}, taking into
account nuclear effects, has recently been made.  The higher twist 
contribution is evaluated  by adding a term $h(x)/Q^2$ to the to the pQCD 
prediction for $xF_3$ at leading twist, for each $x,Q^2$ point.
Agreement of the measurements of $h(x)$ with the form of 
the renormalon prediction is again found. 
\noindent

In conclusion, the present consensus is that one may cut out the need 
for higher twist contributions by cutting out the kinematic region
in which they are important.  The cut has
typically been made such that only data for which $Q^2 \geqsim 4\,$GeV$^2$, 
$W^2 \geqsim 10\,$GeV$^2$ are considered and most analyses have been made
using the structure function $F_2$. Thus we note that: i) 
$F_L$ may be subject to higher twist corrections for all $x$, at 
low $Q^2$ (see Sec.~\ref{sec:hitwist}); ii) the renormalon predictions 
indicate that $xF_3$ may have small higher twist corrections for all $x$;
ii) if the $Q^2$ cut is lowered to $1\,$GeV$^2$, as in the most recent 
parametrizations, one should be more cautious (however since 
most of the data at such low $Q^2$ are also at low $x$,  
higher twist terms are not expected to be very significant). 
Subject to these warnings we may have confidence in leading twist DGLAP 
analyses.
Of course the above considerations refer to the conventional higher twist 
effects discussed in Sec.~\ref{sec:hitwist}. In HERA very low $x$ data we may 
see the effects of unconventional higher twist effects which contribute to 
parton shadowing as discussed in Sec.~\ref{sec:lowx}.
 
When considering the applicability of QCD analyses of DIS data we should also
consider the presence
of diffractive-like events in the deep inelastic scattering
event sample. It has been established that there is a contribution to this 
sample  of approximately $10\%$ from events with a large
rapidity gap towards the proton remnant, with characteristics as
measured in hadronic diffractive exchange~\cite{gallo97,eich97}.
These events may originate from a different production mechanism
as compared to the bulk of the deep inelastic scattering data, for example, 
they include
processes such as exclusive vector meson production.
So far no special account has been taken of this in the extraction of
partons from inclusive $F_2$ measurements.

The most reliable parton distributions are obtained by 
making global analyses of a wide range of data from DIS and other hard 
processes. It is appropriate to first consider the consistency of data sets.

\subsubsection{Treatment of data sets in global analyses}

In addition to the standard corrections described in Sec.~\ref{sec:data}, there
are further corrections which must be made when combining data from different 
experiments.

Firstly, data must be taken on comparable targets. 
A nuclear target cannot simply be treated as an additive
combination of nucleons, and data taken on such a target have to undergo 
corrections for
nuclear shadowing, binding and Fermi motion effects before they  can be
used to extract information on parton distributions within the free 
nucleon. Modern data are taken on hydrogen or deuterium targets to avoid this
problem. However, for precision measurements, the need for correction even 
extends to deuterium targets. For (anti)neutrino beams the event rate on
proton/deuterium targets is very low. Accordingly the CCFR experiment
uses a heavy target and the data are corrected for nuclear effects before 
they are compared to the electroproduction data.  
Nuclear effects were  measured extensively by the EMC, SLAC and NMC 
collaborations in the mid to late eighties~\cite{NMCnuclear}, 
so that reliable corrections are available for charged leptoproduction
experiments.  

Secondly, the treatment of heavy quark thresholds, needs careful consideration.
The theoretical background to this has been given in Sec.~\ref{sec:heavyq}.
 Different authors have used somewhat differing procedures, 
which will be indicated appropriately when their analyses are described.
One should also be careful when comparing data from neutrino and muon/electron
scattering, since the effect of mass thresholds will be different. In 
neutrino scattering charmed quarks can be produced directly from scattering 
off the strange sea, as well as in the boson-gluon fusion 
process (flavour excitation, $W^*s \to  c$, as well as flavour creation,  
$W^*g \to s\bar c$) and the former process is dominant at the energies
at which present day neutrino data have been taken. The neutrino
data are usually corrected with a `slow rescaling' prescription to allow for 
the charm mass threshold in the flavour excitation 
process~\cite{Brock}. However this is not really
adequate when doing a NLO analysis and a full variable flavour treatment which
also includes the flavour creation process
should be performed~\cite{varflav}. 

Thirdly, in the late 1980's 
there were still significant disagreements between various data
sets. The muo-production experiments EMC~\cite{EMC} and BCDMS~\cite{BCDMS} 
disagreed in both normalization and shape, so much so that the 1990 global 
analysis of HMRS~\cite{HMRS}, published two separate sets of parton 
distributions the E set for EMC data and the B set for BCDMS data. 
Similarly  the neutrino production 
data of the CDHSW and CCFR collaborations differed in shape at low $x$.
 A complete re-analysis of the old SLAC experiment was undertaken incorporating
modern understanding of relevant corrections~\cite{whitlow_thesis}, partly in 
order to help to resolve these discrepancies. 
The EMC data have now  been superseded by the increased precision of the NMC 
data, and the CDHSW data have been superseded by the increased precision of the
CCFR data. There is a remaining disagreement between the NMC(97) Muon data 
and the CCFR(97) neutrino data at low $x$, see Fig.~\ref{fig:ccfrdat2} 
and the discussion 
in Sec.~\ref{sec:xf3dat}. Modern global fits deal with this discrepancy 
differently, as specified in the next 
subsection. Differences in normalization
between data sets are accounted for by allowing
the relative normalizations of data sets to float within the 
overall absolute normalization uncertainties given by the experiments. 

A further problem in the combination of data from different experiments 
concerns the treatment of systematic errors.
 Conventionally point to point systematic errors are accounted for by
adding them in quadrature to the statistical errors. Most global fits do not 
take full account of correlations between experimental
systematic errors since this information has not been available for 
all data sets and the correct procedure when combining different data sets is
still a subject of active discussion.

\subsubsection{Global fits: the method of MRS and CTEQ}
\label{sec:mpdf}
During the 1980's and early 1990's many sets of parton distributions 
were developed to try 
to describe the available data~\cite{HMRS,DOEHLQGHR,KMRS,MTB}. 
Today only the parton distributions of the 
MRS~\cite{MRSD,MRSD',MRSA,MRSG,MRS96},   
CTEQ ~\cite{CTEQ1,CTEQ3,CTEQ4} and GRV~\cite{GRV91,GRV94} groups are used, 
since only these
take account of modern data. We first discuss the analyses of MRS and CTEQ, 
which have a common
philosophy. We begin by giving
a general description of the parametrizations, discussing the 
assumptions which are made and the choice of data input to the analyses.
This will be done in detail only for the MRS 
parametrizations, and significant differences for CTEQ will be pointed out
when relevant.

The parton distributions  must be parametrized by a suitably flexible analytic
 form at the starting $Q^2=Q^2_0$.  MRS have used $Q^2_0 = 4\,$GeV$^2$, for all
but the latest MRSR fits, whereas CTEQ have used $Q^2_0 = 2.56\,$GeV$^2$ 
for all
but the CTEQ4LQ parametrization. Both groups  use inputs which extend the
simple forms of Eqs.~\ref{eq:alfbet},~\ref{eq:sea} as follows
{\small \begin{equation}
xu_v = A_u x^{\eta_1}(1 - x)^{\eta_2} P(x,u)
\label{eq:parambeg}
\end{equation}}
{\small \begin{equation}
xd_v = A_d x^{\eta_3}(1 - x)^{\eta_4} P(x,d)
\end{equation}}
{\small \begin{equation}
xS = A_S x^{-\lambda_S}(1 - x)^{\eta_S} P(x,S)
\end{equation}}
{\small \begin{equation}
xg = A_g x^{-\lambda_g}(1 - x)^{\eta_g} P(x,g)
\label{eq:param}
\end{equation}}
\noindent
where $P(x,i) = ( 1 + \epsilon_i \sqrt x + \gamma_i x)$ for MRS and
$P(x,i) = ( 1 + \gamma_i x^{\epsilon_i})$ for the CTEQ analyses 
(although both groups have
used `minimal' parametrizations $P(x,i) = ( 1 + \gamma_i x)$, 
for the gluon distributions
in some of their parametrizations).
Not all of the normalizations $A_i$ are free parameters, $A_u,A_d$ are
determined by the need to satisfy flavour sum rules and $A_g$ is 
determined in terms of the other three by the momentum sum rule.
The distributions are usually defined within the $\MSB$ renormalization and
 factorization scheme and this will be assumed unless otherwise stated. 
The QCD scale parameter $\Lambda$ 
is usually a parameter of the fit. The treatment of
$\Lambda$ across flavour thresholds is as given by Marciano~\cite{Marciano} as
described in Sec.~\ref{sec:heavyq}. Since MRS express their results in terms of 
$\Lambda$ for four flavours, whereas CTEQ express their results
in terms of $\Lambda$ for five flavours, we chose to quote 
$\alpha_s(M_Z^2)$ for both analyses.

Some comments on this choice of parametrization are in order.
The fact that the $u_v$ valence shape is not the same as the 
$d_v$ valence shape has been known since the earliest days of
neutrino scattering when neutrino and antineutrino scattering data
on protons and deuterium were compared~\cite{Parker,Allasia} . However, 
in more recent years, the data
which fix these valence shapes have come from taking the difference and the 
ratios of $F_2^{\mu p},F_2^{\mu n}$ data~\cite{nmcdata3}. At large $x$,
when only valence distributions are significant, one has
{\small \begin{equation}
\frac{F_2^{\mu n}(x)}{F_2^{\mu p}(x)} = \frac{1 + 4d_v(x)/u_v(x)}
{4 + d_v(x)/u_v(x)}
\end{equation}}
\noindent
from the QPM.
The data indicate that $d_v(x)/u_v(x) \sim (1-x)^1 \to 0$ as $x \to 1$, 
as predicted by Field and Feynman~\cite{FF},
rather than the naive result of SU(3) flavour symmetry, 
$d_v(x)/u_v(x) = 1/2$, for 
all $x$. (However, see references~\cite{gurvitz,thom} for an 
alternative view.)

The sea distribution refers to $u,\bar u, d,\bar d, s, \bar s, c, \bar c$.
Since we must have $q = \bar q$ within each flavour of the sea by 
definition, we may express the sea as 
$S = 2(\bar u + \bar d + \bar s +\bar c)$.
Early parametrizations assumed that the $u,d$ content of the sea
is flavour symmetric, but there is no necessity for this and in 1992
NMC gave the first evidence that this is not the case~\cite{NMCGott}. 
The Gottfried sum rule is violated
(see Sec.~\ref{sec:gott} for the latest data). 
If we abandon the 
assumption that $\bar u = \bar d$, which was used in its derivation, we obtain
{\small \begin{equation}
\int^1_0 \frac{dx}{x} (F_2^p - F_2^n) = \frac{1}{3} \int^1_0 dx 
( u_v - d_v) + \frac{2}{3} \int^1_0 dx (\bar u - \bar d) 
\end{equation}}
\noindent
Thus the observation of the value $\simeq 0.23$ for this sum
tells us that $\bar d > \bar u$, as expected from Pauli 
suppression~\cite{feynman}. (For further discussion of the reasons for this
result see refs.~\cite{thomas,kumano,buccella}). 
In the MRS parametrization this
is taken into account by expressing the structure of the sea as
{\small \begin{equation}
 2\bar u = 0.4(1-\delta)S - \Delta,\ 2\bar d = 0.4(1-\delta)S + \Delta,\ \
 2\bar s = 0.2(1-\delta)S,\ 2\bar c = \delta S
\end{equation}}
\noindent
where,
{\small \begin{equation}
x\Delta = x(\bar d - \bar u) = A_{\Delta} x^{\lambda_{\Delta}}(1 - x)^{\eta_S}
P(x,\Delta)
\end{equation}}
\noindent
with $\epsilon_\Delta=0$, and $\lambda_\Delta \sim 0.5$ from considering the breaking of
$\rho/A_2$ Regge trajectory exchange degeneracy.  The CTEQ group account for $u,d$ differences 
in the sea similarly, but allow  more free parameters.

The strange sea is suppressed relative to the $u$ and $d$ seas and the 
charmed sea is 
very suppressed. The origin of this suppression lies in their larger quark
masses, but we may allow for it by introducing a simple suppression factor.
MRS assume that $\bar s = (\bar u + \bar d)/4$, i.e. 
the strange sea is suppressed by $50\%$ compared to the $u$ and $d$ 
sea distributions at $Q^2_0$. The justification for this comes from CCFR opposite sign
dimuon data~\cite{ccfr_dimuon}, which confirm the earlier data of 
CDHSW~\cite{CDHSmu}. Briefly, opposite sign dimuon events dominantly arise in 
(anti)neutrino scattering when the
struck quark is an ($\bar s$)$s$ quark which becomes a ($\bar c$)$c$ quark 
through the flavour changing weak current. This ($\bar c$)$c$ quark then 
decays through the muon channel to ($\bar s \mu^- \nu_{\mu}$) 
$s \mu^+ \nu_{\mu}$ yielding a muon of opposite sign to that at the 
original lepton scattering vertex. The rate for the process thus gives a 
measure of the size of the strange sea, which is consistent with $50\%$
suppression\fnm{o}\fnt{o}{~The description of the dimuon data is better if this
suppression is applied at $Q^2_0 = 1\,$GeV$^2$, 
rather than $Q^2_0 = 4\,$GeV$^2$.}. The strange sea is treated similarly
in the latest CTEQ and MRS parametrizations, although historically 
CTEQ had allowed
it to have an independent parametrization. 
Their early work on extracting the size
of strange sea from the difference 
$xs(x) = \frac{5}{6}F_2^{\nu N}(x) - 3 F_2^{\mu d}(x) $,
which involves combining data from two different experiments, pointed to
a discrepancy between this determination 
and the dimuon results, indicating a difference between NMC muon
data and CCFR neutrino data, in the region $0.01 < x < 0.1$. As we have seen
this discrepancy has still 
not been resolved even when the updated CCFR(97) and NMC(97) data are used 
(see Fig.~\ref{fig:ccfrdat2}).
CTEQ omit these data points from their subsequent analyses,
whereas MRS include both sets of data to achieve a compromise fit.

The charmed sea has also been treated similarly by the two groups. 
Their standard parametrizations use a zero-mass variable 
flavour number (ZM-VFN) scheme, see Sec.~\ref{sec:heavyq}. Basically
one assumes that
$c(x,Q^2) = 0$ for $Q^2 \le m^2_c$. The charm content of the nucleon at
higher $Q^2$ is then generated by the boson-gluon fusion process, as 
embodied in the DGLAP equations for massless partons. The shape of the charm 
distribution resulting
at $Q^2_0 = 4\,$GeV$^2$ is similar to the shape of the overall sea 
distribution.
The magnitude of the charm contribution to the total sea distribution
depends sensitively on $m_c$. Data on $F_2^{c \bar c}$ 
from the EMC collaboration 
have been used to fix $m_c^2 = 2.7\,$GeV$^2$, and this results in $2\%$ of
the total sea content being charmed at 
$Q^2_0 = 4\,$GeV$^2$~\fnm{p}\fnt{p}{~If one is using
$Q^2_0 = 1\,$GeV$^2$, then $m_c^2 = 2.7\,$GeV$^2$ is within the fitted region
and one must deal with the threshold behaviour smoothly. The formalism of 
Georgi and Politzer has been used to suppress the generated charm density by 
smooth factor~\cite{GP}.}. 
The contributions of bottom and top quarks are treated similarly, and they 
turn out to be negligible for present analyses.
Clearly such a treatment of heavy quarks can only be valid far above the
relevant thresholds, $W^2 \gg 4 m^2_q$.
Near threshold a proper treatment of quark mass effects is required as 
discussed in Sec.~\ref{sec:heavyq}. New parton distributions which incorporate a
consistent treatment of heavy quark effects from the threshold region to 
the asymptotic regime, using general-mass variable flavour number 
(GM-VFN) schemes, have been given by Martin et al~\cite{MRRS}
and Lai and Tung~\cite{LaiTung}. These will be discussed further in 
Sec.~\ref{sec:latest}.

The form $x^{-\lambda_g}$ for the gluon parametrization at small $x$
is suggested by Regge behaviour as explained in Sec.~\ref{sec:Thframe}, 
but whereas the conventional Regge exchange is that of the soft Pomeron, 
with $\lambda_g \sim 0.0$, one may also allow for a hard Pomeron, 
with $\lambda_g \sim 0.5$. The reasons for this are discussed in detail 
in Sec.~\ref{sec:lowx}.  In the original MRSD~\cite{MRSD} 
parametrizations these values were fixed as the alternative possibilities 
MRSD$_0$ and MRSD$_-$. However the input of high precision HERA data to the fits has 
allowed  $\lambda_g$ to be left as a free parameter. 

The form $x^{-\lambda_S}$ in the sea quark parametrization comes from similar
considerations since, 
at small $x$, the process $g \to q\bar q$ dominates the evolution
of the sea quarks. Hence the fits to early HERA data have as a constraint 
$\lambda_S=\lambda_g$. However one only expects this once $Q^2$ is large enough for the 
effect of  DGLAP evolution to be seen, hence it may not be a reasonable
constraint at $Q^2 = Q^2_0$. Furthermore, the exact solution of the DGLAP equations 
predicts $\lambda_S = \lambda_g - \epsilon$, see Eq.~\ref{eq:DLLAlam}. 
The data at low $x$ are now of
sufficient precision to allow $\lambda_g$ and $\lambda_S$ to be separate free
parameters, as in the MRSR~\cite{MRS96} and CTEQ4~\cite{CTEQ4} fits.

The relationship between the measured structure functions and the parton 
distributions is not straightforward at NLO since the evolved parton 
distributions must be convoluted with coefficient functions and all types of
parton may contribute to a particular structure function through the evolution.
However, the simple LO formulae give a good guide to the major contributions.
At LO the relationships between the structure functions and parton
distributions (ignoring charm) may be written as
{\small \begin{equation}
F_2^{\mu p} - F_2^{\mu n} = \frac{1}{3} x (u + \bar u - d - \bar d)
\end{equation}}
{\small \begin{equation}
\frac{1}{2}(F_2^{\mu p} + F_2^{\mu n}) = \frac{5}{18} x (u + \bar u + d + 
\bar d + \frac{4}{5}s)
\end{equation}}
{\small \begin{equation}
F_2^{\nu N} = F_2^{\bar \nu N} = x( u + \bar u + d + \bar d + 2 s)
\end{equation}}
{\small \begin{equation}
\frac{1}{2} x ( F_3^{\nu N} + F_3^{\bar \nu N}) = x(u - \bar u + d - \bar d)
\end{equation}}
Thus, even if there were no discrepancies between different data sets, 
deep inelastic data could only determine four quantities (usually taken to be
$ u + \bar u$, $d + \bar d$, $\bar u + \bar d$ and $s$) with any 
precision, on a point by point basis. Thus the global fits
use information from non-DIS processes in order to have
more complete information on parton distributions.

We will first consider data which constrain quark distributions.
Drell Yan dilepton production in the process $p N \to \mu^+ \mu^- X$ can
be a sensitive probe of the sea quark distribution since the dominant
subprocess is $q \bar q \to \gamma^*$, and thus the cross-sections are directly
proportional to the antiquark distributions. The E605 data~\cite{E605} have
 been used to constrain the sea distribution at medium to large $x$ values,
$0.1 < x < 0.6$, yielding $\eta_S \sim 10$. More recently, the CDF 
data~\cite{CDFDY} provide a further constraint at $x \sim 0.01$. 

The NA51 data~\cite{NA51} on the asymmetry, $A_{DY}$, between the differential
cross-sections $\frac{d^2\sigma}{dM dy}$ at $y=0$ for the processes
$p p \to \mu^+ \mu^- X$ and $p n \to \mu^+ \mu^- X$ 
(where  $M$ and $y$ are the invariant 
mass and rapidity of the lepton pair) has given more 
information on the difference $\bar d - \bar u$. The dominant subprocesses are 
$u\bar u, d\bar d \to \gamma^*$, and the partons are to be
evaluated at $x = M/\sqrt s$. Hence the NA51 measurement
serves to fix $(\bar d - \bar u)$ at the $x$ value, $x  \sim 0.18$. The E866
experiment is now taking data which will give further information on the $\bar u/ \bar d$ asymmetry in the $x$ range $0.05 < x < 0.3$~\cite{Towell}.

Data on $W$  production may also be used to investigate quark 
distributions. The processes $p \bar p \to W^+(W^-) X$ proceed via the 
subprocesses $u\bar d \to W^+(d\bar u \to W^-)$, and the cross-sections are thus 
sensitive to the $u$ and the $d$ distributions at $x \sim M_W/\sqrt s$, 
i.e. $x \sim 0.13$ at CERN, where UA2 data are taken, and 
$x \sim 0.05$ at FNAL, where CDF~\cite{CDFW} and D0 data are taken. 
The $W^{\pm}$ charge asymmetry, $A_W(y)$,
probes the slope of the $d/u$ ratio, since the
$u$ quarks carry more momentum on average than the $d$ quarks and so 
the $W^+$ tend to follow the direction of the incoming proton and the 
$W^-$ that of the antiproton. These measurements  give a 
more direct probe of the $d/u$ distribution than the $F_2^n/F_2^p$ ratio at 
intermediate $x$ values, since $W$ production, at Tevatron energies,
 is less sensitive to the contribution of the antiquarks.

The need for additional information from non-DIS processes is more
pronounced for the gluon distribution. Since the gluon does not couple to the 
photon it does not enter the expressions for the structure functions at all 
at LO. It is merely constrained by the momentum sum rule, and by the way 
the DGLAP equations feed its evolution into the sea 
distribution~\fnm{q}\fnt{q}{~At NLO the gluon distribution may contribute to 
$F_2$ depending on the renormalization scheme chosen.}. 
Consequently the gluon is the parton which has the largest uncertainty. 
We devote Sec.~\ref{sec:gluon} to considering the gluon distribution.
We do not confine our discussion to considering the role of the 
gluon distribution in the global fits. 
We consider ways to extract the gluon distribution from DIS 
using the scaling violations of
$F_2$, paying particular attention to the fits done by the individual 
experiments and we consider information which can be gained from the 
measurement of other structure
functions and from non-DIS processes, both at present and in future.

\subsubsection{Methods of extracting the gluon distribution}
\label{sec:gluon}

\begin{figure}[ht]
\vspace*{13pt}
\begin{center}
\psfig{figure=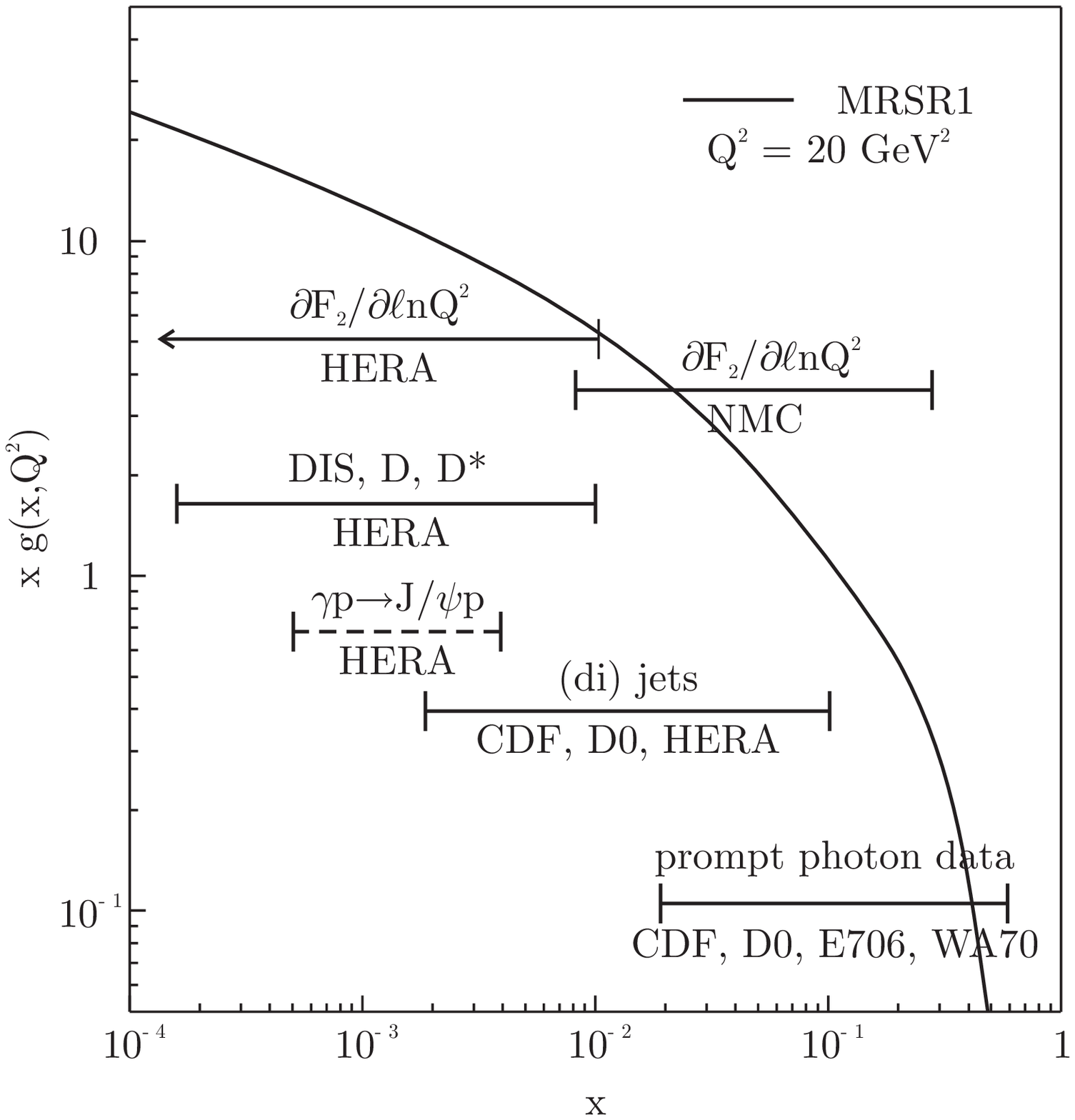,bbllx=-50pt,bblly=150pt,bburx=660pt,bbury=670pt,height=9cm} 
\fcaption{The $x$ intervals in which the gluon may be constrained by various 
sets of data. The MRSR1 gluon distribution is superimposed.}
\label{fig:gluon_over}
\end{center}
\end{figure}

An overview  of the data than can contribute 
to the gluon extraction is given 
in Fig.~\ref{fig:gluon_over} as function of $x$.

We begin by considering the QCD fits made by individual experiments to their own
structure function data. The experimental collaborations have
 the advantage of  knowing the correlations between
their systematic errors and thus can produce reliable error bands on the 
extracted distributions. The usual method follows that of the global fits.
A functional form with few free parameters is assumed for the gluon at a chosen
starting scale, and the coupled singlet 
evolution equations (Eqs.~\ref{eq:DGLAPq},~\ref{eq:DGLAPg}) are solved to NLO
simultaneously (usually in the $\MSB$ scheme). Since the gluon distribution is 
strongly correlated with
the value of $\alpha_s$ (see Sec.~\ref{sec:alphas}), 
a fixed value of $\alpha_s$,
as determined from independent data, is often assumed. Higher 
twist contributions must be included in the formalism if high $x$, low $Q^2$ 
data are included.  

In 1992 Virchaux and Milsztajn~\cite{VirMil} made a fit to
combined SLAC and BCDMS proton and deuteron 
data, which accounted for higher twist terms and for the
correlation with $\alpha_s$. The CDHSW collaboration~\cite{cdhswr} 
made a similar fit to the singlet structure function $F_2$ extracted from 
neutrino scattering on iron.     
These extractions indicated a soft behaviour of the gluon at
high $x$ and a tendency for $xg(x)$ to rise
for $x \leq 0.1$. However, the errors on the gluon distribution were
large, approximately $30\%-50\%$ at $x \sim 0.1$.

\begin{figure}[ht]
\vspace*{13pt}
\begin{center}
\psfig{figure=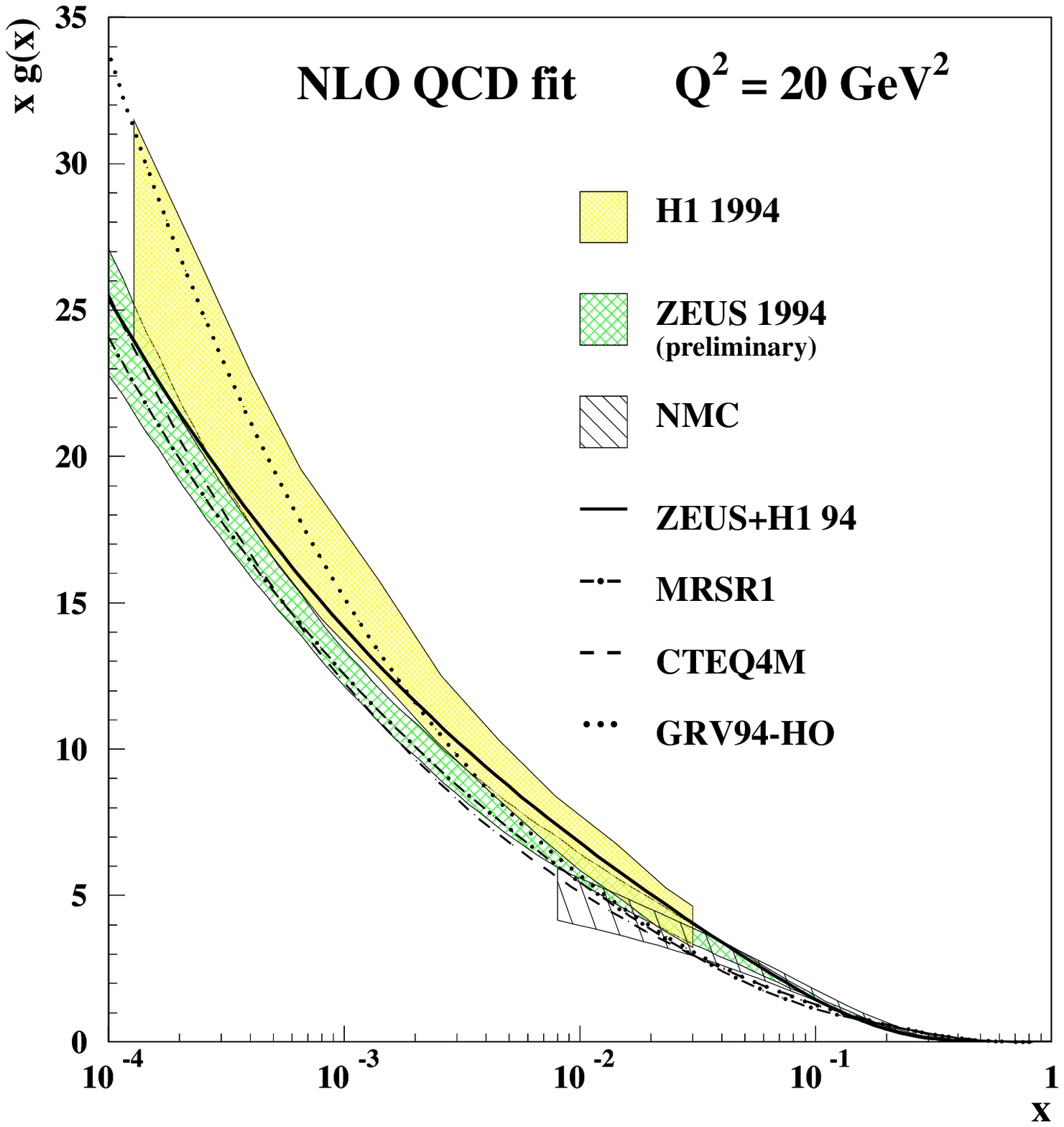,bbllx=-90pt,bblly=150pt,bburx=665pt,bbury=700pt,height=9cm} 
\fcaption{The gluon density $xg(x)$ at 
$Q^2 = $ 20 GeV$^2$ extracted from a NLO DGLAP 
fits, for ZEUS and H1. The NMC fit is also shown at higher $x$. 
The solid line is from a new fit using all HERA(94) data.
The predictions from CTEQ4 (dashed), GRV94 (dotted) and 
MRSR1 (dashed-dotted) are also shown for comparison}
\label{fig:gluon20}
\end{center}
\end{figure}

In 1993 the NMC collaboration used their precise $F_2$ data on proton and 
deuteron targets in the kinematic range
$0.008 \le x \le 0.5$, $1 \le Q^2 \le 48$ GeV$^2$ 
to make an NLO QCD analysis~\cite{qcdnmc}.
The flavour singlet and
non--singlet quark distributions as well as the gluon distribution were
parametrized at the reference scale $Q^2_0= 7$ GeV$^2$ and all  data with $Q^2
\geq $ 1 GeV$^2$ were included in the fit.  Higher twist terms were included 
by using the forms extracted by Virchaux and Milsztajn (see Fig.~\ref{renormalon}) averaged over the proton and
deuteron and suitably extrapolated to lower values of $x$.  
 Their contribution 
was found to be substantial at high $x$ for scales of about 1 GeV$^2$.
The gluon distribution extracted  is shown in Fig.~\ref{fig:gluon20}
for the fixed value of $\alpha_s = 0.113$. The uncertainty on the
distribution is calculated taking into account
the statistical errors, the systematic errors including correlations, the
contribution of the unmeasured region, and  the uncertainties on 
$\alpha_s$ and higher twist contributions. The total uncertainty at 
$x = 10^{-2}$ amounts to $\sim 20\%$.  These data cover the 
region $x > 0.01$ and the gluon distribution shows a tendency to continue to
rise moderately at low $x$. HERA data have now extended our knowledge of 
the gluon distribution down to $x \sim 10^{-4}$.
   
\begin{figure}[ht]
\vspace*{13pt}
\begin{center}
\psfig{figure=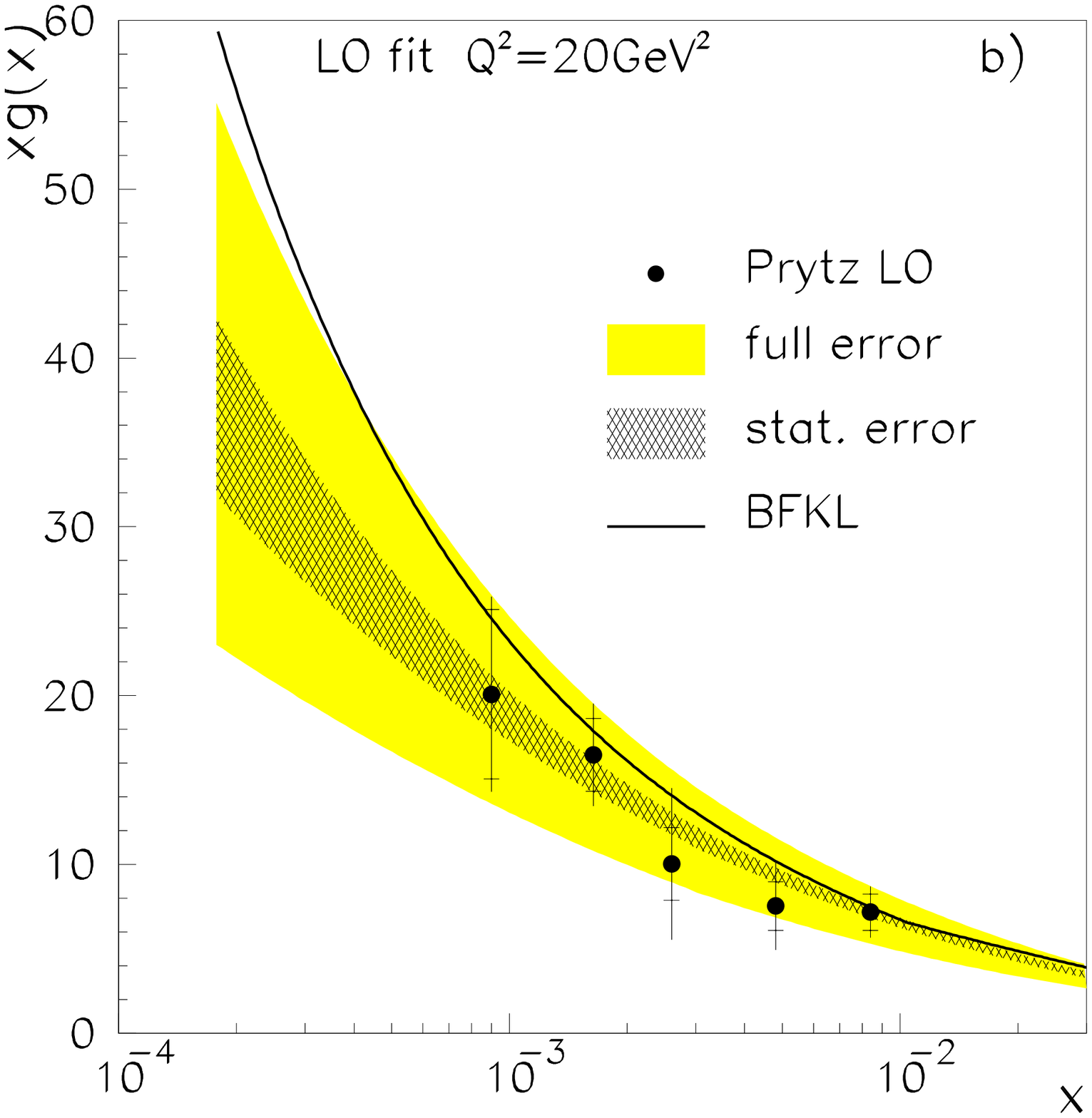,bbllx=-150pt,bblly=100pt,bburx=700pt,bbury=700pt,height=8cm} 
\fcaption{
The gluon density at 20 GeV$^2$ from a LO fit (shaded) and a mixed
DGLAP-BFKL fit (full line). The points are calculated using the method 
of Prytz on H1(93) data. 
The inner error bars (band) represent the statistical errors,
the outer error bars (band) the statistics and systematics added in 
quadrature.}
\label{fig:gluonlo}
\end{center}
\end{figure}

Fig.~\ref{fig:zeusall_q2} 
shows clearly the scaling violations for the measured
low $x$ values at HERA.
Before full  NLO DGLAP QCD fits were applied to extract the gluon 
from the HERA data, several 
approximate methods were used  to deconvolute the gluon density
directly from $F_2$, exploiting the fact that, at low $x$ the gluon is 
by far the most dominant parton and
$F_2$ is essentially given by the singlet sea quark distribution 
which is driven by the gluon (through the $g \to q\bar q$ 
splitting process). These methods gave  a first impression of the behaviour
of the gluon density at low $x$. The simplest method proposed by 
J. Prytz~\cite{PRYTZ} consists of  neglecting the quark contribution 
to scaling violations and performing a Taylor
expansion of the splitting function around $x=\frac{1}{2}$,
 leading to  a very simple LO expression for the
gluon density
\begin{equation}
xg(x,Q^2)\approx\frac{27 \pi}{20 \alpha_s(Q^2)}
 \frac{\partial F_2(\frac{x}{2},Q^2)}{\partial \ln {Q^2}}
\label{eq:kjell}
\end{equation}
This is a crude approximation which 
is expected to hold only to within $20\%$ at $x=10^{-3}$
for a steeply rising gluon~\cite{H1GLU}. 
The results are shown in Fig.~\ref{fig:gluonlo}, using the  H1(93) data.
This demonstrated the strikingly strong rise of the gluon distribution 
at low $x$. Approximate NLO corrections to this approach have 
been calculated~\cite{GOLEC,PRYTZ} and 
several more sophisticated and accurate methods have also been 
proposed~\cite{EKL,gluonprox}. These have usually involved the need for some
assumption about the degree of singularity of the behaviour of $F_2$ or the 
gluon itself as $x \to 0$. Experimentalists have preferred to concentrate
on gluon extractions from full NLO DGLAP fits to their data. 

Before we describe these fits we draw attention to the assumptions
inherent in such extractions.  It is implicitly assumed that the evolution 
equations can be applied in the kinematic range studied. It is not guaranteed 
that this is the case for DGLAP evolution at low $x$, it may be necessary to
consider the BFKL~\cite{BFKL}, or other, evolution equations, 
as will be discussed in Sec.~\ref{sec:lowx}. We illustrate this by comparing
the gluon distribution extracted from a LO DGLAP fit to that extracted from
a fit using somewhat different evolution equations, namely a
hybrid set~\cite{kwiehybrid}, based on the DGLAP and the 
BFKL evolution equations. (The latter was only 
available for the gluon, hence for the evolution of the 
quark distribution the DGLAP equations are used, suitably matched 
to the BFKL evolved gluon distribution). The results are shown in 
Fig.~\ref{fig:gluonlo}. Both fits are made to NMC(95), BCDMS and
H1(93) proton data, where the fixed target data is introduced to constrain
quark densities at high $x$.  The error bands show the statistical and total 
errors (taking into account correlations between the systematic
uncertainties) of the DGLAP fit, which agrees very well with the approximate 
method of Prytz. The full line shows the hybrid fit result. One can see that 
a different, in this case steeper, gluon distribution
can emerge when using different evolution equations even with the same input
data.

Both the H1 and the ZEUS collaborations have made fits of their data
to numerical solutions of the DGLAP equations at NLO in the $\MSB$ scheme. 
Heavy quark contributions have been treated using the three fixed flavour 
number scheme (FFN) given in ref.~\cite{GRS}, and
extended to $\alpha_s^2$ in the coefficient functions according to 
ref.~\cite{nlocc}. Thus there is no charmed parton distribution, charm is 
dynamically generated by the boson-gluon fusion (BGF) process.
 The scale for this process has been taken as
$\sqrt{Q^2+ 4 m_c^2}$, with a charm quark mass of $m_c=1.5$~GeV.  
An uncertainty in the charm quark mass of $\sim0.5$~GeV 
leads to a few percent variation of the gluon density. 
The  small contribution of beauty  quarks
has been  neglected.

Both groups use parametrizations for the input parton distributions which are
very similar to those used by the global fits of MRS and CTEQ, see
Eqs.~\ref{eq:parambeg}-\ref{eq:param}. For the H1 fit~\cite{qcdh1} the
the input scale is taken to be $Q^2_0 = 5$ GeV$^2$ and H1(94) data with 
$Q^2 > 5 $ GeV$^2$ are included in the fit. The function $P(x,i)$ is given by
$P(x,i) = ( 1 + \epsilon_i \sqrt x + \gamma_i x)$ for quark distributions and 
by $P(x,g) = 1$ for the gluon distribution. Since no isoscalar data are 
available yet in the small $x$ domain, $\eta_1=\eta_3$ is assumed. 
The quark and antiquark components of the sea are assumed to be equal,
and the flavour composition of the sea is assumed to be $25\%$ strange 
and equal amounts of $u$ and $d$ flavour (such that the Gottfried sum rule
would not be violated). The normalizations of
the valence quark densities are fixed using the flavour sum rules
and the normalization of the gluon density is determined using
the momentum sum rule. The parameters $\lambda_S$ and $\lambda_g$ which 
describe the low $x$ behaviour of the sea and gluon distributions are free
parameters which are allowed to be different from each other. The value of
$\alpha_s$ is fixed at $\alpha_s = 0.113$, as determined in ref.~\cite{VirMil}. 

The ZEUS fit~\cite{qcdzeus} differs in the following respects. 
The input scale is taken to be $Q^2_0 = 7$ GeV$^2$ but 
ZEUS(94) data for $Q^2 > 1.5 $ GeV$^2$ are included in the fit. The gluon 
distribution uses the function $P(x,g) = (1 + \gamma_g x)$.
Rather than parametrizing the $xu_v$, $xd_v$ and $xS$ distributions, ZEUS chose
to parametrize the singlet quark distribution $x\Sigma$ and the non-singlet 
difference between up and down quarks in the proton $x\Delta_{ud}$ as follows
\begin{eqnarray}\label{input1}
x\Sigma(x)&=& A_{\Sigma}x^{\lambda_{\Sigma}}(1-x)^{\eta_{\Sigma}}(1+\epsilon_{\Sigma}\sqrt{x} + \gamma_{\Sigma}),\nonumber \\
x\Delta_{ud}(x)&=& A_{ud} x^{\eta_{ud1} }(1-x)^{\eta_{ud2}}.\nonumber \\
\end{eqnarray}
The sea quark density is obtained by subtracting the valence distribution
(taken from the MRSD$_-^{\prime}$ parametrization, which is still appropriate
for the higher $x$ valence region) from the singlet distribution, and the 
strange sea is assumed to be $20\%$ of the total sea.

In order to constrain the valence quark densities at high $x$,
 proton and deuteron  results from the BCDMS (H1 fit only) and 
NMC (both H1 (NMC(95)) and ZEUS (NMC(97)) fit) experiments are also used.
In the fitting procedure a term was added to the $\chi^2$
to permit variation of  the relative normalizations of the different data
sets. To avoid possible contributions from higher twist effects, NMC data 
for which $Q^2<4~{\rm GeV}^2$ were excluded from the ZEUS fit, and 
NMC and BCDMS data for which $Q^2<5~{\rm GeV}^2$, and $Q^2<15~{\rm GeV}^2$ if 
$x > 0.5$, were excluded from the H1 fit. 
The $\chi^2$ obtained using uncorrelated systematic errors and the 
$\chi^2$ computed when
considering the full error of each point are given in Table~\ref{tab:tabchi}
for the H1 fit. Only  small adjustments of the relative normalizations 
 are required demonstrating 
 remarkable agreement between these  different experiments.
\begin{table}
\tcaption{The $\chi^2$ values for the H1 NLO QCD fit. 
For each experiment the number of data points used in the QCD fit, 
the $\chi^2$ obtained as described in the text
using only the uncorrelated errors, the $\chi^2$ computed from the same fit
using the full error on each point
and the normalization factors as determined from the fit, are given.
The H1 nominal vertex  and shifted  vertex 
data samples are denoted as
NVX and SVX respectively.}
\centerline{\footnotesize\smalllineskip
\begin{tabular}{cccccccc}\\
\hline
Experiment   &H1(94)   & H1(94) &NMC(95)-p&NMC(95)-D&BCDMS-p&BCDMS-D&total\\
             & NVX &  SVX & & & & & \\
\hline
 data points             & 157 & 16 &  96 &  96 & 174 & 159 & 698 \\
 $\chi^2$ (unco. err.)   & 174 & 13 & 157 & 153 & 222 & 208 & 931 \\
 $\chi^2$ (full error)   &  85 &  6 & 120 & 114 & 122 & 140 & 591 \\
normalization            & 1.00&1.04& 1.00&1.00 & 0.97& 0.97&      \\
\hline\\
\end{tabular}}

\label{tab:tabchi}
\end{table}

The results of the ZEUS fit have been shown in Fig.~\ref{fig:zdat1} and 
Fig.~\ref{fig:zdat2}
versus $x$ and Fig.~\ref{fig:zeusall_q2} versus
$Q^2$. The fit gives a good description of all data used.
The $Q^2$ dependence at fixed $x$ is also described 
well over the nearly four orders of magnitude covered
by the HERA data, see Fig.~\ref{fig:zeusall_q2}.
Adding the statistical and systematic errors in quadrature the quality
of the ZEUS fit is $\chi^2 = 463$ for 408 data points and 13 free parameters.

For ZEUS the parameters describing the low $x$ behaviour of the quark singlet
and gluon distributions at $Q^2_0 = 7$ GeV$^2$ are $\lambda_{\Sigma} = 0.23$ 
and $\lambda_g = 0.24$. For the H1 fit the corresponding dependences of the
sea and the gluon distributions at $Q^2_0 = 5\,$GeV$^2$ are $\lambda_S = 0.19$ 
and $\lambda_g = 0.20$. These powers are similar to those extracted from the
global fits, see Sec.~\ref{sec:latest} and Fig.~\ref{dicklam}. 
Note that there are sizeable
correlations between the fit parameters which were not studied in detail as
the basic aim of these analyses was to extract the gluon density.

\begin{figure}[ht]
\vspace*{13pt}
\begin{center}
\psfig{figure=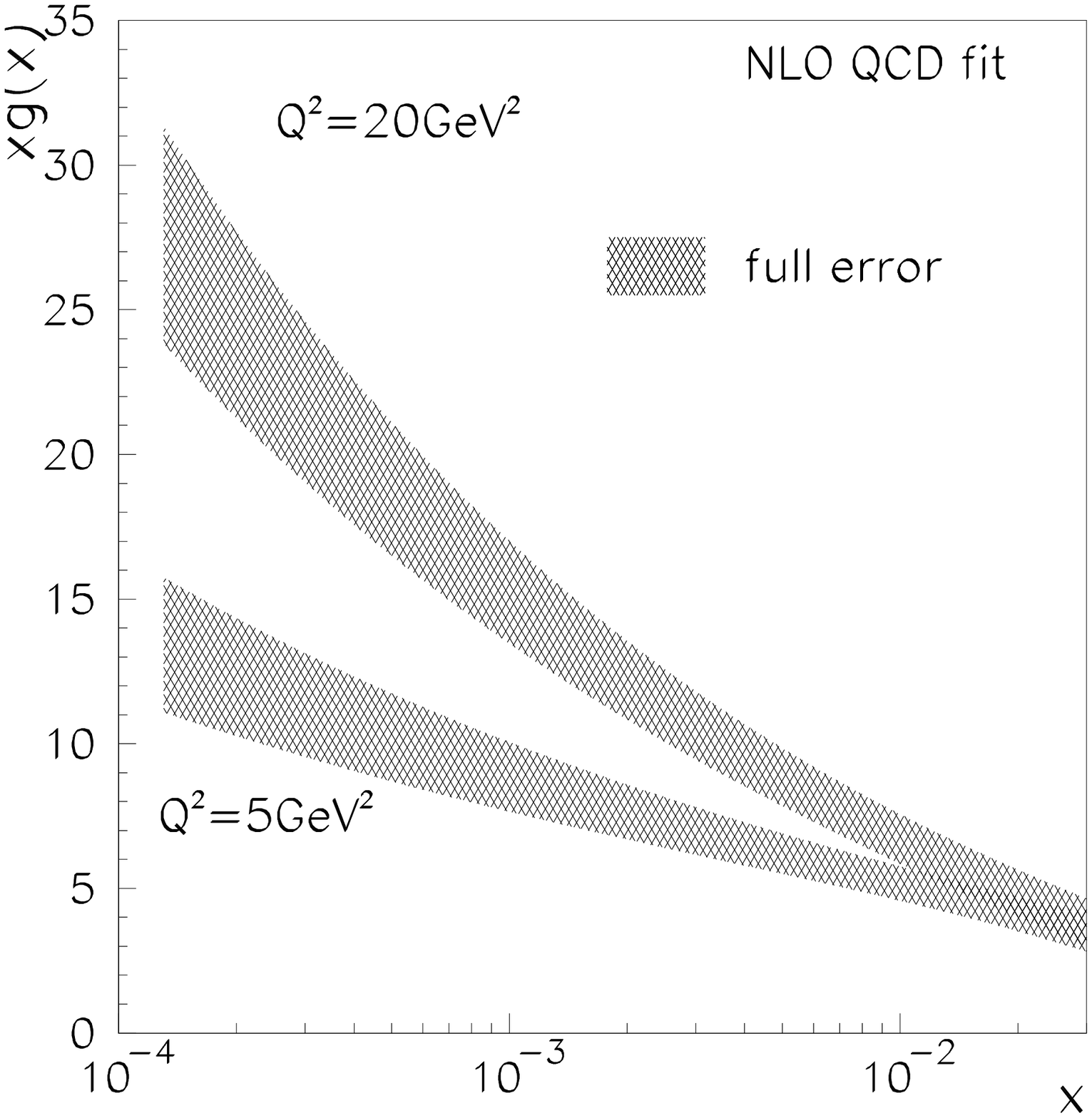,bbllx=-100pt,bblly=100pt,bburx=700pt,bbury=700pt,height=9cm} 
\fcaption{The gluon density $xg(x)$ at $Q^2 = $ 5 GeV$^2$ and 
$Q^2 = $ 20 GeV$^2$ extracted from the H1 NLO DGLAP fit. The error bands 
represent statistical and systematic errors taking into account correlations.}
\label{fig:gluonnlo}
\end{center}
\end{figure}

Fig.~\ref{fig:gluonnlo} shows the NLO gluon
 density $x g(x,Q^2)$ at $Q^2=5~{\rm GeV}^2$ and
$Q^2=20~{\rm GeV}^2$ as extracted from the H1 fit. 
Note that there are no $F_2$ measurements below $5\cdot 10^{-4}
$ at $Q^2= 20 $ GeV$^2$, but  in that region the gluon is constrained by the
data at lower $Q^2$ via the QCD evolution equations.
The error band results from the statistical and systematic errors added
in quadrature, taking into account correlations.
A variation of $\alpha_s$ by $0.005$
would give a  change of 9\%  in the gluon density at 20 GeV$^2$.
The striking rise of the gluon density towards low $x$ is confirmed
and this rise increases with increasing $Q^2$.

Fig.~\ref{fig:gluon20} shows $xg(x)$ at $Q^2 = 20$ GeV$^2$ for both H1 and 
ZEUS. The error bands account for statistical and systematic errors including
correlations, for both fits. The agreement is good, the gluon density 
extracted by ZEUS being somewhat lower than that extracted by H1.
At the lowest $x$ values $\sim 10^{-4}$ the gluon distribution is 
now determined with a precision of about $20\%$, an impressive improvement
in reach and precision compared with the previous fits
shown in Fig.~\ref{fig:gluonlo}.
At around $x= 0.01$ the HERA fits make contact with the fit performed by 
the NMC collaboration on their own data. The agreement with the NMC
result is very good. 

A fit following the H1 prescription (but with fixed $\alpha_s = 0.118$)
was made using both H1(94) and ZEUS(94) data,
as well as NMC(97) and BCDMS fixed target data. 
Only HERA(94) data for which $Q^2 > 5 $ GeV$^2$ were used in this fit.
The result (central value) is also shown in 
Fig.~\ref{fig:gluon20} and is close to the ZEUS fit result.
Fig.~\ref{fig:gluon20} also shows
 a collection of recent determinations from the global analyses of 
CTEQ, GRV and MRS. Although these  analyses include additional data to constrain
quark and gluon distributions, the resulting gluon distributions are 
broadly similar to the HERA fits and to each other in the fitted region. 
However they are starting to deviate from each 
other at the lowest $x$ values.

It is interesting to investigate whether there is any evidence for higher 
twists in the low $x$ HERA data. ZEUS have included BCDMS and SLAC data in 
their fit in order to extend it to lower $Q^2$. Fits are made for various
starting scales down to $Q^2_0 = 0.4\,$GeV$^2$ and all data above the starting 
scale are included in the fit. Higher twist terms of the form $f(x)/Q^2$ are
included in the fit and the value of $f(x)$ is fitted in each $x$ bin. 
It is found that the higher twist contribution is negligible for $x < 10^{-2}$,
even for low $Q^2$ data. The fit is unable to describe the data if a starting
scale $Q^2_0 < 0.8\,$GeV$^2$ is chosen, independent of whether or not such 
higher twist terms are included. This breakdown of the applicability of the
DGLAP equations at low $Q^2$ is discussed further in Sec.~\ref{sec:lowq2}.

A common problem in  extracting the gluon using only
 structure function data is the 
limited sensitivity to the gluon at large $x$. This is demonstrated in a
 fit of the HERA(94) and fixed target
data using a Bayesian treatment of the systematic errors, by 
Alekhin~\cite{alekhin}. The resulting uncertainty on the gluon at $x > 0.3$ 
is essentially $100\%$. Thorne~\cite{thornefl} has shown that 
using information from charm production
in DIS and prompt photon  production in non-DIS processes can  reduce this 
uncertainty significantly, even when these data are only used to impose
weak constraints. The global analyses of MRS, CTEQ and GRV have always     
 included additional data from non-DIS processes in their fits in order
to constrain the gluon.
Data on prompt photon production and single inclusive jet data from hadronic 
collisions have been used to date.

Prompt photon data have been used for the $x$ range: $0.02 < x < 0.5$, see
Fig.\ref{fig:gluon_over}. The process $pN \rightarrow \gamma X$ 
should provide information on the gluon distribution since the dominant
subprocess is $g q \to \gamma q$, at leading order.
However, one must also account for
non-direct $\gamma$ production in fragmentation processes from other 
partons~\cite{gammanlo}. There is also some
uncertainty from factorization and renormalization scale dependence. 
The most reliable information comes
from the lower energy WA70 data on  $pp \rightarrow \gamma X$~\cite{WA70} which
constrains the gluon in the $x$ range, $0.3 \le x \le 0.5$.
There are also newer fixed target data from UA6~\cite{UA6} and E706~\cite{E706}
 which cover the same $x$ range as WA70 and
there are ISR data~\cite{R807}, which extend to lower $x$, ($0.15 \le x \le
0.3$), and data from UA2~\cite{UA2g} and CDF~\cite{CDFg} 
which extend into the medium-small  $x$ region ($0.05 \le x \le 0.15$ and
$0.02 \le x \le 0.1$, respectively). These higher energy data are
sufficiently accurate that one can see that the shape of the $E_T$ distribution
of the photons with respect to the beam axis is steeper than that predicted by
pQCD calculations (even after correction for fragmentation
effects). This discrepancy can be remedied by including the effects of
`intrinsic' $k_T$, i.e. $k_T$ broadening of initial state partons due
to soft gluon radiation~\cite{CTEQ3,MRSG,BaerBlair,Zielinski}. Such 
broadening  would have 
to increase with $\sqrt{s}$ in order to describe all the data~\cite{prompt_kt}.
Until these effects are calculated in QCD it is difficult to use these higher
energy  
data to pin down the gluon~\cite{Durham95,Snomass97,DIS97}. For the lower energy
data of WA70 and E706 the broadening is much less, but even here there are the
scale uncertainties which lead to an uncertainty of $\sim 25\%$ in the gluon
distribution. The MRS and GRV fits use the WA70 data but the recent 
CTEQ analyses consider the uncertainty to be 
sufficiently serious that they either have not used
these data or have introduced a scale parameter to account for this 
uncertainty.

Further information on the gluon may come from 
the single jet inclusive cross-sections for jets with transverse energies 
$E_T \sim 100 $ GeV, since these are dependent on the gluon via  
$gg,gq$ and $g\overline{q}$ initiated subprocesses. 
There are high statistics jet 
measurements from the CDF and D0 collaborations at 
$\sqrt s = 1.8$ TeV~\cite{CDF,D0}.  For jets with transverse energy 
$50 \leqsim E_T \leqsim 200\,$GeV pQCD calculations are considered reliable.
The slope of the $E_T$ distribution
constrains the combination $\alpha_s(\mu^2) g(x,\mu^2)$ at the scale 
$\mu^2 = E_T^2$ and $x = 2E_T/\sqrt s \sim 0.1$. Currently the jet data
are more sensitive to $\alpha_s$ than to the gluon distribution.
However an independent measure of $\alpha_s$, would allow us to use these data 
to constrain the gluon distribution. 
The CTEQ collaboration have included the jet
data in their CTEQ4 fits whereas MRS merely compare their results to the 
jet data.  Finally we note that the CDF jet data for 
$E_T > 200\,$GeV rise above the predictions from most conventional global fits 
such as MRSR and  CTEQ4. This has given rise to many interesting 
suggestions of new physics effects~\cite{NEWPHYS}. However CTEQ have found a 
conventional parametrization (CTEQ4HJ)~\cite{Huston} 
which is able to describe these data.
Another conventional explanation is given by Klasen and Kramer~\cite{KK}.

We now briefly review other processes which may yield information on the gluon
in future. 
Dijet production in high energy $p\bar p$ collisions can give information on the
small $x$ gluon. If the jets are produced with equal transverse momentum and
very forward (with pseudorapidity $\eta >> 1$) then 
the gluon distribution  can be probed down to about $x = 0.005$, for
$\sqrt s = 1.8$ TeV~\cite{glover}. 

Dijet production initiated by the processes
$\gamma g \rightarrow q\overline{q}$ and $ \gamma q \rightarrow
qg$, in DIS at HERA, have also been used to extract the gluon distribution.
A LO extraction has been presented by H1~\cite{h1jetgluon}. The 
NLO corrections are known and the effects of scheme dependence and jet 
algorithms have been
quantified~\cite{dewolf2}. A method to extract the 
gluon density in NLO via an analysis in Mellin space has been 
proposed in ref.~\cite{berger}.
At present there are still some unresolved theoretical questions on the
absolute normalization of the jet rates, but the outlook is promising.

One of the problems with the gluon extraction from dijet production in DIS is 
the background from non-gluon induced processes. 
This is largely avoided by measuring open charm production, which 
 allows one to tag the boson-gluon
fusion process and hence to measure the gluon density directly.
The charm data in DIS have been discussed in Sec.~\ref{sec:cdata}, 
and  the status of the theory in Sec.~\ref{sec:heavyq}. 
 Fig.~\ref{fig:f2cc_all} shows the comparison of the measurements 
of $F^{c\overline{c}}_2$ from ZEUS and
H1 with the prediction using the gluon distribution extracted from the
GRV94 parametrization. The agreement is very good.
Clearly the errors on the measurements are still far too large to 
make a competitive extraction of the gluon from the charm data.
Both high luminosity at HERA and improved experimental 
set-ups (e.g. use of Si vertex detectors) will allow  
considerable improvements in the future such that one should be able to
 measure the gluon density in the $x$ range $0.0001<x<0.01$. 
Open charm production in photoproduction has also been suggested as a means of 
measuring the gluon distribution. However it depends on the gluon density 
via a 
convolution over photon and proton parton densities, and is more likely to shed
light on the former in the short term. 

 Measurement of the longitudinal structure function $F_L$ at HERA
can provide a  handle on the gluon distribution
at low $x$. In Eq.~\ref{eq:flqcd} we gave the relationship between $F_L$ and
$F_2$ and the gluon density, which is valid to order $\alpha_s$ in the
coefficient functions. At $x \leqsim 10^{-3}$ 
the dominant contribution comes from the gluon regardless of the exact shape of 
the gluon distribution and we may write~\cite{AMCS}
{\small \begin{equation}
xg(x,Q^2) = \frac{3}{5}5.8\left[\frac{3\pi}{4\alpha_s}F_L(0.417x,Q^2) - 
\frac{1}{1.97} F_2(0.75x,Q^2)\right]
\label{eq:amcs}
\end{equation}}  
such that an accurate measurement of $F_L$ taken together with present
measurements of $F_2$ should yield an accurate measurement of the gluon.
This approximation is good to $\sim 2\%$, within its own assumptions.
However, heavy quark contributions and higher order corrections complicate
the unfolding of the gluon density from $F_L$. Corrections to 
order $\alpha_s^2$ in the coefficient functions have been given in 
ref.~\cite{nnlofl} and heavy quark contributions have been considered in 
ref.~\cite{hqfl}. It remains true that a measurement of $F_L$ should be
sensitive to the gluon distribution. The present data on $F_L$ have been given
in Sec.~\ref{sec:rdata}. None of the fixed target measurements access small 
enough 
$x$ to be useful in extracting the gluon distribution. The measurement of $F_L$
presented by H1~\cite{h194r} cannot give a measurement of the gluon which is
independent of the H1 gluon distribution as extracted from scaling violations, 
since the method of measurement of $F_L$ essentially represents a consistency
check on the conventional NLO QCD fit~\cite{thornefl}. The prospects for
a future independent measurement of $F_L$ at HERA have been discussed in
Sec.~\ref{sec:rdata}.

Inelastic $J/\psi$ photoproduction at HERA has been suggested as a process 
which 
could allow one to measure the gluon. However it appears that the perturbative
calculation is not well behaved in the limit $p_T(J/\psi) \to 0$, and if
the small $p_T$ region is excluded from the analysis the predictions are not
very sensitive to the small $x$ behaviour of the gluon~\cite{nlojaypsi}.
Elastic (diffractive) $J/\psi$ production in DIS and in photoproduction are 
more promising, since the cross-section depends on $xg(x,Q^2_V)^2$, where
the scale of the process is given by $Q^2_V = M^2_{J/\psi}/4$~\cite{rysbrod}.   
Similarly, diffractive vector meson ($\rho,\phi$) production in DIS, at higher 
$Q^2$, depends on the square of the gluon density. These data could give 
information on the gluon distribution in the region $0.0001 < x < 0.01$.
At the present time the theoretical framework for extracting the gluon
distribution from  these processes is still under development and 
the experimental precision of the data is still fairly low.

In summary, considerable progress has been made in pinning down the gluon 
in a wide kinematic region extending down to $x$ = 10$^{-4}$. 
At present the analyses from the scaling violations of the $F_2$ 
structure function are generally accepted as giving the most reliable 
information, since they are well understood both
experimentally and theoretically (provided that the conventional NLLA is
accepted as appropriate).
However, further progress is expected in determinations from other
processes over the next few years,  such that we may expect to
improve the determination of the gluon distribution over the full $x$ range.

\subsubsection{The MRSR and CTEQ4 parametrizations}
\label{sec:latest}

\begin{table}
\tcaption{A summary of the development of the MRS ad CTEQ PDFs
from 1992 to 1996, specifying the input data sets, the input ($Q^2 = Q^2_0$) 
slope of the gluon ($\lambda_g$) and sea ($\lambda_S$) distributions at low $x$ 
(the notation $\lambda$ is used when $\lambda_g=\lambda_S$) and the value of 
$\alpha_s$ (note this is a parameter of the global fits except for the latest
MRSR sets)}
\centerline{\footnotesize\smalllineskip
\begin{tabular}{ccccc}\\
 \hline
 Year & PDF & Data Sets Used & $Q^2_0$ input & $\alpha_s(M_Z^2)$ \\ 
\hline
 $1992$  & MRSD$^{\prime}$ &$BCDMS-F_2^{\mu p}(F_2^{\mu d})$& $\lambda=0 (D0')$ & $0.112$  \\
       &$Q^2_0=4\,$GeV$^2$  & $NMC(92)-F_2^{\mu p},F_2^{\mu d},F_2^n/F_2^p$ &  $\lambda=0.5 (D-')$ & $$\\
       & CTEQ1& $CCFR(93)-F_2^{\nu N},xF_3^{\nu N}$ & low $x$ flat (1M)&$0.112$    \\
       &$Q^2_0=2.56\,$GeV$^2$  &$WA70-prompt\ \gamma,E605-DY$ & low $x$ singular (1MS)&  \\ 
\hline
 $1993$ & MRSH & $          $ & $\lambda=0.3$ & $0.112$  \\
     & $Q^2_0=4\,$GeV$^2$ & $ ZEUS(92),H1(92)-F_2^{ep}$ &  &  \\
       & CTEQ2 & $ (CDF-DY) $ & $\lambda=0.258$ & $0.110$\\
       &$Q^2_0=2.56\,$GeV$^2$ & & & \\  
\hline
 $1994/5$  & MRSA&$         $& $\lambda=0.3$ & $0.112$ \\
       & MRSA$^{\prime}$ & $ZEUS(93),H1(93)-F_2^{ep}$ &  $\lambda=0.17$& $0.113$\\
       & MRSG & $NA51\ A_{DY}$ &$\lambda_S=0.067,\lambda_g=0.301$ &$0.114$    \\
       &$Q^2_0=4\,$GeV$^2$& $CDF\ A_W$ & & \\
       & CTEQ3 &$$ & $\lambda=0.28$&$0.112$  \\ 
       & $Q^2_0=2.56\,$GeV$^2$& &  & \\ 
\hline
 $1996$ & MRSR& $           $ & $\lambda_S \ne \lambda_g$ (R1) & $0.113$  \\
       &$Q^2_0= 1\,$GeV$^2$  & $ ZEUS(94),H1(94)-F_2^{ep}$ & $ \lambda_S \ne \lambda_g$ (R2) &$0.120$  \\
       & & $E665-F_2^{\mu p}(F_2^{\mu d}),(SLAC-F_2^{ep})$ & $\lambda_S=\lambda_g$ (R3) & $0.113$ \\
       &  & $NMC(95)-F_2^{\mu p},F_2^{\mu d}$ & $\lambda_S=\lambda_g$ (R4) & $0.120$ \\
&CTEQ4& $ (CDF/D0\ jet\ E_T) $ &$ \lambda_S=0.143,\lambda_g=0.206$ & $0.116$\\
 & $Q^2_0=2.56\,$GeV$^2$& & & \\  
\hline\\
\end{tabular}}
\label{tab:pdf}
\end{table}

In Table~\ref{tab:pdf} we summarize the development of the MRS and CTEQ parton 
distributions from 1992 to the 1996 sets. We specify the common data 
sets which were input to the fits
and some of the interesting features of the 
parametrizations, such as the slope of the low $x$ gluon and sea distributions
at $Q^2_0$ (where $\lambda_g = \lambda_S$ is fixed we use the common notation
$\lambda$) and the
value of $\alpha_s(M_Z^2)$. Note that this is a parameter of the global fit
for all but the MRSR PDFs. Only the MRSR and CTEQ4 fits will be discussed in 
any detail.

The data sets originally used in 1992 were the older BCDMS data~\cite{BCDMS} on
$F_2^{\mu p}, F_2^{\mu d}$ (note  MRS only use the proton data), the NMC(92) 
data~\cite{NMCGott} 
on $F_2^{\mu p}, F_2^{\mu d}, F_2^{n/p}$ and CCFR(93) data~\cite{ccfr_lambda} 
on $F_2^{\nu N}, xF_3^{\nu N}$. 
Both groups also supplemented this DIS data by  E605 Drell Yan data and WA70 prompt
photon data. When HERA(92) and HERA(93) data on $F_2^{ep}$ became available 
they were included in the fits 
extending the parton distributions into the low $x$ region. New non-DIS data 
from NA51 on the Drell Yan asymmetry, $A_{DY}$~\cite{NA51}, and from CDF on the
$W^{\pm}$ asymmetry, $A_W$~\cite{CDFW} also became available such that by 1995 
one had the parametrizations MRSA$^{\prime}$, CTEQ3, valid in the regions 
$Q^2 \geqsim 4\,$GeV$^2$, $ x \geqsim 4\times 10^{-4}$. 
These parametrizations still give a reasonable description of data in this 
kinematic region today. 

\begin{figure}[]
\begin{center}
\psfig{figure=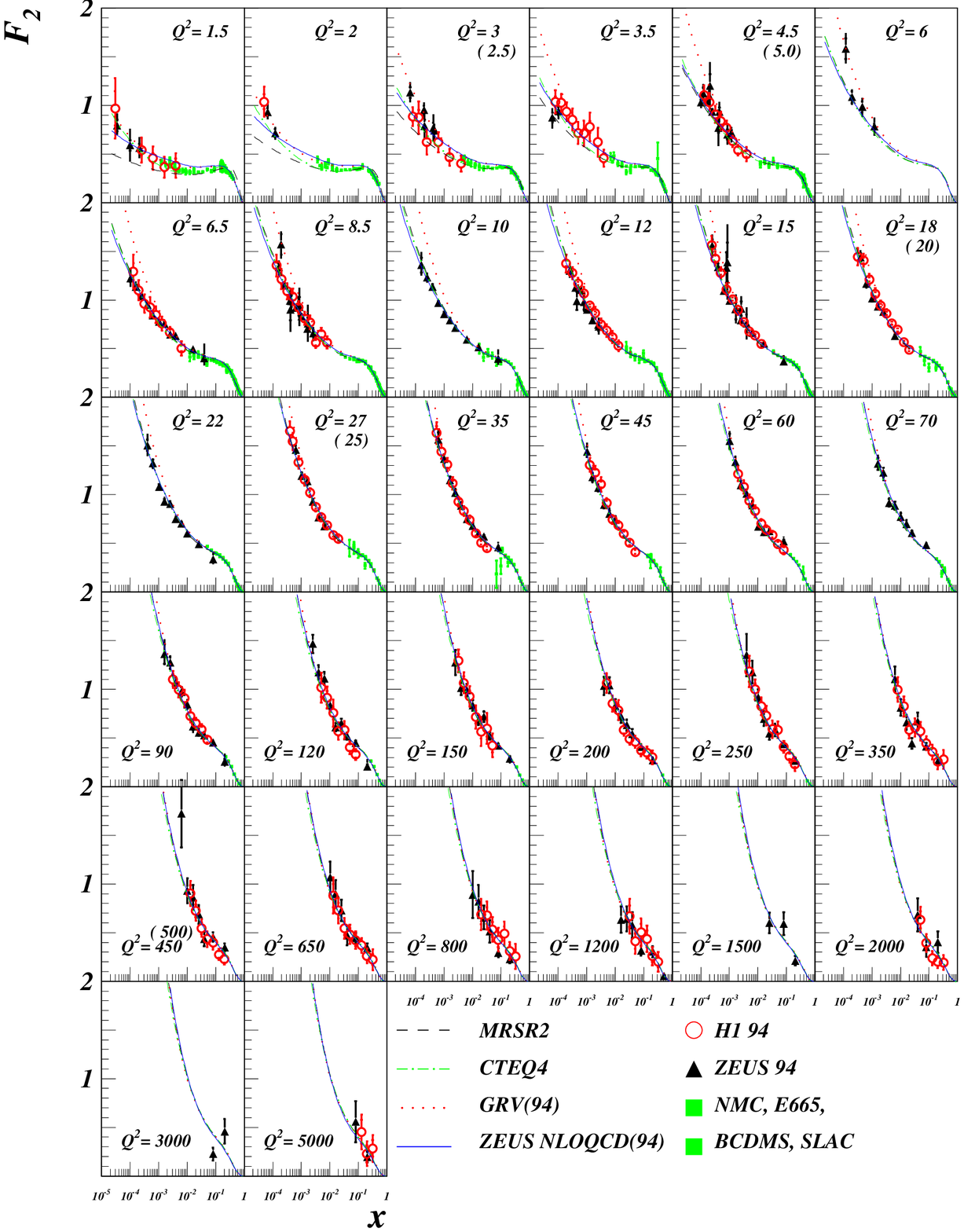,bbllx=60pt,bblly=80pt,bburx=600pt,bbury=820pt,height=0.95\textheight}
\fcaption{ZEUS(94), H1(94), NMC(97), E665, BCDMS and SLAC  
$F_2^p$ data at fixed $Q^2$, 
with the MRSR2, CTEQ4 and GRV94 parton distribution functions and the ZEUS NLO
 QCD fit overlaid. At high $Q^2$ all these parametrizations are 
indistinguishable.}
\label{stamp}
\end{center}
\end{figure}

However when the HERA(94) data were published in 1996 this stimulated new fits by each
team: the MRSR and CTEQ4 series. New data from NMC(95)~\cite{nmcdata1} and 
from E665~\cite{e665data} on $F_2^p$ and $F_2^d$ were also included
in these fits (note MRS do not use E665 $F_2^d$ data). In addition to this 
the CTEQ group include the inclusive jet measurements from CDF~\cite{CDF} and
D0~\cite{D0} in their fit. MRS  compare their fit to these data but they 
do not 
use them in the fit  since the correct treatment 
of systematic errors is not clear. CTEQ drop all the prompt-photon
data since the problems in fitting the $p_T$ spectrum for the CDF data cast doubt on 
our understanding of this process. MRS continue to include the lower energy 
WA70 prompt photon data in their fits. MRS also include SLAC 
data~\cite{whitlow90,whitlow92} on $F_2^{ep}$
in their fits since they drop their starting scale to $Q^2_0 = 1\,$GeV$^2$. 
This change
of input scale was made in response to the fact that HERA(94) data 
extend down to
$Q^2 = 1.5\,$GeV$^2$ and to the work of various 
authors~\cite{BF,GRV,GRV91,GRV94} 
which made it apparent that DGLAP evolution could be 
successfully extended down to
such low $Q^2$ values. We discuss this somewhat surprising observation 
further in Sec.~\ref{sec:lowx}. The quality of fit of the MRSR2 and CTEQ4 
parametrizations to data in the kinematic range $1.5 < Q^2 < 5000\,$GeV$^2$,
$ x > 3\times 10^{-5}$ is illustrated in Fig.~\ref{stamp}.

Both teams have allowed for variations on their basic fits.
There are variants for use with LO calculations and there are variants for 
use in differing renormalization schemes at NLO, such as DIS
rather than $\MSB$. There have also been variants~\cite{MRSAmod} 
which extend below $Q^2 = 1\,$GeV$^2$. Such parametrizations 
are needed for calculating 
radiative corrections to higher $Q^2$ data as well as
for making unfolding corrections to the very low $Q^2$ data described in
Sec.~\ref{sec:f2lepto}. Such variants use phenomenological parametrizations of
low $Q^2$ data adjusted to match smoothly onto the pQCD forms at higher $Q^2$.
Variants which
correspond to differing values of $\alpha_s$ (varying from $0.105$ to $0.130$) 
are also available for use in $\alpha_s$ determinations from data on jet 
production and non-DIS processes~\cite{CTEQ4,MRSalf}.  

\clearpage

Before the HERA data were included in the fits, the parametrizations 
also included variants
which accounted for the possibilities of hard or soft Pomeron exchange. 
As noted above, the inclusion of high precision HERA data has made it possible
to fit the slopes of the low $x$ gluon and sea distributions
 and thus to investigate the nature of the Pomeron
exchange involved. The interesting theoretical possibilities which this opens 
up, and the limitations of conclusions drawn from fits made within the 
framework of conventional pQCD, are discussed in detail in Sec.~\ref{sec:lowx}.
Here we confine ourselves to giving a warning
against over-interpreting the values of $\lambda_g$ and $\lambda_S$ which 
result from the fits to the parton parametrizations.

The exact values of $\lambda_{S,g}$ which 
give the best fit to $F_2$ depend on the $\epsilon_{S,g}$ values. There is a 
correlation between the values of $\lambda$ and $\epsilon$ such that a more 
negative $\epsilon$ reduces the value of $\lambda$ and a more positive value 
increases it. One can fit the same data with different $\epsilon_S$ values
(and $\epsilon_g=0$) such that  $\lambda_g = \lambda_S = 0.2$, 
$\epsilon_S = -3.3$ or $\lambda_g = \lambda_S  = 0.3$, $\epsilon_s = -1.1$ or 
$\lambda_g = \lambda_S = 0.4$, $\epsilon_S = +2.9$ and all these possibilities 
give similar fit quality. 
Hence the same data can appear in need of more or less singular input
depending on the exact form of the parametrization. MRSA has $\lambda_S
=\lambda_g= 0.3, \epsilon_S=-1.1,\epsilon_g=0$, whereas MRSA$^{\prime}$ 
has $\lambda_S=
\lambda_g=0.17,\epsilon_S=-2.55,\epsilon_g=-1.9$. 
The minimal forms which have $\epsilon_{S,g} = 0$ give a 
better guide to the low $x$ behaviour if one only considers the value of 
$\lambda_{S,g}$. 
 
\begin{figure}[ht]

\centerline{\psfig{figure=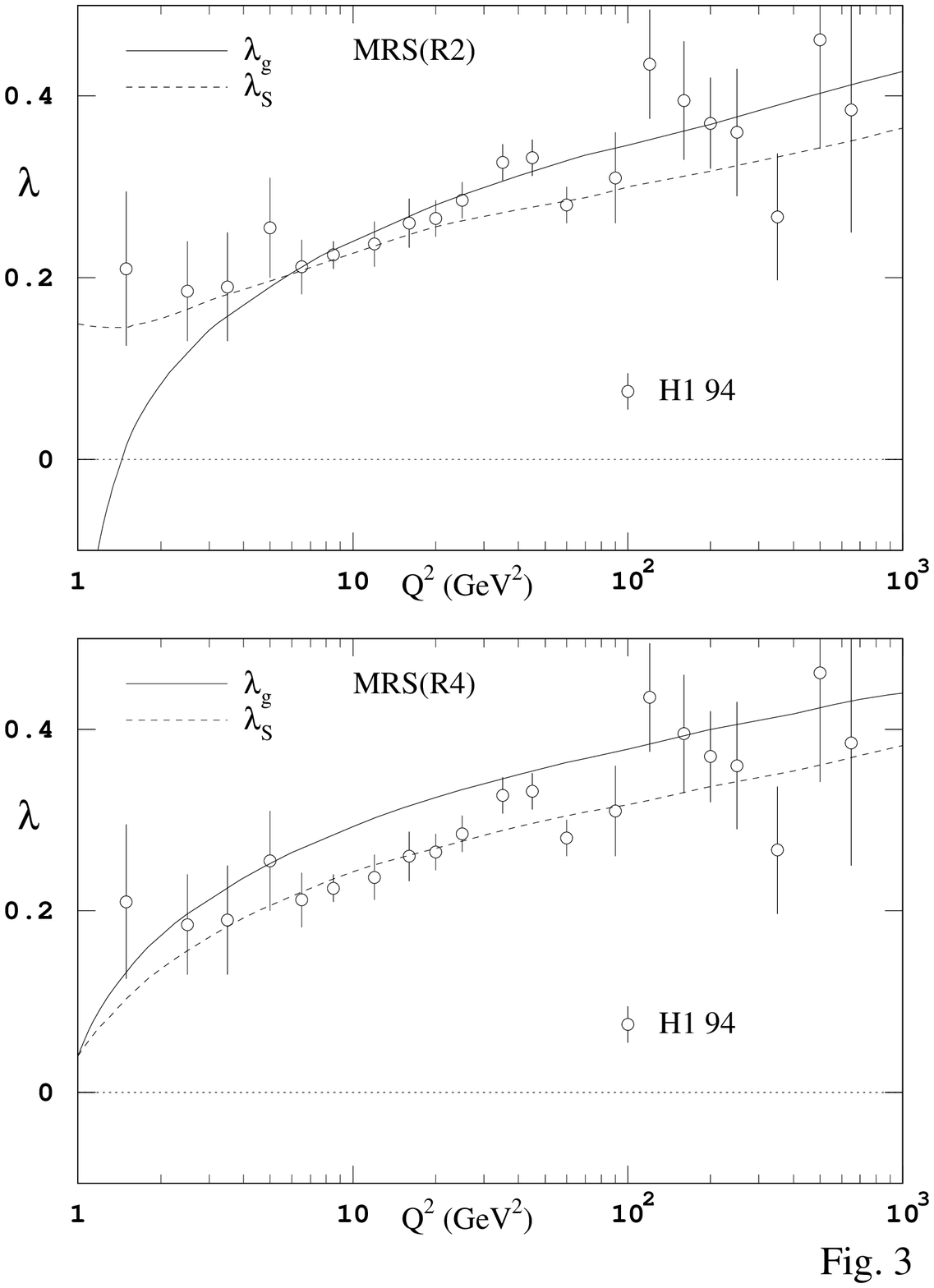,width=6cm,height=9cm}}
\fcaption{The evolution of the parameters $\lambda_S$ and $\lambda_g$ with 
$Q^2$, for the MRSR2 and MRSR4 parametrizations. The evolution of the 
$\lambda$ parameter as evaluated from H1(94) data is also shown. 
Note that these parameters are not evaluated in exactly the same way. For
$Q^2 > Q^2_0$ the MRSR PDFs are not given by the analytic forms 
of Eq.~\ref{eq:param}. Hence to quote a value of $\lambda$ as a function of 
$Q^2$ the authors perform fits of the evolved distributions to forms of this 
type for each value of $Q^2$ which is of interest. The H1 $\lambda$ parameter
is evaluated by the simpler procedure of fitting the form 
$x^{-\lambda}$ to $F_2$ data in the  low $x$ range, $x < 0.1$, 
for various $Q^2$ 
bins. The mean value of $x$ at which this evaluation is made will increase with 
$Q^2$. These different 
procedures this will not give fully comparable $\lambda$ values}
\label{dicklam}
\end{figure}

The MRSR1,2 fits are done for $\lambda_g \neq \lambda_S$ and fixed 
$\alpha_s = 0.113, 0.120$ respectively. The reasons for this choice of 
$\alpha_s$ values is discussed in Sec.~\ref{sec:alphas}.
The best overall fit is MRSR2, which has $\alpha_s=0.120$ and $\lambda_S=0.15,
\epsilon_S=1.1$, $\lambda_g=-0.51,\epsilon_g=-4.2$, at $Q^2_0 = 1\,$GeV$^2$. 
These values
are not directly comparable to the previous MRS $\lambda$ values because of 
the drop in
the starting scale. One notes that the gluon low $x$ slope has become 
valence-like, however this quickly changes as $Q^2$ increases, such that
$\lambda_S$ and $\lambda_g$ become equal (to $\sim 0.2$) at 
$Q^2\sim 5\,$GeV$^2$ 
(as in the  MRSA$^{\prime}$ distributions) and for larger $Q^2$, $\lambda_g >
\lambda_S$, as expected by pQCD, see Fig.~\ref{dicklam}.

\noindent

Table~\ref{tab:chi2} gives $\chi^2$ per data point for the parametrizations 
CTEQ3, CTEQ4, MRSA$^{\prime}$, MRSR1 and  MRSR2  
for the DIS data, and some of the relevant non-DIS data.
 The numbers of parameters
differ between the parametrizations: CTEQ3,
CTEQ4, MRSA$^{\prime}$, MRSR1, MRSR2 have 15,18,17,19,19 shape parameters respectively, 
in addition to the relative normalizations of the data sets, and the value
of $\alpha_s$ when it is fitted rather than fixed. 
We also present the $\chi^2$ per degree of freedom 
for fits to the DIS data 
in the final row so that we can compare the goodness of fit of each of these 
parametrizations to current data.

\begin{table}[ht]
\tcaption{ $\chi^2$ per data point for the parametrizations 
CTEQ3, CTEQ4, MRSA$^{\prime}$, MRSR1 and MRSR2  
for the DIS data, and relevant non-DIS data, including the CDF/D0
jet data, for which systematic errors have not been included. Note that  the 
$\chi^2$ values are worked out for all the data sets using 
the published  forms of the parametrizations and the CTEQ evolution programme
for all the quoted values except those of MRSR, 
which were supplied by the authors. Not all of the data sets 
included were available at the time when the earlier parametrizations were 
issued, the $\chi^2$ are worked out retrospectively. 
The number of data points quoted differs for MRSR1,
MRSR2, since these have a lower starting value $Q^2_0= 1\,$GeV$^2$. The other 
parametrizations have their $\chi^2$ values worked out from the CTEQ 
starting point 
$Q^2=2.56\,$GeV$^2$. Some of the $\chi^2$ values for MRSR fits to non-DIS data
are estimated since 
the authors indicate that they are similar in quality to those 
achieved with MRSA and 
MRSA$^{\prime}$. 
The WA70 $\chi^2$ for MRSA$^{\prime}$ is estimated as equal to that for MRSA.
The $\chi^2$ per degree of freedom given in the final row includes only the
DIS data.}
\centerline{\footnotesize\smalllineskip
\begin{tabular}{cccccc}\\
 \hline
 Data & CTEQ3 & MRSA$^{\prime}$ & CTEQ4 & MRSR1 & MRSR2 \\ 
\hline
 $BCDMS\ F_2^p$&$130/168$&$157/168$&$145/168$&$265/174$&$320/174$  \\
 $BCDMS\ F_2^d$&$187/156$&$214/156$&$186/156$&$$&$$  \\
 $CCFR(93)\ F_2$&$69/63$&$68/63$&$83/63$&$41/66$&$56/66$  \\
 $CCFR(93)\ xF_3$&$41/63$&$54/63$&$47/63$&$51/66$&$47/66$  \\
 $NMC ratio$&$133/89$&$143/89$&$131/89$&$136/85$&$132/85$  \\
 $NMC(95)\ F_2^p$&$147/104$&$129/104$&$97/104$&$155/129$&$147/129$  \\
 $NMC(95)\ F_2^d$&$137/104$&$152/104$&$93/104$&$139/129$&$129/129$ \\
 $ZEUS(94)\ F_2^p$&$549/179$&$369/179$&$243/179$&$326/204$&$308/204$  \\
 $H1(94)\ F_2^p$&$220/172$&$150/172$&$119/172$&$158/193$&$149/193$  \\
 $E665\ F_2^p$&$48/35$&$38/35$&$41/35$&$62/53$&$63/53$  \\
 $E665\ F_2^d$&$45/35$&$29/35$&$32/35$&$$&$$  \\
 $SLAC\ F_2^p$&$$&$$&$$&$108/70$&$95/70$  \\
 $CDF\ A_W$&$3/9$&$3.7/9$&$4.3/9$&$\sim 4/9$&$\sim 4/9$  \\
 $NA51\ A_DY$&$0.4/1$&$0.1/1$&$0.6/1$&$\sim 0.1/1$&$\sim 0.1/1$  \\
 $E605$&$93/119$&$94/119$&$98/119$&$\sim 95/119$&$\sim 95/119$  \\
 $WA70$&$23/39$&$\sim 21/39$&$$&$$&$$  \\ 
 $CDF\ jet$ &$$&$$&$$&$222/24$&$52/24$ \\ 
\hline
 total $\chi^2/ndf$&$1.49$&$1.31$&$1.06$&$1.33$&$1.32$ \\ 
\hline\\
\end{tabular}}
\label{tab:chi2}
\end{table}

When comparing these $\chi^2$ values one should bear in mind that it is not
really possible to define a one standard deviation covariance matrix of 
uncertainties on these parton distribution functions. Firstly, because
the correlations between experimental systematic errors are not easily 
taken into account. Secondly because there are theoretical uncertainties, such
as the treatment of heavy quarks, the treatment of scale uncertainties when
dealing with non-DIS data, the choice of $Q^2_0$ and of the form of the input
parametrization, which are not easy to quantify in terms of a correlated
systematic uncertainty. A related issue arises when using PDFs
as input for calculations of high energy scattering processes in
order to constrain the parameters of the Standard Model, or even to investigate
physics beyond the Standard Model. The systematic uncertainty on a result
coming from the choice of PDF is not easily quantified since the best 
global fits are derived from similar assumptions. We need to consider variants
on the standard PDFs in which the parameters most sensitive to the derived 
result are varied as widely as possible, while maintaining a reasonable global 
fit quality taking into account the systematic errors on the input data.
For an interesting discussion of these issues see 
ref.~\cite{Snomass97,huskuhl,botje}. 

Both the MRS team~\cite{MRRS,Rob97} and the CTEQ team~\cite{LaiTung} 
have recently updated their parton 
distributions to account for heavy quark production consistently from the 
threshold region to the asymptotic region. The new CTEQ parton distributions,
CTEQ4HQ, were produced by fitting to the same data sets as the CTEQ4 series 
to aid comparison. They are similar to the standard CTEQ4M set but
the fit quality is somewhat improved, $\chi^2/ndf = 1.03$, 
especially for low $x$ data. The MRS 
team have included the  new NMC(97) data~\cite{nmcdata2} 
in their (MRRS) analysis~\cite{MRRS}.
The fit quality has improved, $\chi^2/ndf = 1.19$ (compared to
$\chi^2/ndf = 1.25$ for MRSR2 on the same data) especially at low $x$, and the
corresponding value of $\alpha_s$,$\alpha_s(M_Z^2) = 0.118$ also gives a 
better fit to BCDMS data. The resulting description 
of HERA and EMC charm data is good.
More recently~\cite{Rob97} MRS have also included the updated CCFR(97) 
data~\cite{ccfr_seligman}
in their fits (excluding the contentious region $x < 0.1$). 
This does not change the fit quality and fit
parameters significantly compared to MRRS. The resulting parton 
distributions are very similar to those of CTEQ4HQ when evolved to 
$Q^2 = 25\,$GeV$^2$.

\subsubsection{Dynamically generated partons: the GRV parametrization}
\label{sec:GRV}

Whereas the parametrizations provided by the MRS and CTEQ groups depend
crucially on the non-perturbative input parametrization at $Q^2_0$, the
parametrizations of the GRV group~\cite{GRV91,GRV94} are far less dependent
on their inputs.

The original idea behind these parametrizations is that at some VERY LOW
scale $Q^2 = \mu^2$ the nucleon consists only of constituent valence quarks.
As $Q^2$ increases, one generates the gluons and sea quarks in the 
nucleon dynamically from these valence quarks, through the
conventional DGLAP equations for the processes $q \to q g$, $g \to q \bar q$.
It did not prove possible to describe all relevant data (in particular the gluon
distributions required by the prompt photon data) using such a picture, and it
was accordingly modified to include gluons and sea quarks at the starting scale
$\mu^2$, BUT these distributions have a valence-like shape 
(i.e. small at small $x$). One may even 
interpret this picture by saying that the gluons and sea quarks are frozen
upon the valence current quarks for $Q^2 < \mu^2$, and these composite objects
form the constituent quarks~\cite{GRV94}.

The valence quark distributions are fixed by taking the valence 
parametrizations from a conventional fit, such as MRSA, at $Q^2_0 = 4\,$GeV$^2$
and evolving them backwards to $\mu^2$, using the QCD scale parameter 
$\Lambda = 200\,$ MeV, for four flavours, matching $\alpha_s$ at 
flavour thresholds in the usual way~\cite{Marciano}. The valence-like sea and
gluon distributions at $\mu^2$ are parametrized by simple forms like
{\small \begin{equation}
 xg(x,\mu^2) = A x^{\alpha}(1 - x) ^\beta,\ \ x\bar q(x,\mu^2) = A' 
x^{\alpha'} (1 - x) ^{\beta'}
\end{equation}}
\noindent
where the free parameters are set by fitting to fixed target-DIS and prompt
photon data, after $Q^2$ evolution from the starting scale $\mu^2$ back up to
the $Q^2$ values of the relevant data sets has been performed (i.e. the
gluon distribution at any scale $Q^2 > \mu^2$ consists of the original
gluon distribution at $\mu^2$ augmented by the gluons which evolve from the
valence quarks). 
The scale $\mu^2$ is set as the scale at which the gluon distribution is as 
hard as the valence quark distribution $u_v$. The splitting $q \to q g$ 
naturally generates a softer distribution in the split products than that
of the original quark, hence a gluon distribution which is 
harder than the $u_v$ distribution is considered physically unreasonable. 
Thus the
scale $\mu$ at which $g \sim u_v$ is assumed to be the lowest scale
at which the DGLAP equations may be used.

The original predictions of GRV (GRV91)~\cite{GRV91} were a little too steep
to describe the HERA low $x$ data, a better fit was obtained 
(GRV94)~\cite{GRV94}
by modifications which took into account more modern fixed target 
data to fix the
input distributions, and by treating the heavy quark component of $F_2$
in a more sophisticated manner~\cite{GRS}, using a fixed flavour number (FFN) 
scheme. This essentially uses 3-flavours 
of massless parton and generates the charm contribution to $F_2$ through the
boson-gluon fusion process. 

It turns out that $\mu^2 \sim 0.3\,$GeV$^2$, which seems a very low scale
at which to use perturbative QCD. However,  GRV argue that it is
the value of $\alpha_s(\mu^2)/\pi \sim 0.2$, which determines the relative 
size of higher order corrections, not the value of $\alpha_s(\mu^2)$, and 
that their predictions are perturbatively stable between LO and NLO if one
considers measurable quantities like structure functions, rather than parton
distributions. They emphasize that their calculation is only applicable 
to the
leading twist operators of QCD, and clearly at very low $Q^2$ higher twist
operators can be important. Hence GRV state that the calculation can be
made consistently at low $Q^2$, but they only expect the resulting 
parametrization to describe reality for somewhat higher $Q^2$ 
($\geqsim 0.6\,$GeV$^2$). 
The very recent low $Q^2$ data from HERA indicate that this
expectation is correct, GRV predictions begin to describe the
data qualitatively, for $Q^2$ values, $Q^2 \geqsim 0.8\,$GeV$^2$, 
as illustrated in Fig.~\ref{stamp} and Fig.~\ref{fig:f2_lowQ2_all}. 

However one
can also see that the GRV94 parametrization lies systematically above the 
lowest $x$ data for moderate $Q^2$ values.
Note that since HERA data were not input to the GRV94 parametrization 
it represents a prediction for the behaviour of $F_2$ in the HERA region.
Whereas at larger $x$, $x > 0.01$, the approach is
similar to that of MRS and CTEQ, at small $x$, $x < 0.01$, it 
loses sensitivity to the form of the initial distributions at $Q^2 =\mu^2$.
This is because of the long 
evolution length between $\mu^2$ and the $Q^2$ values at which we compare with 
data, taken together with the fact that the initial
distributions are valence-like. 
The  resulting gluon and sea quark distributions 
are very steep at small $x$ for
large $Q^2$, but they flatten as $Q^2$ decreases, and this clearly describes
the trend of the HERA low $x$ data qualitatively. 
Such behaviour is characteristic of the predictions of pQCD
at low $x$. We discuss this in detail in Sec.~\ref{sec:lowx}.

\subsection{Determinations of $\alpha_s(M_Z^2)$}
\label{sec:alphas}
 
The simplest predictions of pQCD concern the 
$Q^2$ evolution of the non-singlet structure functions. In principle one
can extract a value of $\Lambda$, and hence a value of
$\alpha_s$, from such data independent of assumptions as to the shape of
the non-singlet quark distributions.
However the early data on non-singlet quantities (such as $xF_3$ in neutrino
scattering and $F_2^p -F_2^n$ in muon scattering) were insufficiently accurate
for  precision determinations. Hence $\alpha_s$ determinations from deep 
inelastic scattering data were done within the global fits to singlet 
and non-singlet structure functions, in which the parton distributions are 
fitted at the same time as $\alpha_s$. There is then a coupling
between the value of $\Lambda$  extracted and the shape of the 
gluon distribution, which is most easily understood from the LO DGLAP 
equations (Eqs.~\ref{eq:DGLAPq},~\ref{eq:DGLAPg}. Increasing $\Lambda$
increases the negative contribution from the $P_{qq}$ term, but this may be 
compensated by the positive contribution from the $P_{qg}$ term if the 
gluon is made harder. The global fits of the MRS and CTEQ teams 
address this problem by looking at data from other physical 
processes to tie down the gluon distribution.

Before the most recent high precision HERA(94) data were included in these fits
the $\alpha_s$ values extracted were in the region of 
 $\alpha_s \sim 0.113$. This contrasts with the values of 
$\alpha_s \sim 0.120$ extracted from LEP data from the $n$-jet production
rates and the hadronic width of the $Z^0$~\cite{Marti}. 
It also seems that the recent
CDF data on the single jet inclusive $E_T$ distribution is best described by 
$\alpha_s \sim 0.120$. Much thought has gone into
understanding the origins of this discrepancy 
including the suggestion that new physics maybe 
responsible for a process dependent discrepancy. However, when one looks more
closely at the theoretical uncertainties involved in the extractions
(as well as at the experimental statistical and systematic errors) it is not 
clear that there is any significant discrepancy. 
The reviews of Bethke~\cite{Bethke} and Stirling~\cite{Stiralf}
summarize the latest measurements and calculations of theoretical 
uncertainty for $\alpha_s$ extractions from non-DIS processes. 

In the present section we discuss the
uncertainties on extractions of $\alpha_s$ from structure function 
data\fnm{r}\fnt{r}{~We shall 
not discuss the use of $n+1$ jet rate in DIS measurements to measure 
$\alpha_s$, since there are still some unresolved theoretical questions on
the absolute normalizations of the jet rates.
For latest results see ref.~\cite{weber}, for a review of
the technique see ref.~\cite{jetrates}.} paying particular attention to 
the estimation of experimental and theoretical uncertainties.
We first consider the evaluations of $\alpha_s$ from
fixed target data. We concentrate on
analyses of data from individual experiments rather than global fits, since the
systematic errors can then be specified properly. We then consider 
the impact of the HERA data including the further theoretical uncertainties
which are relevant for small $x$ data.
The results for $\alpha_s$ are presented in Table~\ref{tab:alphas}, where 
statistical and systematic errors are combined into a single 
experimental error and an estimate of the theoretical
uncertainty is also given, where available.

\begin{table}
\tcaption{A summary of the results for $\alpha_s(M_Z^2)$ extracted from
structure function data. The experimental errors include both statistical and systematic errors . The treatment of correlated errors and the methods used
by different authors are discussed in the text}
\centerline{\footnotesize\smalllineskip
\begin{tabular}{cccc}\\
 \hline
 Data & $\alpha_s(M_Z^2)$ & Authors & Method  \\ 
\hline
BCDMS/SLAC $F_2$ &$0.113 \pm 0.003(exp) \pm 0.004(th)$  & VM~\cite{VirMil} & DGLAP NLO  \\ 
\hline
 CCFR(93) $xF_3$ &$0.107 \pm 0.006(exp)$  & CCFR~\cite{ccfr_lambda}&  DGLAP NLO  \\
  CCFR(93) $F_2+xF_3$ &$0.111 \pm 0.004(exp)$ &CCFR~\cite{ccfr_lambda}& DGLAP NLO       \\
\hline 
CCFR(97) $xF_3$ & $0.118 \pm 0.006(exp) $ & CCFR~\cite{ccfr_seligman}&  DGLAP NLO   \\
  CCFR(97) $F_2+xF_3$ & $0.119 \pm 0.004(exp)$ &CCFR~\cite{ccfr_seligman}& DGLAP NLO       \\  
 CCFR(97) $F_2+xF_3$ & $0.119 \pm 0.002(exp) \pm 0.004(th)$ &CCFR~\cite{ccfr_seligman}& DGLAP NLO \\
\hline
Global GLS&$ 0.108^{+0.005}_{-0.006}(exp) \pm^{0.004}_{0.006}(th)$ & CCFR~\cite{ccfr_xF3}& sum rule NNNLO HT\\
CCFR GLS(97)&$0.112^{+0.007}_{-0.009}(exp) \pm 0.009(th)$ & CCFR~\cite{Spentzouris} & sum rule NNNLO HT\\
\hline
HERA(94) in global & $0.1146 \pm 0.0036(exp)$ & Alekhin~\cite{alekhin}& Bayesian\\
HERA(93)&$0.120 \pm 0.005(exp) \pm 0.009(th)$ &BF~\cite{BAllalpha}& DGLAP NLO\\
H1(94)&$0.122 \pm 0.004(exp) \pm 0.007(th)$ & BF~\cite{Ballupdate} & DGLAP NLO\\
\hline\\
\end{tabular}}
\label{tab:alphas}
\end{table}

\subsubsection{Determinations of $\alpha_s$ from fixed target data}

Two main methods are used: analysis of scaling violations and of sum rules. 
Analyses of scaling 
violation data are most usually done at NLO by numerical solution of the 
DGLAP equations. The experimental uncertainties on such measurements are now
dominated by the systematic errors coming from energy scale uncertainties.
For a discussion of possible future 
improvements see ref.~\cite{Snomass97}. It is 
clear that taking into account correlations between systematic errors
can lead to a significant reduction in the total systematic error~\cite{botje2}.
However, theoretical uncertainties are just as important. 
Bl\"umlein et al~\cite{Blumalf} have considered some of  the
theoretical uncertainties involved in extracting $\alpha_s$ from such fits. 
 Differences arise from various sources. 
There are different approximations to the solution of Eq.~\ref{eq:alphas2} for
 $\alpha_s$ at second order and there are different ways of dealing with the
behaviour of $\alpha_s$ at the heavy quark thresholds 
(see Sec.~\ref{sec:heavyq}) 
and these choices give an uncertainty of $\pm 0.001$. One
may choose to perform the evolution in Mellin space or in $x$ space and 
this choice leads to
an uncertainty of $\pm 0.003$. Scheme uncertainty seems to give the 
largest contribution. Considering the correct scale for the evaluation of 
$\alpha_s$ to range from  $Q^2/4$ to $4Q^2$ gives an uncertainty of
$^{+0.004}_ {-0.006}$ from renormalization and 
$\pm 0.003$ from mass factorization
even when $Q^2$ is restricted to $Q^2 > 50\,$GeV$^2$. 
The scheme dependence should be considerably reduced if one could work to 
NNLO. Small $x$ data is particularly sensitive to the need for higher order
corrections.
One should also consider the uncertainty coming from restricting the analysis 
to the conventional leading $\ln Q^2$ approximations. 
There can be important corrections of the form $\alpha_s \ln (1-x)$ at high 
$x$ and $\alpha_s \ln (1/x)$ at small $x$. Presently the highest $x$ data come 
from the BCDMS experiment, and their major contribution is in the region
$0.3 < x < 0.5$, such that high $x$ logs are not likely to be important. 
However the HERA data do access small enough $x$ that $\ln (1/x)$ corrections 
could be important.

One of the earliest analyses which accounted for theoretical uncertainties
was given  by Virchaux and Milsztajn~\cite{VirMil} using BCDMS and SLAC $F_2$ 
data. This analysis
accounted for the correlated systematic errors in the BCDMS data, the relative
normalizations of the data sets, and the need for target mass corrections and
higher twist contributions at high $x$, low $Q^2$. Theoretical uncertainties 
from the choice of the renormalization and factorization scales, the position 
of the flavour thresholds, and the value of $R$ were also 
considered. The quoted value
 is very insensitive to the uncertainties on the gluon distribution, since
a substantial part of the data lies at high $x$, $x > 0.25$, where the 
non-singlet structure function dominates.

Accurate data are now available on non-singlet quantities. In 1993
 CCFR gave an NLO QCD analysis of their non-singlet $xF_3$ 
data~\cite{ccfr_lambda} 
for $Q^2 > 10\,$GeV$^2$ -  large enough that higher twist effects 
are considered negligible. The analysis was also done using
$F_2$ data instead of $xF_3$ data for $Q^2 > 15\,$GeV$^2$, $x > 0.5$, 
where $F_2$ is dominantly non-singlet. (However, the increased precision 
gained by using higher statistics $F_2$ data, is partly negated by increased 
theoretical uncertainty in using a quantity which is not strictly a 
non-singlet.) These analyses gave low values of 
$\alpha_s$ in agreement with the BCDMS value.
In 1997 CCFR updated their data~\cite{ccfr_seligman} as detailed in 
Sec.~\ref{sec:xf3dat}. 
Fits have again been made to $xF_3$ and $xF_3 + F_2(x>0.5)$ data under the 
same $Q^2$ cuts as for the 1993 analysis, and the values of $\alpha_s$ 
extracted are both significantly higher. 
If the correlations between systematic errors are included in the fits
then a reduced systematic error results, such that CCFR quote the final value 
$\alpha_s = 0.119 \pm 0.002(exp) \pm 0.004(th)$, where the theoretical 
uncertainty includes accounting for higher twist corrections and 
renormalization and factorization scale uncertainties.
We present the values from both the updated and the older data 
to aid comparison with the analyses of other authors
where the older data was used.

As an illustration of how different analysis techniques and different 
assumptions can affect the results extracted for $\alpha_s$, we briefly 
review some independent analyses of the CCFR data. Kataev et 
al~\cite{kataev0919} (KKSP) have used the alternative technique of 
reconstruction of the structure function from its Mellin moments, using an 
expansion in terms of Jacobi polynomials~\cite{russkys}, 
in the same kinematic regions as defined by CCFR(93).
The main advantage of this method is that 
NNLO corrections can be included. In practice
the inclusion of NNLO correction in this analysis does not change the
central values for $\alpha_s$ significantly (e.g. 
$\alpha_s = 0.109 \pm 0.006(exp) \pm 0.003 (th)$ for the $xF_3$ data) 
although it reduces theoretical
uncertainties due to scheme dependence. Target mass corrections were included
in the KKSP analysis but dynamical higher twist contributions were not.
Sidorov~\cite{Sidarov1} has also used the Jacobi polynomial method to 
analyse the CCFR(93) $xF_3$ data,  including higher twist
terms by adding a term $h(x)/Q^2$ to the leading twist expressions for 
$xF_3$, and including data in the $x,Q^2$ range $0.015 < x < 0.65, 
1.3 < Q^2 < 501\,$GeV$^2$, rather than cutting out the low $Q^2$ data. 
The value of $\alpha_s$ extracted is then even lower than that extracted by
KKSP, $\alpha_s = 0.104^{+0.006}_{-0.008}$. 
This analysis has also been used for a combined fit to
world neutrino data from other collaborations~\cite{Sidarov2}. The data from
CDHS~\cite{CDHSW}, SKAT~\cite{SKAT}, BEBC-WA59~\cite{Varvell}, 
BEBC-GGM~\cite{BEBC_GGM} and the new JINR-IHEP~\cite{JINR} data on $xF_3$ were
analysed in the kinematic region $0.03 < x < 0.80, 0.5 < Q^2 < 196\,$GeV$^2$
(with a cut of $x> 0.35$ on CDHS data to account for known 
problems~\cite{cdhswr}). This global analysis also yields a very low value for 
$\alpha_s$, $\alpha_s = 0.107 \pm 0.003$.
More recently KKSP have updated their NNLO analysis to consider the 
CCFR(97) $xF_3$ data including consideration of higher twist 
effects~\cite{KKSP97,KKSPpade}. Using the renormalon prediction for
the higher twist correction~\cite{das} they obtain 
$\alpha_s = 0.117 \pm 0.006(exp) \pm 0.003(th)$ but note that this should be 
compared with $\alpha_s = 0.121 \pm 0.006(exp) \pm 0.006(th)$ at NLO from their
own NLO analysis, rather than that of CCFR, and that they do not take into 
account correlations between systematic errors.
Tokarev and Sidorov~\cite{Toksid} have also used the 
Jacobi polynomial method to analyse the updated CCFR(97) $xF_3$ data, including
target mass, higher twist and nuclear corrections, and they quote a value of
$\alpha_s \sim 0.112$, much lower than the values from the corresponding
analyses of CCFR or KKSP. In contrast, Shirkov et al~\cite{Shirkov} have made an
analysis of the same data accounting for heavy quark thresholds, and they
obtain, $\alpha_s = 0.122 \pm 0.004$. Thus the theoretical error on 
$\alpha_s$ quoted by the CCFR collaboration from their CCFR(97) analysis may
be somewhat underestimated. 
 
Data on the GLS sum rule can also be used to extract $\alpha_s$. The GLS sum 
rule is a prediction for the $n = 1$ moment of the non-singlet structure 
function $xF_3$ which has been worked out to NNNLO
{\small \begin{equation}
\int_0^1 xF_3(x,Q^2) \frac{dx}{x} = 3 (1 - \frac{\alpha_s(Q^2)}{\pi} - 
a(n_i)\frac{\alpha_s(Q^2)}{\pi}^2 - b(n_i) \frac{\alpha_s(Q^2)}{\pi}^3) 
- \Delta HT
\end{equation}}
where $a,b$ are known calculable functions of the appropriate number of 
flavours $n_i$, and $\Delta HT$ represents the higher twist contribution.
Higher twist contributions to the low $n$ moments are much more reliably
estimated than  contributions to the structure functions. The higher
twist contribution to the GLS sum rule
 has been estimated by several different techniques~\cite{GLSnew} to be 
$0.27\pm\frac{0.14}{Q^2}$.  This constitutes the major theoretical 
uncertainty on the measurement. The major experimental uncertainty comes from
the need to extrapolate to $x \to 0$ in order to perform the integral.
Accordingly, the CCFR collaboration made a 
measurement using their own CCFR(93) data 
and that of BEBC-WA59, WA25, GGM, SKAT and FNAL-E180 (and some SLAC 
data on $F_2$ for $x> 0.5$), to achieve the best possible kinematic 
coverage~\cite{ccfr_xF3}. 
One can measure the value of the sum for various different $Q^2$ values, and 
thus determine $\alpha_s(Q^2)$ as a function of $Q^2$, in the low $Q^2$ range 
where most of the data lie, $1 < Q^2 < 20\,$GeV$^2$. Since $\alpha_s(Q^2)$ is 
changing 
rapidly in this region, such an analysis should be more sensitive than 
those made at higher $Q^2$. The equivalent value of $\alpha_s(M_Z^2)$ 
extracted is low~\fnm{s}\fnt{s}{~A more recent evaluation of the 
GLS sum rule using the new data from the 
JINR-IHEP detector~\cite{JINR} yields a much larger value of 
$\alpha_s(M_Z^2)$, 
$\alpha_s \sim 0.120$, but the statistical and systematic errors on these data 
are still rather large.} (see Table~\ref{tab:alphas}).
However, we note that Chyla and Kataev~\cite{chyla}(CK) 
have made an independent analysis of 
the data on the GLS sum rule from the CCFR(93) data alone~\cite{ccfr_gls}, 
paying particular attention to uncertainties due to scheme dependence, and 
they obtain the higher value $\alpha_s = 0.115 \pm 0.005(exp) \pm 0.003(th)$. 
Finally, CCFR have recently updated their GLS 
analysis~\cite{Spentzouris} to use the updated CCFR(97)data, 
in the region $0.02 < x
< 0.5$. A power law fit is used to make the extrapolations into the remaining 
$x$ region (where other $\nu N$
DIS data was used to determine the parameters of this fit). Their
preliminary value for $\alpha_s$ from this GLS analysis is larger than the 
corresponding result on 93 data (see Table~\ref{tab:alphas}), 
but it is still lower 
than the result from the DGLAP NLO fit to the scaling violations.

\subsubsection{Determinations of $\alpha_s$ from HERA data}

We will now consider how the HERA data have contributed to our knowledge of 
$\alpha_s$. We have already discussed the global fits of MRS and CTEQ 
in Sec.~\ref{sec:pdf}. The
purpose of these fits is to extract the parton distributions rather than to
measure $\alpha_s$. Thus in the latest MRSR fits distributions are given for
fixed values of $\alpha_s$: low, $\alpha_s = 0.113$, and high, 
$\alpha_s = 0.120$. 
We note that in these fits the HERA(94) data prefer the latter value, 
in contrast to the fixed target data 
(not including the CCFR(97) data) which prefer the former. The CTEQ4 
analysis also uses the somewhat higher value of $\alpha_s = 0.116$ 
to accommodate the HERA(94) data. 
However, the correlations between experimental
systematic errors are not accounted for in these global fits.  
Alekhin~\cite{alekhin} has made a new global fit to the high precision DIS 
data from BCDMS, the latest NMC(97) data and H1(94) and ZEUS(94) data, 
for which information on correlations is available. He takes account of
point to point correlations using a Bayesian treatment of systematic errors.
The data is cut such that $W^2 > 16\,$GeV$^2$, $Q^2> 9\,$GeV$^2$ to reduce
sensitivity to higher twist contributions. 
He makes target mass corrections and
accounts for heavy quarks using the GRV formalism.
The fit is made with $\alpha_s$ and the parametrization of the parton 
distributions as free parameters
and the error quoted on $\alpha_s$ fully accounts for its 
correlation to the shape of the gluon distribution. 
The value of $\alpha_s$ extracted is $\alpha_s = 0.1146 \pm 0.0036$.
 
It is also interesting to consider the HERA data alone rather than 
within a global fit. Ball and Forte~\cite{BAllalpha} predict that structure 
functions in the small $x$, high $Q^2$
limit exhibit `Double Asymptotic Scaling' (DAS), which we discuss in 
Sec.~\ref{sec:lowx}. One may  fit directly to the 
DAS predictions for the structure functions in order to 
extract $\alpha_s$, and such a fit to H1(94) data yields 
$\alpha_s = 0.113 \pm 0.002(stat) \pm 0.007(sys)$~\cite{deRoeck}, 
but such a procedure can only be 
approximate since the DAS formulae are merely good approximations to the full
NLO QCD predictions. Ball and Forte have made an NLO QCD fit tailored to
low $x$  HERA data in order to measure $\alpha_s$. They take a set of
parton distributions such as MRSR, known to fit data at high $x$, 
evolve them to a starting scale $Q^2_0$ (which is fixed in any one fit, but 
choices are varied in the range $1 < Q^2_0 < 25\,$GeV$^2$) and then cut off 
their low $x$ tails for $x \leq 0.01$, replacing them with new tails: 
$xq \sim x^{-\lambda_S}$ for singlet quarks (at the low $x$ values considered
the singlet quark distribution may be considered as identical to the sea quark
distribution)
and $xg \sim x^{-\lambda_g}$ for the gluon, where the 
$\lambda$ values are parameters of the fit. These new distributions (together
 with unmodified non-singlet distributions) are then evolved to the $Q^2$
values of the HERA(93) data, using a value of $\alpha_s$ which is also
a parameter of the fit.

The results of the best fit (see Table~\ref{tab:alphas})
show that $\alpha_s$ and the small $x$
shape of the gluon distribution, characterized by the parameter $\lambda_g$, 
are correlated as usual, BUT the shape of the gluon and the singlet quark 
distribution, 
characterized by the parameter $\lambda_S$, are 
also strongly correlated in this kinematic region and the singlet quark
distribution is strongly constrained  directly through the measurement 
of $F_2$. Thus
the errors on $\alpha_s$ are smaller at low $x$ than at moderate $x$ because
at small $x$ the gluon is `driving' the behaviour of the singlet quark 
distribution. 
The theoretical uncertainties from sources such as scheme dependence are 
discussed in some detail, with similar conclusions to those of 
Bl\"umlein~\cite{Blumalf}. However Ball and Forte go further and consider
the need for $\ln (1/x)$ corrections to the conventional leading
$\ln Q^2$ expansion. The necessary modifications to the 
conventional expansion scheme will be discussed in Sec.~\ref{sec:resum},
here we quote only the result that the uncertainty on $\alpha_s$ from this
source is $^{+0.002}_{-0.006}$, asymmetric because the central value of a fit
including $\ln (1/x)$ contributions is $\alpha_s = 0.115$. An update
to this analysis
using the H1(94) data suggests that this uncertainty has now been reduced
since fits including $\ln (1/x)$ corrections are no longer competitive in 
$\chi^2$ to the conventional fits~\cite{Ballupdate}.

However, Thorne~\cite{Thornenew} disagrees with the conclusion 
that $\ln (1/x)$ terms are not necessary to fit the low $x$ data. This will be 
discussed in detail in Sec.~\ref{sec:resum}. 
The consequence for the extraction of
$\alpha_s$ is that low $x$ data cannot give a reliable estimate of $\alpha_s$ 
when treated conventionally. This is because a fully renormalization scheme 
consistent perturbative expansion necessitates the inclusion of 
$\ln (1/x)$ terms. Unfortunately, this can only be done to leading order 
at the moment, 
due to the lack of full knowledge of the $\ln (1/x)$ terms to next-to-leading
order. Thus a fit to $\alpha_s$ at NLO 
cannot be consistently performed in regions of $x$ where $\ln (1/x)$ terms are 
important. He estimates this region as $x \leqsim 0.01$. Thus conventional 
extractions of $\alpha_s$ using the HERA data are unreliable. When the 
next-to-leading
$\ln (1/x)$ terms are fully known an extraction which does not suffer from
uncertainties due to scheme dependence can be made. Thorne's provisional
value for a properly extracted value of $\alpha_s$ is $\sim 0.115$~\cite{priv}.

A further reservation on the use of low $x$ data in conventional fits comes 
from the uncertainties due to the treatment of heavy quark thresholds. None of
the analyses discussed above 
have treated heavy quark production consistently in all kinematic regions.
A correct treatment of charm production will make a significant difference
at small $x$ where $F_2^{c\bar c}$ is a greater fraction of $F_2$.  
If a QCD fit is made without accounting for the charm
threshold correctly then threshold behaviour can fake a stronger dependence of
$F_2$ on $Q^2$ than is truly attributable to QCD scaling violations and thus 
make the value of $\alpha_s$ seem larger. For example, 
the MRRS analysis discussed in Sec.~\ref{sec:latest} attempts to treat
 heavy quark 
production consistently from threshold to the asymptotic region  and the 
corresponding value of $\alpha_s$ is $0.118$, lower than the previous
best fit value of MRSR2 ($\alpha_s = 0.120$). 

To summarize, when one considers the systematic and theoretical uncertainties
 quoted on $\alpha_s(M_Z^2)$ values it is not clear that there is any 
significant difference between values of $\alpha_s$ extracted from lower 
energy DIS processes ($\alpha_s \sim 0.113$) and those extracted from LEP data
($\alpha_s \sim 0.120$), particularly in view of the updated CCFR(97) result 
($\alpha_s\sim 0.119$). 
The values of $\alpha_s$ extracted from HERA data ($\alpha_s \sim 0.120$) 
are quite consistent with those extracted from 
the lower energy, fixed target DIS data and with the LEP data. 
However, a correct treatment of 
the correlations between systematic errors, of the charm 
threshold and of the need to include $\ln (1/x)$ terms at low
 $x$ would all lead to a reduction of $\alpha_s$ compared to that extracted 
from a conventional NLO QCD fit made without these corrections. 
 One awaits an analysis which includes 
all of these considerations before drawing final conclusions. 

\subsection{Models for parton distribution functions}
\label{sec:modelpdfs}

One of the problems encountered in establishing whether DGLAP alone is 
sufficient at low $x$ and $Q^2$ is the flexibility allowed in the
functional form chosen for the shape in $x$ of the PDFs at the starting 
scale $Q^2_0$. The actual forms used by the MRS and CTEQ groups have been
discussed in Sec.~\ref{sec:mpdf} and we noted in Sec.~\ref{sec:pdfhilo} 
the constraints from 
counting rules as $x$ tends to 1 and from Regge theory at low $x$.  
Typically about 15 to 20 shape parameters are used in global fits to describe 
the nucleon PDFs. In this section we review very briefly 
some of the other methods and approaches that have been used to model or 
constrain the input distributions. 

\subsubsection{Quantum statistics}
\label{sec:pdfqstats}
The idea that the Pauli exclusion principle could have an effect on the
parton distributions and in particular force the sea to be flavour 
non-symmetric is almost as old~\cite{FF} as the quark-parton model. 
There is also considerable support from experiment, as we have seen from
 the measured value of the Gottfried sum rule 
and the difference in the Drell Yan asymmetry measured in $pp$ and $pn$ 
scattering. Recently a number of authors have explored the use of 
functional
forms based on Fermi-Dirac statistics for quarks and antiquarks and
on Bose-Einstein statistics for gluons. So, for example, the quark
distributions are of the form
\begin{equation}
xq(x)={f(x)\over \exp({x-x_q\over \bar{x}})+1}
\end{equation}
where $\bar{x}$ is a parameter playing the role of a universal 
temperature,
$x_q$ is the thermodynamic potential of quark species $q$ and
$f(x)$ is the weight function $Ax^\alpha(1-x)^\beta$ allowing
incorporation of the usual counting rule and Regge constraints if
required. 

Bucella, Bourrely, Soffer and coworkers~\cite{buccella1} have
developed a model for polarized parton distributions and have added
phenomenological constraints, specifically the dominance of $u^\uparrow$ 
at large $x$ (from $F_2^n/F_2^p$ and $g_1^p/F_1^p$)
and the approximate equality $u^\downarrow=d/2$ (from QPM
sum rules for $u^{\uparrow,\downarrow}$, the parton distribution
functions for the quark spin parallel and antiparallel to that of
the proton). They also split the quark distribution functions into two
pieces: a `gas' part corresponding to the valence quarks which
is important at large $x$ and a common `liquid' part to model
the rise in the sea at low $x$. Ref.~\cite{buccella1}$^d$ gives the most 
detailed comparison with data. The 8 or so model parameters are
determined at the starting scale $Q^2_0=4\,$GeV$^2$ by fitting
unpolarized CCFR(93) $xF_3$ data and NMC(92) $F^p_2$, $F^n_2$ data.
Standard NLO DGLAP evolution is then used to calculate the structure 
functions at other values of $Q^2$. The resulting
PDFs give a reasonable description of $F^p_2$, $F^n_2$ at higher $Q^2$
(including the early data from HERA) and also the polarized structure
functions $g_1^p$ and $g_1^n$. The model has been refined 
further~\cite{buccella2} to produce various relationships between 
polarized and unpolarized structure functions. Another important outcome
of these ideas is that the discrepancy in the Ellis-Jaffe sum 
rule~\cite{jafell} for
$g_1^p$ may be linked to that in the Gottfried sum rule, but we will
not pursue this further here.

A more formal extension of the use of statistical mechanics to parameterize
the parton distributions is outlined in the paper by Mangano, Miele \& 
Migliore~\cite{mangano}. Ideas from non-equilibrium statistical mechanics
are used to derive modifications to the DGLAP evolution equations to take
account of Pauli blocking for quarks and antiquarks and statistical effects
in gluon emission. Finally Bhalerao~\cite{bhalerao} uses a phenomenological
approach to consider how the statistical distributions are modified by
the finite size of the nucleon. The effect of the finite volume is to
soften the input parton distribution somewhat, which when coupled with
standard DGLAP evolution allows the model to be fit to $F_2$ data quite
successfully.

\subsubsection{The MIT bag model}
\label{sec:pdfbag}

The MIT bag model~\cite{MITbag} was developed to study
the spectroscopy of hadronic states, but it also  
provides hadronic wavefunctions. There have been many attempts
over the last 20 years to use bag model wavefunctions as a basis for
calculating valence quark-parton distributions. 
 In the bag model the quarks are taken to be massless fermions satisfying
the free particle Dirac equation inside a static sphere of fixed radius.
The spectrum of hadronic states is derived by applying the gluon field as
a surface boundary condition on the confined quark fields.
One of the major problems faced is that
in a naive approach the bag model calculations do not give the correct
support for the PDFs (i.e. they are non-zero for $x$ outside [0,1]). 
Procedures for overcoming this difficulty
have been proposed by a number of authors and a fairly recent 
paper~\cite{bagsupport} emphasizes the importance of ensuring the 
correct support. Steffens and Thomas~\cite{NLObag} have taken
bag model parton distributions one step further by providing a 
calculation at next-to-leading order. While the quality of the description
of the structure function data is not much changed, the reliability of
the calculation is greatly increased. In leading order, to get a good
fit the starting scale must be taken very low ($\mu^2=0.0676\,$GeV$^2$) and the
corresponding value of
$\alpha_s$ is larger than 1. In NLO the starting scale becomes 
$\mu^2=0.115\,$GeV$^2$ and $\alpha_s=0.77$. In both LO and NLO, apart from
the starting scale, the bag radius and the scalar and vector di-quark
masses are parameters to be determined from data. In fact they are 
fixed by comparison with the MRSD$_0^{\prime}$ parameterization at 
$Q^2=10\,$GeV$^2$. An alternative approach to the support problem
has been investigated in a recent paper by  Jasiak~\cite{jasiak}. 
His method also allows the parton distributions to be normalized correctly, 
thus addressing a common problem in bag model calculations.
He compares his bag model calculations to the GRV94 parameterization and
finds that it is necessary to use the Politzer scaling variable
(which reduces the effects of target mass at low energies) to get 
approximate agreement at the starting scale.

\subsubsection{Lattice QCD}
\label{sec:pdflattice}

Much progress has been made in recent years in the calculation
of nucleon parton distribution functions from lattice QCD~\cite{gockeler}.
The present situation was summarized by Schierholz~\cite{schierholz}
at the recent DIS97 workshop. The lattice calculations are based on
the operator product expansion for the leading twist moments of the 
structure functions (Sec. 3), which gives
\begin{equation}
\int^1_0dx x^{n-2}F_2(x,Q^2)=
   \sum_{a=u,d,g}C_{2,n}^a(\mu^2/Q^2,\alpha_s(\mu^2))A_n^a(\mu)
\label{eqn:ope}
\end{equation}
where $C_{2,n}$ are Wilson coefficients and $A_n$ are forward 
$\gamma^*p$ matrix elements of various local operators at scale $\mu$. 
In parton model language
the matrix elements are related to the moments of the corresponding
parton distributions, $A_n^a= q_n^a$. On the lattice
the matrix elements are calculated from two and three point correlation
functions, so far in the quenched approximation,\fnm{t}\fnt{t}{~Which means
that the effects of internal quark loops are ignored.}
 and in terms of bare
lattice operators which are functions of the lattice spacing $d$. 
Calculations are done for a range of quark masses,
typically $30-200\,$MeV/c$^2$ so that the chiral limit may be taken 
reliably. To
estimate the physical matrix elements (PDF moments) the operators
must be renormalized at scale $\mu$ as they are divergent in $d$. The
renormalization constants have been calculated perturbatively to one
loop order.  To date results have been obtained for moments up to $n=3$
as this is the maximum possible on a hypercubic lattice. A further constraint
is that the
required computing time increases rapidly with $n$. 
Table~\ref{tab:lattice} shows results from ref.~\cite{schierholz} 
for moments of the unpolarized structure functions compared to those
obtained from the CTEQ3M global fit.

\begin{table}
\tcaption{Lattice QCD results~\cite{schierholz}
 for moments of unpolarized parton distributions compared
with those calculated from the CTEQ3M global fit. The
lattice calculations were done in the quenched approximation
and the figures in brackets give the error.}
\centerline{\footnotesize\smalllineskip
\begin{tabular}{lcc}\\
 \hline
 Moment & Lattice & CTEQ3M \\
    & $\mu^2\approx 5\,$GeV$^2$   & $\mu^2=4\,$GeV$^2$ \\
 \hline
 $\langle x\rangle^u$ & 0.410(34) & 0.284 \\
 $\langle x\rangle^d$ & 0.180(16) & 0.102 \\
 $\langle x^2\rangle^u$ & 0.108(16) & 0.083 \\
 $\langle x^2\rangle^d$ & 0.036(8) & 0.025 \\
 $\langle x^3\rangle^u$ & 0.020(10) & 0.032 \\
 $\langle x^3\rangle^d$ & 0.000(6) & 0.008 \\
 $\langle x\rangle^g$ & 0.53(23) & 0.441 \\
\hline\\
\end{tabular}}
\label{tab:lattice}
\end{table}

The table shows that the lattice results for the lowest moments
are systematically larger
than those from the CTEQ global fit, particularly for the $u$ quark.
Although not the subject of this review, it is worth noting that
the same lattice calculations also give results for the moments
of the polarized structure functions and these are in better agreement
with measurements. The lattice group still have a large number of
systematic effects to study, for example lattice spacing, improved
actions and the effect of the quenched approximation. 
However they do not think that
these will change the pattern of the above results significantly.
One possible explanation is that higher twist effects in the data
are responsible
for the difference. This is an area that is beginning to receive
attention again for a number of reasons, but it will take a lot more
study, both from the lattice and phenomenological sides before this 
explanation is established.

In principle the idea of using lattice QCD to give the parton
distribution functions at the starting scale is a very attractive
as one would then be able to compare data at all $Q^2$ with a complete 
QCD calculation. However, leaving aside the present discrepancy 
in the unpolarized moments, it will be a long time before lattice 
technology is able to provide calculations of other than the lowest
order moments. The challenge of how to use the incomplete information
provided by the low order moments from the lattice has been taken up by
Mankiewicz \& Weigl~\cite{weigl}. Their idea is to use the lattice
moments to give the large $x$ behaviour of the parton distributions
and phenomenological or Regge ideas to fix the small $x$ behaviour.
Technically they do this by using the Ioffe-time distributions~\cite
{Ioffe},
effectively transforming the parton distributions to coordinate space.
To be more specific the parton distribution function $q(x,\mu^2)$
and the Ioffe distribution $Q(z,\mu^2)$ are related by
\begin{eqnarray}
Q(z,\mu^2)&=&\int_0^1 du\ \ q(u,\mu^2)\sin uz,\nonumber \\
q(x,\mu^2)&=&{2\over \pi}\int^\infty_0 dz\ \ Q(z,\mu^2)\sin uz,
\end{eqnarray}
where $z$ is the Ioffe-time variable, which is a Lorentz invariant
measure of the longitudinal distance along the light cone between
the quark fields. $Q(z,\mu^2)$ is the gauge invariant correlation
function of two quark fields on the light cone. The advantage of
using $Q(z,\mu^2)$ is that it allows the separation, in a very clear
way, of the two different regimes (large and small $x$) which 
determine the behaviour of
the PDFs. Behaviour at large $z$ corresponds to small $x$ and this
has to be determined phenomenologically at present. One finds that
if $q(x)\sim x^{-(1+\alpha)}$ then $Q(z)\sim z^\alpha$ and vice-versa.
The authors have used CTEQ3, MRSA and GRV94 to evaluate the singlet 
$Q(z)$ distributions for the $u$ and $d$ quarks at the scale 
$\mu^2=4\,$GeV$^2$. In all cases they find that the large $z$ behaviour
is approximately constant or slowly rising, indicating that $\alpha$ is 
either 0 or a small positive number. The results also indicate that
at large $z>10$ (small $x$ and corresponding to linear distances of
order $2\,$fm) quark-quark correlations at distances larger
than the electromagnetic size of the nucleon are important. The
behaviour at small values of $z$ (large $x$) is almost linear and
in principle well determined by the low order parton distribution
moments from lattice QCD. The two different behaviours have to be matched
smoothly at a transition region around $z=5$. The scheme that Mankiewicz
and Weigl use is a power series interpolation in $z$ with the parameters
determined from the lattice moments at small $z$, and from
the CTEQ NLO fit and the nucleon electromagnetic 
radius at large $z$. The resulting singlet
distribution for the $u$ quark is larger than that from global fits,
which is a reflection of the larger lattice moments. A similar
approach for the non-singlet valence quark distributions gives results
in better agreement with global fits, which indicates that perhaps
the contribution of the $q\bar{q}$ sea at small $x$ is
not properly understood yet.  
 
\subsubsection{Other approaches}
\label{sec:pdfskyrm}

\begin{figure}[ht]
\centerline{\psfig{figure=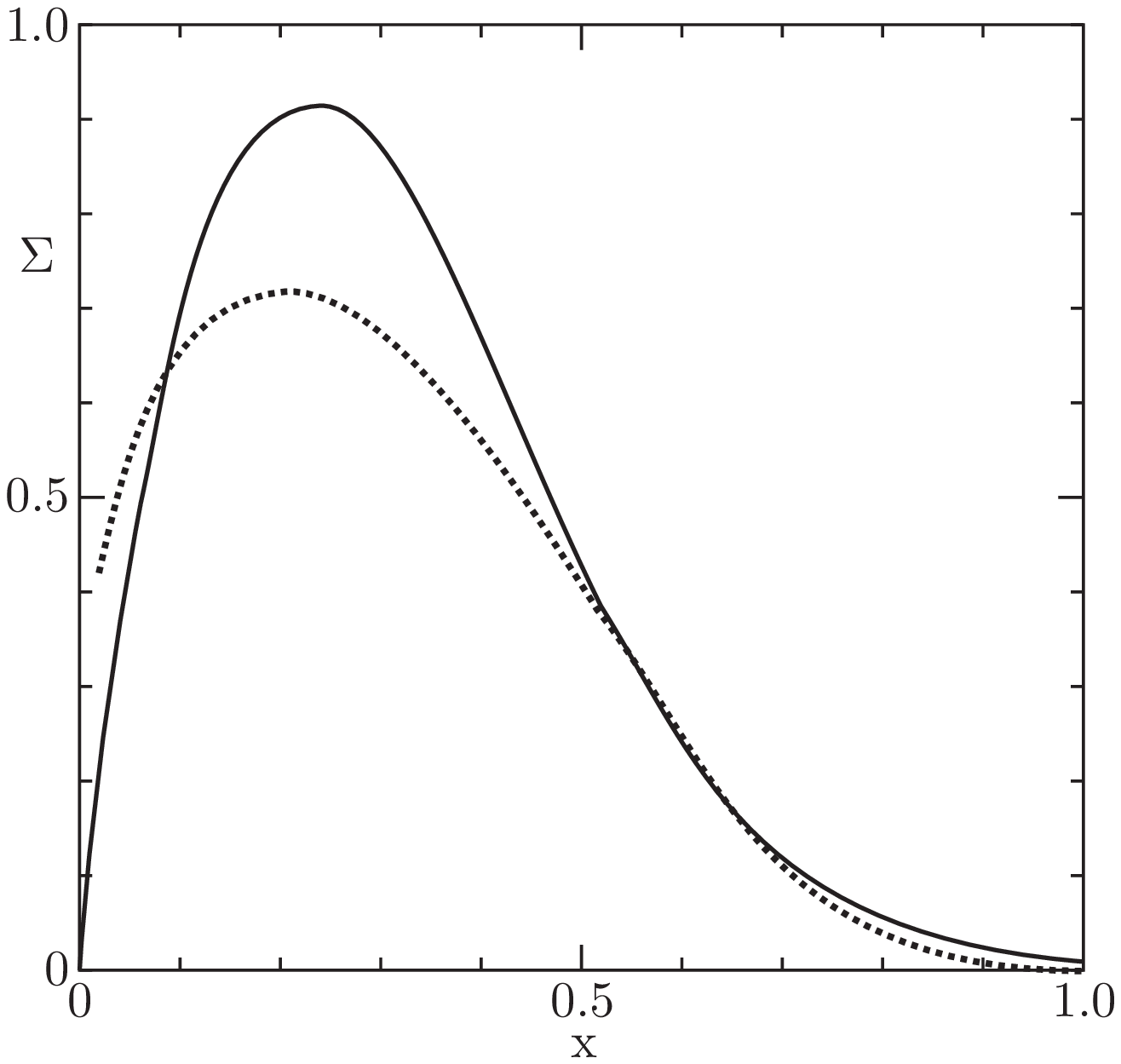,width=6cm,height=5cm}}
\fcaption{The isosinglet unpolarized distribution 
$\Sigma={1\over 2}x[u+d+\bar{u}+\bar{d}]$ from the large $N_c$ chiral
model calculation of Diakonov et al (line) compared to that from GRV94 
(points).}
\label{fig:diakonov}
\end{figure}

It is clear from the discussion above that a different approach is
needed to calculate the small $x$ behaviour of the structure functions.
Recently there has been a revival of interest in non-perturbative methods
that use other than quark degrees of freedom~\cite{bj96}. The ideas
go back to those of Skyrme and more recently the instanton~\cite{shuryak}, 
which are based on classical solutions of Euclidean QCD field equations. 
QCD has chiral symmetry which is spontaneously broken giving rise to the 
almost massless pion as the corresponding Goldstone boson.
Diakonov, Petrov and co-workers~\cite{diakonov} have used these ideas
to calculate the quark momentum distributions in the large-$N_c$ limit
and at a low scale. The basis of their model is that in the large-$N_c$
limit QCD is equivalent to an effective meson theory with baryons as
solitonic excitations. The effective theory is expressed in the form
of a chiral lagrangian for the pion field. There are various parameters in 
the model, such as the constituent quark mass which is taken to be 
$350\,$MeV/c$^2$ and a UV cutoff (technically a Pauli-Villars regularization 
of $600\,$MeV is used in the numerical calculations). The latter
parameter also sets the scale at which the parton distributions are
calculated. The calculated isosinglet singlet unpolarized distributions are 
compared with the NLO GRV94 parameterization  and the isovector
polarized distributions with the corresponding parameterizations from
ref.~\cite{GRSV}. In both cases the agreement between the calculated
and the phenomenological distributions (which are taken as a good 
indication of the behaviour of the data) is quite good in shape but the
normalizations do not quite agree. 
Fig.~\ref{fig:diakonov} shows the comparison for the isosinglet unpolarized
distribution.

\noindent

From the figure it can be seen that the model lies
above the `data' in the isosinglet case. For the isovector it 
is the other way around.
Diakonov et al argue that this is very encouraging as the calculations
can be improved and that they already give a lot of support to the
GRV approach of a low starting scale with `valence-like' input
distributions. A somewhat similar approach based on the Nambu-Jona-Lasino
model is discussed by Gamberg, Reinhardt and Weigl~\cite{gamberg}.

Another recent calculation that gives support to the GRV assumption
of an intrinsic `valence like' gluon momentum distribution 
is that of Hoyer and Roy~\cite{hoyroy}. It is based
on an analysis of the long time-scale structure of the nucleon using 
a Fock state decomposition of the nucleon wave-function proposed some
years ago~\cite{brodsky8081}. The basic idea is that the partons
in the nucleon Fock states ($|qqq\rangle,~|qqqg\rangle,
~|qqqq\bar{q}\rangle,\ldots$) have similar velocities in order to stay 
together over long times. This then gives a probability for an
$n$-parton state proportional to $\displaystyle 
\left(m^2_N-\sum^n_1{m^2_{\bot i}\over x_i}\right)^{-2}$, where
$m^2_{\bot i}=m^2_i+k^2_{\bot i}$ is the squared transverse mass of 
parton $i$. This leads to the probability distribution of a given Fock
state being peaked at 
$\displaystyle x_i=m_{\bot i}/\sum^n_i m_{\bot i}$.
For the 3 quark plus multi-gluon states, assuming a
common transverse mass $m_{\bot i} \approx 0.3 - 0.4\,$GeV, the
probability of the $n$-gluon state is found to be proportional
to $1/n^4$. This then gives $\bar{x_q}=1/4,~\bar{x_g}=1/6$ for the
mean fractional momenta carried by the quarks and gluons and 
$\epsilon_q=3/4,~\epsilon_g=1/4$ for the 
fractions of the nucleon momentum at the starting scale, where both quark and 
gluon states have a valence-like shape.

To summarize this brief survey of non-perturbative models for the PDF input
functions, we have seen that there are some very promising ideas but that
much more work needs to be done before one could use such results directly
in a global fit. A more likely scenario is to use some of the
non-perturbative results as constraints on the input functional forms and
parameters - as Mankiewicz and Weigl have done with the structure function
moments from lattice calculations. It is also interesting to note that a
number of the model approaches do appear to give support to the GRV
valence-like input forms.


\section{pQCD at low $x$}
\label{sec:lowx}

With the advent of data from the HERA collider we are into a new phase of
testing the applicability of pQCD. This time we are interested in testing its
predictions at very low $x$ where we may be moving out of the region where
the conventional approximations (LLA and NLLA) as embodied in the DGLAP
equations are applicable. We outline the main theoretical approaches to
the low $x$ region in Sec.~\ref{sec:thlowx} and we discuss the current 
phenomenology of each of these approaches in the subsequent sections.

\subsection{Theoretical approaches to low $x$ physics}
\label{sec:thlowx}
\subsubsection{Low $x$ solutions of the DGLAP equations}

Consider Eqs.$~\ref{eq:pqq}-\ref{eq:pgg}$ 
for the (LO) splitting functions as $z = x/y \to 0$.
{\small \begin{equation}
 P_{qq} \to \frac{4}{3},\ P_{qg} \to \frac{1}{2},\ P_{gq} \to 
\frac{4}{3}\frac{1}{z},\ P_{gg} \to 6\frac{1}{z}
\label{eq:smallx}
\end{equation}}
\noindent
We see that the gluon splitting functions are singular as $z\to 0$ (and this
result remains true for higher orders). Thus the gluon distribution will
become steep as $x \to 0$, and its contribution to the evolution of the
quark distribution will become dominant, so that the quark singlet 
distributions, and hence the structure function $F_2$, will also become
steep as $x \to 0$. 

Quantitatively,
{\small \begin{equation}
 \frac{dg(x,Q^2)}{d \ln Q^2} \simeq \frac{\alpha_s(Q^2)}{2\pi} \int^1_x
\frac{dy}{y} \frac{6}{z} g(y,Q^2)
\label{eq:glev}
\end{equation}}
\noindent
may be solved subject to the nature of the boundary function $xg(x,Q^2_0)$.
There are  two possibilities, depending on whether this input function is
singular or non-singular. Formally, the distinction is made by considering
whether the singularities of the operator product expansion matrix elements 
lie to the left or the right of those of the anomalous dimensions. In practice,
this means that if the input function is flatter(steeper) than 
$xg(x,Q^2_0) \sim x^{-0.25}$ it is considered non-singular(singular).

The non-singular solution was given (in the DLLA) 
by de Rujula et al~\cite{deRujula} 
{\small \begin{equation}
 xg(x,Q^2) \simeq \exp\left(2 \left[\xi (Q^2_0,Q^2)\ln\frac{1}{x}\right]
^{\frac{1}{2}}\right)
\label{eq:DLLAg}
\end{equation}} 
\noindent
where
{\small \begin{equation}
 \xi(Q^2_0,Q^2) = \int^{Q^2}_{Q^2_0} \frac{dq^2}{q^2} \frac{3\alpha_s(Q^2)}
{\pi}
\label{eq:DLLAxi}
\end{equation}}
\noindent
Given a long enough evolution length from $Q^2_0$ to $Q^2$, this will
generate a steeply rising gluon distribution at small $x$, starting from 
the flattish behaviour (or even a valence-like shape~\cite{GRV}) 
of $xg(x,Q^2)$ at $Q^2 = Q^2_0$. 

\begin{figure}[ht]
\centerline{\psfig{figure=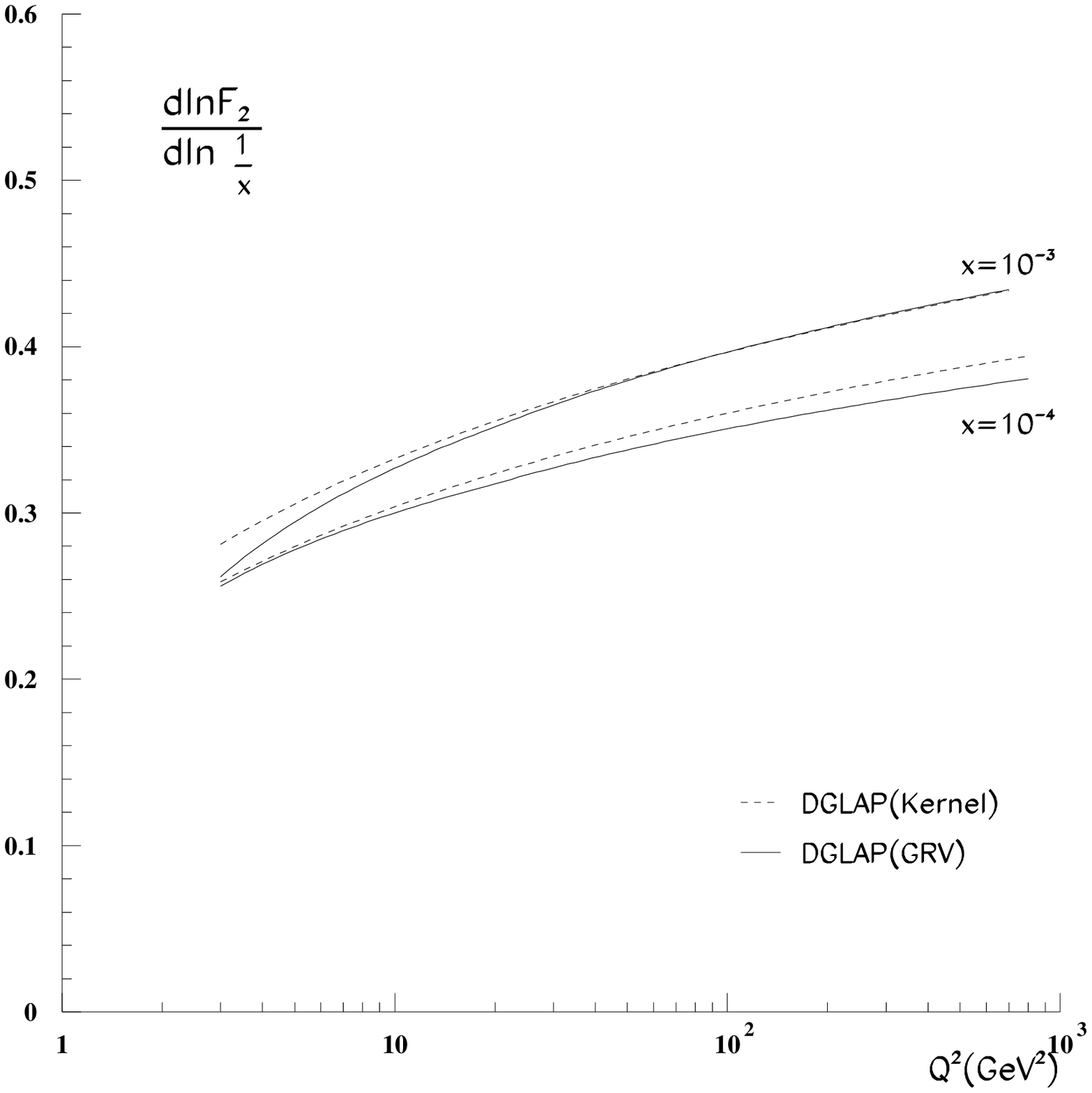,width=9cm,height=9cm}}
\fcaption{The quantity $\lambda = \frac{d\ln F_2}{d\ln(1/x)}$ as a function of 
$Q^2$ for two different $x$ values, as evaluated from the DGLAP 
kernels~\cite{Peslam} and from the parametrization of GRV}
\label{roy1b} 
\end{figure}

Specifically, the gluon distribution   
rises faster than any power of $\ln(1/x)$, but slower than any power of $x$.
However, over the limited $x,Q^2$ range of HERA data it may mimic a power 
behaviour, $xg(x,Q^2) \sim x^{-\lambda_g}$, with 
{\small \begin{equation}
\lambda_g = \left(\frac{12}{\beta_0}\frac{ \ln(t/t_0)}{\ln(1/x)}\right)^{\frac{1}
{2}}
\end{equation}}
\noindent
where $t=\ln(Q^2/\Lambda^2)$, $t_0=\ln(Q_0^2/\Lambda^2)$.
The steep behaviour of the gluon will generate a 
steep behaviour of $F_2$ at small $x$, $F_2 \sim x^{-\lambda_S}$, where
 the DGLAP evolution kernels give~\cite{Peslam}
{\small \begin{equation}
\lambda_S \approx \left(\frac{12}{\beta_0}\frac{\ln(t/t_0)}{\ln(1/x)}\right)
^{\frac{1}{2}} - 
\frac{a}{\ln(1/x)}
\label{eq:DLLAlam}
\end{equation}}
\noindent
where $a$ depends on the input shape of the 
gluon at $t_0$, ($a = 0.75$ for a flat input, $a = 1.25$ for a valence-like
input). 
In Fig.~\ref{roy1b} we illustrate the behaviour of this equation by showing
$\lambda_S =d \ln F_2/d \ln(1/x)$ as a function of 
$Q^2$ for two different $x$ values.

This steep behaviour of $F_2$ at small $x$ contrasts with the Regge 
expectation of a flattish behaviour of $F_2$ coming from the exchange of
the soft Pomeron (see Sec.~\ref{sec:pdfhilo}). 
However, there is not necessarily any contradiction between 
these two predictions, since it is not clear over what range of $Q^2$
the Regge prediction should be applicable. If we assume it is appropriate 
for very low $Q^2$ values, then pQCD will generate a steep shape 
for higher $Q^2$ by its evolution. This is the usual explanation of 
conventional DGLAP evolution
for the steep shape of $F_2$ observed at low $x$ in HERA 
data~\cite{BF,GRV}. The current phenomenology of this 
explanation is explored in Sec.~\ref{sec:conv}. 
\noindent

This approach suggests that we might
extend our concept of the Pomeron. The Regge prediction was based on 
the exchange of the soft Pomeron in the t-channel. 
If we continue this idea into 
the higher $Q^2$ realm it implies that the Pomeron can be identified with
the strongly ordered gluon ladder of Fig.~\ref{ladder}. Retracing the 
argument that a flattish behaviour of $F_2$ as $x\to 0$ 
derives from a flattish behaviour of the $V^*N$ total 
cross-section as $s(V^*N) (= W^2)$ increases,
we see that a steeply rising behaviour of $F_2$ as $x\to 0$ would imply a
steeply rising $V^*N$ total cross-section, and the data for larger $Q^2$ do 
indeed exhibit this behaviour, see Fig.~\ref{fig:sigtotw}.
Hence the corresponding `QCD' Pomeron is hard, i.e. it
has an intercept $\alpha_P$ significantly in excess of unity.

The singular solution to Eq.~\ref{eq:glev} has been given by 
Yndurain et al~\cite{yndurain,Yndurainold,Yndnow}.
 If a steep $\lambda$ value is input 
then it remains stable under $Q^2$ evolution, overriding the prediction of  
de Rujula et al~\cite{deRujula} such that one has~\cite{EKL,BF,KMRS} 
{\small \begin{equation}
F_2 \sim \ln Q^2 x^{-\lambda}
\label{eq:Yndlam}
\end{equation}}
with a fixed value of $\lambda$ equal to the input value, for all $Q^2$. 
Whereas such a solution implies
an unconventional hard Pomeron right from the start, it may be preferred 
theoretically because such solutions are stable against 
the inclusion of higher 
order diagrams, whereas the solutions with flattish (or valence-like) 
input gluons are quite strongly sensitive to the inclusion of NNLO and NNNLO 
terms~\cite{EKL}. The current phenomenology of this solution is explored
further in Sec.~\ref{sec:conv}

\subsubsection{The BFKL equation}

Although it is now evident that
it is possible to fit all data with $Q^2 \geqsim 1.5\,$GeV$^2$ conventionally,
there have been criticisms of this explanation.  Firstly, because 
it is necessary to
begin the $Q^2$ evolution at $Q^2 \simeq 1.0\,$GeV$^2$, or lower, 
and one may doubt 
whether perturbative calculations are reliable for such low $Q^2$ values, 
since
$\alpha_s$ is correspondingly large ($\alpha_s \geqsim 0.4$). Secondly,
in the usual LLA approach we are summing terms which are 
leading in $\ln(1/x)$ only when they are accompanied by leading $\ln Q^2$. 
One might legitimately ask if, in the low $x$ region, it would not also
be appropriate to sum diagrams which are leading in $\ln(1/x)$ independent
of $\ln Q^2$. This is what is done by the BFKL equation~\cite{BFKL}.   
This involves considering the evolution of a gluon distribution which is 
not integrated over $k_T$, since breaking the association to leading 
$\ln Q^2$ implies that the gluon ladder need not be ordered in $k_T$. The 
unintegrated gluon distribution $f(x,k_T^2)$ relates to the more familiar 
gluon distribution as follows
{\small \begin{equation}
 xg(x,Q^2) = \int^{Q^2}_0 \frac{dk_T^2}{k_T^2} f(x,k_T^2)
\label{eq:BFKLg}
\end{equation}}
\noindent
The BFKL equation may then be written
{\small \begin{equation}
 \frac{ df(x,k_T^2)} { d\ln(1/x)}= \int  dk'^2_T K(k_T^2,k'^2_T) 
f(x,k'^2_T)\ =\ K \otimes f =\ \lambda \ f 
\label{eq:BFKL}
\end{equation}}
\noindent
which describes the evolution in $\ln(1/x)$ of the unintegrated gluon density.

The solution of this equation is controlled by the largest eigenvalue 
$\lambda$ of the kernel $K$~\fnm{u}\fnt{u}{~For the form of the BFKL 
kernel and details of the techniques 
for solution see Kwiecinski~\cite{Kwiecinski}.}. To leading order in
$\ln(1/x)$), and fixed $\alpha_s$, we obtain the very steep power law behaviour
{\small \begin{equation}
 xg(x,Q^2) \sim f(Q^2)\ x^{-\lambda_L}
\label{eq:BFKLsol}
\end{equation}}
\noindent 
where 
{\small \begin{equation}
 \lambda_L = \frac{3\alpha_s}{\pi} 4\ln2 \simeq 0.5
\label{eq:BFKLlam}
\end{equation}}
\noindent 
(for $\alpha_s \simeq 0.25$, as appropriate at $Q^2 \sim 4\,$GeV$^2$).
This would lead to very steeply rising $V^*N$ cross-sections, corresponding
to a hard Pomeron of intercept $\alpha_P = 1.5$. This hard Pomeron has been
termed the BFKL Pomeron.

This power law behaviour of the BFKL solution could explain data which
is already steeply rising at moderate $Q^2$, without need of a long 
evolution length from $Q^2_0$ to $Q^2$, and hence without need to use 
perturbative QCD at very low $Q^2$. The power $\lambda_L \simeq 0.5$ is too
steep to fit current data, however the simple derivation just sketched should 
be improved in various ways. One must solve the full equation with an
appropriate boundary conditions rather than
consider just the leading eigenvalue. 
Then one must consider incorporating the running of $\alpha_s$ 
with $Q^2$ (or more appropriately $k_T^2$) and one must consider the 
upper(UV) 
and lower(IR) cut-offs on the $k_T^2$ integration in Eq.~\ref{eq:BFKL} 
and finally
one must bear in mind that the kernel of the equation has only been 
completely calculated to leading order in $\ln(1/x)$ ($LL(1/x)$). 
All component parts for next-to-leading order ($NLL(1/x)$) calculations are now
complete, but it is not yet clear how to implement 
them~\cite{NLOlnx,doksh_dis97,levin97}. The current phenomenology
of the BFKL equation is discussed in Sec.~\ref{sec:BFKL}. 

\subsubsection{Unifying the DGLAP and BFKL equations}

\begin{figure}[ht]
\centerline{\psfig{figure=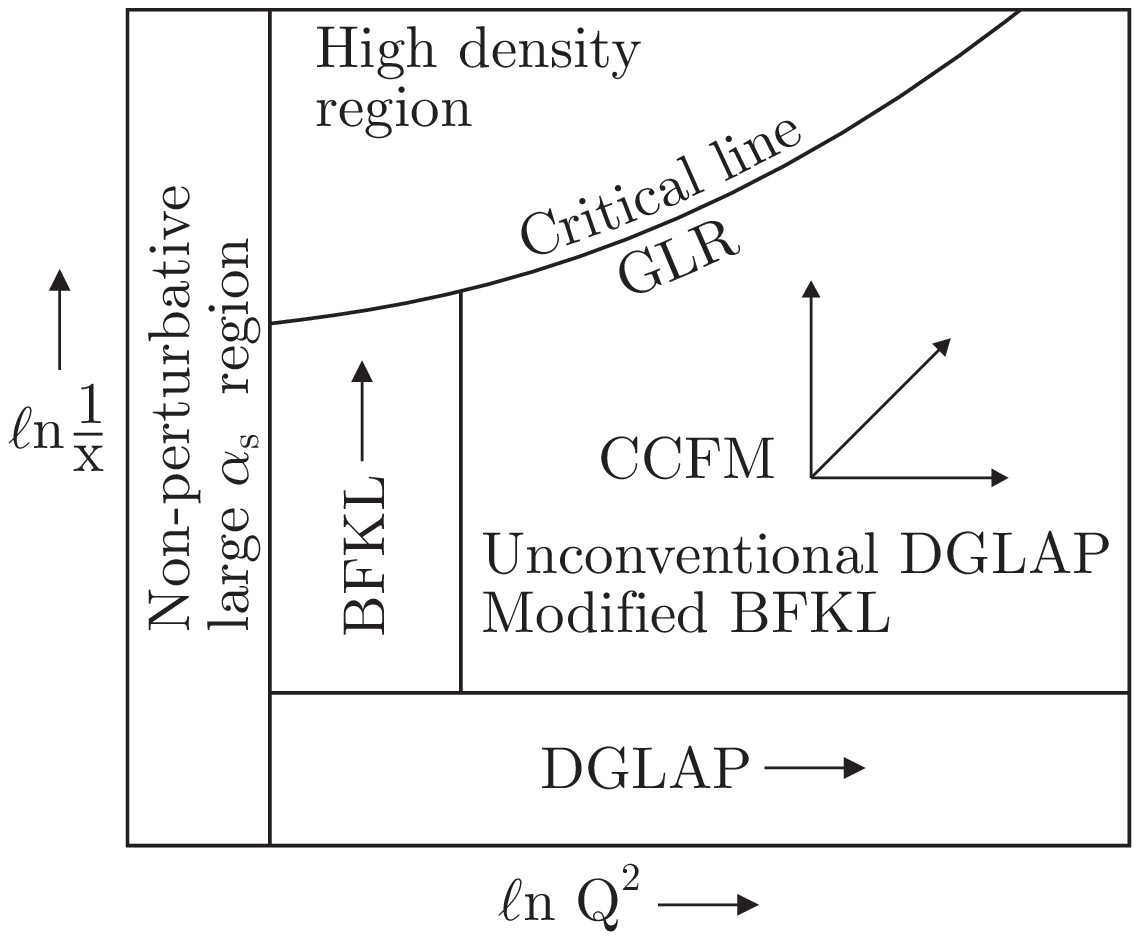}}
\fcaption{Schematic representation of the applicability of various evolution
equations across the $x,Q^2$ plane}
\label{eqplane}
\end{figure}

\begin{figure}[t]
\centerline{\psfig{figure=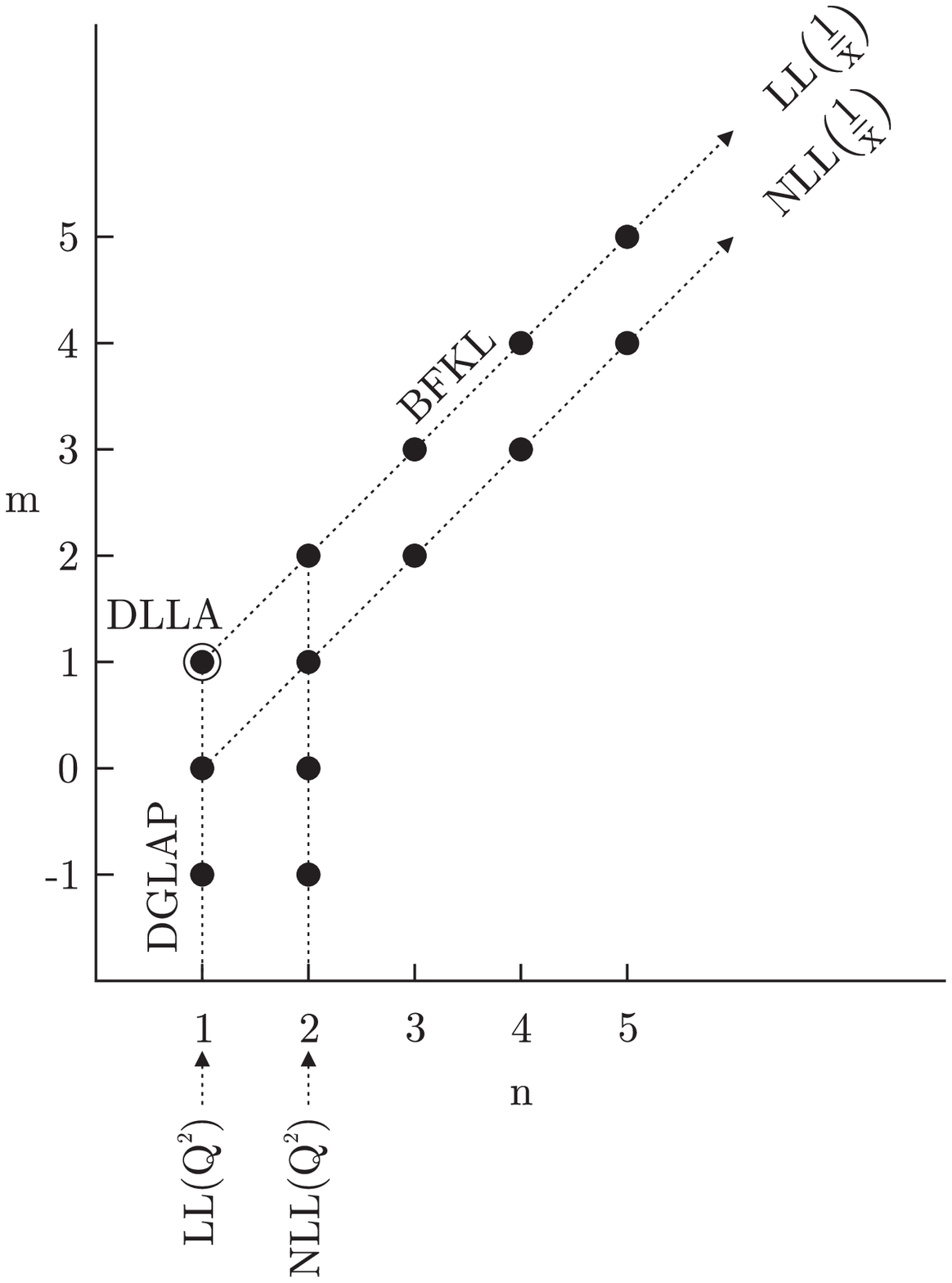,width=7cm,height=9cm}}
\fcaption{The $n,m$ plane with the BFKL and DGLAP summations indicated}
\label{Dick}
\end{figure}

Whereas the conventional DGLAP equations deal with $Q^2$
evolution and may be inadequate at low $x$, the BFKL equation deals with $1/x$
evolution and may be inadequate at high $Q^2$. Fig.~\ref{eqplane} gives a 
diagram of the applicability of various evolution equations across the $x,Q^2$ 
plane.
One would like to have
an approach which can be used throughout the kinematic plane. 
One possible way is to attempt to incorporate extra terms in
$\ln(1/x)$ into the DGLAP equations (`Unconventional' DGLAP). This 
amounts to recalculating the splitting functions and coefficient functions 
to `resum' all $\ln(1/x)$
contributions. If we rewrite Eq.~\ref{eq:DGLAPg} for small $x$, where the
gluon is dominant, as follows
{\small \begin{equation}
\frac{d g(x,Q^2)}{d \ln Q^2} = \int^1_x \frac{dy}{y} P_{gg}(\frac{x}{y},
\alpha_s)\ g(y,Q^2)
\end{equation}}
{\small \begin{equation}
 P(x,\alpha_s) = P^0(x)\ \alpha_s(Q^2) + P^1(x)\ \alpha_s^2(Q^2) +\ldots
\end{equation}}
\noindent
in order to bring out the dependence on $x$ and $Q^2$, then at small $x$ 
we have
{\small \begin{equation}
 P^n(x) = \frac{1}{x} \left[ a_n \ln^n(\frac{1}{x}) + b_n \ln^{n-1}
(\frac{1}{x}) +\ldots\right]
\end{equation}}
\noindent
so that in general
{\small \begin{equation}
xP(x,\alpha_s) = \Sigma_{n=1}^{\infty}\ \Sigma^n_{m=1}
\ A_{nm}\ \alpha_s^n(Q^2)\ \ln^{m-1}(\frac{1}{x})
\end{equation}}
\noindent
and the coefficient functions may be expanded similarly. It is convenient to
write this in terms the anomalous dimensions, as follows
{\small \begin{equation}
\gamma^N(\alpha_s) = \Sigma_{n=1}^{\infty}\ \Sigma^n_{m-\infty}
A_{nm}\ \alpha_s^n\ N^{-m}
\label{eq:resum}
\end{equation}}
\noindent
where the sum over $m$ extends to negative values to represent the 
contribution of the terms which are non-singular as $x \to 0$, and $N$ is the
moment index (see Sec.~\ref{sec:moments}).
Ideally we should sum the whole of $n,m$ space, but in practice we 
make different approximations as to which terms are most important. 
Conventionally we consider terms which are leading in $\ln Q^2$ so we 
sum Eq.~\ref{eq:resum} over $m$ for $n = 1$ in the LLA, and for 
$n = 2$ as well in the NLLA. 
However one could re-order the sum to first sum over terms with
a leading $\ln(1/x)$ ($LL(1/x)$). These are the terms for which $n = m$ and this
approximation gives the BFKL equation at $LL(1/x)$. Terms with $m = n-1$ give
the $NLL(1/x)$ approximation. The common point $n = m = 1$
represents the DLLA. The $n,m$ plane and these differing summations are 
illustrated in Fig.~\ref{Dick}.
\noindent

We are defining different expansion schemes to sum up the contributions of 
different terms to the evolution of the parton distributions, and eventually to
the $\gamma^*-parton$ cross-section. The conventional large $x$ scheme, where
$\ln Q^2$ is considered leading, sums up all logs of the form
{\small \begin{equation}
\alpha_s^p (\ln Q^2)^q (\ln\frac{1}{x})^r
\label{eq:pqr}
\end{equation}}
\noindent
for which $p = q \geq r \geq 0$, at LO. 
At NLO terms for which $p = q + 1 \geq r \geq 0$ are also summed. In
the small $x$ scheme, where  $\ln(1/x)$ is
considered leading, terms for which $p = r \geq q \geq 1$ are summed 
at $LL(1/x)$, and terms for which $p = r + 1 \geq q \geq 1$ are also summed 
at $NLL(1/x)$. This small 
$x$ scheme can be used to supplement the conventional splitting functions 
(and coefficient functions) in order to include $\ln(1/x)$ higher order terms 
appropriately at small $x$ within the framework of the DGLAP equations.
The current phenomenology of this approach
is discussed in Sec.~\ref{sec:resum}

Another way to approach the need for a solution applicable across the whole
kinematic plane is to develop alternative evolution equations which sum more
general classes of diagrams. The CCFM 
equation~\cite{CCFM,Webber} is the best established of such equations.
It is based on the idea of coherent gluon
radiation, which leads to angular ordering of gluon emissions in the gluon 
ladder such that $\theta_i > \theta_{i-1}$, where $\theta_i$ is the angle 
that the $i$th gluon makes to the original direction 
(see Fig.~\ref{ladder}).
Outside this angular region there is destructive interference such that 
multi-gluon contributions vanish to leading order. Angular ordering implies
ordering in the transverse momenta $k_T$ divided by the energies $E$ 
of the gluons on the rungs of the gluon ladder.
An additional scale is necessary to specify the
maximum angle of gluon emission. This extra scale can be taken to be the scale
$Q$ of the probe, such that we deal with a scale dependent unintegrated gluon 
density $f(x,k_T^2,Q^2)$. The integral equation for $f(x,k_T^2,Q^2)$ can be 
approximated by the BFKL equation at small $x$, where $f$ becomes independent
 of $Q^2$ and ordering in $k_T/E$ does not imply $k_T$ ordering. 
However, at moderate $x$, ordering in $k_T$ is implied and one
 recovers the DGLAP equation for the integrated gluon
distribution $g(x,Q^2)$. The CCFM equation takes into account some of the 
effects of the 
next-to-leading order $\ln(1/x)$ terms which have yet to be implemented in the
BFKL equation~\cite{KMS,KMS96,Bottazzi}. 

Li~\cite{Li2,Li3} 
has taken an alternative approach to the unification of the DGLAP
 and BFKL equations. He proposes a modified BFKL equation which
contains an intrinsic $Q^2$ dependence which does not come from a $\ln Q^2$
summation but from consideration of the phase space boundary for radiative
corrections. 

Kwiecinski, Martin and Stasto~\cite{KMStas} have developed a different 
modified BFKL equation
which incorporates both $\ln(1/x)$ and $\ln Q^2$ resummation, such that the
equation maybe used over all $x$ and $Q^2 \geqsim 1\,$GeV$^2$. 
A major uncertainty in the solution of the BFKL equation comes from the 
treatment of the infra-red region. The modified equation 
improves on this by only considering $f(x,k^2_T)$ in the perturbative domain.

The CCFM equation and these modified BFKL equations may be preferable to
the approach which incorporates $\ln(1/x)$ resummation within DGLAP since they
all deal with an unintegrated gluon density and hence preserve more of the
physics of the non-$k_T$ ordered gluon ladder.
The current phenomenology of  the CCFM equation and of modified BFKL equations
is discussed in Sec.~\ref{sec:CCFM} 

\subsubsection{Non-linear effects and higher twist at low $x$}

\begin{figure}[ht]
\centerline{\psfig{figure=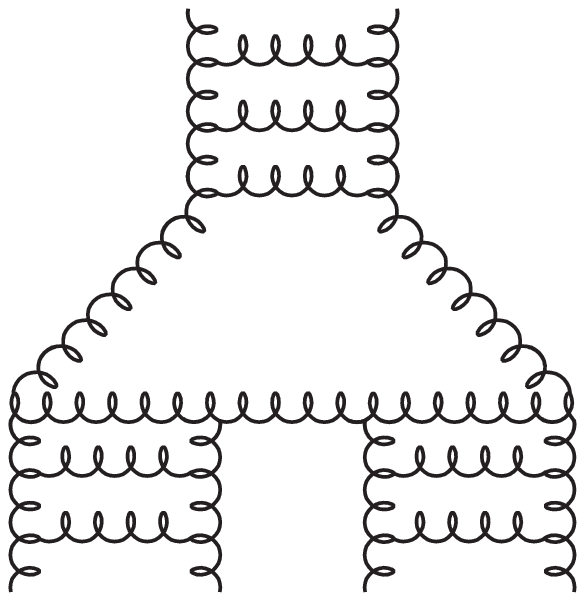,width=5cm,height=4cm}}
\fcaption{An example of a multi-ladder diagram  representing gluon-gluon
interaction}
\label{shadowing}
\end{figure}

Two further, related questions occur when considering the 
implications of a steeply rising gluon density at small $x$.
The first is that a  steeply increasing $V^*N$ cross-section
would eventually violate the Froissart (unitarity) bound.
This bound may, of course, not be applicable in the case of particles 
off-mass shell~\cite{bound}, 
but one can view the problem from a slightly different
perspective which needs consideration independent of such a violation. As
$x \to 0$ the gluon density is becoming very large so that the possibility
of gluon annihilation, or recombination, may compete with the usual
evolution. Another way of expressing this is to say that the gluons 
shadow, or screen, each other from the virtual-boson probe. A possible 
multi-ladder diagram representing such a gluon-gluon interaction is given
in Fig.~\ref{shadowing}. 

\noindent
 
\begin{figure}[ht]
\centerline{\psfig{file=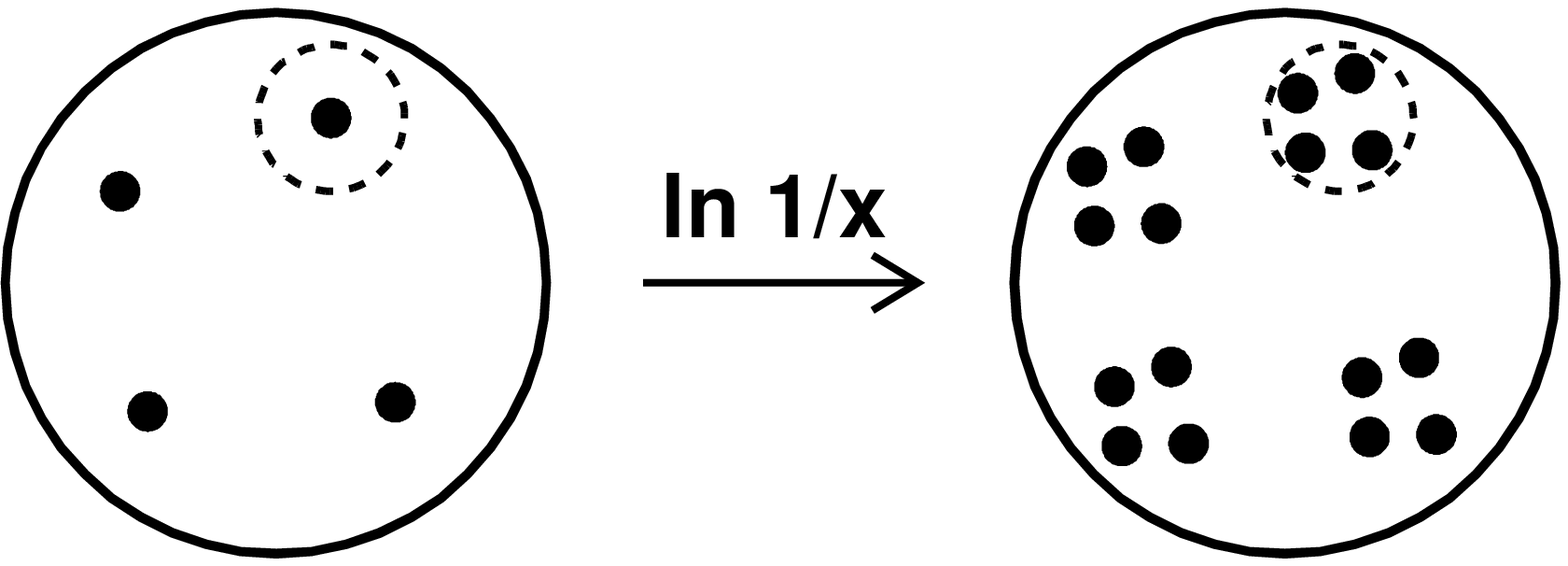,width=8cm,height=4cm}}
\fcaption{Schematic diagram showing the increase in local parton density due
to evolution in $x$, leading to hotspot regions in the proton.}
\label{fig:hotspot}
\end{figure}

An estimate of the size of gluon density necessary for shadowing to
become important may be given as follows.
A measurement at the scale $Q^2$ probes a gluon of transverse size 
$\sim1/Q$, longitudinal size $\sim 1/(xP)$. The number of gluons per unit
rapidity $\ln(1/x)$ which can interact with the probe is $xg(x,Q^2)$, so 
the transverse area which they occupy is $xg(x,Q^2)\ \pi/Q^2$. When this 
is comparable to $\pi R^2$, the transverse area of the nucleon, we should 
expect to get parton shadowing~\cite{Mueller93}, i.e. when
$ xg(x,Q^2) \geqsim R^2 Q^2$. For $R \sim 1$ fm and $Q^2 \sim 10\,$GeV$^2$
we obtain $xg(x,Q^2) \geqsim 250$. However there is a large uncertainty on this
estimate, a more realistic value for the area occupied by the gluons may be
$xg(x,Q^2)\ 4\pi/Q^2$, which would already give shadowing for 
$xg(x,Q^2) \geqsim 60$, at $Q^2 \sim 10\,$GeV$^2$. Such values
of the gluon momentum distribution have not yet been reached even for the
lowest $x$ data at HERA. However the relevant size for $R$ may not be the
nucleon radius, but the radius of a constituent quark ($\sim 0.4$fm) 
if the gluon ladders of Fig.~\ref{shadowing} couple to the same parton 
within the nucleon. In this case shadowing may be expected in HERA data.
Such a scenario is called `hot-spot' shadowing~\cite{hotspots}, see 
Fig.~\ref{fig:hotspot}.
\noindent 

In considering shadowing we are moving towards a new region in the 
$x,Q^2$ plane where the parton density is high (see Fig.~\ref{eqplane})
 and the usual methods of perturbative QCD 
cannot be applied despite the fact that $\alpha_s$ is still quite small. 
We need quantum statistical methods which are in process of development. 
It is useful to define a parameter, 
$ \omega(x,Q^2) = \frac{\alpha_s}{Q^2}\ \rho$, where 
$\rho = \frac{xg(x,Q^2)}{\pi R^2}$
is the gluon density in the transverse plane. Since $\alpha_s/Q^2$ gives 
the cross-section for gluon absorption by a parton in the hadron, $\omega$
represents the probability of gluon recombination (or shadowing).
When $\omega \ll \alpha_s$ shadowing is negligible and we may use linear
evolution equations such as BFKL or DGLAP.  When 
$\omega \gg \alpha_s \leqsim 1$ we have reached the high density region and we
approach the unitarity limit. When $\omega \simeq \alpha_s$ 
shadowing is significant but parton densities are not overwhelmingly 
large. In Fig.~\ref{eqplane} we indicate a `critical line' separating 
this region from 
the very high density region.  (Note that this figure is only schematic and 
the relative positions 
of many of the lines is still in debate, for example Mueller~\cite{Mueller96}
contends that the non-perturbative region extends to higher $Q^2$ as $x$ 
decreases such that it may subsume the high density region (see also 
ref.~\cite{Lev}) and 
Levin~\cite{levin97} contends that the BFKL region may be completely hidden
in the shadowing region).

To the right of this critical line a first attempt to account for 
shadowing was developed in which one adds a non-linear term to the 
usual evolution equations.
The probability of evolution, or parton splitting is
$\propto \alpha_s \rho$, and the probability of recombination is
$\propto \alpha_s^2 \rho^2 / Q^2$, so that the evolution of the gluon 
distribution is given by
{\small \begin{equation}
 \frac{d^2 xg(x,Q^2) }{d\ln Q^2 d\ln(1/x)} = \frac{3 \alpha_s }{\pi}
xg(x,Q^2) - \frac{\alpha_s^2 \gamma}{Q^2 R^2} [xg(x,Q^2)]^2
\end{equation}}
\noindent
where the proportionality factor $\gamma = 81/16$ is calculated from 
Regge theory as the triple Pomeron (or triple gluon ladder) vertex 
(see Fig.~\ref{shadowing}). This equation
is known as the GLR equation~\cite{GLRMQ}.
When $xg(x,Q^2) \sim \pi Q^2 R^2/\alpha_s(Q^2)$ the non-linear 
term cancels the linear term and the 
evolution of $xg(x,Q^2)$ stops. This is known as saturation.
Correspondingly the steep rise in the gluon distribution would flatten off
and the $V^*N$ cross-section would also flatten off, hence there would be 
no violation of the unitarity bound. However we have now moved to the left of 
the critical line and the GLR equation cannot be trusted.
 
There have been some phenomenological attempts to consider shadowing
corrections to both the DGLAP and the BFKL equations~\cite{KMRS,AKMS,GLM} using
this simple picture. Gotsman et al~\cite{GLM} have suggested that such 
corrections are responsible for the reduction of the BFKL slope $\lambda_L$
to match the observed values.
However the GLR equation has been criticized as inadequate 
even in its supposed region of validity. Zhu et al~\cite{whu} have pointed out
that the GLR equation does not conserve the momentum of the gluons and have
proposed a modified GLR equation. Bartels~\cite{multi} has pointed out that 
there may be ladder-ladder interactions before the
recombination and furthermore, considering the recombination of 
just two gluon ladders may not go far enough,  multi-ladder diagrams
may also be important. An extension of the GLR equation to 
account for multi-ladder correlations was developed by Laenen and
Levin~\cite{Laenen} and, more recently, the Geiger-Mueller~\cite{GM} 
approach to shadowing has been developed by Ayala, Gay-Ducati and 
Levin~\cite{AGL} such that results should be applicable closer
to the critical line. This work
has suggested that shadowing effects become significant in the low $x$ 
region before it is necessary to take into account large $\alpha_s \ln(1/x)$ 
terms (see also ref.~\cite{shab}). 

The interpretation of shadowing in terms of the Operator Product
Expansion of operators of different twists is not obvious. A more rigorous
evaluation of higher twist effects at small $x$ is undoubtedly 
necessary. We expect twist-4 gluon operators to be most important at
small $x$~\cite{Bartels}.
The relevant operators are the 2-gluon and 4-gluon operators and these may mix
via a $ 4 \to 2 $ gluon transition vertex. 
Bartels~\cite{newBart} has recently made a phenomenological study of such 
higher twist effects at low $Q^2$. He uses the fact that diffractive vector 
meson production has been established as a twist-4 effect to put a lower limit
on the twist-4 contribution to DIS of a few percent at
$Q^2 \sim 10\,$GeV$^2$, $x \sim 10^{-3}$. The resulting small $x$ behaviour of 
the twist-4 contribution to $F_2$ comes out as negative and its expected 
$1/Q^2$ fall off is compensated by a strong dependence on $Q^2$ of the 
twist-4 scaling violations \fnm{v}\fnt{v}{~In one scenario considered the 
twist-4 contribution has an $x,Q^2$ dependence approximately proportional to
the square of the leading twist contribution to $F_2$. This would imply that 
the usual forms used to describe higher twist contributions in fits to 
structure function data (see Sec.~\ref{sec:hitwist}) are inadequate.}
 such that twist-4 effects may remain important to unexpectedly large $Q^2$.

A new picture of the physics of the BFKL equation has been developed 
by considering the interaction of chains of colour dipoles in impact parameter 
space~\cite{Mueller}. This picture may also enable us to study saturation and
unitarity effects, which occur when the dipole density grows,
in a theoretically consistent manner~\cite{PesSal}.
There have also been studies of the high density region in which 
attempts are made to formulate a Lagrangian for high density 
QCD~\cite{Jalilian}. For example, one may replace 
the complex QCD interaction between partons by the interaction of a parton 
with momentum fraction $x$ with a classical field created by all partons 
with momentum fraction larger than $x$. The DGLAP and BFKL equations
are recovered at the correct low density limits. There are also formal studies
of this region which build on the old idea of Reggeon Calculus, but using
reggeized gluons (i.e. BFKL Pomeron) as the basic object in an effective field
theory~\cite{reggecal}.

Higher twist effects, shadowing, parton saturation and the approach to the 
critical line are also of relevance to the 
second question which arises when considering a steeply rising gluon 
density. Namely, whereas the $V^*N$ cross-section does rise steeply with 
$W^2$ (for small $x$)
the real photon nucleon cross-section does not. We need to 
understand this transition from hard to soft physics, but we will require 
non-perturbative techniques since perturbative QCD cannot be used
when $Q^2$ is small and $\alpha_s$ is correspondingly large. 
Bartels~\cite{newBart} has suggested that the transition may need an 
interplay of higher and leading twist effects. 
Shadowing is another of the possible explanations for the transition. 
The transition region is discussed fully in Sec.~\ref{sec:lowq2}.

\subsubsection{Simple parametrizations}
\label{simple}

We finish this introductory section by mentioning a few simple parametrizations
of $F_2$ at small $x$ which have been inspired by various of the theoretical 
approaches mentioned above, but which find their ultimate justification in the
quality of the fits to the data.

As we have seen, pQCD predicts $F_2 \propto x^{-\lambda}$ at small $x$, where
$\lambda$ is a constant for the naive solution to the BFKL equation 
(Eq.~\ref{eq:BFKLlam}) and for the singular input solution to the DGLAP 
equations (Eq.~\ref{eq:Yndlam}), but $\lambda$ varies with $x$ and $Q^2$ for 
the commonly used non-singular input solution to the DGLAP equations 
in the DLLA 
(Eq.~\ref{eq:DLLAlam}). The latter solution inspired the parametrization
{\small \begin{equation}
 F_2 = n_i x^{-\gamma\sqrt{\frac{T}{\xi}}}
\end{equation}}
of De Roeck, Klein and Naumann~\cite{deRoeck}, where $\xi = \ln(1/x)$ and
$T = \ln(\alpha_s(Q^2_0)/\alpha_s(Q^2))$. This form yields a $\chi^2/ndf$ of
$109/167$ (using correlated systematic errors) 
in a fit to H1(94) data in the kinematic region $x < 0.1$, 
$Q^2 > 1.5\,$GeV$^2$, with two free parameters: $Q^2_0 = 0.365\,$GeV$^2$ and 
$\alpha_s = 0.113$. This parametrization is essentially a simplification
of the 2-loop result of the DAS approach, see Sec.~\ref{sec:DAS}

The simple parametrization
{\small \begin{equation}
 F_2 = a + b \ln(\frac{Q^2}{Q^2_0})\ln(\frac{x_0}{x})
\end{equation}}
has been given by Buchmuller and Haidt~\cite{haidt}, inspired by the search for
a form which would saturate, but not violate, the Froissart bound at very low
$x$. This double logarithmic form also represents the first term of the sum
which produces the non-singular input solution to the DGLAP in the DLLA.
 This form yields a 
$\chi^2/ndf$ of $83/72$ (using uncorrelated systematic errors) 
in a fit to H1(94) data in the kinematic region 
$x < 0.01$, $Q^2 > 5\,$GeV$^2$, with four free parameters 
$Q^2_0 = 0.5\,$GeV$^2$, $x_0 = 0.074$, $a = 0.078$ and $b = 0.364$.
It has proved possible to extend this simple form to describe very low $Q^2$
data~\cite{haidt2}, by making the substitution 
$\ln(Q^2/Q^2_0) \to \ln(1 + Q^2/Q^2_0)$. Both H1(94)  data and H1(95)SVX 
data at very low $Q^2$ ($Q^2$ down to $0.35\,$GeV$^2$) can be well 
described in a fit to all data for which $x < 0.005$. The parameter values are
 $Q^2_0 = 0.55\,$GeV$^2$, $x_0 = 0.04$, with intercept $a \simeq 0$, slope 
$b \simeq 0.45$.   

De Roeck and De Wolf~\cite{deRdeW} have given the parametrization
{\small \begin{equation}
 F_2 = C_0 \Gamma(\delta) (\frac{z}{2})^{1-\delta} I_{\delta+1}(z)
\end{equation}}
where $z = 2 (\frac{12}{\beta_0} \ln(\frac{1}{x}) T^a)^{\frac{1}{2}}$, $\Gamma$
is the Gamma function 
and  $I_{\delta+1}(z)$ is a Bessel function of order $\delta +1$,
where $\delta = (11 + 2n_i/27)/\beta_0$. This parametrization is inspired by 
the similarity between the $x$ dependence of $F_2$ at small $x$ and the 
$\sqrt{s}$ dependence of the average charged multiplicity in $e^+ e^-$ 
collisions. A fit with $\chi^2/ndf = 265/231$ (using correlated systematic
errors) to both H1(94) and ZEUS(94) data in the kinematic region
$x < 0.05$, $5 < Q^2 < 250\,$GeV$^2$ has been obtained, with values of the
two free parameters: $C_0 = 0.389$, $a = 0.708$, ($Q^2 = 1\,$GeV$^2$ and 
$\alpha_s = 0.113$ are fixed). 
The success of the fit suggests that both deep inelastic scattering at small 
$x$ and $e^+e^-$ annihilation can be adequately described by
angular ordered QCD radiation in an essentially free phase space.

\subsection{Phenomenology: conventional DGLAP} 
\label{sec:conv}

The success of the MRSR, CTEQ4 and GRV94 parametrizations already  
 indicates that purely conventional DGLAP
evolution summing only the leading (and next-to-leading) $\ln Q^2$ terms is 
adequate to describe the data down to $Q^2 = 1.5\,$GeV$^2$. 
This is surprising. When the HERA(92) data 
first revealed the steep rise of $F_2$ 
at low $x$ in 1993  this was taken as an indication that 
one was moving beyond conventional DGLAP into the region where the BFKL 
equation is applicable. The rise in $F_2$ at the lowest available $Q^2$ value,
$Q^2 \sim 10\,$GeV$^2$, was too steep to have been generated by DGLAP 
evolution from a conventional flat input at 
the conventional starting points for $Q^2$ evolution 
in the MRS and CTEQ parametrizations ($Q^2_0 = 4\,$GeV$^2$ and 
$Q^2_0 = 2.56\,$GeV$^2$ respectively) and the 
GRV approach starting from a much lower input scale was considered 
controversial. The most popular explanation for the rise of $F_2$ at low $x$
was that given by the prediction of the AKMS group~\cite{AKMS};
 $F_2 \sim A x^{-\lambda_L} Q + F_2^{bg}$, where the first term represents a
solution of the BFKL equation obtained by evolving down in $x$ from $x=0.01$, 
and the second term represents a soft non-perturbative term which describes 
larger $x$ data. Such solutions will be discussed further in 
Sec.~\ref{sec:BFKL}

The MRS team had anticipated a need for non-conventional input
by including a term $x^{-\lambda}$ in 
the gluon and sea distributions input at $Q^2_0$, in the MRSD 
parametrizations~\cite{MRSD}. 
The Regge prediction, $\lambda = 0$, from the soft Pomeron, 
represents the conventional input, but the alternative value, $\lambda = 0.5$,
from the BFKL Pomeron, was also 
tried. The data preferred a value $\lambda \simeq 0.3$, and this was initially
taken as evidence for non-conventional BFKL evolution, since the prediction of
$\lambda = 0.5$ is rather naive, as we have already 
mentioned\fnm{w}\fnt{w}{~Of course if it is really necessary to
use the BFKL equation, one cannot use  parametrizations like MRS or CTEQ
(which are based on the DGLAP equations) at all. However 
Kwiecinski et al~\cite{KMRS} suggested that such a BFKL inspired input 
could represent a first approximation to the correct BFKL treatment.}.

However the need for unconventional behaviour was
rapidly challenged by the following considerations. 
Firstly, the data are only too steep to be generated from a flat input if 
the input scale is as high as $Q^2_0 \simeq 4\,$GeV$^2$. If one lowers the 
input scale to $Q^2_0 = 1\,$GeV$^2$, one can fit the data with conventional
DGLAP evolution and an input which is quite 
compatible with the conventional Regge soft Pomeron. The work of Ball and 
Forte~\cite{BF}, which we discuss below, has been
influential in establishing this idea, but such an approach had 
already been suggested by the success of the GRV91 parametrization which 
has  a flattish shape ($\lambda \simeq 0.1$) at $Q^2 = 1\,$GeV$^2$. The latest
MRSR fits confirm the success of this idea for all data for which 
$Q^2 > 1.5\,$GeV$^2$. For example, MRSR4 has the flat input 
$\lambda_g = \lambda_S = 0.04$ at $Q^2_0=1\,$GeV$^2$.

 The success of these approaches derives from the DLLA 
prediction of de Rujula et al~\cite{deRujula} given 
in Eqs.~\ref{eq:DLLAg},~\ref{eq:DLLAxi}. 
As we have already remarked, this is not strictly 
compatible with a power law dependence of $F_2 \sim x^{-\lambda}$, 
but it can mimic a power law
behaviour, over a limited region of $x,Q^2$ (see  Eq.~\ref{eq:DLLAlam}). 
Of course the exact value of $\lambda$ predicted for any $x,Q^2$ value 
depends on the chosen input
values of $Q^2_0$ and $\Lambda$, but the variation of $\lambda$ with $Q^2$ 
(and $x$) will always have the same general shape (see Fig.~\ref{roy1b}). 
Thus we see that 
if we make parametrizations of the data which start conventional 
DGLAP evolution from a high value of $Q^2_0$  we will need a steeper
input value of $\lambda$ than if we use a lower
value of $Q^2_0$. Freedom to lower the input scale means that the conventional
approach CAN be successful with a flattish input.

Secondly, even if we wish to maintain the somewhat larger input scale of 
$Q^2_0 = 4\,$GeV$^2$, in order to avoid the non-perturbative region, 
the observation that one needs a steep input $\lambda$
at this scale is not necessarily an indication of 
unconventional behaviour. One has inadequate knowledge of the 
non-perturbative physics  which determines the inputs, and hence one
cannot  exclude this as a reasonable input to conventional DGLAP 
evolution. Thus one should consider conventional explanations based on the
solution of Eq.~\ref{eq:glev} applicable when one has a singular input gluon 
distribution: namely Eq.~\ref{eq:Yndlam}. 
Such a solution was chosen by Yndurain et al~\cite{Yndurainold} 
for their early parametrizations and  recent
work, which is discussed further below, suggests that it can still 
provide a reasonable fit to present data~\cite{Yndnow}. 

\subsubsection{Non-singular input gluon: double asymptotic scaling}
\label{sec:DAS}

The work of Ball and Forte~\cite{BF} on Double Asymptotic Scaling 
seeks to  eliminate any dependence of the conventional DGLAP predictions on 
the form of the non-perturbative input. They define 
two new variables $\sigma$ and $\rho$, as follows
{\small \begin{equation}
\sigma = \sqrt {\ln(x_0/x).\ln(t/t_0)};\ \ \rho = \sqrt{\ln(x_0/x)/\ln(t/t_0)}
\end{equation}} 
\noindent
where $Q^2_0, x_0$ are low $Q^2$, high $x$ starting points.
Taking the leading singularities in the splitting functions,
as appropriate at low $x$ and  high $Q^2$ ($Q^2 > Q^2_0, x < x_0$), they 
derive that the gluon distribution, $g(\sigma,\rho)$, 
should rise exponentially with $\sigma$, for
fixed $\rho$, and should be independent of $\rho$, for fixed $\sigma$, 
provided that asymptotic values of $\sigma$ and $\rho$ are reached. 
In other words,
$\ln(g(\sigma,\rho))/\rho$ should scale in both variables asymptotically, and
this property is termed Double Asymptotic Scaling (DAS). Double Asymptotic
Scaling is a property of conventionally evolved pQCD provided that
the gluon distribution at $Q^2_0$ is NOT very singular. This caveat is the
only remaining dependence of this prediction on the input distribution.
The term `very singular' is specified as $\lambda > \gamma/\rho \geqsim 0.3$, 
where $\gamma = 2\sqrt{N_c/\beta_0}$. 

Of course what we actually measure is $F_2(\sigma,\rho)$ rather than 
$g(\sigma,\rho)$, so these predictions need to be extended. 
Rescaling factors $R_F, R_F'$ are defined as 
{\small \begin{equation}
 R_F' = R\ exp\left[\delta \frac{\sigma}{\rho} + \frac{1}{2} \ln(\sigma) + 
\ln(\frac{\rho}{\gamma})\right];\ \ R_F = R_F'\ exp(-2\gamma \sigma)
\end{equation}}
 where $\delta = (11 + 2 n_i/27)/\beta_0$ and 
$R$ is an arbitrary normalization factor. The prediction for $F_2$ is that 
$\ln(R_F' F_2(\sigma,\rho))$ should rise linearly with $\sigma$ 
for fixed $\rho$.
In the rescaling factors, the term in $\ln(\rho/\gamma)$ takes care of the 
transition from the gluon distribution to $F_2$ and the terms in 
$\delta(\sigma/\rho)$ and $\ln(\sigma)$ take care of calculable 
sub-asymptotic
effects, such that $\ln(R_F'F_2)$ versus $\sigma$ at fixed $\rho$ lies on the
SAME straight line independent of $\rho$. The slope of this straight 
line is predicted as $2\gamma=2.4$, at leading order, for $N_c=3 $ colours 
and 
$n_i=4$ flavours. A measurement of this slope is thus a measurement of the
first coefficient of the QCD $\beta$ function $\beta_0$. The final rescaling
 term,  $exp(-2\gamma\sigma)$, which defines $R_F$, removes this slope  
such that $\ln(R_F F_2)$ exhibits scaling in both $\sigma$ and $\rho$ (DAS).
 
\begin{figure}[ht]

\centerline{\psfig{figure=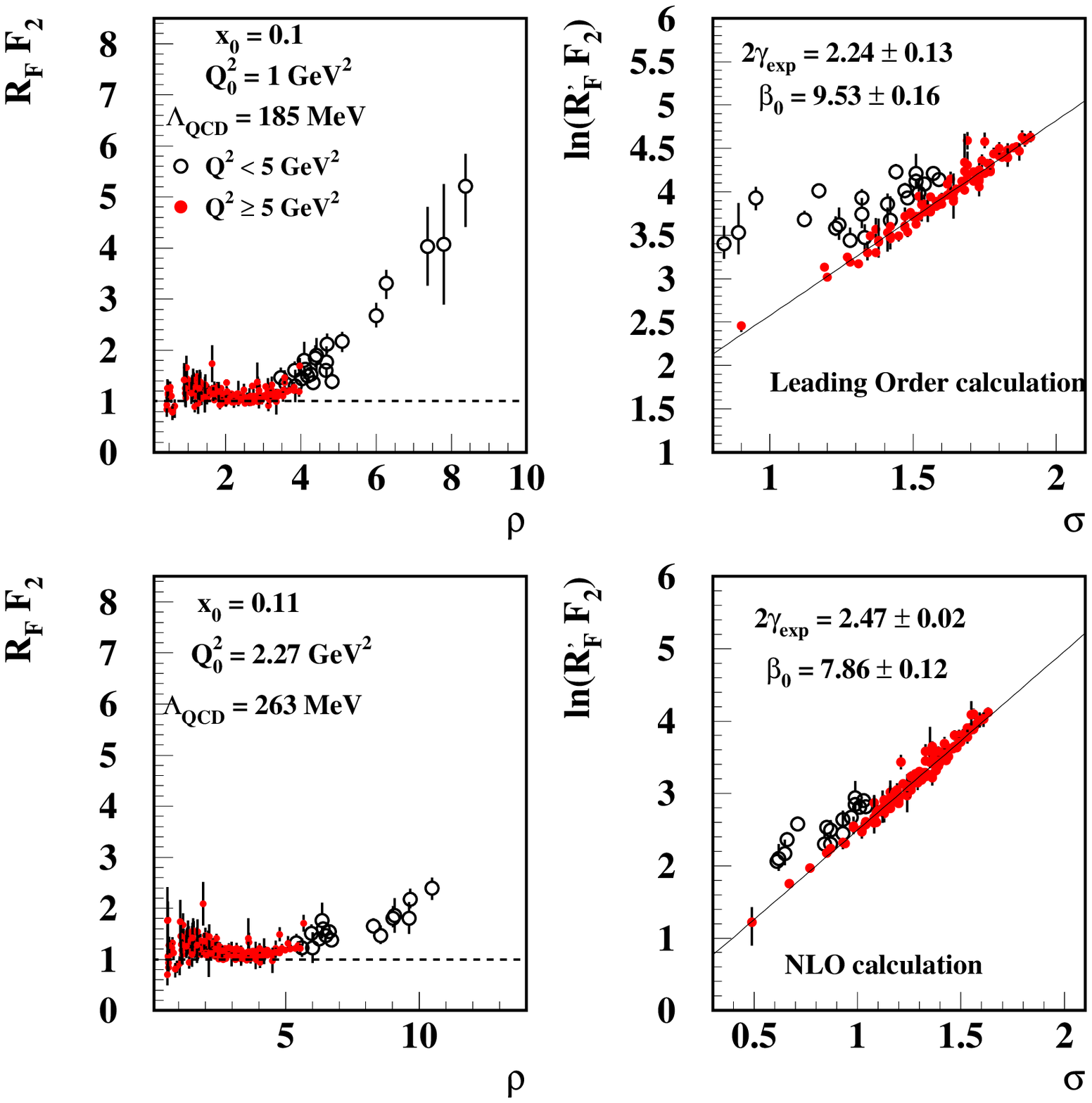,bbllx=10pt,bblly=130pt,bburx=580pt,bbury=690pt,height=11cm}}
\fcaption{The quantities $\ln(R_F' F_2)$ versus $\sigma$ and $R_F F_2$ versus
$\rho$ are illustrated for both LO (top) and NLO (bottom) evaluations. 
The data points are from the ZEUS(94) analysis, points with
$Q^2 < 5\,$GeV$^2$ are indicated separately.}
\label{arnulf}
\end{figure}

The HERA(92) and HERA(93) data confirmed the predictions for a non-singular 
gluon input (and $x_0 = 0.1, Q^2_0 = 1\,$GeV$^2$). 
They exhibit Double Asymptotic Scaling for 
 $\rho$, $\sigma$; $\rho > 1.8$, $\sigma > 1.4$. 
However, the interpretation of the HERA(94) 
data needs a little more care, since these data include points at very
low $Q^2$, $Q^2 < 5\,$GeV$^2$, for which the evolution length from a 
$1\,$GeV$^2$ starting scale may not be sufficient for DAS to be manifest. 
Even if we exclude these low $Q^2$ points, the data
are now sufficiently precise that one can see that leading order DAS is not
perfect, see Fig.~\ref{arnulf}. The slope of $\ln(R_F' F_2)$ versus $\sigma$ is 
somewhat lower than $2\gamma$ and $R_F F_2$ versus $\rho$ is not quite 
flat, it rises as $\rho$ increases.

\noindent

Agreement can be improved somewhat by the inclusion of subleading 
terms~\cite{Mank}, but Ball and Forte find better agreement by extending
their calculation to next-to-leading order~\cite{BFZakopane}. Most of
the NLO correction can be incorporated into a redefinition of the scaling 
variables such that one uses $\ln(\alpha_s(t_o)/\alpha_s(t))$ , rather than
$\ln(t/t_0)$, where $\alpha_s$ is taken at second order. A slight change in the 
definition of the rescaling factor $R_F$ is also necessary
{\small \begin{equation}
R_F^{2-loops} = R_F \left[1 + \frac{\rho}{\gamma}(\epsilon\alpha_s(Q^2) -
\epsilon'\alpha_s(Q^2_0))\right]
\end{equation}}
where  
$\epsilon =\frac{1}{\beta_0\pi}(\frac{103}{27} + \frac{3\beta_1}{\beta_0})$
and $\epsilon' = \epsilon + \frac{78}{\pi\beta_0\gamma^2}$. 
If one plots the data in terms of these new scaling variables one recovers
the prediction that the slope of $\ln(R_F'^{2-loops} F_2)$ versus 
$\sigma$ is equal to
$2\gamma$ (with a somewhat higher input scale $Q^2_0 \sim 2\,$GeV$^2$), but
$\ln(R_F^{2-loops} F_2)$ is still not flat, see Fig.~\ref{arnulf}. 
However, although DAS is not perfect, the shape of $\ln(R_F^{2-loops} F_2)$
CAN be described by a full (conventional) 2-loop calculation at the expense of
reintroducing some dependence on the non-perturbative input. 
Data in the $Q^2$ region, $1.5 < Q^2 < 5\,$GeV$^2$, is also described by 
such a full 2-loop calculation.

The procedure adopted in this NLO fit has already been described in 
Sec.~\ref{sec:alphas}
where it was used to extract $\alpha_s$. A fit to H1(94) data in the $\MSB$ 
scheme produces a $\chi^2$ of $80$ for $169$ data points,
to be compared with the $\chi^2$ values given in Table~\ref{tab:chi2}. 

\subsubsection{Singular input gluon}
\label{sec:ynd}

An alternative view of the predictions of conventional DGLAP is given by 
Yndurain and collaborators~\cite{Yndnow,Barreiro}. 
They use the non-singular input solution of Eq.~\ref{eq:glev}, such that
$F_2 \sim x^{-\lambda}$, with a fixed value of $\lambda$ equal to the
input value, for all $Q^2$.  
The early work of Yndurain et al~\cite{Yndurainold} gave a parametrization of
the structure functions from an approximate 
analytic solution to the DGLAP evolution equations. As we have remarked, an 
exact analytic solution cannot be found for the entire $x,Q^2$ 
plane, but their solution was designed to be
a good approximation in the very large and very small
$x$ regions. More recent work~\cite{Yndnow} has pointed out that the low $x$ 
solution is of relevance to HERA data today. 
At small $x$ they predict (at leading order)
{\small \begin{equation}
 F_2(x,Q^2) \simeq B_S [\alpha_s(Q^2)]^{-d_+(1+\lambda)} x^{-\lambda}
\end{equation}}
\noindent
where the only free parameters are, $B_S$, $\lambda$ and the QCD
scale parameter $\Lambda$, since $d_+$ is specified by the anomalous
dimensions of quark and gluon operators.
This form gives a good fit to the HERA(93) data ($x < 0.01$, $8.5 < Q^2 < 65\,$GeV$^2$) 
with the same sort of values for the
parameters, which were used in the ordinal fits to fixed target DIS data in 
1980, namely $\lambda=0.38$, $B_S = 2.7 \times 10^{-3}$ and $\Lambda
= 200 MeV$ for four flavours.  

When HERA(94) data are considered several further considerations arise. 
One must now fit data across a broad kinematic range.  The form 
$\alpha_s^{-d_+(1+\lambda)}$ ensures that this term grows rapidly
with $Q^2$ producing the strong spiking at low $x$ and high $Q^2$ which is 
seen in the data.  However it also decreases as $Q^2$ decreases and
sub-dominant terms could become important. The dominance of the 
singlet term at low $x$ is only overwhelming if we also have large $Q^2$,
and since the HERA(94) data comprises points at 
$Q^2 \leqsim 5\,$GeV$^2$, it is 
necessary to include the non-singlet contribution to $F_2$.  
The increased precision of the data also requires a fit at next-to-leading
order. Most of the effect can be accounted for by evaluating 
 $\alpha_s$ to second order, but there are some modifications to the 
formulae. The full extended forms are detailed in 
references~\cite{Barreiro,Adel}.
We note that the NLO corrections can be quite large and that the analytic
NLO expressions are not as exact a solution to the NLO DGLAP equations as 
the analytic LO expressions are to the LO DGLAP equations.
Finally, the range of $Q^2$ values is such that one is crossing flavour 
thresholds. The solutions are only meant to hold with respect to a 
constant number of flavours. If separate fits are made in different 
$Q^2$ regions: $n_i = 3, Q^2 < 10\,$GeV$^2$:
\ $n_i=4,\ 10 < Q^2 < 100\,$GeV$^2$: $\ n_i=5,\ Q^2 > 100\,$GeV$^2$, then
the $\chi^2$ values are 
$112/181$ for H1(94) data and $202/175$ for ZEUS(94) data, and the $\lambda$ 
values
are $\lambda = 0.29, 0.33, 0.35$ respectively, for the different $Q^2$ ranges.

\begin{figure}[ht]

\centerline{\psfig{figure=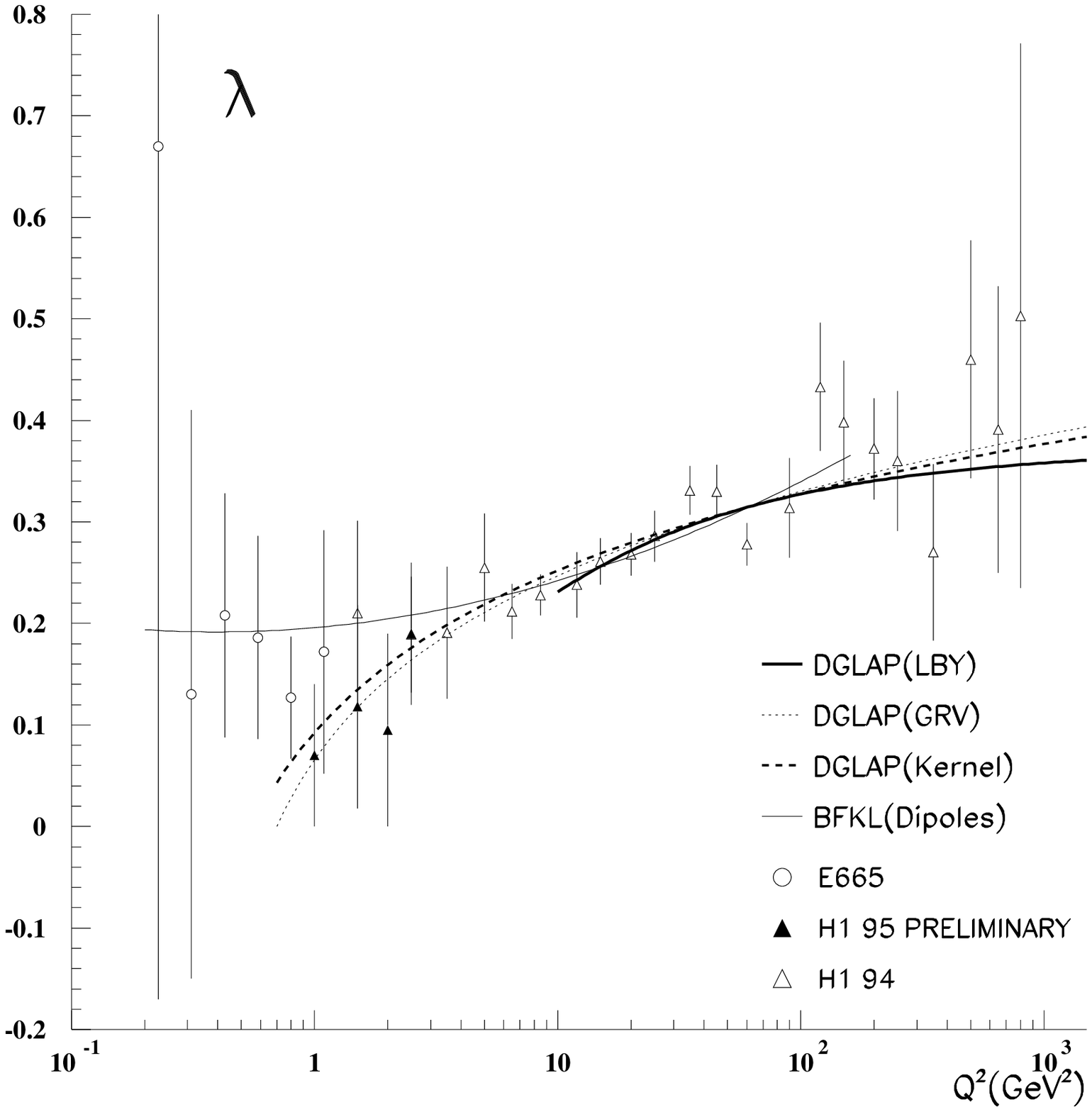,width=9cm,height=9cm}}
\fcaption{The quantity $\lambda = \frac{d\ln F_2}{d\ln(1/x)}$ as a function of 
$Q^2$ averaged over the $x$ range $ 10^{-4} < x < 10^{-2}$, as evaluated from 
the DGLAP kernels, the parametrization of GRV, the parametrization of Lopez,
Barreiro and Yndurain and from the BFKL dipole approach. Data from H1(94) 
and E665 are also shown for comparison.}
\label{roy2}
\end{figure}

In principle, a picture which predicts constant $\lambda$ 
changing only at flavour thresholds should be experimentally 
distinguishable from the non-singular input prediction of a smooth variation
of $\lambda$ with $Q^2$ (and $x$). However one must account for the way
in which $\lambda$
can actually be measured. A fit of the form $x^{-\lambda}$ is made to data
in the region $10^{-4} < x < 10^{-2}$ for each $Q^2$ bin. Such an average 
 $\lambda$ can also be evaluated for the theoretical predictions.
This comparison is illustrated in Fig.~\ref{roy2}
 where one can see that in practice
the predictions for singular and non-singular inputs are hard to 
distinguish. 

\noindent

The fact that a singular input, $F_2 \sim x^{-\lambda}$, $\lambda \geqsim 0.3$,
is not compatible with the conventional  soft Pomeron
is not a problem for the present approach because it should be taken
together with an alternative model of total hadronic cross-sections, 
which was also suggested many years ago~\cite{LYCross}, in which there is 
always a contribution from hard scattering. This work has recently been
updated~\cite{Adel2} to provide a consistent picture of the transition region
from DIS to photo-production, $Q^2 = 8.5 \to 0\,$GeV$^2$, which we discuss in
Sec.~\ref{sec:lowq2}. 
This approach also suggests that the higher $Q^2$ data ,
$Q^2 > 12\,$GeV$^2$, is best fitted by a combination of singular and 
non-singular inputs (corresponding to a soft and hard Pomeron respectively)
at $Q^2_0 \sim 2\,$GeV$^2$, each of which evolves in $Q^2$ in its own 
characteristic way: i.e. 
{\small \begin{equation}
 F_2(x,Q^2_0) = ( B_S x^{-\lambda} + C_S) (1 - x)^\nu
\label{eq:sns}
\end{equation}}
\noindent 
where $B_S$ and $C_S$ are both constants and the larger $x$ data require the
$(1-x)^\nu$ term. The $B_S$ term evolves to become 
$B_S \alpha_s^{-d_+(1 +\lambda_0)}$ (at LO) and the $C_S$ term 
evolves according to the prediction of de Rujula et al. 
A fit of this combined approach to ZEUS data gives $\lambda_0 = 0.43$, larger
than the values obtained from fits to the singular term alone, but 
interestingly consistent with the value $\lambda_0 = 0.47$
which is imposed by the condition that $F_2 \to 0$ as $Q^2 \to 0$~\cite{Adel2}.

\subsection{Phenomenology: BFKL and beyond}
\label{sec:BFKL}

Phenomenological attempts to solve the BFKL equation are still
somewhat underdeveloped because of the problems we referred to earlier
 namely: incorporating the running of $\alpha_s$, 
and introducing IR and UV cut-offs on the $k_T^2$ integrations.
 Many authors ~\cite{FHS,McDermott,Bartels}
have restricted themselves to addressing technical aspects 
and there is general agreement that no definitive work can be done
until the consequences of the $NLL(1/x)$ contributions have been fully 
worked out. Progress is being made on this matter
 at the time of writing~\cite{Gamici3}.

 Much work has  been 
done on the cut-offs and on their relationship to running 
$\alpha_s$. These limits  need more careful consideration for BFKL than 
for DGLAP. Firstly because
the lack of $k_T^2$ ordering means that the BFKL solution may diffuse into the
infra-red region where perturbative calculations are unreliable~\cite{cigar}
(given that $\alpha_s$ in the BFKL kernel is allowed to run, 
$\alpha_s \to \alpha_s(k_T^2)$, and $\alpha_s(k_T^2)$ can be very large for 
low $k_T^2$). Secondly
because the DGLAP formulation ensures energy conservation order by order, but
the BFKL formulation does not. Hence one has to impose it by the choice of 
the UV cut-off when working at finite order.
 
Thus the exact value of the power $\lambda_L$ for the BFKL gluon depends on 
the details of the full solution. Criticisms of the BFKL equation based 
purely on the naive value $\lambda_L = \frac{3\alpha_s}{\pi} \ln 2$ are
correspondingly inappropriate.~\cite{Yndnow,ernst} 
 In some early work it was suggested that the power 
$\lambda_L \simeq 0.5$ would be significantly reduced by 
consideration of the cut-offs~\cite{Collins}. Further 
 work~\cite{AKMS,McDermott,FHS} has modified these conclusions. Askew 
et al~\cite{AKMS} find that the effect of the IR cut-off  alters the
normalization of the resulting gluon but 
does not affect its slope very significantly.
Forshaw et al~\cite{FHS}  suggest that the effect of the UV cut-off 
on the predicted value of the slope is not very significant, 
basically because
the running of $\alpha_s$ weights the integrand in the BFKL equation towards
the infra-red region.

However there is another constraint which may be more significant than that of
 the UV cut-off~\cite{KMS96}.
 Consider a link in the gluon chain where the longitudinal momentum
fraction decreases from $x/z$ to $x$ and the transverse momentum $k_T'$
changes to $k_T$, with emission of a gluon of transverse momentum $q_T$. 
We require 
{\small \begin{equation}
k_T^2/z > q_T^2 
\label{eq:const}
\end{equation}}
\noindent
in order that the virtuality of exchanged 
gluons is controlled by their transverse momenta. This implies that
$k_T^2/z > k_T'^2$, for any given value of $k_T$, and this is a much 
stronger constraint than that due to energy momentum conservation, which
merely imposes, $Q^2/x \sim W^2 > k_T'^2$ , provided that $Q^2 > k_T^2$.
If one considers the effect of this constraint on the solutions for the BFKL
equation one finds that it is very significant for fixed 
$\alpha_s$. It modifies the  asymptotic solution $x^{-\lambda_L}$,
  such that $\lambda_L$ is a far less steep 
function of $\alpha_s$. The usual value quoted $\lambda_L \sim 0.5$ is reduced
to $\lambda_L \sim 0.3$. However, if one considers the case when $\alpha_s$ 
is running, the effect is less extreme because of the weighting of the
integrand towards the infra-red.  
The application of this constraint goes some way towards accounting for 
$NLL(1/x)$ effects. We consider this constraint further
in Sec.~\ref{sec:CCFM} where it is applied to the CCFM and modified BFKL
equations.

Recent work has concentrated on the consequences of incorporating running
$\alpha_s$ into the BFKL equation, and the way in which it exacerbates the 
problem of drift into the non-perturbative infra-red region, such that the
full solution to the BFKL equation may depend crucially on the IR boundary
conditions and be essentially determined by non-perturbative 
physics~\cite{Ross,Gamici}. This implies that it may be essential to include
higher twist terms in our analysis of low $x$ physics as discussed by
Bartels~\cite{newBart}. Indeed, if the transition to low $Q^2$ needs
an interplay of higher and leading twist effects it is also reasonable to
assert that higher twist terms must be non-negligible just above the 
transition region and thus that the success of leading twist DGLAP down to 
very low $Q^2$ is misleading.
Furthermore Mueller~\cite{Mueller96} (see also Lev~\cite{Lev})
has recently warned that
leading twist pQCD may not be usable to very low $x$ unless $Q^2$ is
larger than the mean parton transverse momentum. The OPE may breakdown 
because of diffusion into the infra-red in the non-$k_T^2$ ordered gluon 
ladder and this only gets worse if $\alpha_s$ is running. 
 
Less pessimistically Camici and Ciafaloni~\cite{Gamici} consider that one may
use leading twist pQCD provided that $\alpha_s \ln(1/x)$ is not too 
large. They have begun the work of 
putting together the $NLL(1/x)$ corrections to the BFKL equation in order to
understand running coupling effects~\cite{Gamici3} 
and they conclude that these corrections go in the direction of 
softening the BFKL Pomeron~\cite{ciafsep}. 
Recent work by several other authors points
in the same direction~\cite{Haakman,BF97,doksh_dis97,levin97}. 

\subsubsection{Phenomenology of the BFKL equation}

\begin{figure}[ht]
\centerline{\psfig{figure=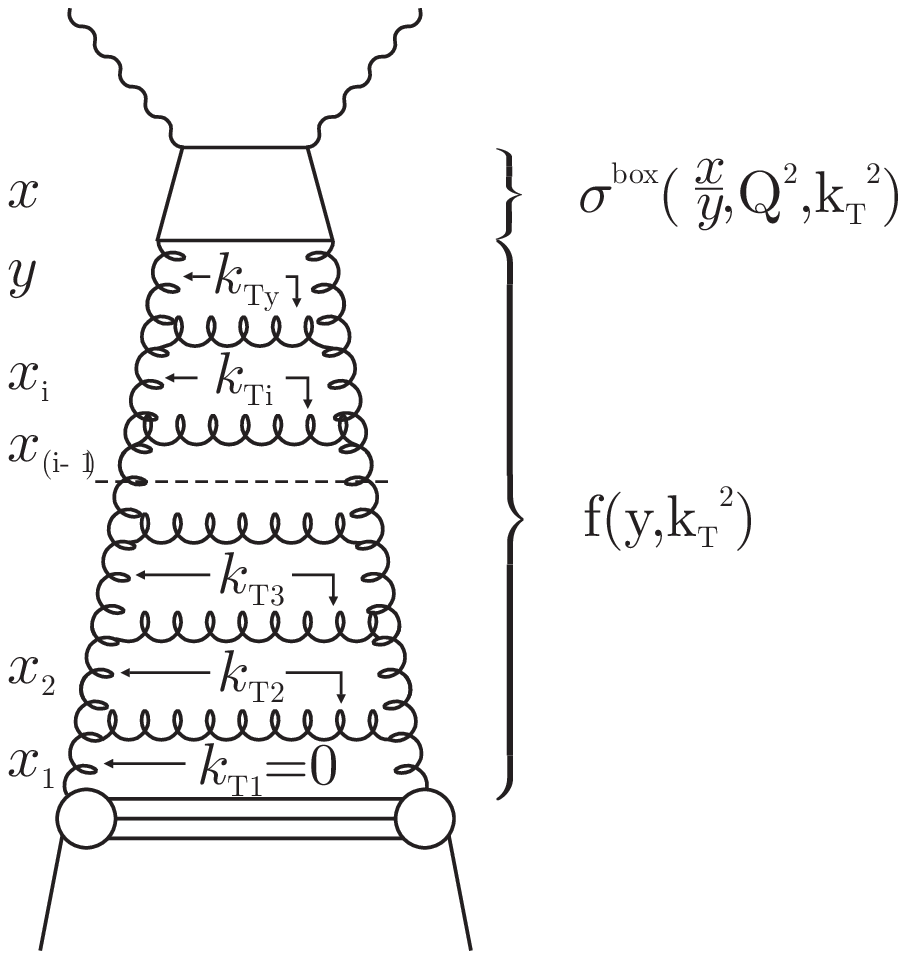,width=7cm,height=7cm}}
\fcaption{Schematic representation of the convolution of the gluon ladder 
with the quark box}
\label{ktladder}
\end{figure}

The practical application of the BFKL equation to predict $F_2$ requires 
convolution of the gluon ladder with the quark box which connects it to the
vector-boson probe, see Fig.~\ref{ktladder}. 
This is done using the $k_T$ factorization formula, which is a generalization 
of collinear factorization,
{\small \begin{equation}
F(x,Q^2) = \int^1_x \frac{dy}{y} \int \frac{dk_T^2}{k_T^2} f(y,k_T^2)\ 
\sigma^{box}(\frac{x}{y},k_T^2,Q^2)
\label{eq:ktfa}
\end{equation}}
\noindent
where $f(y,k_T^2)$ is the unintegrated gluon density at the top of the gluon 
ladder and $\sigma^{box}$ denotes the quark box (and crossed box) contributions
to the boson-gluon subprocess. This convolution requires further consideration
of the limits on the $k_T$ integration. One expects the behaviour of the
unintegrated gluon distribution,
$f(x,k_T^2) \simeq x^{-\lambda_L}$, to feed through into the structure 
function. However, the hard physics which this embodies should be added to a 
background of 
conventional soft processes, which must be there to describe the behaviour 
of $F_2$ at larger $x$. Phenomenologically this gives a description of $F_2$
like that suggested by Eq.~\ref{eq:sns}. 
 The combination of these two terms results in a 
smaller effective power of $\lambda$ for $F_2 \sim x^{-\lambda}$, 
in the region of the present HERA data, than that coming from the BFKL
contribution alone~\cite{AKMS,White}.

An early phenomenological attempt to solve the BFKL equation incorporating
many of the above considerations was given by 
the AKMS~\cite{AKMS} group. 
This picture was successfully confronted with the early HERA 
data~\cite{AKMSGolec}. However it is unable to describe present HERA data
which extend to much lower $Q^2$, since the effective 
$\lambda$ values necessary to describe data with  $Q^2 < 10\,$GeV$^2$ are 
decreasing and the BFKL slope calculated by AKMS 
is simply too steep~\cite{ernst2}. 
A modified BFKL equation which addresses the deficiencies
of the AKMS approach has recently been developed and we discuss this in 
Sec.~\ref{sec:CCFM}.

The colour dipole formulation of the BFKL equation~\cite{Mueller} can be
used to give a phenomenologically successful description of low $x$ $F_2$ data
One considers a $q\bar q$ (onium) state, within the proton, as a colour dipole.
Since the valence quarks of the proton are irrelevant at small $x$, it is
a reasonable physical picture to consider only the onia in the proton, these
represent both sea quarks and gluons since
the wave function of the onium state may evolve as it emits one, or more, soft
gluons. If one pictures this development in impact parameter space, the
original dipole of size $b$ becomes two colour dipoles $qg$ and $g\bar q$ of
smaller size. Each of these dipoles can then branch independently, leading
to a cascade of dipoles as $x$ gets smaller, explaining the rise in the
number of dipoles (or gluons) at small $x$.  To reveal the properties of the
gluon distribution so generated, we must consider an interaction which probes
the onium state. Theoretically it is simplest to consider onium-onium
scattering. One derives a cross-section
{\small \begin{equation}
\sigma(b,b',Y) = \frac{8\pi\alpha_s^2 b b'}{\sqrt{\pi k Y}} exp(\lambda_L Y -
\ln^2(b'/b)/(kY))
\end{equation}}
\noindent
where $b,b'$ are the sizes of the two onia, $Y = \ln(1/x)$ and 
$k = \alpha_s N_c 14\zeta(3)/\pi$ (where the $\zeta$ function gives 
$\zeta(3) = 1.202$). One notes that the high energy (small $x$) behaviour of
the BFKL Pomeron $x^{-\lambda_L}$ is reproduced, and also that the dependence
on the ratio $(b'/b)$ will reproduce the dynamics of the BFKL equation such 
that there is diffusion in $k_T$~\cite{PesSal}.

In order to relate this to the structure functions in DIS we must consider
a photon probe, of transverse size $1/Q$. One first derives an expression for
$F_2$ of an onium state of size $b$, and one then convolutes this with the 
probability to find a dipole of this size in the proton, which is essentially
a non-perturbative function. However, since $b$ can be
regarded as a factorization scale, the final result must be independent of $b$
so that we may set the $b$ dependence of this non-perturbative function to
cancel out that in $F_2^{onium}$. For the detailed formalism see 
reference~\cite{Peschanski}, a simple representation of the solution is 
given by 
{\small \begin{equation}
 F_2 = C a^{\frac{1}{2}} x^{-\lambda_L} \frac{Q}{Q_0} exp\left(
-\frac{a}{2}\ln^2(\frac{Q}{Q_0})\right)
\end{equation}}
where $a = (\alpha_s N_c 7\zeta(3) \ln(1/x)/\pi ) ^{-1}$, $C$ is the
normalization of the non-perturbative function and $Q_0$ is the 
non-perturbative scale ($Q_0 \gg 1/b$). 

Navelet et al~\cite{Peschanski} have used this form to explain the rise of 
$F_2$ at small $x$ seen in the HERA data. They restrict themselves to 
 low $x$ ($x < 5\times 10^{-2}$) and moderate $Q^2$ 
($ 1.5 < Q^2 < 150\,$GeV$^2$) 
data since they are not concerned with conventional physics at larger $x$, and
the BFKL equation does not evolve with $Q^2$. They fit H1(94) data with 3
parameters: the non-perturbative scale $Q_0$, the normalization of the
non-perturbative function which gives the probability of finding an onium
in the proton, and the slope $\lambda_L$. They obtain a $\chi^2$ of 101 for 
130 data points, giving parameter values $Q_0 =0.63$ GeV,  for the 
non-perturbative scale, and $\lambda_L = 0.28$, for the dominant 
contribution to the  steep behaviour of $F_2$ at small $x$. 
Their prediction for the experimentally measurable quantity
$\lambda = d\ln F_2/d\ln(1/x)$ as a function of $Q^2$, averaged over the
$x$ range $10^{-4} < x < 10^{-2}$, 
is illustrated in Fig.~\ref{roy2} along with the measured values,
and the predictions from the DGLAP equations with both singular and 
non-singular inputs. Very accurate data at low $Q^2$ ($1 < Q^2 < 10\,$GeV$^2$)
may be able to  discriminate these theoretical pictures. 

We note that within this picture 
$\lambda_L$ is predicted to be $12\alpha_s \ln2/\pi$ as given in 
Eq.~\ref{eq:BFKLlam},
so that the fitted value of $\lambda = 0.28$ would give an effective value 
of $\alpha_s \simeq 0.11$, which corresponds well with the values derived
from conventional analyses of DIS experiments, at the scale $M_Z^2$. 
However, most
other work on the BFKL equation has assumed that a scale more like
$Q^2 \sim 5\,$GeV$^2$ would be appropriate, so that $\lambda_L \sim 0.5$. The
authors interpret this discrepancy as indicating that the data yield an
effective value of $\lambda_L$ for which the effect of cut-offs, running
$\alpha_s$ and 
$NLL(1/x)$ terms in the BFKL equation are already taken into account.

Alternatively, Nikolaev et al~\cite{Zoller} point out that whereas the 
solution of the BFKL equation
is dominated by the leading eigenvalue, subleading values can be important
non-asymptotically, and introducing a running coupling constant can lead us 
into this region~\cite{Braun}. They have calculated 
the subleading eigenfunctions and singularities of the BFKL equation, 
with a running coupling constant, in the colour 
dipole representation. They describe $F_2(x,Q^2)$ by the simple Regge expansion
{\small \begin{equation}
F_2(x,Q^2) = \sum_n A_n F^n(Q^2)\left(\frac{x_0}{x}\right)^{\lambda_n}
\end{equation}}
\noindent
and find that they can fit data from E665, ZEUS(94) and H1(94), in the
$x,Q^2$ range, $ 10^{-5} < x < 10^{-1}$, $Q^2 = 1\ \to 10^3\,$GeV$^2$, 
with only
3-poles: $\lambda_0 = 0.4,\ \lambda_1 = 0.22,\ \lambda_2 = 0.15$.
The variation $d\ln F_2/d\ln(1/x)$, with $Q^2$ (at small $x$) is also well 
described.

\subsection{Phenomenology: resummation of ln{\boldmath(1/x)} terms}
\label{sec:resum}
 
Much work has been done to establish that the resummation approach
is really equivalent to the BFKL equation at small 
$x$~\cite{Catani,KM,Ciafaloni,Li,Gamici2}. The $LL(1/x)$ solution of the 
BFKL equation (Eqs.~\ref{eq:BFKLsol},~\ref{eq:BFKLlam}) can be rederived
although it is approached only slowly. This approach has further important
features. It introduces running $\alpha_s$ into 
the formalism, and it extends the original BFKL equation into the quark
sector. One is essentially recasting the leading twist part of the BFKL $k_T$
factorization formula into a collinear form with $\ln(1/x)$ terms included.
It will simplify the discussion if we assume that we are working in a scheme
like the DIS scheme, so that we may just consider the modifications to the
splitting functions rather than needing to refer to both splitting functions
and coefficient functions throughout. Most calculations have been
performed in moment space so that it is actually the modified anomalous 
dimensions which are calculated. 

The calculation of the gluon anomalous dimensions rederives the BFKL result
at $LL(1/x)$. However, it turns out that the approach to the steep asymptotic 
behaviour $xg(x)\simeq x^{-\lambda}, \lambda \simeq 0.5$, is rather slow, 
because the gluon anomalous dimension is given by the series
{\small \begin{equation}
 \gamma_{gg,N}(\alpha_s) = \Sigma_{n=1}^{\infty} A_n \left(\frac{\alpha_s}{N}
\right)^n
\end{equation}} 
\noindent
and the coefficients $A_n$ are zero for $n = 2,3,5$  due to strong 
cancellations coming from colour coherence~\cite{Catani95}. Thus, in the 
HERA region, the resummation of $\ln(1/x)$ terms to $LL(1/x)$ in the
gluon anomalous dimension has only a limited impact on predictions for $F_2$.

However, we may also calculate the quark anomalous dimensions with $\ln(1/x)$
terms included. These are zero at leading order $LL(1/x)$. (Recall that at
small $x$ it was only the gluon splitting functions which were becoming 
singular, see Eq.~\ref{eq:smallx}.) 
However the next-to-leading $NLL(1/x)$ contributions to the quark anomalous
dimensions have been fully calculated (it is the $NLL(1/x)$ contributions 
to the gluon anomalous dimensions which are not yet fully understood)
and they are found to be quite significant~\cite{Catani}, 
since all the coefficients $A_n$ in the series 
{\small \begin{equation}
 \gamma_{qg,N}(\alpha_s) = \Sigma_{n=1}^{\infty} A_n \alpha_s\ 
\left(\frac{\alpha_s}{N}\right)^n
\end{equation}}
\noindent 
are positive definite and large. Thus, the splitting function
$P_{qg}$ is much steeper than in conventional DGLAP.
Hence the quark sector, though formally subleading, turns out to have a 
significant impact on the predictions for $F_2$, such that $Q^2$ evolution
from a flat to a steep quark distributions occurs much more quickly. Thus one
may begin evolution at larger $Q^2_0$ values (so that $\alpha_s$ is small 
enough to trust perturbative QCD) and still achieve $F_2$ in agreement with 
HERA data at moderate $Q^2$.

Ellis, Hautmann and Webber~\cite{EHW}  made the first quantitative 
investigations of resummation, giving a fit to HERA(93) data starting from
flat gluon and sea input distributions at a 
scale of $Q^2_0 = 4\,$GeV$^2$. We do not discuss their work in detail since the
HERA(94) data are manifestly not flat at this scale, however 
many of their results have more general relevance. They find 
that a steep behaviour of $F_2$ at low $x$ can be generated after a very short
evolution length (from $Q^2 = 4 \to 8.5\,$GeV$^2$) only if $NLL(1/x)$ 
effects in the quark sector are included, and that the size of this dominant
contribution from the quark sector is very sensitive to the prescription 
chosen to impose momentum conservation. 

Unlike the conventional DGLAP anomalous dimensions, the resummed anomalous
dimensions do not
obey momentum conservation automatically, hence it must be the higher order 
or subleading terms which restore it. Bl\"umlein et al~\cite{BRV} have
used various prescriptions for imposing energy and momentum 
conservation~\fnm{x}\fnt{x}{~Some of their prescriptions are disputed 
by Bojak and Ernst~\cite{ernst2}}
and they conclude that the effects of subleading terms can be very 
significant. These authors have also considered the effects 
of resummation for both singlet and non-singlet partons and for polarized
and unpolarized structure functions. We will only discuss the
unpolarized singlet case here~\fnm{y}\fnt{y}{~Resummation effects for 
unpolarized non-singlet structure functions are estimated to be only at 
the $1\%$ level}.

The input gluon and sea quark distributions are taken to be 
of the form $x^{-0.2}$ at $Q^2_0 = 4\,$GeV$^2$, which gives a 
reasonable fit to HERA(94) data at this low $Q^2$. If a standard 
prescription (prescription $A$ of ref.~\cite{BRV}) 
is used to impose momentum conservation then the effect
of resummation is found to be very large compared to the NLO 
conventional DGLAP calculation. We may quantify this by taking
the ratio of the predictions for the parton distributions 
 when resummation terms are included
to the corresponding predictions including only the conventional NLO terms,
at the point $Q^2 = 10\,$GeV$^2$,\ $x = 10^{-4}$. Including only $LL(1/x)$
resummation one obtains a ratio of $1.3$ for the gluon
distribution and $1.03$ for the sea quark distribution: adding
the $NLL(1/x)$ resummations in the quark sector  changes the sea quark 
ratio dramatically to $3.3$
and consequently the predictions for $F_2$ are strongly increased.

\begin{figure}[ht]
\centerline{\psfig{figure=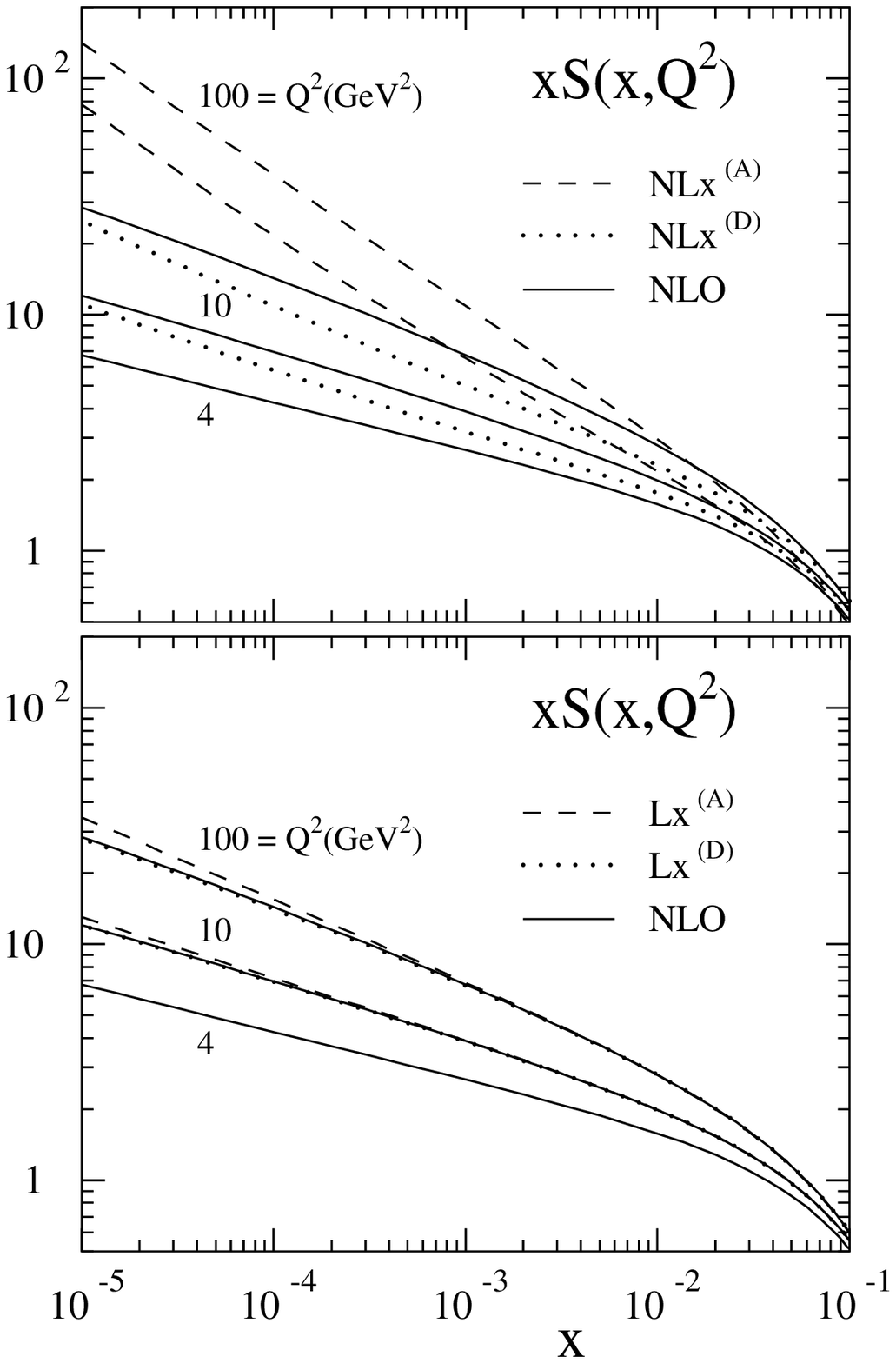,width=6cm,height=10cm}}
\fcaption{Predictions from Bl\"umlein, Riemersma and Vogt 
for the sea quark density as a function of $x$, for 
various different $Q^2$ values, including $LL(1/x)$ resummation in the gluon 
sector (Lx), and $NLL(1/x)$ resummation in the quark sector (NLx), for two 
different methods of implementing energy-momentum conservation labelled as (A) 
and (D)}
\label{vogt}
\end{figure}

However, if a different prescription is used to impose momentum conservation
then the effect of $LL(1/x)$ and $NLL(1/x)$(for quarks) resummation terms is 
completely cancelled by the effective inclusion of higher order terms
which this alternative prescription represents. The final result lies 
slightly below the  conventional NLO prediction. These results are
illustrated for prescriptions $A$ and $D$ of ref.~\cite{BRV} 
in Fig.~\ref{vogt}.
\noindent

Recently these authors have extended this analysis~\cite{BVnew} 
to include the calculation of the $NLL(1/x)$ gluon anomalous dimension 
$\gamma_{gg}$~\cite{NLOlnx}. This has an almost negligible impact on the 
results for $F_2$, but it does soften the predictions for the gluon 
distribution considerably. However, it is not yet clear that the calculations
of $\gamma_{gg}$ are complete to $NLL(1/x)$~\cite{Thorpriv} and the $NLL(1/x)$
expressions for $\gamma_{gq}$ may not have the trivial relationship to 
$\gamma_{gg}$ which obtains at $LL(1/x)$. These questions must be clarified
 before drawing further conclusions~\cite{ballfortemom}.
 
From this discussion we conclude that the conventional BFKL equation, which 
deals only with the gluon sector, may not be very 
important at HERA, but extending the ideas of BFKL, by considering 
$\ln(1/x)$ summation in the quark sector, certainly might be.

\subsubsection{Problems of scheme dependence}

We now consider ambiguities due to scheme dependence.
If we work in renormalization schemes other than the DIS scheme then there is 
freedom in assigning the resummations to the coefficient functions or the
splitting functions according to the renormalization scheme. This choice
does not substantially alter our conclusions. However the choice of 
factorization scheme does. Differences in the speed of 
evolution (in either coefficient functions or splitting functions) can be 
absorbed into a difference in initial conditions, i.e. we have the freedom to
redefine the initial parton distributions~\cite{Ciafaloni,Catani95,Cataninew}.
This is most easily seen by considering that we usually measure only
$F_2$ and its scaling violations $dF_2/d \ln Q^2$. This  does not provide 
enough information to disentangle perturbative from non-perturbative dynamics.
Essentially we have  $F_2 \sim xq$ and $\frac{dF_2}{d \ln Q^2} \sim P_{qg} xg$
(at small $x$), thus steep behaviour of $dF_2/d\ln Q^2$ due to
$\ln(1/x)$ resummation in $P_{qg}$, could be accounted for by a  redefinition 
of the gluon distribution
$xg$, without redefining the quark distribution $xq$ and hence without
affecting the prediction for $F_2$~\cite{Cataninew}.
Redefining the gluon distribution in this way (SDIS scheme) will allow us 
to continue to use conventional DGLAP evolution. This freedom of redefiniton
can only be restricted by knowledge of the 
$NLL(1/x)$ resummation in the gluon anomalous dimensions~\cite{BFscheme} or
by a measurement of the gluon density at small $x$ which does not derive from
the scaling violations of $F_2$, e.g. measurement of $F_L$ of $F_2^{c\bar c}$.
 
This scheme dependence explains why
various authors~\cite{resmmers} have considered resumming $\ln(1/x)$ terms
in calculating the splitting functions and coefficient functions and come 
to somewhat different conclusions as to the importance of such resummation.
There is agreement that resummation effects can be very significant, but it is
unclear that the HERA data require them. One can always modify the input 
distribution shapes and/or the input scale to fit the data. 
Ball and Forte~\cite{BFscheme}
have made fits to HERA data using conventional NLO
DGLAP evolution and compared them with fits including non-conventional
resummation terms, in a variety of renormalization and factorization schemes.
Fits to HERA(93) data with no resummation terms and a long 
evolution length had comparable $\chi^2$ to fits including  resummation terms
with a shorter evolution length. The work of Forshaw et al~\cite{FRT} comes to
a very similar conclusion from a completely different choice of schemes.
This lack of discrimination essentially comes from the fact that, for
$Q^2 \geqsim 10\,$GeV$^2$, resummation corrections affect the predicted size
of $F_2$ much more drastically than its predicted shape. This is no longer 
the case when calculations are extended to lower $Q^2$.

The high precision HERA(94) data, 
including the low $Q^2$, $1.5 < Q^2 < 5\,$GeV$^2$ region, considerably 
reduces the freedom of choice in input shape and input scale.   
Ball and Forte compare the conventional leading $\ln Q^2$ (or large $x$) 
expansion scheme with two other expansion schemes which perform resummation
in slightly different ways: the leading $\ln(1/x)$ (or
 small $x$) scheme and the double leading scheme
in which the two logs are treated symmetrically (in the notation of 
Eq.~\ref{eq:pqr}  all terms
with $p \geq q \geq 1$, $p \geq r \geq 0$ and $q+r \geq p \geq 1$ are summed
at leading order, with an extra power of $\alpha_s$ at next-to-leading order).
The latter two schemes 
are only adequate if $x$ is small, hence perturbative calculations are
made conventionally down to an $x$ value, $x = x_0$, and then these schemes
are used for lower $x$. The NLO fit in the conventional expansion 
scheme and $\MSB$ renormalization/factorization scheme, has already been
described in Sec.~\ref{sec:alphas}. For the non-conventional expansion 
schemes, in addition to the free parameters, 
$\ \lambda_S,\ \lambda_g,\ \alpha_s$, the value of
$x_0$ is also a parameter of the fit.

\begin{figure}[ht]
\centerline{\psfig{figure=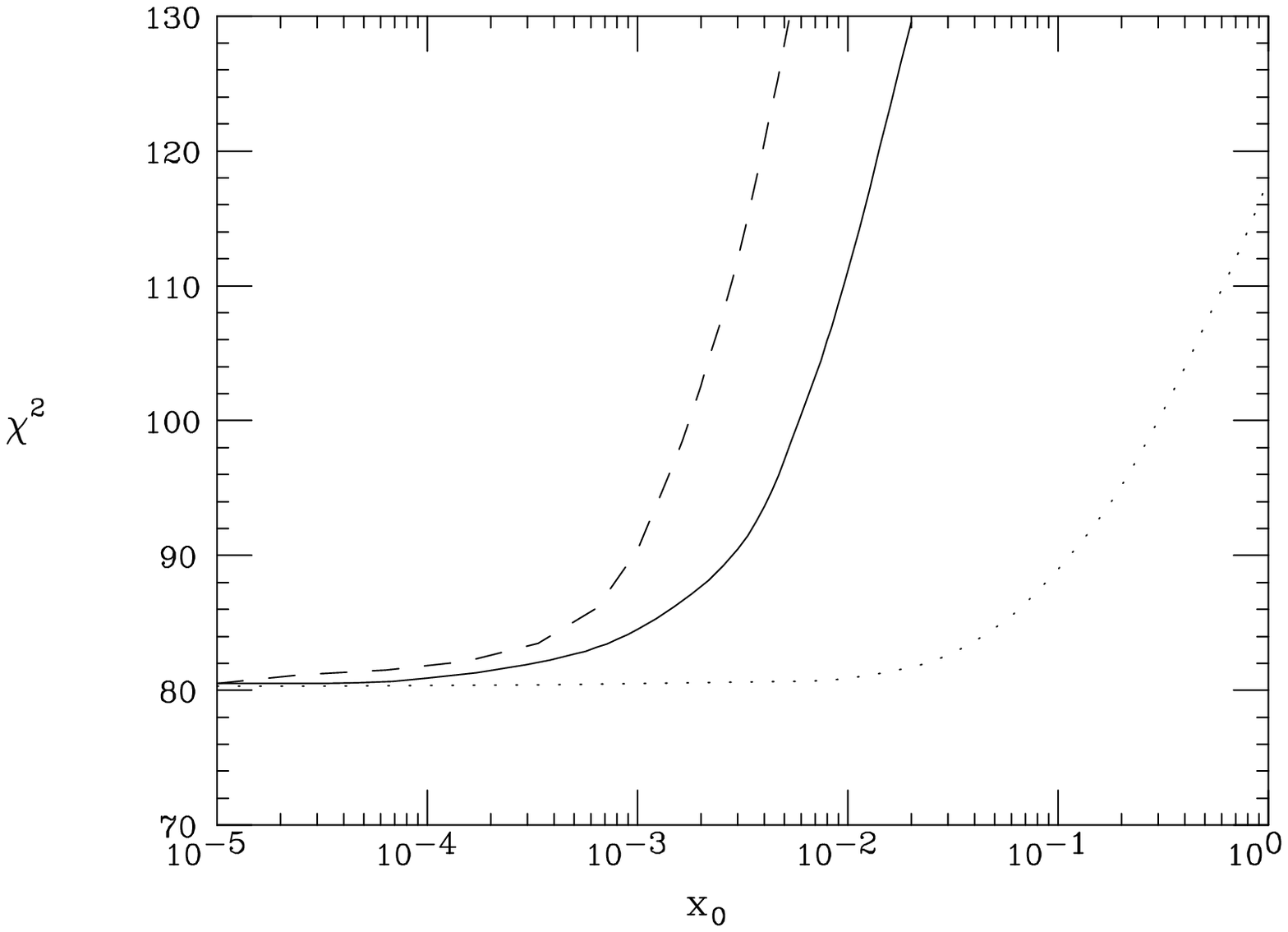,bbllx=30pt,bblly=200pt,bburx=570pt,bbury=630pt,width=8cm,height=7cm}}
\fcaption{The $\chi^2$ of the Ball and Forte fit as a function of $x_0$ in the
$\MSB$ scheme. The different line types correspond to: double leading 
expansion with standard (full) or $Q_0$ (dashed) factorization and small $x$
expansion (dotted).}
\label{x0}
\end{figure}

The detailed results depend on precisely which choice of schemes is made 
 but one conclusion is common. The data do not favour unconventional 
(resummation) expansion schemes  at all. Fig.~\ref{x0} illustrates the variation
of $\chi^2$ with $x_0$ for fits in the small $x$ and double leading expansion 
schemes. There is no minimum in $x_0$
indicating that the data do not need to use the unconventional schemes. The 
best $\chi^2$ is 80 to 169 H1(94) data points, for a conventional NLO fit as 
applied in the $\MSB$ scheme.
The other features of this fit are in broad agreement with the MRSR2 fit
with a high value of $\alpha_s$ ($\alpha_s = 0.122$) and values of
$\lambda_S,\lambda_g$ such that
the gluon distribution is valence-like and the quark distribution is rising 
moderately for low $Q^2$ ($\simeq 1\,$GeV$^2$)~\fnm{z}\fnt{z}{~As usual 
one must beware of making direct comparisons of $\lambda$ values 
which are evaluated somewhat differently by different authors}.

\noindent

A similar but stronger conclusion, that resummation of $\ln(1/x)$ terms 
cannot fit the HERA(94) data (for both ZEUS and H1) for low $Q^2$, ( $1.5 < Q^2
< 5\,$GeV$^2$) is  made by Bojak and Ernst~\cite{Bojak,ernst2}. These authors 
have repeated the analyses of Ellis et al~\cite{EHW} and Forshaw 
et al~\cite{FRT} using  the HERA(94) data, and they have extended the work of
Ball and Forte to use more sophisticated parametrizations in their fits.
However in all cases they conclude that the resummation terms introduce
an $x$ dependence which is simply too steep to fit the lower $Q^2$ data,
for which the slope of $F_2$ is flattening.

\subsubsection{Scheme independent calculations}
 
However these conclusions are disputed by work~\cite{ThorneR,Thornenew} 
which follows the suggestion of Catani~\cite{Cataninew} that
 one should try to formulate the dynamics of scaling violation
entirely in terms of scheme independent quantities such as the measurable 
structure functions rather than parton distributions.
Consider two observables such as $F_2$ and $F_L$ (or $F_2$ and $F_2^{c\bar c}$).
One can write down the evolution equation
{\small \begin{equation}
\frac{dF_2(x,Q^2)}{d \ln Q^2} = \int^1_x \left[ \Gamma_{22}(\frac{x}{y},\alpha_s)
F_2(y,Q^2) + \Gamma_{2L}(\frac{x}{y},\alpha_s) F_L(y,Q^2) \right]
\end{equation}}
where we relate the scaling violations in the observables to the values of the
same observables. The kernels $\Gamma$ are thus observables
themselves. Because of the structure of this equation they may be thought of as
physical anomalous dimensions. They can be related to the usual splitting
factors and coefficient functions as evaluated in any scheme, and it can be 
established that the results turns out to be scheme independent as expected for
an observable. The results do, of course, differ according to order of 
accuracy at which 
the evaluation is made, and in particular evaluations can be made 
conventionally, or including $\ln(1/x)$ resummation terms, such that one may
establish whether such resummation terms are necessary without scheme
ambiguity.

Thorne~\cite{ThorneR,Thornenew} goes further and 
argues that the fact that calculations for $LL(1/x)$ terms are scheme dependent
indicates that the method of incorporating these terms is incorrect. A 
correct, complete,  leading order renormalization scheme consistent
calculation  for an observable quantity
like a structure function (rather than a parton distribution) will  naturally 
include $LL(1/x)$ terms in the form of Catani's physical anomalous
dimensions. Each of our usual expansion schemes (the conventional loop 
expansion in $\alpha_s$ or the small $x$ expansion in $\alpha_s \ln(1/x)$) 
have 
shortcomings because some of the terms appearing at what we call higher orders
are not actually subleading  to terms which have already appeared.
However the full set of terms in the combination of leading order expressions
for both expansion schemes is genuinely leading order, and it is 
renormalization (and hence factorization) scheme independent. He labels this
combined scheme the LORSC scheme~\fnm{aa}\fnt{aa}{~In some of the earlier 
references the LORSC scheme is called the LO(x) scheme.} and it
is in this combined expansion 
scheme that the physical anomalous dimensions appear.

Thus the power counting of small $x$ logarithms is different for the 
physical anomalous dimensions than
for the usual anomalous dimensions (and coefficient functions) and 
terms which are normally considered as classifiable as leading order or 
next-to-leading order get 
somewhat mixed up. An analysis to $LL(1/x)$ in the physical anomalous 
dimensions (LORSC scheme) can be performed consistently using
the $LL(1/x)$ gluon anomalous dimensions $\gamma_{gg}$ and the 
$NLL(1/x)$ quark anomalous
dimensions $\gamma_{qg}$ which have already been calculated. 
An analysis to $NLL(1/x)$ in the physical anomalous dimensions (NLORSC scheme)
is not possible until the $NLL(1/x)$ gluon anomalous dimensions ($\gamma_{gg}$
and $\gamma_{gq}$) have been fully evaluated. 

One further feature of Thorne's LORSC fit is notable. He demands that any 
deviation of the structure function from a flat Regge type behaviour at 
small $x$ must come from perturbative effects. Thus his inputs are 
non-perturbative functions, flat at small $x$, convoluted with functions of 
the physical anomalous dimensions which are
determined by evolution from a non-perturbative scale $A_{LL}$ to the 
input scale $Q^2_0$. Only when $Q^2_0 = A_{LL} \simeq 1\,$GeV$^2$, are the 
inputs purely non-perturbative. 
He then requires insensitivity of the fits to $Q^2_0$. This approach 
predicts a relationship between the small $x$ inputs for $F_2$, $dF_2/d\ln Q^2$
and $F_L$ such that one can no longer redefine inputs freely.

\begin{figure}[ht]

\centerline{\psfig{figure=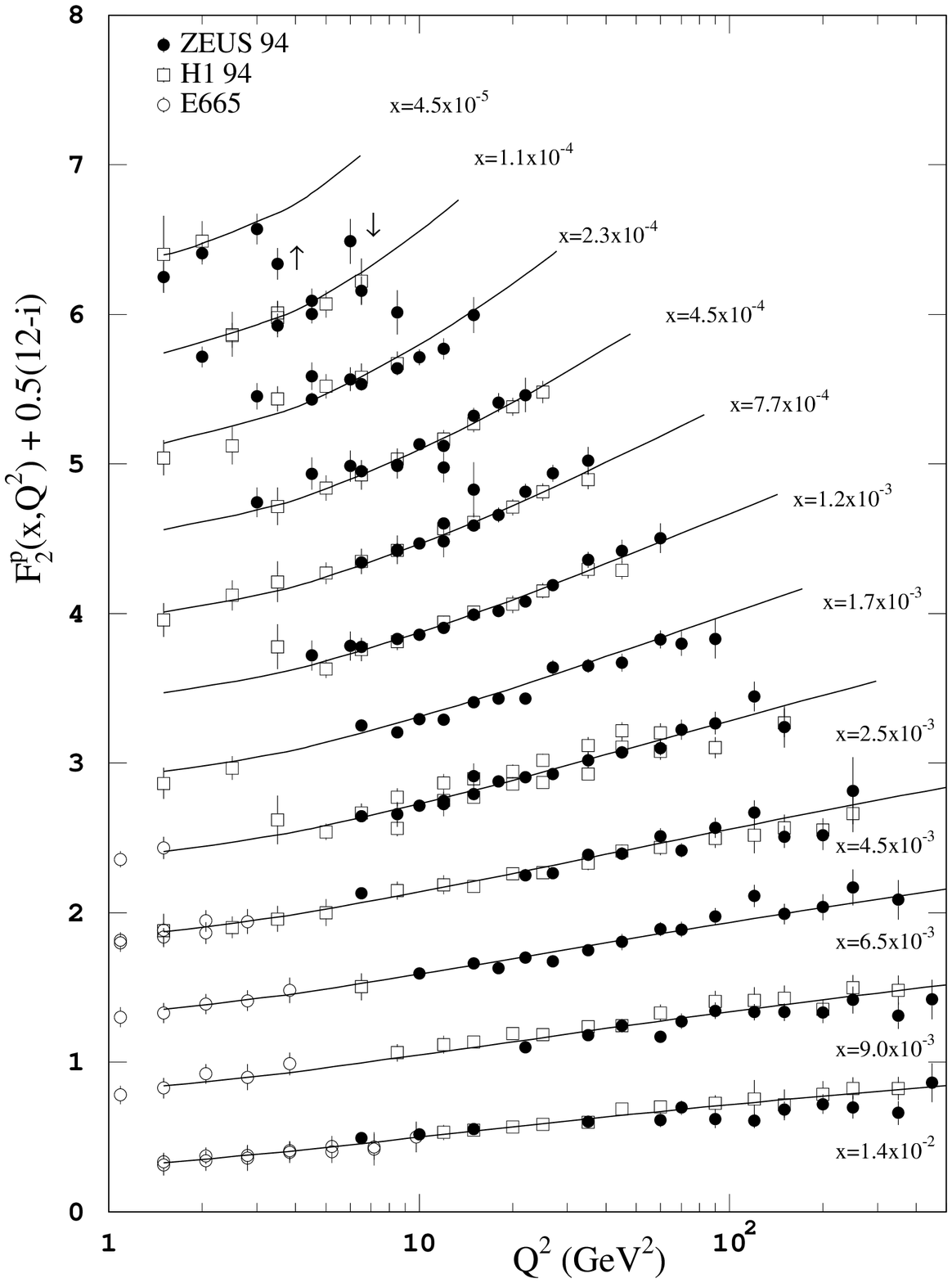,height=9cm}}
\fcaption{The solid line shows the LORSC fit to HERA(94) and E665 data on
$F_2$ as a 
function of $Q^2$ for different $x$ bins. For clarity of display $0.5(12-i)$ 
is added to the value of $F_2$ each time the value of $x$ is decreased, for
$i=1,12$. The ZEUS data are renormalized by 1.015 to produce the best fit.
Data are assigned to the $x$ value closest to the experimental $x$ bin.}
\label{thornefit}
\end{figure}

Thorne~\cite{ThorneR,Thornenew} 
has performed fits to H1(94) and ZEUS(94) data, and 
also to the data of BCDMS, NMC(97), E665 and CCFR(93), 
which establish that one does 
get a better $\chi^2$ working in the LORSC scheme than with conventional NLO
fits such as those of MRSR.  The quality of the fit is shown in 
Fig.~\ref{thornefit} and the $\chi^2$ are presented in 
Table~\ref{tab:thorne}, where
the $\chi^2$ for the MRSR2 fits are also given for comparison. Note
that the $\chi^2$ values for MRSR fits to NMC data differ from those already 
given in Table~\ref{tab:chi2}, since Thorne has used the new NMC(97) 
data~\cite{nmcdata2}, and has
re-evaluated the MRSR $\chi^2$ for these data. A fairer comparison between 
the LORSC scheme and 
the conventional NLO treatment may be obtained by making  
a conventional NLO fit in exactly the same circumstances as the LORSC fit, i.e.
including the same data sets and using the same programme but leaving out
resummation terms. Thorne has done such a fit and it is included in
Table~\ref{tab:thorne} as the `NLO' fit. One sees that the LORSC fit is an 
 improvement over the NLO fit overall. As expected most of the improvement 
is at low $x$:
 if a cut $x<0.1$ is made on all data sets then the LORSC fit gives a 
$\chi^2$ of $483$ for $548$ data points, whereas the NLO fit gives a 
$\chi^2$ of $554$. The NLO fit is superior for higher $x$ data where 
$\alpha_s^2$ terms are important, a more definitive comparison should come
when an NLORSC fit becomes available.
\noindent
\begin{table}
\tcaption{ $\chi^2$ per data point, for QCD fits done to DIS data sets
in the LORSC scheme and in 
the conventional NLO scheme by Thorne~\cite{ThorneR,Thornenew} 
and by MRS~\cite{MRS96}}
\centerline{\footnotesize\smalllineskip
\begin{tabular}{cccc}\\ 
\hline
 $Data$ & LORSC & NLO & MRSR2\\ 
\hline
 $BCDMS\ F_2^p$ & $181/174$ & $218/174$ & $320/174$  \\
 $CCFR(93)\ F_2$& $59/66$ & $48/66$ & $56/66$ \\
 $CCFR(93)\ xF_3$& $48/66$ & $39/66$ & $47/66$  \\
 $NMC\ ratio$& $142/85$ & $137/85$ & $132/85$ \\
 $NMC(97)\ F_2^p$& $122/129$& $131/129$ & $135/129$  \\
 $NMC(97)\  F_2^d$& $114/129$& $107/129$ & $99/129$ \\
 $ZEUS(94)\ F_2^p$ & $253/204$& $281/204$ & $308/204$  \\
 $H1(94)\ F_2^p$&$123/193$ & $145/193$ & $149/193$ \\
 $E665\ F_2^p$&$63/53$ & $63/53$ & $63/53$ \\  
\hline\\
\end{tabular}}
\label{tab:thorne}
\end{table}

The major criticism of the approach of the present section is that it 
deals with an integrated gluon density. One has lost some of the physics 
of the non-integrated
gluon ladder. Thus such an approach cannot address the 
problem that $k_T^2$ values in a non-$k_T$ ordered gluon ladder may drift into 
the infra-red region~\cite{Martin,Gamici}. However Camici and 
Ciafaloni~\cite{Gamici} consider that Thorne's approach should be valid 
provided that $\alpha_s \ln(1/x)$ is not `too' large. 
A related criticism is that one also cannot evaluate the effect 
of the imposition of constraints such as Eq.~\ref{eq:const} and  the approach
does not consider higher twist operators~\cite{KMStas}. 
Thus it is interesting to consider
the CCFM equation or the modified BFKL equations which preserve a richer
physical structure.

\subsection{Phenomenology: CCFM and modified BFKL equations}
\label{sec:CCFM}
\subsubsection{The CCFM equation}

 The CCFM equation~\cite{CCFM} is defined in terms of a scale dependent 
unintegrated gluon density $f(x,k_T^2,Q^2)$ which specifies the chance of 
finding a gluon with longitudinal momentum fraction $x$ and transverse momentum
$k_T$ at the scale $Q^2$.  
The practical application of the CCFM equation to predict $F_2$ involves
many of the same considerations as the application of the BFKL equation. One 
has to incorporate running of $\alpha_s$ and to consider the UV 
and IR cut-offs
on the $k_T$ integration (although this is somewhat more straightforward since
the  angular ordering constraint acts as an
infra-red regulator and we may avoid the severe problems of drift into the
non-perturbative region which beset the implementation of the BFKL 
equation~\cite{Bottazzi}). One also has to convolute the 
CCFM gluon with the quark box at the top of the gluon ladder and
add the result to the soft background appropriate at large $x$. 
 
The phenomenology has been explored by Kwiecinski et
al~\cite{KMS}. They obtain $f(x,k_T^2,Q^2)$ by approximating
 the full CCFM equation appropriately at small $x$. Accordingly, 
the splitting 
functions retain only the $1/x$ terms and the effect of singlet quarks on
the gluon evolution is ignored. Running $\alpha_s$ is incorporated by
using $\alpha_s(k_T^2)$. One must input a non-perturbative gluon 
distribution  to the equation, at a starting scale $Q^2_0$. Kwiecinski et al
chose the form $f(x,k_T^2,Q^2_0) \sim 3(1-x)^5 exp(-k_T^2/Q^2_0)$, i.e. flat 
at low $x$ and with a narrow $k_T$ distribution. The starting scale is chosen 
as $Q^2_0 = 1\,$GeV$^2$. 

The resulting gluon distribution can be compared
with that which would be obtained conventionally from the DGLAP equations,
 or from the BFKL equation, starting from the same input.
The CCFM  unintegrated distribution  has  a $k_T^2$ 
dependence which broadens and develops a significant tail as $x$ decreases, 
however diffusion in $k_T$ is reduced compared to the BFKL equation, and
correspondingly sensitivity to the IR cut-off is reduced.
The angular ordering in CCFM also introduces a dependence of the 
unintegrated gluon on the scale $Q^2$, which is significant at low $Q^2$, 
whereas the BFKL gluon acquires $Q^2$ dependence only from the $k_T^2$ 
integration. Hence the CCFM gluon evolves faster in $Q^2$ than the BFKL gluon.

\begin{figure}[t]

\centerline{\psfig{figure=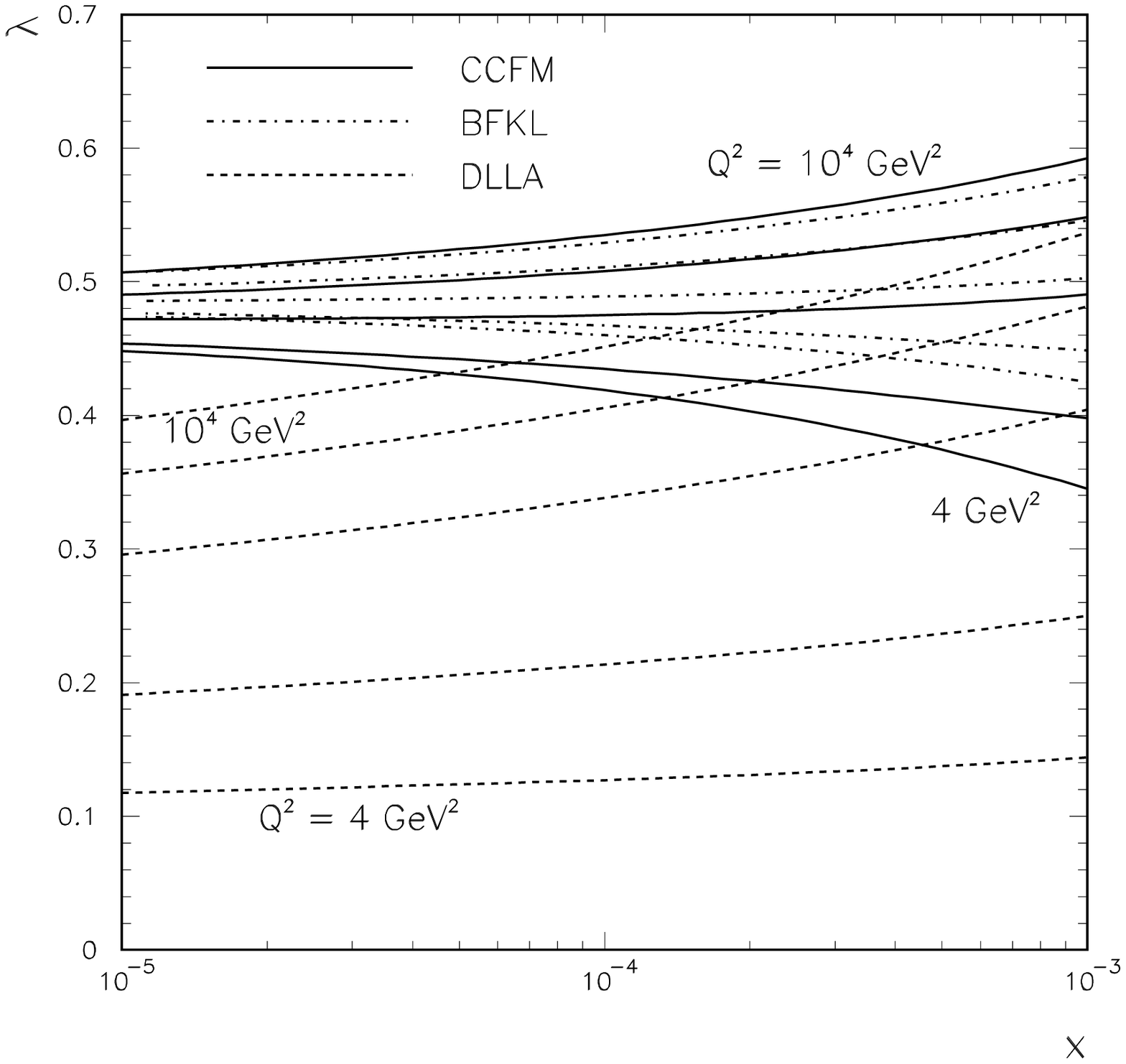,bbllx=30pt,bblly=160pt,bburx=540pt,bbury=680pt,width=9cm}}
\fcaption{The effective values of $\lambda$, $xg \sim x^{-\lambda}$, for CCFM,
BFKL and the DLLA, 
for $Q^2 = 4,10,10^2,10^3,10^4\,$GeV$^2$ versus $x$.}
\label{ccfmlam}
\end{figure}

Thus, if we compare the integrated gluon distributions, the CCFM gluon 
distribution is much steeper at small $x$ than that generated by
conventional DGLAP in the DLLA, 
but less steep than the BFKL gluon at moderate $Q^2$, 
evolving to become very similar at high $Q^2$. These features are illustrated
in Fig.~\ref{ccfmlam}.
Characterizing the steepness of the gluon's slope in terms of the form 
$x^{-\lambda}$, we see that the CCFM slope $\lambda$ 
has a stronger $Q^2$ dependence than that of the BFKL, such that for low 
$Q^2 \sim 10\,$GeV$^2$ it is smaller  by 
$\sim 0.1$, whereas at higher $Q^2$ one has $\lambda \sim 0.5$ just as for
the BFKL result. Kwiecinski et al conclude that the part of the 
$NLL(1/x)$ effects which are included in the CCFM equation 
modify the steep behaviour of the $LL(1/x)$ BFKL result such that 
the onset of this form is more delayed for CCFM. The recent work of
Bottazzi et al comes to similar conclusions~\cite{Bottazzi}. 

\noindent

Kwiecinski et al also use the CCFM gluon to make predictions for 
$F_2$, using the $k_T$ factorization theorem and
imposing an infra-red cut-off, $k_0^2 \sim 1\,$GeV$^2$, on the 
$k_T^2$ integration. The CCFM prediction for $F_2$ is then  added to a 
background from the
conventional soft Pomeron in order to connect 
smoothly to the larger $x$ region.  The result was compared to HERA(93) data 
and it gave a good description, lying between the MRSA$^{\prime}$ and GRV 
predictions.  
Note that it is NOT a fit but a prediction, although there is
some freedom in the choice of the non-perturbative input shape and input 
scale, 
the background shape and the infra-red cut-off value. It is encouraging 
that physically reasonable choices give a reasonable description.

In the preceding discussion no consideration has been given to the 
constraints coming from energy-momentum conservation, which impose a 
UV cut-off on $k_T$ integrations, or to the  constraint embodied in 
Eq.~\ref{eq:const}, which is more restrictive than angular ordering at 
small $z$, since it automatically embodies the angular ordering constraint
($Q^2 (1-z)^2/z^2 > q_t^2$) until $Q^2$ falls
below $k_T^2$. These two constraints
were considered in a separate publication of Kwiecinski et al~\cite{KMS96}
(see also~\cite{dDS}).
As a result the slope $\lambda$ for the CCFM gluon is further reduced by 
$\sim 0.1$, and the predictions for $F_2$ show a reduced slope with 
respect to $Q^2$ as well as with respect to $1/x$. A reasonable description
of HERA(93) data can still be obtained. However, the authors have not 
pursued this
approach to make detailed comparisons with HERA(94) data, preferring to regard
the work as a comparison of the behaviour of the CCFM, BFKL and DGLAP
equations within the present incomplete state of theoretical understanding.

\subsubsection{Modified BFKL equations}

The modified BFKL equation of Li~\cite{Li2} introduces $Q^2$ dependence
into the BFKL equation essentially from applying a UV cut-off at $Q^2$ to
integrals over loop momenta and from introducing a fixed IR cut-off $Q^2_0$
on virtual gluon emission. A $Q^2$ dependent unintegrated gluon distribution 
is obtained which is used to predict $F_2$ using
the $k_T$ factorization theorem and a form of the gluon input which is flat at
$Q^2_0 = 1\,$GeV$^2$, $x_0 =0.1$, with a narrow $k_T$ distribution, similar to 
that used for the CCFM solution discussed above. 
The resulting integrated gluon distribution has a $Q^2$ dependent slope 
at small $x$ given by $xg \sim (\frac{x}{x_0})^{-\alpha_s \ln (Q^2/Q_0^2)}$
and correspondingly the rise in $F_2$ at small $x$ is less steep at smaller
$Q^2$ than that predicted by the unmodified BFKL equation.
Thus reasonable agreement with the data is obtained in the $Q^2$ range
 $8 < Q^2 < 20\,$GeV$^2$ (only 4 flavours are considered). More 
recently~\cite{Li3} this work has been extended to relax the strong ordering
in $x$ usually assumed in the derivation of the BFKL equation, with the
result that the predicted gluon distribution saturates (and thus obeys the
unitarity bound) at small $x$, ($x \leqsim 10^{-4}$).

Kwiecinski, Martin and Stasto~\cite{KMStas} have developed a modified BFKL
equation which is applicable over all $x$ and $Q^2 > 1\,$GeV$^2$.
They develop a unified BFKL and DGLAP equation for the gluon by 
adding to the BFKL equation terms which include leading order DGLAP evolution
(i.e. terms which are leading in $\alpha_s$ but subleading in 
$\alpha_s \ln(1/x)$) as follows
{\small \begin{eqnarray}
f(x,k^2_T) = f^0(x,k^2_T) & + & \frac{3\alpha_s(k^2_T)}{\pi} k^2_T 
                                \int^1_x \frac{dz}{z}
                                \int_{k_0^2} dk'^2_T 
                               K(k'^2_T,k^2_T) f(\frac{x}{z},k'^2_T) \nonumber \\
                          & + & \frac{3\alpha_s(k^2_T)}{\pi}\int^1_x 
                                \frac{dz}{z} (\frac{z}{6} P_{gg}(z) - 1)
                                \int^{k^2_T}_{k^2_0} \frac{dk'^2_T}{k'^2_T} 
                                f(\frac{x}{z},k'^2_T) \\
                          & + & \frac{\alpha_s(k^2_T)}{2\pi} 
                                \int^1_x dz P_{gq}(z) \Sigma(\frac{x}{z},k^2_T)
                                \nonumber
\label{eq:kmstas}
\end{eqnarray}}
\noindent       
The first term is the starting distribution. The second term embodies the
usual BFKL equation integrated from the IR cut-off $k^2_0$ 
 but the  kernel $K(k'^2_T,k^2_T)$ is subject to the 
constraint given in Eq.~\ref{eq:const}. The third term 
 in $P_{gg}$ gives the usual DGLAP convolution
of the gluon density with the splitting function 
but the gluon density has been split into an infra-red piece $\frac{x}{z}
g(\frac{x}{z},k^2_0)$
which is incorporated in the starting distribution and a piece 
which is obtained from integrating over the unintegrated gluon density. The 
$-1$ allows for the contribution already included in the BFKL summation. 
The fourth term in $P_{gq}$ is the usual DGLAP contribution allowing quarks 
to contribute to the evolution of the gluon.
The starting distribution is taken to be
$f^0(x,k^2_T) = \frac{\alpha_s(k^2_T)}{2\pi} \int^1_x dz P_{gg}(z) \frac{x}{z}
g(\frac{x}{z},k^2_0)$, given entirely in terms of a flat non-perturbative
gluon distribution, $\frac{x}{z}g(\frac{x}{z},k^2_0) = N(1-\frac{x}{z})^\beta$
, so that the
rise in $F_2$ at small $x$ is generated entirely by perturbative dynamics
(either from $\ln(1/x)$ or $\ln Q^2$). 

Before we can solve the above equation we must
specify the quark distribution $\Sigma(y,k^2_T)$. This is given by a 
further coupled equation
{\small \begin{eqnarray}
\Sigma(x,k^2_T) = S^a(x) + V(x,k^2_T) & + & \int^{k^2_T}_{k^2_0} 
                                           \frac{dk'^2_T}{k'^2_T}
                                           \frac{\alpha_s(k^2_T)}{2\pi} 
                          \int^1_x dz P_{qq}(z) S(\frac{x}{z},k'^2_T)\nonumber \\
                                      & + & \sum_i \int^\infty_{k^2_0} 
                                          \frac{dk'^2_T}{k'^2_T}
                                          \int^1_x \frac{dz}{z}
                             S^{box}_i(z,k'^2_T,Q^2) f(\frac{x}{z},k'^2_T)\\
                                      & + & \sum_i \int^1_x \frac{dz}{z} 
S^{box}_i(z,k'^2_T=0,Q^2) \frac{x}{z}g(\frac{x}{z},k^2_0)
\nonumber 
\label{eq:kmstasfac}
\end{eqnarray}}
The first, fourth and fifth terms in the above equation express the 
application of the $k_T$ factorization theorem in different kinematic regions.
Recall that this theorem is usually used to predict $F_2$ from the 
unintegrated gluon distribution. However, it is really only 
the sea quark part of $F_2$, which is
predicted since the cross-section $\sigma^{box}$ in Eq.~\ref{eq:ktfa} 
involves only the $g \to q\bar q$ transition. In the present formalism
one wants to evaluate $S(x,k^2_T)$, the sea quark distribution, rather than its
contribution to $F_2$ and this is why the cross-sections are denoted 
as $S^{box}$ rather than $\sigma^{box}$.
This evaluation implicitly involves an integration over the transverse
momentum $\kappa$ of the exchanged quark in the box, thus it involves three
separate contributions according to the relative sizes of $\kappa$, $k_T$ 
and $k_0$. The contribution from the
infra-red region $k^2_T,\kappa^2 < k^2_0$ is parametrized by a form 
appropriate for
light sea quarks $S^a(x) = C_P x^{-0.08} (1-x)^8$ as suggested by the soft
non-perturbative Pomeron. The contribution from the region $k^2_T < k^2_0 <
\kappa^2$ is given by a convolution of the integrated gluon distribution 
$\frac{x}{z}g(\frac{x}{z},k^2_0)$ with $\sigma^{box}$ for each flavour of 
quark (5th term). 
Finally, when $k^2_T,\kappa^2 > k^2_0$, the $k_T$ factorization theorem 
may be used fully perturbatively (4th term). Note that the 
charm component of the sea is evaluated perturbatively in all regions.
Thus far the $k_T$ factorization theorem has been used to play the role of 
$P_{qg} \otimes xg$ convolution in the DGLAP equations. To complete the
evaluation of $\Sigma(x,k^2_T)$ one must also 
add the $P_{qq} \otimes xq$ convolution as given by the third term (for sea 
quarks) and a parametrization of the valence quarks $V(x,k^2_T)$.
(This is  mainly determined by fixed target data and the GRV94 parametrization 
is used). In summary, one no longer just adds a background of conventional
soft processes to the result from the $k_T$ factorization theorem. The physics 
of conventional processes is incorporated in a more thorough manner.

\begin{figure}[htbp]
\vspace*{13pt}
\begin{center}
\begin{tabular}[t]{ll}
\psfig{figure=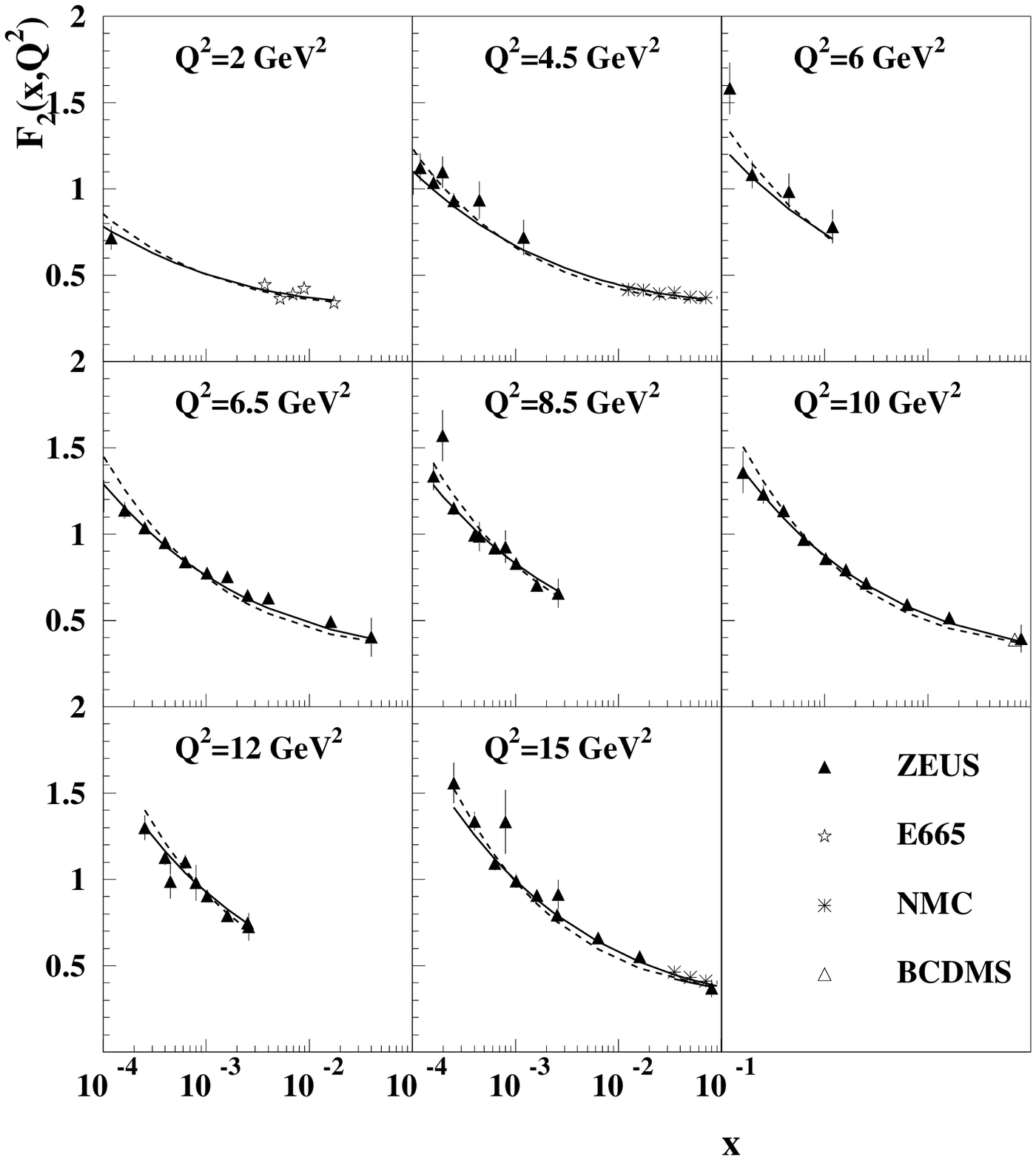,width=.475\textwidth} &
\psfig{figure=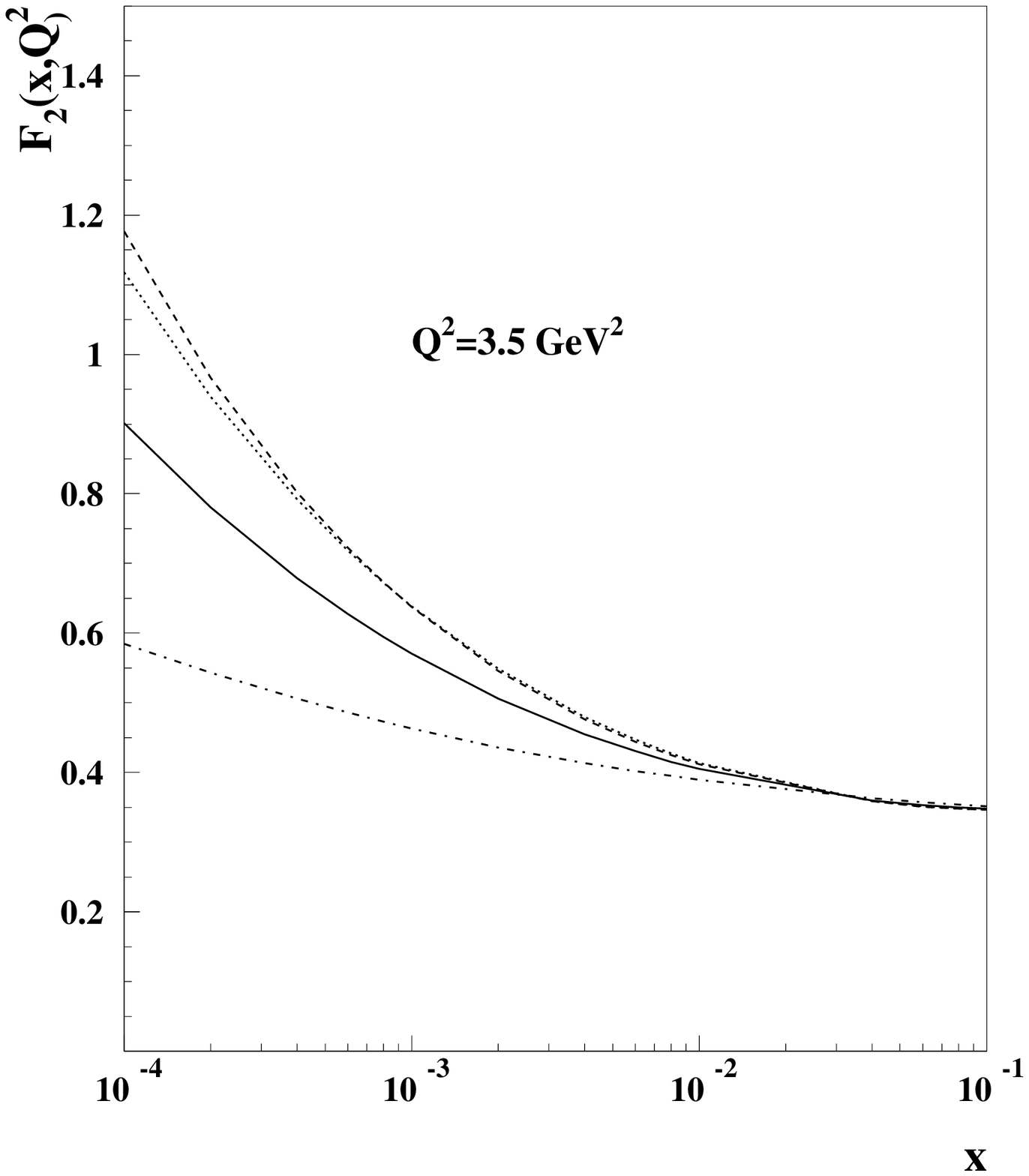,width=.475\textwidth} 
\end{tabular}
\fcaption{$lhs$: the Kwiecinski, Martin and Stasto description 
$F_2$ data (from EMC, BCDMS, NMC(97), ZEUS(94)) at small $x$ 
using $f(x,k^2_T)$ evaluated with (continuous curves) 
and without (dashed curves) the kinematic constraint of Eq.~\ref{eq:const}:
$rhs$: the dotted and dot-dashed curves are obtained using DGLAP 
in the gluon sector and in both the gluon and the quark sectors respectively,
whereas the continuous and dashed curves are as for the $lhs$.}
\label{stasto}
\end{center}
\end{figure}

Thus we have two coupled equations which can be solved for the unintegrated
gluon distribution and the singlet quark distribution. From these $F_2$ (in the
DIS scheme) and the integrated gluon distribution can easily be obtained. 
This gluon distribution exhibits a much less steep behaviour than that 
resulting from the unmodified  BFKL equation.  
In fact it is quite similar to that of the conventional
DGLAP parametrization of MRSR2 and its slope varies with $Q^2$ similarly. The
result for $F_2$ exhibits a rise at small $x$ which softens as $Q^2$ decreases
just as observed. There are only 2 variable parameters  in the fit,
$N$ and $\beta$, which specify the flat non-perturbative gluon input. 
The fits are not sensitive to the valence parameters or indeed to the Regge 
form chosen for the light quark sea. 
The normalization $C_P$ of the light quark sea is determined from the
momentum sum rule and $\alpha_s(M_Z^2) = 0.12$ is kept fixed. 
The input scale is $k_0^2 = 1\,$GeV$^2$.
Fits are made to HERA(94) 
data and to NMC(97), BCDMS, E665 data, with a resulting
 $\chi^2/ndf = 1.07$ and parameters $N=1.57, \beta=2.5$. This fit is an
improvement on the fully conventional fit of the MRSR2 parametrization to 
the same data,
which has $\chi^2/ndf=1.12$. Finally we note that it is important that 
the kinematic constraint of Eq.~\ref{eq:const} be included in the analysis. 
A fit without this constraint
results in too steep a slope of $F_2$ at small $x$ and moderate $Q^2$ and
a much poorer $\chi^2/ndf(=1.8)$. It is also interesting that it is the 
constrained fit which gives the better description of the WA70 prompt photon
data which sample the gluon at high $x$. The quality of the fit and the effect
of the constraint are illustrated in Fig.~\ref{stasto}. 

Fig.~\ref{stasto} also compares the present approach to a 
DGLAP NLO calculation, where the DGLAP calculation replaces the unified
DGLAP/BFKL approach in both the quark and gluon sectors, or only in the gluon 
sector. All the calculations start from the same flat gluon input at 
scale $1\,$GeV$^2$. From this comparison we can see that the use of the 
unified approach is most significant in the quark sector. 
In the previous subsection we found that $\ln(1/x)$ resummation was also most
significant in the quark sector. A further similarity in the non-conventional
approaches of these two subsections is that 
one starts with a non-perturbative flat gluon distribution
 (and flattish sea distribution) at a scale of about $Q^2 \sim 1\,$GeV$^2$ 
and the rise in $F_2$ is generated completely perturbatively, rather than being
input or generated from a much lower starting scale. If we focus
instead on the differences between the current approach and the unmodified
($LL(1/x)$ BFKL equation, we see that the partial inclusion of $NLL(1/x)$
effects, which the introduction of running $\alpha_s$ and the application of
the constraint of Eq.~\ref{eq:const} represent, has lead to a softening of
the BFKL Pomeron.

\subsection{Summary and outlook for pQCD at low $x$}
\label{sec:summary}

To sum up, there has been much debate as to whether  non-conventional QCD 
evolution is required by the HERA data. Good fits to the data can be obtained
from all approaches described in this section if one restricts the $Q^2$ range to
$Q^2 \geqsim 10\,$GeV$^2$. However it is more of a challenge to fit data,
in the $Q^2$ range, $1 < Q^2 < 10 \,$GeV$^2$. The approaches which meet this challenge are: conventional
DGLAP evolution with soft gluon input, BFKL in the colour dipole approach, 
unconventional DGLAP 
including $\ln(1/x)$ resummation worked out in the LORSC scheme
and the modified BFKL equation which incorporates 
$\ln Q^2$ and $\ln(1/x)$ summation. The 
extension of the validity of perturbative QCD into what was previously thought 
of as the non-perturbative region ($Q^2 \leqsim 4\,$GeV$^2$) is surprising, 
the transition region to
non-perturbative physics appears to be very narrow.
However we must be mindful of the warnings issued
by Levin~\cite{levin97} (that non-linear shadowing effects may be a more 
important modification to the conventional evolution equations than 
$\alpha_s \ln(1/x)$ terms),  
Bartels~\cite{newBart} (that higher twist effects should be 
important above the transition region) and Mueller~\cite{Mueller96} 
(that the pQCD may not be 
appropriate at small $x,Q^2$ because of diffusion into the infra-red).
In a recent review of higher order QCD corrections Van Neerven~\cite{Neerven} 
has also emphasized that considering only the most 
singular parts of the $\ln(1/x)$ corrections can give misleading results. 
 It is also possible that 
further higher order and subleading terms (requiring non-perturbative
calculations) could be important~\cite{Gamici}.

The $\chi^2$ of the conventional fits are comparable to those of the 
unconventional fits. Thus it is hard to establish that non-conventional QCD
processes are definitely needed to describe HERA data on $F_2$ and its scaling
violations. However, if they are not
needed, this itself requires some explanation since we have clearly reached
a region where $\alpha_s \ln(1/x) \sim 1$, so that conventional expansions
in $\alpha_s$ should no longer be reliable. 
We gather together here a few possible reasons which we have mentioned in 
passing. It may be because of cancellations between higher order or 
subleading terms and the leading $\ln(1/x)$ contributions:  we already know 
that there are extraordinary cancellations such that there are no 
$\ln^2(1/x)$ terms in the singlet sector~\cite{BF95,BRV}, and that the
coefficients of many of the $LL(1/x)$ terms in the gluon anomalous dimensions
are zero. It may be because the dominant effect of the $\ln(1/x)$ 
resummation is already included in the conventional sum when it is taken to 
NLO such that the term with $n = m = 2$ is included.
It may be because the leading behaviour of the BFKL equation 
(Eqs.~\ref{eq:BFKLsol},~\ref{eq:BFKLlam}) is approached very slowly as
derived in $\ln(1/x)$ resummation calculations. It may be because shadowing
effects mask the effect of $\ln(1/x)$ terms.
It may be because the effect of incorporating running $\alpha_s$ and
higher order effects into the BFKL formalism softens the BFKL 
Pomeron~\cite{ciafsep}. 
Work on the CCFM equation and on modified BFKL equations points 
in this direction. Haakman et al~\cite{Haakman} even contend that the
solution of the BFKL equation is modified so that the BFKL Pomeron behaves 
instead like the Pomeron corresponding to the result of de Rujula et 
al~\cite{deRujula} (see Eq.~\ref{eq:DLLAlam})
which softens as $Q^2$ decreases. 
Ball and Forte~\cite{BF97} contend that if the BFKL equation is derived from
a high energy factorization theorem 
it does not result in a hard Pomeron at all, and the conventional and 
non-conventional sums give results which are very alike for the structure 
function $F_2$. Camici and Ciafaloni~\cite{Gamici3} consider
that a full understanding of $NLL(1/x)$ effects would also lead to an 
understanding of the transition to soft physics as $Q^2 \to 0$.

We should be able to gain some answers by measuring different structure
functions. The freedom to redefine the parton distributions can only be
maintained if we consider only $F_2$ and $dF_2/d\ln Q^2$. 
If we use the parton distributions so determined to predict the high 
energy behaviour of other observables such as $F_L$ or $F_2^{c\bar c}$, 
then these predictions will  differ if the parton densities are 
redefined, and accurate measurements of such quantities could lead to a 
resolution of the ambiguities~\cite{Catani95}. 
For example, both Forshaw et al~\cite{FRT} and Ball and Forte~\cite{BF95} 
found that including the $\ln(1/x)$ resummation
terms alters the relative normalizations of the quark and gluon distributions
One requires a larger gluon to fit the $F_2$ data if 
one does not include resummation terms because the starting scale has to be
lower and there is more time for evolution. Thus if one compares 
predictions for  $F_L$ based on parton distributions extracted from a 
conventional fit to predictions for $F_L$ 
based on parton distributions extracted from a fit including $\ln(1/x)$
resummation terms one will see a significant difference: the conventional fit
will predict a larger $F_L$. 
Within the conventional framework one may relate the gluon distribution at 
small $x$ both to $F_L$ and $F_2$ (Eq.~\ref{eq:amcs}) and to 
$dF_2/d\ln Q^2$ and $F_2$ (Eq.~\ref{eq:kjell}) and refs.~\cite{EKL,gluonprox}.
Thus it is clearly possible to obtain  a set of self consistent relationships
between $F_L$ and $F_2$ and $dF_2/d\ln Q^2$ which will be violated if 
non-conventional effects enter. Such relationships have been given by
in a simple form by Kotikov and Parente~\cite{kotpar} (see also 
ref.~\cite{dekabasu}).

\begin{figure}[ht]

\centerline{\psfig{figure=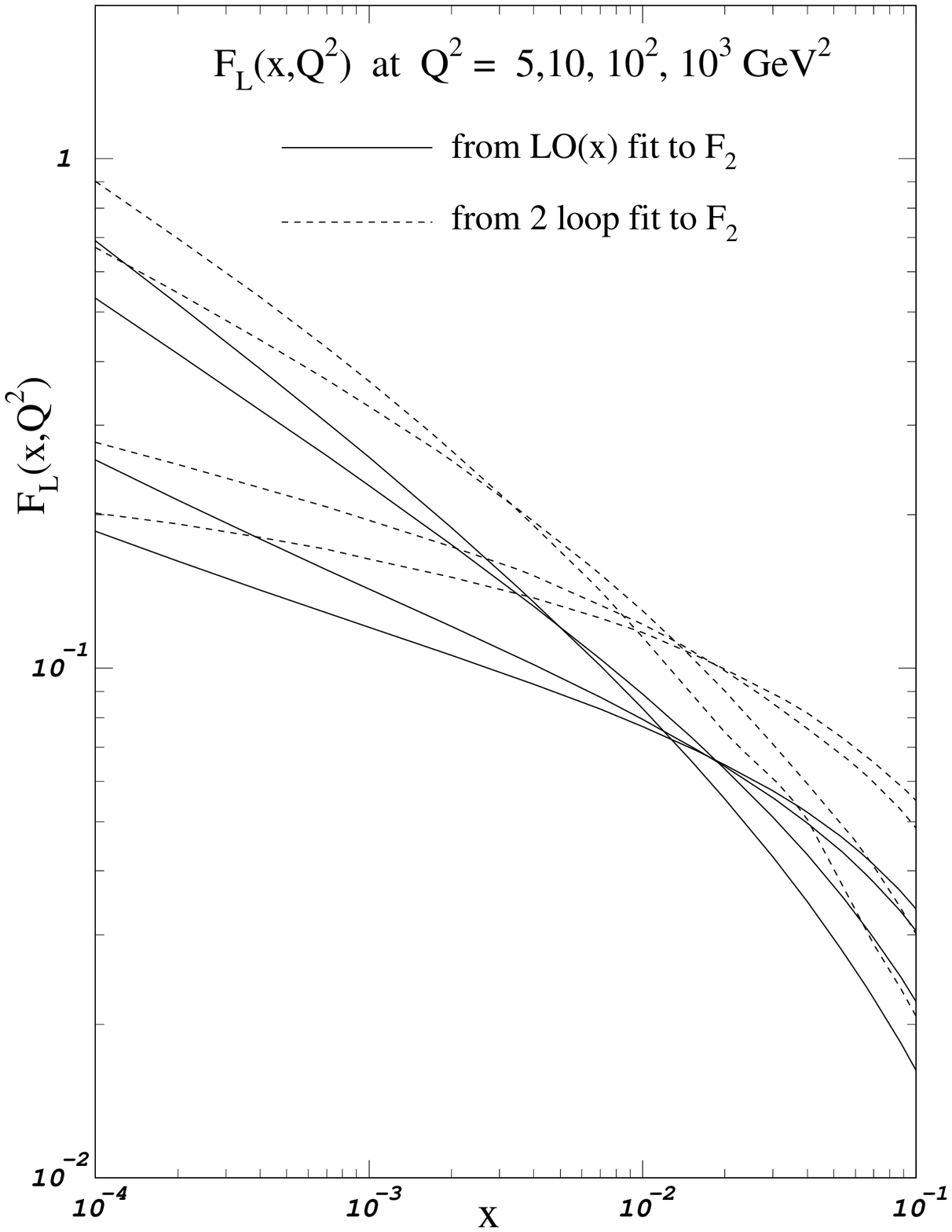,height=9cm}}
\fcaption{The prediction for $F_L$ from the renormalization scheme
consistent calculation including LL(1/x) terms (called LO(x) fit), 
compared to the conventional prediction from an NLO (2-loop) fit.}
\label{flrob}
\end{figure}

\begin{figure}[ht]

\centerline{\psfig{figure=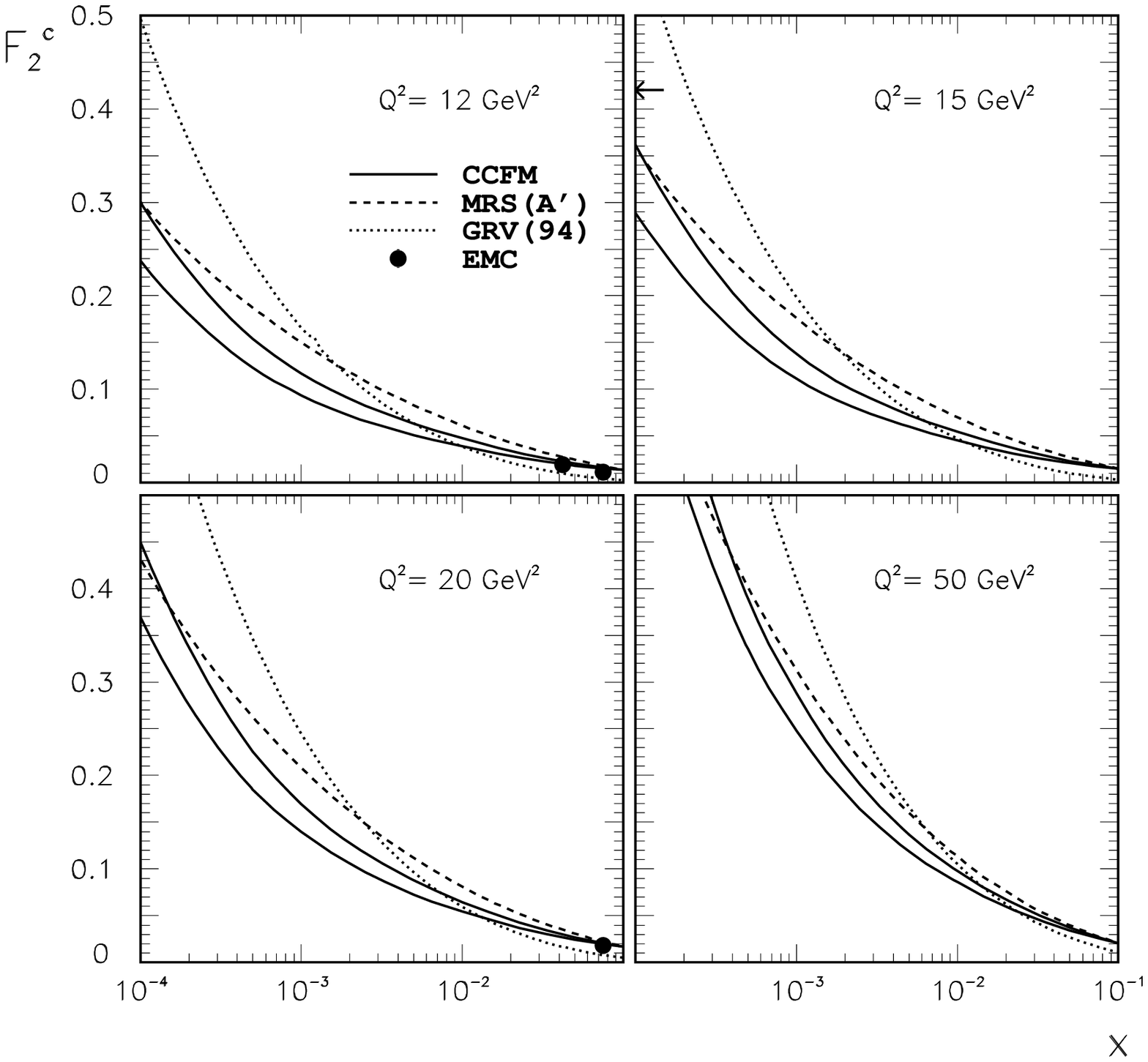,width=8cm,height=6cm}}
\fcaption{The prediction for $F_2^{c\bar c}$ from the CCFM equation
for two values of the charmed quark mass $m_c=1.4$ GeV (upper) and $m_c=1.7$ GeV
(lower) compared to that from the GRV and MRSA$^{\prime}$ parametrizations. 
EMC charm
data are shown for comparison}
\label{ccfmf2c}
\end{figure}

Many of the non-conventional analyses give predictions for $F_L$: 
BFKL dipoles, CCFM and the
modified BFKL equation of Kwiecinski Martin and Stasto, and the 
scheme independent analysis of Thorne. All of these non-conventional 
predictions share the common feature of being lower than the conventional 
predictions. Predictions from the analysis of Thorne are shown
 in Fig.~\ref{flrob}.
There is a similar discrepancy between conventional and
non-conventional predictions for $F_2^{c\bar c}$. The prediction from the CCFM
analysis of Kwiecinski et al is shown in Fig.~\ref{ccfmf2c}.

\noindent

\noindent

Present measurements of $F_L$ and $F_2^{c\bar c}$ were given in 
Secs.~\ref{sec:rdata},~\ref{sec:cdata}.
Most of the measurements of $F_L$ are not in the region of the $x,Q^2$ plane 
which would allow us to use
them to establish or refute the need for $\ln(1/x)$ terms. We need to measure
$F_L$ at HERA energies to reach small $x$ and moderate $Q^2$. We
 note that the glimpse of $F_L$ given by H1~\cite{h194r} is not a model 
independent 
measurement and thus cannot be used to discriminate the need for $\ln(1/x)$ 
terms, see ref.~\cite{thornefl} for a full discussion. 
Thus, as explained in Sec.~\ref{sec:rdata}, 
we need to run HERA at different
beam energies~\cite{AMCS}, or to use the ISR events~\cite{kps,favart}. 
It is unlikely that HERA will run at lower beam energies in the near future.
However, measurement of $F_2^{c\bar c}$ and of $F_L$ using ISR events should
become possible with the proposed upgrades to the HERA machine, which will
yield very large integrated luminosity.

 In view of the difficulty in
distinguishing signals of non-conventional behaviour in current 
structure function measurements it will  also be 
 useful to look for other signals of non-conventional
behaviour, for example in the hadronic final state.

\subsection{Searching for BFKL effects in the hadron final state}
\label{sec:hfs}

Two characteristic features of BFKL dynamics are the absence of strong $k_T$
ordering along the gluon chain\fnm{bb}\fnt{bb}{~However, note that even the 
lack of strong $k_T$ ordering 
which is taken to be a signal for BFKL evolution is already there to some 
extent in conventional NLO, for one pair of gluons in the ladder.}, and the 
consequent growth of the cross-section 
as $exp(\lambda \Delta y)$ where the rapidity interval, 
$\Delta y = \ln (x_1/y)$,
is defined between the gluons at either end of the gluon ladder, with 
longitudinal momentum fractions $x_1$ and $y$ (see Fig.~\ref{ladder}). 
The larger the value of $\Delta y$ the
more dominant are the leading $\ln(1/x)$ contributions. 

Thus for events at low $x$, hadron production in the 
region between the current jet and the proton remnant should be 
sensitive to the difference between BFKL and DGLAP dynamics. 
In the leading log DGLAP scheme 
the parton cascade follows  strong ordering in transverse
momentum $k_{Tn}^2 \gg k_{Tn-1}^2 \gg... \gg k_{T1}^2$, 
whereas there is no ordering in transverse momentum~\cite{muellercarg} for
the BFKL scheme. The transverse momentum follows a kind of random walk 
in $k_{T}$ space: the $k_{Ti}$ value is close to  the $k_{Ti-1}$ value, but it
can be both larger or smaller~\cite{barlott}.
As a consequence, BFKL evolution is expected to produce more gluons  
with substantial transverse energy $E_T$ in the region between the struck 
quark and the proton remnant, for low $x$ events, compared to DGLAP 
evolution~\cite{KMSET}.

\begin{figure}[ht]
\centerline{\psfig{figure=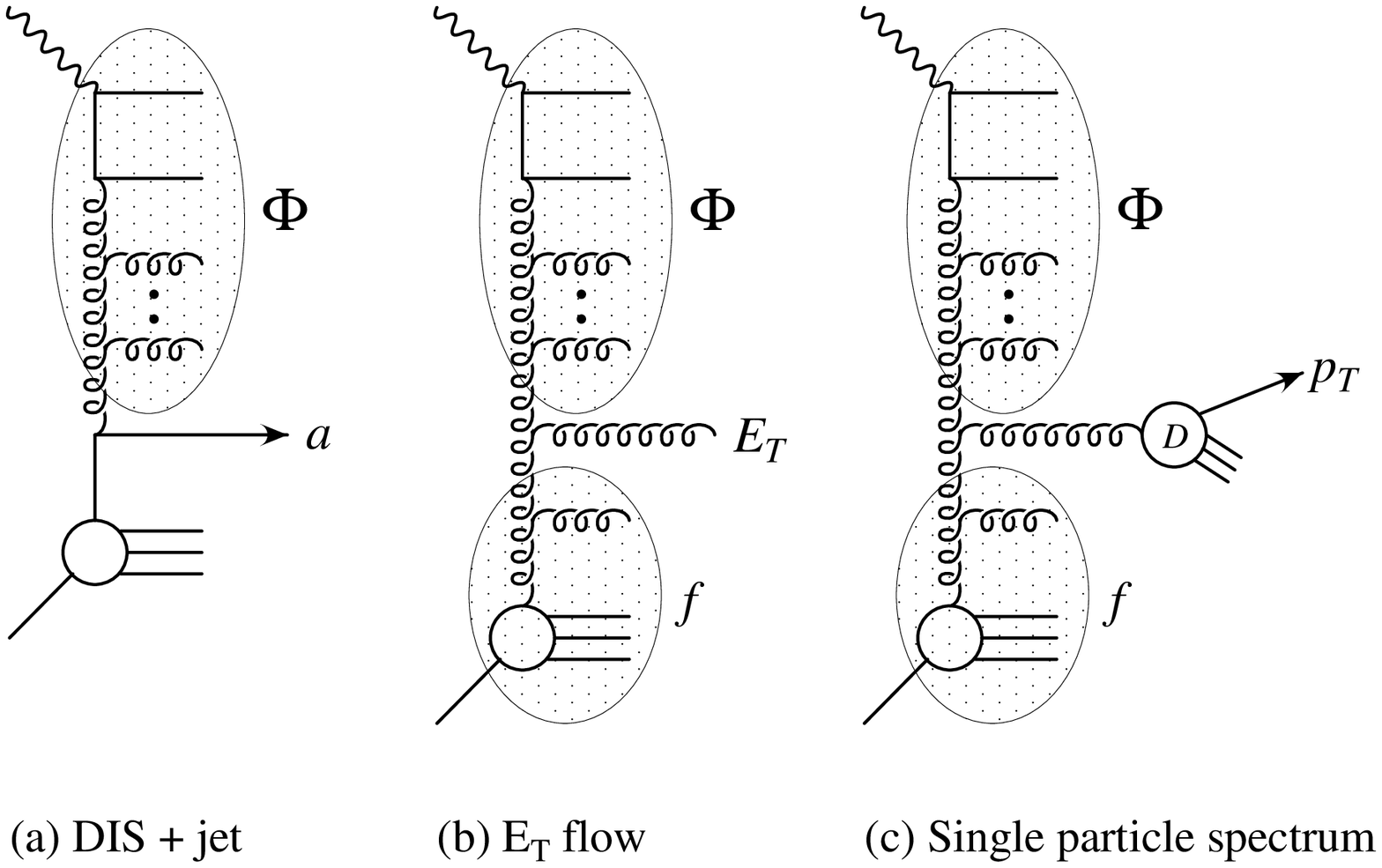,height=8cm}}
\fcaption{Processes which test BFKL dynamics: a)
DIS with a forward jet: b) $E_T$ flow in DIS: 
c) single particle $p_T$ spectra in DIS.}
\label{jan}
\end{figure}

Some processes which
may reveal these effects are: deep inelastic events with a measured forward 
jet; total transverse energy flow and single particle $p_T$ 
distributions in DIS events; production of a pair of 
jets at large rapidity interval in hadronic collisions or in DIS, and the 
azimuthal decorrelations between these jets. The `forward' direction at HERA
is the proton direction. We discuss only the former two
predictions in detail, (see Fig.~\ref{jan}), 
since there is, as yet, no data from DIS on the latter
processes. For a discussion of the interesting effects which may be observed 
in future and possible present indications in non-DIS data, 
see references~\cite{MueNav,AKMGr,Kim,Barjnew}

\noindent

To make comparisons of such predictions with data we must account for 
hadronization of the final state partons and for the experimental acceptance
and resolution. This is usually done within 
Monte Carlo models, based upon QCD phenomenology. We briefly discuss the three
main models used. Firstly,
the LEPTO Monte Carlo incorporates the QCD matrix elements to first order and
takes higher order parton emissions into account approximately by using the 
concept of parton showers i.e. branching processes of partons. This model is 
often called the MEPS model. The branching processes are based on DGLAP 
dynamics and the partons generated in the leading log parton showers are 
strongly ordered in $k_T$. Hadronization is done within the LUND string 
model~\cite{LUNDstr}. Secondly the  HERWIG Monte Carlo has a 
similar QCD treatment with leading log parton showers, but has a different
hadronization model involving a clustering algorithm~\cite{HERWIG}. Thirdly
the ARIADNE Monte Carlo calculates multi-gluon radiation
by emitting gluons from a chain of 
independently radiating colour dipoles spanned by colour connected 
partons~\cite{ariadne}. 
This is often called the CDM model (or CDMBGF since the boson-gluon fusion
process has to be added from the QCD matrix elements). 
It has no strong $k_T$ ordering and should thus give some indication of BFKL 
behaviour, although it does not explicitly use the BFKL equation.
Hadronization is again done by the LUND string model. 
 
Monte Carlo programmes which incorporate BFKL evolution have recently been 
developed by Schmidt~\cite{schmidt} and Orr and Stirling~\cite{stirbfkl}, and a 
Monte Carlo implementation of the CCFM equation in the linked dipole chain
LDC model is under development~\cite{LDC}, but these Monte Carlos
have not yet been used extensively to confront the data.  

Apart from these Monte Carlo implementations of QCD based models, 
there are some analytic calculations, based on (approximate) solutions of the 
BFKL equation. These are generally calculations at the parton level, and
thus they should not really be compared directly with the data. 
Such calculations are used to give some indication of the differences 
between BFKL and conventional dynamics, before hadronization.
Thus they serve to give a first indication of the effect of BFKL dynamics 
on the variable under study.

\subsubsection{Energy flows}

We first discuss transverse energy flow in DIS events. If we suppose that
there is no strong $k_T$ ordering then we expect to find more transverse 
energy, $E_T$, emitted in the central region between the current jet and 
the proton remnant (in the lab frame this is the experimentally problematic
forward region) than would result from conventional evolution. Parton level
calculations incorporating BFKL dynamics yield a fairly flat central plateau 
with $E_T \simeq 2\,$GeV per unit of 
rapidity in the HERA regime~\cite{KMSET}. Early 
observations~\cite{ZHexp} of such a level of $E_T$ were taken as indicating 
a need for BFKL dynamics, particularly since comparison of the data with
LEPTO and ARIADNE favoured the latter. However, these observations led to
a reconsideration of the hadronization 
prescriptions in both Monte Carlos and comparisons with more recent versions 
(LEPTO6.4~\cite{Ing},ARIADNE(4.07)~\cite{Lonn}) show that one can
produce  $E_T$ values $\simeq 2\,$GeV even from conventional dynamics.
In LEPTO the new features included in the model are soft colour
interactions (SCI)~\cite{softcol}, which are intended to describe rapidity 
gap events (usually thought to be of diffractive origin) without introducing 
the concept of a Pomeron and its structure function, 
and a modified sea-quark/remnant treatment (SQT), giving a smoother 
transition from BGF events to events where the photon interacts with a 
sea quark.

\begin{figure}[ht]

\centerline{\psfig{figure=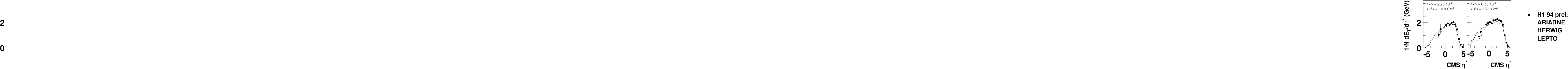,bbllx=0pt,bblly=500pt,bburx=400pt,bbury=720pt,height=8cm}}
\fcaption{The $E_T$ flow versus $\eta$ in the hadronic CMS for H1(94) 
preliminary data.
The proton direction is to the left. The data are compared to the models 
CDM, ARIADNE and HERWIG (see text).}
\label{fig:ethilo_prel}
\end{figure}

\begin{figure}[ht]

\centerline{\psfig{figure=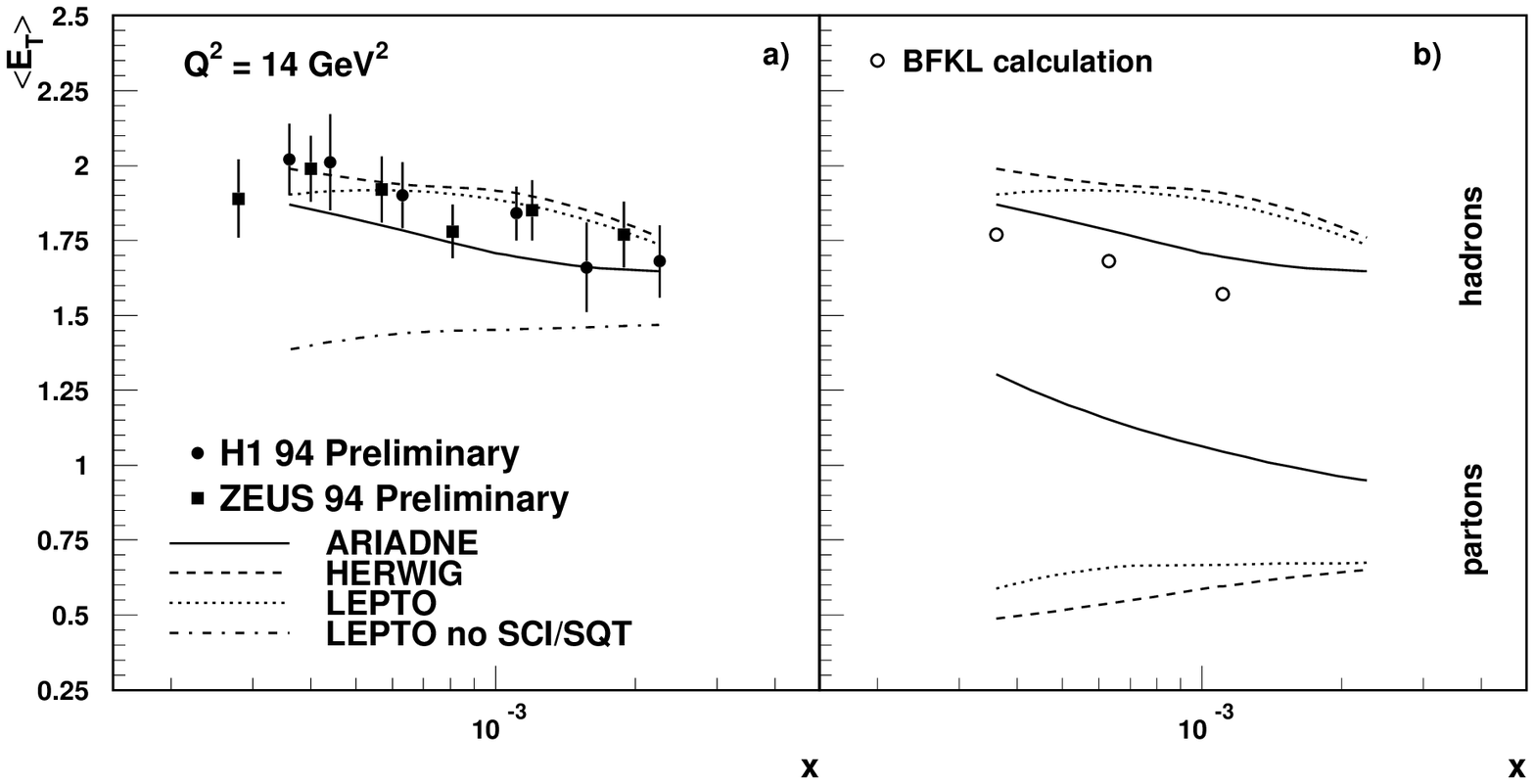,width=9cm,height=7cm}}
\fcaption{Mean transverse energy in the central rapidity region at fixed 
$Q^2$ as functions of $x$; (a) shows H1(94) and ZEUS(94) preliminary data 
compared to the 
ARIADNE, HERWIG and LEPTO model; (b) Compares hadron and parton level 
calculations of these models with a BFKL calculation at the parton level.}
\label{fig:H1et}
\end{figure}

Results for the energy flow versus pseudorapidity $\eta$ are shown in
the hadronic centre of mass system (CMS) in
Fig.~\ref{fig:ethilo_prel}, for two different $x$ bins at a $Q^2 \sim 
14$ GeV$^2$. Overlaid are model predictions, which broadly agree
with the data. It is of interest to check the $x$ dependence of the 
$E_T$ for $\eta < 0$ (the proton direction).     
This is illustrated in Fig.~\ref{fig:H1et} 
where the H1~\cite{eth1}and ZEUS~\cite{etzeus}
 data from the 1994 run
 are displayed. The average value of $E_T$ in the central region 
$(-0.5 < \eta^* < 0.5)$ is plotted as a function of $x$ for 
$Q^2 $= 14 GeV$^2$, and compared with model predictions.
 As $x$ decreases the values of $<E_T>$ increase. This observation 
can be explained by  the Monte Carlo models, 
but in different ways. Models based on DGLAP evolution with strong $k_T$
ordering produce $60-80\%$ of the $E_T$ in hadronization. Hadronization 
compensates for the fact that at the parton level the $<E_T>$ actually 
decreases as $x$ decreases, contrary to the trend in the data. 
The importance of hadronization is demonstrated by the predictions
from LEPTO without the newly introduced ingredients SCI and SQT.
This prediction is much below the data and has clearly a different $x$ 
dependence.

By contrast the CDM model with no $k_T$ ordering
produces more $E_T$ in the central region, with the right 
$x$ dependence,  at parton level, so that only $30-40\%$ extra needs
to be supplied by hadronization. The $x,Q^2$ dependence of the 
$\langle E_T\rangle$ also
agrees  with the analytical 
BFKL calculation~\cite{KMSET}, but this calculation was only 
made at the parton level, thus we see that it leaves room for only $10-20\%$
more of $\langle E_T\rangle$ to come from 
hadronization\fnm{cc}\fnt{cc}{~Even at the parton 
level BFKL calculations are not yet fully reliable since the $NLL(1/x)$ 
corrections have yet to be worked out, for example study of the CCFM 
equation~\cite{Bottazzi} indicates that angular ordering significantly reduces
transverse energy flow.}.

Hence we are now in a situation that all models can account for the data,
despite differences in the underlying parton dynamics: the mean $E_T$ 
can be made to agree by exploiting as yet unconstrained variations in 
hadronization models.
This unsatisfactory situation may be clarified by considering $E_T$
distributions rather than mean $E_T$. 
One expects $E_T$ from hadronization to be limited, whereas 
high $E_T$ should come from partonic activity. Thus the tails of the $E_T$ 
distributions are of interest.  First 
studies indeed show that for fixed $Q^2$ the high $E_T$ tail  gets harder
with decreasing $x$, and similarly for fixed $x$ it gets harder 
with increasing $Q^2$. This is demonstrated
 more clearly when studying $p_T$ spectra
of charged particles as discussed in the next section.

\noindent

\noindent

\subsubsection{Charged particle transverse momentum spectra}

\begin{figure}[ht]
\centerline{\psfig{figure=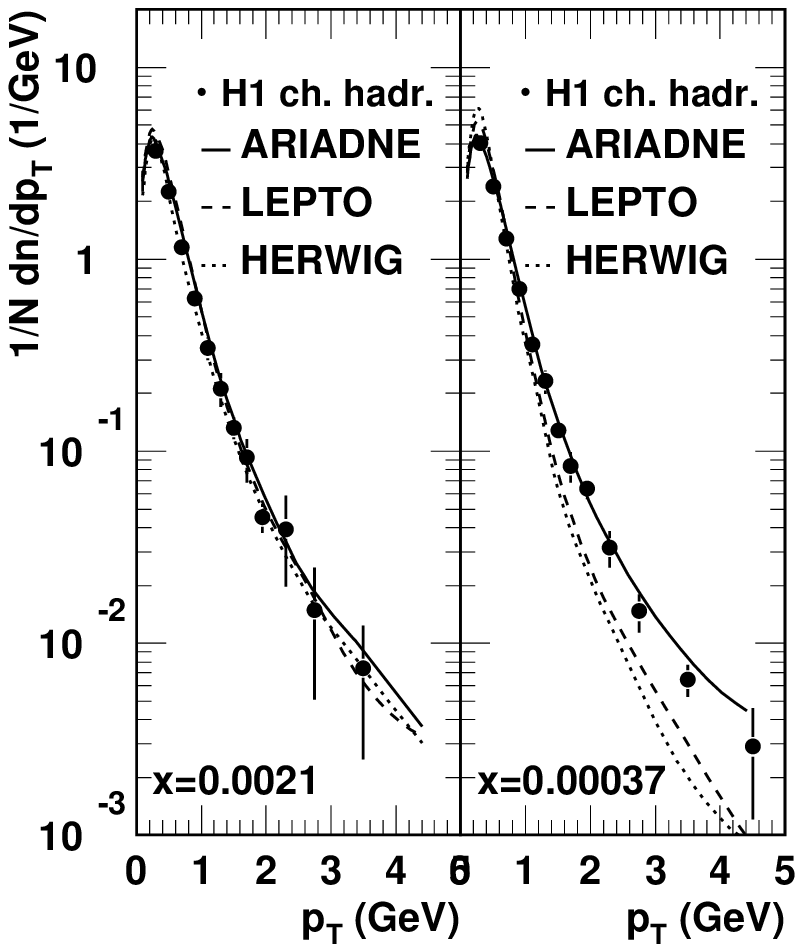,width=9cm,height=9cm}}
\fcaption{The $p_T$ spectra of the charged particles from $0.5 < \eta < 1.5$
(CMS). H1(94) data are shown for two different $x$ bins at $Q^2$ = 14 GeV$^2$.
Overlaid are the models ARIADNE, LEPTO and HERWIG (see text).}
\label{fig:ptdataf}
\end{figure}

It has been demonstrated  with Monte Carlo studies that a 
more direct measurement of the partonic activity in the hadronic final state
could be the measurement of single particle transverse
momentum ($p_T$) spectra~\cite{kuhlen}. A harder $p_T$ tail
is expected when there is unordered partonic activity in the 
ladder. H1 has made a measurement of the charged particle $p_T$
spectra in the CMS region $0.5 < \eta < 1.5$, limited by the acceptance
of the tracker detectors~\cite{h1tracks}.
The result is shown for two $x$ bins at $Q^2 \simeq 14 $ GeV$^2$, together
with model predictions, in Fig.~\ref{fig:ptdataf}.
The errors reflect the statistical and systematic uncertainties.
Clearly at high $x$ all models agree with the data, whereas at small $x$
the ARIADNE prediction is closest to the data. LEPTO and HERWIG produce 
significantly too few large $p_T$ tracks. 

\begin{figure}[ht]
\centerline{\psfig{figure=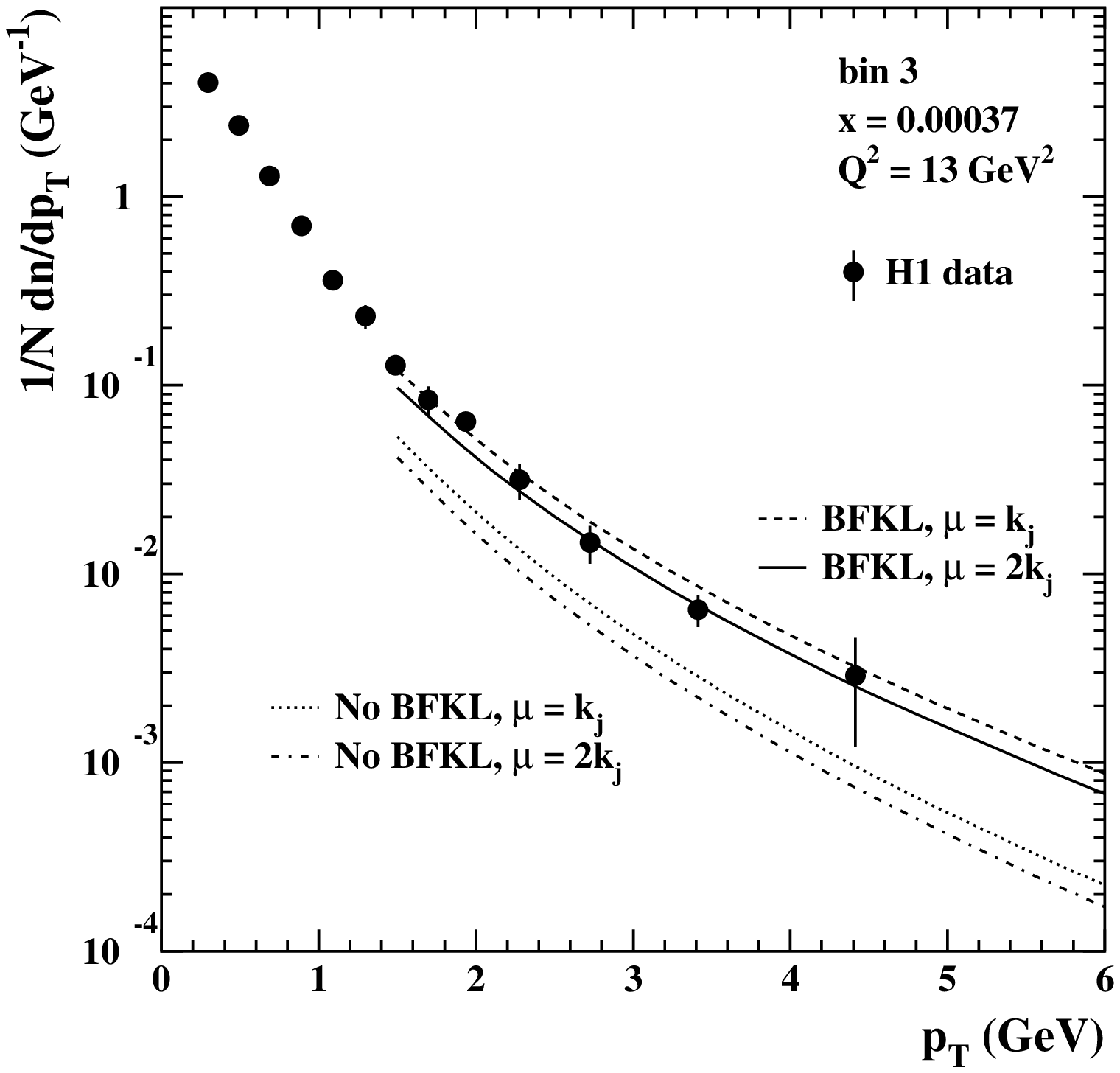,width=8cm,height=8cm}}
\fcaption{Transverse momentum distribution (H1(94) data)
compared with a calculation (including fragmentation) including and
excluding BFKL effects, shown for 2 choices of the scale.}
\label{fig:bfklpt}
\end{figure}

In Fig.~\ref{fig:bfklpt}
the result of a theoretical calculation~\cite{martin_pt}
 with and without BFKL evolution is shown. It consists of a  BFKL
calculation at the parton level folded with 
experimentally measured fragmentation functions.
The normalization of the BFKL calculations has been made using the 
 preliminary forward jet cross-sections (see next subsection). 
Good agreement with the data
is observed for the calculation which includes  BFKL evolution.

\noindent

\subsubsection{Forward jets}

Next we consider DIS events with a measured forward jet.  Jets are 
expected to be less sensitive to hadronization than $\langle E_T\rangle$ 
and should give a more definitive indication of BFKL behaviour. 
Mueller~\cite{Muellerjet} suggested that studying deep inelastic events at 
small $x$, with a measured jet with transverse momentum $k_{Tj}^2 \sim Q^2$ 
and large longitudinal momentum $x_j$, such that $z = x/x_j$ is small 
(see Fig.~\ref{jan}a, where the forward jet is indicated as particle $a$). 
The choice
$k_{Tj}^2 \sim Q^2$ suppresses conventional gluon radiation 
from DGLAP evolution, since there is no room for such evolution if $k_T$ is
strongly ordered, whereas 
BFKL dynamics predicts a cross-section which depends on $z^{-\lambda}$ at 
small $z$. Hence the cross-section for such events should be much
larger for BFKL than for DGLAP dynamics. Moreover the choice 
$k^2_{Tj} \sim Q^2$ also minimizes the problems of 
drift into the infra-red region which beset the BFKL calculation for
$F_2$ (provided $Q^2$ is chosen large enough) and the parton distributions 
need only be used at the large value $x_j$, where they are well 
known~\fnm{dd}\fnt{dd}{~Note that the parton emerging from the proton is 
considered to be collinear with the proton and that it emits a soft t channel
 gluon. Thus the energy of the forward jet is nearly equal to that of 
the parton emerging from the proton so that $x_j$ can be taken to be equal to
the longitudinal momentum fraction of this emergent parton.}. Hence 
the BFKL calculations should be reliable. Such calculations have been 
made by various authors~\cite{Barjold,martin_pt,Barjnew}.

H1~\cite{ZHexp,H1jet} have produced data on this process from both the '93 and 
'94 runs. We discuss both sets of data, since the 
'94 data are still preliminary.
For the '93 data, DIS events are selected by the cuts, 
$E_e > 12\,$ GeV, to minimize backgrounds and, $160^o < \theta_e< 173^o$, 
to concentrate the study at small $x$. These cuts allow a study of  
forward jet production in the region $Q^2 \simeq 20\,$GeV$^2$, and 
$2\times 10^{-4} < x < 2\times 10^{-3}$. A further cut, $y > 0.1$, is 
necessary to ensure separation of  the 
forward jet from the  current jet at the top of the ladder.  
The jets are found by a cone 
algorithm, requiring $E_T > 5\,$GeV in a cone of radius 
$ R = \sqrt{\Delta\eta^2 + \Delta\phi^2 }= 1.0$, in pseudorapidity, $\eta$, 
and
azimuthal angle, $\phi$, in the HERA frame. Jets are defined as forward jets 
if,  $x_j = E_j/E_p > 0.025$, $0.5 < p^2_{Tj}/Q^2 < 4 $,
$6^o < \theta_j < 20^o$, and $p_{Tj} > 5\,$GeV (where $p_{Tj}$ is taken as an
 approximation to $k_{Tj}$). These criteria ensure that $x/x_j \leqsim 0.1$. 
For the 94 data the criteria are modified as follows. The DIS selection cuts
remain almost the same, but the $x$ region is shifted
to $5\times 10^{-4} < x < 3.5\times 10^{-3}$. Jets are found by requiring 
$E_T > 3.5\,$GeV in the cone of radius $R = 1$, and the criteria to select 
forward jets are tightened to: $x_j > 0.035$, $0.5 < p^2_{Tj}/Q^2 < 2 $,
$7^o < \theta_j < 20^o$ and $p_{Tj} > 3.5\,$GeV.
 
ZEUS have produced data on forward jets from their '95 run~\cite{ZEUS95fjet}
with similar DIS selection cuts: $E_e > 10$ GeV, $y > 0.1$: and jet selection
cuts: $x_j > 0.036$, $0.5 < p^2_{Tj}/Q^2 < 2$, $\theta_j > 8.5^o$, 
$p_{Tj} > 5$ GeV, using the cone algorithm with $R = 1$: and a somewhat 
wider kinematic range: $Q^2 \geqsim 15\,$GeV$^2$, $4.5\times 10^{-4} < x <
4.5\times 10^{-2}$.

This is the best that can 
currently be done in fulfilling the requirements $k^2_{Tj} \simeq Q^2$ and
$x/x_j$ small, given the limitations of statistics and the need to separate 
the
forward jet from the proton remnant\fnm{ee}\fnt{ee}{~It has been 
suggested that 
looking for forward $\pi^0$ or prompt $\gamma$ may be 
cleaner~\cite{KLM,K9702213}.}. This requires $x_j \leqsim 0.1$, 
essentially imposed by the lower angular cuts.

\begin{figure}[ht]

\centerline{\psfig{figure=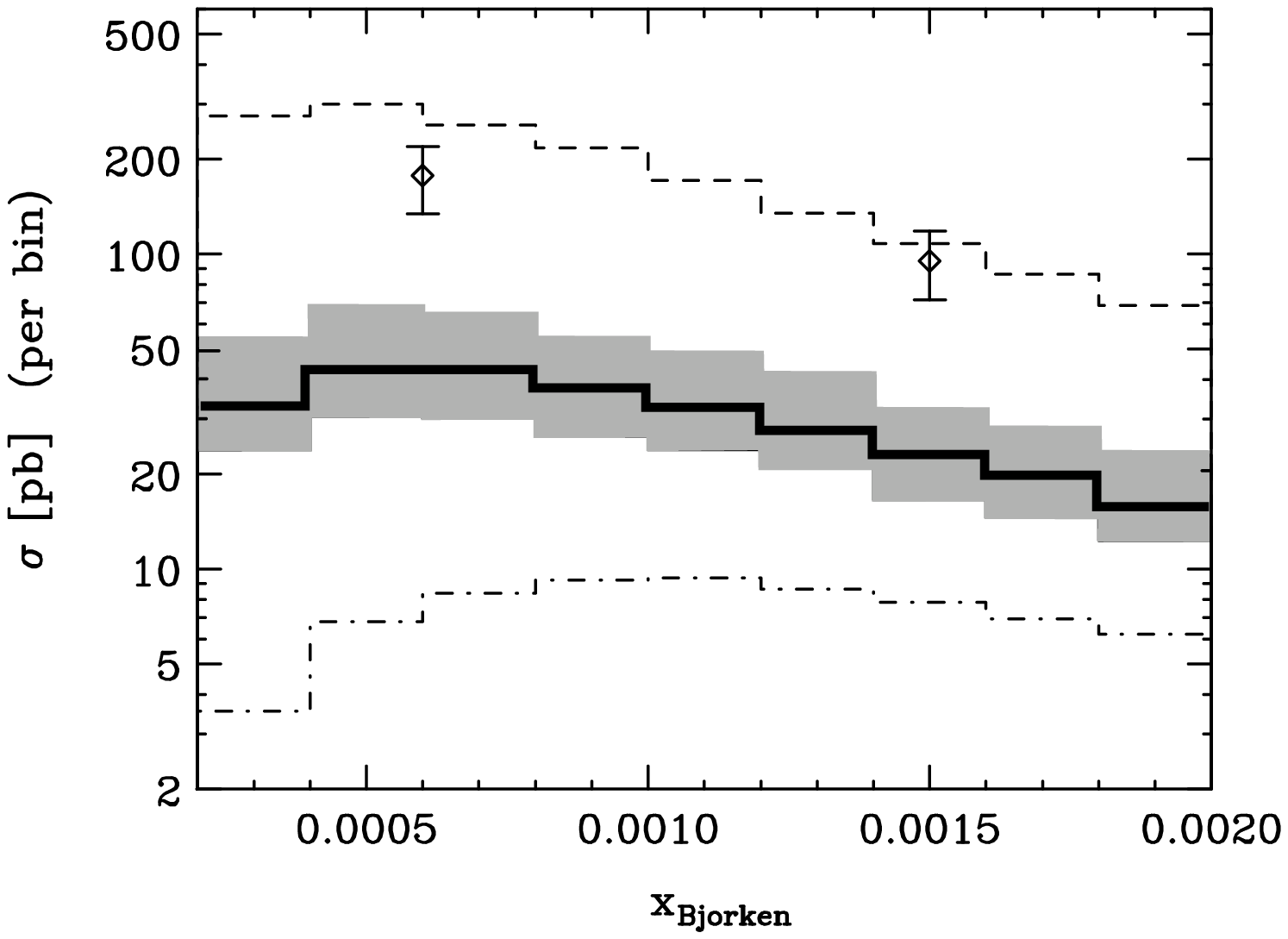,width=8cm,height=6cm}}
\fcaption{Forward jet cross-section as a function of $x$ from the NLO(LO) 
MEPJET calculations given as the solid(dash-dot) histograms. The BFKL 
calculation of Bartels is shown as the dashed histogram. H1(93) data are shown
for comparison}
\label{mirkes}
\end{figure}

For the H1(93) data the number of forward jet events, in two $x$ bins, 
is shown in Fig.~\ref{mirkes} and is compared 
with the two Monte Carlo models LEPTO and ARIADNE in Table~\ref{tab:93jet}. 
We see that the data favour ARIADNE (where there is no $k_T$ ordering) 
even when the latest versions of each Monte Carlo are used. 
\noindent
\begin{table}
\tcaption{Number of observed DIS events with a selected forward jet (H1(93)
selection), compared with predictions from the LEPTO(MEPS) and the ARIADNE(CDM)
Monte Carlos.}
\centerline{\footnotesize\smalllineskip
\begin{tabular}{cccc}\\ 
\hline
 $x\times 10^3$ & Observed events & MEPS & CDM \\ 
\hline
 $0.2 \to 1.0$&$271$& $135$ & $240$  \\
 $1.0 \to 2.0$&$158$& $ 96$ & $121$  \\ 
\hline\\
\end{tabular}}
\label{tab:93jet}
\end{table}

\begin{figure}[ht]
\centerline{\psfig{figure=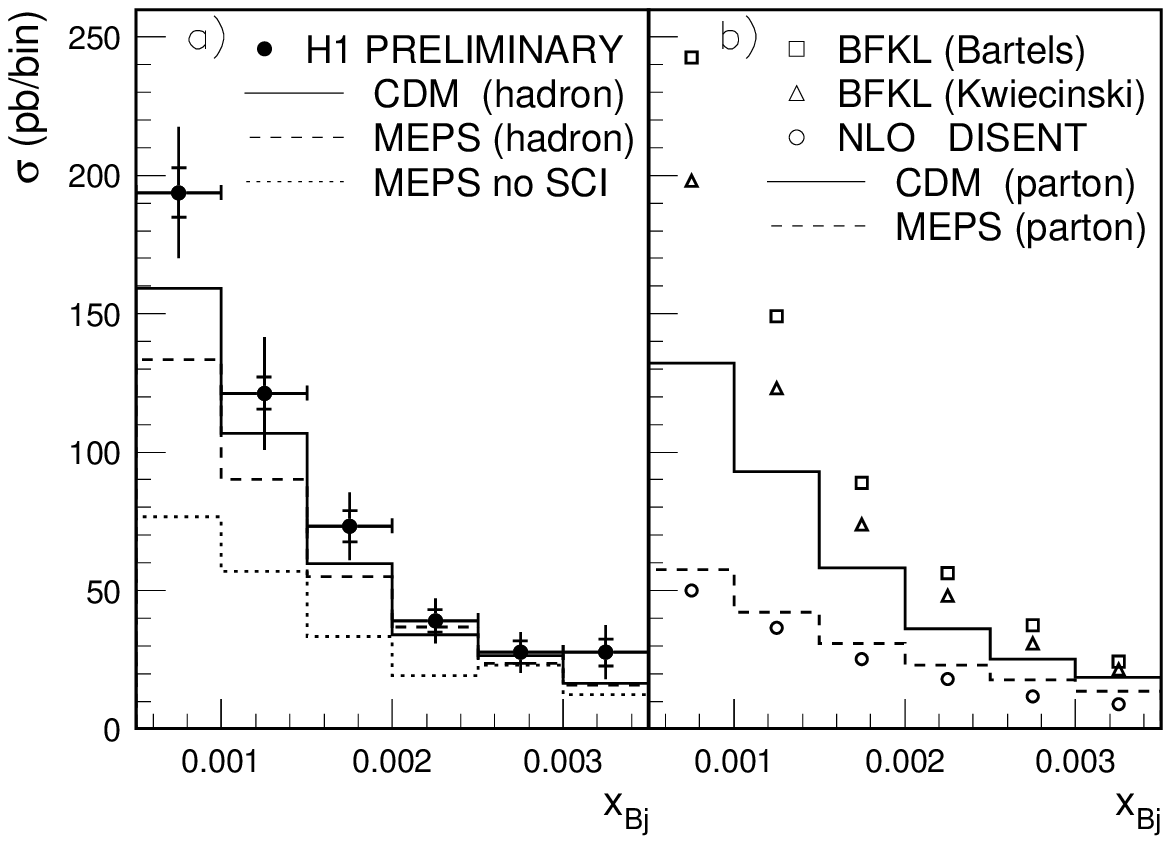,width=9cm,height=7cm}}
\fcaption{a) H1(94) preliminary data on the forward jet cross-section as a 
function of $x$ compared to the hadron level predictions of the ARIADNE(CDM) 
and LEPTO(MEPS) Monte Carlos (both with and without SCI): 
b) parton level predictions from the same Monte Carlos, and from the
NLO Monte Carlo DISENT and the BFKL calculations of Bartels et 
al~\cite{Barjnew} and Kwiecinski et al~\cite{martin_pt}}
\label{fig:fwdjets}
\end{figure}

The H1(94) data have sufficient 
statistics to look at the cross-section as a 
function of $x$ in six $x$ bins. They are shown in Fig.~\ref{fig:fwdjets}a 
together 
with the predictions of the LEPTO and ARIADNE Monte Carlos. 
 ARIADNE gives a reasonable  description of the data, while LEPTO
is somewhat worse.
Furthermore, Fig.~\ref{fig:fwdjets}b shows that for the jets 
selected it appears that the hadronization corrections are again large 
for the LEPTO model, somewhat masking the true parton dynamics.
These effects should become smaller for increased jet $E_T$.
However, it has been shown~\cite{deroeckfj} that, in the kinematic range
considered, increasing $E_T$ reduces the sensitivity to the BFKL effect.
 
\begin{figure}[ht]
\centerline{\psfig{figure=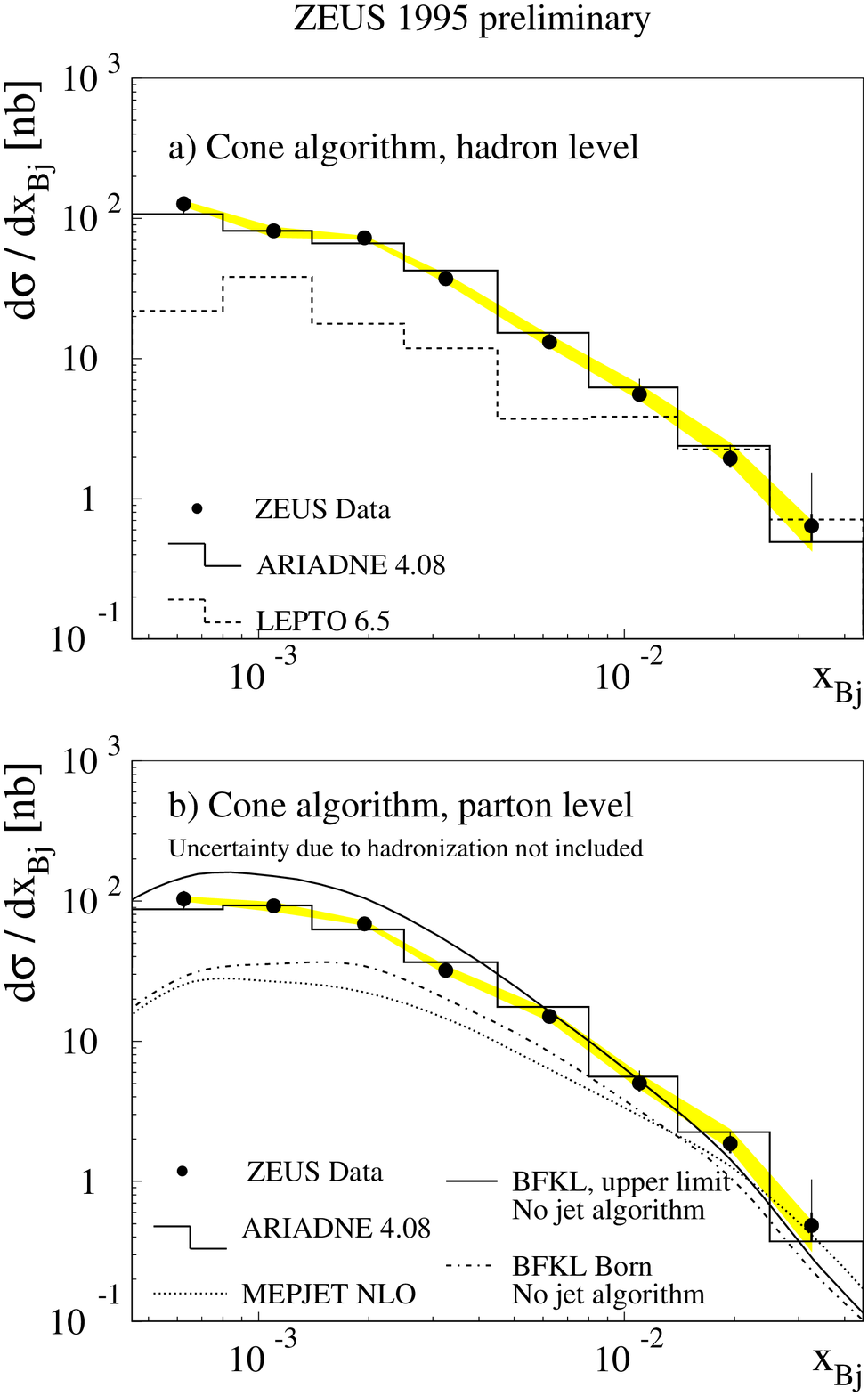,width=7cm,height=10cm}}
\fcaption{ ZEUS 95 preliminary data on the forward jet differential 
cross-section as a function of $x$ compared to the hadron level and parton 
level predictions of the ARIADNE and LEPTO Monte Carlos and to the
NLO Monte Carlo MEPJET and the BFKL calculations of Bartels et 
al~\cite{Barjnew}}
\label{ZEUS95jets}
\end{figure}

The ZEUS '95 data is shown in Fig.~\ref{ZEUS95jets}. The wider kinematic range
allows us to see clearly that the Monte Carlos agree with each other and with
the data at higher $x$ but begin to disagree for $x < 10^{-2}$. ARIADNE
gives a successful description of the data. However,
as remarked earlier, we should not interpret the success of ARIADNE as 
a definitive indication for BFKL dynamics because the ARIADNE Monte 
Carlo does 
not actually incorporate BFKL evolution. Moreover, the current LEPTO/HERWIG
 and ARIADNE Monte Carlos do not
treat the conventional dynamics of jet production exactly beyond LO. We need
calculations of both BFKL and conventional dynamics to higher orders, in order
to make a fair comparison. Such calculations are just becoming available.
\noindent

There are now three different Monte Carlos incorporating conventional
NLO jet calculations, MEPJET by 
Mirkes and Zeppenfeld~\cite{Mirkes}, DISENT by  
Catani and Seymour~\cite{catsey} and DISASTER++ by Graudenz~\cite{disaster}.
Results indicate that  the conventional predictions for the jet cross-sections
 with one of the jets going forward, with large $x_j$ and 
$k^2_{Tj}\sim Q^2$, are very sensitive to the inclusion
of higher orders. When a forward jet is required the NLO (2+1) jet 
cross-section exceeds the LO (2+1) jet
cross-section by a factor of $\sim 4$, and further corrections may be expected
from higher order effects.  However these are not expected to be so dramatic, 
since the main effect of the LO to NLO correction is dominated by the 
special kinematic requirements of these jets.
The NLO calculation of the forward jet cross-section
from MEPJET is shown compared to the H1(93) data in Fig.~\ref{mirkes} and 
compared to the ZEUS 95 data in Fig.~\ref{ZEUS95jets}. The NLO calculation 
from DISENT is shown in 
Fig.~\ref{fig:fwdjets}b. These calculations indicate that conventional 
dynamics at NLO is still inadequate to explain the data.

Bartels et al~\cite{Barjnew} have compared 
a BFKL calculation with an analytic calculation of conventional dynamics at 
the Born level.
For the BFKL calculations one considers the cross-sections for the processes
$g \to g + (n g) + q\bar q$ and $q \to q + (n g) + q\bar q$, where the many 
gluons, $(n g)$, are summed by the BFKL ladder. For the analytic calculation 
one considers the Born cross-sections for these processes without $(n g)$.
The calculations are subject to the experimental cuts used by the  
HERA collaborations, where both the '93 and the '94 data are considered, but
hadronization of the forward parton into a jet is not considered.  
These calculations are illustrated on the H1(93) data in Fig.~\ref{mirkes},
on the H1(94) data in Fig.~\ref{fig:fwdjets}b and on the ZEUS '95 data in
Fig.~\ref{ZEUS95jets}. Kwiecinski et al~\cite{martin_pt} have also made a BFKL
calculation of the forward jet cross-section at the parton level,
which differs from that of 
Bartels et al by including massive charm contributions and by incorporating
the running of $\alpha_s$ numerically. This calculation is compared to the
H1(94) data in Fig.~\ref{fig:fwdjets}b. 
 
These comparisons seem to indicate that BFKL dynamics are
necessary. However, we should be cautious because there are still some 
limitations to the calculations. It has 
been found that conventional NLO calculations also cannot explain the rate 
of centrally produced dijets in HERA data~\cite{dewolf2}. 
However, a study of NLO calculations for jet production in photoproduction
events has shown that the calculations\cite{frix}
 become unreliable in some special
regions of phase space for dijet quantities defined with equal cuts
on the transverse energy of the observed jets, as is usually done
for dijet measurements. This issue needs to be clarified for the DIS case.
The discrepancy between data and conventional NLO calculations could also
be due to the need for conventional higher order 
corrections or it may be attributable to more subtle BFKL effects.  
The BFKL calculations are also not exact: charm production is
not properly treated and hadronization is still not included. We also note 
that Jung has suggested that a resolved photon contribution in DIS events 
could be a possible explanation
for rising forward jet cross-sections at low $x$~\cite{Jung}. 

In conclusion the data from the hadron final states gives tantalizing 
indications that BFKL dynamics are necessary. We await further developments 
both theoretical, in more sophisticated calculations, using the newly 
developed BFKL and CCFM(LDC) Monte Carlos, and 
experimental, in measurement of further interesting processes.
We note that a promising novel way to investigate BFKL dynamics
has been proposed for deep inelastic $\gamma^* \gamma^*$ scattering at LEP
and at future linear $e^+e^-$ colliders
\cite{bartelbfkl,brodsky}. The total cross section for this process
is expected to be strongly affected by BFKL dynamics.
The 1997 LEP data may yield a first
measurement of this process. 

\section{ The low $Q^2$ region}
\label{sec:lowq2}

In Sec.~\ref{sec:data} it was shown that the HERA data now cover the region
down to $Q^2 \sim 0.1 $ GeV$^2$. The major goal for these data 
is the
study of the transition  from deep inelastic scattering to the 
photoproduction $(Q^2 \rightarrow 0)$ limit. At $Q^2 = 0$ the dominant
processes are of non-perturbative nature and are well described by 
Regge phenomenology. As $Q^2$ increases, the exchanged photon is expected to 
shrink and pQCD to take over. The HERA data clearly 
demonstrate that the inclusive cross-sections
 in the  region $Q^2 > 1$  GeV$^2$ can be described by 
pQCD. How far down in $Q^2$ does pQCD continue to 
work? Where exactly is the transition 
region and is this transition smooth?
These are some of the intriguing  questions to be answered by 
these low $Q^2$ data.
The transition region is expected to shed light on the interplay between soft 
and hard interactions. Several phenomenological models attempt to 
describe this region and will be compared to the data.

\begin{figure}[htbp]
\vspace*{13pt}
\begin{center}
\psfig{figure=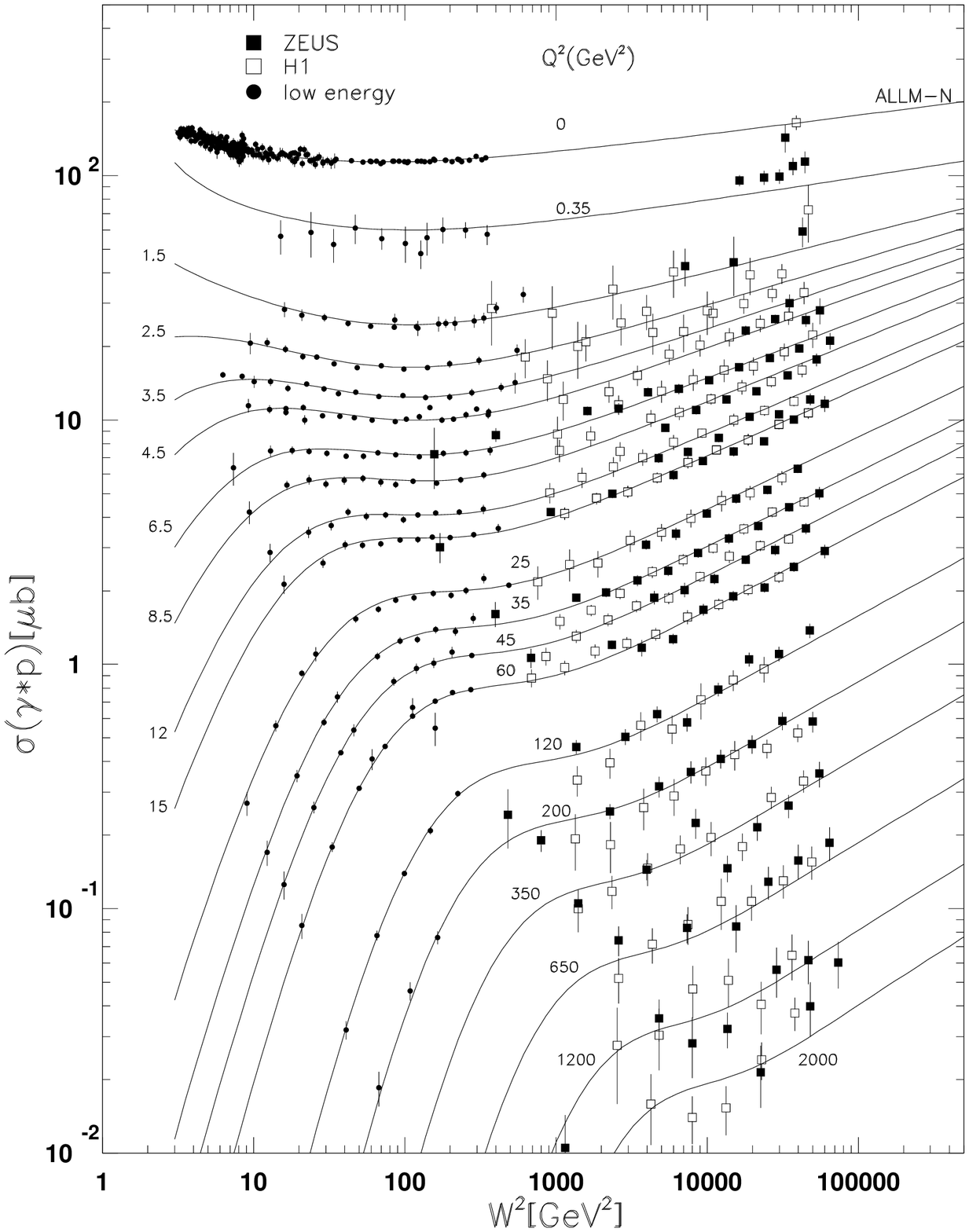,bbllx=0pt,bblly=0pt,bburx=525pt,bbury=800pt,height=.95\textheight} 
\fcaption{The total $\gamma^*p$ cross-section from $F_2$ 
measurements as a function of $W^2$ for different $Q^2$. 
The curves are the model of ALLM~\cite{ALLM}.}
\label{fig:sigtotw}
\end{center}
\end{figure}

It is instructive to remember that the behaviour of $F_2$ at small $x$ 
is related to the behaviour
of $\sigma^{\gamma^*p}$ at large $\gamma^*p$ centre of mass energy $W$. 
Recall Eqs.$~\ref{eq:sigTL}-\ref{eq:epsgam}$ and the small $x$ approximation
Eq.~\ref{eq:relatesim}. For ease of reference we shall write this as
\begin{equation}
\sigma^{\gamma^*p}(W^2,Q^2)\approx{4\pi^2\alpha\over Q^2}F_2(x,Q^2)
\end{equation}
where $\sigma^{\gamma^*p}=\sigma_T+\sigma_L$.
As we discussed in Sec.~\ref{sec:pdfhilo}, Regge theory provides a 
reasonable description of the $s$ dependence of 
the photoproduction cross-section, and we might thus expect it to describe
low $Q^2$ DIS data. We recall that in this picture the $s(\gamma^*p) = W^2$ 
dependence of the cross-section at high energy is given by 
$\sigma^{\gamma^*p} \sim  W^{2\lambda}$, 
where $\lambda$ is a constant determined by the intercept of the soft Pomeron 
trajectory, $\alpha_P = 1 + \lambda$, and $\lambda \simeq 0.08$ gives 
a good fit to hadron-hadron cross-sections~\cite{DoLa}. (We note that Cudell
et al~\cite{cudell} have suggested that $\lambda = 0.096$ gives a better fit.)
Using this idea Donnachie and 
Landshoff (DOLA)~\cite{DoLa2} developed a model which successfully 
described the low $Q^2$ fixed target $F_2$ data (for $Q^2<10\,$GeV$^2$) 
and gave the correct limit as $Q^2\to 0$. 
The structure function is described by a sum of two terms: one for the Pomeron
(which describes the contribution of sea quarks and dominates at high energy) 
and one for a Reggeon of intercept $\alpha_R \simeq 0.5$
(which describes the contribution of valence quarks). Each of these
terms takes the form 
{\small \begin{equation}
  \left(\frac{Q^2}{Q^2 + M^2_{R,P}}\right)^ {\alpha_{R,P}} x^{(1-\alpha_{R,P})} (1 - x)^{\beta_{R,P}}
\label{eq:DOLA}
\end{equation}}
\noindent
By construction the $W$ 
dependence of $\sigma^{\gamma^*p}$ is very weak, as it is for 
hadron-hadron total cross-sections. At HERA it is clear
that this simple picture cannot be adequate as the steep rise of $F_2$ 
as $x\to 0$ becomes a strong rise of $\sigma^{\gamma^*p}$ as $s$ increases.
This is shown in Fig.~\ref{fig:sigtotw} which shows $\sigma^{\gamma^*p}$ as 
a function of $W^2$ for 
different values of $Q^2$ from 0 to $2000\,$GeV$^2$.

\begin{figure}[htbp]
\vspace*{13pt}
\begin{center}
\psfig{figure=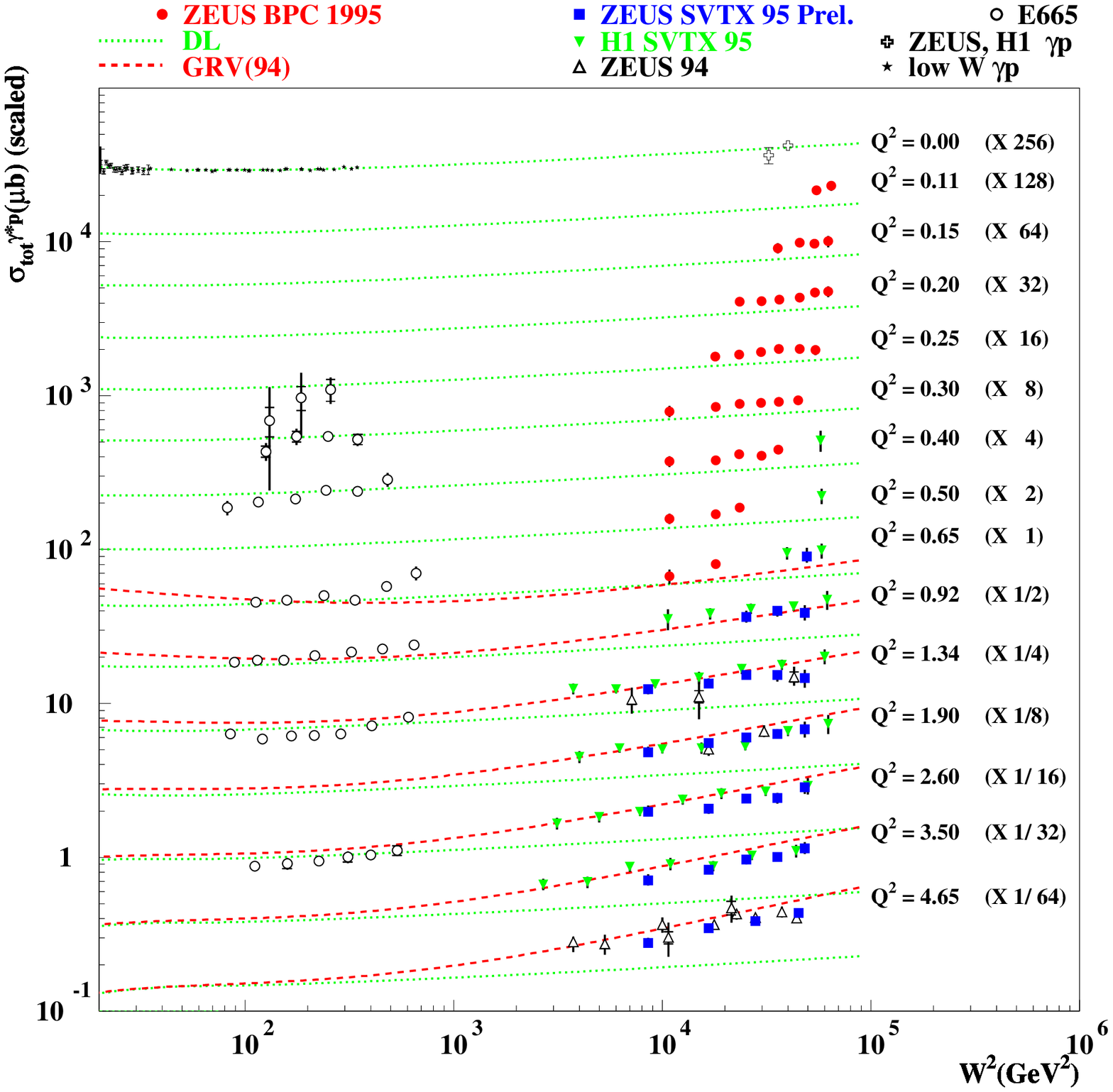,bbllx=30pt,bblly=100pt,bburx=800pt,bbury=770pt,height=.8\textheight} 
\fcaption{The total $\gamma^*p$ cross-section from $F_2$ 
measurements as a function of $W^2$ at low  $Q^2$. 
The curves are the models of DOLA (dotted lines) and GRV (dashed lines).}
\label{fig:sigtotw2}
\end{center}
\end{figure}

\begin{figure}[ht]
\vspace*{13pt}
\begin{center}
\psfig{figure=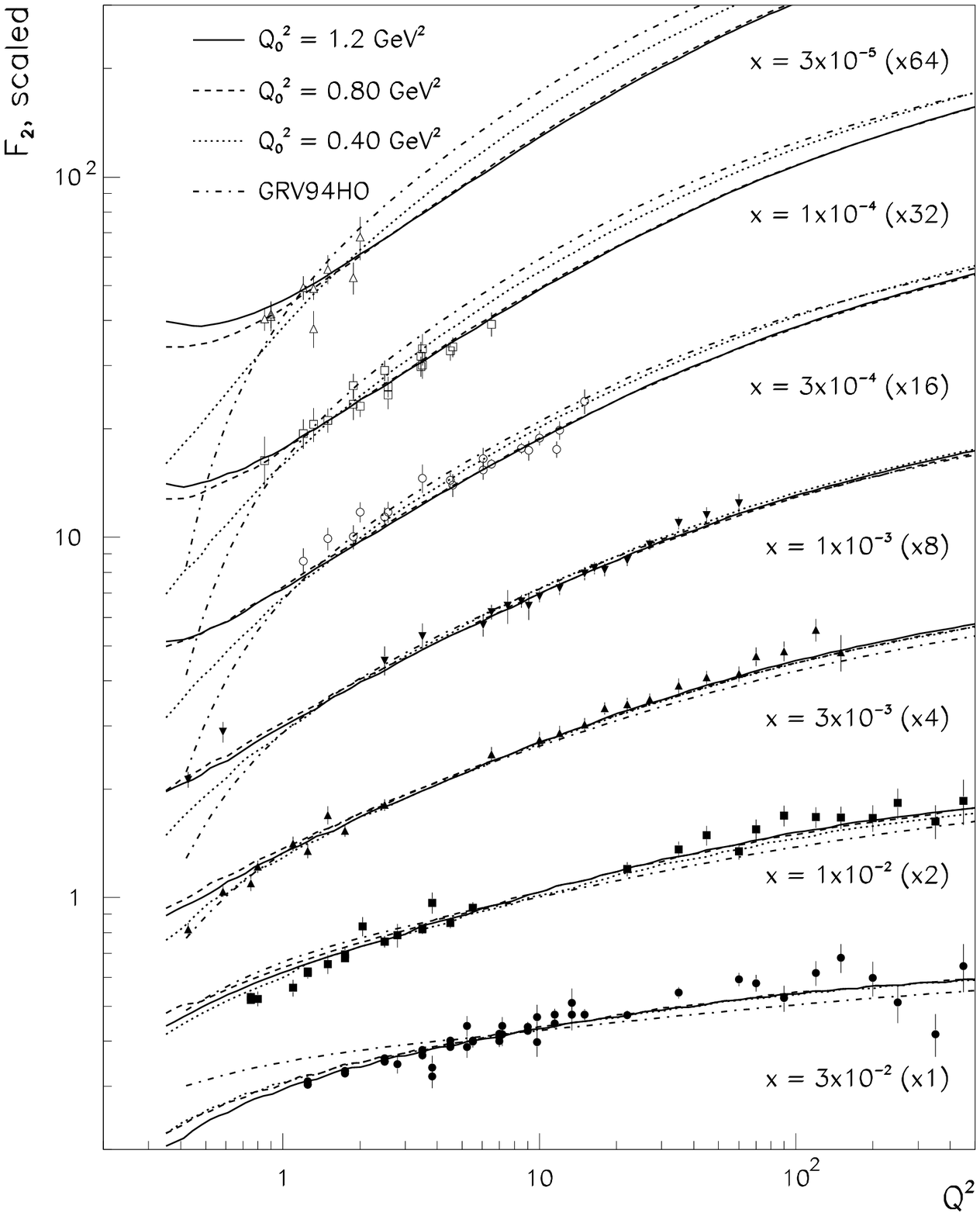,bbllx=-60pt,bblly=80pt,bburx=430pt,bbury=700pt,height=11cm} 
\fcaption{$F_2$ data and results from a perturbative QCD DGLAP global fit
for 3 different starting scales $Q_0^2$ = 0.4, 0.8 and 12. GeV$^2$. 
The GRV prarametrization for $F_2$ is shown for comparison. The data sets used
are ZEUS(94), ZEUS(95)SVX (preliminary) and H1(95)SVX, ZEUS(95)BPC, NMC(97), 
BCDMS, SLAC, E665. The data in each $x$ bin are scaled by the factors 
indicated in brackets}
\label{fig:gluon_scale}
\end{center}
\end{figure}

\begin{figure}[htbp]
\vspace*{13pt}
\begin{center}
\psfig{figure=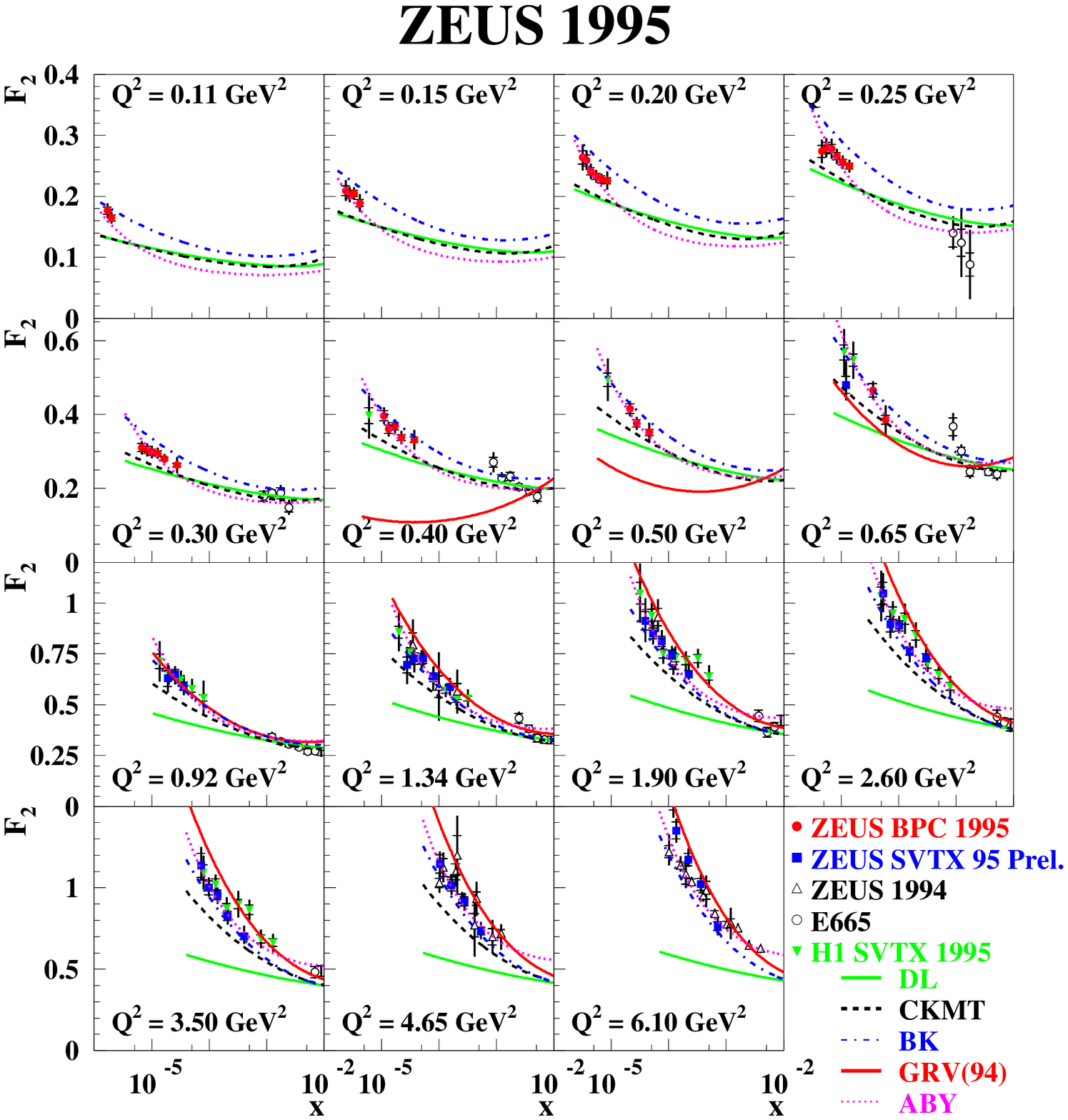,bbllx=50pt,bblly=70pt,bburx=650pt,bbury=770pt,height=.87\textheight}
\fcaption{Measurement of the proton structure function $F_2(x,Q^2)$
in the low $Q^2$ region by H1(95)SVX (full triangles), ZEUS(95)BPC 
(full circles) and ZEUS(95)SVX (preliminary)(full squares) 
 together with previously published results from ZEUS(94) (open triangles) and
E665  (open squares). Various  predictions for $F_2$ are compared with the 
data: the model of DOLA (full line/small), the model of CKMT
(dashed line ), the model of BK (dashed-dotted line),
 the parametrization of GRV (full line/large) and the model 
of ABY (dotted line).}
\label{fig:f2_lowQ2_all}
\end{center}
\end{figure}

It is of interest to check if the
new  low $Q^2$ data from HERA approach the DOLA limit.
Data in the $Q^2$ range, $0.11 < Q^2 < 6.5$ GeV$^2$, are shown in 
Fig.~\ref{fig:sigtotw2} compared with the DOLA and GRV94 
(see Sec.~\ref{sec:GRV}) predictions.
Considering the HERA data together with the lower $W^2$ data from E665, we see
that the energy dependence predicted by the DOLA model is too weak
for $Q^2$ values above $\sim 0.4\,$GeV$^2$. For lower $Q^2$ values we do not
have the benefit of a large $W$ lever arm in the data. The shape 
of the model is compatible with the 
data  in this limited $W$ interval, but the magnitude of the  DOLA
prediction lies significantly below the data.   
We note that the HERA photoproduction data were included in the DOLA fit 
and hence affect the normalization somewhat. 
These photoproduction cross-sections 
have total errors of typically 15\%-20\%~\cite{H1STOT,ZEUSSTOT}.
By contrast we see that the GRV prediction describes the data reasonably
well for higher $Q^2$ and begins to fail for $Q^2 \leqsim 0.8\,$GeV$^2$
(as shown more clearly in Fig.~\ref{fig:f2_lowQ2_all}).
 ZEUS has  studied the effect of lowering the starting scale $Q^2_0$ and 
including lower $Q^2$ data in the NLO DGLAP fit which they used to extract 
the gluon distribution (see Sec.~\ref{sec:gluon}).  
The result is shown in Fig.~\ref{fig:gluon_scale}.
It is found that standard DGLAP works well for data and
 starting scales  $Q_0^2$ as low as 0.8 GeV$^2$,
 but for fits starting from lower $Q^2$ values the scaling
violations become too large at low $x$.
 ZEUS has also introduced higher twist terms of the form 
$f(x)/Q^2$ into this fit and it is found that such terms are negligible for 
$x < 0.02$ and that the fit it still unable to 
describe the data for starting scales lower than $0.8$ GeV$^2$.
This comparison demonstrates that data with $Q^2$ less than about 0.8 GeV$^2$
 approach the pure  soft regime,  
and  can no longer be described by leading twist  pQCD (DGLAP)
calculation, or even by the inclusion of conventional higher twist terms.
Hence these data constitute the effective transition 
region, and the transition from DIS to photoproduction appears to be smooth.


\begin{figure}[ht]
\vspace*{13pt}
\begin{center}
\psfig{figure=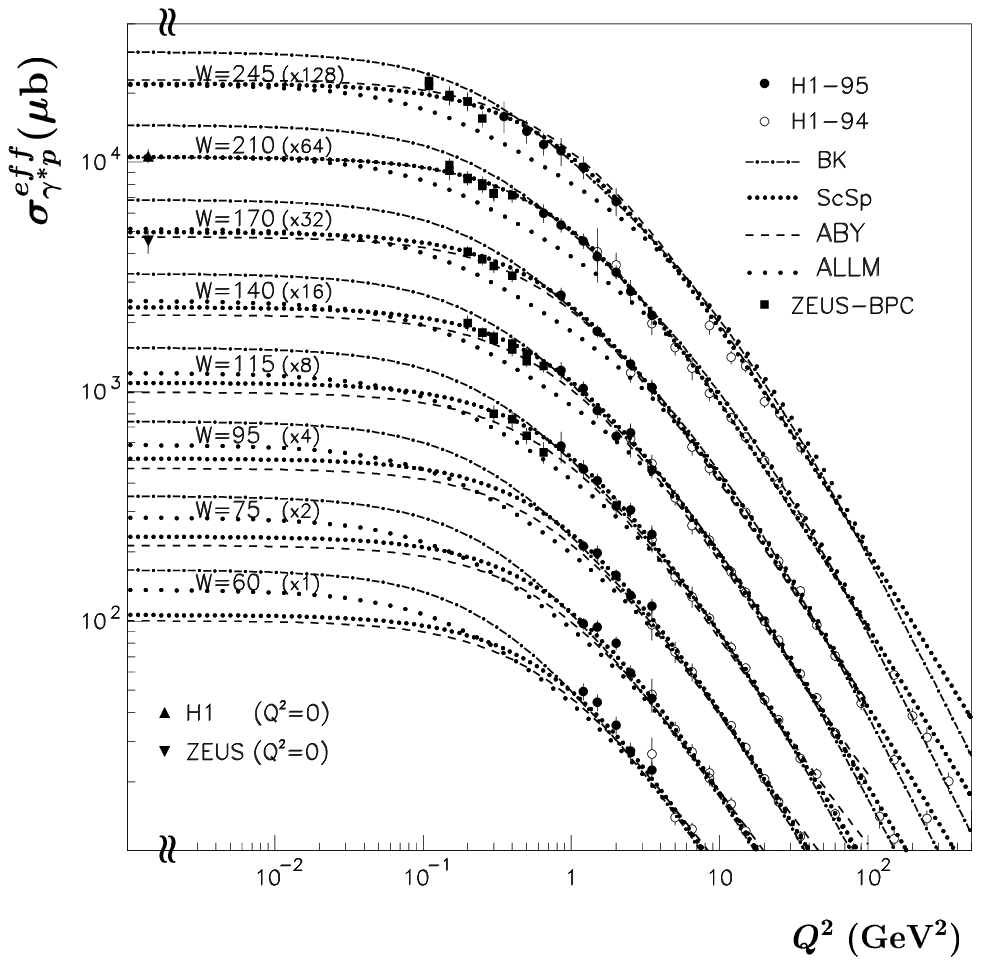,bbllx=160pt,bblly=300pt,bburx=400pt,bbury=600pt,height=10cm}
\fcaption{ Measurement of the virtual photon-proton cross-section
 $\sigma_{\gamma^*p}^{eff}$
    as a function of $Q^2$ at various values of   $W$ (in GeV),
from H1(95)SVX and H1(94) (circles) and ZEUS(95)BPC (squares). 
The cross-sections for consecutive $W$ values are multiplied
with the factors indicated in brackets.
The  errors represent the statistical and systematic errors added 
in quadrature. The photoproduction points (cross: $W= 210 $ GeV, diamond:
$W= 170$ GeV) are from~\cite{H1STOT,ZEUSSTOT}.
Global normalization uncertainties are not included in the errors shown.
The curves represent the  ALLM (dotted/small),  BK
(dashed-dotted line), ScSp (dotted line/large) and ABY (dashed)}
\label{fig:w_all}
\end{center}
\end{figure}

 We now turn  to the phenomenological
developments and models which aim to describe this region.
More details and the formulae for many of the models 
discussed below are 
summarized in ref.~\cite{levy96} and in the very extensive review of low 
$x$, low $Q^2$ electroproduction by Badelek and Kwiecinski~\cite{badelekR}. 
 Comparisons with data are presented  in Fig.~\ref{fig:f2_lowQ2_all},
where $F_2$ is shown as a function of $x$ for the smallest $Q^2$ bins, and in
 Fig.~\ref{fig:w_all}, where the quantity $\sigma^{eff}$ is shown as function 
of $Q^2$ for different $W$ bins. The quantity $\sigma^{eff}$ is the effective 
measured virtual photon-proton  cross-section, $\sigma^{eff} = \sigma_T +
\varepsilon \sigma_L$, for $ep$ collisions in the HERA kinematic range, 
as defined 
by Eqs.~\ref{eq:sigTL},~\ref{eq:epsgam},~\ref{eq:vareps}. It is used because it
 can be determined from the data without assumptions about $R$. Since 
$\varepsilon \geqsim 0.9$, for most of the kinematic range, it also represents
most of the total $\sigma^{\gamma^*p}$ cross-section.

A number of groups have 
modified the Regge approach of DOLA by allowing the parameter 
$\lambda(=\alpha_P - 1)$, which fixes the Pomeron intercept, 
to vary with $Q^2$, for example ALLM~\cite{ALLM} and 
CKMT~\cite{CKMT}. Supposing that the soft Pomeron of DOLA, 
is not a `bare' Pomeron but a shadowed Pomeron, 
then as $Q^2$ increases shadowing gradually disappears and the bare Pomeron 
is revealed to be a harder Pomeron of larger intercept~\cite{CKMT,Levin}. 
In the approach of CKMT the data are fitted to the sum of contributions from
Pomeron exchange and from Reggeon exchange using similar forms to that 
given in Eq.~\ref{eq:DOLA}. 
However, the intercept of the Pomeron Regge trajectory is expressed as
$\alpha_P = 1 + \lambda(Q^2)$ and the form taken for the $Q^2$ dependence is 
$\lambda (Q^2) = \lambda_0(1+2Q^2/(Q^2+d))$, where the parameters $\lambda_0$
and $d$ have been determined to be 
$\lambda_0 = 0.07684$ and $d= 1.117$ GeV$^2$, from a fit to 
 NMC(92) data in the region $1 < Q^2 < 5\,$GeV$^2$ and 
real photoproduction data
from HERA and from lower $W$ fixed target experiments. 
The model assumes that this prescription accounts for
non-perturbative contributions to $F_2$ for $Q^2$ values up to about 
2 GeV$^2$. For higher $Q^2$ the pQCD DGLAP
 evolution equations  are applied to predict the 
$Q^2$ dependence of $F_2$. Thus, since the effective value of 
$\lambda$ quickly becomes larger than that used by DOLA as $Q^2$ increases,
the structure function $F_2$ predicted by CKMT rises faster with decreasing 
$x$ (as $Q^2$ increases)  than the DOLA prediction, as shown in 
Fig.~\ref{fig:f2_lowQ2_all}. However, although the CKMT prediction
is  systematically above that from DOLA, for $Q^2 \geqsim 0.3\,$GeV$^2$
it is still below the data. 

 The model of Abramowicz et al~\cite{ALLM}
(ALLM) is also based on a Regge motivated approach extended into the large 
$Q^2$ regime compatibly with pQCD expectations. 
Contributions from Reggeon and Pomeron exchange, of a similar form to that 
given in Eq.~\ref{eq:DOLA}, are generalized such that the dependence
 $x^{1-\alpha_{R,P}}$ becomes $x^{-\lambda_{R,P}(Q^2)}$ for both the Reggeon 
and the Pomeron terms
 and the parameters $\lambda_{R,P}$ vary with $Q^2$ logarithmically, emulating
pQCD evolution in the high $Q^2$ region. This model gives a parametrization of
the whole $x,Q^2$ phase space. It was fitted to DIS and 
photoproduction data before the final HERA(94) and HERA(95)(low $Q^2$) data
became available. The fitted value of $\lambda_P$ changes smoothly from 
0.08 at $Q^2 = 0$, to about 0.4 at $Q^2 \sim 10^3$ GeV. 
In Fig.~\ref{fig:w_all} one can see that this model has the correct limit at 
$Q^2 = 0$ and that it agrees with data for $Q^2 > 2$ GeV$^2$. However
the prediction is below  the data for small $Q^2$ values, such that the 
transition from the soft to the hard regime is too slow.

The model of  Badelek and Kwiecinski~\cite{badelek} (BK)
 combines the concepts of Generalized Vector Meson Dominance (GVMD) 
with dynamical parton  models such as that  of GRV. 
Here the structure function is 
assumed to be 
the sum of two contributions:  a GVMD model term
$F_2^{VMD}$ 
 and a partonic term $F_2^{part}$. For the latter the 
GRV model was used and it becomes 
dominant above $Q^2\sim$ 1 GeV$^2$. 
In this model  $F_2$ is given by
\begin{equation}
F_2(x,Q^2)= F_2^{VMD}(x,Q^2) + 
\frac{Q^2}{Q^2+Q^2_{VMD}}F_2^{part}(\overline{x},Q^2+Q^2_{VMD}),
\end{equation}
with $\overline{x} = (Q^2+Q^2_{VMD})/(W^2+Q^2+Q^2_{VMD})$.
Vector mesons with mass squared smaller than $Q^2_{VMD}$
are included in the $F_2^{VMD}$ term. The value of  $Q^2_{VMD}$ is the only
parameter of this model and it was chosen, rather than fitted,
to be 1.2 GeV$^2$ before the HERA low $Q^2$ data were available. 
The model has perforce a smooth transition 
from the pQCD region to the real photon limit, but the predictions lie
above the data in the limit $Q^2 \rightarrow 0$, as can be seen in 
Fig.~\ref{fig:w_all}. This figure and Fig.~\ref{fig:f2_lowQ2_all}
illustrate that there is better agreement with HERA data for the BK
model than for the ALLM, however there is a tendency for the BK predictions
to lie too high at the lowest $Q^2$ values, whereas the ALLM predictions lie
too low. Thus it seems that the transition from the hard to the soft regime 
is also too slow in the BK parametrization.

A GVMD inspired approach has been proposed by Schildknecht and Spiesberger 
(ScSp)~\cite{schildknecht} 
to fit the low and medium $Q^2$ HERA data up to $Q^2$ values of 350 GeV$^2$. 
For real photoproduction
the cross-section is almost saturated (up to 78\%~\cite{sakurai}) 
by low mass vector meson states, namely $\rho^0, \omega, \phi$.
For non-zero $Q^2$ these contributions decrease rapidly and, according
to GVMD, their role at small $x$ is taken over by more massive states.
The authors make a 4 parameter fit to all
available $\gamma^*p$ data with $W > 30$ GeV,
and obtain a remarkably good description of the HERA data, as shown in 
Fig.~\ref{fig:w_all}.
A similar approach has been proposed in~\cite{shaw}, where in addition
to the pure GVMD contribution, a region of large masses, $m$, 
and large  coherence lengths, $l =
(Mx(1+m^2/Q^2))^{-1}$, of photon fluctuations is considered. 
In this region the authors expect that 
the hadronic behaviour of the photon  becomes 
a complicated multi-jet state. This model allows for good fits  to 
HERA(94) data and fixed target data, but has not been confronted yet with 
the very low $Q^2$ HERA(95)  data.

A different  approach to  the low $Q^2$ behaviour of $F_2$ in the transition 
region has been presented in~\cite{Adel,Adel2} by Adel, 
Barreiro and Yndurain (ABY). It assumes that perturbative  QCD  evolution
is applicable  to the lowest values of $Q^2$. The strong coupling constant 
is assumed to become independent of $Q^2$ for values below roughly 1 GeV$^2$,
that is $\alpha_s$ `saturates'. There are assumed to be two contributions
to $F_2$, one with a singular and one with a non-singular input 
(see Sec.~\ref{sec:ynd}) to pQCD
evolution. These translate into hard and soft contributions to the
$\gamma^*p$ cross-section respectively. 
The hard contribution prevents  $F_2$ decreasing with decreasing $x$ 
for $Q^2$ values below 1 GeV$^2$ and thus remedies the failure of the
GRV approach. The form of the singular input is $\sim x^{-\lambda_S}$, 
with $\lambda_S = 0.47$ independent of $Q^2$, chosen to ensure that
$F_2 \to 0$ as $Q^2 \to 0$. The form of the non-singular input is 
$\sim Q^2/(Q^2 + M^2)$, $M = 0.87$ GeV, and this does not evolve with $Q^2$
for $Q^2 \leqsim 10\,$GeV$^2$. Above this $Q^2$ a constant term is used for the
non-singular input to the DGLAP equations, and evolution is started from 
$Q^2_0 \sim 2\,$GeV$^2$.  The result of a fit of the ABY
approach to data is compared with the new $F_2$ results
in Fig.~\ref{fig:f2_lowQ2_all} and Fig.~\ref{fig:w_all}. 
Note that this fit used the HERA(94) data ($Q^2 > 1.5 $ GeV$^2$)
and also  a preliminary version of the ZEUS(95)BPC low 
$Q^2$ data, but not the H1(95)SVX low $Q^2$ data. There is 
 good agreement with present data, but one can see that in this approach
the rise of $F_2$ with decreasing $x$
occurs at lower $x$ than for any of the other approaches,
namely for  $x <10^{-4}$. In this sort of approach the singular
(hard) term always dominates as $x \to 0$ ($W$ increases).
Thus it should be easy to check these predictions 
with lower $x$ data.

There are several other models which use both a hard and a soft component to
explain the data in the transition region~\cite{Troshin,Kot,LGM}.
Gotsman et al~\cite{LGM} have
developed an explanation for the transition region using soft and hard 
contributions, which suggests both that the hard contribution can still be
significant at $Q^2 = 0$ and that the soft contribution can be sizeable 
at large $Q^2$. Both ABY and Troshin and 
Tyurin~\cite{Troshin} have observed that the Donnachie Landshoff fit of the
soft Pomeron to high energy hadron-hadron cross-section data using the form,
$\sigma_{tot} \sim s^{0.08}$,
is not the only good fit to available data. The form, $\sigma_{tot} \sim
a + b s^{0.5}$, which combines contributions from a soft and a hard Pomeron, 
also gives a good fit. One may contrast approaches which consider two
Pomerons, hard and soft, with approaches which consider a single Pomeron
whose intercept changes with $Q^2$. It is important to remember that both of 
the forms given above are pre-asymptotic, unitarity requires that 
cross-sections rise as $\ln^2 s$ asymptotically~\fnm{ff}\fnt{ff}{~Strictly one 
should not use the terminology Pomeron in the pre-asymptotic regime, 
but this has now become common practice.}.

\begin{figure}[ht]
\vspace*{13pt}
\begin{center}
\psfig{figure=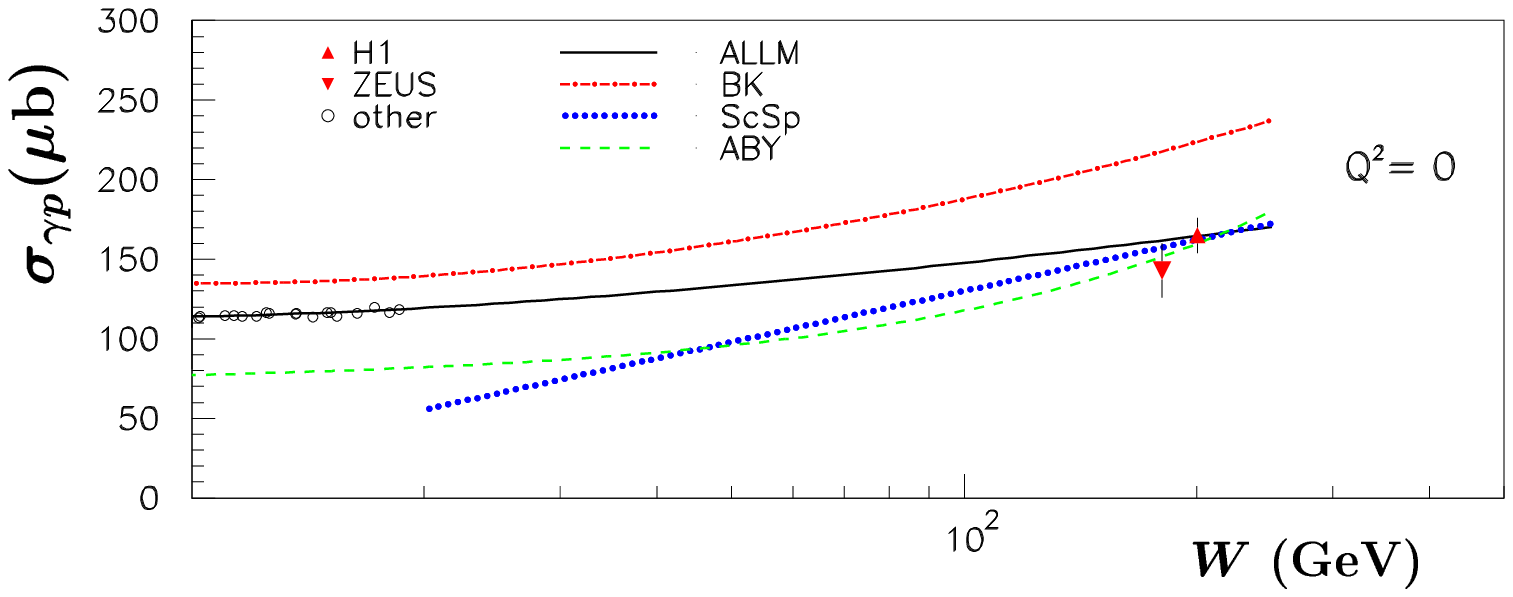,bbllx=20pt,bblly=400pt,bburx=480pt,bbury=600pt,width=10cm} 
\fcaption{The $W$ dependence of the photoproduction cross-section. 
The curves represent the  
 ALLM (full line),  BK
(dashed-dotted line), ScSp (dotted line/large) and 
ABY (dotted line/small)}
\label{fig:w_q20}
\end{center}
\end{figure}

The ABY and GVDM models appear to give a  good description of the data 
shown in Fig.~\ref{fig:w_all}. Can we be satisfied with these models? 
The answer is partially given by checking the 
$W$ dependence of the photoproduction cross-section~\cite{shekelyan}.
This is shown in Fig.~\ref{fig:w_q20}
 for the ALLM, BK, ABY and GVDM models. The ALLM model includes all
these data in their fit, while other models include data above 
a minimum $W$ value (ScSp, ABY), or no data at all (BK).
Consequently ALLM describes the data, while the other models fail.
In particular, naively extrapolating the GVDM prediction to 
low $W$ yields negative cross-sections for $W < 6 $ GeV. It has also been
pointed out that the ratio $R$ for this model is anomalously high in the 
high $Q^2$ region~\cite{martin}.

In summary, it turns out that the region $0.1 < Q^2< 1$ GeV$^2$
spans a kinematic range in which Regge or VMD inspired models describe the 
data at low $Q^2$, and models based on pQCD account well
for the higher $Q^2$ domain (but see the discussion in Sec.~\ref{sec:summary} 
for comments on the reliability of conventional pQCD at such $Q^2$ values). 
The road towards a full understanding of this region lies in finding a proper
marriage between these two limits, as has already been attempted with partial
success, and further work is in progress~\cite{martin}. Clearly 
the HERA data in the low $(x,Q^2)$  region will help to discriminate 
 between different theoretical approaches to low-$Q^2$ dynamics.
Improved measurements of $\sigma^{\gamma p}$ over a wider $W$ range 
from HERA will also be important.

\section{{\boldmath $ep$} cross-sections at very large $Q^2$}
\label{sec:hiq2}

Complete formulae for the NC and CC cross-sections applicable at high $Q^2$ 
have been given 
in Secs.~\ref{sec:NCxsec},~\ref{sec:CCxsec}. At $Q^2$
of order $M_Z^2$ the parity violating structure function $F_3$ gives a
positive contribution to $\sigma^{NC}(e^-p)$ but reduces 
$\sigma^{NC}(e^+p)$. At such large $Q^2$ values, the CC cross-section
is of comparable magnitude to the NC -- a visible manifestation of 
electroweak unification -- with $\sigma^{CC}(e^-p)$ larger than 
$\sigma^{CC}(e^+p)$ again from the $F_3$ contributions.

\begin{figure}[ht]
\centerline{\psfig{figure=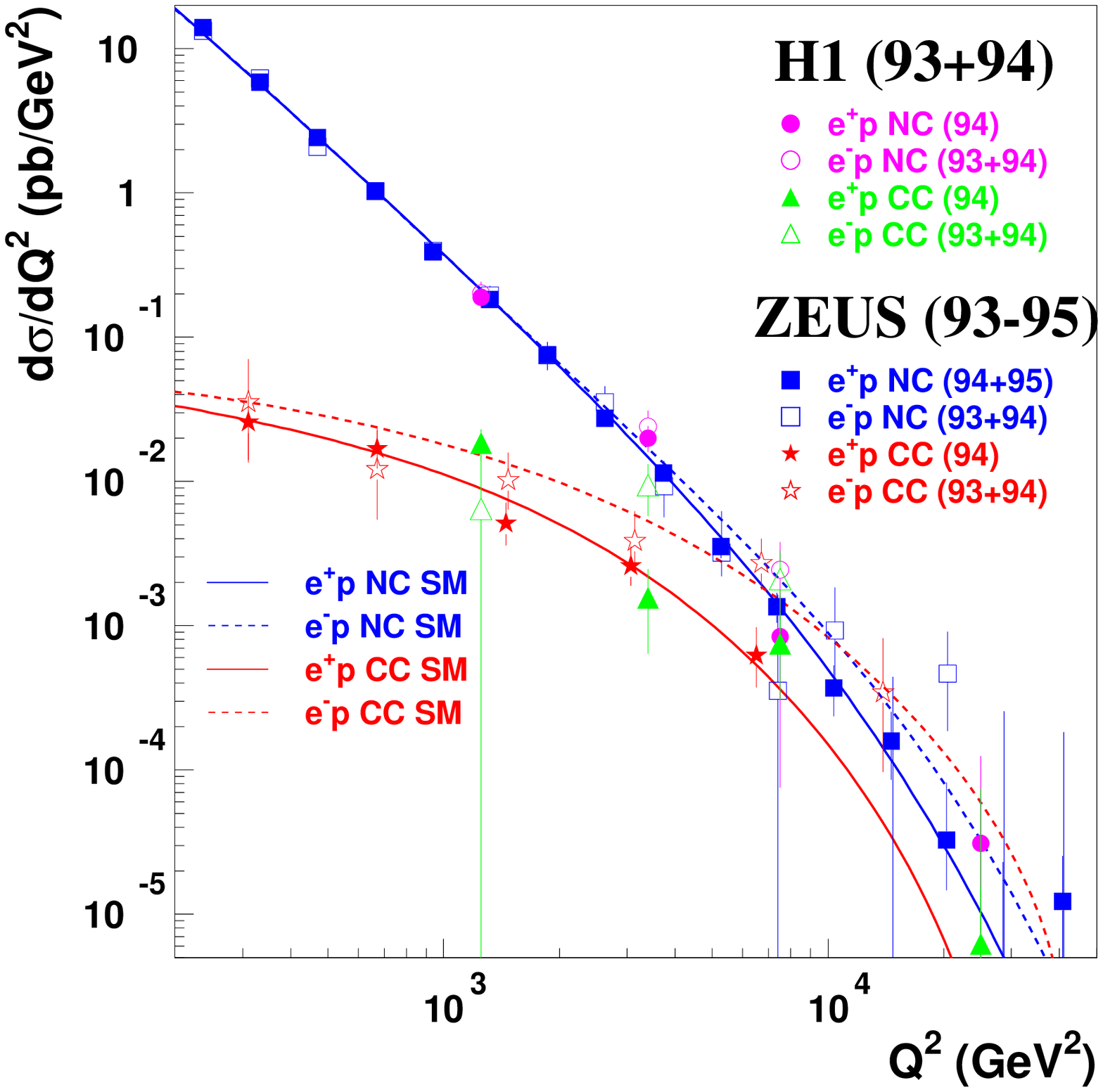,width=8cm,height=12cm}}
\fcaption{High $Q^2$ CC and NC cross-sections measured at HERA compared
to Standard Model predictions.}
\label{fig:hiq2xsecs}
\end{figure}

Fig.~\ref{fig:hiq2xsecs} shows the cross-sections at large $Q^2$ from
HERA data collected in the period 1993-95~\cite{hiq2_ichep96}. 
At $Q^2\sim 100\,$GeV$^2$ 
$\sigma^{NC}$ is dominated by $\gamma$ exchange, is charge 
independent and is two orders of magnitude larger than the CC 
cross-section. For $Q^2\geq 4000\,$GeV$^2$ the measured CC and NC
cross-sections are indeed of comparable magnitude, with some indication
of charge dependence. The precision of the data do not yet permit a
measurement of $xF_3$.

\noindent

Apart from providing a foretaste of future physics at HERA,
the measurement of high $Q^2$ cross-sections has recently been a 
very topical issue. In February 1997 the DESY Laboratory 
announced that both H1 and ZEUS had
observed an excess of NC events above the expectation of the Standard Model
calculation for $Q^2>15000\,$GeV$^2$. Detailed results have
since been published by H1~\cite{h1hiq2} and ZEUS~\cite{zeushiq2}. While it
is not our intention to consider this in any detail, we would like to 
summarise some characteristics of the data, to discuss the 
reliability of the Standard Model (SM) calculation and to summarise very
briefly some of the possible explanations. 

\subsection{The Standard Model calculation}
\label{sec:hiq2sm}

To calculate the Standard Model NC
and CC cross-sections at large $Q^2$ requires knowledge of the parton
distributions at large $x$, for example at $Q^2=5000\,$GeV$^2$ the 
$x$ range is roughly $0.05$ to 1. In this region of $x$ one is relying on 
the high statistics data from SLAC and BCDMS to determine the $x$ dependence.
The valence quarks dominate and their momentum 
distributions are relatively well determined by the global fits. 
At the very largest $Q^2$ values considered, the $u$ quark contribution
 is the most important. The large $x$ region generally is also where 
one has greatest confidence in the DGLAP equations. Higher twist effects 
can be minimised by suitable cuts on the input data.
 All the required Standard Model input couplings and
masses are very well determined~\cite{PDG}, many of them from the LEP1. The
uncertainty on the cross-section comes largely from the systematic uncertainty
in the large $x$ input data and from the uncertainty in $\alpha_s$.
One possible estimate of the uncertainty in the structure functions 
at large $Q^2$ is simply to take recent global fit results, e.g. CTEQ4 
and MRSR and compare 
them. The fitting teams have also considered the effect of varying 
$\alpha_s$ on the resulting PDFs. However this does not give a complete 
estimate of the uncertainty as both CTEQ and MRS use NLO DGLAP evolution,
fit essentially the same data, and use very similar functional forms for 
the input distributions. As we have noted elsewhere, the problem of the proper
inclusion of experimental systematic errors in pQCD fits has been addressed 
by the HERA experimental groups in their own attempts to extract the gluon 
momentum distribution~\cite{fiterr}.
Both H1 and ZEUS have performed NLO QCD fits to
extrapolate the fixed target data with $x>0.1$ from SLAC, BCDMS and NMC(97) to 
the high $Q^2$ region. The systematic errors on the data when propagated
to the HERA kinematic region $x>0.5,~y>0.25$ result in a $6-7\%$ uncertainty.
Varying $\alpha_s(M_Z^2)$ between 0.113 and 0.123 gives an additional $2\%$
uncertainty.

\subsection{The high $Q^2$ data}
\label{sec:hiq2data}

H1 and ZEUS base their analyses on samples of e$^+$p DIS collected in
the period 1994-96. The full details of the event selections are given in
the refs.~\cite{h1hiq2,zeushiq2}. Both experiments identify the large
$Q^2$ events by the presence of a beam positron scattered through a large 
angle, with the primary vertex defined by reconstructed tracks located within
a reasonable distance from the nominal interaction point. As discussed in
Sec.~\ref{sec:heraexp} the HERA detectors enable the struck quark jet to be 
detected as well, thus allowing radiative corrections and photoproduction
background to be limited by cuts on the quantity $\sum(E-p_z)$, summed over
all calorimeter cells. A brief summary of the cuts and the NC high $Q^2$
samples is given in Table~\ref{tab:hiq2tab}. In addition to the cuts listed
H1 also requires $0.1<y<0.9$.

\begin{table}[ht]
\tcaption{The NC high $Q^2$ data samples from H1 and ZEUS. $E^\prime_{min}$
is the minimum scattered electron energy; $z_v$ is the reconstructed 
primary vertex position along the beam line with its allowed range;
the minimum and maximum values of $\sum(E-p_z)$ are given within the brackets.
The overall event selection efficiencies for
both H1 and ZEUS are about 80\% for $Q^2>Q^2_{min}$.}
\centerline{\footnotesize\smalllineskip
\begin{tabular}{cccccccc}\\
 \hline
  & $\int {\cal L}dt$ & $E^\prime_{min}$ & $z_v$ & $\sum(E-p_z)$ & 
  $Q^2_{min}$  & No. & SM  \\
    & [pb$^{-1}$] & [GeV] & [cm] & [GeV] & [GeV$^2$] &    
     events & expect.  \\
 \hline
 ZEUS & 20.1 &  20 & $\pm 50$ & (40,70) & 5000 & 191 & $196\pm 17$\\
  H1 & 14.2 &  25 & $\pm 35$ & (43,63) & 2500 & 443 & $427\pm 38$\\
\hline\\
\end{tabular}}
\label{tab:hiq2tab}
\end{table}

Note that for these samples there is no discrepancy with the 
Standard Model expectations. Also the difference in event numbers
for the two experiments is
explained by the differences in integrated luminosities and cut values,
particularly $Q^2_{min}$. For reconstruction of event kinematics, H1
use the electron method and ZEUS the double-angle method but each
experiment uses the other as a systematic check. For both methods at
large $Q^2$ the most important detector uncertainty comes from the
absolute energy calibration of the calorimeters and for both
experiments it amounts to about 3\%. 
Other systematic uncertainties in the measurement
of the event sample are 2.3\% in the luminosity determination and
about 2\% in the radiative correction estimate. Taking these uncertainties
together with the cross-section uncertainty above gives a total 
systematic error of about 8.5\% for each experiment.
Because the energy scale uncertainty has a bigger effect
at large $Q^2$, the systematic error may reach 30\% for the largest $Q^2_e$,
which is why it is so important that both reconstruction methods be used.
 When the samples 
are studied in more detail it is found that there is an excess of 
events above the Standard Model expectation in the region of 
large $x$ and $y$. 

\begin{figure}[ht]
\centerline{\psfig{figure=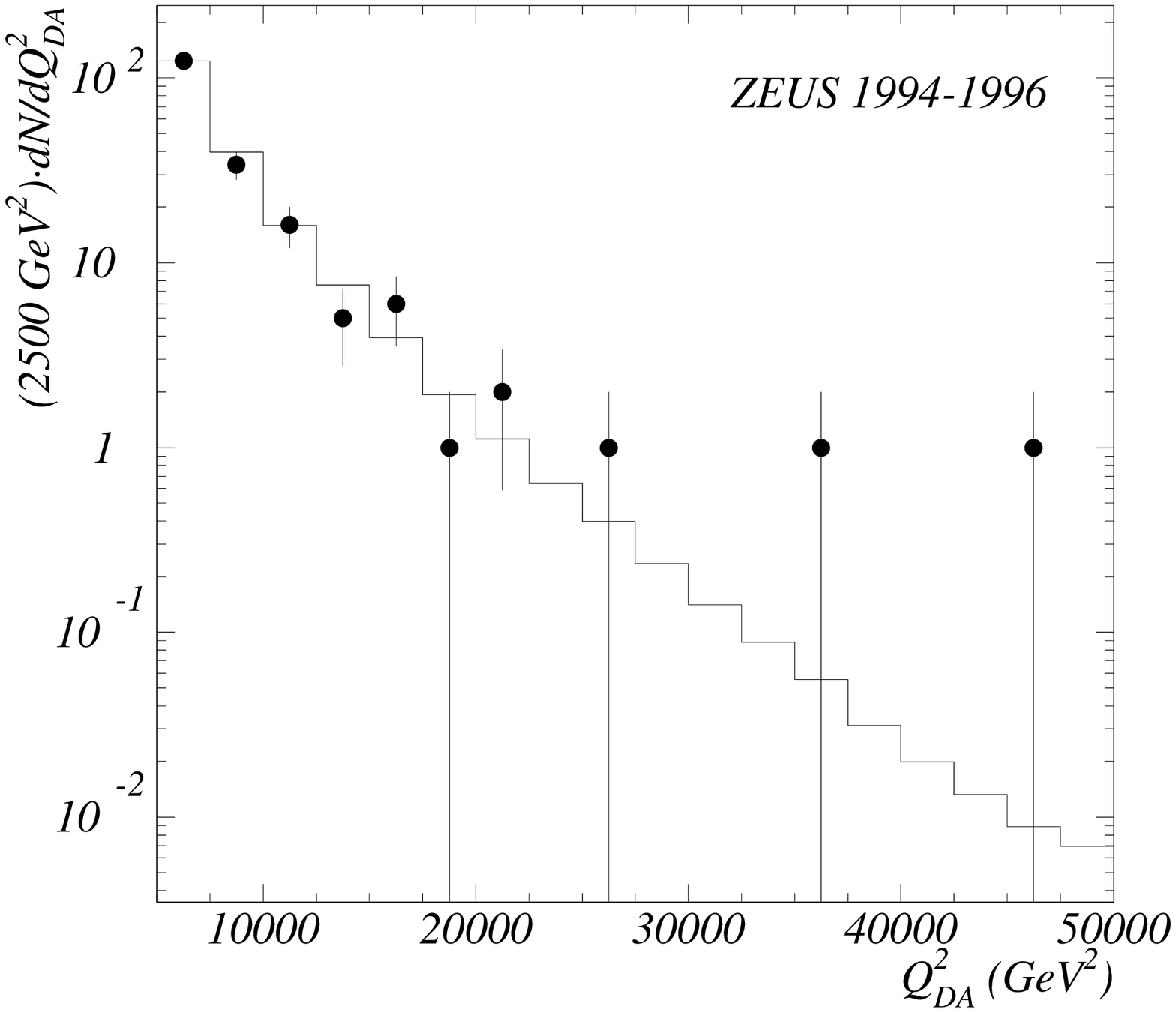,width=10cm,height=10cm}}
\fcaption{The $Q^2$ distribution for the ZEUS high $Q^2$ sample with
$y>0.25$ compared to the SM expectation.}
\label{fig:zhq2dist}
\end{figure}

\begin{figure}[ht]
\centerline{\psfig{figure=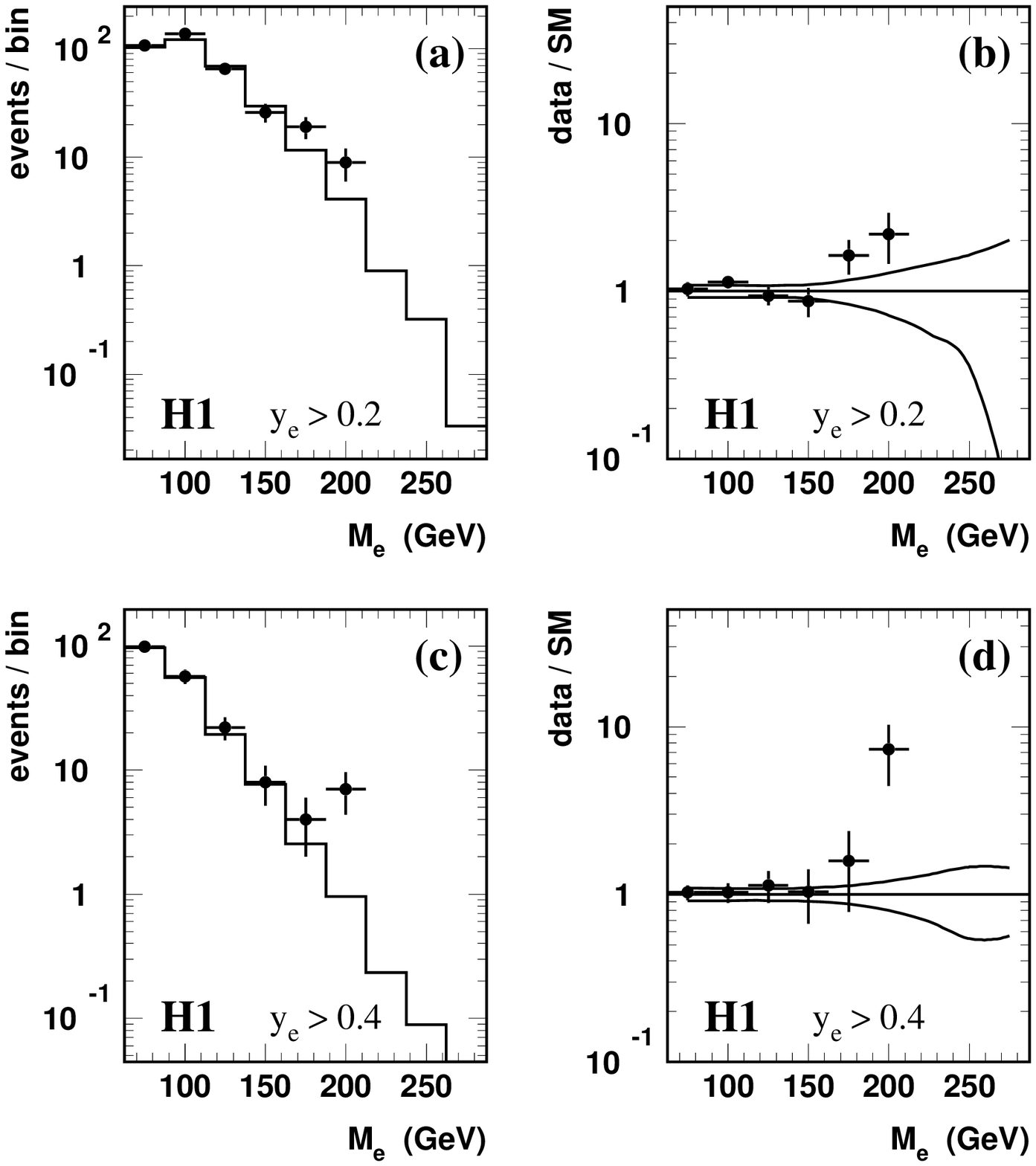,width=10cm,height=12cm}}
\fcaption{The distribution in $M=\sqrt{xs}$ of the H1 high $Q^2$ sample
for $y>0.2$ and $y>0.4$.}
\label{fig:h1mdist}
\end{figure}

The ZEUS group analyse their data using kinematic variables $Q^2,~x,~y$
reconstructed using the double angle method.
Fig.~\ref{fig:zhq2dist} shows the $Q^2$ distribution for the ZEUS data
with $y>0.25$ compared to the SM expectation. An excess of 
events is visible for $Q^2>35000\,$GeV$^2$. H1 use the variables $Q^2$,
$y$ and $M=\sqrt{xs}$ ($M$ is the invariant mass of the $e-q$ system) 
reconstructed using the electron method. Fig.~\ref{fig:h1mdist} shows
the distribution in $M$ of the H1 sample compared to the SM expectation
for $y>0.2$ and $y>0.4$.  There is an excess of events in the bin centered
on $200\,$GeV.

\noindent

\noindent

In more detail, ZEUS observe 2 events with $Q^2>35000\,$GeV$^2$ where 0.15
are expected, giving a Poisson probability of 0.39\% (for 2 or more) and
for $x>0.55$ and $y>0.25$, ZEUS observe 4 events where 0.91 are expected
(probability of 0.60\% for 4 or more). Using simulated experiments ZEUS
find that, for a luminosity of $20.1\,$pb$^{-1}$, there is a 6.0\% probability
to observe a fluctuation of 0.39\% or less in $Q^2$ for any $Q^2$ in the
range. For $Q^2>15000\,$GeV$^2$ H1 observe 12 events where 4.7 are expected
with a probability of 0.6\% (for 12 or more) and for the interval
$187.5<M<212.5\,$GeV with $y>0.4$, 7 events are observed where 0.95 are 
expected with a probability of
0.4\% (for 7 or more). Using simulated experiments for a luminosity of 
$14.2\,$pb$^{-1}$ H1 find that there is a 1\% probability to observe a
fluctuation to the minimum or less (averaged over windows of $25\,$MeV)
for any mass in the $M$ range 80 to 250$\,$GeV. Full details for each of 
the 7 H1 and 4 ZEUS `excess' events are given in Refs.~\cite{h1hiq2,zeushiq2}

H1 have in addition measured the e$^+$p CC cross-section at high $Q^2$.
The crucial event selection cut is to require a total missing transverse
momentum of at least 50$\,$GeV, giving a sample of 31 CC events with
$Q^2>2500\,$GeV$^2$ compared to a SM expectation of 34 events. 
For $Q^2>20000\,$GeV$^2$ H1 find 3 events where 0.74 are expected, with
a probability of 5\%. 
\vfill
\vskip1cm

\subsection{Possible explanations}
\label{sec:hiq2expl}

First could the excess be due to background? Apart from misidentified QED 
Compton scattering and weak boson production, the most dangerous background
processes are all related to photoproduction in which an  
electromagnetic cluster of large energy 
(probably from $\pi^0$ decay) fakes the scattered 
electron. Visually the events in the region of the excess are very clean
with an isolated electron balanced by a single hadronic jet. On a more
quantitative level ZEUS estimate less than 1 event out of 46 observed
in the large $Q^2$ sample (restricted to $x>0.55$) from all background channels
and H1 estimate less than 0.1 event (95\%CL) from all background sources
in their sample of 20 events with $Q^2>10000\,$GeV$^2$.

The effect could be a statistical fluctuation. 
This explanation awaits the increased
statistics expected from the long HERA run in 1997. 

One of the first questions to be addressed is are the effects from the two 
experiments compatible? The answer at present is yes but only just. The
7 excess events from H1 cluster around a mass value of about 200$\,$GeV
($M=\sqrt{xs}$) whereas the 4 ZEUS events have mass values spread between
230 and 250$\,$GeV. The difference in mean mass is larger than the
difference expected from using different methods of reconstruction and
both groups have checked that their results are internally consistent by 
using the other reconstruction method, electron in place of double-angle 
for ZEUS and vice-versa for H1. Another possible cause of spread is 
initial state radiation from the beam positron, though it is unlikely 
that this would affect more than a few of the events. ISR affects the electron 
and double-angle methods differently, and in fact if both methods are
used on the same events the energy of the radiated photon can be
reconstructed, assuming that is collinear with the positron beam. 
The details have been discussed in 
general terms in a recent paper by Wolf~\cite{gwolfisr} and specifically 
for these events by Bassler \& Bernardi~\cite{bbisr}. The latter authors
also propose a modification of the $\Sigma$ method (see Sec.~{\ref{sec:h1dat})
which improves the resolution somewhat at very large $Q^2$. They use the
method on the combined event sample, correcting two events for ISR and
find that although the difference in mean mass between the ZEUS and H1
events is reduced from $26\pm 10$ to $17\pm 7\,$GeV, it is still larger
than one would expect from the decay of a single narrow resonance.

A number of authors have considered various ways in the which the 
parton distributions could be modified at large $x$ without altering
the good description of low $Q^2$ fixed target data. Note that, because
of the feed-down effect in $x$ as one evolves up in $Q^2$, any changes
have to be made well above the $x$ region at which the excess has been
observed at HERA. Kuhlmann, Lai \& Tung~\cite{kuhlmann}, within the 
context of the CTEQ global fit, investigate how large a change would
be required to the valence quark distributions to fit the HERA excess.
They first establish that the effect of changing the parameters of the 
standard $u_v$ distribution to increase its magnitude for $x>0.75$ is
not nearly sufficient. They then show that adding an extra component
to $u_v$ of the form $0.02(1-x)^{0.1}$ at $Q^2_0$ produces an increase 
at $x\approx 0.5-0.6$ of about the right size to explain the number of 
excess events. Of course by its nature such a modification predicts an 
even larger increase in the cross-section at larger $x$ values. They
do not attempt to look for any justification for the extra term and
point out that if the effect is associated with a narrow structure
or structures then this sort of modification would not be appropriate. 
  
Rock and Bosted~\cite{rock} suggest that one might learn something
about $F_2$ in the range $0.7\leq x \leq 0.97$ from SLAC $ep$ data
taken in the $N^*$ resonance region. Their idea is to use Bloom-Gilman
duality~\cite{bgduality} which states that the average of the structure
function in the resonance region is approximately equal to its value in
DIS. They compare SLAC data to the NMC parametrization (Eq.~\ref{eq:nmcfit})
 and to the 
recent CTEQ4M global fit and find that for $x>0.7$ the average resonance
data give a value of $F_2$ about one order of magnitude above the 
conventional fit. This is however still roughly an order of magnitude 
below what is required to explain the excess at high $Q^2$. The authors
warn of the care necessary at large $x$ as $F_2$ is decreasing rapidly,
but argue that the form of the fit parametrization should perhaps
be looked at again including allowance for higher twist terms. 

Gunion \& Vogt~\cite{gunion} point out that
an intrinsic charm component in the proton could give rise to an
excess of events localised in the $e-q$ mass near 200$\,$GeV.  The intrinsic
charm model is formulated in terms of nucleon Fock 
states~\cite{brodsky8081} and in particular the state 
$|uudc\bar{c}\rangle$, will be important at large $x$ values. 
Existing constraints limit the size of this contribution to only 
15\% in the $e^+ p$
NC channel, but could be much larger in the CC channel. So, if the size
of the HERA excess in the NC channel is confirmed, then their 
version of intrinsic charm would not be a viable explanation. 
An alternative version of the intrinsic charm approach has been suggested
by Melnitchouk and Thomas~\cite{mtcharm} in the context of the meson cloud
model for long-range nucleon structure. In this the charm sea component
is taken to arise from quantum fluctuations to identified states - for
example $D^-+\Lambda_c$. Using this approach the authors show that they can
produce a larger effect than that of Gunion \& Vogt. However Melnitchouk
\& Thomas are not claiming that this is the explanation of the HERA excess 
but merely that there is a SM approach with some physical basis 
which could provide a cross-section of about the right magnitude without 
violating the standard PDF description of data at lower $Q^2$. The direct 
investigation of charm
in DIS is very much on the agenda for study at HERA so as luminosity
increases it may be possible to answer some of the questions raised by
these authors directly.

Szczurek and Budzabowski~\cite{szczurek} have studied 
whether the effect of a meson cloud and the 
inclusion of target mass corrections in the description of the 
low $Q^2$ DIS data may, after QCD evolution, influence the large $x$ 
and large $Q^2$ region. 
Meson cloud effects, i.e. the admixture of $\pi N$, $\pi\Delta$ 
and similar components in the Fock expansion of the proton wave function,
have been put forward as a way of understanding 
the Gottfried Sum Rule, the measured 
Drell Yan asymmetries in $pp$ and $pd$
collisions and semi-inclusive production of fast neutrons at HERA.
The inclusion of such effects at
low $Q^2$ leads to an enhancement of the quark distributions at 
large $x$, which survives QCD evolution to 
the HERA high $Q^2$ region. For $Q^2> 10^4\,$GeV$^2$ the predicted 
increase due to these effects is 30\% (100\%) for $x = 0.7$ (0.9). 
When integrating the region $x>0.5$ (0.7) the relative enhancement  
of the cross section due to meson cloud effects and target mass corrections
for $Q^2= 10^4\,$GeV$^2$ is 20\% (60\%), slowly increasing with $Q^2$.
Hence it is unlikely that these mechanisms can provide a complete
explanation of the high $Q^2$ effect, but they need to be considered 
when analysing future high $Q^2$ data at large $x$.

Kochelev~\cite{kochelev} suggests the possibility that the excess of events
at high $Q^2$ is due to an instanton contribution to the proton structure
function. Instantons are 
non-perturbative strong vacuum fluctuations, which describe the tunnelling 
between different gauge rotated vacua in QCD. A characteristic feature is that
they yield a large multi-gluon final state in DIS. The instanton contribution
does not evolve with $Q^2$ at the same rate as the standard PDF contribution
and this leads to a relative increase in the instanton fraction at large $x$
and $Q^2$. First estimates suggest, albeit with large uncertainties, that the
instanton contribution in the region of the high $Q^2$ HERA events can be 
$50\%$ or more, and would therefore increase the SM prediction considerably.
Due to the specific nature of the instanton final state in DIS, which includes
flavour democracy, the study of such final state properties as energy flow and
strange or charm particle abundances could provide decisive information on this
scenario.

The most exciting possibility is that one is seeing the hint of something
beyond the Standard Model. The HERA observations have unleashed a flood 
of theoretical speculation which it would not be appropriate to summarise
here. A compact survey of the field has been given by 
Dokshitzer~\cite{doksh_dis97} in his summary of theoretical developments
at the DIS97 Workshop. The models are in three broad categories:
contact interactions, leptoquarks or an $R$-parity violating 
supersymmetric squark; all are
conveniently summarised in the paper by Altarelli et al~\cite{cern_thlq},
in which existing constraints are also discussed. Contact interactions
are a way of describing the effect of new physics at a scale well below
that of the new force quanta (as the Fermi theory approximates the full
electroweak theory for low $Q^2$ CC interactions). The effective 
Lagrangian for the four-point interaction is of the form
\begin{equation}
{\cal L}_{4}=4\pi\sum_{q=u,d}\sum_{i,j=L,R}{\eta_{ij}\over (\Lambda^2_{ij})^2}
\bar{e_i}\gamma^\mu e_i\bar{q_j}\gamma_\mu q_j,
\end{equation}
where $\Lambda_{ij}$ is the mass scale and coupling parameter and 
$\eta_{ij}=\pm 1$ allows for destructive and constructive interference.
Contact interactions will generate an excess of events
as $Q^2$ increases, but not a structure in $x$. There are of course 
limits on contact interactions from $e^+e^-$ interactions at LEP and
p\=p interactions at the Tevatron, but there are still a few possibilities
which could just about explain the HERA excess with mass scales in the range
$1.7-2.5\,$TeV depending on the sign and helicity
of the coupling. If the effect is concentrated around a specific 
$M=\sqrt{xs}$ value then the most natural
explanation is either a leptoquark (an electron-quark bound state) or
an $R$-parity violating squark coupling to $e-q$. 
There are many possible leptoquarks and they
are classified by spin and isospin quantum numbers. An $e^+ u$ or $e^+d$ 
state with standard Yukawa couplings and a mass of about 200$\,$GeV is
not yet ruled out by Tevatron data but such an exclusion is getting close. 
One would not expect such a scalar leptoquark to decay to the $\nu \bar q$ 
final state, whereas a vector
one could. Perhaps the most favoured explanation is that of an $R$-parity
violating squark coupling to valence $d$ quarks ($e^+ d$, such as stop) or 
one coupling to the $\bar u$ sea ($e^+ \bar u$). When one takes into account
limits from other high energy processes and atomic parity violation 
experiments, the most favoured options are left-handed
scalar charm or top squarks.
 
\subsection{August 1997 update}
\label{sec:hiq2aug97}

For the Hamburg Lepton-Photon Symposium~\cite{straub97} and the Jerusalem
EPS HEP conference~\cite{elsen97}, the HERA high $Q^2$ NC results were
updated with the addition of data collected to the end of June 1997. 
This has increased the accumulated luminosity by roughly 
67\% (to $23.7\,pb^{-1}$
for H1 and  $33.5\,pb^{-1}$ for ZEUS). Unfortunately the new data have
not clarified the experimental picture very much. The measured NC
cross-sections are still in excess of the SM expectation, but the number
of `signal' events has only increased slightly. For H1, in the mass
range $187.5<M<212.5\,$GeV, 8 events are now observed where $1.53\pm 0.29$
are expected with probability 0.3\% (compared to 7 observed and $0.95$ 
expected in the earlier report). For ZEUS, with $x>0.55$ and $y>0.25$,
5 events are observed where $1.51\pm 0.13$ are expected with a probability
of 1.9\% (compared to 4 observed and 0.91 expected before). 
Preliminary CC cross-sections
for the same total luminosities were also presented, the results from
both experiments are above the SM expectation, for $Q^2>10^4\,$GeV$^2$
H1 and ZEUS together observe 28 events where $17.7\pm 4.3$ are expected.
The tentative conclusion that a single narrow resonance structure in $x$ 
cannot explain the data from both experiments has been reinforced. 
From the discussion in the previous section on uncertainties in and novel
models for the PDFs at large $x$, it is also clear that more work needs
to be done on the calculation of the standard model NC and CC cross-sections
at large $x$ and $Q^2$.
Full analyses of all
high $Q^2$ data collected up to the end of 1997 running (estimated to be
totals of 38 and $48\,$pb$^{-1}$ for H1 and ZEUS respectively), which
more than double the statistics of the original publications, 
are expected in early 1998.

\section{Summary and outlook}
\noindent

In this review we have concentrated on the complete data sets from
the latest round of muon beam fixed target experiments, NMC at CERN and
E665 at FNAL, and the first substantial data on proton structure from
the two HERA general purpose experiments, H1 and ZEUS. We have covered
briefly the revised neutrino beam data (on an iron target) from CCFR at
FNAL. The CCFR data are important because they are the most accurate
measurements of $\nu N$ structure functions and they provide a direct 
measurement
of the non-singlet component from which $\alpha_s$ has been determined free 
of the convolution with the gluon momentum distribution. The NMC and E665
data on $F_2^p$ and $F_2^d$ are consistent with each other and with the earlier
high statistics measurements on $eN$ scattering at SLAC and the BCDMS $\mu N$
experiment at CERN. There is still a discrepancy at low $x$ between 
the $F^p_2$ values derived from CCFR data and those of NMC and E665. 
 In the design of both E665 and NMC special attention
was paid to efficient triggering and kinematic reconstruction at very small 
angles in the laboratory allowing the measurement of structure 
functions at small
$x$. The push to small $x$ has seen a natural continuation with the 
data from the HERA collider, in which the $e p$ centre of mass energy has 
increased by an order of magnitude (from $20-30\,$GeV to $300\,$GeV). The
$F_2^p$ data from H1 and ZEUS are consistent with each and match smoothly 
to the fixed target data, though the region of direct overlap is small. At
HERA the measurement of $F_2$, or equivalently  $\sigma^{\gamma^*p}$, at
very small values of $Q^2$ has been performed either by shifting the
$ep$ interaction point and/or by the use of dedicated small electromagnetic
calorimeters near the beam line. The data provides almost complete coverage
from $Q^2=0$ to the `safely' deep inelastic region with $Q^2>4\,$GeV$^2$
and has enabled precision studies of the transition region in which
the description by pQCD breaks down.

New measurements of $R$ have been made by both NMC and CCFR.
If $R$ has not been measured in the experiment then the teams have taken 
a consistent approach on how to handle the correction for $R$ in the 
extraction 
of $F_2$ from the cross-sections. For the fixed target experiments the
SLAC parametrization of $R$ has been used and for HERA experiments $R_{QCD}$
has been used. The systematic errors of the fixed target experiments are
in the range 2-3\% and those from HERA around 5\%.
We now have data on $F_2$ in the ranges $3\cdot 10^{-6}<x<0.9$ and
$0.11< Q^2 < 5000\,$GeV$^2$.

The expansion of the kinematic region to very low $x$ at $Q^2$ values of
at least a few GeV$^2$ has transformed the study of gluon dynamics within
QCD. Two major facts about $F_2$ have emerged from the HERA measurements: 
the first is that $F_2$ rises as $x$ decreases, the rise increasing as $Q^2$ 
increases and the second is that standard NLO QCD appears to describe the
data from the highest $Q^2$ values to the surprisingly low value of 
$1-2\,$GeV$^2$. It is agreed that the rise of $F_2$ at low $x$ is caused
by an increase in the $q \bar q$ sea which in turn is being driven by a rapid 
increase in the gluon density. Much theoretical effort has been expended 
on trying to understand exactly what the dynamical mechanism is. So 
far there is no definitive answer. Early on some argued strongly
that the data would require the use of BFKL inspired multi-gluon dynamics
giving singular parton distributions at 
a conventional starting scale of $Q^2_0\approx 4\,$GeV$^2$, while others 
pointed to the success of the GRV approach using a very low starting scale, 
$\sim 0.3\,$GeV$^2$, with valence-like non-singular input distributions.
The success of Double Asymptotic Scaling as advocated by Ball and
Forte also indicated that conventional DGLAP evolution could describe the 
data. However, as the precision and extent of the data improved 
it became apparent that he GRV94 parametrization itself
cannot describe all the low $x$ data and that Double Asymptotic Scaling needs
corrections within a full NLO calculation. The legacy of the GRV approach is
that most successful fits to data within the conventional framework use a low 
starting scale, $Q^2 \sim 1\,$GeV$^2$, with non-singular inputs. 

However, much work continues on the question whether or not BFKL dynamics
are necessary to describe the HERA data. The phenomenologically successful 
non-conventional approaches incorporating (and going beyond) BFKL dynamics 
are: the resummation of $\ln(1/x)$
terms using `physical' anomalous dimensions as advocated by Catani and 
implemented by Thorne; unified DGLAP-BFKL approaches  based on the 
unintegrated gluon distribution, such as the CCFM equation and modified 
BFKL equations; and 
attempts to solve the BFKL equation within the colour dipole approach. 
Thus we have a situation whereby both conventional and non-conventional 
QCD dynamics can fit the data for $Q^2 > 1\,$GeV$^2$, 
and the difference in $\chi^2$ of the fits to $F_2$ data is not
sufficiently large to allow us to make a clear decision.

We have clearly reached a region 
where $\alpha_s \ln(1/x) \sim 1$ so that conventional expansions in $\alpha_s$
should no longer be reliable. The fact that conventional approaches are 
successful thus needs some explanation. We have summarized some of the 
possible explanations in Sec.~\ref{sec:summary}. Here we recall that recent
theoretical work suggests that a full treatment of $NLL(1/x)$ effects within 
the BFKL equation would lead to a softening of the BFKL Pomeron such that it
becomes more consistent with the behaviour predicted by the conventional
expansions. Camici and Ciafaloni have even suggested that a full understanding
of $NLL(1/x)$ contributions would lead to an understanding of the transition
to soft physics as $Q^2 \to 0$.
It is also possible to imagine a `conspiracy' whereby
higher twist terms and shadowing corrections at low $x$
could be large, even at moderate $Q^2$, hiding BFKL effects. 
Finally there is always the nagging doubt
that the freedom to choose a fairly arbitrary function of $x$ for the
input parton distributions may be hiding a real breakdown of the standard
description. Hence it is very important that the efforts to derive information
about the input parton distributions using non-perturbative techniques, such
as the lattice and chiral dynamics, continue to be pursued vigorously. 
 
Some of these questions should find answers when different variables affected
by non-conventional dynamics are measured. The measurements from the
hadron final state already give tantalizing glimpses of new dynamics. 
We look forward to accurate measurements of structure functions  other than 
$F_2$. Measurements of $F_L$ will be made in the HERA
region, when increased luminosity allows us to exploit the ISR method, or, by 
use of different beam energies.
The period covered by this review has seen the publication of the 
first measurements of the charm structure function, $F_2^{c\bar{c}}$,
at HERA from a DIS event sample containing identified $D$ and $D^*$ mesons, and
we look forward to more accurate measurements in future.
There has also been a major advance in the theoretical treatment of 
heavy flavours
in DIS with the calculation to NLO of the massive quark coefficient
functions. This has in turn stimulated a lively debate on the correct way to 
handle heavy flavours, in particular charm, through the threshold region,
where the mass effects are obviously important, to large $Q^2$, where the
mass is no more important than that of the light quarks. The jury is still out.

In the immediate future we expect to see more data on $F_2$ from the HERA
experiments. The data that we have reported on was collected in 1994
when HERA delivered $6\,$pb$^{-1}$. When quality checks and detailed
cuts are made typically about 50\% of the delivered luminosity is used
for cross-section measurement and structure function determination. In 1995,
1996 and 1997 HERA delivered 12, 17 and  $36\,$pb$^{-1}$, respectively. 
If the full samples from 1995-97 are used then the $Q^2$ at which
systematic errors dominate will increase from around 100$\,$GeV$^2$ now
to above $500\,$GeV$^2$. The aggregated sample should allow of initial studies
of $F_L$ using the ISR method. The understanding of systematic errors will
also improve as detector studies can be made with finer geometrical
binning. The number of identified DIS charm events will also increase
from higher luminosity and from more efficient methods of tagging,
particularly from the use of micro-vertex detectors. H1 has one installed
and has taken some data with it, ZEUS hopes to install one for
operation in the year 2000.
The urgent question of whether the excess of HERA DIS events 
seen at very high $Q^2$ is simply a statistical fluctuation or not
should be resolved. If the anomaly remains it will require much more 
high luminosity at high $Q^2$ for its complete understanding.
Also in the future is the prospect of running with polarized e$^\pm$ 
beams and possibly polarized protons. The option of running with deuterons 
instead of protons is also under consideration. It is planned to upgrade the
luminosity of the HERA collider incrementally between now and the
year 2000, many details of this and the physics that will then become
possible can be found in the report of the 1996 Workshop on Future
Physics at HERA~\cite{heraws96}. 

Apart from HERA we have the prospect of the NuTeV experiment at Fermilab.
Thus the next five years should see a further significant increase in
our understanding of the nature of parton distributions and the forces 
which shape them.
Deep inelastic scattering and hadronic structure
functions are still vibrant and important fields of study and uniquely
important tools in unravelling the complexities of QCD 30 years after such
methods first led us to the dynamical quark-parton model.

\nonumsection{Acknowledgements}
\noindent
Many people have helped us to accumulate information for this review.
For advice on the data and helping to prepare plots we thank:
H. Abramowicz, A. Bodek, G. Contreras, E. Kabuss, M. Kuhlen, F. Lehner, 
A. Levy, A. Quadt, E. Rondio, W. Seligman, 
W. Smith, B. Surrow, J. Tickner, U. Yang.
For many discussion on the complexities of QCD and small $x$ physics
we thank R. Ball, A. D. Martin, R. G. Roberts, R. Thorne and for supplying
original figures we thank B. R. Webber, Ch. Royon, P. Sutton.
We thank M. Lancaster for reading and commenting on an early version 
of the manuscript.

\nonumsection{References}


\noindent
\begin{thebibliography}{000}
%
%
\bibitem{lpham97} {\em Symposium on Lepton and Photon Interactions, Hamburg}
July 1997, {\tt http://www.desy.de/lp97/}.
\bibitem{jer97} {\em EPS HEP97 Conference, Jerusalem}, August 1997, 
{\tt http://www.cern.ch/hep97/}.
\bibitem{gallo97}E. Gallo, invited talk at {\em Proc.
of the 1997 Symposium on Lepton Photon Interactions, 
Hamburg}, July 1997, eds. A. De Roeck and A. Wagner, {\tt hep-ex/9710013}.
\bibitem{eich97}R. Eichler, invited talk at {\em EPS/HEP Conference 97, 
Jerusalem}, August 1997, {\tt http://www.cern.ch/hep97/}.
\bibitem{Halzen} F.\ Halzen and A.D.\ Martin, {\em Quarks and Leptons}, John
Wiley, 1984.
\bibitem{roberts} R.G. Roberts, {\em The Structure of the Proton}, 
CUP 1990.
\bibitem{cg} C.G.\ Callan and D.\ Gross, \Journal{\PRL}{22}{156}{1969}.
\bibitem{feynman} R.P. Feynman, \Journal{\PRL}{23}{1415}{1969};\\
            {\em Photon Hadron Interactions}, Benjamin N.Y. 1972.
\bibitem{bjorken} J.D. Bjorken, \Journal{\em Phys. Rev.}{179}{1547}{1969}.
%
\bibitem{nobel1990} J.I. Friedman, H.W. Kendall \& R.E. Taylor, 
          \Journal{\RMP}{63}{573,597,615}{1991}.
\bibitem{fisk_sciulli} H.E. Fisk \& F. Sciulli, 
     \Journal{\em Ann. Rev. Nucl. Part. Sci.}{32}{499}{1982}.
\bibitem{GLS} D.J.\ Gross and C.H.\ Llewellyn-Smith, 
               \Journal{\NPB}{14}{337}{1969}.
\bibitem{Adler} S.L.\ Adler, \Journal{\em Phys. Rev.}{143}{1144}{1966}.
\bibitem{Gottfried} K.\ Gottfried, \Journal{\PRL}{18}{1154}{1967}.
\bibitem{Abramowicz} H.\ Abramowicz et al, \Journal{\ZPC}{17}{283}{1983}.
\bibitem{Farrar} S.J.\ Brodsky and G.\ Farrar, 
                 \Journal{\PRL}{31}{1153}{1973}.
\bibitem{DoLa} A.\ Donnachie and P.V.\ Landshoff, 
\Journal{\PLB}{296}{257}{1992}.
\bibitem{ForRoss} J.R. Forshaw and D.A. Ross, {\em QCD and the Pomeron}, CUP
1997.
\bibitem{Buras} see for example, A.Buras, \Journal{\em  Rev.Mod.Phys}{52}
{199}{1982}.
\bibitem{PMS} M.R.\ Pennington, \Journal{\PRD}{26}{2048}{1982};\\
P.M.\ Stevenson, \Journal{\PRD}{23}{2916}{1981}.
\bibitem{DGLAP}Yu.\ Dokshitzer, \Journal{\JETP}{46}{641}{1977};\\
  V.N.\ Gribov and L.N.\ Lipatov, \Journal{\SJNP}{15}{438,675}{1972};\\ 
  L.N.\ Lipatov,\Journal{\SJNP}{20}{95}{1975};\\
  G.\ Altarelli and G.\ Parisi, \Journal{\NPB}{126}{298}{1977}.
\bibitem{Renton} P. Renton, {\em Electroweak Interactions}, CUP, 1990.
\bibitem{AMCS} A.M. Cooper-Sarkar et al, \Journal{\ZPC}{39}{281}{1988};\\
A.M. Cooper-Sarkar et al, {\em Proc.
of the Workshop on Physics at HERA, DESY 1991}, 
eds W. Buchm\"uller \& G. Ingelman, Vol I, p155.
\bibitem{yndurain} F.J. Yndurain, {\em Quantum Chromodynamics}, 
Springer-Verlag 1983.
\bibitem{tm} H.\ Georgi and H.D.\ Politzer, \Journal{\PRD}{14}{1829}{1976};\\
O.\ Nachtmann, \Journal{\NPB}{63}{237}{1973}.
\bibitem{1987} {\em High $p_t$ physics and higher twists}, 
\Journal{Nucl.Phys.Proc.Supp}{7B}{}{1989}. 
\bibitem{Miramontes}J.L.\ Miramontes, M.A.\ Miramontes and J.\ 
Sanchez-Guillen, \\ \Journal{\PRD}{40}{2184}{1989};\\
R.K.\ Ellis, W.\ Furmanski and R.\ Petronzio, \Journal{\NPB}{212}{29}{1980};\\
E.V. Shuryak and A.I. Vainshtein, \Journal{\NPB}{199}{451}{1982}.
\bibitem{GLSnew}V.M. Braun and A.V. Kolesnichenko, 
\Journal{\NPB}{283}{723}{1987};\\ G.G. Ross and R.G. Roberts, 
\Journal{\PLB}{322}{425}{1994};\\
I.I. Balitsky, V.M. Braun and A.V. Kolesnichenko,
 \Journal{\PLB}{242}{245}{1990}, and \Journal{\PLB}{318}{648}{1993}.
\bibitem{Harindranath}A. Harindranath et al, {\tt hep-ph/9711298}.
\bibitem{webberon}B.R. Webber, {\em Proc. of DIS96, Rome}, 
Eds G. D'Agostini \& A. Nigro, \\ World Scientific 1997, p77.
\bibitem{das}M. Dasgupta and B.R. Webber, \Journal{\PLB}{382}{273}{1996};\\
M. Maul et al, {\tt hep-ph/9710531}.
\bibitem{Ster97} G. Sterman, {\em Summary talk at DIS97, Chicago} \\
{\tt http://www.hep.anl.gov/dis97/}. 
\bibitem{Marciano} W.J. Marciano, \Journal{PRD}{29}{5801}{1984}.
\bibitem{hws_cc} E. Laenen et al, {\em Proc. of the Workshop on Future 
Physics at HERA, DESY 1996}, eds G. Ingelman, 
A. De Roeck \& R. Klanner, Vol. I, p393;\\
Wu-Ki Tung, {\em Proc. of DIS97, Chicago},
{\tt http://www.hep.anl.gov/dis97/};\\
M.Buza et al, {\tt hep-ph/9707263};\\
W.L.van-Neerven, {\tt hep-ph/9708452};\\
J.C.Collins, M.G. Ryskin and A.D. Martin, {\tt hep-ph/9709440}.
\bibitem{MRSD} A.D. Martin, R.G. Roberts and W.J.Stirling, 
\Journal{\PRD}{47}{867}{1993}.
\bibitem{MRSA} A.D. Martin, R.G. Roberts and W.J.Stirling, \Journal{\PRD}{50}
{6734}{1994}.
\bibitem{MRS96} A.D. Martin, R.G. Roberts and W.J.Stirling, 
\Journal{\PLB}{387}{419}{1996}.
\bibitem{GRV91} M. Gluck, E. Reya and A Vogt, \Journal{\ZPC}{48}{471}{1990},
\Journal{\ZPC}{53}{127}{1992}.
\bibitem{GRV94} M. Gluck, E. Reya and A. Vogt, \Journal{\ZPC}{67}{433}{1995}.
\bibitem{GRS} M. Gluck, E. Reya and M.\ Stratmann, 
\Journal{\NPB}{422}{37}{1994}.
\bibitem{bgf_lw} J.P Leveille \& T. Weiler, \Journal{\NPB}{147}{147}{1979}
\bibitem{nlocc} E. Laenen et al, \Journal{\NPB}{291}{325}{1992};
             \Journal{\em ibid}{392}{162,229}{1993};\\
S. Riemersma, J. Smith \& W.L. van Neerven,
            \Journal{\PLB}{347}{143}{1995}.
\bibitem{vogt_cc} A. Vogt, {\em Proc. of DIS96, Rome}, 
Eds G. D'Agostini \& A. Nigro, World Scientific 1997, p254.
\bibitem{LaiTung} H. L. Lai and W.K. Tung, \Journal{\ZPC}{74}{463}{1997}.
\bibitem{MRRS} A.D. Martin et al, {\tt hep-ph/9612449}.
\bibitem{varflav} M.A.G. Aivazis et al, \Journal{\PRD}{50}{3102}{1994}.
\bibitem{ScmidtC} C.R. Schmidt, {\em Proc. of DIS97, Chicago}, 
{\tt hep-ph/9706496}.
\bibitem{Buza} M. Buza et al, {\tt hep-ph/9612398}.
\bibitem{Olsca} F.I. Olness and R.J. Scalise, {\em Proc. of DIS97, Chicago},
{\tt hep-ph/9707459}.
\bibitem{DickRob}R.G. Roberts and R.S. Thorne, {\tt hep-ph/9709442},
{\tt hep-ph/9711223}.
%
%
\bibitem{mishra_sciulli} S.R. Mishra \& F. Sciulli, 
     \Journal{\em Ann. Rev. Nucl. Part. Sci.}{39}{259}{1989}.
\bibitem{sloan} T. Sloan, G. Smadja \& R. Voss, 
         \Journal{\em Phys. Rep.}{162}{45}{1988}. 
\bibitem{nmcdet} NMC, P. Amaudruz et al, \Journal{\NPB}{371}{3}{1992};\\
              I.G. Bird, PhD Thesis, Free University of Amsterdam (1992).
\bibitem{mount} R.P. Mount, \Journal{\NIM}{187}{401}{1981}.
\bibitem{e665det} E665, M.R. Adams et al, \Journal{\NIMA}{291}{533}{1990}.
\bibitem{ccfrdet} W.K. Sakumoto et al, \Journal{\NIMA}{294}{179}{1990};\\
                  B.J. King et al, \Journal{\NIMA}{302}{254}{1991}.
\bibitem{ccfr_norm} CCFR, R. Blair et al, \Journal{\PRL}{51}{343}{1983};\\
P. S. Auchinloss et al, \Journal{\ZPC}{48}{411}{1990};\\
CDHSW, P. Berge et al, \Journal{\ZPC}{49}{187}{1991}.
\bibitem{bek} S. Bentvelsen, J. Engelen \& P. Kooijman, {\em Proc.
of the Workshop on Physics at HERA, DESY 1991}, 
eds W. Buchm\"uller \& G. Ingelman, Vol I, 23;\\
S. Bentvelsen, PhD Thesis, University of Amsterdam 1994.
\bibitem{jb} F. Jacquet \& A. Blondel, {\em Proc. of the Study of an ep
facility for Europe}, ed U Amaldi, DESY 79/48, p391.                  
\bibitem{h1det} H1, I. Abt et al, \Journal{NIMA}{386}{310}{1997}.
\bibitem{h1up} H1, {\em Technical proposal for the upgrade of the rear region
of the H1 detector}, DESY Internal Report PRC-93/02.
\bibitem{h1n94} H1, S. Aid et al, \Journal{\NPB}{470}{3}{1996}.
\bibitem{h1v95} H1, C. Adloff et al, DESY preprint DESY 97-042.
\bibitem{sigma} U. Bassler \& G. Bernardi, \Journal{\NIMA}{361}{197}{1995}.
\bibitem{zeusdet} ZEUS, M. Derrick et al, \Journal{\PLB}{293}{465}{1992};\\
               The ZEUS Detector Status Report 1993, DESY 1993;\\
               M. Derrick et al, \Journal{\ZPC}{63}{391}{1994}.
\bibitem{zcalflt} W.H. Smith et al, \Journal{\NIMA}{355}{278}{1995}.
\bibitem{zn94} ZEUS, M. Derrick et al, \Journal{\ZPC}{72}{399}{1996}.
\bibitem{whitlow90} L.W. Whitlow et al, \Journal{\PLB}{250}{193}{1990}.
\bibitem{motsai} L.W. Mo \& Y.S. Tsai, \Journal{\RMP}{41}{205}{1969};\\
                 Y.S. Tsai, SLAC preprint SLAC-PUB-848, 1971.
\bibitem{dubna} A.A. Akhundov et al, \Journal{\SJNP}{26}{660}{1977};\\
       A.A. Akhundov et al, {\em Proc. 1992 Zeuthen Workshop
on Elementary Particle Theory}, eds J. Bl\"umlein \& T. Riemann, p209
and references therein.
\bibitem{radcor} Report of the Working Group on Radiative Corrections,
{\em Proc. of the Workshop on Physics at HERA, DESY 1991}, eds 
W. Buchm\"uller \& G. Ingelman, vol II, 787.
\bibitem{hector} A. Arbuzov et al, \Journal{\CPC}{94}{128}{1996}.
\bibitem{helios} J. Bl\"umlein, 
{\em Proc. of the Workshop on Physics at HERA, DESY 1991}, eds 
W. Buchm\"uller \& G. Ingelman, vol III, 1272.
\bibitem{terad91} A.A. Akhundov et al, {\em ibid}, 1285.
\bibitem{heracles} K. Kwiatkowski, H. Spiesberger \& H-J. M\"ohring,
{\em ibid}, 1294.
\bibitem{django} G.A Schuler \& H. Spiesberger, {\em ibid}, 1419.
\bibitem{lepto} G. Ingelman, 
{\em Proc. of the Workshop on Physics at HERA, DESY 1991}, eds 
W. Buchm\"uller \& G. Ingelman, vol III, 1366.
\bibitem{ariadne} L. L\"onnblad, \Journal{\CPC}{71}{15}{1992};
                 \Journal{\ZPC}{65}{285}{1995}.
\bibitem{bbkk} B. Badelek et al, \Journal{\JPG}{19}{1671}{1993}.
\bibitem{kotwal} A. V. Kotwal, Harvard Thesis 1995.
\bibitem{bardin} D. Bardin et al, {\em Proc. of the Workshop on
Future Physics at HERA, DESY 1996}, eds G. Ingleman, A. De Roeck \&
R. Klanner, Vol I, p13. 
\bibitem{blobel} V. Blobel, {\em Proc. 1984 CERN School of Computing},
Aiguablava, Spain, \\ CERN 85-09, 88, (1985). 
\bibitem{dag1} G. D'Agostini, \Journal{\NIMA}{362}{487}{1995}.
\bibitem{zech} G. Zech, DESY preprint DESY 95-113, (1995).
\bibitem{aqthesis} A. Quadt, Oxford Thesis 1997.
\bibitem{gmuon} I.G. Bird, {\em NMC Internal Report} NMC/91/1, (1991).
\bibitem{syserrs} D.E. Soper \& J.C. Collins, {\em Issues in the
determination of PDFs}, CTEQ Note 94/01, {\tt hep-ph/9411214};\\
M. Botje, M. Klein \& C. Pascaud, {\em Proc. of the Workshop on
Future Physics at HERA, DESY 1996}, eds G. Ingleman, A. De Roeck \&
R. Klanner, Vol I, p33;\\
P. Burrows et al, {\em Proc. of  Snowmass Workshop 1996}, 
{\tt hep-ex/9612012}. 
\bibitem{emc_rev} T. Sloan, K. Bazizi, S.J. Wimpenny, {\em Proc. ICHEP90,
Singapore}, edited by K.K.Phua and Y. Yamaguchi, World Scientific 1990, p639.
\bibitem{whitlow92} L.W. Whitlow et al, \Journal{\PLB}{282}{475}{1992}.
\bibitem{whitlow_thesis} L.W. Whitlow, Ph.D. Thesis Stanford University,
SLAC Report 357 (1990).
\bibitem{BCDMS} A Benvenuti et al, \Journal{\PLB}{223}{485}{1989},
              \Journal{\em ibid}{237}{592}{1990}.
\bibitem{bcdmserr} BCDMS, A.C. Benvenuti et al, 
\Journal{\PLB}{195}{91}{1987};\\ A. Ouraou, Th\`{e}se, Universit\'{e}
Paris VII (1988).
\bibitem{nmcdata1} NMC, M. Arneodo et al, \Journal{\PLB}{364}{107}{1995}.
\bibitem{kabuss}  NMC, E.-M. Kabuss, {\em Proc. of DIS96, Rome},  
Eds G. D'Agostini \& A. Nigro, World Scientific 1997, p.136.
\bibitem{nmcdata2} NMC, M. Arneodo et al, \Journal{\NPB}{483}{3}{1997}.
\bibitem{milsztajn} A. Milsztajn et al, \Journal{\ZPC}{49}{527}{1991}.
\bibitem{e665data} E665, M.R. Adams et al, \Journal{\PRD}{54}{3006}{1996}.
\bibitem{hera92} H1, I. Abt et al, \Journal{\NPB}{407}{515}{1993};\\
                ZEUS, M. Derrick et al, \Journal{\PLB}{316}{412}{1993}.
\bibitem{hera93} H1, T. Ahmed et al, \Journal{\NPB}{439}{471}{1995};\\
                ZEUS, M. Derrick et al, \Journal{\ZPC}{65}{379}{1995}.
\bibitem{zv94} ZEUS, M. Derrick et al, \Journal{\ZPC}{69}{607}{1996}.
\bibitem{h194r} H1, C. Adloff et al, \Journal{\PLB}{393}{452}{1997}.
\bibitem{zbpc95} ZEUS, J. Breitweg et al, DESY preprint DESY 97-135.
\bibitem{zsvx95} ZEUS, B. Surrow, {\em Proc. of DIS97, Chicago},
{\tt http://www.hep.anl.gov/dis97/}.
\bibitem{ccfr_lambda} CCFR, P.Z. Quintas et al, \Journal{\PRL}{71}{1307}{1993}.
\bibitem{ccfr_gls} CCFR, W.C. Leung et al, \Journal{\PLB}{317}{655}{1993};\\
D. Harris et al, Fermilab-Conf-95-114.
\bibitem{ccfr_seligman} CCFR, W. G. Seligman et al, {\tt hep-ex/9701017}, 
\Journal{PRL}{79}{1213}{1997};\\ 
W. G. Seligman, Ph. D. thesis, Columbia University 1997, Nevis Report 292.
\bibitem{ccfr_radcor1} A. De R\'{u}jula et al, \Journal{\NPB}{154}{394}{1979}.
\bibitem{ccfr_dimuon} CCFR, A. O. Bazarko et al, \Journal{\ZPC}{65}{189}{1995}.
\bibitem{brodsky-ma} S. Brodsky \& B. Ma, \Journal{\PLB}{381}{317}{1996}. 
\bibitem{CCFRth} S.P. Mishra, {\em Proc. of Lepton-Photon Symposium and 
Europhysics Conf, Geneva 1991}, edited by S. Hegarty, K. Potter and 
E. Quercigh, p135.
\bibitem{GKR} M. Gluck et al, \Journal{\PLB}{398}{381}{1997}.
\bibitem{caldwell97} A. Caldwell, {\em Unpublished talk at the 1997 DESY Theory
Workshop},\\ Hamburg, September 1997.
\bibitem{barabash} IHEP-JINR, L. S. Barabash et al, {\tt hep-ex/9611012},
submitted to \PLB.
\bibitem{hand} L.N. Hand {\it Phys Rev }{\bf 129} (1963) 1834.
\bibitem{Gilman}F.J.Gilman, {\it Phys Rev }{\bf 167} (1968) 1366.
\bibitem{emcr} EMC, J.J. Aubert et al, \Journal{\NPB}{259}{189}{1985};
\Journal{\em ibid}{293}{740}{1987}.
\bibitem{cdhswr} CDHSW, P. Berge et al, \Journal{\ZPC}{49}{187}{1991}.
\bibitem{e140x} E140X, L.H. Tao et al, \Journal{\ZPC}{70}{387}{1996}.
\bibitem{ccfr_r} CCFR, U.K. Yang et al, \Journal{\JPG}{22}{775}{1996};\\
A. Bodek, {\em Proc. of DIS96, Rome} Eds G. D'Agostini \& A. Nigro,\\
World Scientific 1997, p213.
\bibitem{nmc_r} NMC, A.J. Milsztajn, {\em ibid}, p220.
\bibitem{bry} A. Bodek, S. Rock \& U.K. Yang, Univ. 
Rochester preprint UR-1355, 1995.
\bibitem{nnlor} E. Zijlstra \& W.L. van Neerven, 
\Journal{\NPB}{383}{525}{1992}.
\bibitem{royon} H. Navelet et al., \Journal{\PLB}{364}{357}{1996}.
\bibitem{dewolf} A. De Roeck and E. De Wolf, \Journal{\PLB}{388}{843}{1996}.
\bibitem{thornefl} R.S. Thorne, {\tt hep-ph/9708302}.
\bibitem{kps} M.W. Krasny, W. Placzek \& H. Spieseberger, 
\Journal{\ZPC}{53}{687}{1992};\\ W. Placzek, Ph.D. Thesis, 
\Journal{Acta Physica Polonica}{B24}{1229}{1993}.
\bibitem{frey} A. Frey, Ph.D. Thesis, University of Bonn, BONN-IR-96-03 (1996).
\bibitem{favart} L. Favart et al, \Journal{\ZPC}{72}{425}{1996}.
\bibitem{bauerdick} L. Bauerdick, A Glazov and M. Klein,
{\em Proc. of the Workshop on Future Physics at HERA, DESY 1996}, 
editors G. Ingleman, A. De Roeck and R. Klanner, Vol I, p77.
\bibitem{bfp} BFP, A.R. Clark et al, \Journal{\PRL}{45}{1465}{1980};\\
             G.D. Gollin et al, \Journal{\PRD}{24}{55}{1981}.
\bibitem{emcf2cc} EMC, J.J. Aubert et al, \Journal{\NPB}{213}{31}{1983}.
\bibitem{fastbgf} S. Riemersma, J. Smith \& W.L. van Neerven,
            \Journal{\PLB}{347}{143}{1995}.
\bibitem{nlocc_pty} B.W. Harris \& J. Smith, \Journal{\NPB}{452}{109}{1995};\\
              \Journal{\PLB}{353}{535}{1995}.
\bibitem{pdg96} Particle Data Group, R. M. Barnett et al, 
\Journal{\PRD}{54}{1}{1996}.
\bibitem{h1_d*} H1, C. Adloff et al, \Journal{\ZPC}{72}{593}{1996}.
\bibitem{aroma} G. Ingelman, J. Rathsman \& G.A. Schuler, AROMA2.1 DESY 
ISSN 0418-9833 (1995).
\bibitem{z_d*} ZEUS, M. Derrick et al, DESY-97-089, 
\Journal{\PLB}{407}{402}{1997}.
\bibitem{nmcdata3} NMC, M. Arneodo et al, \Journal{\NPB}{487}{3}{1997}.
\bibitem{e665_ratio} E665, M. R. Adams et al, \Journal{\PRL}{75}{1466}{1995}.
\bibitem{nmc_gsr} NMC, M. Arneodo et al, \Journal{\PRD}{50}{R1}{1994}.
%
%
%
\bibitem{esw} R.K. Ellis, W.J. Stirling \& B.R. Webber,
{\em QCD and Collider Physics}, CUP 1996.
\bibitem{BurGam} A.\ Buras and K. Gaemers, \Journal{\NPB}{132}{249}{1978}.
\bibitem{BB} A. Buras and  Bialas, \Journal{\PRD}{21}{1825}{1980}.
\bibitem{OR} J.F.\ Owens and K. Reya, \Journal{\PRD}{17}{3003}{1979}.
\bibitem{Yndurainold}C.\ Lopez and F.J.\ Yndurain, 
\Journal{\NPB}{171}{231}{1980}, \Journal{\NPB}{183}{157}{1981};
A. Gonzalez-Arroyo, et al, \Journal{\NPB}{153}{161}{1979},
\Journal{\NPB}{174}{474}{1980}.
\bibitem{PDFLIB} H. Plothow-Besch, 
\Journal{\em Comp.Phys.Comm}{75}{396}{1993}. 
\bibitem{VirMil}M.\ Virchaux and A.\ Milsztajn, 
\Journal{\PLB}{274}{221}{1992}.
\bibitem{NMCrat} P.Amaudruz et al, \Journal{\NPB}{371}{3}{1992}.
\bibitem{EMC} J.J. Aubert etal, \Journal{\NPB}{293}{740}{1987}, 
\Journal{\NPB}{213}{31}{1983}.
\bibitem{Varvell} K. Varvell et al, \Journal{\ZPC}{36}{1}{1987}.
\bibitem{Sidarov1}A.V. Sidorov, \Journal{\PLB}{389}{379}{1996}.
\bibitem{KKSP97} A.L.Kataev et al, {\tt hep-ph/9706534}.
\bibitem{Sidarov2}A.V. Sidorov, {\tt hep-ph/9609345}, 
JINR Rapid Comm.80,11 (1996).
\bibitem{NMCnuclear} J. Gomez et al, \Journal{PRD}{49}{4348}{1994}:\\
R.G. Arnold et al, \Journal{\PRL}{52}{727}{1984};\\
M. Arneodo et al, \Journal{\NPB}{331}{1}{1990};\\
J. Ashman et al, \Journal{\ZPC}{57}{211}{1993};\\
M. Arneodo et al, \Journal{\NPB}{481}{3}{1997}, \Journal{\NPB}{481}{23}{1997}.
\bibitem{Brock} R. Brock, \Journal{PRL}{44}{1027}{1980}, {\em Proc. of 
Physics at Fermilab in the 1990's,
Breckenridge, Colorado}, ed. D. Green and H.Lubatti, World Scientific 1990, 
p358;\\
J. Kaplan and F. Martin, \Journal{\NPB}{115}{333}{1976};\\
H. Georgi and H.D. Politzer, \Journal{\PRD}{14}{1829}{1976}.
\bibitem{HMRS}P.N.\ Harriman et al., \Journal{\PRD}{42}{798}{1990}.
\bibitem{DOEHLQGHR} J.F.Owens and W.K.Tung, \Journal{\em Ann. Rev. Nucl. Sci.}
 {42}{291}{1992}.
\bibitem{KMRS} J.\ Kwiecinski et al., \Journal{\PRD}{42}{3645}{1990}.
\bibitem{MTB} J.G. Morfin and W.K.Tung, \Journal{\ZPC}{52}{13}{1991}.
\bibitem{MRSD'} A.D. Martin, R.G.Roberts and W.J. Stirling, 
 \Journal{\PLB}{30}{145}{1993}.
\bibitem{MRSG} A.D. Martin, R.G. Roberts and W.J. Stirling, \Journal{\PLB}{354}
{155}{1995}.
\bibitem{CTEQ1} J. Botts et al, \Journal{\PLB}{304}{159}{1995}.
\bibitem{CTEQ3} H. Lai et al, \Journal{\PRD}{51}{4763}{1995}.
\bibitem{CTEQ4} H. Lai et al, \Journal{\PRD}{55}{1280}{1997}.
\bibitem{Parker} M.A. Parker et al, \Journal{\NPB}{232}{1}{1984}.
\bibitem{Allasia} D. Allasia et al, \Journal{\NPB}{239}{301}{1984},
\Journal{\PLB}{135}{231}{1984},\\ \Journal{\ZPC}{28}{321}{1985}.
\bibitem{FF} R. Field and R.P.Feynman, \Journal{\PRD}{15}{2590}{1977}.
\bibitem{gurvitz} S.A. Gurvitz, {\tt nucl-th/9701014}.
\bibitem{thom}A.W. Thomas and W. Melnitchouk, {\tt hep-ph/9708484}.
\bibitem{NMCGott} P. Amaudruz et al, \Journal{\PLB}{295}{159}{1992},
\Journal{\PRL}{66}{2712}{1991}.
\bibitem{thomas} F.M. Steffens and A.W. Thomas, \Journal{\PRC}{55}{900}{1997}. 
\bibitem{kumano} S. Kumano, {\tt hep-ph/9707524}.
\bibitem{buccella} F. Buccella et al, {\tt hep-ph/9705475}.
\bibitem{CDHSmu} H. Abramowicz et al, \Journal{\ZPC}{15}{19}{1982}.
\bibitem{GP} H. Georgi and H.D. Politzer, \Journal{\PRD}{14}{1829}{1976}.
\bibitem{E605} G. Moreno et al, \Journal{\PRD}{43}{2815}{1991};\\
D. Malde et al, \Journal{\PRL}{64}{2479}{1990};\\
C.N. Brown et al, \Journal{\PRL}{63}{2537}{1989}.
\bibitem{CDFDY} F. Abe et al,\Journal{\PRD}{49}{1}{1994}.
\bibitem{NA51} A. Baldit et al, \Journal{\PLB}{332}{244}{1994}.
\bibitem{Towell} R. Towell, {\em Proc. of DIS97, Chicago},
{\tt http://www.hep.anl.gov/dis97/}.
\bibitem{CDFW} F. Abe et al, \Journal{\PRL}{74}{850}{1995}, 
\Journal{\PRL}{69}{28}{1992}.
%
\bibitem{qcdnmc} NMC, M. Arneodo et al.,
\Journal{\PLB}{309}{222}{1993}.
\bibitem{PRYTZ} K. Prytz, \Journal{\PLB}{311}{286}{1993};\\
 K. Prytz, Rutherford-Appleton Laboratory
                           preprint : RAL-94-036.
\bibitem{H1GLU} H1,
 I. Abt  et al., Phys. Lett. \Journal{\PLB}{321}{161}{1994}.
\bibitem{GOLEC} K. Golec-Biernat, \Journal{\PLB}{328}{495}{1994}.
\bibitem{EKL} R.K.\ Ellis, Z.\ Kunst and E.M.\ Levin, \Journal{\NPB}{420}
{517}{1994}.
\bibitem{gluonprox} K. Bora nd D.K.Choudury, \Journal{\PLB}{354}{151}{1995};\\
A.V.Kotikov and G.Parente, \Journal{\PLB}{379}{195}{1996};\\
M.B. Gay-Ducati and V.P.B. Goncalves, \Journal{\PLB}{390}{401}{1997}.
\bibitem{BFKL} E.A.\ Kuraev, L.N.\ Lipatov, V. \ Fadin, \Journal{\JETP}{45}
{199}{1977};\\
Ya.Ya.\ Balitsky and L.N.\ Lipatov, \Journal{\SJNP}{28}{822}{1978};\\
L.N.\ Lipatov, \Journal{\SJNP}{63}{904}{1980};\\
L.N.\ Lipatov,{\em Perturbative QCD}, editor A. Mueller,
(World Scientific, 1989), p411.
\bibitem{kwiehybrid} J. Kwiecinski,  \Journal{\ZPC}{29}{561}{1985}.
\bibitem{qcdh1} H1, S. Aid et al., \Journal{\NPB}{470}{3}{1996}.
\bibitem{qcdzeus}ZEUS, paper N-647, submitted to
{\em EPS HEP97 Conference},\\ Jerusalem, August 1997.
\bibitem{alekhin} S. Alekhin, IHEP 96-79, {\tt hep-ph/9611213}.
\bibitem{gammanlo} P. Aurenche et al., \Journal{\NPB}{399}{34}{1993}.
\bibitem{WA70} M. Bonesini et al, \Journal{\ZPC}{38}{371}{1988}.
\bibitem{UA6} A. Bernasconi et al, \Journal{\PLB}{206}{163}{1988};\\
G. Sozzi et al,\Journal{\PLB}{317}{243}{1993}.
\bibitem{E706} G. Alversen et al, \Journal{\PRD}{48}{5}{1993};\\
M. Zielinski, {\em Proc. of DIS97, Chicago},\\ 
{\tt http://www.hep.anl.gov/dis97/}, {\tt hep-ex/9711017}.
\bibitem{R807} T. Akesson et al, \Journal{\SJNP}{51}{836}{1990}.
\bibitem{UA2g} J. Alitti et al, \Journal{\PLB}{299}{174}{1993}.
\bibitem{CDFg} F. Abe et al, \Journal{\PRL}{73}{2662}{1994}.
\bibitem{BaerBlair} H. Baer and D. Reno, {\em Proc. of DIS97, Chicago};\\
R.Blair, {\em Proc. of DIS97, Chicago}, {\tt http://www.hep.anl.gov/dis97/}.
\bibitem{Zielinski} M. Zielinski, {\em Proc. of DIS97, Chicago},
{\tt http://www.hep.anl.gov/dis97/}.
\bibitem{prompt_kt} J. Huston et al., \Journal{\PRD}{51}{6139}{1995}.
\bibitem{Durham95} R.G. Roberts, \Journal{\JPG}{22}{675}{1996}.
\bibitem{Snomass97}M. Albrow et al, {\tt hep-ph/9706470}, {\em Summary of 
Structure Function Subgroup, Proc. of Snowmass Workshop 1996}.
\bibitem{DIS97} J. Bl\"umlein et al, {\em Summary of Working Group I,  
Proc. of DIS97, Chicago}, {\tt hep-ph/9707420}.

\bibitem{CDF} F Abe et al, \Journal{\PRL}{77}{438}{1996} \\
F. Chlebana, {\em Proc. of DIS97, Chicago}, {\tt http://www.hep.anl.gov/dis97/}
\bibitem{D0} G.C. Blazey, {\em Proc. of XXXI Rencontre de Moriond}, 
March 1996, p155;\\
R. Hirosky, {\em Proc. of DIS97, Chicago}, {\tt http://www.hep.anl.gov/dis97/}.
\bibitem{NEWPHYS} V. Barger et al, \Journal{\PLB}{382}{178}{1996};\\
G. Altarelli etal, \Journal{\PLB}{375}{292}{1996};\\
P. Chiappetta et al, \Journal{\PRD}{54}{789}{1996};\\
J. Ellis and D. Ross, \Journal{\PLB}{383}{187}{1996};\\
R.S. Chivukula et al, \Journal{\PLB}{380}{92}{1996};\\
M. Bander, \Journal{\PRL}{77}{601}{1996}.
\bibitem{Huston}J. Huston, \Journal{\PRL}{77}{444}{1996}.
\bibitem{KK} M. Klasen and G. Kramer, \Journal{\PLB}{386}{384}{1996}.
\bibitem{glover} W. Giele et al, \Journal{\PRD}{53}{120}{1996}.
\bibitem{h1jetgluon} H1, S. Aid et al.,
\Journal{\NPB}{354}{494}{1995}.
\bibitem{dewolf2} E.A.DeWolf et al, {\em Summary of Hadron Final States Working
Group, Proc. of DIS97, Chicago}, {\tt http://www.hep.anl.gov/dis97/}.
\bibitem{frix} G. Frixione and G. Ridolfi, {\tt hep-ph/9707345}.
\bibitem{berger} D. Graudenz et al.,  \Journal{\ZPC}{70}{77}{1996}.
\bibitem{nnlofl} E.Zijlstra and W.L.vanNeerven, 
\Journal{\NPB}{383}{525}{1992};\\
S. Larin and J. Vermaseren, \Journal{\ZPC}{57}{93}{1993}
\bibitem{hqfl}E. Witten, \Journal{\NPB}{104}{445}{1976};\\
L. Orr and W.J.Stirling, \Journal{\PRL}{66}{1673}{1991};\\
E. Laenen et al, \Journal{\PLB}{291}{325}{1992}, 
\Journal{\NPB}{392}{162}{1993}, \Journal{\PLB}{392}{229}{1993}.
\bibitem{nlojaypsi} M. Kramer et al, \Journal{\PLB}{348}{657}{1995},
\Journal{\NPB}{459}{3}{1996}. 
\bibitem{rysbrod} M.G.Ryskin, \Journal{\ZPC}{57}{89}{1993};\\
M.G. Ryskin et al, \Journal{\ZPC}{76}{231}{1997};\\
S.J.Brodsky et al, \Journal{\PRD}{50}{3134}{1994}.
\bibitem{BF}R.D.\ Ball and S.\ Forte, \Journal{\PLB}{335}{77}{1994}, and
\Journal{\PLB}{336}{77}{1994}.
\bibitem{GRV}M.\ Gluck, E.\ Reya and A.\ Vogt, 
\Journal{\PLB}{306}{391}{1993}, \Journal{\NPB}{130}{76}{1992}, 
\Journal{\ZPC}{53}{433}{1995}.
\bibitem{MRSAmod} A.D. Martin, R.g.Roberts and W.J.Stirling, 
\Journal{\PRD}{51}{4763}{1995}.
\bibitem{MRSalf}A.D. Martin et al, \Journal{\PLB}{356}{89}{1995}.
\bibitem{huskuhl} J. Huston, {\em Proc. of DIS97, Chicago}, \\
S. Kuhlmann et al, {\em Proc. of DIS97, Chicago},
{\tt http://www.hep.anl.gov/dis97/}.
\bibitem{botje} M. Botje, {\em Proc. of DIS97, Chicago}, {\tt hep-ph/9707289}.
\bibitem{Rob97} R.G. Roberts, {\em Proc. of DIS97, Chicago}, 
{\tt hep-ph/9706269}.
\bibitem{Marti} S. Marti, {\em Proc. of DIS97, Chicago},
{\tt http://www.hep.anl.gov/dis97/}.
\bibitem{Bethke}S. Bethke, \Journal{Nucl.Phys.Proc.Supp.}{54A}{314}{1997}.
\bibitem{Stiralf}W.J. Stirling, {\tt hep-ph/9709429}.
\bibitem{weber} M. Weber, {\em Proc. of DIS97, Chicago},
{\tt http://www.hep.anl.gov/dis97/}.
\bibitem{jetrates} D. Graudenz, {\tt hep-ph/9708362}.
\bibitem{botje2} M. Botje, {\em Proc. of DIS97, Chicago}, {\tt hep-ph/9707292}.
\bibitem{Blumalf}J. Bl\"umlein et al, {\em Proc. of the Workshop on 
Future Physics at HERA, DESY 96}, eds G. Ingelman, 
A. De Roeck \& R. Klanner, Vol I, p52.
\bibitem{kataev0919} A.L. Kataev et al, \Journal{\PLB}{388}{179}{1996}.
\bibitem{russkys}J. Chyla and J.Rames, \Journal{\ZPC}{31}{151}{1986}:\\
V.G. Krivokhizkhin et al, \Journal{\ZPC}{36}{51}{1987},\Journal{\ZPC}{48}{347}
{1990};\\
A.V. Kotikov, G. Parente and J. Sanchez-Guillen, 
\Journal{\ZPC}{58}{465}{1993};\\
A.L. Kataev and A.V. Sidorov, \Journal{\PLB}{331}{179}{1994};\\
J. Chyla and J. Rames, \Journal{\PLB}{343}{351}{1995}
\bibitem{CDHSW} H. Abramowicz et al, \Journal{\PRL}{57}{298}{1986},
\Journal{\ZPC}{28}{51}{1985};\\
J. Berge et al, \Journal{\ZPC}{49}{187}{1991}.
\bibitem{SKAT} V.V. Ammsov et al, \Journal{\ZPC}{30}{175}{1986}.
\bibitem{BEBC_GGM} P.C. Bosetti et al, \Journal{\NPB}{203}{362}{1982}.
\bibitem{JINR}L.S. Barabash et al., JINR-E1-96-308, {\tt hep-ex/9611012}.
\bibitem{KKSPpade} A.L. Kataev, {\tt hep-ph/9709509}.
\bibitem{Toksid} M.V. Tokarev and A.V.Sidorov, {\tt hep-ph/9707438}.
\bibitem{Shirkov} D.V. Shirkov et al, {\tt hep-ph/9707514}.
\bibitem{ccfr_xF3}M. Shaevitz et al, 
\Journal{\em Nucl. Phys. Proc. Supp.}{B38}{188}{1995}.
\bibitem{chyla}J. Chyla and A. L. Kataev,\Journal{\PLB}{297}{385}{1992}.

\bibitem{Spentzouris} P. Spentzouris, {\em Proc. of DIS97, Chicago},
{\tt http://www.hep.anl.gov/dis97/}.
\bibitem{BAllalpha} R.D. Ball and S. Forte, \Journal{\PLB}{358}{365}{1995}. 
\bibitem{deRoeck} A. De Roeck et al, \Journal{\PLB}{385}{411}{1996}.
\bibitem{Ballupdate}R.D. Ball and S. Forte, {\em Proc. of DIS96, Rome}, 
Eds G. D'Agostini \& A. Nigro, World Scientific 1997, p208.
\bibitem{Thornenew} R.S. Thorne, \Journal{\PLB}{392}{463}{1997}, 
{\tt hep-ph/9701241}, {\tt hep-ph/9710541}, {\em Proc. of DIS97, Chicago},
{\tt hep-ph/9706233}.
\bibitem{priv}R.S. Thorne, private communication.
\bibitem{buccella1} (a) F. Buccella \& J. Soffer, 
\Journal{\em Mod. Phys. Lett.}{A8}{225}{1993};\\
(b) C. Bourrely et al, \Journal{\ZPC}{62}{431}{1994};\\
(c) C. Bourrely \& J. Soffer, \Journal{\PRD}{51}{2108}{1995};\\
(d) C. Bourrely \& J. Soffer, \Journal{\NPB}{445}{341}{1995}.
\bibitem{buccella2} C. Bourrely \& J. Soffer, \Journal{\PRD}{53}{4067}{1996};
F. Buccella et al, {\em Proc. of HADRONS96, Ukraine}, p130; G. Miele, 
{\em Cont. to Lepton-Photon Symposium, Beijing 1995}, {\tt hep-ph/9508203};
F. Buccella et al, {\tt hep-ph/9705475}.
\bibitem{jafell} J. Ellis and R.L. Jaffe, \Journal{\PRD}{9}{1444}{1974},
\Journal{\PRD}{10}{1669}{1974}.
\bibitem{mangano} G. Mangano, G. Miele \& G. Migliore, 
\Journal{\em Nuovo Cim. A}{108}{867}{1995}.
\bibitem{bhalerao} R. S. Bhalerao, \Journal{\PLB}{380}{1}{1996};
erratum \Journal{\em ibid}{B387}{881}{1996}.
%
\bibitem{MITbag} A. Chodos et al, \Journal{\PRD}{10}{2599}{1974}.
\bibitem{bagsupport} F.M. Steffens \& A.W. Thomas, 
\Journal{\NPA}{568}{798}{1994}.
\bibitem{NLObag} F.M. Steffens \& A.W. Thomas,
\Journal{\em Prog. Th. Phys. Suppl.}{120}{145}{1995}.
\bibitem{jasiak} J. Jasiak, {\tt hep-ph/9706503}.
%
\bibitem{gockeler} M. G\"ockeler et al, \Journal{\PRD}{53}{2317}{1996}.
\bibitem{schierholz} C. Best et al, Minireview, 
{\em Proc. of DIS97, Chicago}, {\tt hep-ph/9706502}.
\bibitem{weigl} L. Mankiewicz \& T. Weigl, \Journal{\PLB}{380}{134}{1996};\\
\Journal{\PLB}{389}{334}{1996}.
\bibitem{Ioffe} B.L. Ioffe, V.A. Khoze \& L.N. Lipatov, 
{\em Hard Processes}, North-Holland (1984), Ch. 3.
%
\bibitem{bj96} J.D. Bjorken, {\em Talk at the SLAC Summer Institute 1996},
{\tt hep-ph/9611421}.
\bibitem{shuryak} T. Sch\"afer \& E.V. Shuryak,
{\tt hep-ph/9610451}.
\bibitem{diakonov} D.I. Diakonov et al, \Journal{\NPB}{480}{341}{1996};
\Journal{\PRD}{56}{4069}{1997}.
\bibitem{GRSV} M. Gl\"uck, E. Reya, M. Stratmann \& W. Vogelsang,
\Journal{\PRD}{53}{4775}{1996}.
\bibitem{gamberg} L. Gamberg, H. Reinhardt \& H. Weigl, talk at the
{\em MENU97 Conference, Vancouver}, July 1997, {\tt hep-ph/9708266}.
\bibitem{hoyroy} P. Hoyer \& D.P. Roy, {\tt hep-ph/9705273}
\bibitem{brodsky8081} S.J. Brodsky et al, \Journal{\PLB}{93}{451}{1980};
\Journal{\PRD}{23}{2745}{1981}.
\bibitem{deRujula} A.\ de Rujula et al., \Journal{\PRD}{10}{1649}{1974}.
\bibitem{Peslam} H. Navelet et al, \Journal{\MPLA}{12}{857}{1997}.
\bibitem{Yndnow} F.J. Yndurain, {\tt hep-ph/9604263}, {\tt hep-ph/9605265}.
\bibitem{Kwiecinski} J.\ Kwiecinski, \Journal{\JPG}{19}{1443}{1993}.
\bibitem{NLOlnx} V.\ Fadin and L.N.\ Lipatov, 
\Journal{\NPB}{477}{767}{1996};\\
V.S. Fadin, R.Fiore and A.Quartarolo, \Journal{\PRD}{53}{2729}{1996};\\
V.S. Fadin, R.Fiore and M.I.Kotsky, \Journal{\PLB}{359}{181}{1995}, 
\Journal{\PLB}{389}{737}{1996}, \Journal{\PLB}{387}{593}{1996}, 
hep-ph/9704267;\\
V.Del Duca, \Journal{\PRD}{54}{989}{1996}, \Journal{\PRD}{54}{4474}{1996};\\
G. Camici and M. Ciafaloni, \Journal{\PLB}{386}{341}{1996}, 
\Journal{\NPB}{496}{305}{1997}, {\tt hep-ph/9707390};\\
A.Stepanova and F.V. Tkachov, {\tt hep-ph/9710242};\\
V.S. Fadin et al, {\tt hep-ph/9711427}.
\bibitem{doksh_dis97} Y.L. Dokshitzer, {\em Summary talk, Proc. of  DIS97, 
Chicago}, {\tt hep-ph/9706375}.
\bibitem{levin97} E.M. Levin, {\em Summary talk, Proc. of DIS97, Chicago},
{\tt hep-ph/9706341}, {\tt hep-ph/9706448}.
\bibitem{CCFM} M.\ Ciafaloni, \Journal{\NPB}{296}{49}{1988};\\
S.\ Catani, F.\ Fiorini and G.\ Marchesini, \Journal{\PLB}{234}{339}{1990};
\Journal{\NPB}{336}{18}{1990}.
\bibitem{Webber} G.\ Marchesini and B.R.\ Webber, \Journal{\NPB}{349}
{617}{1991};\\
G.\ Marchesini, {\em Proc. of 1990 Workshop on QCD at 200 TeV}, edited by
L.\ Cifarelli and Yu.\ Dokshitzer,(N.Y.Plenum, 1990)p.183
\bibitem{KMS}J.\ Kwiecinski, A.D.\ Martin and P.J.\ Sutton, \Journal{\PRD}{52}
{1445}{1995}, \Journal{\PRD}{53}{6094}{1996}.
\bibitem{KMS96}J.\ Kwiecinski, A.D.\ Martin and P.J.\ Sutton, \Journal{\ZPC}
{71}{585}{1996};\\
B.\ Andersson, G.\ Gustafson and J.\ Samuleson, 
\Journal{\NPB}{463}{217}{1996};\\
M.\ Ciafaloni, \Journal{\NPB}{296}{49}{1988}.
\bibitem{Bottazzi} G. Bottazzi et al, {\tt hep-ph/9702418}; 
{\tt hep-ph/9708474}; \\
G. Salam, {\em Proc. of DIS97, Chicago}, {\tt hep-ph/9705233}; 
{\tt hep-ph/9707383};\\
M. Scorletti, {\tt hep-ph/9710559}.
\bibitem{Li2} H.\ Li, {\tt hep-ph/9703328}.
\bibitem{Li3} H.\ Li, {\tt hep-ph/9709236}.
\bibitem{KMStas}J. Kwiecinski, A.D. Martin and A.M. Stasto,
\Journal{\PRD}{56}{3991}{1997}.
\bibitem{bound}A.L.Ayala et al, \Journal{\PLB}{388}{188}{1996};\\
S.M.Troshin and N.E.Tyurin, \Journal{\em Europhys.Lett.}{37}{239}{1997}.
\bibitem{Mueller93} A.H.\ Mueller, \Journal{\JPG}{19}{1463}{1993},\\
{\it Nucl. Phys.} {\bf B} (Proc. Suppl.)
18C (1990) 125.
\bibitem{hotspots}M.G.\ Ryskin and E.M. Levin, \Journal{\em Phys.Rep.}
 {189}{267}{1990};\\
A.H.\ Mueller, \Journal{\NPB}{335}{115}{1990};\\
W.Zhu et al, \Journal{\PRD}{54}{847}{1996}. 
\bibitem{Mueller96} A.H. Mueller, \Journal{\PLB}{396}{251}{1997}.
\bibitem{Lev} F.M. Lev, {\tt hep-ph/9606204}, {\tt hep-ph/9606334}.
\bibitem{GLRMQ} L.V.\ Gribov, E.M. Levin, M.G.\ Ryskin, \Journal{\em 
Phys.Rep}{100}{1}{1983};\Journal{\NPB}{188}{155}{1981}\
A.H.\ Mueller and J.\ Qiu, \Journal{\NPB}{260}{427}{1986}.
\bibitem{AKMS} J.\ Kwiecinski et al, \Journal{\PRD}{44}{2640}{1991},
\Journal{\PRD}{46}{921}{1992}; 
A.J.\ Askew et al, \Journal{\PRD}{47}{3775}{1993}, 
\Journal{\NPB}{416}{739}{1994}, \Journal{\PRD}{49}{4402}{1994},
\Journal{\PLB}{325}{212}{1994}, \Journal{\MPLA}{8}{3813}{1993}.
\bibitem{GLM}E.\ Gotsman, E.M.\ Levin and U.\ Maor, 
\Journal{\PLB}{379}{186}{1996}.
\bibitem{whu} W.Zhu et al, \Journal{|NPB}{449}{183}{1995}
\bibitem{multi} J.\ Bartels, \Journal{\ZPC}{60}{4171}{1993}, 
\Journal{\PLB}{298}{204}{1993};\\
E.M.\ Levin,M.G.\ Ryskin and A.G.\ Shuvaev, \Journal{\NPB}{387}{519}{1992};\\
E.\ Laenen, E.M.\ Levin and A.G. Shuvaev, \Journal{\NPB}{419}{38}{1994};\\
A.\ Martin, \Journal{\JPG}{19}{1604}{1993}.
\bibitem{Laenen} E.\ Laenen and E.M.\ Levin, \Journal{\NPB}{451}{207}{1995}.
\bibitem{GM} see for example, E.M. Levin, {\tt hep-ph/9710546}.
\bibitem{AGL} A.L. Ayala, M. B. Gay-Ducati and E.M. Levin, \Journal{\NPB}{493}
{305}{1997}, {\tt hep-ph/9706347}, {\tt hep-ph/9710539}.
\bibitem{shab} Yu. M. Shabelski and D.Treleani, \Journal{\PLB}{403}{364}{1977}.
\bibitem{Bartels} J.\ Bartels, \Journal{\PLB}{298}{204}{1993}, \Journal{\ZPC}
{60}{471}{1993};\\
E.M. Levin, M.G. Ryskin and A .G. Shuvaev, \Journal{\NPB}{387}{580}{1992};\\
A.P.\ Bukhvostov et al, \Journal{\NPB}{258}{601}{1985}.
\bibitem{newBart} J. Bartels and C. Bontus, {\em Proc. of DIS97, Chicago},\\
{\tt http://www.hep.anl.gov/dis97/}.
\bibitem{Mueller} A.H.\ Mueller, \Journal{\NPB}{415}{373}{1994};\\
A.H.\ Mueller and B.\ Patel, \Journal{\NPB}{425}{471}{1995};\\ 
A.H.\ Mueller, \Journal{\NPB}{437}{107}{1995};\\
Chen Zhang and A.H.\ Mueller, \Journal{\NPB}{451}{579}{1995}
\bibitem{PesSal} R. Peschanski and G. Salam, {\em Proc. of the Workshop on 
Future Physics at HERA, DESY 1996}, eds G. Ingelman, 
A. De Roeck \& R. Klanner, Vol. I, p110.
\bibitem{Jalilian} J. Jalilian-Marian et al, 
\Journal{\PRD}{55}{5414}{1997}, 
{\tt hep-ph/9701284}, {\tt hep-ph/9706377}, {\tt hep-ph/9709432};\\
L. McLerran and R. Venugopalan, \Journal{\PRD}{49}{2233,3352}{1994},
 \Journal{\PRD}{50}{2225}{1994}, \Journal{\PRD}{53}{458}{1996}
\bibitem{reggecal}A.R. White, {\em Proc. of DIS97, Chicago}, 
{\tt hep-ph/9705297};\\
C. Ewerz, {\tt hep-ph/9707257}.
\bibitem{haidt}W. Buchmuller and D. Haidt, {\tt hep-ph/9605428}.
\bibitem{haidt2} D. Haidt, {\em Proc. of DIS97, Chicago},
{\tt http://www.hep.anl.gov/dis97/}.
\bibitem{deRdeW} A. De Roeck and E. de Wolf, \Journal{\PLB}{388}{843}{1996}.
\bibitem{Mank}L. Mankiewicz et al, \Journal{\PLB}{393}{175}{1997}.
\bibitem{BFZakopane}R.D. Ball and S. Forte, 
\Journal{\em Acta Physica Polonica}{B26}{2097}{1995}
\bibitem{Barreiro}C. Lopez, F. Barreiro and F.J. Yndurain, \Journal{\ZPC}{72}
{561}{1996}.
\bibitem{Adel}K. Adel, F. Barreiro and F.J.Yndurain, \Journal{\NPB}{495}{221}
{1997}.
\bibitem{LYCross}C. Lopez and F.J. Yndurain, \Journal{\PRL}{44}{1118}{1980}.
\bibitem{Adel2}K. Adel et al, {\tt hep-ph/9612469}. 
\bibitem{FHS} J.R.\ Forshaw, P.N.\ Harriman and P.J.\ Sutton, 
\Journal{\JPG}{19}{1616}{1993}.
\bibitem{McDermott}M.F.\ McDermott, J.R.\ Forshaw and G.G.\ Ross, 
\Journal{\PLB}{349}{189}{1995};\\
M.F.\ McDermott and J.R.\ Forshaw, \Journal{\NPB}{484}{283}{1997}.
\bibitem{Gamici3} G. Camici and M. Ciafaloni, {\tt hep-ph/9707390}, 
\Journal{\PLB}{395}{118}{1997}.
\bibitem{cigar}J.\ Bartels, \Journal{\JPG}{19}{1611}{1993};\\
J.\ Bartels, H.\ Lotter and M.\ Vogt, \Journal{\PLB}{373}{215}{1996}.
\bibitem{ernst}I. Bojak and M. Ernst, \Journal{\PRD}{53}{80}{1996}.
\bibitem{Collins}
J.C.\ Collins and P.V.\ Landshoff, \Journal{\PLB}{276}{196}{1992}.
\bibitem{Ross} K.D. Anderson et al, {\em Proc. of DIS97, Chicago},
{\tt hep-ph/9706215}; {\tt hep-ph/9705466}.
\bibitem{Gamici} G. Camici and M. Ciafaloni, \Journal{\PLB}
{386}{341}{1996}.
\bibitem{ciafsep} M. Ciafaloni, {\tt hep-ph/9709390}.
\bibitem{Haakman} L.P.A. Haakman et al, {\tt hep-ph/9707262}.
\bibitem{BF97}R.D. Ball and S. Forte, \Journal{\PLB}{405}{317}{1997},
{\em Proc. of DIS97, Chicago}, {\tt hep-ph/9706459}.
\bibitem{White} A.R. White, {\em Proc. of AUP Workshop on QCD, Paris 1996}, 
p283.
\bibitem{AKMSGolec} Askew et al, \Journal{\PLB}{325}{212}{1994}.
\bibitem{ernst2} I. Bojak and M.Ernst, {\tt hep-ph/9702282}.
\bibitem{Peschanski}H.\ Navelet, R.\ Peschanski, Ch.\ Royon, S.\ Wallon,
\Journal{\PLB}{385}{357}{1996}.
\bibitem{Zoller}V.R. Zoller, {\em Proc. of DIS97,Chicago}, 
{\tt hep-ph/9706370};\\
N.N.Nikolaev et al, \Journal{\em JETP, Lett.}{66}{138}{1997}.
\bibitem{Braun} M. Braun and G.P.Vacca, {\tt hep-ph/9706314}.
\bibitem{Catani} S.\ Catani and F.\ Hautmann, \\ 
\Journal{\NPB}{427}{475}{1995}; \Journal{\PLB}{315}{517}{1993};\\
S.\ Catani. M.\ Ciafaloni and F.\ Hautmann, \\
\Journal{\NPB}{366}{135}{!991}; \Journal{\PLB}{242}{97}{1990};\\
J.\ Collins and R.K.\ Ellis, \Journal{\NPB}{360}{3}{1991}.
\bibitem{KM}J.\ Kwiecinski and A.D.\ Martin, \Journal{\PLB}{353}{123}{1995}.
\bibitem{Ciafaloni} M.\ Ciafaloni, \Journal{\PLB}{356}{74}{1995};\\
S.\ Catani, {\em Proc. of DIS95, Paris} 
Eds. J.F. Laporte and Y. Sirois, p281;\\
R.D.\ Ball and S.\ Forte, \Journal{\PLB}{358}{365}{1995},
{\em Proc. of DIS95, Paris}, Eds. J.F. Laporte and Y. Sirois, p273.
\bibitem{Li} H. Li, \Journal{\PLB}{405}{347}{1997}. 
\bibitem{Gamici2} G. Camici and M. Ciafaloni, \Journal{\NPB}
{496}{305}{1997}.
\bibitem{Catani95} S.\ Catani, \Journal{\ZPC}{70}{263}{1996}. 
\bibitem{EHW} R.K. Ellis, F. Hautmann and B.R. Webber, \Journal{\PLB}
{348}{582}{1995}.
\bibitem{BRV}J. Bl\"umlein, S. Riemersma, A.Vogt, 
\Journal{\em Acta Physica Polonica}{B28}{211}{1997}; 
{\em Proc. of SPIN96, Amsterdam} p218; 
\Journal{\em Nucl. Phys. Proc. Supp.}{51C}{30}{1996}; 
{\em Proc. of DIS96, Rome}, 
Eds G. D'Agostini \& A. Nigro, World Scientific 1997, p572.
\bibitem{BVnew} J. Bl\"umlein and A. Vogt, {\em Proc. of DIS97, Chicago},
 {\tt hep-ph/9706371}; {\tt hep-ph/9707488}.
\bibitem{Thorpriv}R.S. Thorne, private communication.
\bibitem{ballfortemom}R.D.\ Ball and S.\ Forte, 
\Journal{\PLB}{359}{362}{1995}.
\bibitem{Cataninew} S. Catani, \Journal{\ZPC}{75}{665}{1997}.
\bibitem{BFscheme} R.D. Ball and S. Forte, {\em Proc. of DIS95, Paris}, p273.
\bibitem{resmmers} R.K.\ Ellis, F.\ Hautmann and B.R.\ Webber, \Journal{\PLB}
{348}{582}{1995};\\
R.D.\ Ball and S.\ Forte, \Journal{\PLB}{351}{513}{1995}; {\em Proc. of XXX
Recontres de Moriond, 1995}, p143;
\Journal{\em Acta Physica Polonica}{B26}{2097}{1995}; 
{\em Proc. of DIS96, Rome}, 
Eds G. D'Agostini \& A. Nigro, World Scientific 1997, p172;\\
J.R.\ Forshaw, R.G.\ Roberts and R.S.\ Thorne, \Journal{\PLB}{356}{79}{1995};\\
R.S.\ Thorne, {\em Proc. of EPS Conference, Brussels 1995}, edited by
J. Lemone, C. VanderVelde and F. Verbeure, p123.
\bibitem{FRT} J.R.\ Forshaw, R.G.\ Roberts and R.S.\ Thorne, 
\Journal{\PLB}{356}{79}{1995}.
\bibitem{Bojak}I. Bojak and M. Ernst, \Journal{\PLB}{397}{296}{1997}.
\bibitem{ThorneR} R.S. Thorne, {\em Proc. of DIS96, Rome}, 
Eds G. D'Agostini \& A. Nigro, World Scientific 1997, p234; 
{\em Proc. of the Workshop on Future Physics 
at HERA, DESY 1996} eds G. Ingelman, 
A. De Roeck \& R. Klanner, Vol I, p107; {\em Proc. of DIS97, Chicago}, 
{\tt hep-ph/9706233}.
\bibitem{Martin} A.D.\ Martin, 
\Journal{\em Acta Physica Polonica}{B27}{1287}{1996};\\
M.G.\ Ryskin, Yu.M.\ Shabelskii, \Journal{\em Phys. Atom. Nucl}
{58}{1782}{1995}.
\bibitem{dDS} V. del Duca et al, \Journal{\PRD}{51}{4756}{1995}.
\bibitem{Neerven} W.L. vanNeerven, {\em Proc. of the Workshop on 
future Physics at HERA, DESY 1996} eds G. Ingelman, 
A. De Roeck \& R. Klanner, Vol I, p56.
\bibitem{BF95} R.D. Ball and S. Forte, \Journal{\PLB}{351}{313}{1995}.
\bibitem{kotpar} A.V. Kotikov and G. Parente, {\tt hep-ph/9609439},
{\em Proc. of DIS96, Rome}, 
Eds G. D'Agostini \& A. Nigro, World Scientific 1997, p237;,
{\tt hep-ph/9710252}. 
\bibitem{dekabasu}R. Deka and D.K.Choudury, \Journal{\ZPC}{75}{679}{1997};\\
T. Sarkar and R. Basu, {\tt hep-ph/9607232}.
\bibitem{muellercarg}
  A.H. Mueller, Columbia preprint CU-TP-658 (1994).
\bibitem{barlott} J. Bartels, H. Lotter,   \Journal{\PRL}{309}{400}{1993}.
\bibitem{KMSET} J. Kwiecinski et al, \Journal{\PRD}{50}{217}{1994};\\
K. Golec-Biernat et al, \Journal{\PLB}{335}{220}{1994}.
\bibitem{MueNav} A.H. Mueller, {\it Nucl. Phys.} {\bf B} (Proc. Suppl.)
18C (1990) 125; \\ {\it J. Phys.} G17 (1991) 1443;\\
A.H. Mueller and H. Navelet, \Journal{\NPB}{282}{727}{1987};\\
 V. del Duca and C. R, Schmidt, \Journal{\PRD}{49}{4510}{1994};\\
W.J.Stirling, \Journal{\NPB}{423}{56}{1994};\\
V. del Duca, {\tt hep-ph/9707348};\\
 J. Bartels et al, \Journal{\ZPC}{76}{75}{1997}.
\bibitem{Barjold}J. Bartels, A. De Roeck and M. Loewe, \Journal{\ZPC}{54}{635}
{1992};\\
J. Kwiecinski, A.D. Martin and P.J. Sutton, \Journal{\PLB}{287}{254}{1992},\\
\Journal{\PRD}{46}{921}{1992};\\
W.K. Tang, \Journal{\PLB}{278}{363}{1991}.
\bibitem{AKMGr} A.J. Askew et al, \Journal{\PLB}{338}{92}{1994};\\
 J. Kwiecinski et al., \Journal{\PRD}{54}{1874}{1996};\\
 J. Kwiecinski et al., {\tt hep-ph/9707375}.
\bibitem{Kim} V.T.Kim and G.B. Pivovarov, {\tt hep-ph/9709433}.
\bibitem{Barjnew} J. Bartels et al, \Journal{\PLB}{384}{300}{1996}.
\bibitem{LUNDstr} B. Andersson et al, \Journal{\em Phys.Rep.}{97}{31}{1983};\\
 T. Sj\"ostrand, {\it Comp. Phys. Comm.} 39 (1986) 347; \\
 T. Sj\"ostrand and M. Bengtsson, {\it Comp. Phys. Comm.} 43 (1987) 367; \\
 T. Sj\"ostrand, CERN-TH-6488-92 (1992).
\bibitem{HERWIG} G. Marchesini et al., {\it Comp. Phys. Comm.} 
67 (1992) 46;\\
HERWIG MC Version 5.9, G. Marchesini et al, {\tt hep-ph/9607393}.
\bibitem{schmidt}C.R. Schmidt, \Journal{\PRL}{78}{4531}{1997}.
\bibitem{stirbfkl} L.H. Orr and W.J. Stirling, {\tt hep-ph/9706529}.
\bibitem{LDC} H. Kharraziha and L. L\"onnblad, {\tt hep-ph/9709424};\\
B. Anersson et al, {\tt hep-ph/9711403}
\bibitem{ZHexp} H1, S. Aid et al., \Journal{\PLB}{356}{118}{1995}.
\bibitem{Ing} G. Ingelman, A. Edin, J. Rathsman, {\it Comp. Phys. Comm.} 101
(1997) 108;\\
A. Edin et al, \Journal{\JPG}{22}{943}{1996}.
\bibitem{Lonn} L. L\"onnblad, \Journal{\JPG}{22}{947}{1996}.
\bibitem{softcol} A. Edin, G. Ingelman and J. Rathsman,
\Journal{\ZPC}{75}{57}{1997},\\ \Journal{\PLB}{336}{371}{1996},
\bibitem{eth1} H1, C. Adloff et al., paper 
pa02-73, submitted to {\em ICHEP 96, Warsaw}, July 1996.
\bibitem{etzeus} N. A. Pavel, {\em Proc. of DIS96, Rome}, 
Ed.s G. D'Agostini, A. Nigro, (1997) 502.
\bibitem{kuhlen} M. Kuhlen \Journal{\PLB}{382}{441}{1996}.
\bibitem{h1tracks} H1,  C. Adloff et al.,
\Journal{\NPB}{485}{3}{1997}.
\bibitem{martin_pt}  J. Kwiecinski, S.C. Lang, A.D. Martin,
DTP-97-56, {\tt hep-ph/9707240}.
\bibitem{Muellerjet} A.H. Mueller, \Journal{\JPG}{17}{1443}{1991};\\
J. Bartels et al, \Journal{\ZPC}{54}{635}{1992};\\
W.K. Tung, \Journal{\PLB}{278}{363}{1992};\\
J. Kwiecinski et al, \Journal{\PRD}{46}{921}{1992}, \Journal{\PLB}{287}{254}
{1992}.

\bibitem{H1jet}H1, paper pa03-049, submitted to {\em ICHEP 96, Warsaw},
Poland, July 1996.
\bibitem{ZEUS95fjet}ZEUS, paper N-659, submitted to {\em EPS/HEP 
Conference 97, Jerusalem}, August 1997.
\bibitem{KLM} J. Kwiecinski et al, \Journal{\PRD}{54}{1874}{1996}, 
\Journal{\PRD}{55}{1273}{1997}.
\bibitem{K9702213}J. Kwiecinski et al, \Journal{\em Acta Physica Polonica}
{B27}{3455}{1996}.
\bibitem{deroeckfj} J. Bartels, A. De Roeck and M. W\"usthoff, 
{\em Proc. of the Workshop on Future Physics at HERA, DESY 1996}, 
Eds. G. Ingelman,.A. De Roeck, R. Klanner, p598.
\bibitem{Mirkes} E. Mirkes and D. Zeppenfeld , 
\Journal{Acta Physica Polonica}{B27}{1323}{1996}, 
\Journal{\PRL}{78}{428}{1997}, 
{\em Proc. of the Workshop on Future Physics at HERA, DESY 1996},
eds G. Ingelman, A. De Roeck \& R. Klanner, p588, {\tt hep-ph/9706437},
{\tt hep-ph/9711224}.
\bibitem{catsey} S. Catani and M.H. Seymour, \Journal{\PLB}{378}{287}{1996},
\Journal{\NPB}{485}{291}{1997}, \Journal{\PRD}{55}{6189}{1997}.
\bibitem{disaster} D. Graudenz, {\tt hep-ph/9710244}.
\bibitem{Jung} H. Jung, {\tt hep-ph/9709425}.
\bibitem{bartelbfkl} J. Bartels, A. De Roeck, H. Lotter,
\Journal{\PLB}{389}{742}{1996}.
\bibitem{brodsky} S.J. Brodsky, F. Hautmann, D.E. Soper,
\Journal{\PRL}{78}{803}{1997}.

\bibitem{cudell} J.R. Cudell et al, {\tt hep-ph/9601336}, {\tt hep-ph/9701312}.
\bibitem{DoLa2}A. Donnachie and P.V. Landshoff, \Journal{\ZPC}{61}{139}{1994}.
\bibitem{H1STOT}T. Aid  et al, \Journal{\ZPC}{69}{27}{1995}.
\bibitem{ZEUSSTOT}M. Derrick et al, \Journal{\ZPC}{63}{408}{1995}.
\bibitem{levy96} A. Levy, {\em Proc. of DIS96, Rome}
Eds. A. Negri and G. D'Agostini, and Tel-Aviv preprint TAUP-2349-96 (1996);\\
A. Levy, DESY preprint DESY 97-013 (1997). 
\bibitem{badelekR} B. Badelek, J. Kwiecinski, 
{\it Rev. Mod. Phys.} {\bf 68} (1996) 445. 
\bibitem{ALLM} H. Abramowicz, E.M. Levin, A. Levy and U. Maor,
\Journal{\PLB}{269}{465}{1991}.\\
We show the parametrizations as presented in:
A. Marcus PhD. thesis Tel-Aviv University TAUP~2350-96~(1996) (unpublished).
\bibitem{CKMT} A.\ Capella et al., \Journal{\PLB}{337}{358}{1994};\\
E.\ Gotsman, E.M.\ Levin and U.\ Maor, \Journal{\PLB}{379}{186}{1996};\\
J.R.Cudell, P.V.\ Landshoff and A.\ Donnachie, 
\Journal{\NPB}{482}{241}{1996}.
\bibitem{Levin}E.M.\ Levin, CBPF-NF-010/95, {\em Lectures to the Gleb 
Wataghiin School, Campinas, Brazil, 1994}, p158.
\bibitem{badelek} B. Badelek and J. Kwiecinski, \Journal{\PLB}{295}{263}{1992},
 \Journal{\ZPC}{43}{251}{1989}.
\bibitem{schildknecht}  D. Schildknecht and H. Spiesberger
Bielefeld Uni. preprint BI-TP-97-25 (1997).
\bibitem{sakurai} J.D. Sakurai and D. Schildknecht, 
\Journal{\PLB}{40}{121}{1972};\\
B. Gorczyca and D. Schildknecht, \Journal{\PLB}{47}{71}{1973}.
\bibitem{shaw} G. Kerley and G. Shaw, Manchester University preprint
MC-TH-97-12 (1997).
\bibitem{Troshin}S.M. Troshin and N.E.Tyurin, \Journal{\PRD}{55}{7305}{1997}.
\bibitem{Kot}A.V. Kotikov, {\tt hep-ph/9507320}.
\bibitem{LGM}E. Gotsman, E.M. Levin and U. Maor, {\tt hep-ph/9708275}.
\bibitem{shekelyan} V. Shekelyan, to be published in {\em Proc. of the 1997 
Symposium on Lepton and Photon Interactions, Hamburg}, 
July 1997, eds. A. De Roeck and A. Wagner.
\bibitem{martin} A. D. Martin, private communication and 
\Journal{\NPB}{392}{162}{1993}.
%
\bibitem{hiq2_ichep96} D. Acosta, U. Bassler, invited talks at 
{\em EPS HEP97 Conference}, Jerusalem, August 1997, 
{\tt http://www.cern.ch/hep97/}.
\bibitem{h1hiq2} H1, C. Adloff et al,
\Journal{\ZPC}{74}{191}{1997}.
\bibitem{zeushiq2} ZEUS, J. Breitweg et al,
\Journal{\ZPC}{74}{207}{1997}.
\bibitem{PDG} Particle Data Group, R.M. Barnett et al, 
\Journal{\PRD}{54}{1}{1996}.
\bibitem{fiterr} C. Pascaud \& F. Zomer, preprint LAL 95-05 (1995);\\
M. Botje, M. Klein \& C. Pascaud, {\em Proc. of the Workshop on Future 
Physics at HERA, DESY 1996},
eds G. Ingelman, A. De Roeck \& R. Klanner, Vol.1 p33 (1996);\\
M. Botje, {\em Proc. of DIS97, Chicago}, {\tt hep-ph/9707289}.
\bibitem{gwolfisr} G. Wolf, DESY preprint DESY 97-047, {\tt hep-ex/9704006}.
\bibitem{bbisr} U. Bassler \& G. Bernardi, DESY preprint DESY 97-136.
%
\bibitem{kuhlmann} S. Kuhlmann, H. L. Lai \& W. K. Tung, 
\Journal{\PLB}{409}{271}{1997}.
\bibitem{rock} S. Rock \& P. Bosted, {\tt hep-ph/9706436}.
\bibitem{bgduality} E.D. Bloom \& F.J. Gilman, 
\Journal{\PRL}{25}{1140}{1970}.
\bibitem{gunion} J.F. Gunion \& R. Vogt, {\tt hep-ph/9706252}. 
\bibitem{mtcharm} W. Melnitchouk \& A.W. Thomas, {\tt hep-ph/9707387}.
\bibitem{szczurek} A. Szczurek and A. Budzabowski,
\Journal{\PLB}{408}{275}{1997}.
\bibitem{kochelev} N.I. Kochelev, {\tt hep-ph/9710540}.
\bibitem{cern_thlq} G. Altarelli et al, CERN preprint CERN-TH/97-40,
{\tt hep-ph/9703276}.
\bibitem{straub97} B. Straub, to be published in {\em Proc.
of the 1997 Symposium on Lepton Photon Interactions, 
Hamburg}, July 1997, eds. A. De Roeck and A. Wagner,
{\tt http://www.desy.de/lp97/}.
\bibitem{elsen97} E. Elsen, to be published in {\em Proc. of
EPS/HEP Conference 97, Jerusalem}, August 1997, 
{\tt http://www.cern.ch/hep97/}.
\bibitem{heraws96} {\em Proc. of the Workshop on Future Physics at HERA,
DESY 1996}, eds G. Ingelman, A. De Roeck \& R. Klanner, 2 Vols. DESY 1996. 

\end{thebibliography}
\end{document}